\documentclass[11pt]{article}

\usepackage{amsmath,amssymb,amstext,amsthm}
\usepackage{amsfonts}
\usepackage{graphicx,epstopdf,epsfig,multirow,epic,bm}
\usepackage{color}
\usepackage{multicol}
\usepackage{algorithm}
\usepackage{algorithmic}
\usepackage{CJK}
\usepackage{paralist}

\theoremstyle{definition}
\newtheorem{thm}{Theorem}
\newtheorem{cor}[thm]{Corollary}
\newtheorem{lem}[thm]{Lemma}
\newtheorem{prop}[thm]{Proposition}
\theoremstyle{definition}
\newtheorem{defn}{Definition}
\theoremstyle{definition}
\newtheorem{rem}{Remark}

\theoremstyle{definition}
\newtheorem{problem}{Problem}

\theoremstyle{definition}
\newtheorem{conj}{Conjecture}
\theoremstyle{definition}

\theoremstyle{definition}

\DeclareGraphicsExtensions{.eps,.eps.gz}
\normalsize \rm
\usepackage{titlesec}

\newcommand{\qqed}{\hfill $\square$}
\newcommand{\paralled}{\hfill $\parallel$}

\usepackage{indentfirst}
\begin{document}

\thispagestyle{empty}

\begin{CJK}{GBK}{song}

\begin{center}
{\huge \textbf{Recent Colorings And Labelings\\[12pt] In Topological Coding}}\\[14pt]
{\large Bing \textsc{Yao}$^{1}$, \quad Hongyu \textsc{Wang}$^{2,3}$}\\[8pt]
(\today)\\[16pt]
\end{center}
{\small 1. College of Mathematics and Statistics, Northwest Normal University, Lanzhou, 730070 CHINA\\
2. National Computer Network Emergency Response Technical Team/Coordination Center of China, Beijing, 100029, CHINA\\
3. School of Electronics Engineering and Computer Science, Peking University, Beijing, 100871, CHINA}

\vskip 1cm

\pagenumbering{roman}
\tableofcontents

\newpage

\setcounter{page}{1}
\pagenumbering{arabic}

\thispagestyle{empty}

\begin{center}
{\huge \textbf{Recent Colorings And Labelings\\[12pt] In Topological Coding}}\\[14pt]
{\large Bing \textsc{Yao}$^{1,\dagger}$, \quad Hongyu \textsc{Wang}$^{2,3,\ddagger}$}\\[8pt]
(\today)\\[14pt]
{\small 1. College of Mathematics and Statistics, Northwest Normal University, Lanzhou, 730070 CHINA\\
2. National Computer Network Emergency Response Technical Team/Coordination Center of China, Beijing, 100029, CHINA\\
3. School of Electronics Engineering and Computer Science, Peking University, Beijing, 100871, CHINA\\
$^{\dagger}$ yybb918@163.com;\quad $^{\ddagger}$ why200904@163.com}
\end{center}

\vskip 1cm

\begin{quote}
\textbf{Abstract:} Topological Coding consists of two different kinds of mathematics: topological structure and mathematical relation. The colorings and labelings of graph theory are main techniques in topological coding applied  in asymmetric encryption system. Topsnut-gpws (also, colored graphs) have the following advantages: (1) Run fast in communication networks because they are saved in computer by popular matrices rather than pictures. (2) Produce easily text-based (number-based) strings for encrypt files. (3) Diversity of asymmetric ciphers, one public-key corresponds to more private-keys, or more public-keys correspond more private-keys. (4) Irreversibility, Topsnut-gpws can generate quickly text-based (number-based) strings with bytes as long as desired, but these strings can not reconstruct the original Topsnut-gpws. (5) Computational security, since there are many non-polynomial (NP-complete, NP-hard) algorithms in creating Topsnut-gpws. (6) Provable security, since there are many mathematical conjectures (open problems) in graph labelings and graph colorings. We are committed to create more kinds of new Topsnut-gpws to approximate practical applications and anti-quantum computation, and try to use algebraic method and Topsnut-gpws to establish graphic group, graphic lattice, graph homomorphism \emph{etc}. 
\end{quote}

\newpage


\section{Research background and preliminary}

\subsection{Research background}

The authors in \cite{Bernstein-Buchmann-Dahmen-Quantum-2009} point: ``There are many important classes of cryptographic systems beyond RSA and DSA and ECDSA, and they are believed to resist classical computers and quantum computers, such as Hash-based cryptography, Code-based cryptography, Lattice-based cryptography, Multivariate-quadratic-equations cryptography, Secret-key cryptography''. Notice that the lattice difficulty problem is not only a classical \emph{number theory}, but also an important research topic of computational complexity theory. Researchers have found that \emph{lattice theory} has a wide range of applications in cryptanalysis and design. Many difficult problems in lattice have been proved to be NP-hard. So, this kind of cryptosystems is generally considered to have the characteristics of quantum attack resistance \cite{Wang-Xiao-Yun-Liu-2014}.

We started learning \emph{graph theory} from a famous book ``Graph Theory With Application'' written by Prof. Bondy and Prof. Murty in 1976, And they presented us a good book ``Graph Theory'' with more algorithms for modern graph theory in 2008 \cite{Bondy-2008}. Many of our research works of graph labelings were based on a famous survey article contributed by Gallian in \cite{Gallian2020}, see ``\emph{A survey: recent results, conjectures and open problems on labeling graphs}'' in \cite{Gallian-1989}. Prof. Gallian introduces over 200 graph labelings up to 2019 after more than ten years of unremitting collection, his survey has collected over 2830 articles all over the world. It must be admitted that Prof. Gallian's comprehensive article has greatly helped us to learn new graph labelings quickly, and promotion for the study and research of graph labelings, and powerful tools to carry out effective research works, and obtain our own encouraging achievements.

\emph{Topological Coding} is a combinatoric subbranch of \emph{graph theory} and \emph{cryptography}. Topological coding is made up of two different kinds of mathematics: topological structure and mathematical relation. The colorings and labelings of graph theory are main techniques in topological coding. The domain of topological coding involves millions of things, and a graph of topological coding connects things together to form a complete ``story'' under certain constraints in asymmetric encryption system \cite{Bing-Yao-2020arXiv}. It is known that there is no polynomial quantum algorithm to solve some lattice problems, so is the graph isomorphism problem \cite{Bernstein-Buchmann-Dahmen-Quantum-2009}.

Encrypting a network wholly to resist full-scale attacks and sabotage by attackers is needed extensively today. On October 23, 2019, according to an article published in Nature, the 53 bit quantum computer ``Sycamore'' developed by Google is to complete a specific problem computing in 200 seconds, the problem will take 10,000 years for the world's fastest supercomputer. So, network security will be challenged tremendously in the near future.

By aiming to researching on topics of topological coding we have proposed new graph labelings and have designed new colorings by combining traditional colorings and traditional labelings together. These new colorings and labelings help greatly us to design text-based strings, number-based strings and graphic passwords (Topsnut-gpws) made by topological structures and number theory, graphic groups, Topcode-matrices, graphic lattices, graph homomorphism lattices, and so on. Topcode-matrices can be used to describe topological graphic passwords used in information security and graph connected properties for solving some problems coming in the investigation of Graph Networks and Graph Neural Networks proposed by GoogleBrain and DeepMind \cite{Battaglia-27-authors-arXiv1806-01261v2}.

Since the colorings and labelings mentioned here are selected from many of articles, for the convenience of research, we have retained their names and notations as that in the original articles, and some of individual colorings, or individual labelings may appear in the different definitions of this article.

\subsection{Terminology and graph operations}

Standard terminology and notation of graphs and digraphs used here are cited from \cite{Bang-Jensen-Gutin-digraphs-2007}, \cite{Bondy-2008} and \cite{Gallian2020}. All graphs are \emph{simple} and no \emph{loops}, unless otherwise stated. The following notation and terminology will be used in the whole article:
\begin{asparaenum}[$\bullet$ ]
\item All non-negative integers are collected in the set $Z^0$, and all integers are in the set $Z$, so $Z^+=Z^0\setminus \{0\}$.
\item A $(p,q)$-graph $G$ is a graph having $p$ vertices and $q$ edges, and $G$ has no multiple edge and directed-edge; and $G^c$ is the complementary graph of the $(p,q)$-graph $G$.
\item The \emph{cardinality} of a set $X$ is denoted as $|X|$, so the \emph{degree} of a vertex $x$ in a $(p,q)$-graph $G$ is denoted as $\textrm{deg}_G(x)=|N(x)|$, where $N(x)$ is the set of neighbors of the vertex $x$.
\item A vertex $x$ is called a \emph{leaf} if $\textrm{deg}_G(x)=1$.
\item Adding the new edges of an edge set $E^*$ to a graph $G$ produces a new graph, denoted as $G+E^*$, where $E^*\cap E(G)=\emptyset$.
\item A \emph{caterpillar} is a tree $T$, such that the deletion of all leaves of $T$ produces a path. A \emph{lobster} is a tree $H$, such that the deletion of all leaves of $H$ produces a caterpillar.
\item A symbol $[a,b]$ stands for an integer set $\{a,a+1,a+2,\dots, b\}$ with two integers $a,b$ subject to $a<b$, and $[a,b]^o$ denotes an \emph{odd-set} $\{a,a+2,\dots, b\}$ with odd integers $a,b$ holding $1\leq a<b$ true, and $[\alpha,\beta]^e$ is an \emph{even-set} $\{\alpha,\alpha+2,\dots, \beta\}$ with even integers $\alpha,\beta$ and $\alpha<\beta$.
\item The symbol $[a,b]^r$ is an \emph{interval} of real numbers.
\item A \emph{text-based password} is abbreviated as \emph{TB-paw} and is made by 52 English characters and numbers of $[0,9]$.
\item A \emph{text-based string} $D=t_1t_2\cdots t_m$ has its own \emph{reciprocal text string} defined by $D^{-1}=t_mt_{m-1}\cdots t_2t_1$, also, we say that $D$ and $D^{-1}$ match with each other, where $m$ is the \emph{number} of the text-based string $D$.
\item A \emph{number-based string} $S(n)=c_1c_2\cdots c_n$ with $c_j\in [0,9]$, where $n$ is the \emph{number} (length) of the number-based string $S(n)$.
\item Let $S$ be a set. The set of all subsets is denoted as $S^2=\{X:~X\subseteq S\}$, called the \emph{power set} of $S$, and $S^2$ contains no empty set at all. For example, for a set $S=\{a,b,c,d,e\}$, so $S^2$ has its own elements $\{a\}$, $\{b\}$, $\{c\}$, $\{d\}$, $\{e\}$, $\{a,b\}$, $\{a,c\}$, $\{a,d\}$, $\{a,e\}$, $\{b,c\}$, $\{b,d\}$, $\{b,e\}$, $\{c, d\}$, $\{c, e\}$, $\{d,e\}$, $\{a,b,c\}$, $\{a,b,d\}$, $\{a,b,e\}$, $\{a,c,d\}$, $\{a,c,e\}$, $\{a,d,e\}$, $\{b,c,d\}$, $\{b,c,e\}$, $\{a,b,c,d\}$, $\{a,b,c,e\}$, $\{a,c,d,e\}$, $\{b,c,d,e\}$ and $\{a,b,c,d,e\}$.
\item $[a,b]^2$ is the set of all subsets of an integer set $[a,b]$ for $a<b$ and $a,b\in Z^0$.
\item Let $S(k_i,d_i)^q_1=\{k_i, k_i + d_i, \dots , k_i +(q-1)d_i\}$ be an \emph{integer set} for integers $k_i\geq 1$ and $d_i\geq 1$, and let $\{(k_i,d_i)\}^m_1=\{(k_1,d_1), (k_2,d_2),\dots ,(k_m,d_m)\}$ be a \emph{matching-pair sequence}. In some articles, $S_{q-1,k,d}=\{k,k+d,k+2d,\dots, k+(q-1)d\}$ for integers $q\geq 1$, $k\geq 0$ and $d\geq 1$.
\item \textbf{Topsnut-gpws}. Topsnut-gpws is the abbreviation of ``graphical passwords based on topological structure and number theory'' in Topological Coding, a mixture subbranch of graph theory and cryptography (Ref. \cite{Yao-1909-01587-2019}, \cite{Wang-Xu-Yao-2016}, \cite{Wang-Xu-Yao-Key-models-Lock-models-2016} and \cite{Yao-Sun-Zhang-Li-Zhang-Xie-Yao-2017-Tianjin-University}).
\item A \emph{sequence} $\textbf{\textrm{d}}=(m_1,m_2,\dots,m_n)=(m_k)^n_{k=1}$ consists of positive integers $m_1, m_2, \dots , m_n$. If a graph $G$ has its \emph{degree-sequence} ${\textrm{deg}}(G)=\textbf{\textrm{d}}$, then $\textbf{\textrm{d}}$ is \emph{graphical}, we call $\textbf{\textrm{d}}$ \emph{degree-sequence}, each $m_i$ a \emph{degree component}, and $n=L_{ength}(\textbf{\textrm{d}})$ the \emph{length} of $\textbf{\textrm{d}}$.
\item A \emph{caterpillar} $T$ is a tree, after removing all of leaves of the caterpillar $T$, the remainder is just a \emph{path}; a \emph{lobster} is tree, and the deletion of all leaves of the lobster produces a caterpillar.
\item $\textrm{gcd}(a,b)$ is the \emph{maximal common factor} between two positive integers $a$ and $b$.
\end{asparaenum}

\begin{lem} \label{thm:basic-sequence-lemma}
\cite{Bondy-2008} (Erd\"{o}s-Galia Theorem) A sequence $\textbf{\textrm{d}}=(m_k)^n_{k=1}$ with $m_{i}\geq m_{i+1}\geq 0$ to be the \emph{degree-sequence} of a $(p,q)$-graph graph $G$ if and only if the sum $2q=\sum^n_{i=1}m_i$ and
\begin{equation}\label{eqa:555555}
\sum^k_{i=1}m_i\leq k(k-1)+\sum ^n_{j=k+1}\min\{k,m_j\}.
\end{equation}
\end{lem}

\begin{figure}[h]
\centering
\includegraphics[width=15.6cm]{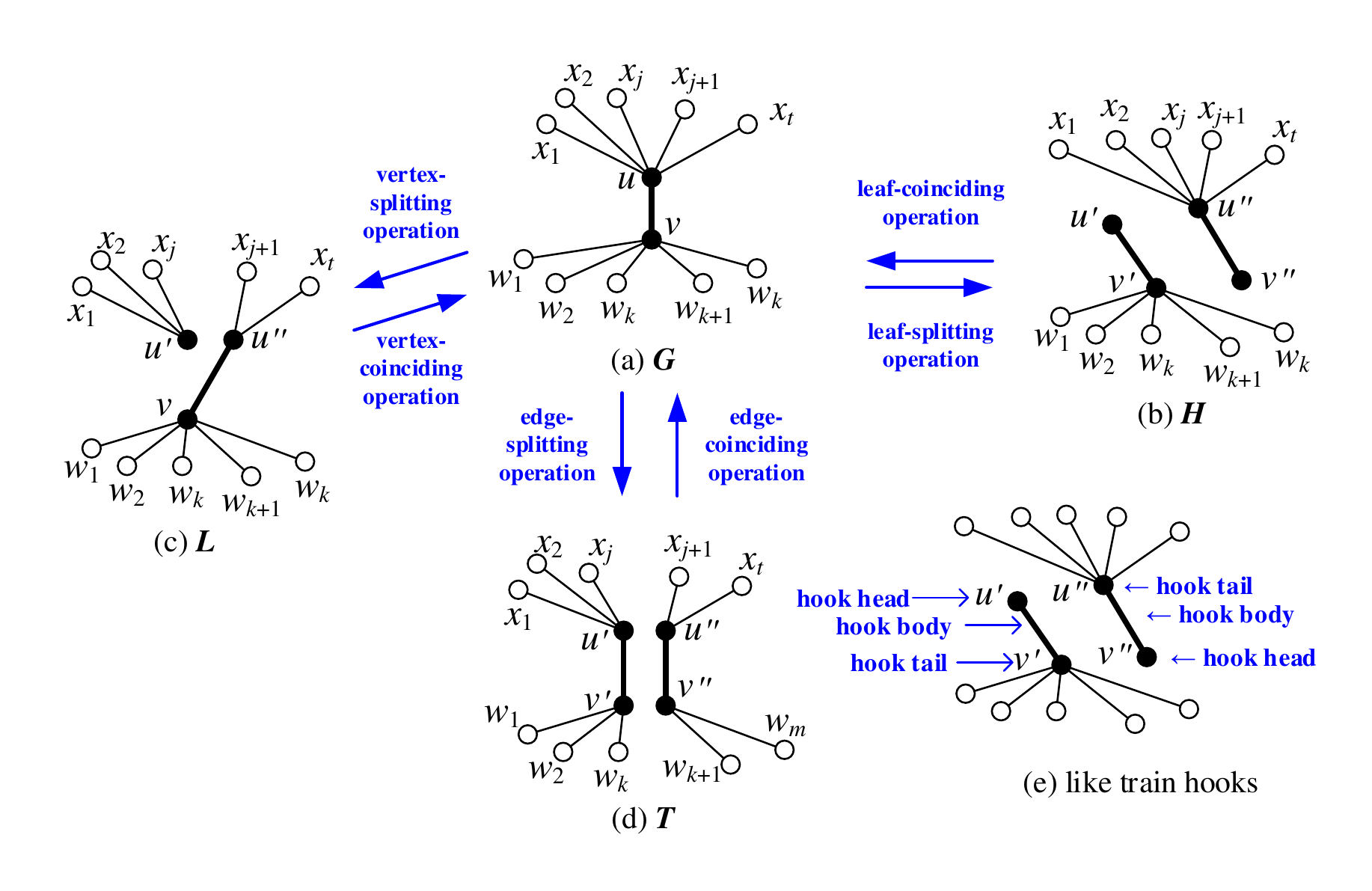}\\
\caption{\label{fig:6-operations}{\small Six graph operations.}}
\end{figure}

\begin{defn}\label{defn:six-splitting-coinciding-operations}
Let $G$ be a $(p,q)$-graph. There are the following basic graph operations:
\begin{asparaenum}[\textrm{Opera}-1. ]
\item \cite{Yao-Sun-Zhang-Mu-Sun-Wang-Su-Zhang-Yang-Yang-2018arXiv} A \emph{vertex-splitting operation} is defined in the way: Vertex-split a vertex $u$ of $G$ into two vertices $u\,',u\,''$ such that $N(u\,')=\{x_1,x_2,\dots ,x_j\}$ and $N(u\,'')=\{v,x_{j+1},x_{j+2},\dots ,x_t\}$ with $N(u\,')\cap N(u\,'')=\emptyset$ and $N(u)=N(u\,')\cup N(u\,'')$; the resultant graph is written as $G\wedge x$, named as a \emph{vertex-split graph}, see Fig.\ref{fig:6-operations} from (a) to (c), and moreover $|V(G\wedge x)|=|V(G)|+1$, $|E(G\wedge x)|=|E(G)|$.
\item \cite{Yao-Sun-Zhang-Mu-Sun-Wang-Su-Zhang-Yang-Yang-2018arXiv} A \emph{(non-common neighbor) vertex-coinciding operation} is defined by coinciding two vertices $u\,'$ and $u\,''$ in to one $x=u\,'\odot u\,''$ if $N(u\,')\cap N(u\,'')=\emptyset$ such that $N(x)=N(u\,')\cup N(u\,'')$; the resultant graph is written as $G(u\,'\odot u\,'')$, called \emph{vertex-coincided graph}, see Fig.\ref{fig:6-operations} from (c) to (a), and moreover $|V(G(u\,'\odot u\,''))|=|V(G)|-1$, $|E(G(u\,'\odot u\,''))|=|E(G)|$.
\item \cite{Bing-Yao-2020arXiv} A \emph{leaf-splitting operation} is defined as: Let $uv$ be an edge of the graph $G$ with complex degrees $\textrm{deg}_G(u)\geq 1$ and $\textrm{deg}_G(v)\geq 1$. A \emph{leaf-splitting operation} is defined as: Remove the edge $uv$ from $G$, and add a new leaf $v\,''$, next join it with the vertex $u~(=u\,'')$ of the graph $G-uv$ by a new edge $uv\,'$, and then add another new leaf $u\,'$ to join it with the vertex $v~(=v\,'')$ of $G-uv$ by another new edge $vu\,'$, the resultant graph is written as $G\overline{\wedge} uv$, see Fig.\ref{fig:6-operations} from (a) to (b) for understanding the leaf-splitting operation. And $|V(G\overline{\wedge} uv)|=|V(G)|+2$, $|E(G\overline{\wedge} uv)|=|E(G)|+1$.
\item \cite{Bing-Yao-2020arXiv} A \emph{leaf-coinciding operation} is defined as: For a graph graph $H$ having two leaf-edges $uv\,'$ and $u\,'v$ with degrees $\textrm{deg}_H(u)\geq 2$ and $\textrm{deg}_H(v\,')=1$, as well as $\textrm{deg}_H(u\,')=1$ and $\textrm{deg}_H(v)\geq 2$, if $[N(u)\setminus \{v\,'\}]\cap [N(v)\setminus \{u\,'\}]$, we coincide two edges $uv\,'$ and $u\,'v$ into one edge $uv=uv\,'\overline{\ominus} u\,'v$ with $u=u\odot u\,'$ and $v=v\odot v\,'$. The new graph is written as $H(uv\,'\overline{\ominus} vu\,')$, and the process of obtaining $H(uv\,'\overline{\ominus} vu\,')$ is called a \emph{leaf-coinciding operation}, see Fig.\ref{fig:6-operations} from (b) to (a), and moreover $|V(H(uv\,'\overline{\ominus} vu\,'))|=|V(H)|-2$, $|E(H(uv\,'\overline{\ominus} vu\,'))|=|E(H)|-1$. This operation is very similar with the connection of two train hooks, see Fig.\ref{fig:6-operations} (e).
\item \cite{Yao-Sun-Zhang-Mu-Sun-Wang-Su-Zhang-Yang-Yang-2018arXiv} An \emph{edge-splitting operation} is defined as: Split the edge $uv$ of $G$ into two edges $u\,'v\,'$ and $u\,''v\,''$ such that neighbor sets $N(u\,')=\{v\,',x_1,x_2,\dots ,x_j\}$ and $N(u\,'')=\{v\,'',x_{j+1},x_{j+2},\dots ,x_t\}$ with $N(u)=N(u\,')\cup N(u\,'')$, $N(v\,')=\{u\,',w_1,w_2,\dots ,w_k\}$ and $N(v\,'')=\{u\,'',w_{k+1},w_{k+2},\dots ,w_{m}\}$ with $N(v)=N(v\,')\cup N(v\,'')$; the resultant graph is written as $G\wedge uv$, named as an \emph{edge-split graph}, with $N(v\,')\cap N(u\,')=\emptyset$, $N(v\,')\cap N(u\,'')=\emptyset$, and $N(u\,'')\cap N(v\,')=\emptyset$, $N(u\,')\cap N(v\,'')=\emptyset$, see Fig.\ref{fig:6-operations} from (a)$G$ to (d)$T$, and moreover $|V(G\wedge uv)|=2+|V(G)|$, $|E(G\wedge uv)|=|E(G)|+1$. Especially, if two neighbor sets $N(u\,'')=\{v\,''\}$ and $N(v\,'')=\{u\,'',w_1,w_2,\dots ,w_{m}\}$, as well as two neighbor sets $N(v\,')=\{u\,'\}$ and $N(u\,')=\{v\,',x_1,x_2,\dots ,x_t\}$, we say $G\wedge uv$ is the result of doing a leaf-splitting operation, here $\textrm{deg}(u\,'')=1$ and $\textrm{deg}(v\,')=1$.
\item \cite{Yao-Sun-Zhang-Mu-Sun-Wang-Su-Zhang-Yang-Yang-2018arXiv} An \emph{edge-coinciding operation} is defined by coinciding two edges $u\,'v\,'$ and $u\,''v\,''$ of $G$ into one edge $uv=u\,'v\,'\overline{\ominus} u\,''v\,''$, when $N(u\,')\cap N(u\,'')=\emptyset$ and $N(v\,')\cap N(v\,'')=\emptyset$, such that $N(u)=N(u\,')\cup N(u\,'')\cup \{v=(v\,'\odot v\,'')\}$ and $N(v)=N(v\,')\cup N(v\,'')\cup \{u=(u\,'\odot u\,'')\}$; the resultant graph is written as $G(u\,'v\,'\overline{\ominus} u\,''v\,'')$, called an \emph{edge-coincided graph}, see Fig.\ref{fig:6-operations} from (d) to (a), and moreover $|V(G(u\,'v\,'\overline{\ominus} u\,''v\,'')|=|V(G)|-2$, $|E(G(u\,'v\,'\overline{\ominus} u\,''v\,'')|=|E(G)|-1$.
\item If two vertex-disjoint colored graphs $G$ and $H$ have $k$ pairs of vertices with each pair of vertices colored with the same colors, then we, by the vertex-coinciding operation, vertex-coincide each pair of vertices from $G$ and $H$ into one, the resultant graph is denoted as $\odot_k \langle G,H\rangle$ (called \emph{vertex-coincided graph}), and moreover $|V(\odot_k \langle G,H\rangle)|=|V(G)|+|V(H)|-k$, $|E(\odot_k \langle G,H\rangle)|=|E(G)|+|E(H)|$. Clearly, $\odot_k \langle G,H\rangle$ holds for vertex-disjoint uncolored graphs $G$ and $H$ too.\qqed
\end{asparaenum}
\end{defn}

\begin{thm}\label{thm:vertex-splitting-leaf-splitting-connectivity}
\cite{Wang-Zhang-Mei-Yao2018-Split} For the vertex-splitting connectivity $\kappa_{split}(G)$ and the leaf-splitting connectivity $\kappa^{leaf}_{split}(G)$ of a connected graph $G$, we have
\begin{equation}\label{eqa:555555}
\kappa_{split}(G)=\kappa(G)\quad \textrm{and}\quad \kappa^{leaf}_{split}(G)=\kappa\,'(G),
\end{equation}
where $\kappa(G)$ is the connectivity and $\kappa\,'(G)$ is the edge connectivity of the connected graph $G$ \cite{D-B-West1996, Bondy-2008}.
\end{thm}

\begin{thm}\label{thm:2-vertex-split-graphs-isomorphic}
$^*$ Suppose that two connected graphs $G$ and $H$ admit a mapping $f:V(G)\rightarrow V(H)$. In general, a vertex-split graph $G\wedge u$ with $\textrm{deg}_G(u)\geq 2$ is not unique, so we have a vertex-split graph set $S_G(u)=\{G\wedge u\}$, similarly, we have another vertex-split graph set $S_H(f(u))=\{H\wedge f(u)\}$. If each vertex-split graph $L\in S_G(u)$ corresponds another vertex-split graph $T\in S_H(f(u))$ such that $L\cong T$, and vice versa, we write this fact as
\begin{equation}\label{eqa:2-vertex-split-graphs-isomorphic}
G\wedge u\cong H\wedge f(u),
\end{equation} thereby, $G$ is \emph{isomorphic} to $H$, namely, $G\cong H$.\paralled
\end{thm}

\begin{defn} \label{defn:11-lattice-integer-lattice}
\cite{Bernstein-Buchmann-Dahmen-Quantum-2009} A \emph{lattice} is a set
\begin{equation}\label{eqa:555555}
\textrm{\textbf{L}}(\textbf{B}) =\left \{\sum^m_{i=1}x_i\textbf{b}_i :~x_i \in Z, 1 \leq i \leq n \right \}
\end{equation} based on a \emph{lattice base} $\textbf{B}=(\textbf{b}_1, \textbf{b}_2,\dots , \textbf{b}_n)$ consisted of linear independent vectors $\textbf{b}_1, \textbf{b}_2,\dots , \textbf{b}_n$ in $R^m$ with $n\leq m$. If each component of each vector $\textbf{b}_i$ in the base $\textbf{B}$ is an integer, we call $\textrm{\textbf{L}}(\textbf{B})$ an \emph{integer lattice}, denoted as $\textrm{\textbf{L}}(\textbf{ZB})$ for distinguishing.\qqed
\end{defn}

\subsection{Basic colorings and labelings, Conjectures}

All \emph{colorings} and \emph{labelings} mentioned in this article are on graphs of graph theory, very often, people say \emph{graph colorings} and \emph{graph labelings} \cite{Gallian2020}.

\subsubsection{Basic labelings}

\begin{defn}\label{defn:totally-normal-labeling}
\cite{su-yan-yao-2018} A $(p,q)$-graph $G$ admits a bijection $f:V(G)\rightarrow [m,n]$ or $f: V(G)\cup E(G)\rightarrow [m,n]$, we denoted the set of colored vertices of $G$ as $f(V(G))=\{f(x):x\in V(G)\}$, the set of colored edges of graph $G$ as $f(E(G))=\{f(uv):uv\in E(G)\}$. If $|f(V(G))|=p$, then $f$ is called a \emph{vertex normal labeling} of $G$; when as $|f(E(G))|=q$, $f$ is called an \emph{edge normal labeling} of $G$; and as $|f(V(G)\cup E(G))|=p+q$, then $f$ is called a \emph{totally normal labeling}.\qqed
\end{defn}

\begin{defn}\label{defn:universal-label-set}
\cite{Bing-Yao-Cheng-Yao-Zhao2009} A graph $G$ admits a coloring $h:S\subseteq V(G)\cup E(G)\rightarrow [a,b]$, and write the color set $h(S)=\{h(x): x\in S\}$. The \emph{dual coloring} $h\,'$ of the coloring $h$ is defined as: $h\,'(z)=\max h(S)+\min h(S)-h(z)$ for $z\in S$. Moreover, $h(S)$ is called the \emph{vertex color set} if $S=V(G)$, $h(S)$ the \emph{edge color set} if $S=E(G)$, and $h(S)$ a \emph{universal color set} if $S=V(G)\cup E(G)$. Furthermore, if $G$ is a bipartite graph with its vertex set bipartition $(X,Y)$, and holds $\max h(X)<\min h(Y)$ true, we call $h$ a \emph{set-ordered coloring} (resp. \emph{set-ordered labeling}) of $G$.\qqed
\end{defn}

We restate basic $W$-type labelings as follows:

\begin{defn} \label{defn:basic-W-type-labelings}
\cite{Gallian2020, Yao-Sun-Zhang-Mu-Sun-Wang-Su-Zhang-Yang-Yang-2018arXiv, Bing-Yao-Cheng-Yao-Zhao2009, Zhou-Yao-Chen-Tao2012} Suppose that a connected $(p,q)$-graph $G$ admits a mapping $\theta:V(G)\rightarrow \{0,1,2,\dots \}$. For each edge $xy\in E(G)$, the induced edge color is defined as $\theta(xy)=|\theta(x)-\theta(y)|$. Write vertex color set by $\theta(V(G))=\{\theta(u):u\in V(G)\}$, and edge color set by
$\theta(E(G))=\{\theta(xy):xy\in E(G)\}$. There are the following constraint conditions:

B-1. $|\theta(V(G))|=p$;

B-2. $\theta(V(G))\subseteq [0,q]$, $\min \theta(V(G))=0$;

B-3. $\theta(V(G))\subset [0,2q-1]$, $\min \theta(V(G))=0$;

B-4. $\theta(E(G))=\{\theta(xy):xy\in E(G)\}=[1,q]$;

B-5. $\theta(E(G))=\{\theta(xy):xy\in E(G)\}=[1,2q-1]^o$;

B-6. $G$ is a bipartite graph with the bipartition $(X,Y)$ such that $\max\{\theta(x):x\in X\}< \min\{\theta(y):y\in Y\}$ ($\max \theta(X)<\min \theta(Y)$ for short);

B-7. $G$ is a tree having a perfect matching $M$ such that $\theta(x)+\theta(y)=q$ for each matching edge $xy\in M$; and

B-8. $G$ is a tree having a perfect matching $M$ such that $\theta(x)+\theta(y)=2q-1$ for each matching edge $xy\in M$.

\noindent \textbf{Then}:
\begin{asparaenum}[\textrm{Blab}-1. ]
\item A \emph{graceful labeling} $\theta$ satisfies B-1, B-2 and B-4.
\item A \emph{set-ordered graceful labeling} $\theta$ holds B-1, B-2, B-4 and B-6 true.
\item A \emph{strongly graceful labeling} $\theta$ holds B-1, B-2, B-4 and
B-7 true.
\item A \emph{set-ordered strongly graceful labeling} $\theta$ holds B-1, B-2, B-4, B-6 and B-7 true.
\item An \emph{odd-graceful labeling} $\theta$ holds B-1, B-3 and B-5 true.
\item A \emph{set-ordered odd-graceful labeling} $\theta$ holds B-1, B-3, B-5
and B-6 true.
\item A \emph{strongly odd-graceful labeling} $\theta$ holds B-1, B-3,
B-5 and B-8 true.
\item A \emph{set-ordered strongly odd-graceful labeling} $\theta$ holds B-1, B-3, B-5, B-6 and B-8 true.\qqed
\end{asparaenum}
\end{defn}

\begin{figure}[h]
\centering
\includegraphics[width=16.4cm]{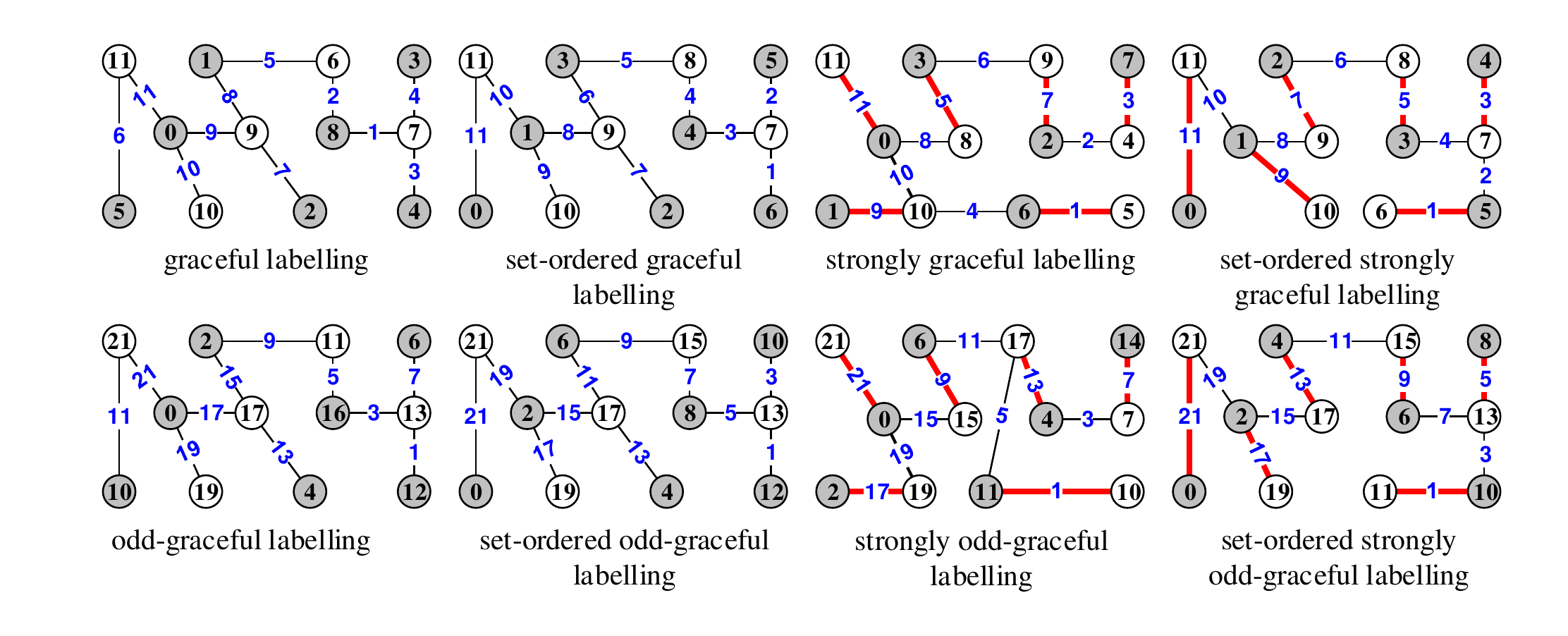}
\caption{\label{fig:graceful-odd-graceful}{\small Examples for understanding Definition \ref{defn:basic-W-type-labelings}, the red edges are the matching edges.}}
\end{figure}

\begin{defn} \label{defn:11-old-labelings-Gallian}
Let $G$ be a $(p,q)$-graph.

(1) \cite{Gallian2020} An \emph{edge-magic total labeling} $f$ of $G$ holds $f(V(G)\cup E(G))=[1,p+q]$ such that for any edge $uv\in E(G)$, $f(u)+f(uv)+f(v)=c$, where the magic constant $c$ is a fixed positive integer; and furthermore $f$ is \emph{a super edge-magic total labeling} if $f(V(G))=[1,p]$.

(2) \cite{Gallian2020} A \emph{felicitous labeling} $f$ of $G$ holds $f(V(G))\subset [0,q]$, $f(u)\neq f(v)$ for distinct $u,v\in V(G)$, and all edge colors $f(uv)=f(u)+f(v)~(\bmod~q)$ for $uv\in E(G)$ are distinct from each other; and furthermore $f$ is \emph{q super felicitous labeling} if $f(V(G))=[1,p]$.

(3) \cite{Zhou-Yao-Chen2013} An \emph{odd-elegant labeling} $f$ of $G$ holds $f(V(G))\subset [0,2q-1]$, $f(u)\neq f(v)$ for distinct $u,v\in V(G)$, and $f(E(G))=\{f(uv)=f(u)+f(v)~(\bmod~2q):uv\in E(G)\}=[1,2q-1]^o$.

(4) \cite{Gallian2020} A \emph{$k$-graceful labeling} $f$ of $G$ holds $f(V(G))\subset [0,q+k-1]$, $|f(V(G))|=p$ and $f(E(G))=\{f(uv)=|f(u)-f(v)|:uv\in E(G)\}=[k,q+k-1]$ true.\qqed
\end{defn}

\begin{defn} \label{defn:3-parameter-labelings}
\cite{Gallian2020} Let $G_i$ be a $(p,q)$-graph for $i\in [1,m]$, and integers $k_i\geq 1$ and $d_i\geq 1$.

(1) A \emph{$(k_i,d_i)$-graceful labeling} $f$ of $G_i$ hold $f(V(G_i))\subseteq [0, k_i + (q-1)d_i]$, $f(x)\neq f(y)$ for distinct $x,y\in V(G_i)$ and $\pi(E(G_i))=\{|\pi(u)-\pi(v)|:\ uv\in
E(G_i)\}=S(k_i,d_i)^q_1$.

(2) A labeling $f$ of $G_i$ is said to be $(k_i,d_i)$-\emph{arithmetic} if $f(V(G_i))\subseteq [0, k_i+(q-1)d_i]$, $f(x)\neq f(y)$ for distinct vertices $x,y\in V(G_i)$ and $\{f(u)+f(v): uv\in E(G_i)\}=S(k_i,d_i)^q_1$.

(3) A $(k_i,d_i)$-\emph{edge antimagic total labeling} $f$ of $G_i$ hold $f(V(G_i)\cup E(G_i))=[1,p+q]$ and $\{f(u)+f(v)+f(uv): uv\in E(G_i)\}=S(k_i,d_i)^q_1$, and furthermore $f$ is \emph{super} if $f(V(G_i))=[1,p]$.

(4) A \emph{$(k_i,d_i)$-harmonious labeling} of a $(p,q)$-graph $G_i$ is defined by a mapping $h:V(G)\rightarrow [0,k+(q-1)d_i]$ with $k_i,d_i\geq 1$, such that $f(x)\neq f(y)$ for any pair of vertices $x,y$ of $G$, $h(u)+h(v)(\bmod^*~qd_i)$ means that $h(uv)-k=[h(u)+h(v)-k](\bmod~qd_i)$ for each edge $uv\in E(G)$, and the edge color set $h(E(G))=S(k_i,d_i)^q_1$ holds true.

(5) In \cite{Acharya-Hegde-Arithmetic-1990}, a labeling $f$ of $G$ is said to be \emph{$(k,d)$-arithmetic} if the vertex labels are distinct nonnegative integers and the edge labels induced by $f(x) + f(y)$ for each edge $xy$ are $k, k + d,k + 2d,\dots , k + (q-1)d$.\qqed
\end{defn}

\begin{rem}\label{rem:3-parameter-labelings}
A \emph{Topsnut-gpw sequence} $\{G_{(k_i,d_i)}\}^m_1$ is made by an integer sequence $\{(k_i,d_i)\}^m_1$ and a $(p, q)$-graph $G$, where each Topsnut-gpw $G_{(k_i,d_i)}\cong G$, and each Topsnut-gpw $G_{(k_i,d_i)}\in \{G_{(k_i,d_i)}\}^m_1$ admits one labeling of four parameterized labelings defined in Definition \ref{defn:3-parameter-labelings}. A Topsnut-gpw sequence $\{G_{(k_i,d_i)}\}^m_1$ has the following properties:

(i) $\{(k_i,d_i)\}^m_1$ is a random sequence holding a group of \emph{restrictions}.

(ii) $G_{(k_i,d_i)}\cong G$ is a \emph{structural regularity}.

(iii) Each $G_{(k_i,d_i)}\in \{G_{(k_i,d_i)}\}^m_1$ admits randomly one labeling defined in Definition \ref{defn:3-parameter-labelings}.

(iv) Each $G_{(k_i,d_i)}\in \{G_{(k_i,d_i)}\}^m_1$ has its matching $H_{(k_i,d_i)}\in \{H_{(k_i,d_i)}\}^m_1$ under the meaning of dual labeling, image-labeling, inverse labeling and twin labeling, and so on.\paralled
\end{rem}

\begin{defn}\label{defn:various-k-rotatable-labeling}
(\cite{Yao-Zhang-Sun-Mu-Sun-Wang-Wang-Ma-Su-Yang-Yang-Zhang-2018arXiv, Yao-Sun-Zhang-Mu-Sun-Wang-Su-Zhang-Yang-Yang-2018arXiv, Wang-Xu-Yao-2016, Wang-Xu-Yao-Key-models-Lock-models-2016, Wang-Xu-Yao-2017-Twin}) A $(p,q)$-graph $G$ is \emph{$W$-type $k$-rotatable} for a constant $k\in [a,b]$ if any vertex $u$ of $G$ can be colored as $f(u)=k$ by some $W$-type labeling $f$, where $W$-type $\in \{$graceful, odd-graceful, twin odd-graceful, elegant, odd-elegant, twin odd-elegant, felicitous, edge-magic total, edge-magic total graceful, 6C, \emph{etc.}$\}$.\qqed
\end{defn}

\subsubsection{Basic colorings}

\begin{defn}\label{defn:labeling-coloring}
$^*$ For a $(p,q)$-graph $G$, let $f:V(G)\rightarrow [a,b]$ and $g:E(G)\rightarrow [a,b]$ be two mappings of $G$, and write colored vertex set as $f(V(G))=\{f(w):w\in V(G)\}$ and colored edge set $g(E(G))=\{g(uv):uv\in E(G)\}$, and moreover let $h=(f,g)$ be a \emph{compound mapping} induced by both $f$ and $g$, write $h(V(G)\cup E(G))=f(V(G))\cup g(E(G))$. We have the following constraint conditions:
\begin{asparaenum}[\textrm{Cond}-1. ]
\item \label{labeling-v} $|f(V(G))|=p$;
\item \label{labeling-e} $|g(E(G))|=q$;
\item \label{coloring-v} $|f(V(G))|<p$;
\item \label{coloring-e} $|g(E(G))|<q$;
\item \label{proper-v} $f(u)\neq f(v)$ for any edge $uv\in E(G)$; and
\item \label{proper-e} $g(uv)\neq g(uw)$ for any pair of adjacent edges $uv,uw\in E(G)$.
\end{asparaenum}
\textbf{Then}:
\begin{asparaenum}[\textrm{CL}-1. ]
\item $G$ admits a \emph{labeling} $f$ if Cond-\ref{labeling-v} holds true, namely, $f(x)\neq f(y)$ for any pair of distinct vertices $x,y\in V(G)$.
\item $G$ admits a \emph{coloring} $f$ if Cond-\ref{coloring-v} holds true, namely, $f(u)= f(w)$ for some two distinct vertices $u,w\in V(G)$.
\item $G$ admits a \emph{proper coloring} $f$ if Cond-\ref{coloring-v} and Cond-\ref{proper-v} hold true.
\item $G$ admits a \emph{proper edge coloring} $g$ if Cond-\ref{coloring-e} and Cond-\ref{proper-e} hold true.

\item $G$ admits a \emph{total labeling} $h=(f,g)$ if both Cond-\ref{labeling-v} and Cond-\ref{labeling-e} hold true.
\item $G$ admits a \emph{total coloring} $h=(f,g)$ if one of Cond-\ref{coloring-v} and Cond-\ref{coloring-e} holds true.
\item $G$ admits a \emph{proper total coloring} $h=(f,g)$ if one of Cond-\ref{coloring-v} and Cond-\ref{coloring-e} holds true, and both Cond-\ref{proper-v} and Cond-\ref{proper-e} hold true.

\item $G$ admits a \emph{v-coloring e-labeling} $h=(f,g)$ if Cond-\ref{coloring-v} and Cond-\ref{labeling-e} hold true.
\item $G$ admits an \emph{e-coloring v-labeling} $h=(f,g)$ if Cond-\ref{labeling-v} and Cond-\ref{coloring-e} hold true.

\item $G$ admits a \emph{v-proper-coloring e-labeling} $h=(f,g)$ if Cond-\ref{coloring-v}, Cond-\ref{labeling-e} and Cond-\ref{proper-v} hold true.
\item $G$ admits an \emph{e-proper-coloring v-labeling} $h=(f,g)$ if Cond-\ref{labeling-v}, Cond-\ref{coloring-e} and Cond-\ref{proper-e} hold true.\qqed
\end{asparaenum}
\end{defn}

\begin{rem}\label{rem:various-labeling}
A total labeling can be transformed into a proper total coloring through a series of v-coloring e-labelings, e-coloring v-labelings, v-proper-coloring e-labelings and e-proper-coloring v-labelings. In fact, a proper total coloring is a composition of v-proper-coloring and e-proper-coloring. There are three parameters for various proper colorings: the number $\chi(G)=\min_f\{|b-a|:f(V(G))\subseteq [a,b]\}$ over all proper colorings $f$ of a graph $G$ is called \emph{chromatic number}; the number $\chi\,'(G)=\min_g\{|b-a|:g(E(G))\subseteq [a,b]\}$ over all proper edge colorings $g$ of $G$ is called \emph{chromatic index}; and the number $\chi\,''(G)=\min_h\{|b-a|:f(V(G))\subseteq [a,b], g(E(G))\subseteq [a,b]\}$ over all proper total colorings $h=(f,g)$ of $G$ is called \emph{total chromatic number}.

Since there are hundreds of mutually different labelings and mutually different colorings, the sentence ``a $W$-type coloring'' or ``a $W$-type labeling'' stands for one of the existing colorings and labelings of graph theory hereafter.\paralled
\end{rem}

\begin{defn}\label{defn:Marumuthu-edge-magic-graceful-labeling}
\cite{Marumuthu-G-2015} If there exists a constant $k\geq 0$, such that a $(p, q)$-graph $G$ admits a total labeling $f:V(G)\cup E(G)\rightarrow [1, p+q]$, each edge $uv\in E(G)$ holds
$|f(u)+f(v)-f(uv)|=k$, and $f(V(G)\cup E(G))=[1, p+q]$ true, we call $f$ an \emph{edge-magic graceful labeling} of $G$, and $k$ a \emph{magic constant}. Moreover, $f$ is called a \emph{super edge-magic graceful labeling} if $f(V(G))=[1, p]$.\qqed
\end{defn}

\begin{defn}\label{defn:topcode-matrix-definition}
\cite{Yao-Sun-Zhao-Li-Yan-ITNEC-2017} A \emph{Topcode-matrix} (or \emph{topological coding matrix}) is defined as
\begin{equation}\label{eqa:Topcode-matrix}
\centering
{
\begin{split}
T_{code}= \left(
\begin{array}{ccccc}
x_{1} & x_{2} & \cdots & x_{q}\\
e_{1} & e_{2} & \cdots & e_{q}\\
y_{1} & y_{2} & \cdots & y_{q}
\end{array}
\right)_{3\times q}=
\left(\begin{array}{c}
X\\
E\\
Y
\end{array} \right)=(X,~E,~Y)^{T}
\end{split}}
\end{equation}\\
where \emph{v-vector} $X=(x_1, x_2, \cdots, x_q)$, \emph{e-vector} $E=(e_1$, $e_2 $, $ \cdots $, $e_q)$, and \emph{v-vector} $Y=(y_1, y_2, \cdots, y_q)$ consist of non-negative integers $e_i$, $x_i$ and $y_i$ for $i\in [1,q]$. We say $T_{code}$ to be \emph{evaluated} if there exists a function $f$ such that $e_i=f(x_i,y_i)$ for $i\in [1,q]$, and call $x_i$ and $y_i$ to be the \emph{ends} of $e_i$, and $q$ is the \emph{size} of $T_{code}$.\qqed
\end{defn}

\subsection{Famous conjectures}

A good conjecture has something like a magnetic pull for the mind of a mathematician. At its best, a mathematical conjecture states something extremely profound in an extremely precise and succinct way, crying out for proof or disproof. Mordechai Rorvig pointed: ``\emph{Given more conjectures, there is more grist for the mill of the mathematical mind, more for mathematicians to prove and explain}''.
The following long-standing conjectures can be found in \cite{Bondy-2008}, \cite{Gallian2020} and the books of graph theory.

\begin{conj} \label{conj:c5-Graceful-Tree-Conjecture00}
(Alexander Rosa, 1966) Each tree admits a graceful labeling.
\end{conj}

\begin{conj} \label{conj:c5-Graceful-Cahit-1994}
(Cahit, 1994) Each tree admits an ordered graceful labeling.
\end{conj}

\begin{conj} \label{conj:c5-strongly-Graceful}
(H. J. Broersma and C. Hoede, 1999) Every tree containing a perfect matching admits a strongly
graceful labeling.
\end{conj}

\begin{conj} \label{conj:c5-Bermond-1979}
(Bermond, 1979) Every lobster admits a graceful labeling.
\end{conj}

\begin{conj} \label{conj:c5-Graceful-Truszczynski}
(Truszczy\'{n}ski) All unicyclic connected graphs, except $C_n$, $n = 1$ or $2$ $(\emph{mod}\ 4)$, admit graceful labelings.
\end{conj}

\begin{conj} \label{conj:box}
(R.B. Gnanajothi, 1991) Any tree admits an odd-graceful labeling.
\end{conj}

\begin{conj} \label{conj:Kelly-Ulam-conjecture}
(Kelly-Ulam's Reconstruction Conjecture, 1942) Let both $G$ and $H$ be graphs with $n$ vertices. If there is a bijection $f$: $V(G)\rightarrow V(H)$ such that $G-u\cong H-f(u)$ for each vertex $u\in V(G)$, then $G\cong H$.
\end{conj}

\begin{conj} \label{conj:c2-KT-conjecture}
(Gy\'{a}r\'{a}s and Lehel, 1978; B\'{e}la Bollob\'{a}s, 1995) For integer $n\geq 3$, given $n$ vertex-disjoint trees $T_k$ of $k$ vertices with respect to $1\leq k\leq n$. Then complete graph $K_n$ can be decomposed into the union of $n$ edge-disjoint trees $H_k$, namely $K_n=\bigcup ^{n}_{k=1}H_k$, such that $T_k\cong H_k$ whenever $1\leq k\leq n$. Also, write this case as $\langle T_1,T_2,\dots, T_m\mid K_n\rangle$.
\end{conj}

\begin{conj} \label{conj:box}
(A. G. Chetwynd and A. J. W. Hilton, 1985) Let $G$ be a graph on $2m$ vertices. If $G$ is a regular graph with degree at
least $m$ (when $m$ is odd) or $m-1$ (when $m$ is even), then $G$ is $1$-factorable, as the \textbf{$1$-Factorization
Conjecture}.
\end{conj}

\begin{conj} \label{conj:chapter3-perfect-1-factor}
(Anton Kotzig, 1964; Perfect 1-Factorization Conjecture) For integer $n \geq 2$, complete graph $K_{2n}$ can be decomposed into $2n-1$ perfect matchings such that the union of any two matchings forms a hamiltonian cycle of $K_{2n}$.
\end{conj}

\begin{conj} \label{conj:c4-Hajos-1961}
(Haj\'{o}s, 1961) If $\chi(G)=m$, then $G$ contains a subdivision of $K_m$.
\end{conj}

\begin{conj} \label{conj:c4-Hadwiger-1943}
(Hadwiger, 1943) If $\chi(G)=m$, then $G$ is ``contractible'' to a graph which contains $K_m$.
\end{conj}

\begin{conj} \label{conj:c4-Reed-B-conjecture-X}
(Bruce Reed, 1998) $ \chi(G)\leq \big \lceil \frac{1}{2}(\Delta(G)+1+K(G))\big \rceil$, where $K(G)$ is the clique number of $G$, that is, it is maximal one of vertex numbers of complete graphs contained by $G$.
\end{conj}

\begin{figure}[h]
\centering
\includegraphics[width=16.4cm]{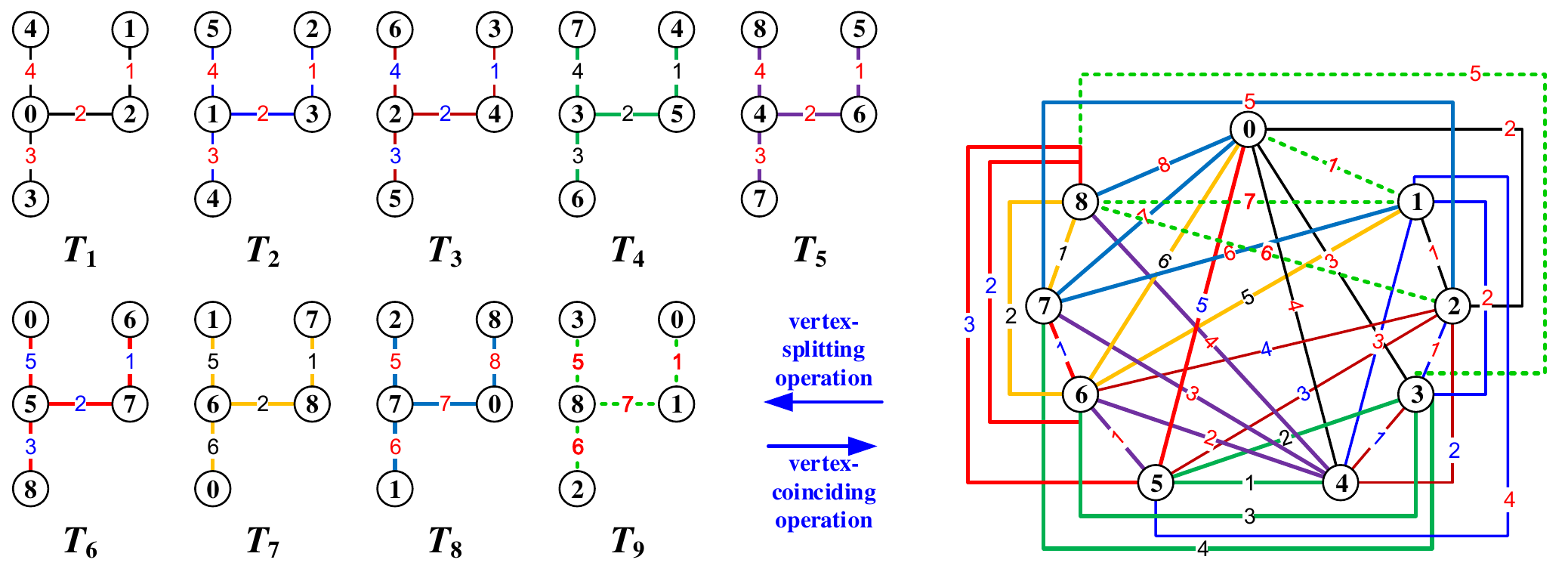}
\caption{\label{fig:Ringel-Kotzig}{\small An example for illustrating the Ringel-Kotzig Decomposition Conjecture, $E(K_9)=\bigcup ^9_{k=1}E(T_k)$ with $E(T_i)\cap E(T_j)=\emptyset$ if $i\neq j$.}}
\end{figure}

\begin{rem}\label{rem:333333}
Rosa \cite{A-Rosa2} said: ``\emph{When does a combinatorial problem become a disease? Certainly the extreme ease of formulating the problem has something to do with it: most identified ``diseases'' are understandable to undergraduates or even to good high school students. They are highly contagious, and so they attract the attention of not only professional mathematicians but also of scores of layman mathematicians}.''

The origins of graceful labelings lie in the problem of packing isomorphic copies of a given tree into a complete graph (Gerhard Ringel and Anton Kotzig, 1963; Alexander Rosa, 1967). If each tree admits a graceful labeling, then this will settle a longstanding and well-known Ringel-Kotzig Decomposition Conjecture \cite{Ringel-G}: ``\emph{A complete graph $K_{2n+1}$ can be decomposed into $2n+1$ subgraphs that are all isomorphic with a given tree having $n$ edges.}'' (see Fig.\ref{fig:Ringel-Kotzig}, $E(K_9)=\bigcup ^9_{k=1}E(T_k)$)

In Wikipedia (TheFreeDictionary.com mirror): (1) Deciding whether the clique number of a graph is greater than or equal to $k$ for $k \geq 3$ is NP-complete.
(2) Deciding subgraph is NP-complete. (3) Others' NP-complete problems are: Complete coloring (\cite{Garey-Johnson-1979-NP-C, Manlove-McDiarmid-1995-trees-NP-C}), $b$-coloring, Edge-coloring \cite{I-Holyer-SIAM-1981}, Strong edge coloring \cite{M-Mahdian-NP-C-2002}, Graph isomorphism, Cut (graph theory), Maximum common subgraph isomorphism problem, K-edge-connected graph, Cubic graph, Clique (graph theory), Vertex cover problem, Graph isomorphism, Hamiltonian path problem, and so on.\paralled
\end{rem}

\section{Recent graph labelings}

\begin{defn}\label{defn:perfect-graph-vs-subgraph-labelings}
\cite{Yao-Zhang-Sun-Mu-Sun-Wang-Wang-Ma-Su-Yang-Yang-Zhang-2018arXiv} Suppose that a connected graph $G$ admit a $\theta$-labeling (resp. $\theta$-coloring), where $\theta$-labeling (resp. $\theta$-coloring) is one of the existing graph labelings (resp. the existing graph colorings). If every connected proper subgraph of $G$ also admits this type of $\theta$-labeling (resp. $\theta$-coloring), then we call $G$ a \emph{perfect $\theta$-labeling graph} (resp. perfect $\theta$-coloring graph).\qqed
\end{defn}

\begin{rem}\label{rem:333333}
Since a set-ordered graceful labelings is equivalent to a group of labelings on trees admitting set-ordered graceful labelings (Ref. \cite{SH-ZXH-YB-2018-5, Su-Wang-Yao-Image-labelings-2021-MDPI, Yao-Liu-Yao-2017, Yao-Yao-Hui-Cheng-2012-Malaysian}), so each caterpillar is a perfect $\theta$-labeling graph for this group of labelings. So, we can conjecture: Each tree is a perfect $\theta$-labeling graph for this group of labelings (resp. the existing graph colorings).\paralled
\end{rem}

\subsection{Total graceful labelings}

\begin{defn} \label{defn:generalized-total-graceful-labelings}
\cite{Wang-Xu-Yao-Ars-2018, Wang-Xu-Yao-2019} Suppose that a $(p,q)$-graph $G$ admits a \emph{labeling} $\theta:V(G)\cup E(G)\rightarrow [1,M]$, and write
$\theta(V(G))=\{\theta(u):u\in V(G)\}$,
$\theta(E(G))=\{\theta(xy):xy\in E(G)\}$. There are the following
constraint conditions:
\begin{asparaenum}[(a)]
\item \label{11-Proper01} $|\theta(V(G))|=p$, $|\theta(E(G))|=q$ and $\theta(xy)=|\theta(x)-\theta(y)|$ for every edge $xy\in E(G)$;
\item \label{ve-total-magic} $\theta(V(G))\cup \theta(E(G))= [1,p+q]$;
\item \label{11-super-tgraceful} $\theta(E(G))=[1,q]$;
\item \label{set-set-ordered} $G$ is a bipartite graph with the bipartition $(X,Y)$ such that $\max\{\theta(x):x\in X\}< \min\{\theta(y):y\in Y\}$ ($\max \theta(X)<\min \theta(Y)$ for short); and
\item \label{11-general-tgraceful-001} $\theta(V(G))\cup \theta(E(G))\subset [1,M]$ with $M\geq 2q+1$ and $\max [\theta(V(G))\cup \theta(E(G))]=M$.
\end{asparaenum}
\textbf{Then}:
\begin{asparaenum}[\textrm{Tog}-1.\,]
\item A \emph{total graceful labeling} $\theta$ holds (\ref{11-Proper01}) and (\ref{ve-total-magic}) true.
\item A \emph{super total graceful labeling} $\theta$ holds (\ref{11-Proper01}), (\ref{ve-total-magic}) and (\ref{11-super-tgraceful}) true.
\item A \emph{set-ordered total graceful labeling} $\theta$ holds (\ref{11-Proper01}), (\ref{ve-total-magic}) and (\ref{set-set-ordered}) true.
\item A \emph{super set-ordered total graceful labeling} $\theta$ holds (\ref{11-Proper01}), (\ref{ve-total-magic}), (\ref{11-super-tgraceful}) and (\ref{set-set-ordered}) true.
\item A \emph{generalized total graceful labeling} $\theta$ holds (\ref{11-Proper01}) and (\ref{11-general-tgraceful-001}) true.
\item A \emph{super generalized total graceful labeling} $\theta$ holds (\ref{11-Proper01}), (\ref{11-general-tgraceful-001}) and (\ref{11-super-tgraceful}) true.
\item A \emph{set-ordered generalized total graceful labeling} $\theta$ holds (\ref{11-Proper01}),
(\ref{11-general-tgraceful-001}) and (\ref{set-set-ordered}) true.
\item A \emph{super set-ordered generalized total graceful labeling} $\theta$ holds (\ref{11-Proper01}), (\ref{11-general-tgraceful-001}), (\ref{11-super-tgraceful}) and (\ref{set-set-ordered}) true.\qqed
\end{asparaenum}
\end{defn}

\begin{defn} \label{defn:generalized-total-odd-graceful-labelings}
\cite{Wang-Xu-Yao-Ars-2018, Wang-Xu-Yao-2019} In Definition \ref{defn:generalized-total-graceful-labelings}, we substitute the constraint conditions ``$\theta(V(G))\cup \theta(E(G))= [1,p+q]$ and $\theta(E(G))=[1,q]$'' by ``$\theta(V(G))\cup \theta(E(G))\subseteq [1,p+2q-1]$ and $\theta(E(G))=[1,2q-1]^o$'', then we get a total odd-graceful labeling, and moreover we have \emph{$W$-type total odd-graceful labelings} for $W$-type$~\in\{$super, set-ordered, super set-ordered, generalized, super generalized, set-ordered generalized, super set-ordered generalized$\}$, respectively.\qqed
\end{defn}

\subsection{Twin-type of labelings}

\begin{defn}\label{defn:22-twin-odd-graceful-labeling}
\cite{Wang-Xu-Yao-2017-Twin} For two connected $(p_i,q)$-graphs $G_i$ with $i=1,2$, and let $p=p_1+p_2-2$, if a $(p,q)$-graph $G=\odot \langle G_1, G_2\rangle$ admits a vertex labeling $f$: $V(G)\rightarrow [0, q]$ such that

(i) $f$ is just an odd-graceful labeling of $G_1$, so $f(E(G_1))=\{f(uv)=|f(u)-f(v)|: uv\in E(G_1)\}=[1, 2q-1]^o$;

(ii) $f(E(G_2))=\{f(uv)=|f(u)-f(v)|: uv\in E(G_2)\}=[1,2q-1]^o$; and

(iii) $|f(V(G_1))\cap f(V(G_2))|=k\geq 0$ and $f(V(G_1))\cup f(V(G_2))\subseteq [0, 2q-1]$.\\
Then $f$ is called a \emph{twin odd-graceful labeling} (Tog-labeling) of $G$.\qqed
\end{defn}

\begin{figure}[h]
\centering
\includegraphics[width=16cm]{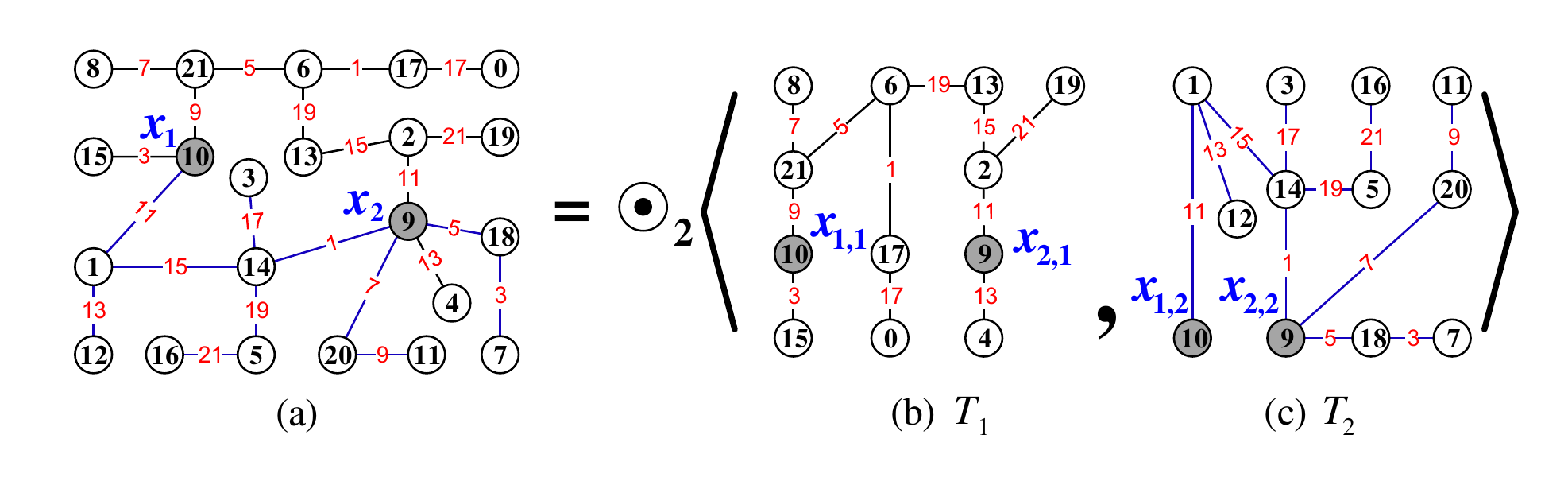}
\caption{\label{fig:odd-elegant-labeling-why}{\small An example for understanding Definition \ref{defn:22-twin-odd-graceful-labeling}, cited from \cite{Wang-Xu-Yao-2017-Twin}.}}
\end{figure}

\begin{defn}\label{defn:twin-odd-elegant-graph}
\cite{Wang-Xu-Yao-2017} For two connected $(p_i,q)$-graphs $G_i$ with $i=1,2$, and let $p=p_1+p_2-2$, if a $(p,q)$-graph $G=\odot \langle G_1, G_2\rangle$ admits a vertex labeling $f$: $V(G)\rightarrow [0, q-1]$ such that

(i) $f$ is just an odd-elegant labeling of $G_i$ with $i=1,2$;

(ii) $|f(V(G_1))\cap f(V(G_2))|=k\geq 0$ and $f(V(G_1))\cup f(V(G_2))\subseteq [0, q-1]$.\\
Then $f$ is called a \emph{twin odd-elegant labeling} (Toe-labeling) of $G$, and $G=\odot \langle G_1, G_2\rangle$ is called a \emph{Toe-matching partition}, where $G_1$ is the \emph{Toe-source} and $G_2$ is a \emph{Toe-association}.\qqed
\end{defn}

\begin{figure}[h]
\centering
\includegraphics[width=16cm]{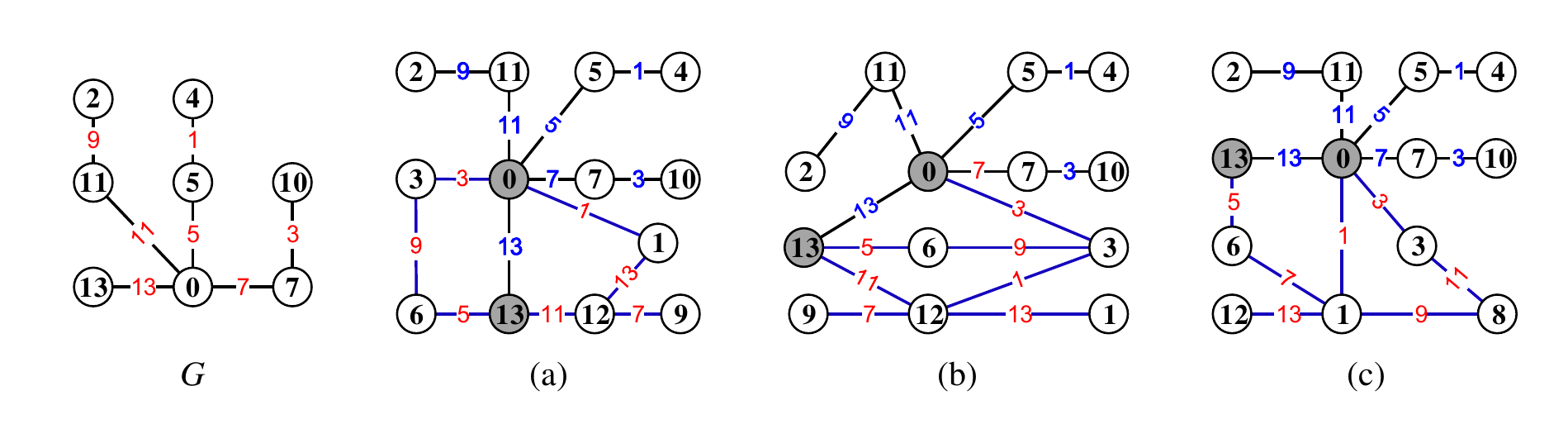}
\caption{\label{fig:2-odd-graceful-vs-elegant}{\small A scheme for illustrating Definition \ref{defn:twin-odd-elegant-graph}, cited from \cite{Wang-Xu-Yao-2017}.}}
\end{figure}

In Fig.\ref{fig:2-odd-graceful-vs-elegant}, a $(8,7)$-graph $G$ admits an odd-graceful labeling $f$, each graph $H_{\textrm{(k)}}$ in $\odot_2 \langle G,H_{\textrm{(k)}}\rangle$ admits an odd-elegant labeling $g_{\textrm{(k)}}$ for k$=$a, b, c; and furthermore both labelings $f$ and $g_{\textrm{(k)}}$ are a \emph{twin (odd-graceful, odd-elegant) labeling}. Clearly, $H_{\textrm{(k)}}\not\cong H_{\textrm{(i)}}$ for $\textrm{(k)}\neq \textrm{(i)}$.

\begin{defn}\label{defn:twin-odd-graceful-labelings}
\cite{Wang-Xu-Yao-2017-Twin, Yao-Zhang-Sun-Mu-Sun-Wang-Wang-Ma-Su-Yang-Yang-Zhang-2018arXiv} Suppose $f:V(G)\rightarrow [0,2q-1]$ is an odd-graceful labeling of a $(p,q)$-graph $G$, and $g:V(H)\rightarrow [1,2q]$ is a labeling of another $(p\,',q\,')$-graph $H$ such that each edge $uv\in E(H)$ has its own color defined as $g(uv)=|g(u)-g(v)|$ and the edge color set $g(E(H))=[1,2q-1]^o$. We call both labelings $f$ and $g$ a \emph{twin odd-graceful labeling} when $f(V(G))\cup g(V(H))\subseteq [0,2q]$, both graphs $G$ and $H$ are a matching of \emph{twin odd-graceful graphs}. \qqed
\end{defn}

\begin{defn}\label{defn:2odd2-matching-labeling}
\cite{Wang-Xu-Yao-2017} For two connected $(p_i,q)$-graphs $G_i$ with $i=1,2$, and let $p=p_1+p_2-2$, if a $(p,q)$-graph $G=\odot \langle G_1, G_2\rangle$ admits a vertex labeling $f$: $V(G)\rightarrow [0, q]$ such that

(i) $f$ is an odd-graceful labeling of $G_1$;

(ii) $f:V(G_2)\rightarrow [0,2|E(G_2)|]$ holding $f(E(G_2))=\{f(uv)=f(u)+f(v)~(\bmod~2|E(G_2)|): uv\in E(G_2)\}=[1, 2|E(G_2)|-1]^{o}$ true.

Then $f$ is called a \emph{2-odd graceful-elegant labeling} (a 2odd2-labeling) of $G$ (called a \emph{2odd2-graph}), and $\odot \langle G_1, G_2\rangle$ is called a \emph{2odd2-matching partition}.\qqed
\end{defn}

\begin{defn} \label{defn:twin-k-d-harmonious-labelings}
\cite{Yao-Zhang-Sun-Mu-Sun-Wang-Wang-Ma-Su-Yang-Yang-Zhang-2018arXiv} A $(p,q)$-graph $G$ admits a $(k,d)$-labeling $f$, and another $(p\,',q\,')$-graph $H$ admits another $(k,d)$-labeling $g$. If $(X_0\cup S_{q-1,k,d})\setminus f(V(G)\cup E(G))=g(V(H)\cup E(H))$ with $X_0=\{0,d,2d, \dots ,(q-1)d\}$ and $S_{q-1,k,d}$, then $g$ is called a \emph{complementary $(k,d)$-labeling} of $f$, and both $f$ and $g$ are a \emph{twin $(k,d)$-labeling}.\qqed
\end{defn}

\begin{defn} \label{defn:twin-colorings-general}
\cite{yao-sun-su-wang-matching-groups-zhao-2020} If a $(p,q)$-graph $G$ admits a $W_f$-type coloring $f:V(G)\cup E(G)\rightarrow [a,A]$, and another $(p\,',q\,')$-graph $H$ admits a $W_g$-type coloring $g:V(H)\cup E(H)\rightarrow [a,A]$, such that $f(V(G))\cup f(V(H))\subseteq [a,A]$, and $k=|f(V(G))\cap f(V(H))|$, we call both colorings $f$ and $g$ a \emph{twin ($W_f$, $W_g$)-type coloring}, and $k$ the \emph{coinciding rank}. Moreover, we coincide a vertex $x_i$ of $G$ with a vertex $u_i$ of $H$ if $f(x_i)=g(u_i)$ for $i\in [1,k]$, such that the resultant graph has no two vertices colored with the same color, denoted this graph admitting the $(W_f,W_g)$-type coloring as $\odot_k \langle G,H\rangle$.\qqed
\end{defn}

\begin{figure}[h]
\centering
\includegraphics[width=16.4cm]{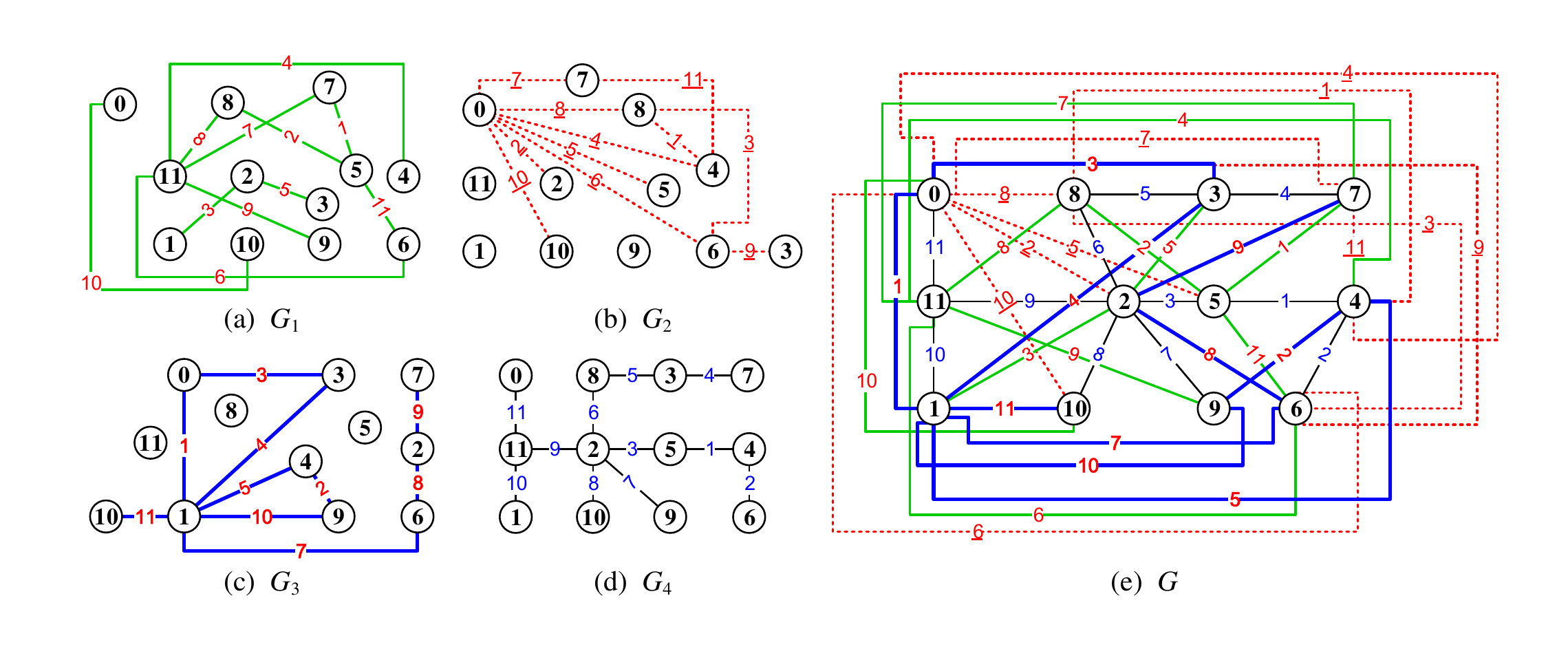}\\
\caption{\label{fig:multiple-graph-matching}{\small A multiple-graph matching partition $G=\odot_f\langle G_i\rangle ^4_1$ for understanding Definition \ref{defn:multiple-trees-labelings}, cited from \cite{Yao-Sun-Zhang-Mu-Sun-Wang-Su-Zhang-Yang-Yang-2018arXiv}.}}
\end{figure}

\begin{defn}\label{defn:multiple-trees-labelings}
\cite{Yao-Sun-Zhang-Mu-Sun-Wang-Su-Zhang-Yang-Yang-2018arXiv} If a $(p,q)$-graph $G$ admits an e-set v-proper $W$-type coloring (resp. labeling) $(F,f)$ defined by $f: V(G)\rightarrow [0, a(p,q)]$ and $F: E(G)\rightarrow [0, b(p,q)]^2$, where $a(p,q)$ and $b(p,q)$ are linear functions of $p$ and $q$, such that $G$ can be decomposed into spanning trees $T_1,T_2,\dots ,T_m$ with $m\geq 2$ and $E(G)=\bigcup^m_{i=1}E(T_i)$ (allow $E(T_i)\cap E(T_j)\neq \emptyset $ for some $i\neq j$), and each spanning tree $T_i$ admits a $W$-type coloring (resp. labeling) $f_i$ induced by $(F,f)$. We call $G$ a \emph{multiple-tree matching partition}, denoted as $G=\oplus_F\langle T_i\rangle ^m_1$. (see Fig.\ref{fig:multiple-graph-matching})\qqed
\end{defn}

\begin{defn}\label{defn:multiple-graph-matching}
\cite{Yao-Sun-Zhang-Mu-Sun-Wang-Su-Zhang-Yang-Yang-2018arXiv} If a $(p,q)$-graph $G$ admits a labeling $f: V(G)\rightarrow [0, p-1]$, such that $G$ can be vertex-split into (spanning) graphs $G_1,G_2,\dots ,G_m$ with $m\geq 2$ and $E(G)=\bigcup^m_{i=1}E(G_i)$ with $E(G_i)\cap E(G_j)=\emptyset $ for $i\neq j$, and each graph $G_i$ admits a $W_i$-type labeling $f_i$ induced by $f$. We call $G$ a \emph{multiple-graph matching partition}, denoted as $G=\odot_f\langle G_i\rangle ^m_1$ (see an example shown in Fig.\ref{fig:multiple-graph-matching}).\qqed
\end{defn}

\begin{thm}\label{thm:set-ordered-matchings-10-labelings}
\cite{Yao-Sun-Zhang-Mu-Sun-Wang-Su-Zhang-Yang-Yang-2018arXiv} If a tree $T$ admits a set-ordered graceful labeling $f$, then $T$ matches with a multiple-tree matching partition $\oplus_f\langle T_i\rangle ^m_1$ with $m\geq 10$ (see Definition \ref{defn:multiple-graph-matching}).
\end{thm}

\begin{defn} \label{defn:matchable-trees}
\cite{Yao-Xiangqian-Zhou-Chen-Cheng-2013, Zhang-Yao-Wang-Wang-Yang-Yang-2013} Let $(X,Y)$ be the bipartition of a complete bipartite graph $K_{n,n}$, where $X=\{u_i:i\in [1,n]\}$, $Y=\{v_j:j\in [1,n]\}$ with $n\geq 2$. For integers $k,p\geq 2$, $K_{n,n}$ has a labeling $\pi _k$ defined as: $\pi _k(u_i)=(k-1)n +i-1$ and $\pi _k(v_i)=(p-k)n+i-1$ for $i\in [1,n]$ with $2k<p+1$. Call two vertices $u_{t_0}\in X$, $v_{t_0}\in Y$ \emph{roots} for a fixed $t_0\in [1,n]$. A pair of vertex-disjoint \emph{rooted trees} $T(u_{t_0}),T(v_{t_0})\subset K_{n,n}$ is called a \emph{matchable pair}, denoted as $M\langle T(u_{t_0}),T(v_{t_0})\mid k,n,p\rangle $, if the following facts hold:

(1) $u_{t_0}\in V(T(u_{t_0}))$ and $v_{t_0}\in V(T(v_{t_0}))$, and $|T(u_{t_0})|+|T(v_{t_0})|=2n$;

(2) $\pi _k(V(T(u_{t_0}))\cup V(T(v_{t_0})))=[n(k-1),nk-1]\cup [n(p-k),n(p-k+1)-1]$; and

(3) $\pi _k(E(T(u_{t_0}))\cup E(T(v_{t_0})))=[n(p-2k)+1,n(p-2k+2)-1]\setminus \{n(p-2k+1)\}$. \qqed
\end{defn}

\begin{figure}[h]
\centering
\includegraphics[width=16.2cm]{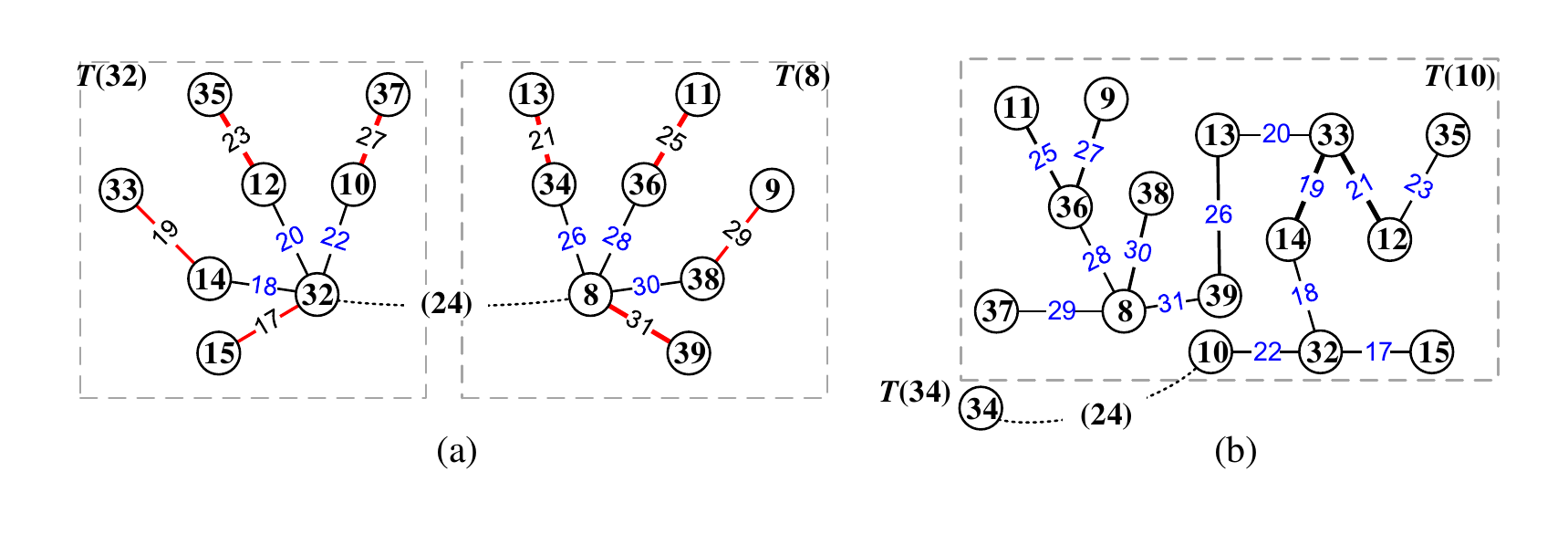}
\caption{\label{fig:Plant-trees-00-eps-converted-to}{\small Two illustrative examples of Definition \ref{defn:matchable-trees}.}}
\end{figure}

Two matchable pairs shown in Fig.\ref{fig:Plant-trees-00-eps-converted-to} are for understanding Definition \ref{defn:matchable-trees}. Fig.\ref{fig:Plant-trees-00-eps-converted-to}(a) is a matchable pair
$M\langle T(8),T(32)\mid 2,8,6\rangle $, where $|T(8)|+|T(32)|=16$, $\pi(V(T(8))\cup V(T(32)))=[8,15]\cup [32,39]$ and $\pi(E(T(8))\cup E(T(32)))=[17,31]\setminus \{24\}$. Fig.\ref{fig:Plant-trees-00-eps-converted-to}(b) shows a matchable pair $M\langle T(10),T(34)\mid 2,8,6\rangle $, where $|T(10)|+|T(34)|=16$, $\pi(V(T(10))\cup V(T(34)))=[8,15]\cup
[32,39]$ and $\pi(E(T(10))\cup E(T(34)))=[17,31]\setminus \{24\}$.

\begin{thm} \label{thm:add-biprtitefraceful-trees-to-each}
Let $G$ be a tree with vertices $x_1,x_2,\dots,x_p$, and $\pi$ be a (set-ordered) graceful labeling of $G$ with $\pi(x_i)<\pi(x_{i+1})$ for $i\in [1,p-1]$. Then

$(i)$ For even $p$ and a collection of $\frac{p}{2}$ matchable pairs $M\langle T(u_{t_0}),T(v_{t_0})\mid k,n,p\rangle $ with $k\in [1,\frac{p}{2}]$, we have a (strongly, set-ordered) graceful tree obtained by identifying the roots $u_{t_0}$, $v_{t_0}$ of each matchable pair $M\langle T(u_{t_0}),T(v_{t_0})\mid k,n,p\rangle $ with the vertices $x_k$, $x_l$ of $G$ respectively, where $k+l=p+1$ and $k\in [1,\frac{p}{2}]$.

$(ii)$ For odd $p$ and a collection of $\frac{p-1}{2}$ matchable pairs $M\langle T(u_{t_0}),T(v_{t_0})\mid k,n,p\rangle $ with $k\in [1,\frac{p-1}{2}]$, and a (strongly, set-ordered) graceful tree $T\,^*$ on $n$ vertices, we have a (strongly, set-ordered) graceful tree obtained by identifying the roots $u_{t_0}$, $v_{t_0}$ of each pair $M\langle T(u_{t_0}),T(v_{t_0})\mid k,n,p\rangle $ with the vertices $x_k$, $x_l$ of $G$ respectively, where $k+l=p+1$ and $k\in [1,\frac{p-1}{2}]$, and identifying a vertex of $T\,^*$ with the vertex $x_{(p+1)/2}$ of $G$.
\end{thm}

\begin{figure}[h]
\centering
\includegraphics[width=14.4cm]{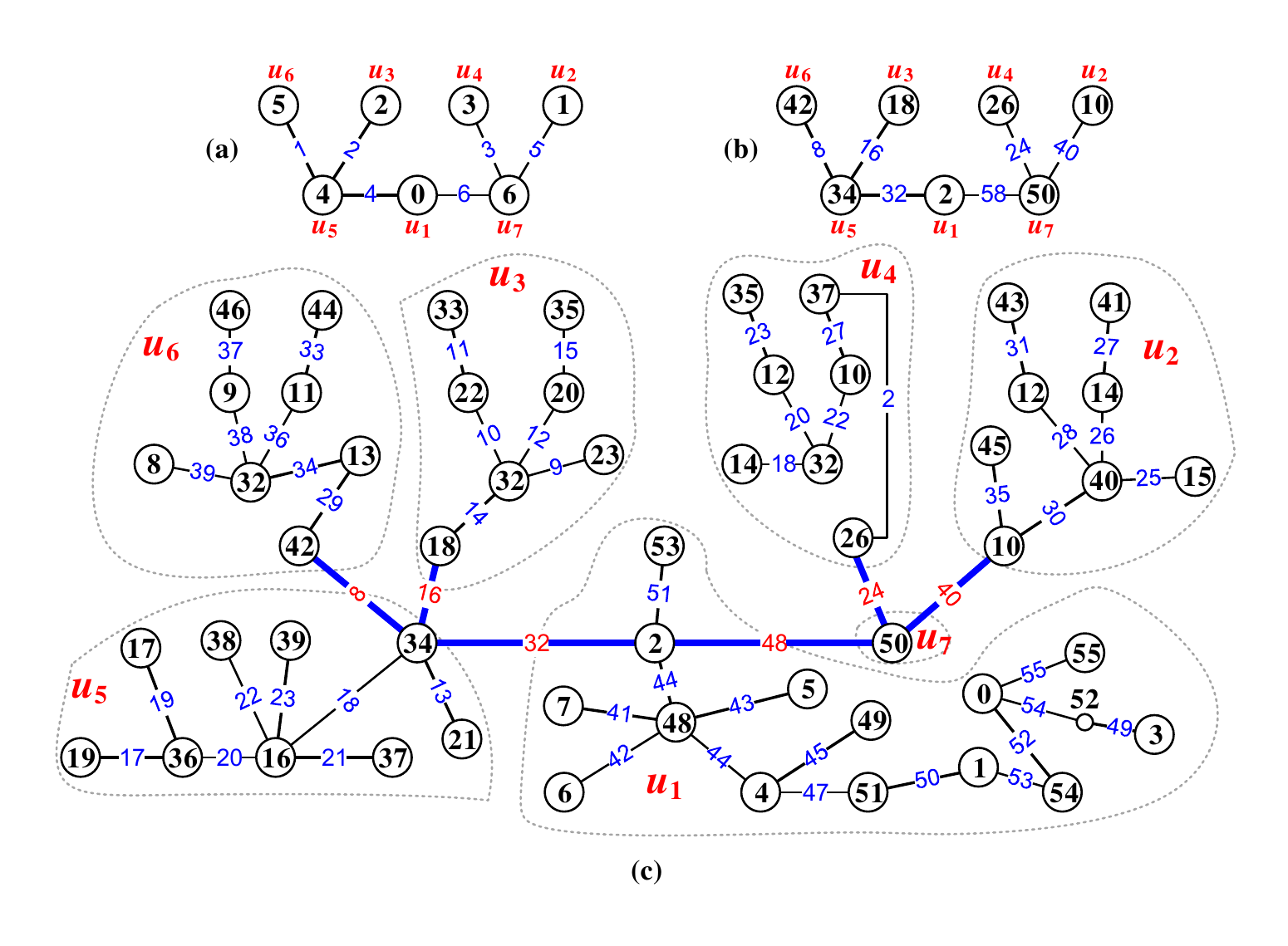}
\caption{\label{fig:plant-matching-trees}{\small An example for illustrating the assertion $(ii)$ of Theorem \ref{thm:add-biprtitefraceful-trees-to-each}, where $M\langle T(2),T(50)\mid 1,8,7\rangle$, $M\langle T(10),T(42)\mid 2,8,7\rangle$, $M\langle T(18),T(34)\mid 3,8,7\rangle$ and $T\,^*=T(26)$, and the black vertices are the roots.}}
\end{figure}

\subsection{Magic-type of labelings}

\subsubsection{Edge-magic labelings}

\begin{defn} \label{defn:super-set-ordered-edge-magic-total-labeling}
\cite{Wang-Yao-Yao-2014Information} Let $G$ be a bipartite $(p,q)$-graph with bipartition $(X, Y)$, and let $G$ admit an \emph{edge-magic total labeling} $f$ defined in Definition \ref{defn:11-old-labelings-Gallian}. There are two constraints:

(C1) $f(V(G))=[1,p]$; and

(C2) $\max \{f(x): x\in X\}<\min\{f(y): y\in Y\}$ ($\max f(X)<\min f(Y)$).\\
\textbf{We call $f$}:
\begin{asparaenum}[(i) ]
\item A \emph{super edge-magic total labeling} of $G$ if $f$ holds (C1) true.
\item A \emph{set-ordered edge-magic total labeling} of $G$ if $f$ holds (C2) true.
\item A \emph{super set-ordered edge-magic total labeling} (super-so-edge-magic total labeling) of $G$ if $f$ holds both (C1) and (C2) true.\qqed
\end{asparaenum}
\end{defn}

\begin{defn} \label{defn:generalized-super-edge-magic-total-labeling}
\cite{Wang-Yao-Yao-2014Information} Let $G$ be a $(p,q)$-graph. If there exist a constant $\mu$ and a mapping $f:V(G)\cup E(G)\rightarrow [1, 2q+1]$ such that $f(u)+f(v)+f(uv)=\mu$ for every edge $uv\in E$, then we call $f$ a \emph{generalized edge-magic total labeling} of $G$, and $\mu$ a \emph{generalized magic constant}. Furthermore, if $G$ is a bipartite graph with bipartition $(X,Y)$, and the labeling $f$ holds $f(V(G))=[1,q+1]$ and $\max f(X)<\min f(Y)$ true, we call $f$ a \emph{generalized super set-ordered edge-magic total labeling}.\qqed
\end{defn}

\begin{defn} \label{defn:anti-edge-magic-total-abelling}
\cite{Wang-Yao-Yang-Yang-Chen-2013-Advanced-Materials} Let $G$ be a $(p,q)$-graph. If there exists a bijection $f : V(G)\cup E(G)\rightarrow [1, p+q]$ such that $\{f(u)+f(v)+f(uv):uv\in E(G)\}=\{k, k+d, k+2d, \cdots, k+(q-1)d\}$ for some values $k,d\in Z^0$, then we call $f$ an \emph{anti-edge-magic total labeling} of $G$. Furthermore, if $G$ is a bipartite graph with bipartition $(X, Y)$, and $f$ holds $f(V(G))=[1,p]$
and $\max f(X)<\min f(Y)$, we call $f$ a \emph{super set-ordered anti-edge-magic total labeling}.\qqed
\end{defn}

\begin{defn} \label{defn:k-d-edge-magic-total-labeling}
\cite{Wang-Yao-Yao-2014Information} A $(p, q)$-graph $G$ admits a bijection $f:V(G)\cup E(G)\rightarrow \{d,2d, \dots, \mu d, k+(\mu+1)d, k+(p+q-1)d\}$ with $\mu\in[1,p+q-1]$, such that $f(u)+f(v)+f(uv)=\lambda$ for each edge $uv\in E(G)$, we call $f$ a \emph{$(k,d)$-edge-magic total labeling}, $\lambda$ a \emph{magic constant}. Moreover, if $G$ is a bipartite graph with bipartition $(X, Y)$, and $f$ holds $f(X)=\{ d,2d, \dots, |X|d\}$, $f(Y)=\{k+|X|d, k+(|X|+1)d, \dots, k+(|X|+|Y|-1)d$, we call $f$ a \emph{super set-ordered $(k,d)$-edge-magic total labeling} of $G$.\qqed
\end{defn}

\begin{defn}\label{defn:edge-magic-total-graceful-labeling}
\cite{Yao-Mu-Sun-Zhang-Wang-Su-2018} An \emph{edge-magic total graceful labeling} $g$ of a $(p,q)$-graph $G$ is defined as: $g: V(G)\cup E(G)\rightarrow [1,p+q]$ such that $g(x)\neq g(y)$ for any two elements $x,y\in V(G)\cup E(G)$, and each edge $uv\in E(G)$ holds $g(uv)+|g(u)-g(v)|=k$ with a constant $k$. Moreover, $g$ is \emph{super} if $\max g(E(G))<\min g(V(G))$ (or $\max g(V(G))<\min g(E(G))$).\qqed
\end{defn}

\begin{defn}\label{defn:magic-graceful-labeling}
\cite{Marumuthu-G-2015} If there exists a constant $k\geq 0$, a $(p, q)$-graph $G$ admits a total labeling $f:V(G)\cup E(G)\rightarrow [1, p+q]$, such that each edge $uv\in E(G)$ holds $|f(u)+f(v)-f(uv)|=k$ and $f(V(G)\cup E(G))=[1, p+q]$ true, we call $f$ an \emph{edge-magic graceful labeling} of $G$, and $k$ a \emph{magic constant}. Moreover, $f$ is called a \emph{super edge-magic graceful $k$-labeling} if $f(V(G))=[1, p]$.\qqed
\end{defn}

\begin{defn}\label{defn:ve-exchanged-labeling}
\cite{Yao-Mu-Sun-Zhang-Wang-Su-Ma-2018} A \emph{ve-exchanged matching labeling} $h$ of an edge-magic graceful labeling $f$ of a $(p, q)$-graph $G$ is defined as: $h:V(G)\cup E(G)\rightarrow [1, p+q]$, each edge $uv\in E(G)$ holds $h(uv)=|h(u)-h(v)|$ true ($h(uv)+|h(u)-h(v)|=k$, or $h(uv)=h(u)+h(v)~(\bmod~q)$, or $|h(u)+h(v)-h(uv)|=k$, or $h(u)+h(uv)+h(v)=k$), such that $h(V(G)\cup E(G))=[1, p+q]$, $h(V(G))\setminus \{a_0\}=f(E(G))$ and $h(E(G))=f(V(G))\setminus \{a_0\}$, where $a_0=\lfloor \frac{p+q+1}{2}\rfloor $ is the singularity of two labelings $f$ and $h$. \qqed
\end{defn}

\begin{defn}\label{defn:ve-inverse-edge-magic-graceful-labeling}
\cite{Yao-Zhang-Sun-Mu-Sun-Wang-Wang-Ma-Su-Yang-Yang-Zhang-2018arXiv} Suppose that a $(p,q)$-graph $G$ admits an \emph{edge-magic graceful labeling} $f$, and a $(q,p)$-graph $H$ admits another \emph{edge-magic graceful labeling} $g$. If $f(E(G))=g(V(H))$ and $f(V(G))=g(E(H))$, we say that both labelings $f$ and $g$ are \emph{ve-inverse} from each other, and both graphs $G$ and $H$ are \emph{inverse} from each other under the edge-magic graceful labelings.\qqed
\end{defn}

\begin{defn}\label{defn:general-ve-inverse-labeling}
$^*$ Suppose that a graph $G$ admits a $W_f$-type labeling (resp. coloring) $f$, and another graph $H$ admits a $W_g$-type labeling (resp. coloring) $g$. If $f(E(G))=g(V(H))$ and $f(V(G))=g(E(H))$, we call $(f,g)$ a \emph{ve-inverse $(W_f,W_g)$-type labeling} (resp. coloring).\qqed
\end{defn}

\begin{defn}\label{defn:edge-magic-total-graceful-labeling}
\cite{Yao-Zhang-Sun-Mu-Sun-Wang-Wang-Ma-Su-Yang-Yang-Zhang-2018arXiv} Suppose that a $(p,q)$-graph $L$ admits an \emph{edge-magic total graceful labeling} $\theta$ defined in Definition \ref{defn:magic-graceful-labeling}, and a $(q,p)$-graph $S$ admits an \emph{edge-magic total graceful labeling} $\varphi$. If $\theta(E(L))=\varphi(V(S))$ and $\theta(V(L))=\varphi(E(S))$, we say that both labelings $\theta$ and $\varphi$ are \emph{inverse} from each other, and moreover both graphs $L$ and $S$ are \emph{inverse} from each other under the edge-magic total graceful labeling.\qqed
\end{defn}

\begin{defn}\label{defn:li-yi-chun-sun-yao-2018east-china}
\cite{li-yi-chun-sun-yao-2018east-china} A $(p,q)$-graph $G$ admits a mapping $f:V(G) \cup E(G)\rightarrow [0, 2(p+q-1)]^e$. If there exists a constant $k$, such that $\mid f(u)+f(v)-f(uv)\mid =k$ for each edge $uv\in E(G)$, we call $f$ an \emph{edge-magic even-graceful labeling}, and $G$ an \emph{edge-magic even-graceful graph}.\qqed
\end{defn}

\begin{defn}\label{defn:wang-yaru-yao-2017}
\cite{wang-yaru-yao-2017} An \emph{odd-even separable edge-magic (total) labeling} $f$ of a $(p,q)$-graph $G$ is an edge-magic (total) labeling and holds $f(V(G))=[1,2L+1]^o$ and $f(E(G))=[2,2M]^e$.\qqed
\end{defn}

\begin{defn}\label{defn:odd-totally-edge-magic-graceful}
\cite{li-yi-chun-sun-yao-2018east-china} If a $(p,q)$-graph $G$ admits a bijection $f : V(G)\cup E(G)\rightarrow [1, 2(p+q)+1]$, and there exists an odd constant $k$ such that $|f(u)+f(v)-f(uv)|=k$ for each edge $uv\in E(G)$, we say $f$ an \emph{odd-totally edge-magic graceful labeling}, and $k$ an \emph{odd magic constant}.\qqed
\end{defn}

\subsubsection{Parameterized magic-type total labelings}

\begin{defn} \label{defn:k-lambda-magically-total-labeling}
\cite{Yao-Chen-Yao-Cheng2013JCMCC} Let $G$ be a connected $(p,q)$-graph. If there are integers $k$ and $\lambda~(\neq 0)$ such that a proper total labeling $f$ of $G$ from $V(G)\cup E(G)$ to $[1,p+q]$ satisfies $f(u)+f(v)=k+\lambda f(uv)$ whenever $ uv\in E(G)$. Then $f$ is called a $(k,\lambda )$-\emph{magically total labeling} (or $(k,\lambda )$-\emph{mtl} for short) of $G$, $k$ and $\lambda $ are called a \emph{magical constant} and a \emph{balanced number}, respectively, and moreover $f$ is \emph{super} if the vertex color set $f(V(G))=[1,p]$, and $G$ is $f$-\emph{saturated} if $f$ is no longer a $(k,\lambda )$-\emph{mtl} of $G+uv$ for any $uv\in E(\overline{G})$, where $\overline{G}$ is the complement of $G$.\qqed
\end{defn}

\begin{lem} \label{thm:complementary-labeling}
Let $G$ be a connected $(p,q)$-graph admitting a $(k,\lambda)$-magically total labeling $f$ defined in Definition \ref{defn:k-lambda-magically-total-labeling}.

$(i)$ Then the complementary labeling $h$ of the labeling $f$ is a $(k\,',\lambda)$-magically total labeling of $G$ where the magical constant $k\,'=(2-\lambda )(p+q+1)-k$; and

$(ii)$ If $f$ is supper, then the partial complementary labeling $h$ of $f$ is a supper $(k\,'',\lambda)$-magically total labeling of $G$ where the magical constant $k\,''=2(p+1)-k-\lambda (2p+q+1)$.
\end{lem}

\begin{lem} \label{thm:theorem-necessary-sufficient}
A connected graph $G$ admits a $(k,\lambda )$-magically total labeling $f$ defined in Definition \ref{defn:k-lambda-magically-total-labeling} if and only if any two incident edges $uv$ and $vw$ of $G$ hold $f(u)-f(w)=\lambda [f(uv)-f(vw)]$ true.
\end{lem}

\begin{lem} \label{thm:parameter-relationship}
Let $G$ be a connected $(p,q)$-graph admitting a supper $(k,\lambda)$-magically total labeling $f$ defined in Definition \ref{defn:k-lambda-magically-total-labeling}. We have:

$(i)$ If $k>0$, then $\lambda \leq -1$;

$(ii)$ Let $\Delta$ be the maximal degree of $G$, then
$\lambda \Delta\leq p-1$;

$(iii)$ $3-\lambda (p+q)\leq k\leq 2(p+q)-\lambda -1$; and

$(iv)$ If $G$ is regular, then
$2k=2(p+1)-\lambda(2p+q+1)$.
\end{lem}

\begin{thm} \label{thm:supper-add-a-vertex}
Let $G$ be a connected $(p,q)$-graph possessing a supper $(k,\lambda)$-magically total labeling $f$ defined in Definition \ref{defn:k-lambda-magically-total-labeling}. If $\lambda \leq 2$, there then is a graph, $H=G+uw$, obtained by adding a new vertex $u$ to $G$ and adjoining $u$ and a certain vertex $w$ of $G$, such that it admits a supper $(k-\lambda,\lambda)$-magically total labeling.
\end{thm}

\begin{thm} \label{thm:c5-positive-negative-relation}
Let $\lambda \geq 1$. A graph $G$ admits a supper
$(k,\lambda)$-magically total labeling $f$ defined in Definition \ref{defn:k-lambda-magically-total-labeling} if and only if it admits a supper $(k\,',-\lambda)$-magically total labeling $h$ where the magical constant $k\,'=k+\lambda (M+m)$ for $m=\min\{f(uv): uv\in E(G)\}$ and $M=\max\{f(uv): uv\in E(G)\}$, And $f$ is supper if and only if $h$ is supper too.
\end{thm}

\begin{thm} \label{thm:supper-property}
Let $G$ be a connected $(p,q)$-graph admitting a supper $(k,\lambda)$-magically total labeling $f$ defined in Definition \ref{defn:k-lambda-magically-total-labeling}. Then $q\leq \lceil (2p-3)/\lambda \rceil$, the equality holds if and only if $G$ is $f$-saturated.
\end{thm}

\begin{thm} \label{thm:catepillar-supper-magic}
Every caterpillar admits a supper $(k,\lambda)$-magically total labeling defined in Definition \ref{defn:k-lambda-magically-total-labeling}.
\end{thm}

\begin{defn} \label{defn:k-sedge-difference-magically-total-labeling}
$^*$ Let $G$ be a connected $(p,q)$-graph. If there are integers $k$ and $\lambda~(\neq 0)$ such that $G$ admits a proper total labeling $h:V(G)\cup E(G)\rightarrow [1,p+q]$ holding $|h(u)-h(v)|=k+\lambda h(uv)$ for each edge $ uv\in E(G)$ true, then we call $h$ a $(k,\lambda)$-\emph{edge-difference magically total labeling} of $G$, $k$ a \emph{magical constant} and $\lambda$ a \emph{balanced number}, and moreover $f$ is \emph{super} if the vertex color set $f(V(G))=[1,p]$.\qqed
\end{defn}

\begin{rem}\label{rem:edge-difference-magically-total}
It is noticeable: (i) a $(k,\lambda )$-magically total labeling defined in Definition \ref{defn:k-lambda-magically-total-labeling} implies some \emph{edge-magic total labeling} defined in Definition \ref{defn:11-old-labelings-Gallian} and some \emph{felicitous-difference total coloring} defined in Definition \ref{defn:new-graceful-strongly-colorings}, since $f(u)+f(v)=k+\lambda f(uv)$ is related with $f(u)+f(uv)+f(v)=k$ and $f(u)+f(v)-f(uv)=k$ as $\lambda=-1$.

(ii) A $(k,\lambda)$-edge-difference magically total labeling defined in Definition \ref{defn:k-sedge-difference-magically-total-labeling} implies some edge-magic graceful total labeling defined in Definition \ref{defn:edge-magic-total-even-odd-graceful-labeling}, some edge-difference magic total coloring and some graceful ev-difference magic total coloring defined in Definition \ref{defn:new-graceful-strongly-colorings}, since $|h(u)-h(v)|=k+\lambda h(uv)$ is related with $h(uv)+|h(u)-h(v)|=k$ and $\big |h(uv)-|h(u)-h(v)|\big |=k$ as $\lambda=-1$.\paralled
\end{rem}

\subsubsection{Couple edge-magic total labelings}

Let $G\,'$ be a copy of $G$, and join a vertex $u$ of $G$ with a vertex $u\,'$ of $G\,'$ by an edge together, the resultant graph is denoted by $\widehat{GG\,'}$, called an \emph{edge-symmetric graph} since $\widehat{GG\,'}-uu\,'$ contains just two components $G$ and $G\,'$ holding $G\cong G\,'$ true. We call $\widehat{GG\,'}$ a \emph{formally edge-symmetric graph} if $u\,'$ is just the imagine vertex of $u$. An \emph{pseudo-edge-symmetric graph} $\widehat{GH}$ is obtained by joining a vertex $u$ of $G$ with a vertex $v$ of $H$ by an edge if $G\not \cong H$.

\begin{defn}\label{defn:CEMTL}
\cite{Yao-Sun-Zhang-Zhang-Yang-Wang-Zhang-ICISCE2017} If there are two constants $k^+>0$ and $k^-\geq 0$, a $(p,q)$-graph $G$ admits a bijection $f :V(G)\cup E(G)\rightarrow [1, p+q]$ such that every edge $uv\in E(G)$ holds one of $f(u)+f(v)+f(uv)=k^+$ and $|f(u)+f(v)-f(uv)|=k^-$ true, then we call $f$ a \emph{$(k^+,k^-)$-couple edge-magic total labeling} ($(k^+,k^-)$-CEMTL for short).\qqed
\end{defn}

\begin{figure}[h]
\centering
\includegraphics[width=16cm]{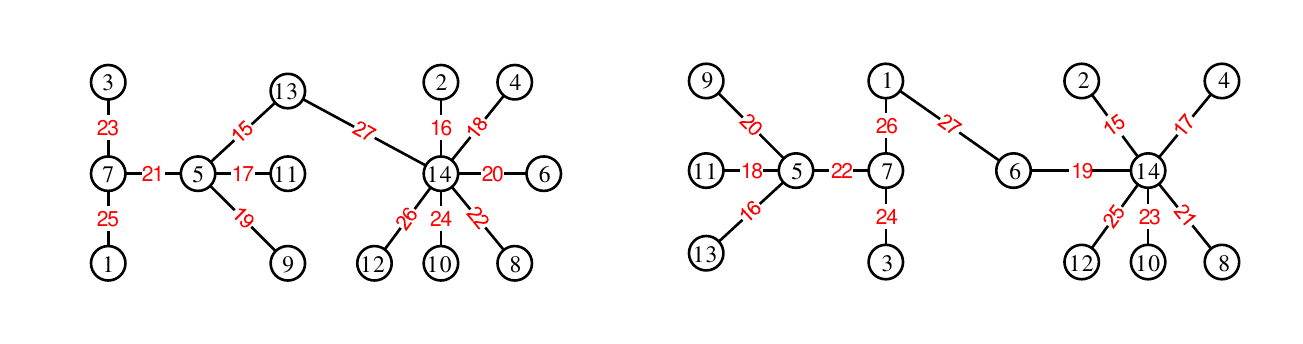}\\
\caption{\label{fig:two-trees-CEMTL}{\small Left tree admits an $(e>v)$-$(33,0)$-CEMTL $f$, and Right tree admits a $(v>e)$-$(34,1)$-CEMTL.}}
\end{figure}

\begin{defn}\label{defn:111111}
\cite{Yao-Sun-Zhang-Zhang-Yang-Wang-Zhang-ICISCE2017} The \emph{dual labeling} $g$ of a labeling $f$ defined in Definition \ref{defn:CEMTL} is a $(v>e)$-$(k^+_{new},k^-_{new})$-CEMTL (resp. a $(e>v)$-$(k^+_{new},k^-_{new})$-CEMTL) if and only if $f$ is an $(e>v)$-$(k^+,k^-)$-CEMTL (resp. a $(v>e)$-$(k^+,k^-)$-CEMTL), where $k^+_{new}=3(p+q+1)-k^+$ and $k^-_{new}=p+q+1+k^-$ (see Fig.\ref{fig:two-trees-CEMTL}).
\end{defn}

\begin{lem} \label{thm:construction-1}
\cite{Yao-Sun-Zhang-Zhang-Yang-Wang-Zhang-ICISCE2017} Suppose that $T$ is a tree admitting set-ordered graceful labelings. Then there exists an edge-symmetric tree $\widehat{TT\,'}$ admitting a super $(k^+,k^-)$-CEMTL.
\end{lem}

\begin{lem} \label{lem:construction-2}
\cite{Yao-Sun-Zhang-Zhang-Yang-Wang-Zhang-ICISCE2017} Suppose that a connected $(p,q)$-graph $G$ admits an $(e>v)$-EMGL $f$ with a magic constant $k$, there are two vertices $w,w\,'$ holding $f(w)+f(w\,')=p+q+1-k$. Then there exists an edge-symmetric graph $\widehat{GG\,'}$ admitting a $(k^+,k^-)$-CEMTL.
\end{lem}

\begin{thm} \label{thm:theorem-1}
\cite{Yao-Sun-Zhang-Zhang-Yang-Wang-Zhang-ICISCE2017} Let $T$ be a graceful tree. Then there exists an edge-symmetric tree $\widehat{HH\,'}$ admitting a super $(k^+,k^-)$-CEMTL, where $H=\widehat{TT\,'}$, where $T\,'$ is a copy of $T$ and $H\,'$ is a copy of $H$.
\end{thm}

\begin{thm} \label{thm:theorem-2}
\cite{Yao-Sun-Zhang-Zhang-Yang-Wang-Zhang-ICISCE2017} Suppose that both trees $T$ and $H$ have the same vertex number, $T$ admits an EMGL, $H$ admits a $(v>e)$-EMGL (resp. $(e>v)$-EMGL). Then the pseudo-edge-symmetric tree $\widehat{TH}$ admits a $(k^+,k^-)$-CEMTL.
\end{thm}

\begin{thm} \label{thm:theorem-3}
\cite{Yao-Sun-Zhang-Zhang-Yang-Wang-Zhang-ICISCE2017} Suppose that both graphs $G$ and $H$ have $p$ vertices and $q$ edges. $G$ admits an EMGL $f$ with magic constant $k_1$, and $H$ admits an $(e>v)$-EMGL $g$ with magic constant $k_2$. If there exist vertices $w\in V(G)$ and $w\,'\in V(H)$ such that $f(w)+g(w\,')=p+q+1-k_1$ (resp. $f(w)+g(w\,')=p+q+1+k_1$), then the pseudo-edge-symmetric graph $\widehat{GH}$ obtained by joining $w$ with $w\,'$ by an edge admits a $(k^+,k^-)$-CEMTL, where $k^-=2k_1$, $k^+=2(2p+q)-1-2k_2$.
\end{thm}

\begin{rem}\label{rem:CEMTLs}
Based on our new labelings, we have applied the labeling orders, that is, $(e>v)$ and $(v>e)$ here in our algorithms for building up network models admitting $(k^+,k^-)$-CEMTLs ($(k^+,k^-)$-couple edge-magic total labelings). However, many EMGLs (edge-magic graceful labelings) and $(k^+,k^-)$-CEMTLs have no dual labelings to be EMGLs, or $(k^+,k^-)$-CEMTLs. We point out that there exist graphs admitting no $(k^+,k^-)$-CEMTLs defined in Definition \ref{defn:CEMTL}.\paralled
\end{rem}

So we generalize Definition \ref{defn:CEMTL} in the following definition:

\begin{defn}\label{defn:generalized-CEMTL}
\cite{Yao-Sun-Zhang-Zhang-Yang-Wang-Zhang-ICISCE2017} If there are two sequences $(k^+_i)^s_1=(k^+_1,k^+_2,\dots, k^+_s)$ and $(k^-_j)^t_1=(k^-_1,k^-_2,\dots $, $k^-_t)$, a $(p,q)$-graph $G$ admits a bijection $f :V(G)\cup E(G)\rightarrow [1, p+q]$ such that every edge $uv\in E(G)$ holds $f(u)+f(v)+f(uv)=k^+_i$ or $|f(u)+f(v)-f(uv)|=k^-_j$ for some $k^+_i,k^-_j$, then we call $f$ a \emph{$\langle (k^+_i)^s_1, (k^-_j)^t_1\rangle$-couple edge-magic total labeling}, rewrite $\langle (k^+_i)^s_1, (k^-_j)^t_1\rangle$-CEMTL for short.\qqed
\end{defn}

\begin{rem}\label{rem:EMGs}
In Definition \ref{defn:generalized-CEMTL}, we can classify the edges of the $(p,q)$-graph $G$ into two sets $E^+=\{uv: f(u)+f(v)+f(uv)=k^+_i\}$ and $E^-=\{xy: |f(x)+f(y)-f(xy)|=k^-_j\}$ by the labeling $f$. So, $f$ is called a super $\langle (k^+_i)^s_1, (k^-_j)^t_1\rangle$-CEMTL if $f(V(G))=[1,p]$, $f$ an $(e>v)$-$\langle (k^+_i)^s_1, (k^-_j)^t_1\rangle$-CEMTL if $f(uv)-[f(u)+f(v)]=k^-_j>0$ for each edge $uv\in E^-$ (resp. a $(v>e)$-$\langle (k^+_i)^s_1, (k^-_j)^t_1\rangle$-CEMTL if $[f(u)+f(v)]-f(uv)=k^-_j>0$ for each edge $uv\in E^-$). We have some particular cases: If $t=0$ (resp. $s=0$), $f$ is a $(k^+_i)^s_1$-CEMTL (resp. a $(k^-_j)^t_1$-CEMTL); if $t=0$ and $s=1$, $f$ is an EMTL (\emph{edge-magic total labeling}); if $t=1$ and $s=0$, $f$ is an EMGL (\emph{edge-magic graceful labeling}); if $t=1$ and $s=1$, $f$ is a $(k^+,k^-)$-CEMTL.

Let $C_{emtl}(G)=\min\{s+t\}$ over all $\langle (k^+_i)^s_1, (k^-_j)^t_1\rangle$-CEMTLs of $G$. We can confirm: ``\emph{Every simple graph admits a $\langle (k^+_i)^s_1, (k^-_j)^t_1\rangle$-CEMTL}''. However, it seems that the work for determining $\langle (k^+_i)^s_1, (k^-_j)^t_1\rangle$-CEMTLs of simple graphs is not slight, so is for $C_{emtl}(G)$.\paralled
\end{rem}

\begin{figure}[h]
\centering
\includegraphics[width=12cm]{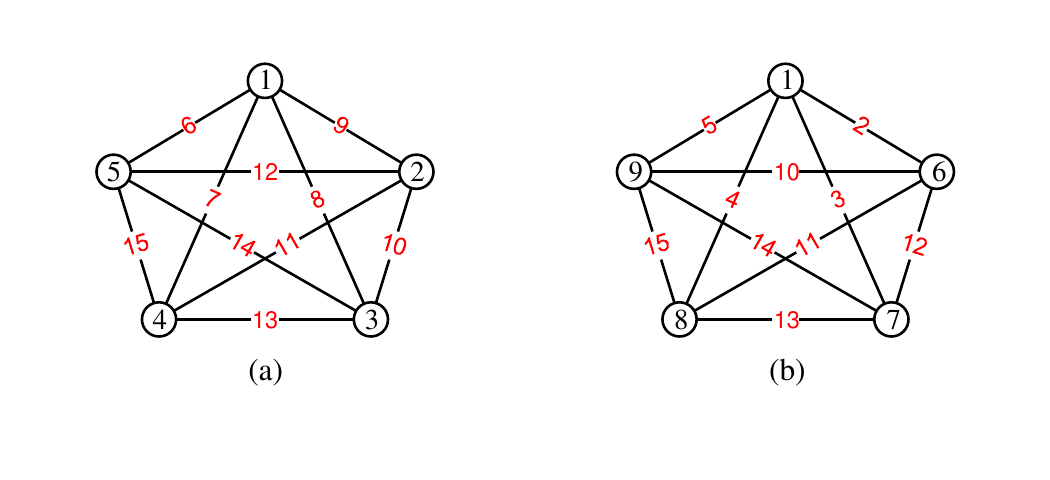}\\
\caption{\label{fig:two-complete-graphs}{\small The complete graph $K_5$ admits two $\langle (k^+_i)^s_1, (k^-_j)^t_1\rangle$-CEMTLs for illustrating Definition \ref{defn:generalized-CEMTL}, and $C_{emtl}(K_5)=3$, where (a) $\langle (k^+_i)^s_1, (k^-_j)^t_1\rangle=\langle (12), (5,6)\rangle$ and (b) $\langle (k^+_i)^s_1, (k^-_j)^t_1\rangle=\langle (25), (2,5)\rangle$.}}
\end{figure}

\begin{defn}\label{defn:sequce-graceful-labeling-k-f}
\cite{Yao-Sun-Zhang-Zhang-Yang-Wang-Zhang-ICISCE2017} If a $(p,q)$-graph $H$ admits a mapping $f:V(H)\rightarrow [0,q]$ such that $f(x)\neq f(y)$ for any pair of distinct vertices $x,y$, and $E_i=\{uv\in E(H):~|f(u)-f(v)|=i\}$, $k_i=|E_i|$ with $\sum^q_{i=1}k_i=q$, we call $f$ a \emph{$(k_i)^q_1$-graceful labeling}, where $(k_i)^q_1=(k_1,k_2,\dots ,k_q)$. Let $K_f=|\{k_i\neq 0:~i\in [1,q]\}|$ and $G_{race}(H)=\max\{K_f:~\textrm{all $(k_i)^q_1$-graceful labelings $f$ of $H$}\}$. And if $K_f=q$, $f$ is the proper graceful labeling introduced in \cite{Gallian2020}; if $K_f<q$, $f$ is a \emph{pseudo-graceful labeling}.
\end{defn}

\subsubsection{Magic-type of labelings, Magic-type of matching labelings}

\begin{defn}\label{defn:Wang-Xu-Yao-Shanghai-2017-00}
\cite{Wang-Xu-Yao-Shanghai-2017} If a $(p,q)$-graph $G$ admits a proper vertex labeling $f : V(G)\rightarrow [0,q]$ and a proper edge labeling $f : E(G)\rightarrow [1, q]$, and there exist two constants $k_1$ and $k_2$ such that $|f(u)+f(v)-f(uv)|=k_1$ or $|f(u)+f(v)-f(uv)|=k_2$ for any edge $uv\in E(G)$, and $f(E(G))=[1,q]$, we call $f$ a \emph{$(k_1, k_2)$-edge-magic graceful pseudo-labeling} (abbreviated as $(k_1, k_2)$-EMG pseudo-labeling).\qqed
\end{defn}

\begin{defn}\label{defn:Wang-Xu-Yao-Shanghai-2017-11}
\cite{Wang-Xu-Yao-Shanghai-2017} If a $(p,q)$-graph $G$ admits a proper total labeling $f : V(G)\cup E(G)\rightarrow [1, p+q]$, and there exist two non-negative constants $k_1$ and $k_2$ with respect to $k_1<k_2$, such that each edge $uv\in E(G)$ satisfies $|f(u)+f(v)-f(uv)|=k_1$ or $|f(u)+f(v)-f(uv)|=k_2$, we call $f$ a \emph{$(k_1, k_2)$-edge-magic graceful total labeling} (abbreviated as $(k_1, k_2)$-EMGTL).\qqed
\end{defn}

\begin{defn}\label{defn:EMGTLs}
\cite{Wang-Xu-Yao-Shanghai-2017} If a $(p,q)$-graph $G$ admits a proper total labeling $f : V(G)\cup E(G)\rightarrow [1, p+q]$, and there exist different non-negative integers $k_i\geq 0$ with $i\in [1,m]$, such that every edge $uv\in E(G)$ satisfies $|f(u)+f(v)-f(uv)|=k_i$ for some $k_i$, we call $f$ a \emph{$(k_i)^m_1$-edge-magic graceful total labeling} ($(k_i)^m_1$-EMGTL for short). \qqed
\end{defn}

\begin{defn}\label{defn:generalized-edge-magic}
\cite{su-yan-yao-2018} For a connected $(p,q)$-graph $G$, if (i) there are different integers $k_{i}\geq0$ with $i\in[1,m]$, such that $0\leq k_{1}< k_{2}< \cdots <k_{m}$; (ii) $G$ admits a total labeling $f$: $V(G)\cup E(G) \rightarrow [1, p+q]$, such that each edge $uv\in E(G)$ corresponds a magic constant $k_{i}$ and holds $|f(u)+f(v)-f(uv)|=k_{i}$ true. Then we call $f$ a $(k_{i})_{1}^{m}$-\emph{edge-magic graceful total labeling} of $G$, also, called a \emph{generalized edge-magic total labeling} ($(k_{i})_{1}^{m}$-EMGTL for short). The number $E_{mgtl}(G)=\min \{ m\}$ over all $(k_{i})_{1}^{m}$-edge-magic graceful total labelings of $G$ is called \emph{edge-magic total labeling number}, and write the number $K_{mgtl}(G)=\min_f \{k_{m_{0}}\}$ over all $(k_{i})_{1}^{m_{0}}$-edge-magic graceful total labelings $f$ of $G$ with $m_{0}=E_{mgtl}(G)$.\qqed
\end{defn}

\begin{defn}\label{defn:relaxed-Emt-labeling}
\cite{Yao-Sun-Zhang-Mu-Sun-Wang-Su-Zhang-Yang-Yang-2018arXiv} Let $f:V(G)\cup E(G)\rightarrow [1,p+q]$ be a total labeling of a $(p,q)$-graph $G$. If there is a constant $k$ such that $f(u)+f(uv)+f(v)=k$, and each edge $uv$ corresponds another edge $xy$ holding $f(uv)=|f(x)-f(y)|$ true, then we call $f$ a \emph{relaxed edge-magic total labeling}.\qqed
\end{defn}

\begin{defn}\label{defn:Oemt-labeling}
\cite{Yao-Sun-Zhang-Mu-Sun-Wang-Su-Zhang-Yang-Yang-2018arXiv} Suppose that a $(p,q)$-graph $G$ admits a vertex labeling $f:V(G) \rightarrow [0,2q-1]$ and an edge labeling $g:E(G)\rightarrow [1,2p-1]^o$. If there is a constant $k$ such that $f(u)+g(uv)+f(v)=k$ for each edge $uv\in E(G)$, and $g(E(G))=[1,2p-1]^o$, then we call $(f,g)$ an \emph{odd-edge-magic matching labeling}. \qqed
\end{defn}

\begin{defn}\label{defn:relaxed-Oemt-labeling}
\cite{Yao-Sun-Zhang-Mu-Sun-Wang-Su-Zhang-Yang-Yang-2018arXiv} Suppose that a $(p,q)$-graph $G$ admits a vertex labeling $f:V(G)\rightarrow [0,2q-1]$ and an edge labeling $g:E(G)\rightarrow [1,2q-1]^o$, and let $s(uv)=|f(u)-f(v)|-f(uv)$ for $uv\in E(G)$. If

(i) each edge $uv$ corresponds an edge $u\,'v\,'$ such that $g(uv)=|f(u\,')-f(v\,')|$;

(ii) and there exists a constant $k\,'$ such that each edge $xy$ has a matching edge $x\,'y\,'$ holding $s(xy)+s(x\,'y\,')=k\,'$ true;

(iii) there exists a constant $k$ such that $f(uv)+|f(u)-f(v)|=k$ for each edge $uv\in E(G)$. \\
Then we call $(f,g)$ an \emph{ee-difference odd-edge-magic matching labeling}.\qqed
\end{defn}

\begin{defn}\label{defn:Dgemm-labeling}
\cite{Yao-Sun-Zhang-Mu-Sun-Wang-Su-Zhang-Yang-Yang-2018arXiv} Suppose that a $(p,q)$-graph $G$ admits a vertex labeling $f:V(G)\rightarrow [0,p-1]$ and an edge labeling $g:E(G)\rightarrow [1,q]$, and let $s(uv)=|f(u)-f(v)|-g(uv)$ for $uv\in E(G)$. If there are:

(i) each edge $uv$ corresponds an edge $u\,'v\,'$ such that $g(uv)=|f(u\,')-f(v\,')|$ (resp. $g(uv)=p-|f(u\,')-f(v\,')|$);

(ii) and there exists a constant $k\,''$ such that each edge $xy$ has a matching edge $x\,'y\,'$ holding $s(xy)+s(x\,'y\,')=k\,''$ true;

(iii) there exists a constant $k$ such that $|f(u)-f(v)|+f(uv)=k$ for each edge $uv\in E(G)$;

(iv) there exists a constant $k\,'$ such that each edge $uv$ matches with one vertex $w$ such that $f(uv)+f(w)=k\,'$, and each vertex $z$ matches with one edge $xy$ such that $f(z)+f(xy)=k\,'$, except the \emph{singularity} $f(x_0)=0$.\\
Then we name $(f,g)$ an \emph{ee-difference graceful-magic matching labeling} (Dgemm-labeling) of $G$, and $G$ is called a \emph{Dgemm-graph}.\qqed
\end{defn}

\begin{defn}\label{defn:22-self-matchings}
\cite{Yao-Sun-Zhang-Mu-Sun-Wang-Su-Zhang-Yang-Yang-2018arXiv} Let $U$ be a universal graph, and its two subgraphs $G,H$ of $U$ hold $E(G)\cap E(H)=\emptyset $ and $V(U)=V(G)\cup V(H)$ true. If $E(U)=E(G)\cup E(H)$, both $G$ and $H$ are \emph{$U$-complementary} from each other, and moreover we call $G$ to be \emph{self-matching} (resp. \emph{self-complementary}) if $G$ is isomorphic to $H$, that is, $G\cong H$.\qqed
\end{defn}

\begin{rem}\label{rem:ABC-conjecture}
In Definition \ref{defn:22-self-matchings}, $G$ is a \emph{public-key}, and $H$ is a \emph{private-key}, $U$ is a topological authentication is $G$ and $H$ both are $U$-complementary from each other. We say that $G$ and $H$ both are \emph{$k$-vertex-coinciding} for $k=V(G)\cap V(H)$.\paralled
\end{rem}

\subsubsection{Edge-magic total-even/total-odd graceful labelings}

\begin{defn} \label{defn:edge-magic-total-even-odd-graceful-labeling}
\cite{Yao-Sun-Zhang-Yao-bi-labeling-2018} Let $G$ be a connected $(p,q)$-graph. If there are a fixed even-magic constant $k$ and a fixed odd-magic constant $k^*$:

(i) $G$ admits a labeling $f:~V(G)\cup E(G)\rightarrow [0,2(p+q-1)]^e$, such that each edge $uv\in E(G)$ holds $|f(u)+f(v)-f(uv)|=k$ true, then $f$ is called an \emph{edge-magic total-even graceful labeling};

(ii) $G$ admits a labeling $\alpha: V(G)\cup E(G)\rightarrow [1,2(p+q)-1]^o$, such that each edge $uv\in E(G)$ holds $|\alpha(x)+\alpha(y)-\alpha(xy)|=k^*$ true, then $f$ is called an \emph{edge-magic total-odd graceful labeling}.

Moreover, if $G$ is a bipartite graph with bipartition $(X,Y)$, such that $\max f(V(G))<\min f(E(G))$ and $\max f(X)<\min f(Y)$, we call $f$ a \emph{set-ordered super edge-magic total-even (resp. total-odd) graceful labeling}. \qqed
\end{defn}

\begin{defn} \label{defn:odd-even-edge-bi-labeling}
\cite{Yao-Sun-Zhang-Yao-bi-labeling-2018} Suppose that $E(H)=E(G)\cup E(T)$ with $E(G)\cap E(T)=\emptyset$ for a $(p,q)$-graph $H$ and its two subgraphs $G$ and $T$. If the $(p,q)$-graph $H$ admits a labeling $f:~V(H)\cup E(H)\rightarrow [0,2(q-1)]$ such that

(i) there exists a constant $k$, $|f(u)+f(v)-f(uv)|=k$ for each edge $uv\in E(G)$, and both $f(u)$ and $f(v)$ are even;

(ii) each edge $xy\in E(T)$ induces an odd $f^*(xy)=|f(x)-f(y)|$;

(iii) $|f(E(G))|+|f^*(E(T))|=q$. \\
Then we call $f$ an \emph{odd-even-edge bi-labeling}, and $k$ an \emph{odd-even-edge even-magic constant}.\qqed
\end{defn}

\begin{defn} \label{defn:00lingfang-jiang-magic-total-labeling}
\cite{Lingfang-Jiang-Yao-Xie-2010} A generalized $d_m$-uniformly lobster $T$ is a lobster and has a \emph{main path} $P=a_0a_1a_2\cdots a_n$, such that the distance $d(x, P)$ between any leaf $x$ and the main path in $T$ is just $m$, that is, $d(x, P)=m$. Let $p=|V(T)|$ and $q=|E(T)|$, so $q=p-1$. If there are two positive integers $\lambda, \mu$, $T$ admits a total labeling $f:V(T)\cup E(T) \rightarrow [0, p+q-1]$ holding (1) $f(a_i)+f(a_ia_{i+1})+f(a_{i+1})=\lambda$ for $a_ia_{i+1}\in E(P)$; and (2) for each path $P_m=a_i x_1 x_2\cdots x_m \subset T-E(P)$, we have
$$f(a_i)+f(a_ix_1)+\sum^m_{i=1} f(x_i)+\sum^{m-1}_{i=1} f(x_ix_{i+1})=\mu$$
then we call $f$ a $(\lambda, \mu)$-\emph{magic total labeling} of the generalized $d_m$-uniformly lobster $T$.\qqed
\end{defn}

\begin{thm} \label{defn:jiang-magic-total-labeling-result}
\cite{Lingfang-Jiang-Yao-Xie-2010} By Definition \ref{defn:00lingfang-jiang-magic-total-labeling}, each generalized $d_m$-uniformly lobster admits a \emph{$(\lambda, \mu)$-magic total labeling} for some pair of two positive integers $\lambda$ and $\mu$.
\end{thm}

\subsection{Parameterized harmonious labeling}

Gallian \cite{Gallian2020} extended the notion of harmoniousness to arbitrary finite Abelian groups as follows. Let $G$ be a $(p,q)$-graph and let $H$ be a finite Abelian group (under addition) of order $q$. Define $G$ to be \emph{$H$-harmonious} if there is an injection $f:V(G)\rightarrow H$ such that the resulting edge colors $f(xy)=f(x)+f(y)$ for edge $xy\in E(G)$ are distinct. Motivated from Gallian's $H$-harmonious labeling, we define:

\begin{defn} \label{defn:k-d-harmonious}
\cite{Yao-Zhang-Sun-Mu-Wang-Zhang2018} A \emph{$(k,d)$-harmonious labeling} of a $(p,q)$-graph $G$ is defined by a mapping $h:V(G)\rightarrow [0,k+(q-1)d]$ with $k,d\geq 1$, such that $f(x)\neq f(y)$ for any pair of vertices $x,y$ of $G$, $h(u)+h(v)~(\bmod^* qd)$ means that $h(uv)-k=[h(u)+h(v)-k](\bmod ~qd)$ for each edge $uv\in E(G)$, and the edge color set $h(E(G))=\{k,k+d,\dots ,k+(q-1)d\}$.\qqed
\end{defn}

\begin{figure}[h]
\centering
\includegraphics[width=16cm]{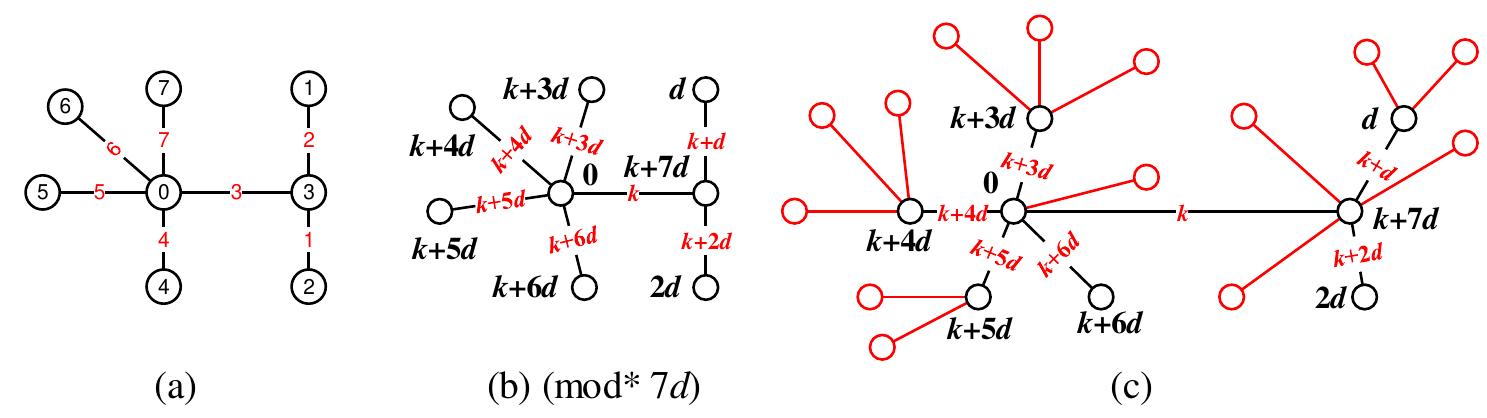}\\
\caption{\label{fig:example-a1}{\small (a) and (b) are a procedure of obtaining a $(k,d)$-harmonious labeling of a caterpillar $T$; (c) adding leaves to the caterpillar $T$.}}
\end{figure}

\begin{figure}[h]
\centering
\includegraphics[width=13.2cm]{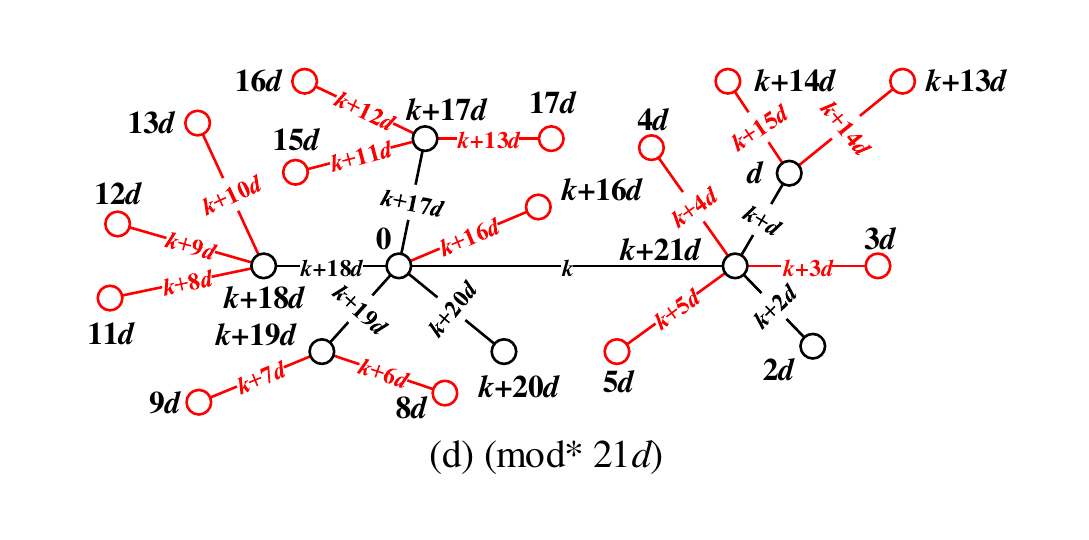}\\
\caption{\label{fig:example-a2}{\small A lobster $H$ obtained by adding leaves randomly to the tree $T$ shown in Fig.\ref{fig:example-a1} admits a $(k,d)$-harmonious labeling, cited from \cite{Yao-Zhang-Sun-Mu-Wang-Zhang2018}. }}
\end{figure}

\begin{rem}\label{rem:333333}
Harmonious graphs naturally arose in the study by Graham and Sloane \cite{Graham-Sloane1980} of modular versions of additive bases problems stemming from error-correcting codes.\paralled
\end{rem}

\begin{defn} \label{defn:more-harmonious-labelings}
\cite{Gallian2020} For a $(p,q)$-graph $G$ admitting a labeling $f$, we have $f(V(G))=\{f(u):u\in V(G)\}$ with $f(x)\neq f(y)$ for distinct vertices $x,y\in V(G)$ and $f(E(G))=\{f(uv):uv\in E(G)\}$. There are the following constraint conditions:
\begin{asparaenum}[(C-1) ]
\item \label{group:harmonious-V} $f(V(G))\subseteq [0, q-1]$;
\item \label{group:harmonious-V1} $f(V(G))\subseteq [0, q]$;
\item \label{group:harmonious-E} $f(uv)=f(u)+f(v)~ (\bmod ~q)$ and $f(E(G))=[0,q-1]$;
\item \label{group:harmonious-sequential} $f(uv)=f(u)+f(v)$ and $f(E(G))=[c, c+q-1]$;
\item \label{group:harmonious-v2q} $f(V(G))\subseteq [k-1,k+2q-1]$ with $k\geq 1$;
\item \label{group:harmonious-vv2q} $f(V(G))\subseteq [k-1,k+2q-2]$ with $k\geq 1$;
\item \label{group:harmonious-odd-even} $f(uv) = f(u) + f(v)+\epsilon$, where $\epsilon=0$ for even $f(u)+f(v)$, and $\epsilon=1$ for odd $f(u)+f(v)$;
\item \label{group:harmonious-e2q} $f(uv)=f(u)+f(v)~(\bmod ~2qk)$ with $k\geq 1$;
\item \label{group:harmonious-e-k-2q} $f(E(G))=[2k,2k+2q-2]^e$ with $k\geq 1$; and
\item \label{group:harmonious-o2q} $f(E(G))=[2k-1,2k+2q-3]^o$ with $k\geq 1$.
\end{asparaenum}
\textbf{We call $f$}:
\begin{asparaenum}[\textrm{Harmo}-1. ]
\item A \emph{harmonious labeling} if (C-\ref{group:harmonious-V}) and (C-\ref{group:harmonious-E}) hold true, it allows that there exists at most an edge $uv$ satisfies $f(u)=f(v)$ if $G$ is a tree.
\item A \emph{strongly c-harmonious labeling} if (C-\ref{group:harmonious-V}) and (C-\ref{group:harmonious-sequential}) hold true.
\item An \emph{even sequential harmonious labeling} if (C-\ref{group:harmonious-v2q}) with $k=1$, (C-\ref{group:harmonious-odd-even}), (C-\ref{group:harmonious-e2q}) with $k=1$ and (C-\ref{group:harmonious-e-k-2q}) with $k=1$ hold true.
\item A \emph{$k$-even sequential harmonious} if (C-\ref{group:harmonious-v2q}), (C-\ref{group:harmonious-odd-even}), (C-\ref{group:harmonious-e2q}) with $k\geq 1$ and (C-\ref{group:harmonious-e-k-2q}) with $k\geq 1$ hold true.
\item An \emph{odd-harmonious labeling} if (C-\ref{group:harmonious-vv2q}) with $k=1$ and (C-\ref{group:harmonious-o2q}) with $k=1$ hold true.
\item A \emph{strongly odd-harmonious labeling} if (C-\ref{group:harmonious-V1}) and (C-\ref{group:harmonious-o2q}) with $k=1$ hold true.
\item An \emph{even-harmonious labeling} if (C-\ref{group:harmonious-v2q}) with $k=1$ and (C-\ref{group:harmonious-e-k-2q}) with $k=0$ hold true.\qqed
\end{asparaenum}
\end{defn}

\begin{defn} \label{defn:even-odd-harmonious-matching-pair}
\cite{Yao-Zhang-Sun-Mu-Wang-Zhang2018} Suppose that each graph $T_i$ of $q$ edges with $i=1,2$ admits an (resp. even)odd-harmonious labeling $f_i$ defined in Definition \ref{defn:more-harmonious-labelings}. If $f_1(V(T_1))\cup f_2(V(T_2))=[0,2q]$, and $|f_1(V(T_1))\cap f_2(V(T_2))|=s$, we say $T_1$ and $T_2$ to be an \emph{(resp. even)odd-harmonious $s$-matching pair}, the vertex-coinciding graph $\odot_s\langle T_1,T_2\rangle$ is a \emph{topological authentication}.\qqed
\end{defn}

\begin{thm} \label{defn:various-harmonious}
\cite{Yao-Zhang-Sun-Mu-Wang-Zhang2018} A tree $T$ has its own vertex set bipartition $V(T)=X\cup Y$ with $X\cap Y=\emptyset$, where $X=\{x_1,x_2,\dots x_s\}$ and $Y=\{y_1,y_2,\dots y_t\}$, $p=|V(T)|=s+t$. Suppose that $T$ admits a set-ordered graceful labeling $f$ such that $f(x_i)=i-1$ with $i\in [1,s]$, and $f(y_j)=s-1+j$ with $j\in [1,t]$, so $\max f(X)<\min f(Y)$. Hence, this tree $T$ admits each of the following harmonious labelings:
\begin{asparaenum}[(E-1) ]
\item A \emph{harmonious labeling} $g_1$ is defined as: $g_1(x_i)=f(x_i)$ with $i\in [1,s]$, and $g_1(y_j)=f(y_{t-j+1})$ with $j\in [1,t]$, and $g_1(x_iy_j)=g_1(x_i)+g_1(y_j)=f(x_i)+f(y_{t-j+1})=f(y_j)-f(x_i)+2f(x_i)+f(y_{t-j+1})-f(y_j)$.

\item An \emph{even-harmonious labeling} $g_2$ is defined in the way: $g_2(x_i)=2f(x_i)$ with $i\in [1,s]$, and $g_2(y_j)=2f(y_{t-j+1})$ with $j\in [1,t]$.

\item An \emph{odd-harmonious labeling} $g_3$ is defined as: $g_3(x_i)=2f(x_i)$ with $i\in [1,s]$, and $g_3(y_j)=2f(y_{t-j+1})-1$ with $j\in [1,t]$.

\item A \emph{$k$-even sequential harmonious labeling} $g_4$ is defined by: For $k\geq 1$, $g_4(x_i)=2k\cdot f(x_i)$ with $i\in [1,s]$, and $g_4(y_j)=2k\cdot f(y_{t-j+1})$ with $j\in [1,t]$.

\item A \emph{strongly c-harmonious labeling} $g_5$ is defined by $g_5(x_i)=f(x_i)$ with $i\in [1,s]$, and $g_5(y_j)=f(y_{t-j+1})$ with $j\in [1,t]$.

\item Under the condition $|s-t|=1$, a \emph{strongly odd-harmonious labeling} $g_6$ is defined as: $g_6(x_i)=2f(x_i)$ with $i\in [1,s]$, and $g_6(y_j)=2f(y_{t-j+1})-1$ with $j\in [1,t]$.

\item A \emph{$(k,d)$-harmonious labeling} $g_7$ is defined by $g_7(x_i)=f(x_i)\cdot d$ with $i\in [1,s]$, and $g_7(y_j)=k+f(y_{t-j+1})\cdot d$ with $j\in [1,t]$.\qqed
\end{asparaenum}
\end{thm}

\begin{defn} \label{defn:k-d-harmonious}
\cite{Yao-Zhang-Sun-Mu-Wang-Zhang2018} A \emph{$(k,d)$-harmonious labeling} $h$ of a $(p,q)$-graph $G$ is defined by a mapping $h:V(G)\rightarrow [0,k+(q-1)d]$ with $k,d\geq 1$, such that $h(x)\neq h(y)$ for any pair of vertices $x,y$ of $G$, $h(u)+h(v)(\bmod^* ~qd)$ means that $h(uv)-k=[h(u)+h(v)-k](\bmod ~qd)$ for each edge $uv\in E(G)$, and the edge color set $h(E(G))=\{k,k+d,\dots, k+(q-1)d\}$.\qqed
\end{defn}

\subsection{Topsnut-matchings}

Let $G$ be a $(p,q)$-graph, and let $f$ be a coloring (resp. labeling) defined as $f:X\rightarrow M$, where $X$ is a subset of $V(G)\cup E(G)$, and $M\subset Z^0$. We call this colored graph $G$ a \emph{Topsnut-gpw}. Let $O_r$ be an operation defined on graphs. If each graph $G_{n}$ of a graph sequence $\{G_{n}\}^m_{n=0}$ can be produced from $G_{n-1}$ by the operation $O_r$, and $G_0$ is not generated by doing the operation $O_r$ to some graph, then we call $G_{n}$ a \emph{recursive $O_r$-graph} and $G_0$ a \emph{$O_r$-root}, as well as $O_r$ a \emph{recursive operation}. Let $l$ stand up a lock (authentication), $k$ be a key (password), and $h$ be the password rule (a procedure of authentication). The function $l=h(k)$ represents the state of ``a key $k$ opens a lock $l$ through the password rule $h$, and call directly $l=h(k)$ a \emph{Topsnut-gpw}.

It is not hard to show that each lobster admits a $(k,d)$-harmonious labeling with $k\geq 1$ and $d\geq 2$. Thereby, we can use $(k,d)$-harmonious labelings of graphs to design more complicated Topsnut-gpws. Let $\{(k_i,d_i)\}_1^n$ be a sequence obtained from two sequences $\{k_1,k_2,\dots ,k_n\}$ and $\{d_1,d_2,\dots ,d_n\}$. We have a Topsnut-gpw set $$F(k_i,d_i)=\{G:~G\textrm{ admits a $(k_i,d_i)$-harmonious labeling}\}$$ for each matching $(k_i,d_i)$ with $i\in [1,n]$. There are random \emph{Topsnut-chains} $G_{i_1},G_{i_2}, \dots ,G_{i_m}$ with $G_{i_j}\in \bigcup ^n_{i=1}F(k_i,d_i)$, such that we can apply them to encrypt more objects once time.

One want to look for all odd-graceful labelings of a $(p,q)$-graph $G$ admitting an odd-graceful labeling. An odd-graceful labeling of the Topsnut-gpw $G$ may determine some matching Topsnut-gpws $H$ admitting matching odd-graceful labelings:

\begin{defn}\label{defn:matching-odd-graceful-labeling}
\cite{Yao-Zhang-Sun-Mu-Wang-Zhang2018} A \emph{$k$-matching odd-graceful labeling} $g$ of an odd-graceful labeling $f$ of a $(p,q)$-graph $G$ is a vertex labeling defined on another graph $H$ as: $g:V(H) \rightarrow [0,2q-1]$, every edge $uv\in E(H)$ has its color $g(uv)=|g(u)-g(v)|$ holding $g(E(H))=[1, 2q-1]^o$ true, such that $f(V(G))\cup g(V(H))=[0,2q-1]$ with $|f(V(G))\cap g(V(H))|=k\geq 0$. We call $H$ admitting a $k$-matching odd-graceful labeling as a \emph{Topsnut-matching} of $G$. \qqed
\end{defn}

\begin{figure}[h]
\centering
\includegraphics[width=16.4cm]{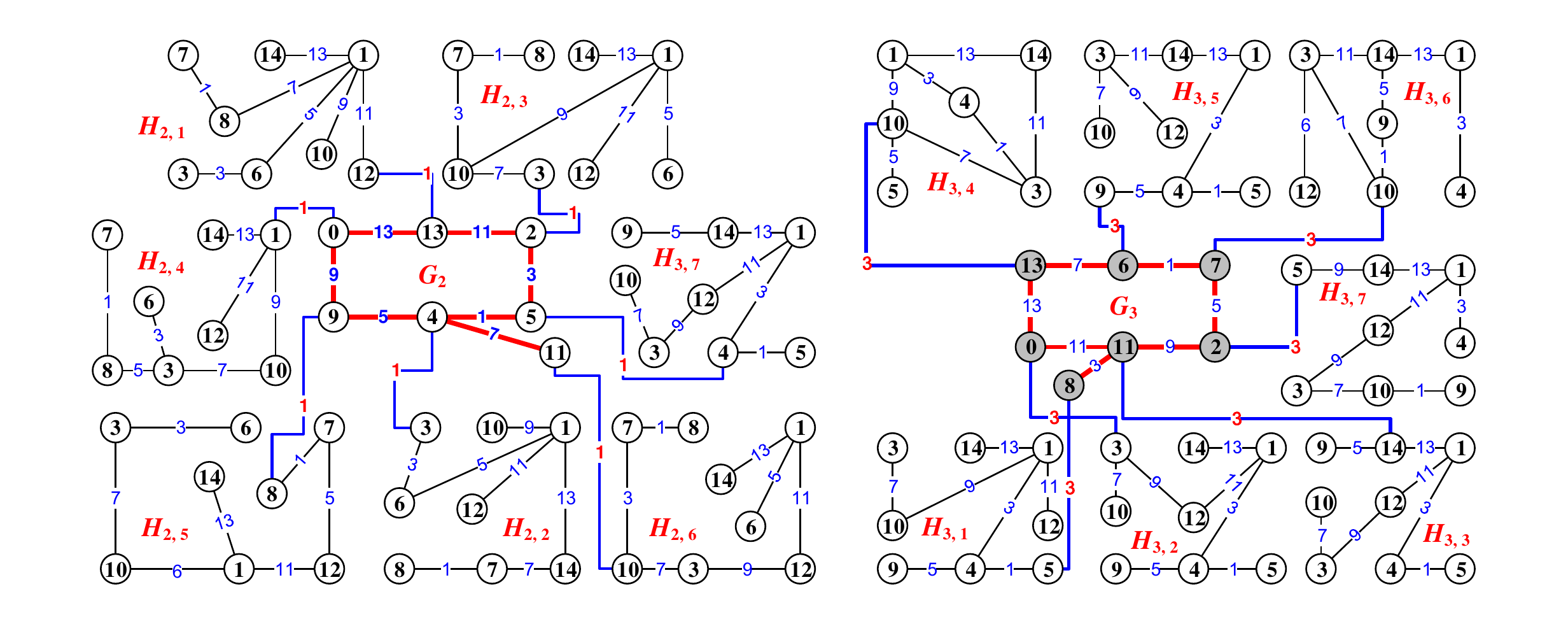}\\
\caption{\label{fig:odd-every-vertex-matchings}{\small Two odd-graceful Topsnut-matching teams $\odot_1\langle G_2,H_{2,i}\rangle ^7_1$ and $\odot_1\langle G_3,H_{3,i}\rangle ^7_1$, cited from \cite{Wang-Yao-2021-SCI}.}}
\end{figure}

In general, a Topsnut-gpw $G$ made by an odd-graceful labeling may have two or more Topsnut-matchings. Let $G$ be an (a pan-)odd-graceful graph with vertices $v_1,v_2,\dots, v_p$. If each vertex $v_i$ of $G$ matches with an (a pan-)odd-graceful graph $H_i$ with $i\in [1,p]$ such that there exists an odd-graceful Topsnut-matching $\odot_1\langle G,H_i\rangle ^p_1$ obtained by vertex-coinciding the vertex $v_i$ with some vertex of $H_i$ into one, these two vertices have been labeled with the same labels. We say $\odot_1\langle G,H_i\rangle ^p_1$ a \emph{(pan-)odd-graceful Topsnut-matching team} (see an example shown in Fig.\ref{fig:odd-every-vertex-matchings}). Moreover, $\odot_1\langle G,H_i\rangle ^p_1$ is called a \emph{uniformly (pan-)odd-gracefully Topsnut-matching team} if $H_i\cong H_j$ for $i\neq j$.

\begin{thm} \label{thm:Topsnut-matching-team}
\cite{Wang-Yao-2021-SCI} Each caterpillar $G$ of $p$ vertices has an (a \emph{pan}-)\emph{odd-graceful Topsnut-matching team} $\odot_1\langle G,H_i\rangle ^p_1$.
\end{thm}

\begin{defn}\label{defn:equential-odd-graceful-labeling}
\cite{Yao-Zhang-Sun-Mu-Wang-Zhang2018} A \emph{$k$-sequential odd-graceful labeling} $h: V(G)\rightarrow [k,2q-1+k]$ such that each induced edge color $h(uv)=|h(u)-h(v)|$ for $uv \in E(G)$ holds $h(E(G))=[1,2q-1]^o$.\qqed
\end{defn}

\begin{thm}\label{thm:two-labelings-magic}
\cite{Yao-Liu-Yao-2017} Let $G$ be a bipartite graph with the bipartition $(X,Y)$, and let $f$ be a labeling $V(G)\rightarrow \{0,1,2,\dots\}$ such that $f(u)\neq f(v)$ for all distinct vertices $u,v\in V(G)$, and $f(xy)=f(y)-f(x)\geq 1$ for each edge $xy\in E(G)$ with $x\in X$ and $y\in Y$. Write $f(V(G))=\{f(w):w\in V(G)\}$. Then we have

$(i)$ The bipartite graph $G$ admits a labeling $g_1$ induced by $f$ such that $g_1(u)\neq g_1(v)$ for distinct vertices $u,v\in V(G)$, and $g_1(x)+g_1(y)=\max f(V(G))+\min f(V(G))-f(xy)$ for each edge $xy\in E(G)$ with $x\in X$ and $y\in Y$.

$(ii)$ For all values of positive integers $d$ and $k$, the bipartite graph $G$ admits a labeling $g_2$ such that $g_2(u)\neq g_2(v)$ for all distinct vertices $u,v\in V(G)$, and $g_2(y)-g_2(x)=k+d\cdot f(xy)$ for each edge $xy\in E(G)$ with $x\in X$ and $y\in Y$.

$(iii)$ There are a labeling $g_3$ and a constant $\lambda>0$ such that $g_3(u)\neq g_3(v)$ for all distinct vertices $u,v\in V(G)$, and $g_3(x)+g_3(xy)+g_3(y)=\lambda$ for each edge $xy\in E(G)$.

$(iv)$ The bipartite graph $G$ admits a labeling $g_4$ such that $g_4(u)\neq g_4(v)$ for all distinct vertices $u,v\in V(G)$, and $g_4(x)+g_4(y)=k+d\cdot \big [\max f(V(G))+\min f(V(G))-f(xy)\big ]$ for each edge $xy\in E(G)$ with $x\in X$ and $y\in Y$.
\end{thm}

\begin{rem}\label{rem:vertex-identified-graph}
Suppose that a $(p,q)$-graph $G$ contains two subgraphs $G_1$ and $G_2$ such that $V(G_1)\cap V(G_2)=\{w_1,w_2, \dots, w_n\}$ and $E(G)=E(G_1)\cup E(G_2)$ with $E(G_1)\cap E(G_2)=\emptyset$, we by $G=\odot_n \langle G_1,G_2\rangle $ denote $G$, and call $G$ a \emph{$n$-vertex-coincided graph} ($n$-vi-$(p,q)$-graph). Moreover, we call $G$ a \emph{uniformly $n$-vertex-coincided $(p,q)$-graph} (uniformly $n$-vi-$(p,q)$-graph) if $|E(G_1)|=|E(G_2)|$ in \cite{Wang-Xu-Yao-Complementary-2017}. Obviously, every tree is a $1$-vi-tree.\paralled
\end{rem}

\begin{defn}\label{defn:CEOG-labeling}
\cite{Wang-Xu-Yao-Complementary-2017} Suppose that a $1$-vi-$(p,q)$-graph $G=\odot_1 \langle G_1,G_2\rangle $ admits a mapping $f$: $V(G)\rightarrow [0, q]$ such that

(i) $f(x)\neq f(y)$ for any pair of vertices $x,y\in V(G)$;

(ii) $f$ is an odd-graceful labeling of $G_1$;

(iii) $f:V(G_2)\rightarrow [0,2|E(G_2)|]$ such that the edge color set $f(E(G_2))=\{f(uv)=f(u)+f(v)~(\bmod~2|E(G_2)|):~uv\in E(G_2)\}=[1, 2|E(G_2)|-1]^{o}$.\\
Then $G$ is called a \emph{complementary edge-odd-graceful $1$-vi-$(p,q)$-graph} (a CEOG $1$-vi-graph for short), $f$ a \emph{complementary edge-odd-graceful labeling} (a CEOG-labeling) of $G$.\qqed
\end{defn}

\begin{rem}\label{rem:333333}
We have the following particular graphs:
\begin{asparaenum}[CEOG-1. ]
\item \cite{Wang-Xu-Yao-Complementary-2017} If a uniformly $1$-vi-$(p,q)$-graph $G=\odot_1 \langle G_1,G_2\rangle $ admits a CEOG-labeling $f$, where $G_1$ is an odd-graceful graph. Then $G_1$ is called an \emph{E-lock-graph}, and $G_2$ is called an \emph{E-key-graph} of $G_1$.

\item \cite{Wang-Xu-Yao-Complementary-2017} Let $G_j$ be a $(p_j,q_j)$-graph with $j=1,2$. A graph $G$ is obtained by identifying a vertex $x_{i,1}$ of $G_1$ with another vertex $x_{i,2}$ of $G_2$ into one vertex $x_i$ (denoted as $x_i=x_{i,1}\odot x_{i,2}$ hereafter) for $i\in [1,m]$, and call these vertices $x_1$, $x_2$, $\dots$, $x_m$ the \emph{identification-vertices}. Write $G=\odot_m \langle G_1, G_2\rangle$, and called it an \emph{$m$-identification graph}. So, $G=\odot_m \langle G_1, G_2\rangle$ has $p_1+p_2-m$ vertices and $q_1+q_2$ edges. Conversely, we can vertex-split the identification-vertex $x_i$ into two vertices $x_{i,1}$ and $x_{i,2}$ (called \emph{splitting-vertices}), so $G$ can be split into two vertex-disjoint subgraphs $G_1$ and $G_2$.
\item \cite{Wang-Xu-Yao-Complementary-2017} For two connected $(p_i,q)$-graphs $G_i$ with $i=1,2$, and $p=p_1+p_2-2$, if the $2$-identification $(p,q)$-graph $G=\odot_2 \langle G_1, G_2\rangle$ admits a mapping $f$: $V(G)\rightarrow [0, q-1]$ such that

\quad (i) $f(x)\neq f(y)$ for any pair of vertices $x,y\in V(G)$;

\quad (ii) $f$ is just an odd-elegant labeling of $G_i$ with $i=1,2$.

\quad Then we say $G$ a \emph{twin odd-elegant graph} (a TOE-graph), $f$ a \emph{TOE-labeling}, $G_1$ a \emph{TOE-source graph}, $G_2$ a \emph{TOE-associated graph}, and $(G_1, G_2)$ a \emph{TOE-matching pair}.
\item \cite{Wang-Xu-Yao-Complementary-2017} Let $T_i$ be a tree of order $n$ with $i=1,2$. If the 2-identification tree $G_T=\odot_2 \langle T_1, T_2\rangle$ admits a mapping $f:$ $V(G_T)\rightarrow [0, 2n-3]$ such that

\quad (i) $f(x)\neq f(y)$ for any pair of vertices $x,y\in V(G_T)$;

\quad (ii) $f$ is an odd-elegant labeling of $T_i$ with $i=1,2$.

\quad Then $G_T$ is called a \emph{twin odd-elegant tree} (a TOE-tree), $f$ a \emph{TOE-labeling}, and $T_1$ a \emph{TOE-source tree}, and $T_2$ a \emph{TOE-associated tree} of $T_1$. Furthermore, we say $T_1$ a \emph{TOE-source self-associated tree} if $T_1$ is isomorphic to $T_2$.\paralled
\end{asparaenum}
\end{rem}

\begin{defn}\label{defn:Module-k-super-graceful}
\cite{Zhang-Sun-Bing-Yao-IMCEC-2018} Suppose that a $(p,q)$-graph $G$ admits a mapping $f:V(G)\cup E(G) \rightarrow [1,p + q]$ such that $f(u)\neq f(v)$ for distinct vertices $u,v\in V(G)$, and each edge color $f(uv)$ for $uv\in E(G)$ is defined as $f(uv)=|f(u)-f(v)|+k~(\bmod~p+q)$ for some integer $k\in Z^0\setminus \{0\}$, and the set of all edge colors is equal to $[1, p + q]$, we call $f$ a \emph{Module-$k$ super graceful labeling}.\qqed
\end{defn}

\begin{figure}[h]
\centering
\includegraphics[width=12cm]{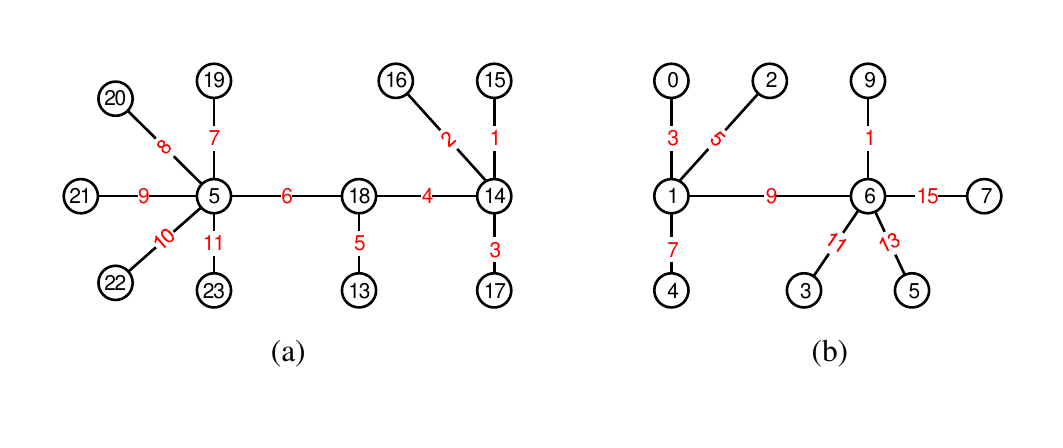}
\caption{\label{fig:module-k}{\small (a) A module-$0$ super-graceful labeling; (b) a module-$2$ odd-elegant labeling.}}
\end{figure}

\begin{defn}\label{defn:Module-k-odd-elegant}
\cite{Zhang-Sun-Bing-Yao-IMCEC-2018} Suppose that a $(p,q)$-graph $G$ admits a mapping $f:V(G) \rightarrow [0,2q-1]$ such that $f(u)\neq f(v)$ for distinct vertices $u,v\in V(G)$, and each edge color $f(uv)$ for $uv\in E(G)$ is defined as $f(uv)=f(u)+f(v)+k~(\bmod~2q)$ for some integer $k\in Z^0\setminus \{0\}$, and the set of all edge colors is equal to $[1, 2q-1]^{o}$ or $[2, 2q]^{e}$, we call $f$ a \emph{Module-$k$ odd-elegant labeling} and $G$ an \emph{Module-$k$ odd-elegant graph}. Moreover, if $(V_1,V_2)$ is the bipartition of $G$, and the Module-$k$ odd-elegant labeling $f$ obeys $\max\{f(u):~u\in V_1\}<\min\{f(v):~v\in V_2\}$, then we call $f$ a \emph{set-ordered Module-$k$ odd-elegant labeling}, and write this case as $\max f(V_1)<\min f(V_2)$.\qqed
\end{defn}

\begin{defn}\label{defn:edge-module-k-odd-graceful}
\cite{Zhang-Sun-Bing-Yao-IMCEC-2018} Let $G$ be a $(p,q)$-graph and admit a mapping $g:V(G) \rightarrow [0,2q-1]$ such that $g(u)\neq g(v)$ for distinct vertices $u,v\in V(G)$, and each edge color $g(uv)$ for $uv\in E(G)$ is defined as $g(uv)=|g(u)-g(v)|+k~(\bmod~2q)$ for even $k$.

(1) If the edge color set $g(E(G))=\{g(uv):~uv\in E(G)\}=[1, 2q-1]^{o}$, we call $g$ an \emph{edge module-k odd-graceful labeling} (Em(k)og-labeling) and $G$ an \emph{edge module-k odd-graceful graph} (Em(k)og-graph). Especially, $G$ is \emph{odd-graceful} when $k=0$.

(2) If $G$ is a bipartite graph with its bipartition $(V_1,V_2)$, such that an \emph{Em(k)og-labeling} $g$ holding $\max\{g(u):~u\in
V_1\}<\min\{g(v):~v\in V_2\}$, then we call $g$ a \emph{set-ordered Em(k)og-labeling}, and write this case as $\max g(V_1)<\min g(V_2)$.\qqed
\end{defn}

\begin{defn}\label{defn:module-k-twin-odd-graceful-graph}
\cite{Zhang-Sun-Bing-Yao-IMCEC-2018} Let $G_j$ be a $(p_j,q_j)$-graph with $j=1,2$, and let $p=p_1+p_2-1$ and $q=q_1+q_2$. If the $(p,q)$-graph $G=\odot \langle G_1,G_2\rangle$ admits a mapping $f$: $V(G)\rightarrow [0, q]$ such that

(i) $f(x)\neq f(y)$ for any pair of vertices $x,y\in V(G)$;

(ii) $f$ is an Em(k)og-labeling of $G_1$;

(iii) the edge color set $$f(E(G_i))=\{f(uv)=|f(u)-f(v)|+k~(\bmod~2q):~ uv\in E(G_i)\}=[1, 2q-1]^{o}$$ for some even integer $k$ and $i=1,2$.\\
Then we call $G$ a \emph{twin edge module-k odd-graceful graph} (twin Em(k)og-graph), $f$ an \emph{Em(k)og-labeling}, $G_1$ a \emph{source graph}, and $G_2$ an \emph{associated graph}.\qqed
\end{defn}

\begin{defn}\label{defn:edge-ordered-graceful-labeling}
\cite{yarong-mu-bing-yao-IMCEC-2018} Suppose that a connected $(p,q)$-graph $G$ admits a set-ordered graceful labeling $f$ defined in Definition \ref{defn:basic-W-type-labelings} and $V(G)=X\cup Y$ such that there exist an integer $M>0$ and another labeling $g$ defined by $g(x)=f(x)$ for $x\in X$ and $g(y)=f(y)+M$ with $y\in Y$. If $g(E(G))=[1,s]\cup [s+1+M,q]$, then $f$ is called an \emph{edge-ordered graceful labeling} of $G$, and $G$ an \emph{edge-ordered graceful graph}.\qqed
\end{defn}

\begin{defn}\label{defn:edge-ordered-odd-graceful-labeling}
\cite{yarong-mu-bing-yao-IMCEC-2018} Suppose that a connected $(p,q)$-graph $G$ admits a set-ordered graceful labeling $f$ defined in Definition \ref{defn:basic-W-type-labelings} and $V(G)=X\cup Y$ such that there exist an integer $M>0$ and another labeling $g$ defined by $g(x)=f(x)$ for $x\in X$ and $g(y)=f(y)+M$ with $y\in Y$. If $g(E(G))=[1,2s-1]\cup [2(s+M)-1,2q-1]$, then we call $f$ an \emph{edge-ordered odd-graceful labeling} of $G$, and $G$ an \emph{edge-ordered odd-graceful graph}.\qqed
\end{defn}

\begin{defn} \label{defn:3-new-all-odd-graceful}
\cite{SH-ZXH-YB-2017-4, SH-ZXH-YB-2018-5} Let $G$ be a $(p, q)$-graph.
\begin{asparaenum}[(1) ]
\item If there is an injection of~$f:V (G) \rightarrow [0,2q+1]^o$ such that the edge color set $\{f(uv) =|f(u)-f(v)|-1:~ uv\in E(G)\}=[1, 2q-1]^o$, $f$ is called an \emph{all odd-graceful labeling} of $G$.

\item $G$ is called a \emph{super odd-graceful graph} if there is a bijective function $f$ with $f(V(G)\cup E(G))=[1,2(p+q)-1]^o$, such that edge color $f(uv) =|f(u)-f(v)|-1$ for each edge $uv\in E(G)$.

\item $G$ is called a \emph{super odd-elegant graph} if there is a bijective function $f$ with $f(V(G)\cup E(G))=[1,2(p+q)-1]^o$, such that each edge $uv\in E(G)$ holds $f(uv) =f(u)+f(v)~(\bmod~M$) with $M=2p+3q-1$.\qqed
\end{asparaenum}
\end{defn}

\begin{defn}\label{defn:new-edge-odd-graceful-combinator}
\cite{Hongyu-Wang-Bing-Yao-Multiple-2020} Let $G$ be a $(p, q)$-graph.
\begin{asparaenum}[\textrm{New}-1. ]
\item \label{defn:edge-odd-graceful-labeling} Suppose that $G$ admits a mapping $f: V(G)\rightarrow[0,2q]$ such that $f(u)\neq f(v)$ for distinct vertices $u,v\in V(G)$, and the edge color set $\{f(uv)=|f(u)-f(v)|: uv\in E(G)\}=[1, 2q-1]^o$, then we call $f$ an \emph{edge-odd-graceful labeling}, and $G$ an \emph{edge-odd-graceful graph}.
\item \label{defn:edge-odd-multiple-labeling-EOM-Private-key} Suppose that the graph $G=H\odot \langle G_k\rangle^n_{k=1}$ obtained by coinciding a vertex $u_k$ of each graph $G_k$ with a vertex $x_k\in V(H)=\{x_1,x_2,\dots ,x_n\}$ is a multiple-vertices graph and admits a mapping $f:V(G)\rightarrow[0,q]$, such that

(i) $f(x)\neq f(y)$ for any pair of vertices $x, y\in V(G)$;

(ii) $f$ is an edge-odd-graceful labeling of $H$;

(iii) $f$ is an edge-odd-graceful labeling of $G_k$ for $k\in[1,n]$.

\quad Then we call $G$ an \emph{edge-odd-multiple graph}, $f$ an \emph{edge-odd multiple labeling}, and $H$ a \emph{base}, and each $G_k$ with $k\in [1,n]$ a \emph{seed}.\qqed
\end{asparaenum}
\end{defn}

\begin{defn}\label{defn:edge-odd-graceful-labeling-arxiv}
\cite{Yao-Zhang-Sun-Mu-Sun-Wang-Wang-Ma-Su-Yang-Yang-Zhang-2018arXiv} A $(p,q)$-graph $G$ admits an \emph{edge-odd-graceful total labeling} $h$ when $h:V(G)\rightarrow [0,q-1]$ and $h:E(G)\rightarrow [1,2q-1]^o$ such that $\{h(u)+h(uv)+h(v):uv\in E(G)\}=[a,b]$ with $b-a+1=q$.\qqed
\end{defn}

\begin{defn}\label{defn:set-ordered-edge-odd-multiple}
(i) \cite{Hongyu-Wang-Bing-Yao-Multiple-2020} If $f$ holds the edge color set $\{f(uv)=|f(u)-f(v)|: uv\in E(G)\}\neq [1, 2q-1]^o$ in Definition \ref{defn:edge-odd-graceful-labeling-arxiv}, then we call $f$ a \emph{blemished (edge-)odd-graceful labeling}, and $G$ a \emph{blemished (edge-)odd-graceful graph}.

(ii) \cite{Hongyu-Wang-Bing-Yao-Multiple-2020} In Definition \ref{defn:new-edge-odd-graceful-combinator}, if each graph $G_k$ with $k\in [1, |V(H)|]$ admits a set-ordered edge-odd-graceful labeling, then $G=H\odot \langle G_k\rangle^n_{k=1}$ with $n=|V(H)|$ is called an \emph{set-ordered multiple-vertices $(p,q)$-graph}, so we call $G$ a \emph{set-ordered-uniformly mv-$(p,q)$-graph}, $f$ a \emph{set-ordered edge-odd multiple labeling} of $G$.
\end{defn}

\subsection{Labelings and perfect matchings}

\subsubsection{Labelings on graphs having perfect matchings}

\begin{defn} \label{defn:New-graphical-passwords}
\cite{SH-ZXH-YB-2018-5} Let $G$ ba a $(p, q)$-graph with a perfect matching $M$, and ``AM''=arithmetic matching.
\begin{asparaenum}[M-1. ]
\item If $G$ admits a super graceful labeling $h$ holding $h(u)+h(v)=3q+2$ for each matching edge $uv \in M$, then $h$ is called an \emph{AM-super graceful labeling}.

\item If there is an injection $h:V (G) \rightarrow [0,2q+1]^o$ such that the edge color set $\{h(uv) =|h(u)-h(v)|-1: uv\in E(G)\}=[1, 2q-1]^o$, then $h$ is called an \emph{all-odd graceful labeling} of $G$. Moreover, if $h$ holds $h(u)+h(v)=2(q+1)$ for each matching edge $uv \in M$, we call $h$ an \emph{AM-all-odd graceful labeling}.

 \item $G$ is called a \emph{super odd-graceful labeling} if $h(V(G)\cup E(G))=[1,2(4q+1)]^o$ with $h(uv) =|h(u)-h(v)|-1$ for each edge $uv\in E(G)$. Moreover, if $h$ holds $h(u)+h(v)=2(3q+1)$ for each matching edge $uv \in M$, we call $h$ an \emph{AM-super odd-graceful labeling}.

 \item If $G$ is a bipartite graph with the bipartition $(X,Y)$, and $G$ is said to be a \emph{super odd-elegant graph} if there is a mapping $h$ holding $h(V(T)\cup E(T))=[1,2(4q+1)]^o$ with $h(uv) =h(u)+h(v)$ ($\bmod~M^*$) for $uv\in E(T)$, where the constant $M^*=2p+3q+(|X|-|Y|)$. Moreover, $h$ is called an \emph{AM-super odd-elegant labeling} if $h$ holds $|h(u)-h(v)|=q+1$ for each matching edge $uv \in M$.

 \item If $G$ admits one modulo three graceful labeling $h$ holding $h(u)+h(v)=3q-2$ for each matching edge $uv \in M$, then we call $h$ an \emph{AM-one modulo three graceful labeling}.

 \item If $G$ admits a Skolem-graceful labeling $h$ holding $h(u)+h(v)=q+2$ for each matching edge $uv \in M$, then $h$ is called an \emph{AM-Skolem-graceful labeling}.

 \item If $G$ admits an odd-even graceful labeling $h$ holding $h(u)+h(v)=2(q+1)$ for each matching edge $uv \in M$, then $h$ is called an \emph{AM-odd-even graceful labeling}.

 \item If $G$ admits a $k$-graceful labeling $h$ holding $h(u)+h(v)=k+q-1$ for each edge $uv \in M$, then $h$ is called an \emph{AM-$k$-graceful labeling}.\qqed
\end{asparaenum}
\end{defn}

\begin{figure}[h]
\centering
\includegraphics[width=16.4cm]{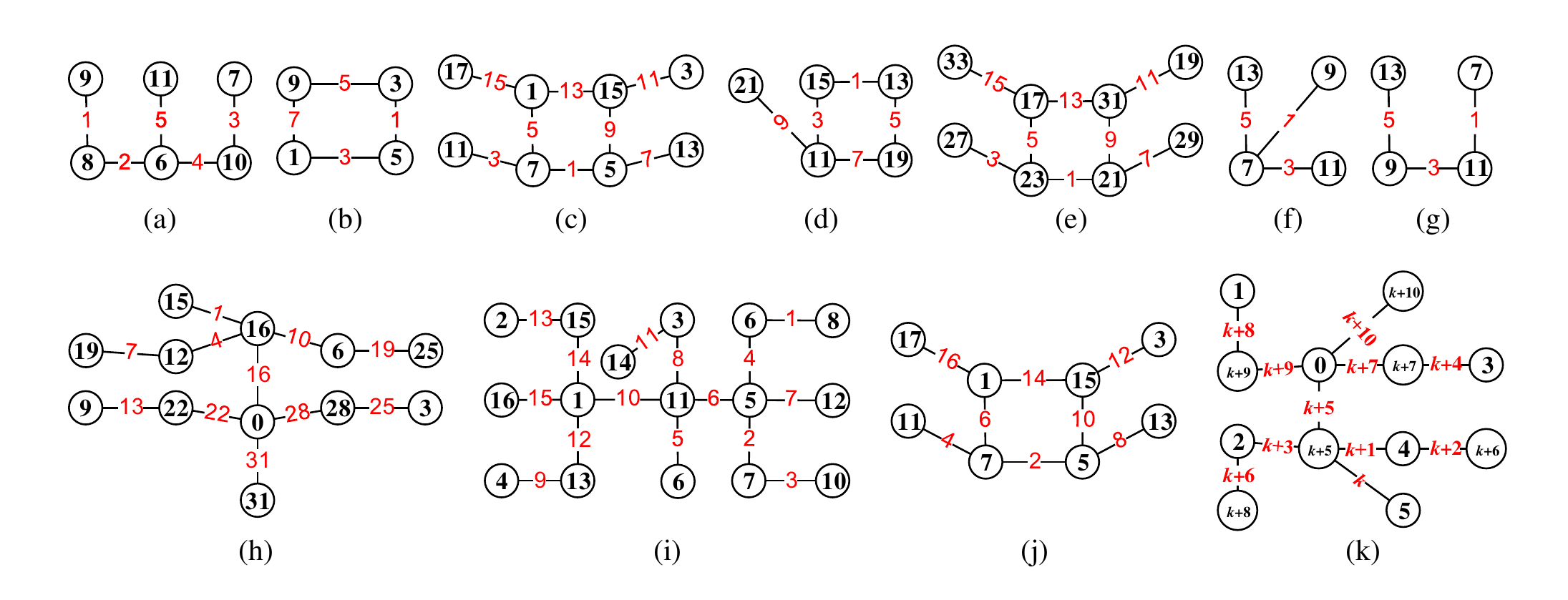}\\
\caption{\label{fig:new-definitionss}{\small Examples for understanding Definition \ref{defn:New-graphical-passwords}, cited from \cite{SH-ZXH-YB-2018-5}.}}
\end{figure}

There are examples shown in Fig.\ref{fig:new-definitionss} for understanding Definition \ref{defn:New-graphical-passwords}: (a) An AM-super graceful labeling; (b) an all-odd graceful labeling; (c) an AM-all-odd graceful labelinga; (d) a super odd-graceful labeling; (e) an AM-super odd-graceful labeling; (f)a super odd-elegant labeling; (g) an AM-super odd-elegant labeling; (h) a one modulo three graceful labeling; (i) an AM-Skolem-graceful labeling; (j) an AM-odd-even graceful labeling; (k) an AM-$k$-graceful labeling.

\begin{figure}[h]
\centering
\includegraphics[width=16.4cm]{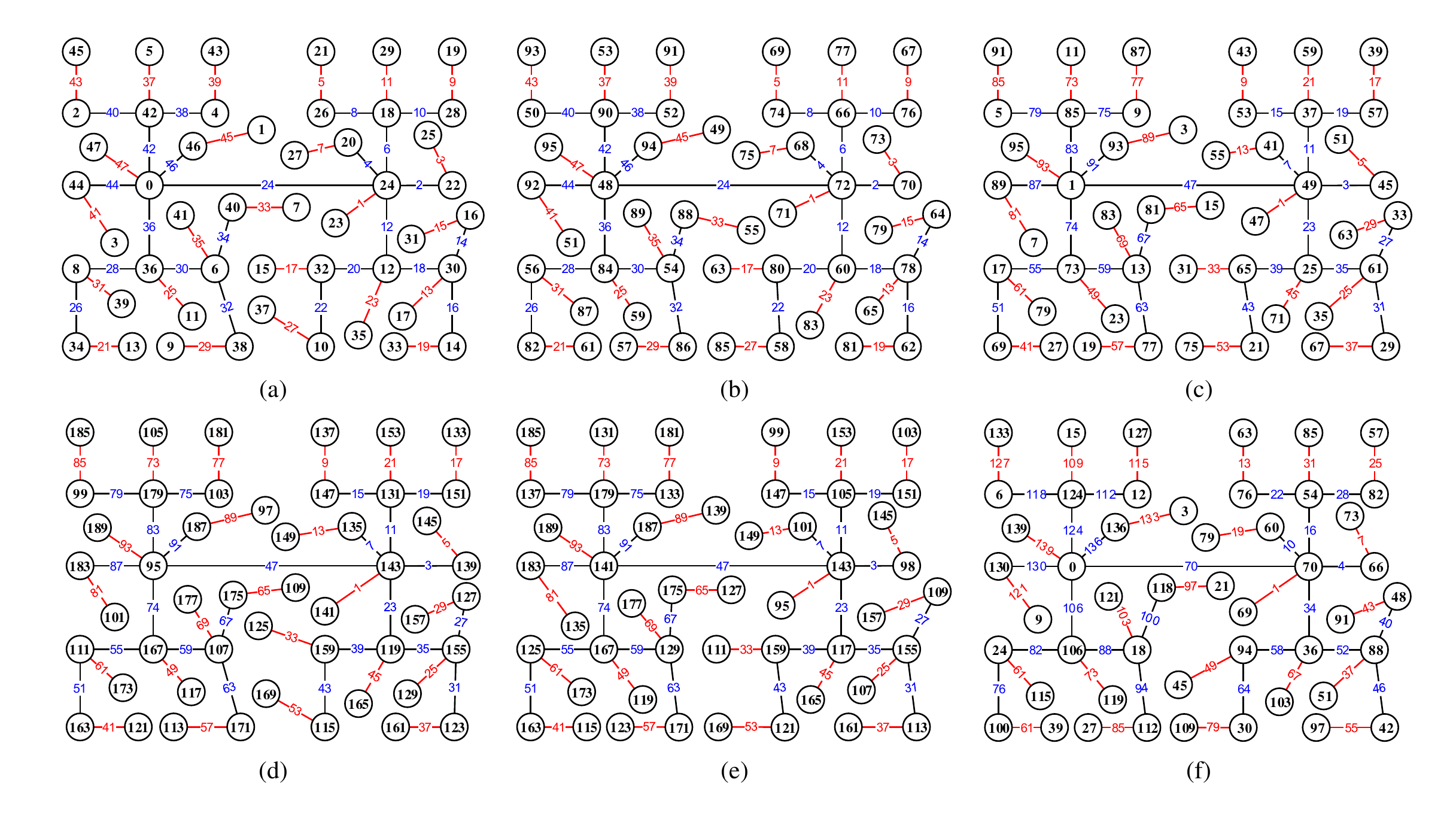}\\
\caption{\label{fig:AM-series-0}{\small More examples for understanding Definition \ref{defn:New-graphical-passwords}, cited from \cite{SH-ZXH-YB-2018-5}.}}
\end{figure}

\begin{figure}[h]
\centering
\includegraphics[width=16.4cm]{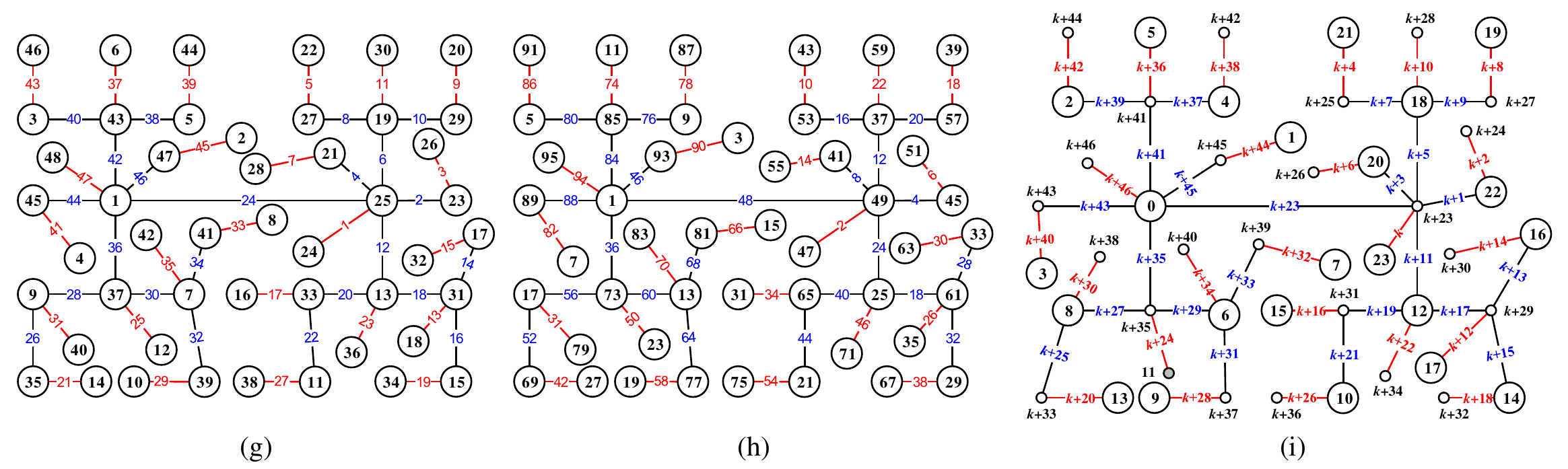}\\
\caption{\label{fig:AM-series}{\small Another group of examples for understanding Definition \ref{defn:New-graphical-passwords}, cited from \cite{SH-ZXH-YB-2018-5}.}}
\end{figure}

In Fig.\ref{fig:AM-series-0} and Fig.\ref{fig:AM-series}, (a) a strongly set-ordered graceful labeling; (b) an AM-set-ordered super graceful labeling; (c) an AM-set-ordered all-odd graceful labeling; (d) an AM-set-ordered super odd-graceful labeling; (e) an AM-set-ordered super odd-elegant labeling; (f) an AM-set-ordered one modulo three graceful labeling; (g) an AM-set-ordered Skolem-graceful labeling; (h) an AM-set-ordered odd-even graceful labeling; (i) an AM-$k$-graceful labeling.

\begin{thm} \label{them:relationships}
\cite{SH-ZXH-YB-2018-5} A \emph{leaf-matching} $(p,q)$-\emph{tree} $T$ contains a perfect matching $M$ such that each matching edge has an end of degree one. Then we have the following pairwise equivalent assertions:
\begin{asparaenum}[(i) ]
\item $T$ admits a strongly set-ordered graceful labeling;
\item $T$ admits an AM-set-ordered super graceful labeling;
\item $T$ admits an AM-set-ordered all-odd graceful labeling.
\item $T$ admits an AM-set-ordered super odd-graceful labeling;
\item $T$ admits an AM-set-ordered super odd-elegant labeling;
\item $T$ admits an AM-set-ordered one modulo three graceful labeling;
\item $T$ admits an AM-set-ordered Skolem-graceful labeling;
\item $T$ admits an AM-set-ordered odd-even graceful labeling; and
\item $T$ admits an AM-$k$-graceful labeling.
\end{asparaenum}
\end{thm}

\begin{figure}[h]
\centering
\includegraphics[width=15cm]{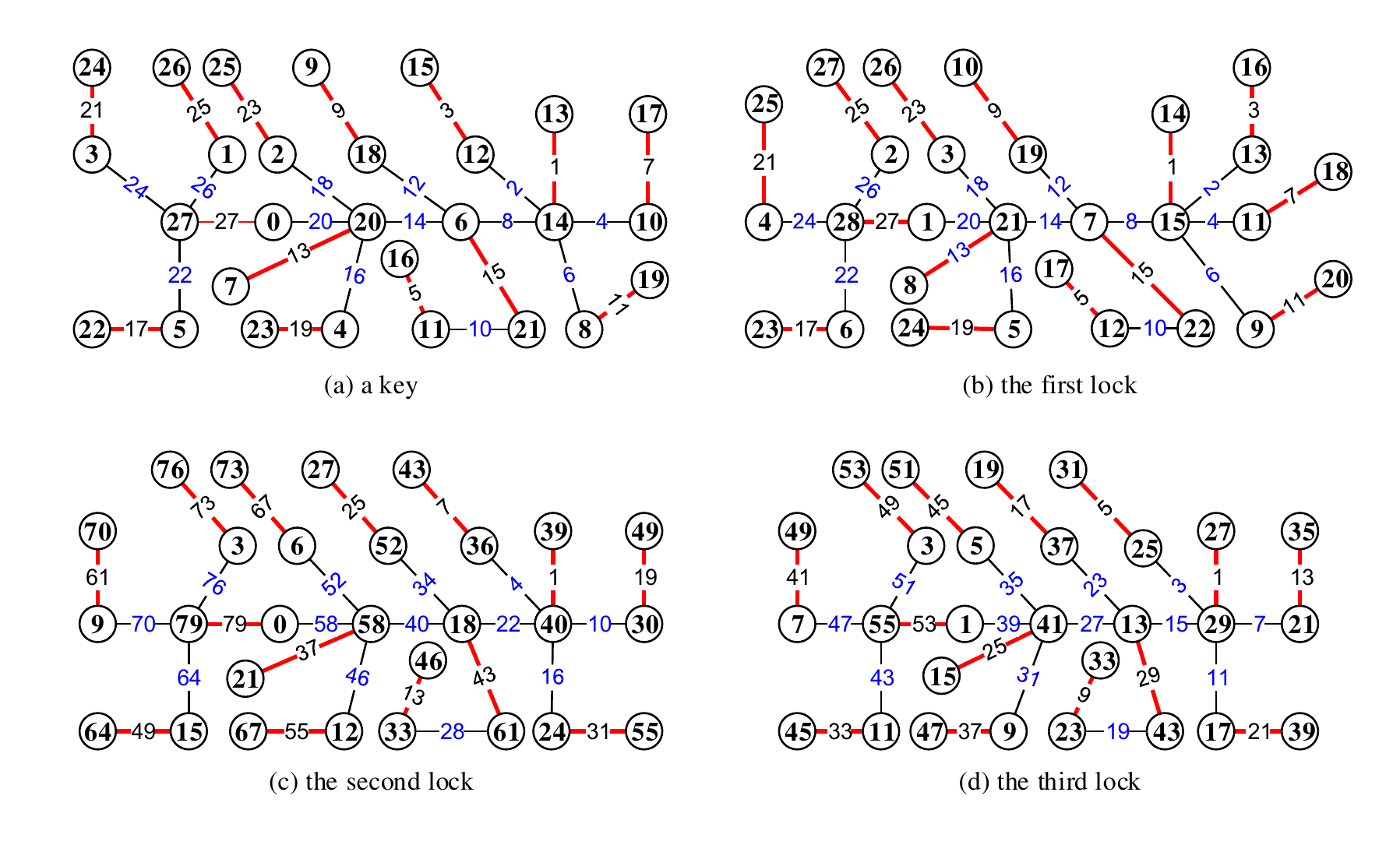}\\
\caption{\label{fig:key-lock-new}{\small A scheme for illustrating the non-symmetric authentication and Theorem \ref{them:relationships}, cited from \cite{SH-ZXH-YB-2018-5}.}}
\end{figure}

\subsubsection{Parameterized labelings on trees having perfect matchings}

\begin{defn} \label{defn:AM-odd-elegant-perfect-matchings}
\cite{SH-ZXH-YB-2017-1, SH-ZXH-YB-2017-2} Let $T$ be a $(p,q)$-tree having a perfect matching $M$, and let $S_{q-1,k,d}=\{k, k + d, \dots , k +(q-1)d\}$ for integers $k\geq 1$ and $d\geq 1$. Suppose that each leaf of $T$ is an end of some matching edge of $M$, and $M(L)$ is the set of all leaves of $T$, and furthermore the graph $T-M(L)$ obtained by deleting all leaves of $T$ is just a tree having $k=|V(T-M(L))|$ vertices.

(1) \cite{SH-ZXH-YB-2017-1} Suppose that $T$ admits an odd-elegant labeling~$f_1$. If $|f_1(u)-f_1(v)|=1$ for each matching edge $uv \in M$ holds true, and the elements of $f_1(E(M))$ form an arithmetic progression having $|M|$ terms with the first term $1$ and the common difference 4. Then we say $f_1$ an \emph{arithmetic matching odd-elegant labeling} (\emph{AM-odd-elegant labeling}) of $T$, and $T$ is an \emph{AM-odd-elegant graph}.

(2) \cite{SH-ZXH-YB-2017-1} Suppose that $T$ admits a super felicitous labeling~$f_2$. If $|f_2(u)-f_2(v)|=k$ for every matching edge $uv \in M$, we call $f_2$ an \emph{arithmetic matching super felicitous labeling} (\emph{AM-super felicitous labeling}) of $T$, and $T$ an \emph{AM-super felicitous graph}.

(3) \cite{SH-ZXH-YB-2017-1} Suppose that $T$ admits a super edge-magic graceful labeling~$f_3$. If $|f_3(u)-f_3(v)|=k$ for each matching edge $uv \in M$, and the elements of $f_3(E(T))$ produce an arithmetic progression having $|M|$ terms with the first term $p+1$ and the common difference $2$. Then we call $f_3$ an \emph{arithmetic matching super edge-magic graceful labeling} (\emph{AM-super edge-magic graceful labeling}) of $T$, and $T$ an \emph{AM-super edge-magic graceful graph}.

(4) \cite{SH-ZXH-YB-2017-2} Suppose that $T$ admits a super edge-magic total labeling~$f_4$. If $|f_4(u)-f_4(v)|=k$ for each matching edge $uv \in M$, and the elements of $f_4(E(T))$ yield an arithmetic progression having $|M|$ terms with the first term $2p-1$ and the common difference $-2$. Then we call $f_4$ an \emph{arithmetic matching super edge-magic total labeling} (\emph{AM-super edge-magic total labeling}) of $T$, and $T$ an \emph{AM-super edge-magic total graph}.\qqed
\end{defn}

\begin{defn} \label{defn:am-parameterized-matchings}
\cite{SH-ZXH-YB-2017-2} Let integers $k,d\geq 0$.

(1) Suppose that $T$ admits a super $(k,d)$-edge antimagic labeling $g$ defined in Definition \ref{defn:3-parameter-labelings}. If $|g(u)-g(v)|=k$ for each matching edge $uv \in M$, and

(i) the elements of $g(E(T))$ produce an arithmetic progression having $|M|$ terms with the first term $p+1$ and the common difference $2$;

(ii) $g(u)+g(v)=b_i$ forms a sequence $\{b_i\}$ being an arithmetic progression having $|M|$ terms with the first term $k+2$ and the common difference $2$, where $n=|M|$. \\
Then we call $g$ an \emph{arithmetic matching super $(k,d)$-edge antimagic total labeling} (\emph{AM-super $(k,d)$-edge antimagic total labeling}) of $T$, and $T$ an \emph{AM-super $(k,d)$-edge antimagic total labeling graph}.

(2) Suppose that $T$ admits a $(k,d)$-arithmetic labeling $h$ defined in Definition \ref{defn:3-parameter-labelings}. If $|h(u)-h(v)|=k$ for each matching edge $uv \in M$, and the elements of $h(E(T))$ produce an arithmetic progression having $|M|$ terms with the first term $k$ and the common difference $2d$. Then we call $h$ an \emph{arithmetic matching $(k,d)$-arithmetic labeling} (\emph{AM-$(k,d)$-arithmetic labeling}) of $T$, and $T$ an \emph{AM-$(k,d)$-arithmetic graph}.\qqed
\end{defn}

\begin{figure}[h]
\centering
\includegraphics[width=16.4cm]{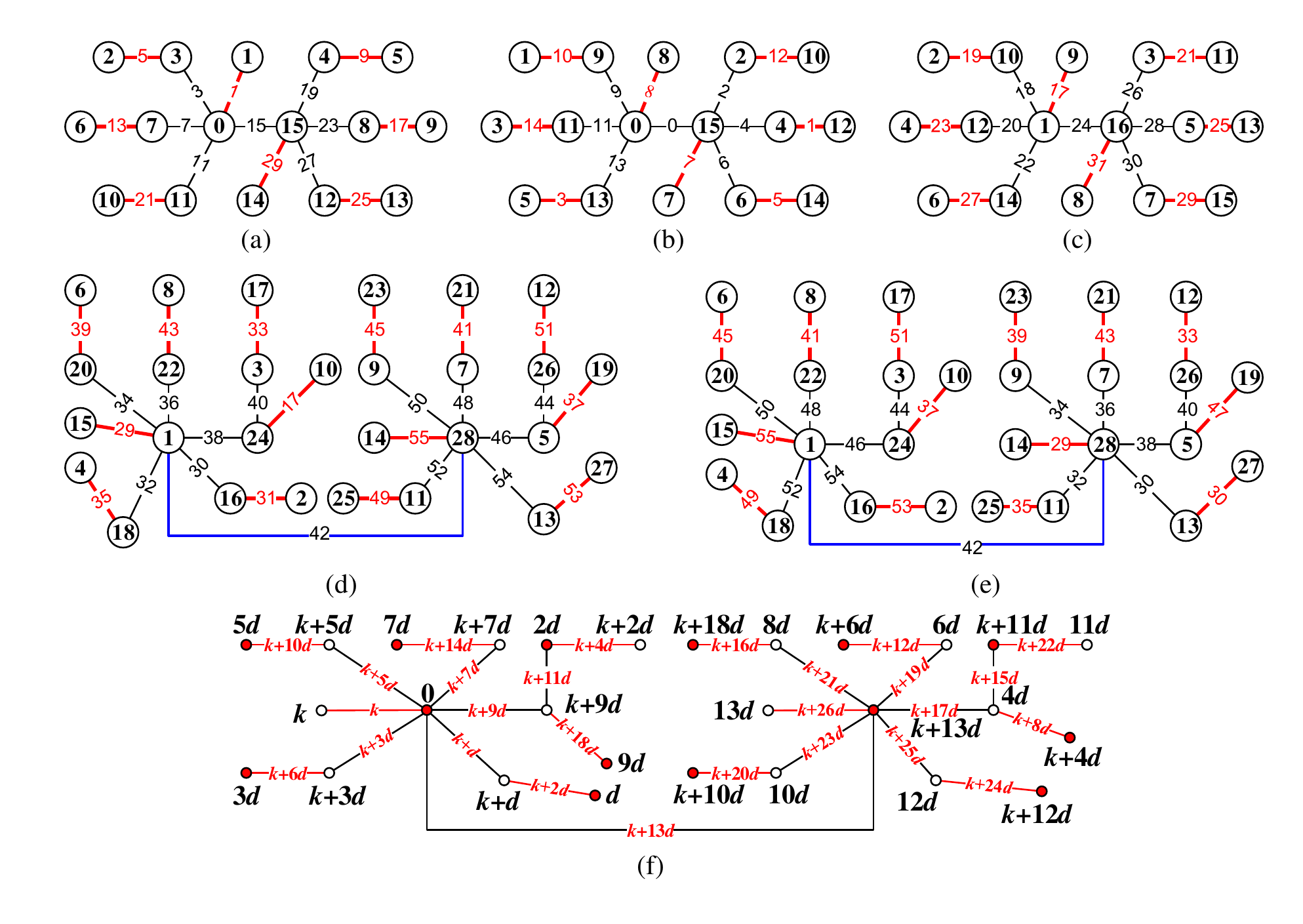}
\caption{\label{fig:for-illustrating-de-59}{\small For illustrating Definition \ref{defn:AM-odd-elegant-perfect-matchings} and Definition \ref{defn:am-parameterized-matchings}: (a) An AM-odd-elegant labeling; (b) an AM-super felicitous labeling; (c) an AM-super edge-magic graceful labeling; (d) an AM-super $(|X|+p+3,2)$-edge antimagic total labeling; (e) an AM-super edge-total graceful labeling; (f) an AM-$(k,d)$-arithmetic labeling.}}
\end{figure}

\begin{defn} \label{defn:totally-odd-even-graceful-labeling}
\cite{Sun-Zhang-Yao-ICMITE-2017} Let $T$ be a $(p,q)$-tree having a perfect matching $M$, and let $S_{q-1,k,d}=\{k, k + d, \dots , k +(q-1)d\}$, $S\,'_{k,d}=\{k, k + 2d, \dots , k +2(q-1)d\}$, $S\,''_{k,d}=\{k, k + d, \dots , k +(p+q-1)d\}$ for integers $k\geq 1$ and $d\geq 1$. Suppose that each leaf of $T$ is an end of some matching edge of $M$, and $M(L)$ is the set of all leaves of $T$, and furthermore the graph $T-M(L)$ obtained by deleting all leaves of $T$ is a tree having $c=|V(T-M(L))|$ vertices. AM is the abbreviation of ``arithmetic matching''.
\begin{asparaenum}[S-1. ]
 \item An \emph{AM-$(k,d)$-totally odd (resp. even) graceful labeling} $f$ of $G$ holds $f(V(G))\subseteq [0, k + 2(q-1)d]$, $f(x)\neq f(y)$ for distinct vertices $x,y\in V(G)$ and $f(E(G))=\{|f(u)-f(v)|:\ uv\in E(G)\}=S\,'_{k,d}$. Meanwhile, $f(u)+f(v)=k+2(q-1)d$ for each matching edge $uv \in M$.

 \item An \emph{AM-$(k,d)$-felicitous labeling} $f$ of $G$ holds $f(V(G))\subseteq [0, k + (q-1)d]$, $f(x)\neq f(y)$ for distinct vertices $x,y\in V(G)$ and $f(E(G))=\{f(uv)=f(u)+f(v)~ (\bmod~qd): uv\in E(G)\}=S_{q-1,k,d}$; and furthermore, $f$ is \emph{super} if $f(V(G))=[0,(\frac{p}{2}-1)d]\cup[k+pd/2,k +(p-1)d]$. Meanwhile, $|f(u)-f(v)|=k+cd$ for each matching edge $uv \in M$.

 \item An \emph{AM-$(k,d)$-edge-magic total labeling} $f$ of $G$ holds $f(V(G)\cup E(G))\subseteq [0, k + (p+q-1)d]$ such that $f(u)+f(v)+f(uv)=a$ for any edge $uv\in E(G)$, where the magic constant $a$ is a fixed integer; and furthermore, $f$ is \emph{super} if $f(V(G))=[1,pd/2]\cup[k+pd/2,k +qd]$. Meanwhile, $|f(u)-f(v)|=k+(c-1)d$ for each matching edge $uv \in M$ and the elements of $f(E(T))$ yield an arithmetic progression having $|M|$ terms with the first term $k+2(p-1)d$ and the common difference $-2d$.

 \item An \emph{AM-$(k,d)$-edge-magic graceful labeling} $f$ of $G$ holds $f(V(G)\cup E(G))\subseteq [0, k + (p+q-1)d]$ such that $\mid f(u)+f(v)-f(uv)\mid=a$ for any edge $uv\in E(G)$, where the magic constant $a$ is a fixed integer; and furthermore, $f$ is \emph{super} if $f(V(G))=[1,pd/2]\cup[k+pd/2,k+qd]$. Meanwhile, $|f(u)-f(v)|=k+(c-1)d$ for each matching edge $uv \in M$ and the elements of $f(E(T))$ yield an arithmetic progression having $|M|$ terms with the first term $k+pd$ and the common difference $2d$.

 \item Suppose that $T$ admits a $(k,d)$-arithmetic~$f$. If $|f(u)-f(v)|=k$ for each matching edge $uv \in M$, and the elements of $f(E(T))$ produce an arithmetic progression having $|M|$ terms with the first term $k$ and the common difference $2d$. Then we call $f$ an \emph{AM-$(k,d)$-arithmetic labeling} of $T$.

 \item An \emph{AM-$(k,d)$-totally odd (resp. even)-elegant labeling} $f$ of $G$ holds $f(V(G))\subset [0,k + 2(q-1)d]$, $f(u)\neq f(v)$ for distinct vertices $u,v\in V(G)$, and $f(E(G))=\{f(uv)=f(u)+f(v)~(\bmod~k+2qd):uv\in E(G)\}=S\,'_{k,d}$. Meanwhile, $|f(u)-f(v)|=k$ for each matching edge $uv \in M$ and the elements of $f(E(T))$ yield an arithmetic progression having $|M|$ terms with the first term $k$ and the common difference $4d$.\qqed
\end{asparaenum}
\end{defn}

\begin{figure}[h]
\centering
\includegraphics[width=14.6cm]{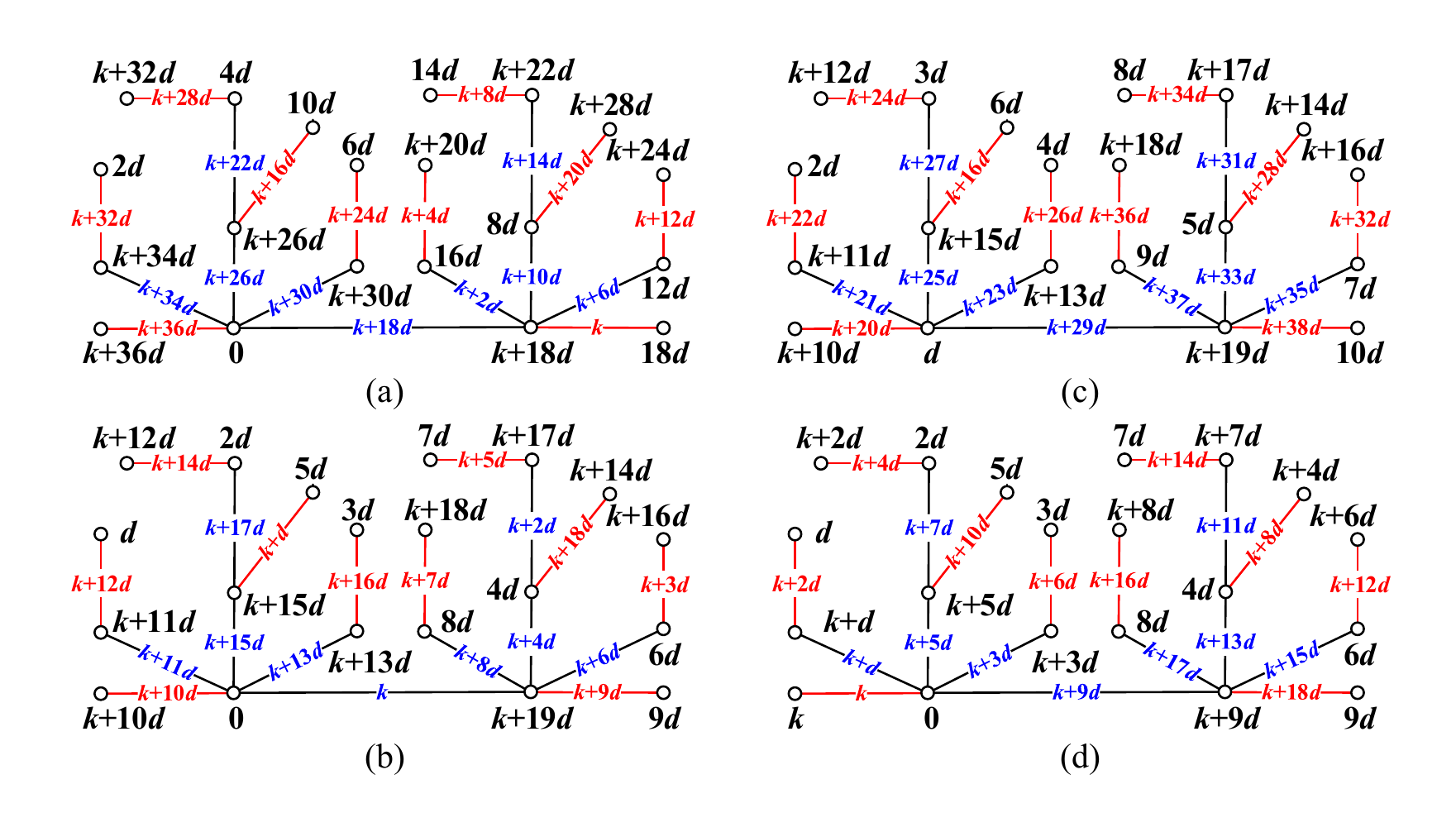}
\caption{\label{fig:matching-labelings-k-d}{\small For illustrating Definition \ref{defn:totally-odd-even-graceful-labeling}: (a) An AM-$(k,d)$-totally odd (resp. even) graceful labeling; (b) an AM-$(k,d)$-super felicitous labeling; (c) an AM-$(k,d)$-super edge-magic total labeling; (d) an AM-$(k,d)$-super edge-magic graceful labeling.}}
\end{figure}

\begin{figure}[h]
\centering
\includegraphics[width=16.4cm]{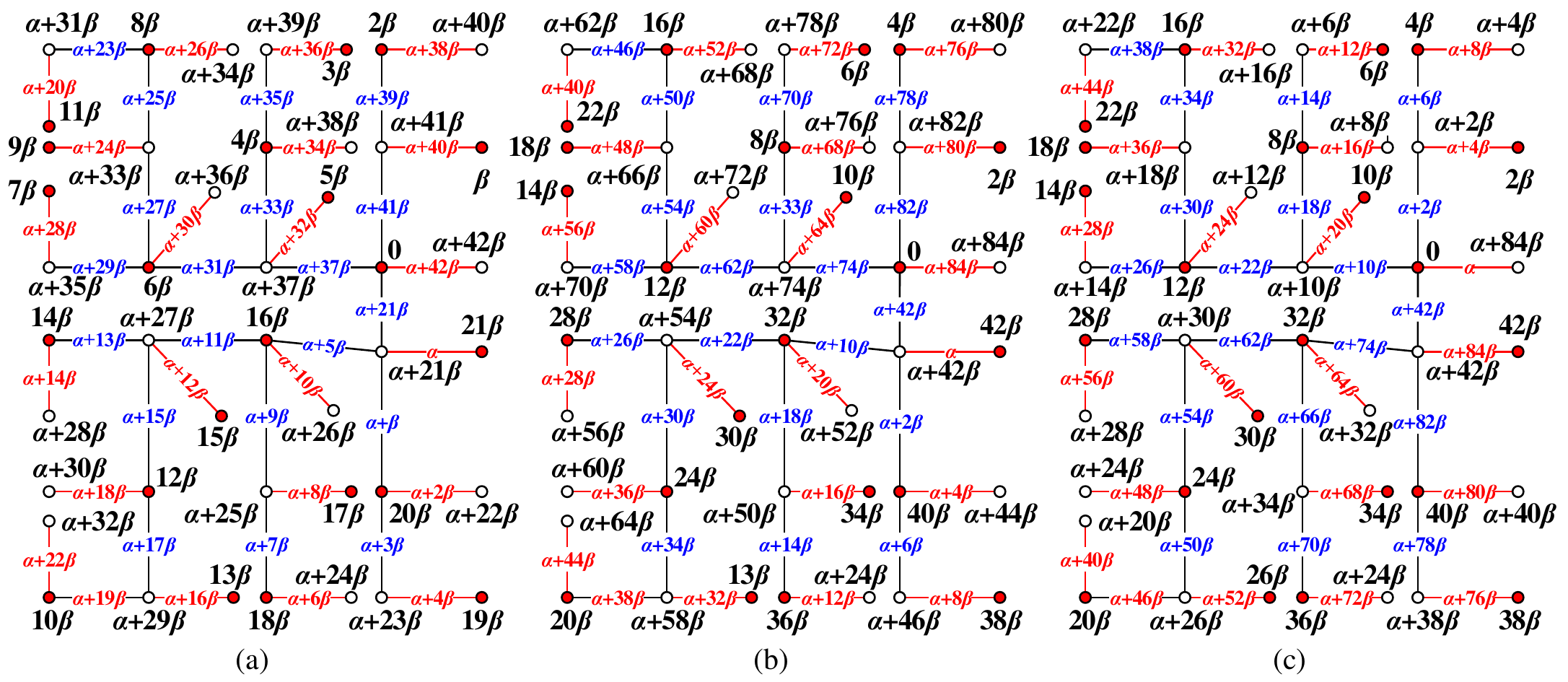}
\caption{\label{fig:1-definition}{\small $p=44$, $q=43$ and $|M|=22$ in Definition \ref{defn:perfect-matching-definitions}: (a) A strongly set-ordered $(\alpha,\beta)$-graceful labeling; (b) an AM-$(\alpha,\beta)$ totally odd/even-graceful labeling; (c) an AM-$(\alpha,\beta)$ totally odd/even-elegant labeling, cited from \cite{Sun-Wang-Yao-2020}.}}
\end{figure}

\begin{figure}[h]
\centering
\includegraphics[width=16.4cm]{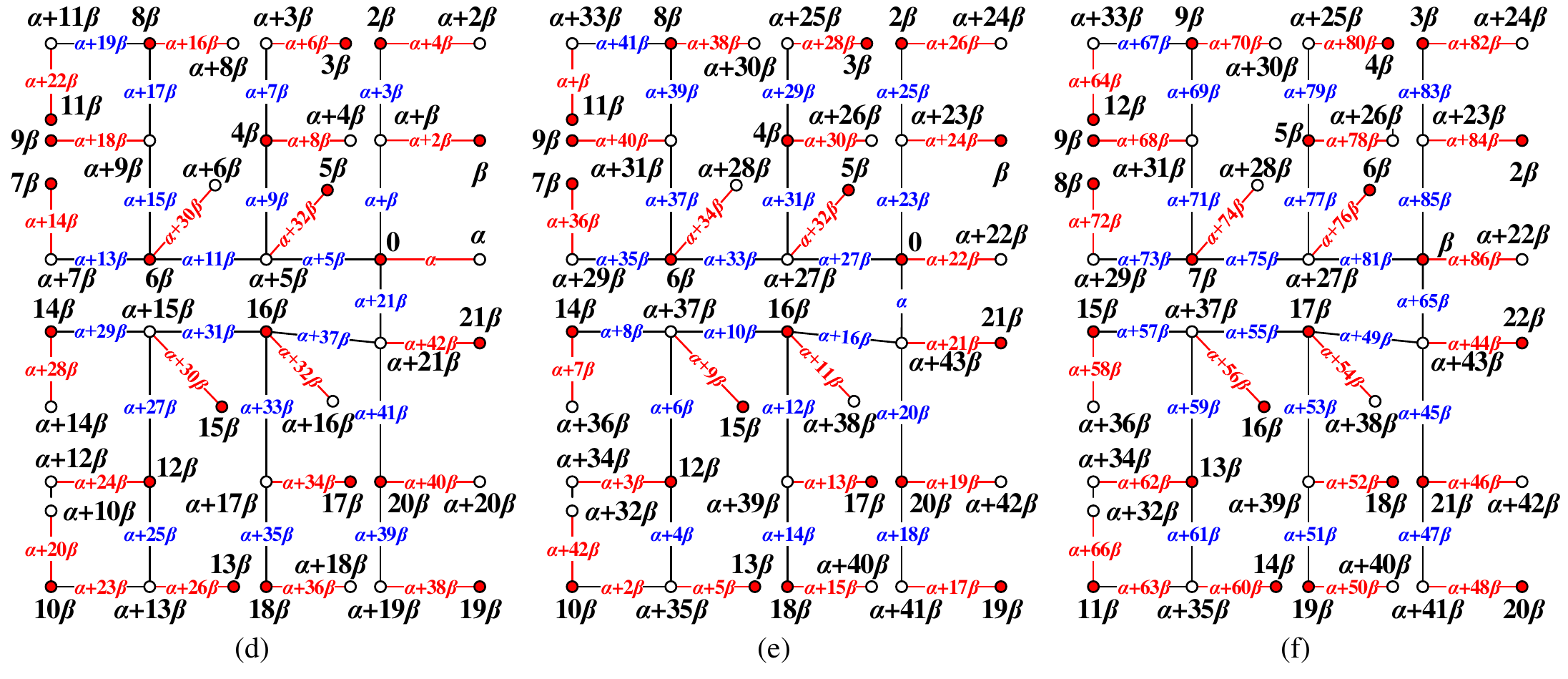}
\caption{\label{fig:2-definition}{\small $p=44$, $q=43$ and $|M|=22$ in Definition \ref{defn:perfect-matching-definitions}: (d) An AM-$(\alpha,\beta)$ arithmetic labeling; (e) a super AM-$(\alpha,\beta)$ felicitous labeling; (f) a super AM-$(\alpha,\beta)$ edge-magic total labeling, cited from \cite{Sun-Wang-Yao-2020}.}}
\end{figure}

\begin{defn} \label{defn:perfect-matching-definitions}
\cite{Sun-Wang-Yao-2020} Suppose that $T$ is a tree having $p$ vertices, $q$ edges and a perfect matching $M$, and its vertex bipartition $V(T)=X\cup Y$ with $X\cap Y=\emptyset$ such that each edge $xy\in E(T)$ holds $x\in X$ and $y\in Y$. Let $S_{m,\alpha,k\beta}=\{\alpha, \alpha + k\beta, \dots , \alpha +k(m-1)\beta\}$ with integers $k\geq 1$, $m\geq 2$, $\alpha\geq 0$ and $\beta\geq 1$, and let $c$ be a constant. There are the following constraint conditions:
\begin{asparaenum}[C-1. ]
\item \label{parameter:totally-odd-even-graceful-v} $\sigma: V(T)\rightarrow S_{p-1,0,\beta}\cup S_{2p-3,\alpha,\beta}$, and $\sigma(x)\neq \sigma(w)$ for any pair of vertices $x,w\in V(T)$;
\item \label{parameter:felicitous-v} $\sigma: V(T)\rightarrow S_{\frac{p}{2},0,\beta}\cup S_{p,\alpha,\beta}$, and $\sigma(x)\neq \sigma(w)$ for any pair of vertices $x,w\in V(T)$;
\item \label{parameter:totally-odd-even-elegant} $\sigma(E(T))=\{\sigma(xy)=\sigma(x)+\sigma(y):xy\in E(T)\}=S_{q,\alpha,2\beta}$;
\item \label{parameter:arithmetic-labeling-e} $\sigma(E(T))=\{\sigma(xy)=\sigma(x)+\sigma(y):xy\in E(T)\}=S_{q,\alpha,\beta}$;
\item \label{parameter:totally-odd-even-graceful-e} induced edge color $\sigma(xy)=|\sigma(x)-\sigma(y)|$ for each edge $xy\in E(T)$, and $\sigma(E(T))=\{\sigma(xy):xy\in E(T)\}=S_{q,\alpha,\beta}$;
\item \label{parameter:totally-odd-even-graceful-2e} induced edge color $\sigma(xy)=|\sigma(x)-\sigma(y)|$ for each edge $xy\in E(T)$, and $\sigma(E(T))=\{\sigma(xy):xy\in E(T)\}=S_{q,\alpha,2\beta}$;
\item \label{parameter:felicitous-e} induced edge color $\sigma(xy)-\alpha=\sigma(x)+\sigma(y)-\alpha~ (\bmod ~q\beta)$ for each edge $xy\in E(T)$, and $\sigma(E(T))=\{\sigma(xy):xy\in E(T)\}=S_{q,\alpha,\beta}$;
\item \label{parameter:edge-magic-total} $\sigma: V(T)\cup E(T)\rightarrow S_{\frac{p}{2}+1,0,\beta}\cup S_{p+q,\alpha,\beta}$, and $\sigma(x)\neq \sigma(w)$ for any pair of vertices $x,w\in V(T)$;
\item \label{parameter:strongly-grace}$\sigma(X)=S_{\frac{p}{2},0,\beta}$ and $\sigma(Y)=S_{p-1,\alpha,\beta}\setminus S_{\frac{p}{2}-1,\alpha,\beta}$;
\item \label{parameter:super-0}$\sigma(X)=S_{\frac{p}{2},0,\beta}$ and $\sigma(Y)=S_{p,\alpha,\beta}\setminus S_{\frac{p}{2},\alpha,\beta}$;
\item \label{parameter:super} $\sigma(X)=S_{\frac{p}{2},0,\beta}\setminus \{0\}$ and $\sigma(Y)=S_{p,\alpha,\beta}\setminus S_{\frac{p}{2},\alpha,\beta}$;

------------ \emph{matching-set}

\item \label{parameter:felicitous-matching} $|\sigma(x)-\sigma(y)|=\alpha+p \beta/2$ for each matching edge $xy \in M$;
\item \label{parameter:edge-magic-total-matching} $|\sigma(x)-\sigma(y)|=\alpha+(p-2)\beta/2$ for each matching edge $xy \in M$;
\item \label{parameter:totally-odd-even-graceful-matching} $\sigma(x)+\sigma(y)=\alpha+2(q-1)\beta$ for each matching edge $xy \in M$;
\item \label{parameter:totally-odd-even-elegant-matching} $|\sigma(x)-\sigma(y)|=\alpha$ for each matching edge $xy \in M$;
\item \label{parameter:edge-difference-matching} $|\sigma(x)-\sigma(y)|=\alpha+(q-1)\beta$ for each matching edge $xy \in M$;

\item \label{parameter:arithmetic-matching-set} $\sigma(M)=S_{|M|,\alpha,2\beta}$;
\item \label{parameter:matching-set-2} $\sigma(M)=S_{|M|,\alpha,4\beta}$;
\item \label{parameter:matching-set-0}$\sigma(M)=\{\alpha+2(p-1)\beta, \alpha+2(p-2)\beta,\dots ,\alpha+2(p-|M|)\beta\}$;
\item \label{parameter:matching-set-1} $\sigma(M)=\{\alpha+p\beta, \alpha+(2+p)\beta,\dots ,\alpha+[2(|M|-1)+p]\beta\}$;

------------ \emph{edge-magic}

\item \label{parameter:edge-magic-total-number} $\sigma(x)+\sigma(y)+\sigma(xy)=c$ for each edge $xy\in E(T)$;
\item \label{parameter:edge-difference} $\sigma(xy)+|\sigma(x)-\sigma(y)|=c$ for each edge $xy\in E(T)$;
\item \label{parameter:edge-magic-total-graceful} $|\sigma(x)+\sigma(y)-\sigma(xy)|=c$ for each edge $xy\in E(T)$; and
\item \label{parameter:graceful-difference} $\big |\sigma(xy)-|\sigma(x)-\sigma(y)|\big |=c$ for each edge $xy\in E(T)$.
\end{asparaenum}
\noindent \textbf{We call $\sigma$}:
\begin{asparaenum}[\textrm{Label}-1. ]
\item A \emph{strongly} \emph{$(\alpha,\beta)$-graceful labeling} if C-\ref{parameter:felicitous-v}, C-\ref{parameter:totally-odd-even-graceful-e} and C-\ref{parameter:edge-difference-matching} hold true.
\item A \emph{set-ordered strongly $(\alpha,\beta)$-graceful labeling} if C-\ref{parameter:felicitous-v}, C-\ref{parameter:totally-odd-even-graceful-e}, C-\ref{parameter:edge-difference-matching} and C-\ref{parameter:strongly-grace} hold true (see an example shown in Fig.\ref{fig:1-definition} (a)).
\item An \emph{$(\alpha,\beta)$-totally odd/even-graceful labeling} if C-\ref{parameter:totally-odd-even-graceful-v}, C-\ref{parameter:totally-odd-even-graceful-2e} and C-\ref{parameter:totally-odd-even-graceful-matching} hold true (see an example shown in Fig.\ref{fig:1-definition} (b)).

\item An \emph{AM-$(\alpha,\beta)$ totally odd/even-elegant labeling} if C-\ref{parameter:totally-odd-even-graceful-v}, C-\ref{parameter:totally-odd-even-elegant}, C-\ref{parameter:totally-odd-even-elegant-matching} and C-\ref{parameter:matching-set-2} hold true (see an example shown in Fig.\ref{fig:1-definition} (c)).
\item An \emph{AM-$(\alpha,\beta)$ felicitous labeling} if C-\ref{parameter:felicitous-v}, C-\ref{parameter:felicitous-e} and C-\ref{parameter:felicitous-matching} hold true.
\item A \emph{super AM-$(\alpha,\beta)$ felicitous labeling} if C-\ref{parameter:felicitous-v}, C-\ref{parameter:felicitous-e}, C-\ref{parameter:felicitous-matching} and C-\ref{parameter:super-0} hold true (see an example shown in Fig.\ref{fig:2-definition} (e)).

\item An \emph{AM-$(\alpha,\beta)$ arithmetic labeling} if C-\ref{parameter:edge-magic-total}, C-\ref{parameter:totally-odd-even-elegant-matching}, C-\ref{parameter:arithmetic-labeling-e} and C-\ref{parameter:arithmetic-matching-set} hold true (see an example shown in Fig.\ref{fig:2-definition} (d)).

------------ \emph{edge-magic}

\item An \emph{AM-$(\alpha,\beta)$ edge-magic total labeling} if C-\ref{parameter:edge-magic-total}, C-\ref{parameter:edge-magic-total-number}, C-\ref{parameter:edge-magic-total-matching} and C-\ref{parameter:matching-set-0} hold true.
\item A \emph{super AM-$(\alpha,\beta)$ edge-magic total labeling} if C-\ref{parameter:edge-magic-total}, C-\ref{parameter:edge-magic-total-number}, C-\ref{parameter:edge-magic-total-matching}, C-\ref{parameter:super} and C-\ref{parameter:matching-set-0} hold true (see an example shown in Fig.\ref{fig:2-definition} (f)).
\item an \emph{AM-$(\alpha,\beta)$ edge-magic graceful labeling} if C-\ref{parameter:edge-magic-total}, C-\ref{parameter:edge-magic-total-matching}, C-\ref{parameter:edge-magic-total-graceful} and C-\ref{parameter:matching-set-1} hold true.
\item A \emph{super AM-$(\alpha,\beta)$ edge-magic graceful labeling} if C-\ref{parameter:edge-magic-total}, C-\ref{parameter:edge-magic-total-matching}, C-\ref{parameter:edge-magic-total-graceful}, C-\ref{parameter:super} and C-\ref{parameter:matching-set-1} hold true (see an example shown in Fig.\ref{fig:3-definition} (g)).

\item An \emph{AM-$(\alpha,\beta)$ edge-difference labeling} if C-\ref{parameter:totally-odd-even-graceful-v}, C-\ref{parameter:edge-difference} and C-\ref{parameter:edge-difference-matching} hold true.
\item A \emph{super AM-$(\alpha,\beta)$ edge-difference labeling} if C-\ref{parameter:totally-odd-even-graceful-v}, C-\ref{parameter:edge-difference}, C-\ref{parameter:edge-difference-matching} and C-\ref{parameter:strongly-grace} hold true (see an example shown in Fig.\ref{fig:3-definition} (h)).

\item An \emph{AM-$(\alpha,\beta)$ graceful-difference labeling} if C-\ref{parameter:totally-odd-even-graceful-v}, C-\ref{parameter:graceful-difference} and C-\ref{parameter:edge-difference-matching} hold true.
\item A \emph{super AM-$(\alpha,\beta)$ graceful-difference labeling} if C-\ref{parameter:totally-odd-even-graceful-v}, C-\ref{parameter:graceful-difference}, C-\ref{parameter:edge-difference-matching} and C-\ref{parameter:strongly-grace} hold true (see an example shown in Fig.\ref{fig:3-definition} (i)).\qqed
\end{asparaenum}
\end{defn}

\begin{defn}\label{defn:2q-modular-odd-graceful-labeling}
\cite{Yao-Sun-Zhao-Li-Yan-ITNEC-2017, Zhao-Ming-Yao-Bing-Yao-2020, Zhao-Yao-Zhang-Sun-Bing-Yao-2019} A labeling $h$ of a $(p,q)$-graph $G$ is called a \emph{2q-modular odd-graceful labeling} if $h : V (G) \rightarrow [0, 2q-1]$, and $h(E(G)) = \{|h(u)-h(v)|: uv\in E(G)\}\cup \{2q-|h(u)-h(v)|: uv\in E(G)\} = [1, 2q-1]^o$.\qqed
\end{defn}

\begin{figure}[h]
\centering
\includegraphics[width=16.4cm]{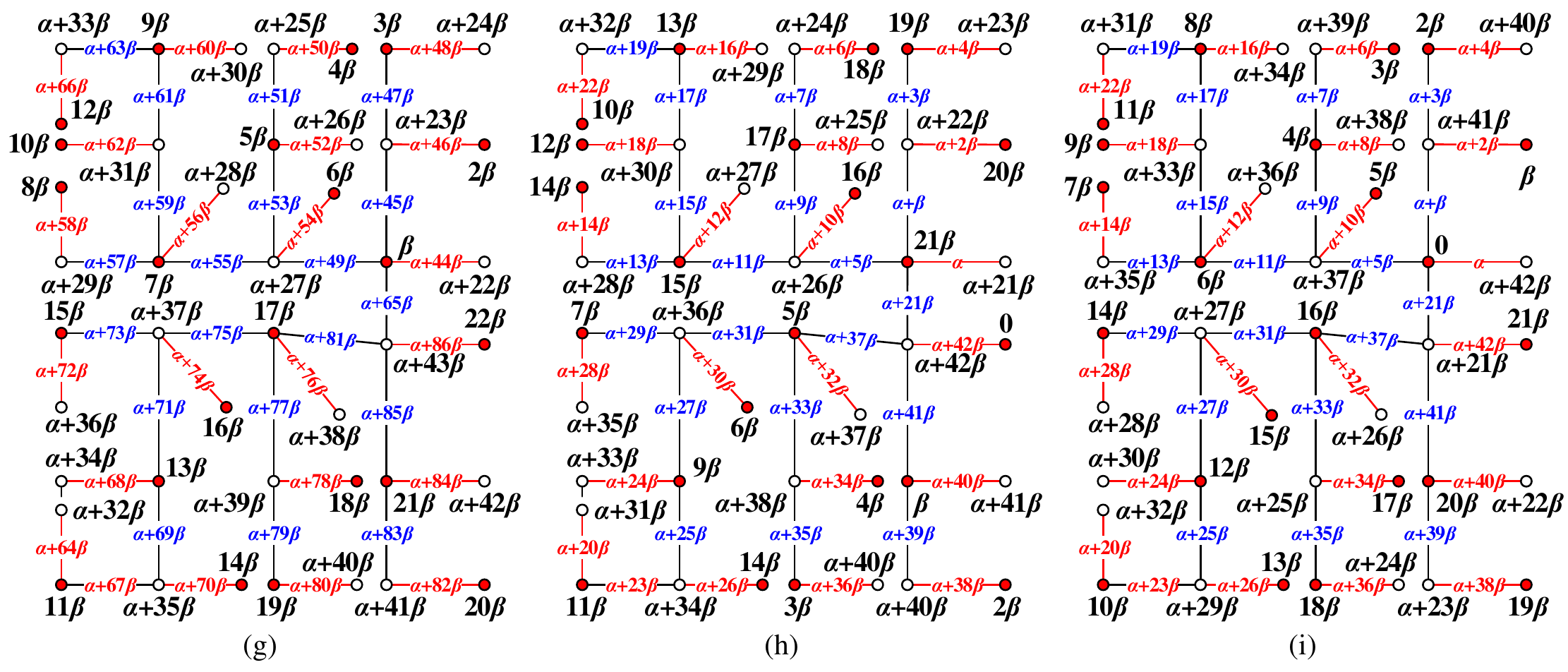}
\caption{\label{fig:3-definition}{\small $p=44$, $q=43$ and $|M|=22$ in Definition \ref{defn:perfect-matching-definitions}: (g) a super AM-$(\alpha,\beta)$ edge-magic graceful labeling; (h) a super AM-$(\alpha,\beta)$ graceful-difference labeling; (i) a super AM-$(\alpha,\beta)$ edge-difference labeling, cited from \cite{Sun-Wang-Yao-2020}.}}
\end{figure}

\begin{rem}\label{rem:333333}
Let $f$ be a strongly set-ordered $(\alpha,\beta)$-graceful labeling of a tree $T$, we have the following equivalent transformations:

(1) An AM-$(\alpha,\beta)$ graceful-difference labeling $\sigma$ is defined by $\sigma(w)=f(w)$ for $w\in V(T)$, $\sigma(xy)=2\alpha+(q-1)\beta-f(xy)$ for $xy\in E(T)$.

(2) An AM-$(\alpha,\beta)$ edge-difference labeling $\sigma$ is defined by $\sigma(x)=\max f(X)+\min f(X)-f(x)$ for $x\in X$, $\sigma(y)=\max f(Y)+\min f(Y)-f(y)$ for $y\in Y$, and $\sigma(xy)=\max f(E(T))+\min f(E(T))-f(xy)$ for $xy\in E(T)$.\paralled
\end{rem}

Before to define six new labelings on trees, we introduce the concept of a \emph{leaf-matching tree}. Let $M(L)$ be the set of all leaves of $(p,q)$-tree $T$, if a tree $T$ with a perfect matching $M$ satisfies that one end of each matching edge of $M$ is just an leaf of $T$, and the remainder $T-M(L)$ is a tree of $p=|V(T-M(L))|$ vertices after removing all leaves of $T$, so call $T$ a \emph{leaf-matching tree}. Furthermore, ''arithmetic matching'' and ''arithmetic progression'' are abbreviated as ``AM'' and ``arp'', respectively.

\begin{defn} \label{defn:parameted-perfect-matchings}
\cite{Sun-Wang-Yao-2020} Suppose that $T$ is a leaf-matching tree on $p$ vertices and $q$ edges, and $M$ is a perfect matching set of $T$.
\begin{asparaenum}[(1)]
\item A labeling $\sigma: V(T)\rightarrow [0, \alpha + 2(q-1)\beta]$ is called an \emph{AM-2-parameter totally odd/even-graceful labeling} if $\sigma(V(T))\subseteq [0, \alpha + 2(q-1)\beta]$, $\sigma(E(T))=S\,'_{\alpha,\beta}$ and $|\sigma(x)-\sigma(y)|$ with each edge $xy\in E(T)$. Meanwhile, each matching edge $xy \in M$ holds $\sigma(x)+\sigma(y)=\alpha+2(q-1)\beta$ true.
\item A labeling $\sigma: V(T)\rightarrow [0, \alpha + (q-1)\beta]$ is called an \emph{AM-2-parameter felicitous labeling} if the vertex color set $\sigma(V(T))\subseteq [0, \alpha + (q-1)\beta]$, $\sigma(E(T))=S_{\alpha,\beta}$ and $\sigma(xy)=\sigma(x)+\sigma(y)~ (\bmod ~q\beta)$ with edge $xy\in E(T)$, and furthermore, $\sigma$ is \emph{super} if $\sigma(V(T))=[0,(\frac{p}{2}-1)\beta]\cup[\alpha+\frac{p}{2}\beta,\alpha +(p-1)\beta]$. Moreover, each matching edge $xy \in M$ holds $|\sigma(x)-\sigma(y)|=\alpha+\rho \beta$ true.
\item A labeling $\sigma: V(T)\rightarrow [0, \alpha + (p+q-1)\beta]$ is called an \emph{AM-2-parameter edge-magic total labeling} if the total color set $\sigma(V(T)\cup E(T))\subseteq [0, \alpha + (p+q-1)\beta]$ such that $\sigma(x)+\sigma(y)+\sigma(xy)=a$ with $xy\in E(T)$, where the $a$ is a magic constant. Then we say $\sigma$ is \emph{super} if $\sigma(V(T))=[1,\frac{p}{2}\beta]\cup[\alpha+\frac{p}{2}\beta,\alpha +q\beta]$. Moreover, each matching edge $xy \in M$ holds $|\sigma(x)-\sigma(y)|=\alpha+(\rho-1)\beta$; and the set $\sigma(M)$ has $|M|$ elements in total and is obtained by an arp with the first term $\alpha+2(p-1)\beta$ and the common difference $-2\beta$.
\item A labeling $\sigma: V(T)\rightarrow [0, \alpha + (p+q-1)\beta]$ is called an \emph{AM-2-parameter edge-magic graceful labeling} if $\sigma(V(T)\cup E(T))\subseteq [0, \alpha + (p+q-1)\beta]$ such that $|\sigma(x)+\sigma(y)-\sigma(xy)|=a$ when $xy\in E(T)$, where the $a$ is a magic constant and, we say $\sigma$ to be \emph{super} if $\sigma(V(T))=[1,p\beta/2]\cup[\alpha+p\beta/2,\alpha+q\beta]$. Furthermore, each matching edge $xy \in M$ holds $|\sigma(x)-\sigma(y)|=\alpha+(\rho-1)\beta$ true; and the set $\sigma(M)$ has $|M|$ elements in total and its elements are obtained by an arp having the first term $\alpha+p\beta$ and the common difference $2\beta$.
\item A labeling $\sigma: V(T)\rightarrow [0, \alpha + (p+q-1)\beta]$ is called an \emph{AM-2-parameter arithmetic labeling} if each matching edge $xy \in M$ holds $|\sigma(x)-\sigma(y)|=\alpha$ true, and the set $\sigma(M)$ has $|M|$ elements in total and its elements yield an arp with the first term $\alpha$ and the common difference $2\beta$.
\item A labeling $\sigma: V(T)\rightarrow [0, \alpha +2(q-1)\beta]$ is called an \emph{AM-2-parameter totally odd/even-elegant labeling} if $\sigma(V(T))\subset [0,\alpha + 2(q-1)\beta]$ and $\sigma(E(T))=\{\sigma(xy)=\sigma(x)+\sigma(y)(\bmod~\alpha+2q\beta):xy\in E(T)\}=S\,'_{\alpha,\beta}$. Furthermore, each matching edge $xy \in M$ holds $|\sigma(x)-\sigma(y)|=\alpha$ true, and the elements of $\sigma(M)$ form an arp having $|M|$ terms, the first term $\alpha$ and the common difference $4\beta$.\qqed
\end{asparaenum}
\end{defn}

\begin{rem}\label{rem:ABC-conjecture}
Notice that the parity property of value of $\alpha$ will determine an AM-2-parameter totally odd-graceful (odd-elegant) labeling if $\alpha=$odd, and an AM-2-parameter totally even-graceful (even-elegant) labeling when $\alpha=$ even. Some examples for illustrating Definition \ref{defn:parameted-perfect-matchings} can be found in Fig.\ref{fig:1-definition}, Fig.\ref{fig:2-definition} and Fig.\ref{fig:3-definition}.\paralled
\end{rem}

\begin{figure}[h]
\centering
\includegraphics[width=14cm]{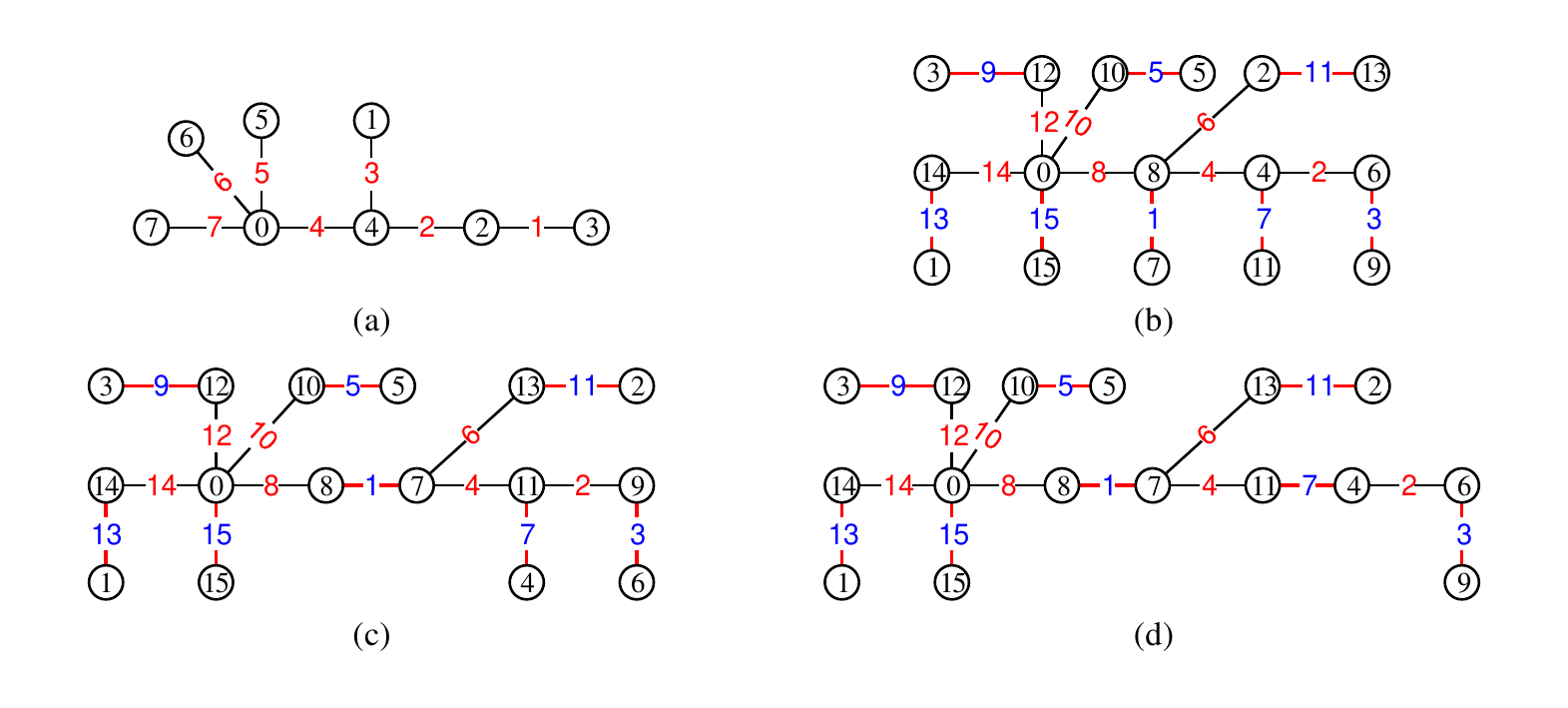}\\
\caption{\label{fig:matching-tree-leaf-matching}{\small A process of obtaining a \emph{leaf-matching tree} shown in (b), cited from \cite{Sun-Wang-Yao-2020}.}}
\end{figure}

\begin{lem} \label{them:parameted-particular-trees-perfect-matching}
\cite{Sun-Wang-Yao-2020} For the following eight labelings:

(1) strongly set-ordered graceful labeling;

(2) strongly $(\alpha,\beta)$-graceful labeling;

(3) AM-2-parameter totally odd/even-graceful labeling;

(4) AM-2-parameter totally odd/even-elegant labeling;

(5) AM-2-parameter arithmetic labeling;

(6) AM-2-parameter super felicitous labeling;

(7) AM-2-parameter super edge-magic total labeling; and

(8) AM-2-parameter super edge-magic graceful labeling.

Then a \emph{leaf-matching tree} $T$ (see Fig.\ref{fig:matching-tree-leaf-matching}) admits an $(i)$-labeling if and only if $T$ admits a $(j)$-labeling too for $i,j\in [1,8]$ and $i\neq j$.
\end{lem}

\begin{lem} \label{thm:parameted-exchange-labels-in-perfect-matching}
\cite{Sun-Wang-Yao-2020} Suppose that a tree $T$ has a perfect matching $M(T)$ and admits
a strongly graceful labeling $\phi$. There are three operations:

$(a)$ \textbf{Exchanging-edge operation}: A new strongly
graceful labeling $\varphi$ of $T$ holding $\varphi(x)=\phi(y)$ and $\varphi(y)=\phi(x)$ for
each matching edge $xy\in M(T)$.

$(b)$ \textbf{Decreasing-leaf operation}: If a matching edge $xy\in M(T)$ satisfies the neighborhood $N(x)=\{y,x_i:i\in [1,d_x-1]\}$ with $d_x=\textrm{deg}_T(x)\geq 3$ and $deg_T(y)=1$. Deleting the edges $xx_j$ and joins $y$ with $x_j$ for $j\in [s+1,d_x-1]$ with $0\leq s<d_x-1$ produces a new tree $H$ such that $H$ is also strongly graceful, and its perfect matching $M(H)$ keeping $|L(T)\cap V(M(T))|-1=|L(H)\cap V(M(H))|$.

$(c)$ \textbf{Increasing-leaf operation}: If a matching edge $xy\in M(T)$ satisfies $deg_T(x)\geq 2$ and $deg_T(y)\geq 2$. A process of (i) moving the edge $xy$ from $T$; (ii) identifying the vertex $x$ with the vertex $y$ into one denoted as $x_0$; and (iii) adding a new vertex $y_0$ to join with $x_0$ will build up another tree $T\,'$ admitting a strongly graceful labeling. Furthermore, the perfect matching $M(T\,')=\{x_0y_0\}\cup \big (M(T)\setminus\{xy\}\big )$ golding $|L(T)\cap V(M(T))|+1=|L(T\,')\cap V(M(T\,'))|$ true.
\end{lem}

By Lemma \ref{them:parameted-particular-trees-perfect-matching} and Lemma \ref{thm:parameted-exchange-labels-in-perfect-matching} we have

\begin{thm} \label{thm:parameted-main-result}
\cite{Sun-Wang-Yao-2020} For a tree $T$ having a perfect matching $M$ and admitting a set-ordered graceful labeling defined in Definition \ref{defn:basic-W-type-labelings}, there exist eight strongly 2-parameter graph labelings defined in Definition \ref{defn:parameted-perfect-matchings}, such that they are equivalent from each other.
\end{thm}

\subsection{Meta-type of labelings}

\begin{defn} \label{defn:proper-bipartite-labeling-ongraphs}
\cite{Yao-Zhang-Hongyu-Wang-Yao-2013} Suppose that a connected $(p,q)$-graph $G$ admits a mapping $\pi:V(G)\rightarrow \{0,1,2,\dots \}$. The induced edge colors of edges $xy\in E(G)$ are defined as $\pi(xy)=|\pi(x)-\pi(y)|$. Write $\pi(V(G))=\{\pi(u):u\in V(G)\}$ and $\pi(E(G))=\{\pi(xy):xy\in E(G)\}$, and there are constraints:
\begin{asparaenum}[(a)]
\item \label{Proper01} $|\pi(V(G))|=p$;
\item \label{Proper02} $|\pi(E(G))|=q$;
\item \label{Graceful-001} $\pi(V(G))\subseteq [0,q]$, $\min \pi(V(G))=0$;
\item \label{Odd-graceful-001} $\pi(V(G))\subset [0,2q-1]$, $\min \pi(V(G))=0$;
\item \label{Graceful-002} $\pi(E(G))=\{\pi(xy):xy\in E(G)\}=[1,q]$;
\item \label{Odd-graceful-002} $\pi(E(G))=\{\pi(xy):xy\in E(G)\}=[1,2q-1]^{o}$;
\item \label{Set-ordered} $G$ is a bipartite graph with bipartition $(X,Y)$ such that $\max\{\pi(x):x\in X\}<\min \{\pi(y):y\in Y\}$,
and write this case as $\pi(X)<\pi(Y)$ for short;
\item \label{Vertex-ordered} Each $u\in V(G)$ holds $\pi(x)<\pi(u)$ for $x\in N(u)$ (resp. $\pi(z)>\pi(u)$ for $z\in N(u)$) true;
\item \label{Meta-vertex-ordered} $G$ is a bipartite graph with bipartition $(X,Y)$ such that every vertex $y\in Y$ holds $\pi(x)<\pi(y)$ when $x\in N(y)$;
\item \label{Graceful-matching} $G$ has a perfect matching $M$ such that $\pi(x)+\pi(y)=q$ for $xy\in M$; and
\item \label{Odd-graceful-matching} $G$ has a perfect matching $M$ such that $\pi(x)+\pi(y)=2q-1$ for $xy\in M$.
\end{asparaenum}
\noindent \textbf{We have}: A \emph{proper labeling} $\pi$ holds (\ref{Proper01}) true and, a \emph{proper total labeling} $\pi$ holds (\ref{Proper01}) and
(\ref{Proper02}) true.

\textbf{Graceful class}: (1) a \emph{graceful labeling} $\pi$ holds (\ref{Proper01}), (\ref{Graceful-001}) and (\ref{Graceful-002}) true; (2) a \emph{set-ordered graceful labeling} $\pi$ holds (\ref{Proper01}),
(\ref{Graceful-001}), (\ref{Graceful-002}) and (\ref{Set-ordered}) true; (3) a \emph{strongly graceful labeling} $\pi$ holds (\ref{Proper01}), (\ref{Graceful-001}), (\ref{Graceful-002}) and (\ref{Graceful-matching}) true; (4) a \emph{strongly set-ordered graceful labeling} $\pi$ holds (\ref{Proper01}), (\ref{Graceful-001}), (\ref{Graceful-002}), (\ref{Set-ordered}) and (\ref{Graceful-matching}) true; (5) a \emph{vertex-ordered graceful
labeling} $\pi$ holds (\ref{Proper01}), (\ref{Graceful-001}), (\ref{Graceful-002}) and (\ref{Vertex-ordered}) true; (6) a \emph{meta-set-ordered graceful labeling} $\pi$ holds (\ref{Proper01}), (\ref{Graceful-001}), (\ref{Graceful-002}) and (\ref{Meta-vertex-ordered}) true; (7) a \emph{strongly meta-set-ordered graceful labeling} $\pi$ holds (\ref{Proper01}), (\ref{Graceful-001}), (\ref{Graceful-002}), (\ref{Meta-vertex-ordered}) and (\ref{Graceful-matching}) true.

\textbf{Odd-graceful class}: (1) an \emph{odd-graceful labeling} $\pi$ holds (\ref{Proper01}), (\ref{Odd-graceful-001}) and (\ref{Odd-graceful-002}) true; (2) a \emph{set-ordered odd-graceful labeling} $\pi$ holds (\ref{Proper01}), (\ref{Odd-graceful-001}), (\ref{Odd-graceful-002}) and (\ref{Set-ordered}) true; (3) a \emph{strongly odd-graceful labeling} $\pi$ holds (\ref{Proper01}), (\ref{Odd-graceful-001}), (\ref{Odd-graceful-002}) and
(\ref{Odd-graceful-matching}) true; (4) a \emph{strongly set-ordered odd-graceful labeling} $\pi$ holds (\ref{Proper01}), (\ref{Odd-graceful-001}), (\ref{Odd-graceful-002}), (\ref{Set-ordered}) and (\ref{Odd-graceful-matching}) true; (5) a \emph{vertex-ordered odd-graceful labeling} $\pi$ holds (\ref{Proper01}), (\ref{Odd-graceful-001}), (\ref{Odd-graceful-002}) and (\ref{Vertex-ordered}) true; (6) a \emph{meta-set-ordered odd-graceful labeling} $\pi$ holds (\ref{Proper01}), (\ref{Odd-graceful-001}), (\ref{Odd-graceful-002}) and (\ref{Meta-vertex-ordered}) true; (7) a \emph{strongly meta-set-ordered odd-graceful labeling} $\pi$ holds (\ref{Proper01}), (\ref{Odd-graceful-001}),
(\ref{Odd-graceful-002}), (\ref{Meta-vertex-ordered}) and (\ref{Odd-graceful-matching}) true.\qqed
\end{defn}

\begin{rem} \label{thm:000000}
The condition `set-ordered' is stronger than `meta-set-ordered', and furthermore `meta-set-ordered' is stronger than `vertex-ordered' in Definition
\ref{defn:proper-bipartite-labeling-ongraphs}. Clearly, a set-ordered graceful labeling is a meta-set-ordered graceful labeling, also, a vertex-ordered graceful labeling.\paralled
\end{rem}

\begin{defn} \label{defn:dual-other-labelings}
\cite{Yao-Zhang-Hongyu-Wang-Yao-2013} Let $G$ be a bipartite $(p,q)$-graph with bipartition $(X,Y)$, where $X=\{x_i:i\in [1,s]\}$ and $Y=\{y_j:j\in [1,t]\}$. We have the
following labelings generated from a proper labeling $\pi$ with $\pi(x_{i})<\pi(x_{i+1})$ for $i\in [1,s-1]$ and $\pi(y_{j})<\pi(y_{j+1})$ for $j\in [1,t-1]$.

$(i)$ The \emph{dual labeling} $\pi\,'$ of $\pi$ is defined as $\pi\,'(u)=\max\pi(V(G))+\min\pi(V(G))-\pi(u)$ for $u\in V(G)$.

$(ii)$ The \emph{partially $X$-dual labeling} $\phi$ of $\pi$ is defined as $\phi(x)=\pi(x_s)+\pi(x_1)-\pi(x)$ for $x\in X$, and $\phi(y)=\pi(y)$ for $y\in Y$; and the \emph{partially $Y$-dual
labeling} $\varphi$ of $\pi$ defined as $\varphi(y)=\pi(y_t)+\pi(y_1)-\pi(y)$ for $y\in Y$, and $\varphi(x)=\pi(x)$ for $x\in X$.

$(iii)$ The \emph{partially $X$-reciprocal transformation} $\pi^{-1}_X$ of $\pi$ is defined as $\pi^{-1}_X(x_i)=\pi(x_{s-i+1})$ for $x_i\in X$, and $\pi^{-1}_X(y_j)=\pi(y_j)$ for $y_j\in Y$. The \emph{partially $Y$-reciprocal transformation} $\pi^{-1}_Y$ of $\pi$ is defined as $\pi^{-1}_Y(y_j)=\pi(y_{t-j+1})$ for $y_j\in Y$, and $\pi^{-1}_Y(x_i)=\pi(x_i)$ for $x_i\in X$.

$(iv)$ A \emph{$(a,b)$-linear transformation} $\psi$ of $\pi$ is defined as $\psi(x)=a\cdot \pi(x)$ for $x\in X$ and $\psi(y)=b+a\cdot \pi(y)$ for $y\in Y$, where integers $b\geq 0$ and $a\geq 1$.\qqed
\end{defn}

\begin{rem}\label{rem:333333}
By the above Definitions \ref{defn:proper-bipartite-labeling-ongraphs} and \ref{defn:dual-other-labelings} we have: Suppose that a proper labeling $\pi$ of a bipartite graph $G$ with the bipartition $(X,Y)$ holds $\pi(X)<\pi(Y)$ true, where $G$ and $\pi$ are defined in Definition \ref{defn:proper-bipartite-labeling-ongraphs}. Then

(i) $\pi\,'(X)<\pi\,'(Y)$ for the dual labeling $\pi\,'$ of $\pi$;

(ii) $\phi(X)<\phi(Y)$ for the partially $X$-dual labeling $\phi$ of $\pi$; and $\varphi(X)<\varphi(Y)$ for the partially $Y$-dual labeling $\varphi$ of $\pi$; and

(iii) $\max\{\psi(x):x\in X\}<\min\{\psi(y)-b:y\in Y\}$ for a $(a,b)$-linear transformation $\psi$ of $\pi$, where integers $b\geq 0$ and $a\geq 1$.\paralled
\end{rem}

\begin{defn} \label{defn:k-d-graceful-elegant-labelings}
\cite{Yao-Zhang-Hongyu-Wang-Yao-2013} Let $Ao(q;k,d)^o=\{k+d,k+3d,\dots ,k+(2q-1)d\}$ for integers $q,d\geq 1$ and $k\geq 0$. Suppose a $(p,q)$-graph $G$ admits a proper labeling
$\pi:V(G)\rightarrow [0,k+(2q-1)d]$ for some integers $d\geq 1$ and
$k\geq 0$.

$(i)$ If the induced edge color set $\pi(E(G))=\{\pi(uv)=|\pi(u)-\pi(v)|:uv\in E(G)\}=Ao(q;k,d)^o$, we say $G$ to be \emph{$(k,d)$-odd-graceful}, and call $\pi$ a \emph{$(k,d)$-odd-graceful labeling} of $G$.

$(ii)$ If the induced edge color set $\{\pi(u)+\pi(v)-k~(\bmod~2qd):uv\in E(G)\}=Ao(q;0,d)^o$, we say $G$ to be \emph{$(k,d)$-odd-elegant}, and call $\pi$ a \emph{$(k,d)$-odd-elegant labeling} of $G$.\qqed
\end{defn}

\begin{defn} \label{defn:111111}
\cite{Yao-Zhang-Hongyu-Wang-Yao-2013} If the graph $G$ in Definition \ref{defn:k-d-graceful-elegant-labelings} has a perfect matching $M$ such that $\pi(u)+\pi(v)=k+(2q-1)d$ for every edge $uv\in M$, then $\pi$ is called a \emph{strongly $(k,d)$-odd-graceful labeling}. Similarly, we can define \emph{strongly set-ordered $(k,d)$-odd-graceful labelings}. Suppose that a bipartite $(p,q)$-graph $H$ with bipartition $(X,Y)$ admits a $(k,d)$-odd-graceful labeling $\pi$ such that $\pi(x)=i_xd$ for $x\in X$ and $\pi(y)=k+i_yd$ for $y\in Y$.

(1) If $\max\{\pi(x):x\in X\}<\min \{\pi(y)-k:y\in Y\}$, we call $\pi$ a \emph{set-ordered $(k,d)$-odd-graceful labeling} (resp. \emph{set-ordered $(k,d)$-odd-elegant labeling}).

(2) $\pi$ is called a \emph{meta-set-ordered $(k,d)$-odd-graceful labeling} (resp. meta-set-ordered $(k,d)$-odd-elegant labeling) if there are some $x_0\in X$ and $y_0\in Y$ such that $\pi(x_0)>\pi(y_0)-k$, and $\pi(u)-k>\max \{\pi(v):v\in N(u)\}$ for each vertex $u\in Y$.

(3) $\pi$ is called a \emph{separately meta-set-ordered $(k,d)$-odd-graceful labeling} (resp. separately meta-set-ordered $(k,d)$-odd-elegant labeling) if there are some $x_0\in X$ and $y_0\in Y$ such that $\pi(x_0)>\pi(y_0)-k$, and every vertex $u\in Y$ holds $\pi(u)-k>\max \{\pi(v):v\in N(u)\}$ and $\pi(u)-k+d\neq \pi(x)$ for every vertex $x\in X$.\qqed
\end{defn}

\begin{rem} \label{thm:aaaaaa}
In Definition \ref{defn:k-d-graceful-elegant-labelings}, a proper labeling $\pi$ means that integers $d\geq 1$ and $k\geq 0$ together hold $\pi(u)\neq \pi(v)$ for distinct vertices $u,v\in V(G)$. A separately
meta-set-ordered $(0,1)^o$-graceful labeling is a meta-set-ordered odd-graceful labeling, and a strongly separated and meta-set-ordered $(0,1)^o$-graceful labeling is a strongly meta-set-ordered odd-graceful labeling. A $(0,1)^o$-elegant labeling is an odd-elegant labeling defined in \cite{xiangqian-zhou2012}.\paralled
\end{rem}

\begin{figure}[h]
\centering
\includegraphics[width=15cm]{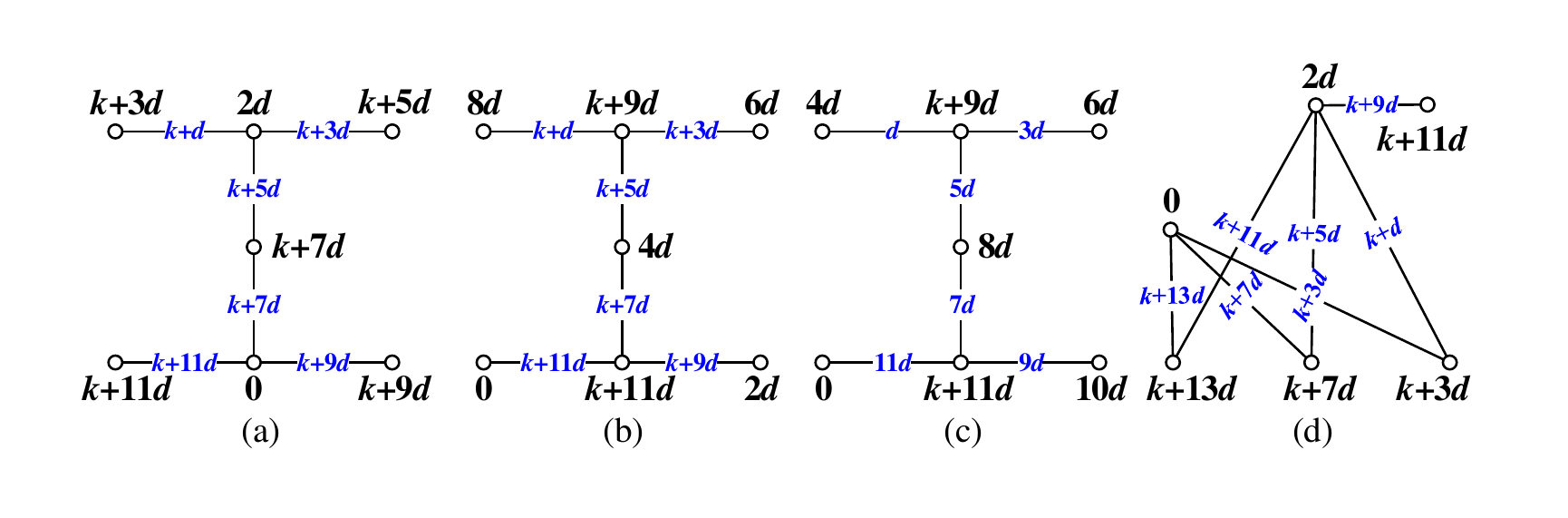}
\caption{\label{fig:k-d-Odd-type-000}{\small For illustrating Definition \ref{defn:k-d-graceful-elegant-labelings}: (a) A set-ordered $(k,d)$-odd-graceful labeling $f$ for all integers $d\geq 1$ and $k\geq 0$; (b) the dual labeling of $f$ is set-ordered $(k,d)$-odd-graceful for all integers $d\geq 1$ and $k\geq 0$; (c) a meta-set-ordered $(k,d)$-odd-elegant labeling for $k\neq d$; (d) a graph $G$ admitting a set-ordered $(k,d)$-odd-graceful labeling for all integers $d\geq 1$ and $k\geq 0$, cited from \cite{Yao-Zhang-Hongyu-Wang-Yao-2013}.}}
\end{figure}

\begin{figure}[h]
\centering
\includegraphics[width=14.2cm]{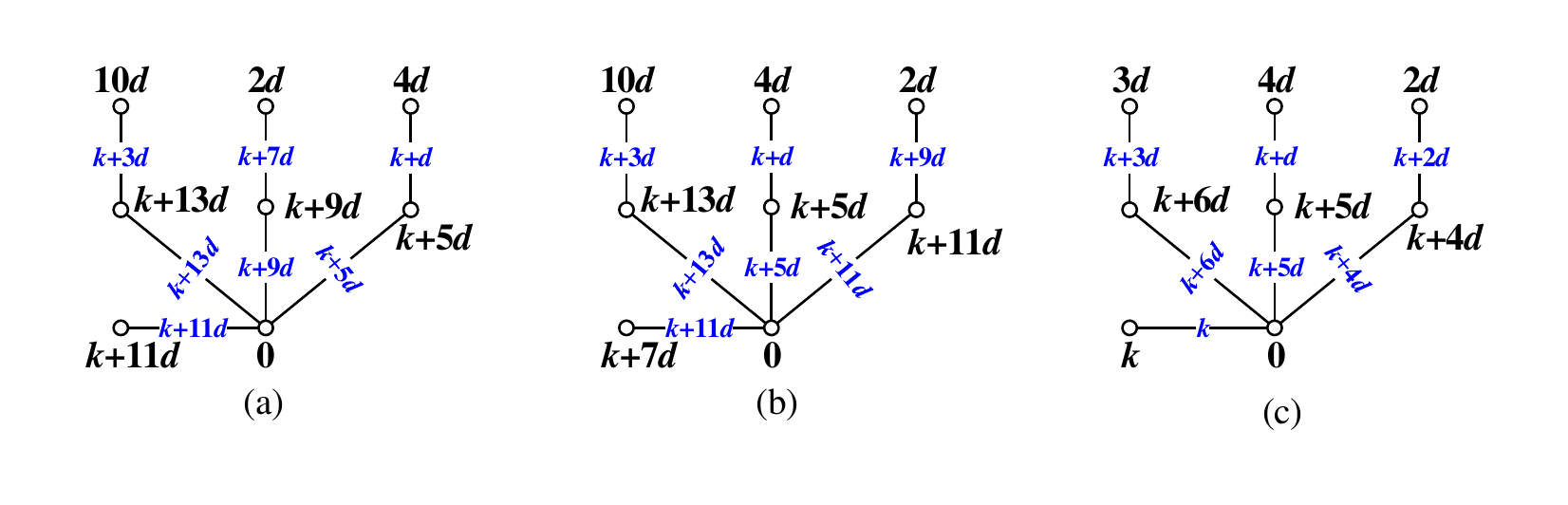}
\caption{\label{fig:k-d-Odd-type-001}{\small For illustrating Definition \ref{defn:k-d-graceful-elegant-labelings}, a tree $T$ admits: (a) a meta-set-ordered $(k,d)$-odd-graceful labeling for $(k,d)\not\in\{ (d,d),(5d,d)\}$; (b) a separately meta-set-ordered $(k,d)$-odd-graceful labeling for $k\neq 3d,5d$; (c) a $(k,d)$-graceful labeling for $k\not\in \{2d,3d,4d\}$ is balanced if $k>m=4d$, cited from \cite{Yao-Zhang-Hongyu-Wang-Yao-2013}.}}
\end{figure}

\subsection{Labelings with more restrictions}

\begin{defn}\label{defn:6C-labeling}
\cite{Yao-Sun-Zhang-Mu-Sun-Wang-Su-Zhang-Yang-Yang-2018arXiv} A total labeling $f:V(G)\cup E(G)\rightarrow [1,p+q]$ for a bipartite $(p,q)$-graph $G$ is a bijection and holds:

(i) (e-magic) $f(uv)+|f(u)-f(v)|=k$;

(ii) (ee-difference) each edge $uv$ matches with another edge $xy$ holding one of $f(uv)=|f(x)-f(y)|$ and $f(uv)=2(p+q)-|f(x)-f(y)|$ true;

(iii) (ee-balanced) let $s(uv)=|f(u)-f(v)|-f(uv)$ for $uv\in E(G)$, then there exists a constant $k\,'$ such that each edge $uv$ matches with another edge $u\,'v\,'$ holding one of $s(uv)+s(u\,'v\,')=k\,'$ and $2(p+q)+s(uv)+s(u\,'v\,')=k\,'$ true;

(iv) (EV-ordered) $\min f(V(G))>\max f(E(G))$, or $\max f(V(G))<\min f(E(G))$, or $f(V(G))\subseteq f(E(G))$, or $f(E(G))$ $\subseteq f(V(G))$, or $f(V(G))$ is an odd-set and $f(E(G))$ is an even-set;

(v) (ve-matching) there exists a constant $k\,''$ such that each edge $uv$ matches with one vertex $w$ such that $f(uv)+f(w)=k\,''$, and each vertex $z$ matches with one edge $xy$ such that $f(z)+f(xy)=k\,''$, except the \emph{singularity} $f(x_0)=\lfloor \frac{p+q+1}{2}\rfloor $;

(vi) (set-ordered) $\max f(X)<\min f(Y)$ (resp. $\min f(X)>\max f(Y)$) for the bipartition $(X,Y)$ of $V(G)$.

We refer to $f$ as a \emph{6C-labeling} of $G$.\qqed
\end{defn}

\begin{defn}\label{defn:odd-6C-labeling}
\cite{Yao-Zhang-Sun-Mu-Sun-Wang-Wang-Ma-Su-Yang-Yang-Zhang-2018arXiv} A $(p,q)$-graph $G$ admits a total labeling $f:V(G)\cup E(G)\rightarrow [1,4q-1]$. If this labeling $f$ holds:

(i) (e-magic) $f(uv)+|f(u)-f(v)|=k$, and $f(uv)$ is odd;

(ii) (ee-difference) each edge $uv$ matches with another edge $xy$ holding $f(uv)=2q+|f(x)-f(y)|$;

(iii) (ee-balanced) let $s(uv)=|f(u)-f(v)|-f(uv)$ for $uv\in E(G)$, then there exists a constant $k\,'$ such that each edge $uv$ matches with another edge $u\,'v\,'$ holding one of $s(uv)+s(u\,'v\,')=k\,'$ and $(p+q+1)+s(uv)+s(u\,'v\,')=k\,'$ true;

(iv) (EV-ordered) $\max f(V(G))<\min f(E(G))$, and $\{|a-b|:a,b\in f(V(G))\}=[1,2q-1]^o$;

(v) (ve-matching) there exists two constant $k_1,k_2$ such that each edge $uv$ matches with one vertex $w$ such that $f(uv)+f(w)=k_1$ or $k_2$;

(vi) (set-ordered) $\max f(X)<\min f(Y)$ for the bipartition $(X,Y)$ of $V(G)$.

We call $f$ an \emph{odd-6C-labeling} of $G$.\qqed
\end{defn}

\begin{defn}\label{defn:odd-even-separable-6C-labeling}
\cite{Yao-Sun-Zhang-Mu-Sun-Wang-Su-Zhang-Yang-Yang-2018arXiv} A total labeling $f:V(G)\cup E(G)\rightarrow [1,p+q]$ for a bipartite $(p,q)$-graph $G$ is a bijection and holds:

(i) (e-magic) $f(uv)+|f(u)-f(v)|=k$;

(ii) (ee-difference) each edge $uv$ matches with another edge $xy$ holding one of $f(uv)=|f(x)-f(y)|$ and $f(uv)=2(p+q)-|f(x)-f(y)|$ true;

(iii) (ee-balanced) let $s(uv)=|f(u)-f(v)|-f(uv)$ for $uv\in E(G)$, then there exists a constant $k\,'$ such that each edge $uv$ matches with another edge $u\,'v\,'$ holding one of $s(uv)+s(u\,'v\,')=k\,'$ and $2(p+q)+s(uv)+s(u\,'v\,')=k\,'$ true;

(iv) (EV-ordered) $\min f(V(G))>\max f(E(G))$ (resp. $\max f(V(G))<\min f(E(G))$, or $f(V(G))\subseteq f(E(G))$, or $f(E(G))$ $\subseteq f(V(G))$, or $f(V(G))$ is an odd-set and $f(E(G))$ is an even-set);

(v) (ve-matching) there exists a constant $k\,''$ such that each edge $uv$ matches with one vertex $w$ such that $f(uv)+f(w)=k\,''$, and each vertex $z$ matches with one edge $xy$ such that $f(z)+f(xy)=k\,''$, except the \emph{singularity} $f(x_0)=\lfloor \frac{p+q+1}{2}\rfloor $;

(vi) (set-ordered) $\max f(X)<\min f(Y)$ (resp. $\min f(X)>\max f(Y)$) for the bipartition $(X,Y)$ of $V(G)$.

(vii) (odd-even separable) $f(V(G))$ is an odd-set containing only odd numbers, as well as $f(E(G))$ is an even-set containing only even numbers.

We refer to $f$ as an \emph{odd-even separable 6C-labeling}.\qqed
\end{defn}

\begin{defn}\label{defn:multiple-meanings-vertex-labeling}
\cite{Yao-Zhang-Sun-Mu-Sun-Wang-Wang-Ma-Su-Yang-Yang-Zhang-2018arXiv, Yao-Wang-Su-Ma-Wang-Sun-ITNEC2020} A $(p,q)$-graph $G$ admits a \emph{multiple edge-meaning vertex labeling} $f:V(G)\cup E(G)\rightarrow [0,p+q]$ if

(1) $f(E(G))=[1,q]$ and $f(u)+f(uv)+f(v)=k$;

(2) $f(E(G))=[p,p+q-1]$ and $f(u)+f(uv)+f(v)=k\,'$;

(3) $f(E(G))=[0,q-1]$ and $f(uv)=f(u)+f(v)~(\bmod~q)$;

(4) $f(E(G))=[1,q]$ and $|f(u)+f(v)-f(uv)|=$a constant $k\,''$; and

(5) $f(uv)=$an odd number for each edge $uv\in E(G)$ holding $f(E(G))=[1,2q-1]^o$, and $\{f(u)+f(uv)+f(v):uv\in E(T)\}=[a,b]$ with $b-a+1=q$.\qqed
\end{defn}

\begin{defn}\label{defn:perfect-odd-graceful-labeling}
\cite{Yao-Zhang-Sun-Mu-Sun-Wang-Wang-Ma-Su-Yang-Yang-Zhang-2018arXiv} Let $f$ be an odd-graceful labeling of a $(p,q)$-graph $G$ defined in Definition \ref{defn:basic-W-type-labelings}, such that $f(V(G))\subset [0,2q-1]^o$ and $f(E(G))=[1,2q-1]^o$. If $\{|a-b|:~a,b\in f(V(G))\}=[1,p]$, then $f$ is called a \emph{perfect odd-graceful labeling} of $G$.\qqed
\end{defn}

\begin{defn}\label{defn:perfect-varepsilon-labeling}
\cite{Yao-Zhang-Sun-Mu-Sun-Wang-Wang-Ma-Su-Yang-Yang-Zhang-2018arXiv} Suppose that a $(p,q)$-graph $G$ admits an $\varepsilon$-labeling $h: V(G)\rightarrow S\subseteq [0,p+q]$. If $\{|a-b|:~a,b\in f(V(G))\}=[1,p]$, we call $f$ a \emph{perfect $\varepsilon$-labeling} of $G$.\qqed
\end{defn}

\begin{defn}\label{defn:6C-complementary-matching}
\cite{Yao-Sun-Zhang-Mu-Sun-Wang-Su-Zhang-Yang-Yang-2018arXiv} For a given $(p,q)$-tree $G$ admitting a 6C-labeling $f$ defined in Definition \ref{defn:6C-labeling}, and another $(p,q)$-tree $H$ admits a 6C-labeling $g$, if both graphs $G$ and $H$ hold $f(V(G))\setminus X^*=g(E(H))$, $f(E(G))=g(V(H))\setminus X^*$ and $f(V(G))\cap g(V(H))=X^*=\{z_0\}$ with $z_0=\lfloor \frac{p+q+1}{2}\rfloor $, then both labelings $f$ and $g$ are pairwise \emph{reciprocal-inverse}. The vertex-coincided graph $\odot_1\langle G,H \rangle $ obtained by vertex-coinciding the vertex $x_0$ of $G$ having $f(x_0)=z_0$ with the vertex $w_0$ of $H$ having $g(w_0)=z_0 $ is called a \emph{6C-complementary matching}.\qqed
\end{defn}

\begin{defn}\label{defn:reciprocal-inverse}
\cite{Yao-Sun-Zhang-Mu-Sun-Wang-Su-Zhang-Yang-Yang-2018arXiv} Suppose that a $(p,q)$-graph $G$ admits a $W$-type labeling $f:V(G)\cup E(G)\rightarrow [1,p+q]$, and a $(q,p)$-graph $H$ admits another $W$-type labeling $g:V(H)\cup E(H)\rightarrow [1,p+q]$. If $f(E(G))=g(V(H))\setminus X^*$ and $f(V(G))\setminus X^*=g(E(H))$ for $X^*=f(V(G))\cap g(V(H))\neq \emptyset$, then both $W$-type labelings $f$ and $g$ are \emph{reciprocal-inverse} (resp. \emph{reciprocal complementary}) from each other, and $H$ (resp. $G$) is an \emph{inverse matching} of $G$ (resp. $H$).\qqed
\end{defn}

\subsection{Image-type of labelings}

\subsubsection{Normal image-type of labelings}

\begin{defn}\label{defn:total-image-dual}
\cite{Yao-Zhang-Sun-Mu-Sun-Wang-Wang-Ma-Su-Yang-Yang-Zhang-2018arXiv} Suppose that a connected graph $G$ admits two $W_i$-type labelings $f_i$ for $i=1,2$. There are constants $M,k$ and $k\,'$ in the following constraint conditions:
\begin{asparaenum}[\textrm{C}-1. ]
\item \label{tt-edge-dual} $f_1(uv)+f_2(uv)=M_{edge}$ for each edge $uv\in E(G)$, where $M_{edge}=\max \{f_1(uv):uv\in E(G)\}+\min \{f_1(uv):uv\in E(G)\}$;
\item \label{tt-vertex-dual} $f_1(x)+f_2(x)=M_{vertex}$ for each vertex $x\in V(G)$, where $M_{vertex}=\max \{f_1(x):x\in V(G)\}+\min \{f_1(x):x\in V(G)\}$;
\item \label{tt-total-dual} $f_1(w)+f_2(w)=M$ for $w\in V(G)\cup E(G)$;
\item \label{tt-vertex-image} $f_1(x)+f_2(x)=k$ for each vertex $x\in V(G)$; and
\item \label{tt-edge-image} $f_1(uv)+f_2(uv)=k\,'$ for each edge $uv\in E(G)$.
\end{asparaenum}
\textbf{Then we have}:
\begin{asparaenum}[(i) ]
\item $f_2$ is an \emph{edge-dual labeling} of $f_{1}$ if C-\ref{tt-edge-dual} holds true.
\item $f_2$ is a \emph{vertex-dual labeling} of $f_{1}$ if C-\ref{tt-vertex-dual} holds true.
\item $f_i$ is a \emph{total dual labeling} of $f_{3-i}$ for $i=1,2$ if C-\ref{tt-total-dual} holds true.
\item $f_i$ is a \emph{vertex-image labeling} of $f_{3-i}$ for $i=1,2$ if C-\ref{tt-vertex-image} holds true.
\item $f_i$ is an \emph{edge-image labeling} of $f_{3-i}$ for $i=1,2$ if C-\ref{tt-edge-image} holds true.
\item $f_i$ is a \emph{totally image labeling} of $f_{3-i}$ for $i=1,2$ if C-\ref{tt-vertex-image} and C-\ref{tt-edge-image} hold true.\qqed
\end{asparaenum}
\end{defn}

\begin{figure}[h]
\centering
\includegraphics[width=16.4cm]{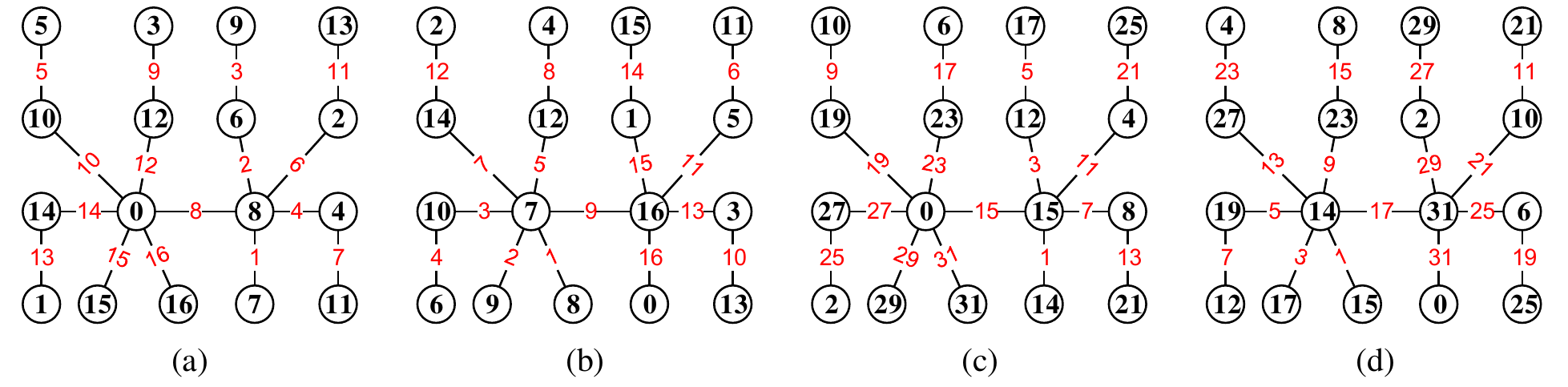}
\caption{\label{fig:graceful-image-aa}{\small Both (a) and (b) are a matching of \emph{set-ordered graceful image-labelings} with $f_a(uv)+h_b(uv)=17$; both (c) and (d) are a matching of \emph{set-ordered odd-graceful image-labelings} with $f_c(uv)+h_d(uv)=32$, cited from \cite{Yao-Zhang-Sun-Mu-Sun-Wang-Wang-Ma-Su-Yang-Yang-Zhang-2018arXiv}.}}
\end{figure}

\begin{defn}\label{defn:mirror-image-labeling}
\cite{Yao-Zhang-Sun-Mu-Sun-Wang-Wang-Ma-Su-Yang-Yang-Zhang-2018arXiv} Let $f_i:V(G)\rightarrow [a,b]$ be a labeling of a $(p,q)$-graph $G$ and let each edge $uv\in E(G)$ have its own color as $f_i(uv)=|f_i(u)-f_i(v)|$ with $i=1,2$. If each edge $uv\in E(G)$ holds $f_1(uv)+f_2(uv)=k$ true, where $k$ is a positive constant, we call both labelings $f_1$ and $f_2$ are a pair of \emph{image-labelings}, and $f_i$ a \emph{mirror-image} of $f_{3-i}$ with $i=1,2$ (see Fig.\ref{fig:graceful-image-aa}).\qqed
\end{defn}

\begin{defn}\label{defn:sujing-image-labelling7}
\cite{Su-Wang-Yao-Image-labelings-2021-MDPI} Let $f:V(G)\cup E(G)\rightarrow [a,b]$ and $g:V(G)\cup E(G)\rightarrow [a\,',b\,']$ be two labelings of a $(p,q)$-graph $G$, integers $a, b, a\,', b\,'$ satisfy $0\leq a<b$ and $0\leq a\,'<b\,'$.

(1) An equation $f(v)+g(v)=k\,'$ holds true for each vertex $v\in V(G)$, where $k\,'$ is a positive constant and it is called \emph{vertex-image coefficient}, then both labelings $f$ and $g$ are called a matching of \emph{vertex image-labelings} (abbreviated as \emph{v-image-labelings});

(2) An equation $f(uv)+g(uv)=k\,''$ holds true for every edge $uv\in E(G)$, and $k\,''$ is a positive constant, called \emph{edge-image coefficient}, then both labelings $f$ and $g$ are called a matching of \emph{edge image-labelings} (abbreviated as \emph{e-image-labelings}).\qqed
\end{defn}

\begin{figure}[h]
\centering
\includegraphics[width=13.4cm]{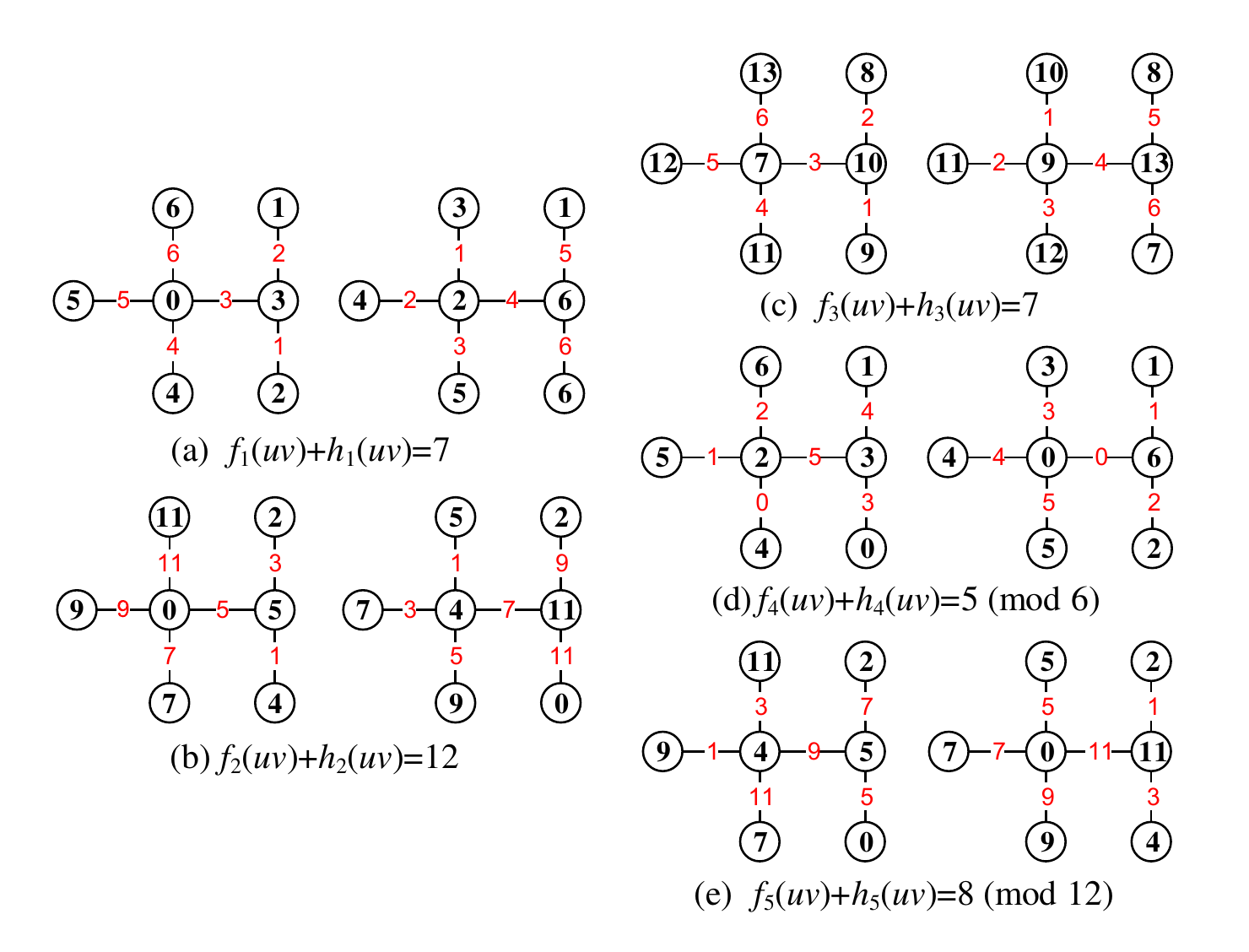}
\caption{\label{fig:graceful-image-11}{\small A tree $T$ admits (Ref. \cite{Gallian2020, Zhou-Yao-Chen-Tao2012, Zhou-Yao-Chen2013}): (a) a matching of set-ordered graceful image-labelings $f_1$ and $h_1$; (b) a matching of set-ordered odd-graceful image-labelings $f_2$ and $h_2$; (c) a matching of edge-magic graceful image-labelings $f_3$ and $h_3$; (d) a matching of set-ordered felicitous image-labelings $f_4$ and $h_4$; (e) a matching of set-ordered odd-elegant image-labelings $f_5$ and $h_5$.}}
\end{figure}

\begin{defn} \label{defn:twin-k-d-harmonious-image-labelings}
\cite{Yao-Zhang-Sun-Mu-Sun-Wang-Wang-Ma-Su-Yang-Yang-Zhang-2018arXiv} A $(p,q)$-graph $G$ admits two $(k,d)$-harmonious labelings $f_i:V(G)\rightarrow X_0\cup S_{q-1,k,d}$ with $i=1,2$, where $X_0=\{0,d,2d, \dots ,(q-1)d\}$ and $S_{q-1,k,d}=\{k,k+d,k+2d,\dots, k+(q-1)d\}$, such that each edge $uv\in E(G)$ is colored with $f_i(uv)-k=[f_i(uv)+f_i(uv)-k~(\textrm{mod}~qd)]$ with $i=1,2$. If $f_1(uv)+f_2(uv)=2k+(q-1)d$, we call both labelings $f_1$ and $f_2$ a \emph{matching $(k,d)$-harmonious image-labelings} of $G$.\qqed
\end{defn}

\begin{figure}[h]
\centering
\includegraphics[width=16.4cm]{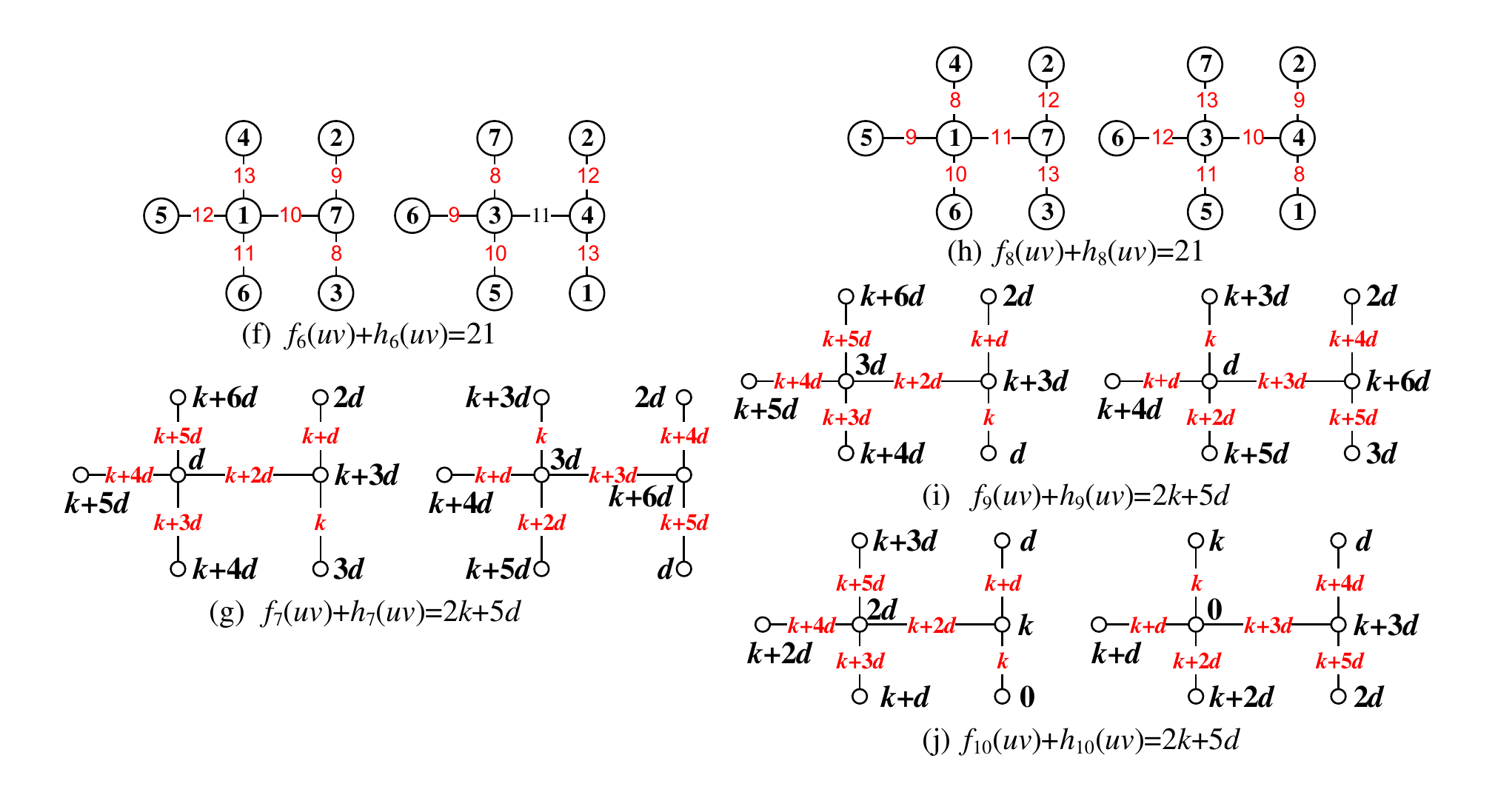}
\caption{\label{fig:graceful-image-parameter}{\small A tree $T$ admits (Ref. \cite{Gallian2020, Zhou-Yao-Chen-Tao2012, Zhou-Yao-Chen2013}): (f) a matching of super set-ordered edge-magic total image-labelings $f_6$ and $h_6$; (g) a matching of set-ordered $(k,d)$-graceful image-labelings $f_7$ and $h_7$; (h) a matching of super set-ordered edge-antimagic total image-labelings $f_8$ and $h_8$; (i) a matching of $(k,d)$-edge antimagic total image-labelings $f_9$ and $h_9$; (j) a matching of $(k,d)$-arithmetic image-labelings $f_{10}$ and $h_{10}$.}}
\end{figure}

\begin{thm} \label{thm:1}
\cite{Su-Wang-Yao-Image-labelings-2021-MDPI} Let $T$ be a tree with $p$ vertices and $q$ edges, its vertex partition is $(X,Y)$, where $|X|=s$ and $|Y|=t$. By Definition \ref{defn:sujing-image-labelling7}, if $T$ admits a set-ordered graceful labeling, the following assertions are mutually equivalent:

$(1)$ $T$ admits a \emph{matching of graceful $v$-image-labelings} with $k=p-1$.

$(2)$ $T$ admits a \emph{matching of odd-graceful $v$-image-labelings} with $k=2p-3$.

$(3)$ $T$ admits a \emph{matching of felicitous $v$-image-labelings} with $k=p-1$.

$(4)$ $T$ admits a \emph{matching of odd-elegant $v$-image-labelings} with $k=p$.

$(5)$ $T$ admits a \emph{matching of super edge-magic total $v$-image-labelings} with $k=p+1$.

$(6)$ $T$ admits a \emph{matching of set-ordered $(k,d)$-graceful $v$-image-labelings} with $k=k\,''+(q-1)d$.

$(7)$ $T$ admits a \emph{super $(s+p+3,2)$-edge anti-magic total v-image-laeblling} and a \emph{super $(s+p+4,2)$-edge anti-magic total labeling}, they are a \emph{matching of $v$-image-labelings} with $k=p+1$.

$(8)$ $T$ admits a \emph{matching of $(k,2)$-arithmetic total $v$-image-labelings} with $k=k\,''+2(t-1)$.
\end{thm}

\begin{cor} \label{cor:connections-several-labelings}
\cite{Su-Wang-Yao-Image-labelings-2021-MDPI} By Definition \ref{defn:sujing-image-labelling7}, if a tree $T$ admits a set-ordered graceful labeling, then the following propositions hold and are equivalent from each other.

$(1)$ $T$ admits a \emph{matching of edge-magic total v$e$-image-labelings} with $k=p+1$ and $k\,'=2p+q+1$.

$(2)$ $T$ admits a $(s+p+3,2)$-edge anti-magic total labeling and a \emph{$(s+p+4,2)$-edge anti-magic total labeling} such that they are a \emph{matching of v$e$-image-labelings} with $k=p+1$ and $k\,'=2p+q+1$.

$(3)$ $T$ admits a \emph{matching of $(k,2)$-arithmetic total v$e$-image-labelings} with $k=k\,''+2(t-1)$ and $k\,'=2(k\,''+2t-2)$.
\end{cor}

\begin{thm} \label{thm:2}
\cite{Su-Wang-Yao-Image-labelings-2021-MDPI} By Definition \ref{defn:sujing-image-labelling7}, if a tree $T$ admits a set-ordered graceful labeling, then the following assertions are mutually equivalent:

$(1)$ $T$ admits a \emph{matching of set-ordered graceful $e$-image-labelings} with $k\,'=q+1$.

$(2)$ $T$ admits a \emph{matching of set-ordered odd-graceful $e$-image-labelings} with $k\,'=2q$.

$(3)$ $T$ admits a \emph{matching of set-ordered felicitous $e$-image-labelings} with $k\,'=2s-1$.

$(4)$ $T$ admits a \emph{matching of odd-elegant $e$-image-labelings} with $k\,'=2q$.

$(5)$ $T$ admits a \emph{matching of super set-ordered edge-magic total $e$-image-labelings} with $k\,'=3(q+1)$.

$(6)$ $T$ admits a \emph{matching of set-ordered $(k,d)$-graceful $e$-image-labelings} with $k\,'=2k\,''+(q+1)d$.

$(7)$ $T$ admits a \emph{matching of set-ordered $(s+p+3,2)$-edge anti-magic total $e$-image-labelings} with $k\,'=3(q+1)$.

$(8)$ $T$ admits a \emph{matching of $(k,2)$-arithmetic total $e$-image-labelings} with $k\,'=k\,''+2q+1$.
\end{thm}

\begin{cor} \label{cor:connections-several-labelings}
\cite{Su-Wang-Yao-Image-labelings-2021-MDPI} By Definition \ref{defn:sujing-image-labelling7}, if a tree $T$ admit a matching of $v$-image-labelings $f,g$ and a matching of $e$-image-labelings $f,h$, then the following propositions hold, and

(1) the set $\{f(v)+g(v)+h(v): v\in V(T)\}$ is an arithmetic sequence if and only if $\{f(uv)+g(uv)+h(uv): uv\in E(T)\}$ is an arithmetic sequence, and the tolerances are equal;

(2) the set $\{f(v)+g(v)+h(v): v\in V(T)\}$ contains two arithmetic sequences if and only if $\{f(uv)+g(uv)+h(uv): uv\in E(T)\}$ is an arithmetic sequences with a tolerance of 2.
\end{cor}

\begin{thm} \label{thm:3}
\cite{Su-Wang-Yao-Image-labelings-2021-MDPI} By Definition \ref{defn:sujing-image-labelling7}, if a tree $T$ admits a graceful labeling $f$, then it admits an \emph{edge-magic total labeling} $g$, so that both labelings $f$ and $g$ are a \emph{matching of $e$-image-labelings} with $k\,'=p+q+1$.
\end{thm}

\begin{thm} \label{thm:4}
\cite{Su-Wang-Yao-Image-labelings-2021-MDPI} If a tree $T$ admits a graceful labeling $f$, then it admits an \emph{set-ordered $(s+p+3,2)$-edge anti-magic total labeling} $g$, so that both labelings $f$ and $g$ are a \emph{matching of $e$-image-labelings} with $k\,'=p+q+1$.
\end{thm}

\subsubsection{Duality image-type of labelings}

Let $\theta:V(L)\cup E(L)\rightarrow [1,M]$ be a labeling (resp. coloring) of a graph $L$, so we have three parameters:
\begin{equation}\label{eqa:dual-image}
{
\begin{split}
M^v_m(L,\theta)=&\max \theta(V(L))+\min \theta(V(L))\\
M^e_m(L,\theta)=&\max \theta(E(L))+\min \theta(E(L))\\
M^{ve}_m(L,\theta)=&\max \theta(V(L)\cup E(L))+\min \theta(V(L)\cup E(L))
\end{split}}
\end{equation}

We show the following duality-image $(W_f,W_g)$-type labelings (resp. colorings) including homomorphic labelings and homomorphic colorings:

\begin{defn}\label{defn:new-image-labeling-colorings}
\cite{Yao-Wang-Su-Sun-image-Type-2020} Suppose that a connected $(p,q)$-graph $G$ admits a $W_f$-type labeling (resp. coloring) $f$, and another connected $(p\,',q\,')$-graph $H$ admits a $W_g$-type labeling (resp. coloring) $g$. There is a bijection $\varphi: V(G)\cup E(G)\rightarrow V(H)\cup E(H)$ such that the image edge $\varphi(u)\varphi(v)\in E(H)$ for each edge $uv\in E(G)$. By the parameters defined in (\ref{eqa:dual-image}), there are the following constraint conditions:
\begin{asparaenum}[(\textrm{C}-1) ]
\item \label{asp:v-dual-f-g} $g(\varphi(x))=M^v_m(G,f)-f(x)$ for $x\in V(G)$;
\item \label{asp:e-dual-f-g} $g(\varphi(u)\varphi(v))=M^e_m(G,f)-f(uv)$ for $uv\in E(G)$;
\item \label{asp:ve-dual-f-g} $g(\varphi(w))=M^{ve}_m(G,f)-f(w)$ for $w\in V(G)\cup E(G)$ and $\varphi(w)\in V(H)\cup E(H)$;

\item \label{asp:v-dual-g-f} $f(\varphi^{-1}(y))=M^v_m(H,g)-g(y)$ for $y\in V(H)$ and $\varphi^{-1}(y)\in V(G)$;
\item \label{asp:e-dual-g-f} $f(\varphi^{-1}(w)\varphi^{-1}(z))=M^e_m(H,g)-g(wz)$ for $wz\in E(H)$ and $\varphi^{-1}(w)\varphi^{-1}(z)\in E(G)$;
\item \label{asp:ve-dual-g-f} $f(\varphi^{-1}(z))=M^{ve}_m(H,g)-g(z)$ for $z\in V(H)\cup E(H)$ and $\varphi^{-1}(z)\in V(G)\cup E(G)$;

\item \label{asp:vertex-image} $f(x)+g(\varphi(x))=A_v$ for $x\in V(G)$;
\item \label{asp:edge-image} $f(x)+g(\varphi(u)\varphi(v))=A_e$ for $uv\in E(G)$; and
\item \label{asp:ve-image} $f(w)+g(\varphi(w))=A_{ve}$ for $w\in V(G)\cup E(G)$ and $\varphi(w)\in V(H)\cup E(H)$.
\end{asparaenum}
\textbf{Then}
\begin{asparaenum}[\textrm{Labe}-1. ]
\item $g$ (resp. $f$) is a \emph{vertex-dual labeling (resp. coloring)} of $f$ (resp. $g$) if (C-\ref{asp:v-dual-f-g}) (resp. (C-\ref{asp:v-dual-g-f})) holds true.
\item $g$ (resp. $f$) is an \emph{edge-dual labeling (resp. coloring)} of $f$ (resp. $g$) if (C-\ref{asp:e-dual-f-g}) (resp. (C-\ref{asp:e-dual-g-f})) holds true.

\item $g$ is a \emph{mixed-dual labeling (resp. coloring)} of $f$ if (C-\ref{asp:ve-dual-f-g}) holds true.
\item $f$ is a \emph{mixed-dual labeling (resp. coloring)} of $g$ if (C-\ref{asp:ve-dual-g-f}) holds true.
\item $f$ is a \emph{mixed-image labeling (resp. coloring)} of $g$ if (C-\ref{asp:ve-image}) holds true.

\item Both $f$ and $g$ are a matching of \emph{totally-dual $(W_f,W_g)$-type labelings (resp. colorings)} if (C-\ref{asp:v-dual-f-g}), (C-\ref{asp:e-dual-f-g}), (C-\ref{asp:v-dual-g-f}) and (C-\ref{asp:e-dual-g-f}) hold true.
\item Both $f$ and $g$ are a matching of \emph{$(W_f,W_g)$-type vertex-image labelings (resp. colorings)} if (C-\ref{asp:vertex-image}) holds true.
\item Both $f$ and $g$ are a matching of \emph{$(W_f,W_g)$-type edge-image labelings (resp. colorings)} if (C-\ref{asp:edge-image}) holds true.
\item Both $f$ and $g$ are a matching of \emph{$(W_f,W_g)$-type total-image labelings (resp. colorings)} if (C-\ref{asp:vertex-image}) and (C-\ref{asp:edge-image}) hold true.

\item If (C-\ref{asp:v-dual-f-g}) and (C-\ref{asp:v-dual-g-f}) hold true, then both $f$ and $g$ are a matching of \emph{vertex-dual $(W_f,W_g)$-type homomorphic labelings (resp. colorings)}.
\item If (C-\ref{asp:e-dual-f-g}) and (C-\ref{asp:e-dual-g-f}) hold true, then both $f$ and $g$ are a matching of \emph{edge-dual $(W_f,W_g)$-type homomorphic labelings (resp. colorings)}.
\item If (C-\ref{asp:v-dual-f-g}), (C-\ref{asp:v-dual-g-f}) and (C-\ref{asp:edge-image}) hold true, then both $f$ and $g$ are a matching of \emph{vertex-dual $(W_f,W_g)$-type edge-image homomorphic labelings (resp. colorings)}.

\item If (C-\ref{asp:e-dual-f-g}), (C-\ref{asp:e-dual-g-f}) and (C-\ref{asp:vertex-image}) hold true, then both $f$ and $g$ are a matching of \emph{edge-dual $(W_f,W_g)$-type vertex-image homomorphic labelings (resp. colorings)}.
\item \label{exa:totally-dual-image-homomorphic} both $f$ and $g$ are a matching of \emph{totally-dual $(W_f,W_g)$-type total-image homomorphic labelings (resp. colorings)} if (C-\ref{asp:v-dual-f-g}), (C-\ref{asp:e-dual-f-g}), (C-\ref{asp:vertex-image}) and (C-\ref{asp:edge-image}) hold true.

\item If (C-\ref{asp:v-dual-f-g}) and (C-\ref{asp:e-dual-f-g}) hold true, then $g$ is a \emph{totally-dual homomorphic labeling (resp. coloring)} of $f$.
\item If (C-\ref{asp:v-dual-g-f}) and (C-\ref{asp:e-dual-g-f}) hold true, then $f$ is a \emph{totally-dual homomorphic labeling (resp. coloring)} of $g$.

\item \label{exa:totally-dual-image-homomorphic} Both $f$ and $g$ are a matching of \emph{uniformly dual-image $(W_f,W_g)$-type homomorphic labelings (resp. colorings)} if (C-\ref{asp:ve-dual-f-g}), (C-\ref{asp:ve-dual-g-f}) and (C-\ref{asp:ve-image}) hold true.\qqed
\end{asparaenum}
\end{defn}

\subsection{Improper-type of labelings}
\begin{defn} \label{defn:2-improper-graceful-labeling}
\cite{Hongyu-Wang-Bing-Yao-2013} Let $G$ be a $(p,q)$-graph, and $f$ be a mapping from $V(G)$ to $\{0,1,2,\dots \}$ such that $\min f(V(G))=0$ and $\max f(V(G))=m$, where $f(V(G))=\{f(x):x\in V(G)\}$. Write $f(E(G))=\{f(uv)=|f(u)-f(v)|:uv\in E(G)\}$. There are four restrictions:

C1. $f(E(G))=[1,q]$ and $m\leq q$;

C2. $f(E(G))=[1,q]$ and $m>q$;

C3. $f(E(G))=[1,2q-1]^o$ and $m\leq 2q-1$;

C4. $f(E(G))=[1,2q-1]^o$ and $m>2q-1$.\\
Then, $f$ is called an \emph{in-improper graceful labeling} (in-imgl) if $f$ holds C1, an \emph{out-improper graceful labeling} (out-imgl) if $f$ holds C2, an \emph{in-improper odd-graceful labeling} (in-imoddgl) if $f$ holds C3, and an \emph{out-improper odd-graceful labeling} (out-imoddgl) if $f$ holds C4.
\end{defn}

\begin{thm} \label{lemma:basic-theorem-000}
Every tree admits $(i)$ \emph{in-imgls} or \emph{out-imgls}; and $(ii)$ \emph{in-imoddgls} or \emph{out-imoddgls} (refer to Definition \ref{defn:2-improper-graceful-labeling}).
\end{thm}

The term ``adding leaves to a $(p,q)$-graph $G$'' means that we join new vertices $u_{i,1},u_{i,2},\dots ,u_{i,m_i}$ to each vertex $u_i$ of $G$, here, some $m_i$ may be zero for $i\in [1,p]$. So $u_{i,1},u_{i,2},\dots ,u_{i,m_i}$ for $i\in [1,p]$ are leaves in the resultant graph.

\begin{thm} \label{them:basic-theorem}
Suppose that a tree $T$ admit set-ordered graceful/odd-graceful labelings. Then adding leaves to $T$ produces a tree that admits \emph{in-imgls} (refer to Definition \ref{defn:2-improper-graceful-labeling}) and proper \emph{odd-graceful labelings}.
\end{thm}

\begin{thm} \label{thm:Corollary11}
Every lobster admits \emph{in-imgls} (refer to Definition \ref{defn:2-improper-graceful-labeling}) and proper \emph{odd-graceful labelings}.
\end{thm}

\begin{thm} \label{thm:1233333}
Let $G$ be a connected bipartite graph having vertices $w_1,w_2,\dots ,w_n$. For each fixed integer $i\in [1,n]$, identifying a certain vertex of every connected bipartite graph $G_{i,j}$ for $j\in [1,m_i]$ (it is allowed some $m_i=0$) with vertex $w_i$ into one vertex produces a connected graph $H^*$ admitting \emph{in-imgls} (resp. \emph{in-imoddgls}, refer to Definition \ref{defn:2-improper-graceful-labeling}) if $G$ and every $G_{i,j}$ admit set-ordered graceful labelings (resp. set-ordered odd-graceful labelings).
\end{thm}

\subsection{Pan-labelings}

\subsubsection{Pan-graceful labelings}

\begin{defn} \label{defn:general graceful labeling}
\cite{Sun-Zhang-Yao-IMCEC-2018} For a $(p, q)$-graph $G$, if there is a labeling $f:V (G) \rightarrow \{0,a_1,a_2,\dots $, $a_q\}$ with $0< a_i<a_j$ and $1\leq i<j\leq q$ such that the edge color set $\{f(uv) =|f(u)-f(v)|: uv\in E(G)\}=\{a_1,a_2,\cdots,a_q\}$, then we call $f$ a \emph{pan-graceful labeling} of $G$.\qqed
\end{defn}

\subsubsection{Pan-type mean labelings}

\begin{defn}\label{defn:pan-odd-graceful-mean-labeling-00}
$^*$ A $(p,q)$-graph $G$ admits a labeling $f:V(G)\rightarrow [0,M_f]$, and the induced edge color $f(uv)=\big \lceil \frac{f(u)+f(v)}{2}\big \rceil $. There are the following constraint conditions:
\begin{asparaenum}[\textrm{Cs}-1. ]
\item \label{V-mean-01} $f(V(G))\subseteq [0,q]$;
\item \label{V-mean-02} $f(V(G))\subset [0,2q-1]$;
\item \label{E-mean-01} $f(E(G))=[1,q]$; and
\item \label{E-mean-02} $f(E(G))=[1,2q-1]^o$.
\end{asparaenum}
\textbf{We call $f$}:
\begin{asparaenum}[\textrm{MeanL}-1. ]
\item A \emph{graceful mean labeling} if Cs-\ref{V-mean-01} and Cs-\ref{E-mean-01} hold true.
\item A \emph{pan-graceful mean labeling} if Cs-\ref{E-mean-01} holds true.
\item An \emph{odd-graceful mean labeling} if Cs-\ref{V-mean-02} and Cs-\ref{E-mean-02} hold true.
\item A \emph{pan-odd-graceful mean labeling} if Cs-\ref{E-mean-02} holds true.\qqed
\end{asparaenum}
\end{defn}

\begin{figure}[h]
\centering
\includegraphics[width=13.4cm]{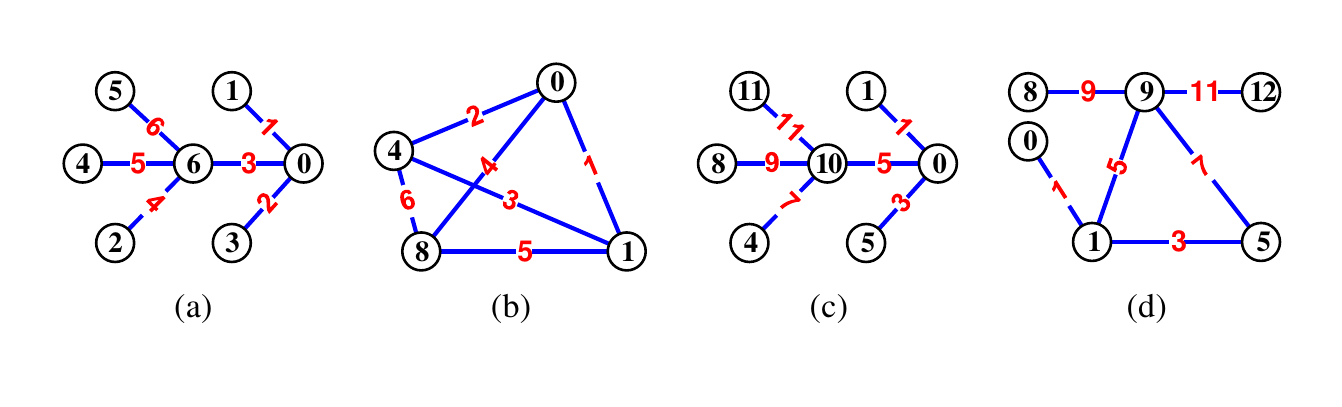}\\
\caption{\label{fig:pan-graceful-mean-00}{\small (a) A graceful mean labeling; (b) a pan-graceful mean labeling; (c) an odd-graceful mean labeling; (d) a graph having a triangle does not admit any odd-graceful labeling, but it admits an odd-graceful mean labeling.}}
\end{figure}

For understanding Definition \ref{defn:pan-odd-graceful-mean-labeling-00}, see Fig.\ref{fig:pan-graceful-mean-00}.

\begin{defn}\label{defn:pan-edge-magic-graceful-labeling-1}
$^*$ In Definition \ref{defn:Marumuthu-edge-magic-graceful-labeling} of an edge-magic graceful labeling, we substitute the condition ``$|f(u)+f(v)-f(uv)|=k$ and $f(V(G)\cup E(G))=[1, p+q]$'' by one of ``$|f(u)+f(v)-f(uv)|=k$ or $(p+q)-|f(u)+f(v)-f(uv)|=k$ such that $f(V(G)\cup E(G))=[1, p+q]$''. The resulting labeling is called a \emph{pan-edge-magic graceful labeling}.\qqed
\end{defn}

\begin{defn}\label{defn:pan-edge-magic-graceful-labeling}
$^*$ A \emph{pan-edge-magic total graceful labeling} $g$ is obtained by replacing the restriction ``$g(uv)+|g(u)-g(v)|=k$'' by one of $g(uv)+|g(u)-g(v)|=k$ and $g(uv)+(p+q)-|g(u)-g(v)|=k$ in Definition \ref{defn:edge-magic-total-graceful-labeling}.\qqed
\end{defn}

\begin{defn}\label{defn:pan-odd-edge-magic-graceful-labeling}
$^*$ A \emph{pan-odd-graceful labeling} is obtained by replacing the restriction ``$f(uv)=|f(u)-f(v)|$ and $f(E(G))=[1, 2q-1]^o$'' by a new one of $f(uv)=|f(u)-f(v)|$, $f(uv)=2q-1-|f(u)-f(v)|$ and $f(E(G))=[1, 2q-1]^o$ Definition \ref{defn:basic-W-type-labelings}.\qqed
\end{defn}

\begin{figure}[h]
\centering
\includegraphics[width=16cm]{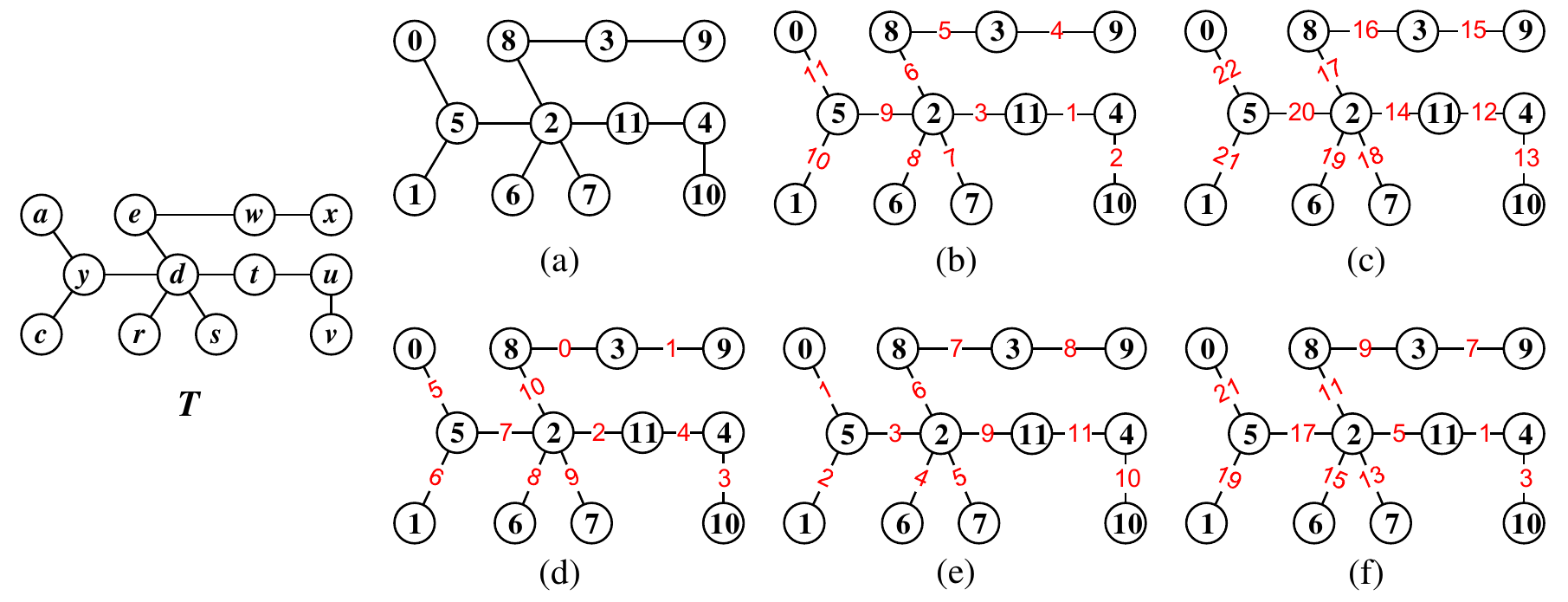}
\caption{\label{fig:a-labeling-many-meanings}{\small A lobster $T$ admits a vertex labeling $f$ shown in (a), and $f$ induces other labelings from (b) to (f).}}
\end{figure}

As an example, a lobster $T$ demonstrated in Fig.\ref{fig:a-labeling-many-meanings} admits:

(1) a pan-edge-magic total labeling $f_1$ such that $f_1(u)+f_1(uv)+f_1(v)=16$;

(2) a pan-edge-magic total labeling $f_2$ holding $f_2(u)+f_2(uv)+f_2(v)=27$ true;

(3) a felicitous labeling $f_3$ satisfying $f_3(uv)=f_3(u)+f_3(v)~(\bmod~11)$;

(4) an edge-magic graceful labeling $f_4$ such that $|f_4(u)+f_4(v)-f_4(uv)|=4$;

(5) an edge-odd-graceful labeling $f_5$ keeping $f_5(uv)=$an odd number, $\{f_5(u)+f_5(uv)+f_5(v):uv\in E(T)\}=[16,26]$.

So, $T$ admits an \emph{e-set v-proper labeling} (see Definition \ref{defn:55-set-labeling}) defined as follows: Let $f:V(T)\rightarrow [0,11]$ with $f(a)=0$, $f(c)=1$, $f(d)=2$, $f(w)=3$, $f(u)=4$, $f(y)=5$, $f(r)=6$, $f(s)=7$, $f(e)=8$, $f(x)=9$, $f(v)=10$, $f(t)=11$. And each edge of $T$ has its own color set as follows:
$$
\begin{array}{lll}
f'(ay)=\{1,5,11,21,22\},&f'(cy)=\{2,6,10,19,21\},&f'(de)=\{6,10,11,17\},\\
f'(dy)=\{3,7,9,17,20\},&f'(dr)=\{4,8,15,19\},&f'(ds)=\{5,7,9,13,18\},\\
f'(dt)=\{2,3,5,9,14\}, &f'(ew)=\{0,5,7,9,16\}, &f'(ut)=\{1,4,11,12\},\\
f'(xw)=\{1,4,7,8,15\},&f'(uv)=\{2,3,10,13\}.
\end{array}
$$

\subsection{Sum-type of labelings}

\begin{defn}\label{defn:difference-sum-labeling}
\cite{Yao-Zhang-Sun-Mu-Sun-Wang-Wang-Ma-Su-Yang-Yang-Zhang-2018arXiv} Let $f:V(G)\rightarrow [0,q]$ be a proper labeling of a $(p,q)$-graph $G$, and let
$$S_{um}(G,f)=\sum_{uv\in E(G)}|f(u)-f(v)|,$$ we call $f$ a \emph{difference-sum labeling}. Find two extremal numbers $\max_f S_{um}(G,f)$ and $\min_f S_{um}(G,f)$ over all difference-sum labelings of $G$.\qqed
\end{defn}

\begin{defn}\label{defn:felicitous-sum-labeling}
\cite{Yao-Zhang-Sun-Mu-Sun-Wang-Wang-Ma-Su-Yang-Yang-Zhang-2018arXiv} Let $f:V(G)\rightarrow [0,q]$ be a proper labeling of a $(p,q)$-graph $G$, and let $$F_{um}(G,f)=\sum_{uv\in E(G)}[f(u)+f(v)]~(\bmod ~q+1),$$ we call $f$ a \emph{felicitous-sum labeling}. Find two extremal numbers $\max_f F_{um}(G,f)$ and $\min_f F_{um}(G,f)$ over all felicitous-sum labelings of $G$.\qqed
\end{defn}

\begin{defn}\label{defn:three-new-sum-labeling}
\cite{Yao-Mu-Sun-Sun-Zhang-Wang-Su-Zhang-Yang-Zhao-Wang-Ma-Yao-Yang-Xie2019} A connected $(p,q)$-graph $G$ admits a labeling $f:V(G)\cup E(G)\rightarrow [1,p+q]$, such that $f(x)\neq f(w)$ for any pair of elements $x,w\in V(G)\cup E(G)$. We have the following sums:
\begin{equation}\label{eqa:ve-11}
F_{ve}(G,f)=\sum _{uv\in E(G)}\big [f(uv)+|f(u)-f(v)|\big ],
\end{equation}
\begin{equation}\label{eqa:ve-22}
F_{|ve|}(G,f)=\sum _{uv\in E(G)}\big |f(u)+f(v)-f(uv)\big |,
\end{equation}
and
\begin{equation}\label{eqa:ve-magic}
F_{magic}(G,f)=\sum _{uv\in E(G)}\big [f(u)+f(uv)+f(v)\big ]=kq.
\end{equation}
\textbf{We call $f$}:
\begin{asparaenum}[(i)]
\item A \emph{ve-sum-difference labeling} of $G$ if it holds (\ref{eqa:ve-11}) true.
\item A \emph{ve-difference labeling} of $G$ if it holds (\ref{eqa:ve-22}) true.
\item A \emph{k-edge-average labeling} of $G$ if it holds (\ref{eqa:ve-magic}) true.\qqed
\end{asparaenum}
\end{defn}

\begin{figure}[h]
\centering
\includegraphics[width=16.4cm]{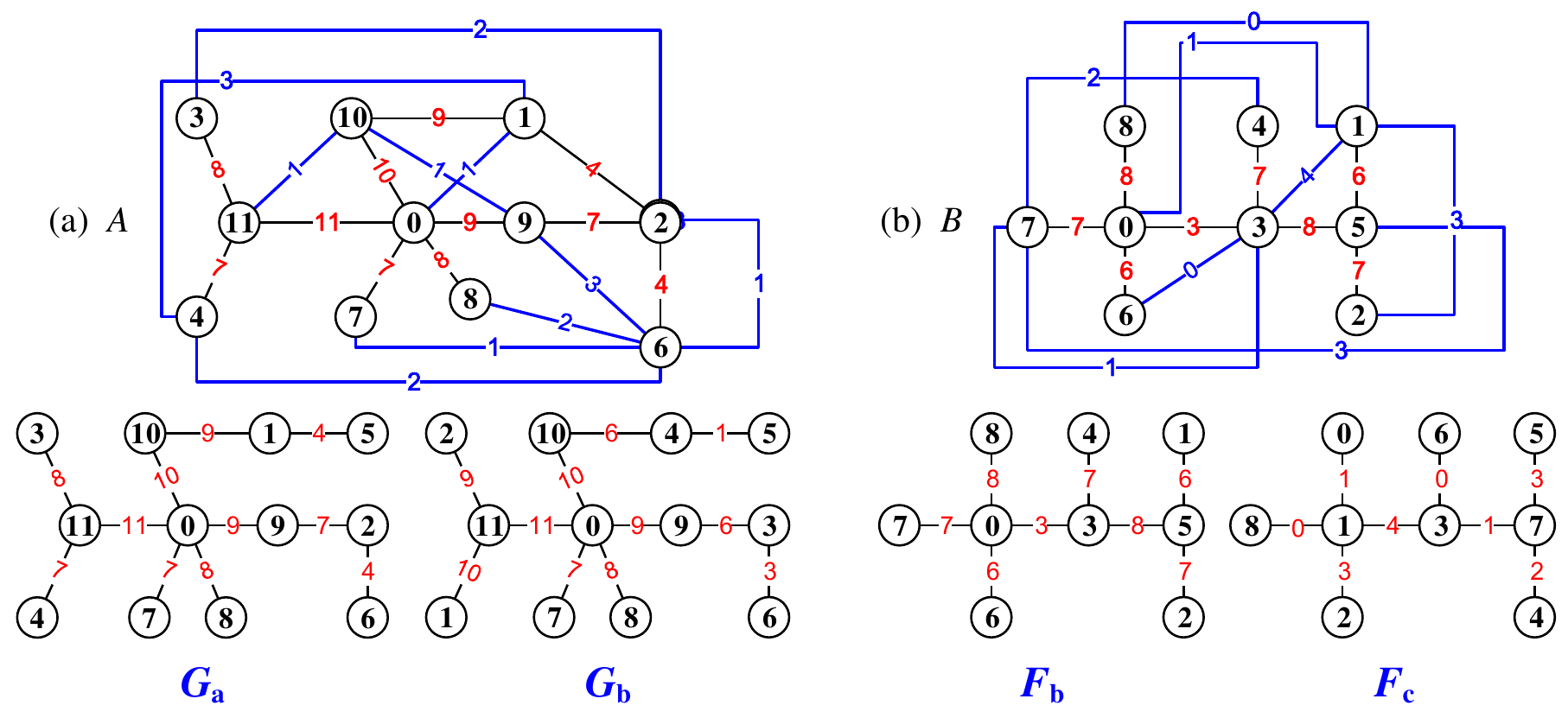}\\
\caption{\label{fig:felicitous-sum-matching}{\small (a) A perfect Max-Min difference-sum matching partition $A=\odot_{12}\langle G_a,G_d\rangle $; (b) a perfect Max-Min felicitous-sum matching partition $B=\odot_{9}\langle F_b,F_c\rangle $, cited from \cite{Yao-Mu-Sun-Sun-Zhang-Wang-Su-Zhang-Yang-Zhao-Wang-Ma-Yao-Yang-Xie2019}.}}
\end{figure}

\begin{defn}\label{defn:difference-sum-matching}
\cite{Yao-Mu-Sun-Sun-Zhang-Wang-Su-Zhang-Yang-Zhao-Wang-Ma-Yao-Yang-Xie2019} Suppose that $G_M$ and $G_m$ are two copies of a $(p,q)$-graph $G$, and $G_M$ admits a difference-sum labeling $f_M$ holding $S_{um}(G_M,f_M)=\max_f S_{um}(G,f)$ true, and $G_m$ admits another difference-sum labeling $f_m$ holding $S_{um}(G_m,f_m)=\min_f S_{um}(G,f)$ true. The identifying graph $H=\odot\langle G_M,G_m\rangle $ is called a \emph{Max-Min difference-sum matching partition}, moreover if $H=\odot_p\langle G_M,G_m\rangle $ holds $E(G_M)\cap E(G_m)=\emptyset$ true, we call $H$ a \emph{perfect Max-Min difference-sum matching partition}. (see a perfect Max-Min difference-sum matching partition shown in Fig.\ref{fig:felicitous-sum-matching}(a))\qqed
\end{defn}

\begin{defn}\label{defn:felicitous-sum-matching}
\cite{Yao-Mu-Sun-Sun-Zhang-Wang-Su-Zhang-Yang-Zhao-Wang-Ma-Yao-Yang-Xie2019} Suppose that $H_M$ and $H_m$ are two copies of a $(p,q)$-graph $H$, and $H_M$ admits a felicitous-sum labeling $h_M$ holding $F_{um}(H_M,h_M)=\max_f F_{um}(G,f)$ true, and $H_m$ admits another felicitous-sum labeling $h_m$ holding $F_{um}(H_m,h_m)=\min_f F_{um}(G,f)$ true. The identifying graph $G=\odot\langle H_M,H_m\rangle $ is called a \emph{Max-Min felicitous-sum matching partition}, and furthermore we call $G$ a \emph{perfect Max-Min felicitous-sum matching partition} if $G=\odot_p\langle H_M,H_m\rangle $ holds $E(H_M)\cap E(H_m)=\emptyset$ true. (see a perfect Max-Min felicitous-sum matching partition shown in Fig.\ref{fig:felicitous-sum-matching}(b))\qqed
\end{defn}

\subsection{Maximal common factor labelings (gcd-labelings)}

\subsubsection{Graceful-type of gcd-labelings}

\begin{defn}\label{defn:graceful-gcd-colorings}
\cite{Su-Hongyu-Wang-Yao-Common-Factor-2021} A $(p,q)$-graph $G$ admits a coloring $f:V(G)\rightarrow [1,M]$ such that each edge $xy\in E(G)$ is colored with $f(xy)=\textrm{gcd}(f(u), f(v))$. If the edge color set $f(E(G))=\{f(xy):xy\in E(G)\}=[1,q]$, we call $f$ a \emph{graceful gcd-coloring}, and $f$ an \emph{odd-graceful gcd-coloring} if $f(E(G))=\{f(xy):xy\in E(G)\}=[1,2q-1]^o$.\qqed
\end{defn}

\begin{figure}[h]
\centering
\includegraphics[width=13.4cm]{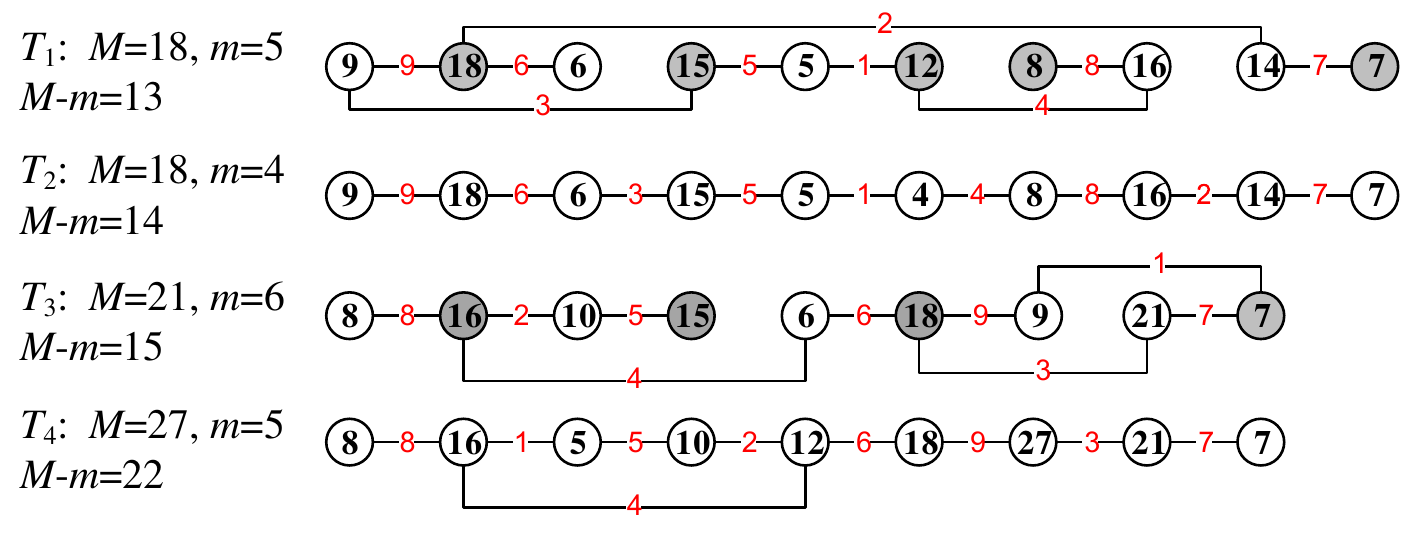}\\
\caption{\label{fig:graceful-gcd}{\small Examples for knowing Definition \ref{defn:graceful-gcd-colorings}.}}
\end{figure}

\begin{rem}\label{rem:graceful-gcd-colorings}
In Definition \ref{defn:graceful-gcd-colorings}, the number
$$
G_{gcd}(G)=\min_f\{\max f(V(G)):~f\textrm{ is a graceful gcd-labeling of $G$}\}
$$ is called a \emph{graceful gcd-labeling number} of the $(p,q)$-graph $G$. Similarly, the parameter
$$OG_{gcd}(G)=\min_f\{\max f(V(G)):~f\textrm{ is an odd-graceful gcd-labeling of $G$}\}$$ is called an \emph{odd-graceful gcd-labeling number} of the $(p,q)$-graph $G$.

Each graph $T_i$ of $9$ edges shown in Fig.\ref{fig:graceful-gcd} admits a graceful gcd-coloring $f_i$ for $i\in [1,4]$. We define a parameter $M_{mgrac}(T_i,f_i)=\max f_i(V(T_i))-\min f_i(V(T_i))$ for $i\in [1,4]$, so
$$M_{mgrac}(T_1)<M_{mgrac}(T_2)<M_{mgrac}(T_3)<M_{mgrac}(T_4).$$
We call $M_{mgrac}(T_i)=\min _{f}\{M_{mgrac}(T_i,f)\}$ the \emph{graceful gcd-coloring M-m-number} (resp. the \emph{odd-graceful gcd-coloring M-m-number} $M_{moddgra}(T_i,f)=\min _{f}\{M_{mgrac}(T_i,f)\}$).\paralled
\end{rem}

\begin{defn}\label{defn:22-graceful-gcd-labelings}
$^*$ For a $(p,q)$-graph $G$, we have:

(1) A \emph{graceful gcd-labeling} is defined as: A $(p,q)$-graph $G$ admits a \emph{proper labeling} $f:V(G)\rightarrow [1,M]$ such that each edge $xy\in E(G)$ is colored with $f(xy)=\textrm{gcd}(f(u), f(v))$ and $f(E(G))=\{f(xy):xy\in E(G)\}=[1,q]$, and $M_{mgral}(G,f)=\max f(V(G))-\min f(V(G))$.

(2) An \emph{odd-graceful gcd-labeling} is defined as: A $(p,q)$-graph $G$ admits a \emph{proper labeling} $f:V(G)\rightarrow [1,M]$ such that each edge $xy\in E(G)$ is colored with $f(xy)=\textrm{gcd}(f(u), f(v))$ and $f(E(G))=\{f(xy):xy\in E(G)\}=[1,2q-1]^o$, and $M_{moddgral}(G,f)=\max f(V(G))-\min f(V(G))$.

Accordingly, there are two extremal numbers: \emph{graceful gcd-labeling M-m-number} $M_{mgral}(G)=\min _{f}\{M_{mgral}(G,f)\}$, \emph{odd-graceful gcd-labeling M-m-number} $M_{moddgral}(G)=\min _{f}\{M_{moddgral}(G,f)\}$.
\end{defn}

\begin{defn}\label{defn:22edge-prime-labelings}
\cite{Su-Hongyu-Wang-Yao-Common-Factor-2021} A $(p,q)$-graph $G$ admits a proper labeling $f:V(G)\rightarrow [1,M]$, such that $\textrm{gcd}(f(u),f(v))$ is just a \emph{prime} for each edge $uv\in E(G)$, we call $f$ a \emph{prime labeling} of $G$. Moreover, we call $f$ an \emph{edge-prime labeling} if $|f(E(G))|=|E(G)|$.\qqed
\end{defn}

\begin{rem}\label{rem:22edge-prime-labelings}
In Definition \ref{defn:22edge-prime-labelings}, the extremal number $P_{rime}(G)=\min_f\{M\}$ over all prime labelings of the $(p,q)$-graph $G$ is called a \emph{prime gcd-labeling number}; and the extremal number $P\,'_{rime}(G)=\min_f\{M\}$ over all edge-prime labelings of the $(p,q)$-graph $G$ is called an \emph{edge-prime gcd-labeling number}.\paralled
\end{rem}

\subsubsection{Distinguishing gcd-labelings}

\begin{defn}\label{defn:distinguishing-gcd-labeling}
$^*$ Suppose that a $(p,q)$-graph $G$ admits a mapping $f:V(G)\rightarrow [1,M_f]$, such that $f(u)\neq f(v)$ for distinct vertices $u,v\in V(G)$, and $\textrm{gcd}(f(u),f(v))\neq \textrm{gcd}(f(u),f(w))$ for any pair of adjacent edges $uv,uw\in E(G)$. Let $F_N(x)=\{\textrm{gcd}(f(x),f(y)):~y\in N(x)\}$ for the neighbor set $N(x)$ of each vertex $x\in V(G)$. We have the following vertex distinguishing common divisor labeling (distinguishing gcd-labeling):

(1) $f$ is called an \emph{adjacent distinguishing gcd-labeling} if $F_N(u)\neq F_N(v)$ for any edge $uv\in E(G)$.

(2) $f$ is called a \emph{distinguishing gcd-labeling} if $F_N(w)\neq F_N(z)$ for distinct vertices $w,z\in V(G)$.

For all distinguishing gcd-labelings $f:V(G)\rightarrow [1,M_f]$, the minimum number of all $M_f$ is called the \emph{distinguishing gcd-number} of $G$, denoted as $D_{gcd}(G)$. For all adjacent distinguishing gcd-labelings $g:V(G)\rightarrow [1,M_g]$, the minimum number of all $M_g$ is called the \emph{adjacent distinguishing gcd-number} of $G$, denoted as $D_{agcd}(G)$. If $D_{gcd}(G)=p$ (resp. $D_{agcd}(G)=p$), we call $f$ a \emph{standard (resp. adjacent) distinguishing gcd-labelings}.\qqed
\end{defn}

\begin{defn}\label{defn:distinguishing-prime-labeling}
$^*$ Suppose that a $(p,q)$-graph $G$ admits a mapping $f:V(G)\rightarrow [1,M_f]$, such that $f(u)\neq f(v)$ for distinct vertices $u,v\in V(G)$, and $\textrm{gcd}(f(u),f(v))\neq \textrm{gcd}(f(u),f(w))$ for any pair of adjacent edges $uv,uw\in E(G)$. Let $F_N(x)=\{\textrm{gcd}(f(x),f(y)):~y\in N(x)\}$, and add a restriction ``$\textrm{gcd}(f(u),f(v))$ to be prime'' for any edge $uv\in E(G)$, we get:

(i) $f$ is called an \emph{adjacent distinguishing prime-labeling} if $F_N(u)\neq F_N(v)$ for any edge $uv\in E(G)$.

(ii) $f$ is called a \emph{distinguishing prime-labeling} if $F_N(w)\neq F_N(z)$ for distinct vertices $w,z\in V(G)$.

For all distinguishing prime-labelings $f:V(G)\rightarrow [1,M_f]$, the minimum number of all $M_f$ is called the \emph{distinguishing prime-number} of $G$, denoted as $D_{pri}(G)$. For all adjacent distinguishing prime-labelings $g:V(G)\rightarrow [1,M_g]$, the minimum number of all $M_g$ is called the \emph{adjacent distinguishing prime-number} of $G$, denoted as $D_{apri}(G)$. If $D_{pri}(G)=p$, or $D_{apri}(G)=p$, we call $f$ \emph{standard (adjacent) distinguishing prime-labelings}.\qqed
\end{defn}

\subsection{Induced total labelings}

A \emph{proper} total coloring of a graph $G$ is defined on a subset of $V(G)\cup E(G)$ such that two adjacent vertices or incident edges or incident of edges and vertices are colored with different colors. Similarly, a mapping $f$ from the vertex set $V(G)$ to $[0,q]$ is a \emph{proper vertex labeling} of $G$ if $f(u)\neq f(v)$ for each edge $uv\in E(G)$. Write $f(V(G))=\{f(u): u\in V(G)\}$ and $f(E(G))=\{f(uv): uv\in E(G)\}$, or $f(V)$ and $f(E)$ if there is no confliction.

\begin{defn}\label{defn:induced-labelingss}
\cite{Yao-Zhang-Zhou-Chen-Zhang-Yao-Li2013} For a mapping $f$ from the vertex set $V(G)$ to $[0,q]$, let $k=\max f(V(G))$ and let $s$ indicate the number of distinct colors used in $f(V(G))$, we call $f$ a \emph{\textbf{proper $(k,s)$-vertex labeling}} of a $(p,q)$-graph $G$. For a $(p,q)$-graph $G$ admitting a $(k,s)$-vertex labeling $f$: $V(G)\rightarrow [0,q]$, an \emph{\textbf{induced difference edge-labeling}} $i^-_f$ of $f$ in $G$ is defined as $i^-_f(uv)=|f(u)-f(v)|$ for each edge $uv\in E(G)$; and an \emph{\textbf{induced harmonious edge-labeling}} $i^+_f$ of $f$ in $G$ is defined as $i^+_f(uv)=f(u)+f(v)$ (mod $q$) for each edge $uv\in E(G)$..\qqed
\end{defn}

\begin{defn} \label{defn:c5-IDT-labeling}
\cite{Yao-Zhang-Zhou-Chen-Zhang-Yao-Li2013} A proper $(k,s)$-vertex labeling $f$ of a $(p,q)$-graph $G$, from $V(G)$ to $[0,q]$, is an \emph{\textbf{induced difference total labeling}} (IDT labeling) of $G$ if its induced difference edge-labeling $i^-_f$ is a proper edge coloring of $G$. We say $f$ to be a $(k,s,t)$-IDT labeling of $G$, where $t=|i^-_f(E)|$. The least number of $k$ spanning over all $(k,s,t)$-IDT labelings of
$G$, denoted as $\chi ^-(G)$, is called the IDT-\emph{\textbf{chromatic number}}.\qqed
\end{defn}

\begin{defn} \label{defn:c5-IHT-labeling}
\cite{Yao-Zhang-Zhou-Chen-Zhang-Yao-Li2013} A proper $(k,s)$-vertex labeling $h$ of a $(p,q)$-graph $G$, from $V(G)$ to $[0,q]$, is an \emph{\textbf{induced harmonious total labeling}} (IHT labeling) of $G$ if its induced harmonious edge-labeling $i^+_h$ is a proper edge coloring of $G$. Let $t=|i^+_h(E)|$, $f$ is also called a $(k,s,t)$-IHT labeling of $G$. The IHT-\emph{\textbf{chromatic number}} of $G$, denoted as $\chi^+(G)$, is the smallest number of $k$ spanning over all $(k,s,t)$-IHT labelings of $G$.\qqed
\end{defn}

\begin{rem}\label{rem:333333}
From Definition \ref{defn:c5-IDT-labeling} and Definition \ref{defn:c5-IHT-labeling}, an IDT-type labeling $f$ of a $(p,q)$-graph $G$ is \emph{\textbf{consecutive}} if $f(V)=[0,p-1]$. Analogously, an IHT-type labeling $h$ of $G$ is \emph{\textbf{consecutive}} if $h(V)=[0,p-1]$. Clearly, every graph admits a $(k,s,t)$-IDT
labeling and a $(k,s,t)$-IHT labeling. A $(k,s,t)$-IDT labeling of $G$ is \emph{\textbf{strong}} if it satisfies two additional requirements that $k=\chi^-(G)$ and $t$ is least, that is to say, for any $(k\,',s\,',t\,')$-IDT labeling of $G$ there is $t\leq t\,'$. A consecutive $(k,s,t)$-IDT labeling of a graph $G$ may be a \emph{graceful labeling} or a \emph{sequential labeling} of a $(p,q)$-graph $G$ if $t=q$, but a \emph{bandwidth labeling} of $G$, such a counterexample is a star. A $(k,s,t)$-IHT labeling of $G$ may be a \emph{harmonious labeling} or a \emph{felicitous labeling} if $t=q$ (\cite{Gallian2020}, \cite{Yao-Yao-Cheng-Li-Xie-Zhang2009}).\paralled
\end{rem}

Let $P_4=uvwx$ denote a path on $4$ vertices. A consecutive $(3,4,3)$-IDT labeling $f$ of $P_4$ can be represented as ${\Large \textcircled{\small 3}}3{\Large \textcircled{\small 0}}2{\Large \textcircled{\small 2}}1{\Large \textcircled{\small 1}}$, where $f(u)={\Large \textcircled{\small 3}}$, $i_f(uv)=3$, $f(v)={\Large \textcircled{\small 0}}$, $i_f(vw)=2$, $f(w)={\Large \textcircled{\small 2}}$, $i_f(wx)=1$, $f(x)={\Large \textcircled{\small 1}}$. Again, $P_4$ admits a $(3,3,3)$-IDT labeling ${\Large \textcircled{\small 3}}3{\Large \textcircled{\small 0}}1{\Large \textcircled{\small 1}}2{\Large \textcircled{\small 3}}$ and a consecutive $(3,3,2)$-IDT labeling ${\Large
\textcircled{\small 2}}2{\Large \textcircled{\small 0}}1{\Large \textcircled{\small 1}}2{\Large \textcircled{\small 3}}$. We have that $\chi^-(P_4)=2$ since $P_4$ admits a strong $(2,3,2)$-IDT labeling ${\Large \textcircled{\small 1}}1{\Large \textcircled{\small 0}}2{\Large \textcircled{\small 2}}1{\Large \textcircled{\small 1}}$.

\subsection{Digraceful labelings}

\begin{defn}\label{defn:digraceful-labeling-digraph}
\cite{Yao-Yao-Hui-Cheng-2012-Malaysian} Let $f$ be a labeling of a digraph $D$ from its vertex set $V(D)$ to $[0, |A(D)|]$, where $A(D)$ is the arc set of $D$, such that for each arc $\overrightarrow{uv} \in A(D)$, the induced are color $f(\overrightarrow{uv}) = f(u)-f(v)$ if $f(u) > f(v)$, otherwise $f(\overrightarrow{uv}) =|A(D)|+1+ f(u)-f(v)$. We call $f$ a \emph{digraceful labeling} of $D$ if the \emph{arc color set} $\{f(\overrightarrow{uv}) : \overrightarrow{uv}\in A(D)\}= [1, |A(D)|]$, and $D$ is called a \emph{digraceful digraph}. \qqed
\end{defn}

\begin{figure}[h]
\centering
\includegraphics[width=16cm]{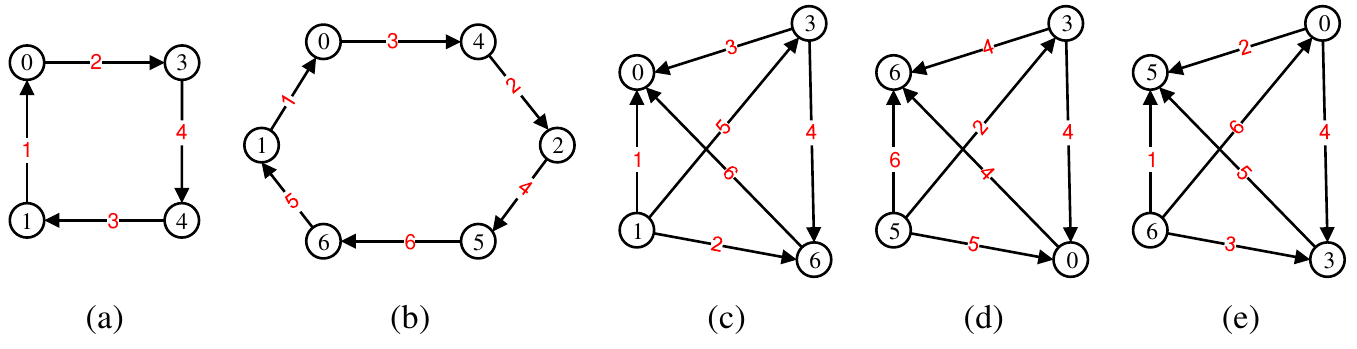}
\caption{\label{fig:Malaysian-11}{\small (a) A digraceful orientation of $C_4$; (b) a digraceful orientation of $C_6$; (c) a digraceful orientation of $K_4$ with a digraceful labeling $f$, and $Fa(f) = Ba(f) = 3$; (d) a digraceful orientation of $K_4$ with the dually digraceful labeling $f^*$ of the digraceful labeling $f$ defined in (c); (e) the conversely digraceful labeling $h^*$ of the converse of an orientation of $K_4$ shown in (c), cited from \cite{Yao-Yao-Hui-Cheng-2012-Malaysian}.}}
\end{figure}

\begin{figure}[h]
\centering
\includegraphics[width=16cm]{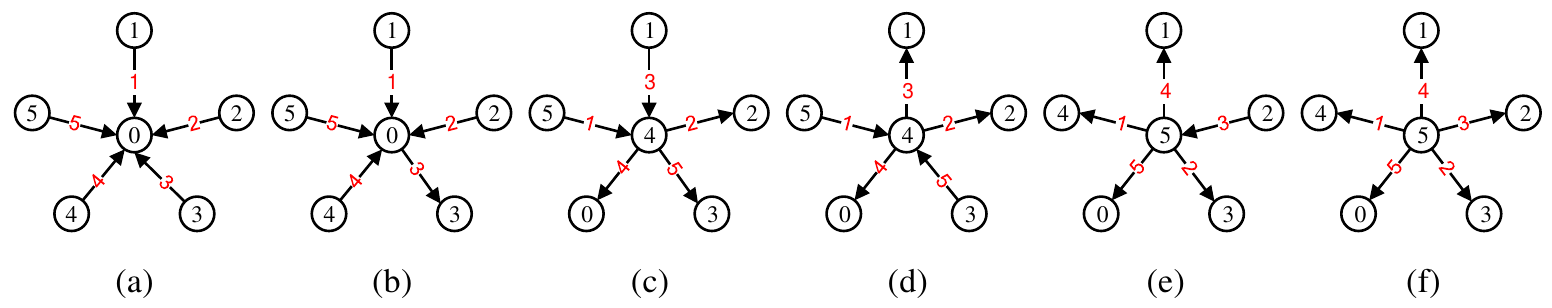}
\caption{\label{fig:Malaysian-22}{\small All digraceful orientations of $K_{1,5}$, cited from \cite{Yao-Yao-Hui-Cheng-2012-Malaysian}.}}
\end{figure}

\begin{rem}\label{rem:333333}
\cite{Yao-Yao-Hui-Cheng-2012-Malaysian} (1) It is not hard to understand that a digraceful labeling $f$ of a digraph $D$ is equivalent to a certain graceful labeling $\theta$ of $D$ defied in \cite{G-S-Bloom-D-F-Hsu-1985} since $f(u) =\theta(u)$ for $u \in V (D)$ and $f(\overrightarrow{uv}) =|A(D)|+1-\theta(\overrightarrow{uv})$ for $\overrightarrow{uv}\in A(D)$, and moreover $f$ is the \emph{arc-dual labeling} of $\theta$ (see Fig.\ref{fig:Malaysian-11}).

(2) Let $f$ be a digraceful labeling of a digraph $D$. An arc $\overrightarrow{uv}$ is called a \emph{forward-arc} if its color $f(\overrightarrow{uv}) = f(u)-f(v)$, otherwise, a \emph{backward-arc}. The number of forward-arcs and the number of backward-arcs are denoted by $Fa_{D}(f)$ and $Ba_{D}(f)$, respectively. We call number $\max_f \{Fa_{D}(f)\}$ spanning over all digraceful labelings $f$ of $D$ the \emph{optimal digraceful number} of $D$, denoted by $Odn(D)$. Furthermore, if $Fa_D(f) = Odn(D) = |A(D)|$, then $f$ is a graceful labeling of the underlying graph $UG(D)$. Clearly, an arc $\overrightarrow{uv}$, an out-degree $d^+(u)$, an arc color $f(\overrightarrow{uv}) = f(u)-f(v)$ and a forward-arc are in the same direction, from the left to the right.

(3) The \emph{converse} $H$ of a digraph $D$ is the digraph obtained from $D$ by conversing all arcs of $D$ by reversing the arc $\overrightarrow{uv}$, it means that we replace the arc $\overrightarrow{uv}$ by the arc $\overrightarrow{vu}$. Notice that $V (H) = V (D)$ and $|A(H)|= |A(D)|$. The conversely digraceful labeling $h^*$ of a digraceful labeling $f$ of $D$ is defined by $h^*(x) = |A(D)|-f(x)$ for all $x \in V (H) = V (D)$.

(4) An undirected graph $G$ has a \emph{digraceful orientation} if there exists an orientation of $G$ which admits a \emph{digraceful labeling}. All orientations of the star $K_{1,5}$ are digraceful, see Fig.\ref{fig:Malaysian-22}. Each vertex of degree one in an undirected tree is called a \emph{leaf}. A caterpillar is an undirected tree $T$ such that the resultant graph obtained by deleting all leaves of $T$ is just a path. A \emph{lobster} $H$ is an undirected tree such that the graph obtained by deleting all leaves from $H$ is just a caterpillar. An in-zero-out ditree is a ditree if, for its every vertex $u$, one of out-degree $d^+(u)$ and in-degree $d^-(u)$ is always equal to zero. A \emph{rooted ditree} $T$ at a fixed vertex $u$ satisfies that in-degree $d^-(u)=0$, out-degree $d^+(u)\geq 1$, and in-degrees $d^-(u)=1$ for all $x \in V (T)\setminus\{u\}$. A vertex $u$ of a ditree $T$ is called an in-leaf if $d^+_T(u)=0$ and $d^-_T(u)=1$, and an out-leaf as if $d^-_T(u)=0$ and $d^+_T(u)=1$.\paralled
\end{rem}

\section{Graph colorings with labeling-type restrictions}

We combine traditional proper total colorings with some graph labelings to build up new colorings for topological coding science. In general, we have the following definition of splitting $\epsilon$-colorings:
\begin{defn}\label{defn:splitting-(odd)graceful-coloring}
(\cite{Yao-1909-01587-2019, Yao-Zhao-Mu-Sun-Zhang-Zhang-Yang-IAEAC-2019}) A connected $(p,q)$-graph $G$ admits a coloring $f:S \rightarrow [a,b]$, where $S\subseteq V(G)\cup E(G)$, and there exists $f(u)=f(v)$ for some distinct vertices $u,v\in V(G)$, and the edge color set $f(E(G))$ holds an $\epsilon$-condition true, so we call $f$ a \emph{splitting $\epsilon$-coloring} of $G$. \qqed
\end{defn}

\subsection{Total colorings with labeling-type of restrictions}

\begin{defn}\label{defn:splitting-(odd)graceful-coloring}
\cite{Yao-Mu-Sun-Zhang-Yang-Wang-Wang-Su-Ma-Sun-2019} Suppose that a connected $(p,q)$-graph $G$ admits a coloring $f:V(G) \rightarrow [0,q]$ (resp. $[0,2q-1]$), such that $f(u)=f(v)$ for some pairs of vertices $u,v\in V(G)$, and the edge color set $f(E(G))=\{f(uv)=|f(u)-f(v)|: ~uv\in E(G)\}=[1,~q]$ (resp. $[1,2q-1]^o$), then we call $f$ a \emph{splitting gracefully total coloring} (resp. \emph{splitting odd-gracefully total coloring}). \qqed
\end{defn}

\subsubsection{A general definition of new colorings}

A group of new colorings, in which some are very similar with that in \cite{Gallian2020, Yao-Zhang-Sun-Mu-Sun-Wang-Wang-Ma-Su-Yang-Yang-Zhang-2018arXiv, Zhou-Yao-Chen-Tao2012, Zhou-Yao-Chen2013}, is as follows:

\begin{defn} \label{defn:new-graceful-strongly-colorings}
\cite{Bing-Yao-2020arXiv} Suppose that a connected $(p,q)$-graph $G$ ($\neq K_p$) admits a total coloring $f:V(G)\cup E(G)\rightarrow [1,M]$, and there are $f(x)=f(y)$ for some pairs of vertices $x,y\in V(G)$. Write $f(S)=\{f(w):w\in S\}$ for a non-empty set $S\subseteq V(G)\cup E(G)$ and let $k$ be a fixed positive integer. There are the following constraint conditions:
\begin{asparaenum}[(1$^\circ$)]
\item \label{27TProper01} $|f(V(G))|< p$;
\item \label{27TProper02} $|f(E(G))|=q$;
\item \label{27TGraceful-001} $f(V(G))\subseteq [1,M]$, $\min f(V(G))=1$;
\item \label{27TOdd-graceful-001} $f(V(G))\subset [1,2q+1]$, $\min f(V(G))=1$;
\item \label{27TGraceful-002} $f(E(G))=[1,q]$;
\item \label{27Tmodulo-01} $f(E(G))=[0,q-1]$;
\item \label{27TOdd-graceful-002} $f(E(G))=[1,2q-1]^o$;
\item \label{27Teven-edge-set-11} $f(E(G))=[2, 2q]^e$;
\item \label{27Tsequential-edge-set} $f(E(G))=[c,c+q-1]$;
\item \label{27Tgraceful-002} $f(uv)=|f(u)-f(v)|$ for each edge $uv\in E(G)$;
\item \label{27Tfelicitous-002} $f(uv)=f(u)+f(v)$ for each edge $uv\in E(G)$;
\item \label{27Tedge-labels-even-odd} For each edge $uv\in E(G)$, $f(uv)=f(u)+f(v)$ when $f(u)+f(v)$ is even, and $f(uv)=f(u)+f(v)+1$ when $f(u)+f(v)$ is odd;
\item \label{27Tmodulo-00} $f(uv)=f(u)+f(v)~(\textrm{mod}~q)$ for each edge $uv\in E(G)$;
\item \label{27Tmodulo-11} $f(uv)=f(u)+f(v)~(\textrm{mod}~2q)$ for each edge $uv\in E(G)$;
\item \label{27Tedge-difference} $f(uv)+|f(u)-f(v)|=k$ for each edge $uv\in E(G)$;
\item \label{27Tgraceful-difference} $\big |f(uv)-|f(u)-f(v)|\big |=k$ for each edge $uv\in E(G)$;
\item \label{27Tfelicitous-differences} $|f(u)+f(v)-f(uv)|=k$ for each edge $uv\in E(G)$;
\item \label{27Tedge-magic} $f(u)+f(uv)+f(v)=k$ for each edge $uv\in E(G)$;
\item \label{27Tmodulo-ordered} There exists an integer $k$ so that $\min \{f(u),f(v)\}\leq k <\max\{f(u),f(v)\}$ for each edge $uv\in E(G)$; and
\item \label{27TSet-ordered} $(X,Y)$ is the bipartition of a bipartite graph $G$ such that $\max f(X)< \min f(Y)$.
\end{asparaenum}
\textbf{A \emph{$W$-type} coloring $f$ is one of the following colorings}:
\begin{asparaenum}[\textrm{TCL}-1. ]
\item An \emph{edge-gracefully total coloring} if (\ref{27TProper01}$^\circ$), (\ref{27TProper02}$^\circ$) and (\ref{27TGraceful-002}$^\circ$) hold true.
\item A \emph{set-ordered edge-gracefully total coloring} if (\ref{27TProper01}$^\circ$), (\ref{27TProper02}$^\circ$), (\ref{27TGraceful-002}$^\circ$) and (\ref{27TSet-ordered}$^\circ$) hold true.

\item An \emph{edge-odd-gracefully total coloring} if (\ref{27TProper01}$^\circ$), (\ref{27TOdd-graceful-001}$^\circ$) and (\ref{27TOdd-graceful-002}$^\circ$) hold true.
\item A \emph{set-ordered edge-odd-gracefully total coloring} if (\ref{27TProper01}$^\circ$), (\ref{27TOdd-graceful-001}$^\circ$), (\ref{27TOdd-graceful-002}$^\circ$) and (\ref{27TSet-ordered}$^\circ$) hold true.

\item A \emph{gracefully total coloring} if (\ref{27TProper01}$^\circ$), (\ref{27TGraceful-001}$^\circ$), (\ref{27TGraceful-002}$^\circ$) and (\ref{27Tgraceful-002}$^\circ$) hold true.
\item A \emph{set-ordered
gracefully total coloring} if (\ref{27TProper01}$^\circ$), (\ref{27TGraceful-001}$^\circ$), (\ref{27TGraceful-002}$^\circ$), (\ref{27Tgraceful-002}$^\circ$) and (\ref{27TSet-ordered}$^\circ$) hold true.
\item An \emph{odd-gracefully total coloring} if (\ref{27TProper01}$^\circ$), (\ref{27TOdd-graceful-001}$^\circ$), (\ref{27TOdd-graceful-002}$^\circ$) and (\ref{27Tgraceful-002}$^\circ$) hold true.
\item A \emph{set-ordered odd-gracefully total coloring} if (\ref{27TProper01}$^\circ$), (\ref{27TOdd-graceful-001}$^\circ$), (\ref{27TOdd-graceful-002}$^\circ$), (\ref{27Tgraceful-002}$^\circ$) and (\ref{27TSet-ordered}$^\circ$) hold true.
\item A \emph{felicitous total coloring} if (\ref{27TGraceful-001}$^\circ$), (\ref{27Tmodulo-00}$^\circ$) and (\ref{27Tmodulo-01}$^\circ$) hold true.
\item A \emph{set-ordered felicitous total coloring} if (\ref{27TGraceful-001}$^\circ$),(\ref{27Tmodulo-00}$^\circ$), (\ref{27Tmodulo-01}$^\circ$) and (\ref{27TSet-ordered}$^\circ$) hold true.
\item An \emph{odd-elegant total coloring} if (\ref{27TOdd-graceful-001}$^\circ$), (\ref{27Tmodulo-11}$^\circ$) and (\ref{27TOdd-graceful-002}$^\circ$) hold true.
\item A \emph{set-ordered odd-elegant total coloring} if (\ref{27TOdd-graceful-001}$^\circ$), (\ref{27Tmodulo-11}$^\circ$), (\ref{27TOdd-graceful-002}$^\circ$) and (\ref{27TSet-ordered}$^\circ$) hold true.
\item A \emph{harmonious total coloring} if (\ref{27TGraceful-001}$^\circ$), (\ref{27Tmodulo-00}$^\circ$) and (\ref{27Tmodulo-01}$^\circ$) hold true, and when $G$ is a tree, exactly one edge color may be used on two vertices.
\item A \emph{set-ordered harmonious total coloring} if (\ref{27TGraceful-001}$^\circ$), (\ref{27Tmodulo-00}$^\circ$), (\ref{27Tmodulo-01}$^\circ$) and (\ref{27TSet-ordered}$^\circ$) hold true.
\item A \emph{strongly harmonious total coloring} if (\ref{27TGraceful-001}$^\circ$), (\ref{27Tmodulo-00}$^\circ$), (\ref{27Tmodulo-01}$^\circ$) and (\ref{27Tmodulo-ordered}$^\circ$) hold true.
\item A \emph{properly even harmonious total coloring} if (\ref{27TOdd-graceful-001}$^\circ$), (\ref{27Teven-edge-set-11}$^\circ$) and (\ref{27Tmodulo-11}$^\circ$) hold true.
\item A \emph{$c$-harmonious total coloring} if (\ref{27TGraceful-001}$^\circ$), (\ref{27Tfelicitous-002}$^\circ$) and (\ref{27Tsequential-edge-set}$^\circ$) hold true.
\item An \emph{even sequential harmonious total coloring} if (\ref{27TOdd-graceful-001}$^\circ$), (\ref{27Tedge-labels-even-odd}$^\circ$) and (\ref{27Teven-edge-set-11}$^\circ$) hold true.
\item A \emph{pan-harmonious total coloring} if (\ref{27TProper02}$^\circ$) and (\ref{27Tfelicitous-002}$^\circ$) hold true.
\item An \emph{edge-magic total coloring} if (\ref{27Tedge-magic}$^\circ$) holds true.
\item A \emph{set-ordered edge-magic total coloring} if (\ref{27Tedge-magic}$^\circ$) and (\ref{27TSet-ordered}$^\circ$) hold true.
\item A \emph{gracefully edge-magic total coloring} if (\ref{27TGraceful-002}$^\circ$) and (\ref{27Tedge-magic}$^\circ$) hold true.
\item A \emph{set-ordered graceful edge-magic total coloring} if (\ref{27TGraceful-002}$^\circ$), (\ref{27Tedge-magic}$^\circ$) and (\ref{27TSet-ordered}$^\circ$) hold true.
\item An \emph{edge-difference magic total coloring} if (\ref{27Tedge-difference}$^\circ$) holds true.
\item A \emph{set-ordered edge-difference magic total coloring} if (\ref{27Tedge-difference}$^\circ$) and (\ref{27TSet-ordered}$^\circ$) hold true.
\item A \emph{graceful edge-difference magic total coloring} if (\ref{27TGraceful-002}$^\circ$) and (\ref{27Tedge-difference}$^\circ$) hold true.
\item A \emph{set-ordered graceful edge-difference magic total coloring} if (\ref{27TGraceful-002}$^\circ$), (\ref{27Tedge-difference}$^\circ$) and (\ref{27TSet-ordered}$^\circ$) hold true.
\item A \emph{felicitous-difference total coloring} if (\ref{27TProper01}$^\circ$), (\ref{27TProper02}$^\circ$) and (\ref{27Tfelicitous-differences}$^\circ$) hold true.
\item A \emph{set-ordered felicitous-difference total coloring} if (\ref{27TProper01}$^\circ$), (\ref{27TProper02}$^\circ$), (\ref{27Tfelicitous-differences}$^\circ$) and (\ref{27TSet-ordered}$^\circ$) hold true.
\item An \emph{ev-difference magic total coloring} if (\ref{27Tgraceful-difference}$^\circ$) holds true.
\item A \emph{set-ordered ev-difference magic total coloring} if (\ref{27Tgraceful-difference}$^\circ$) and (\ref{27TSet-ordered}$^\circ$) hold true.
\item A \emph{graceful ev-difference magic total coloring} if (\ref{27TGraceful-002}$^\circ$) and (\ref{27Tgraceful-difference}$^\circ$) hold true.
\item A \emph{set-ordered graceful ev-difference magic total coloring} if (\ref{27TGraceful-002}$^\circ$), (\ref{27Tgraceful-difference}$^\circ$) and (\ref{27TSet-ordered}$^\circ$) hold true.\qqed
\end{asparaenum}
\end{defn}

\begin{figure}[h]
\centering
\includegraphics[width=16.4cm]{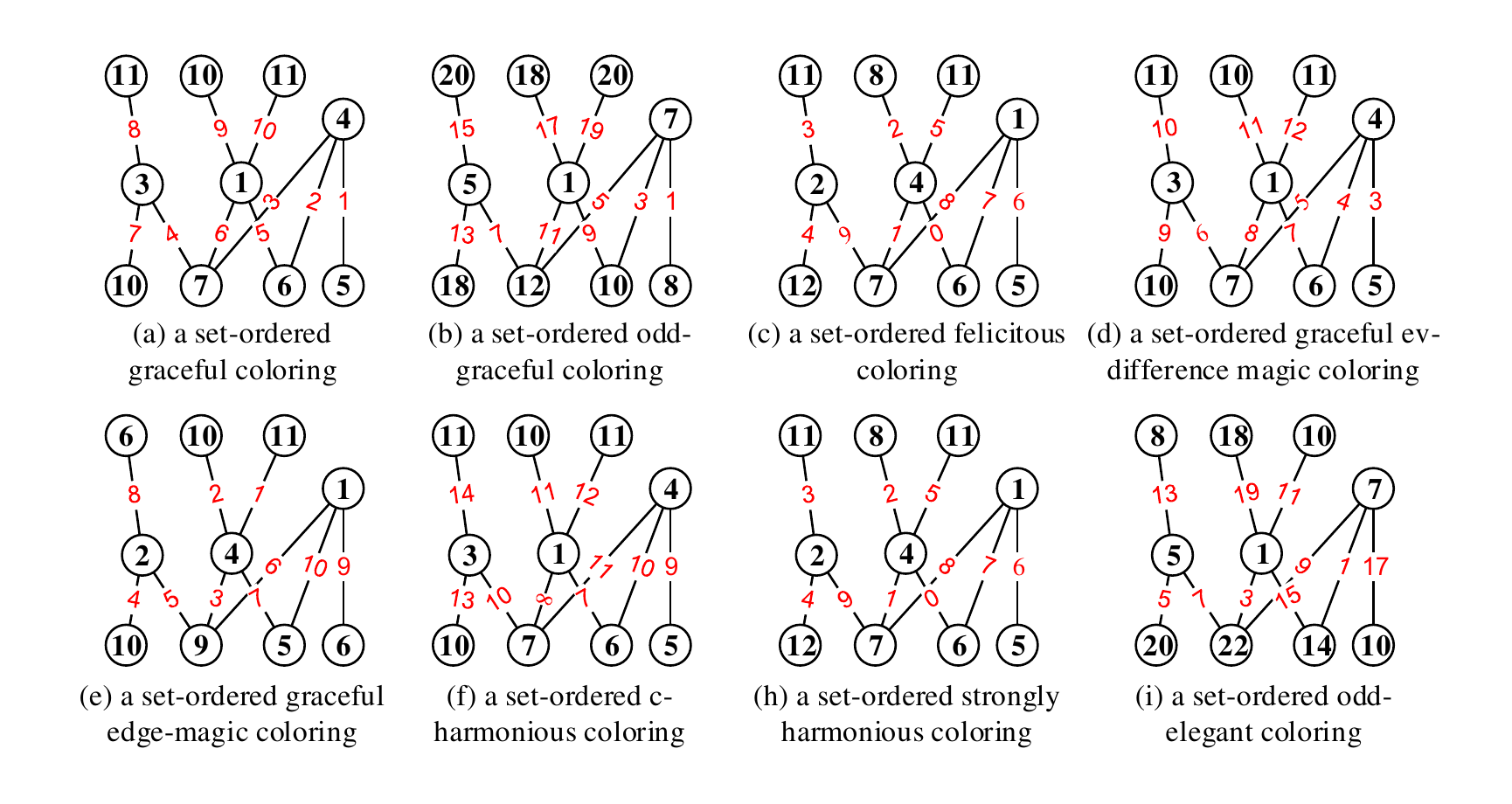}\\
\caption{\label{fig:new-total-colorings-0}{\small Part of examples for understanding Definition \ref{defn:new-graceful-strongly-colorings}, cited from \cite{Bing-Yao-2020arXiv}.}}
\end{figure}

\begin{figure}[h]
\centering
\includegraphics[width=16.2cm]{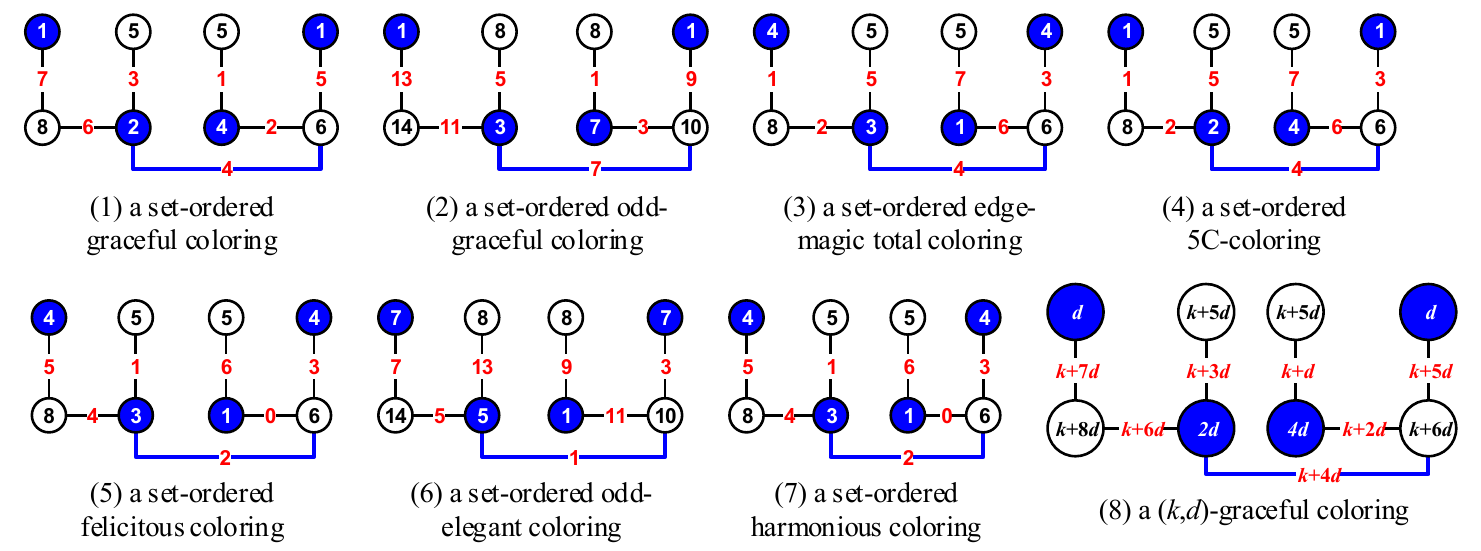}\\
\caption{\label{fig:equivalent-colorings}{\small Examples for showing equivalent colorings in Definition \ref{defn:new-graceful-strongly-colorings}, cited from \cite{Bing-Yao-2020arXiv}.}}
\end{figure}

\begin{rem}\label{rem:333333}
In Definition \ref{defn:new-graceful-strongly-colorings}:

(1) The number $\chi\,''_{W,M}(G)=\min_f\{M:~f(V(G))\subseteq [1,M]\}$ over all $W$-type total colorings $f$ of $G$ for a fixed $W\in \{\textrm{TCL}-1,\textrm{TCL}-2, \dots , \textrm{TCL}-33\}$ is called the \emph{$W$-type total chromatic number} of $G$, and the number $v_{W}(G)=\min_f\{|f(V(G))|\}$ over all $W$-type colorings $f$ of $G$ as \emph{$W$-type total splitting number}. Clearly, $\chi\,''(G)\leq \chi\,''_{W,M}(G)$. Fig.\ref{fig:new-total-colorings-0} shows part of examples for knowing Definition \ref{defn:new-graceful-strongly-colorings}, and the colorings of the examples shown in Fig.\ref{fig:equivalent-colorings} are equivalent from each other.

(2) Substituting ``total coloring'' by ``proper total coloring'' will get another group of $W$-type total colorings that are similar with and stricter than that defined in Definition \ref{defn:new-graceful-strongly-colorings}.\paralled
\end{rem}

\subsubsection{Matching difference total coloring}

\begin{defn}\label{defn:ve-difference-TC}
\cite{Su-Jin-Xu-Yao-New-Total-Colorings-2019} Let $f:V(G)\cup E(G)\rightarrow [1,k]$ be a proper total coloring of a graph $G$. If each edge $uv$ of $E(G)$ holds $f(uv)=|f(u)-f(v)|$ true, we call $f$ a \emph{ve-matching difference total $k$-coloring} of $G$, and call the smallest number of $k$ over all ve-matching difference total $k$-colorings \emph{ve-matching difference total chromatic number}, denoted as $\chi\,''_{ved}(G)$. \qqed
\end{defn}

\begin{defn}\label{defn:ve-sum-TC}
\cite{Su-Jin-Xu-Yao-New-Total-Colorings-2019} For a proper total coloring $g:V(G)\cup E(G)\rightarrow [1,m]$ on a graph $G$, if each edge $uv$ of $E(G)$ holds $g(uv)=g(u)+g(v)$ true, we call $g$ a \emph{ve-matching sum total $m$-coloring} of $G$. The minimal number of $m$ over all ve-matching sum total $m$-colorings is denoted as $\chi\,''_{ves}(G)$, called the \emph{ve-matching sum total chromatic number}. \qqed
\end{defn}

\begin{figure}[h]
\centering
\includegraphics[width=15cm]{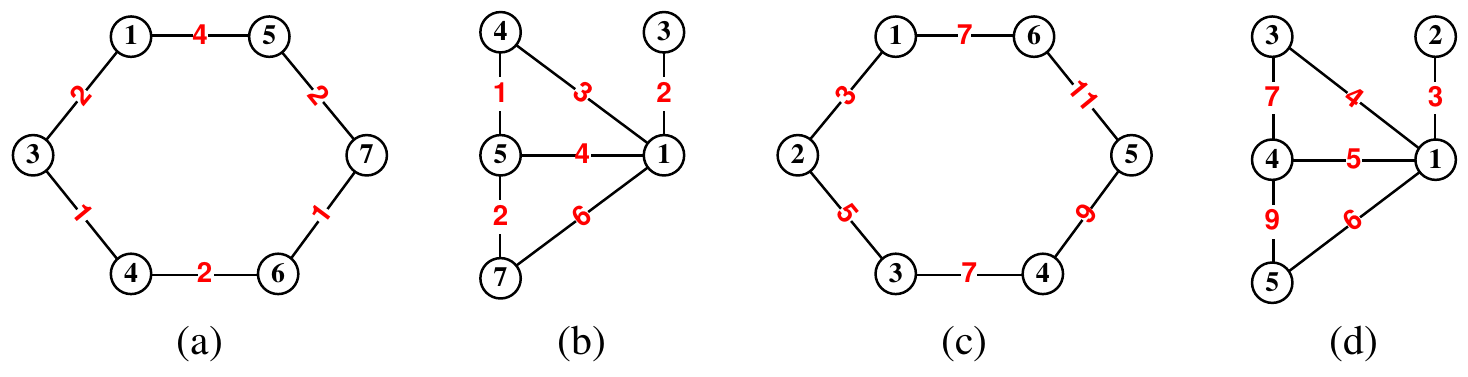}\\
\caption{\label{fig:induce-total-coloring-ss}{\small (a) and (b) are two ve-matching difference total colorings, (c) and (d) are two ve-matching sum total colorings, cited from \cite{Su-Jin-Xu-Yao-New-Total-Colorings-2019}.}}
\end{figure}

See two ve-matching difference total colorings shown in Fig.\ref{fig:induce-total-coloring-ss}(a) and (b) for understanding Definition \ref{defn:ve-difference-TC}. We can see $\chi\,''(G)\leq \chi\,''_{ved}(G)$, in general. Moreover, two ve-matching sum total colorings are shown in Fig.\ref{fig:induce-total-coloring-ss}(c) and (d), clearly, $\chi\,''(G)\leq \chi\,''_{ves}(G)$.

\begin{thm} \label{thm:ved-ves-trees-lemma}
\cite{Su-Jin-Xu-Yao-New-Total-Colorings-2019} About $\chi\,''_{ved}(G)$ defined in Definition \ref{defn:ve-difference-TC} and $\chi\,''_{ves}(G)$ defined in Definition \ref{defn:ve-sum-TC}, we have:
\begin{asparaenum}[\textrm{Result}-1.]
\item Suppose that $H$ is a connected subgraph of a connected graph $G$, then $\chi\,''_{ved}(H)\leq \chi\,''_{ved}(G)$ and $\chi\,''_{ves}(H)\leq \chi\,''_{ves}(G)$.
\item Each tree $T$ admits a ve-matching difference total coloring. When $\Delta(T)$ is odd, then $\Delta(T)+2 \leq\chi\,''_{ved}(T)\leq \frac{3}{2}(\Delta(T)+1)$; When $\Delta(T)$ is even, $\Delta(T)+1 \leq\chi\,''_{ved}(T)\leq \frac{3}{2}\Delta(T)+1$.
\item Each tree $T$ admits a ve-matching sum total coloring and satisfies inequalities $\Delta(T)+2 \leq\chi\,''_{ves}(T)\leq 2\Delta(T)+1$.
\item Every complete bipartite graph $K_{m,n}$ with $n\geq m$ admits a ve-matching sum total coloring and satisfies $\chi\,''_{ves}(K_{m,n})\leq m+2n$.
\item Every complete bipartite graph $K_{m,n}$ with $n\geq m$ admits a ve-matching difference total coloring and satisfies $\chi\,''_{ved}(K_{m,n})\leq 2n+\lceil \frac{n}{2}\rceil$.
\item Each complete graph $K_{n}$ admits a ve-matching sum total coloring and holds $\chi\,''_{ves}(K_{n})\leq2n-1$ true.
\end{asparaenum}
\end{thm}

\subsection{Graceful extremum numbers and gracefully critical authentications}

In \cite{Bing-Yao-2020arXiv}, each bipartite complete graph $K_{m,n}$ does not admit a gracefully total coloring $g$ with $g(x)=g(y)$ for some two distinct vertices $x,y\in V(K_{m,n})$, meanwhile, $K_{m,n}$ admits a graceful labeling $f$ with $f(u)\neq f(w)$ for any pair of vertices $u,w\in V(K_{m,n})$.

\begin{defn}\label{defn:gracefully-critical-graph}
\cite{Bing-Yao-2020arXiv} If a connected $(p,q)$-graph $H^{+}$ admits a gracefully total coloring, and adding some new edge $e\not \in E(H^{+})$ to $H^{+}$ makes a new graph $H^{+}+e$ such that the edge-added graph $H^{+}+e$ does not admit a gracefully total coloring, we call $H^{+}$ a \emph{gracefully$^{+}$ critical graph}. If another connected $(s,t)$-graph $H^{-}$ does not admit a gracefully total coloring, but removing some edge $e\,'\in E(H^{-})$ from $H^{-}$ produces a new graph $H^{-}-e\,'$ admitting a gracefully total coloring, we refer to $H^{-}$ a \emph{gracefully$^{-}$ critical graph}.

A \emph{$(p,s)$-gracefully total number} $R_{grace}(p,s)$ is an extremal number, such that any red-blue edge-coloring of a complete graph $K_{m}$ of $m$ $(\geq R_{grace}(p,s))$ vertices induces a copy $H^{+}_i$ of the gracefully$^{+}$ critical graph $H^{+}$ of $p$ vertices, or a copy $H^{-}_j$ of the gracefully$^{-}$ critical graph $H^{-}$ of $s$ vertices, where each edge of $H^{+}_i$ is red and each edge of $H^{-}_j$ is blue. And, some red-blue edge-coloring of a complete graph $K_{n}$ with $n\leq R_{grace}(p,s)-1$ does not induce any one of copies of the gracefully$^{+}$ critical graph $H^{+}$ of $p$ vertices and copies of the gracefully$^{-}$ critical graph $H^{-}$ of $s$ vertices. \qqed
\end{defn}

\begin{defn}\label{defn:gracefully-total-authentication}
\cite{Bing-Yao-2020arXiv} If a connected graph $G$ contains a copy $T$ of a gracefully$^{+}$ critical graph $H^{+}$ of $p$ vertices and a copy $L$ of a gracefully$^{-}$ critical graph $H^{-}$ of $s$ vertices, and both critical graphs $T$ and $L$ are edge-disjoint in $G$, so that $E(G)=E(T)\cup E(L)$ with $E(T)\cap E(L)=\emptyset $, and $V(G)=V(T)\cup V(L)$ with $|V(V^{+})\cap V(L)|\geq 1$, then we call $G$ a \emph{$(p,s)$-gracefully critical authentication}, and $(T,L)$ a \emph{$(p,s)$-gracefully graph-critical matching}. Moreover, a $(p,s)$-gracefully critical authentication $G^*$ is \emph{optimal} if $|V(G^*)|\leq |V(J)|$ for any $(p,s)$-gracefully critical authentication $J$. \qqed
\end{defn}

\begin{figure}[h]
\centering
\includegraphics[width=16cm]{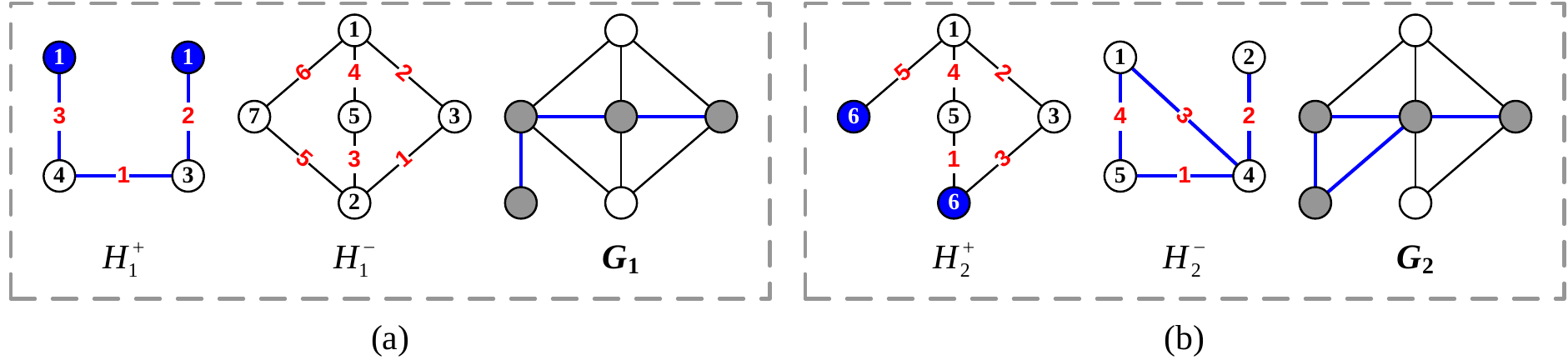}
\caption{\label{fig:extremum-11}{\small (a) $(H^{+}_1,H^{-}_1)$ is a $(4,5)$-gracefully graph-critical matching, and $G_1$ is a $(4,5)$-gracefully critical authentication; (b) $(H^{+}_2,H^{-}_2)$ is a $(5,4)$-gracefully graph-critical matching, and $G_2$ is a $(5,4)$-gracefully critical authentication, cited from \cite{Bing-Yao-2020arXiv}.}}
\end{figure}

\begin{figure}[h]
\centering
\includegraphics[width=16cm]{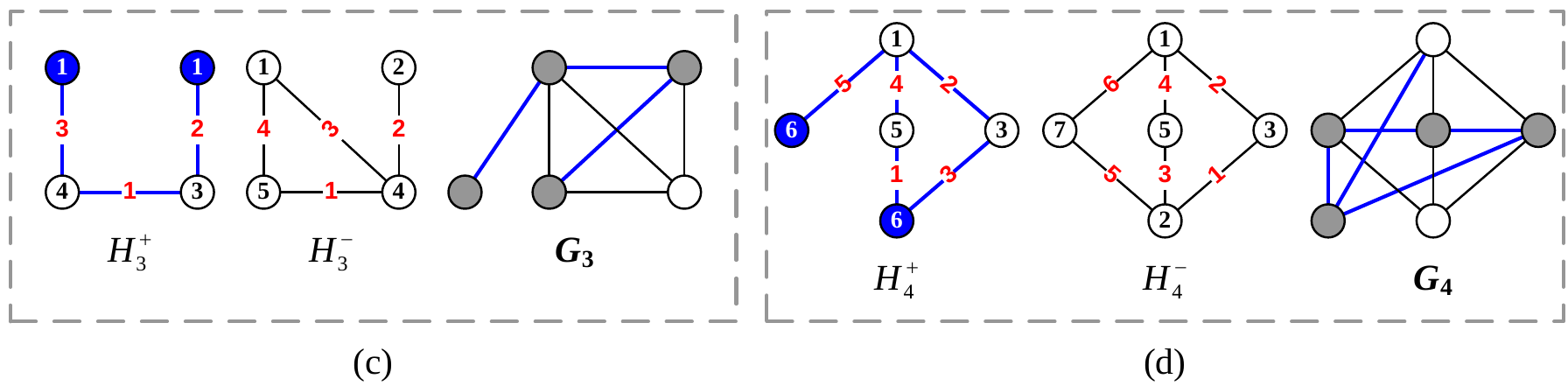}
\caption{\label{fig:extremum-22}{\small (c) $(H^{+}_3,H^{-}_3)$ is a $(4,4)$-gracefully graph-critical matching, and $G_3$ is a $(4,4)$-gracefully critical authentication; (d) $(H^{+}_4,H^{-}_4)$ is a $(5,5)$-gracefully graph-critical matching, and $G_4$ is a $(5,5)$-gracefully critical authentication, cited from \cite{Bing-Yao-2020arXiv}.}}
\end{figure}

See Fig.\ref{fig:extremum-11} and Fig.\ref{fig:extremum-22} for understanding the gracefully$^{+}$ critical graph, the gracefully$^{-}$ critical graph and the $(p,s)$-gracefully critical authentication defined in Definition \ref{defn:gracefully-critical-graph} and Definition \ref{defn:gracefully-total-authentication}. Four graphs $G_1,G_2,G_3 ,G_4$ shown in Fig.\ref{fig:extremum-11} and Fig.\ref{fig:extremum-22} are optimal.

\begin{rem}\label{rem:333333}
In fact, a $(p,s)$-gracefully total number $R_{grace}(p,s)$ is a graph Ramsey. In his book \cite{D-B-West1996}, West defined: Given simple graphs $G_1,G_2,\dots ,G_k$, the (graph) Ramsey number $R(G_1,G_2$, $\dots $, $G_k)$ is the smallest integer $n$ such that every $k$-coloring of $E(K_n)$ contains a copy of $G_i$ in color $i$ for some $i$.\paralled
\end{rem}

\begin{prop} \label{thm:gracefully-total-numbers-graphs} \cite{Bing-Yao-2020arXiv}
(1) If $(H^{+},H^{-})$ is a $(p,s)$-gracefully graph-critical matching, so are $(H^{+}+e,H^{-}-e')$, $(H^{+},H^{+}+e)$ and $(H^{-},H^{-}-e')$ too.

(2) $R_{grace}(p,s)=R_{grace}(s,p)$ with $p\geq 4$ and $s\geq 4$.
\end{prop}

\begin{figure}[h]
\centering
\includegraphics[width=14.4cm]{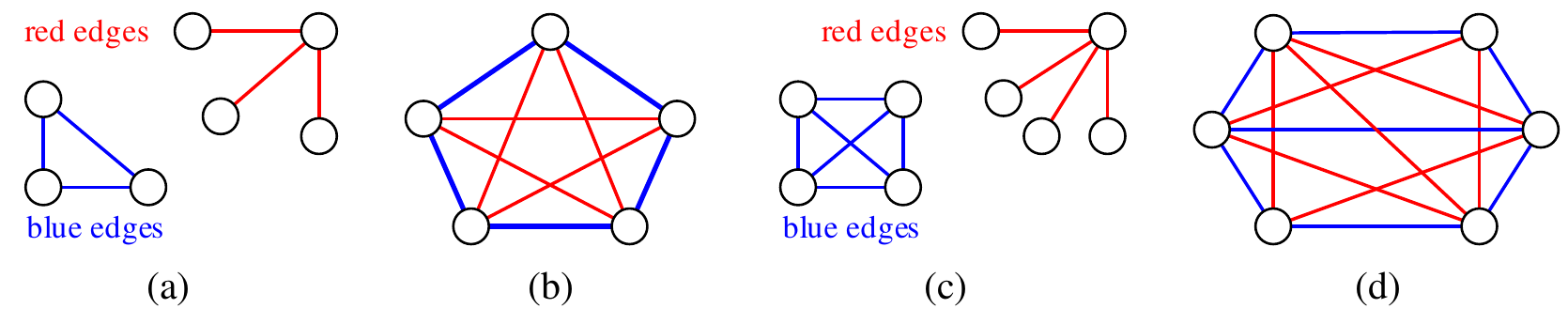}
\caption{\label{fig:extremum-33}{\small (b) A red-blue edge-coloring of $K_5$; (c) a red-blue edge-coloring of $K_6$, cited from \cite{Bing-Yao-2020arXiv}.}}
\end{figure}

In Fig.\ref{fig:extremum-33}, we can see the following facts:

(1) In Fig.\ref{fig:extremum-33}(a), a red-blue edge-coloring of $K_4$ does not induce any copy of a gracefully$^{-}$ critical graph $H^{-}$ of four vertices and any copy of a gracefully$^{+}$ critical graph $H^{+}$ of four vertices defined in Definition \ref{defn:gracefully-total-authentication}, so the $(4,4)$-gracefully total number $R_{grace}(4,4)=5$.

(2) A red-blue edge-coloring of $K_5$ does not induce a gracefully$^{-}$ critical graph $H^{-}$ of five vertices and a gracefully$^{+}$ critical graph $H^{+}$ of five vertices, such that each edge of $H^{+}$ is red and each edge of $H^{-}$ is blue. So, the $(5,5)$-gracefully total number $R_{grace}(5,5)=6$; and

(3) $R_{grace}(4,5)\geq 6$.

\begin{rem}\label{rem:333333}
Let $K_{2,n}$ (as a gracefully$^+$ critical graph $H^{+}$) be a complete bipartite graph with its vertex set $V(K_{2,n})=\{x_1,x_2\}\cup \{y_1,y_2,\dots ,y_n\}$ and edge set $E(K_{2,n})=\{x_iy_j:~i\in[1,2],j\in [1,n]\}$. We define a graceful labeling $f$ for $K_{2,n}$ in the way: $f(x_i)=i$ for $i\in[1,2]$, and $f(y_j)=2j+1$ for $j\in [1,n]$; next we set $f(x_iy_j)=f(y_j)-f(x_i)=2j+1-i$, which deduces $f(E(K_{2,n}))=[1,2n]$. Clearly, $K_{2,n}$ does not admit a gracefully total coloring $g$ with $g(x)=g(y)$ for some distinct two vertices $x,y\in V(K_{2,n})$.

We remove an edge $x_2y_n$ from $K_{2,n}$ to obtain a connected bipartite graph $K_{2,n}-x_2y_n$ (as a gracefully$^+$ critical graph $H^{+}$), and define a total coloring $g$ as: $g(x_1)=f(x_1)=1$, $g(x_2)=2n$, $g(y_n)=2n$, and $g(y_j)=f(y_j)$ for $j\in [1,2n-1]$; set $g(x_iy_j)=g(y_j)-g(x_i)$. So, $g(E(K_{2,n}-x_2y_n))=[1,2n-1]$, and $g(x_2)=2n=g(y_n)$. We claim that $g$ is a gracefully total coloring of the graph $K_{2,n}-x_2y_n$.

We add a new vertex $w$ to $K_{2,n}$, and add new edges $wy_k$ with $k\in [2,n]$, and add edges $wx_1$ and $y_1y_k$ with $k\in [2,n]$ to $K_{2,n}+w$, the resultant graph is denoted as $G$ with $(n+3)$ vertices. It is not hard to see that these new edges induce just a connected bipartite graph $K_{2,n}-x_2y_n$. Thereby, $G$ contains a gracefully$^{+}$ critical $(p,q)$-graph $H^{+}=K_{2,n}-x_2y_n$ and a gracefully$^{-}$ critical $(s,t)$-graph $H^{-}=K_{2,n}$, and both critical graphs $K_{2,n}-x_2y_n$ and $K_{2,n}$ are edge-disjoint in $G$. Thereby, $G$ is a smallest $(n+2,n+2)$-gracefully total authentication \cite{Bing-Yao-2020arXiv}.\paralled
\end{rem}

\begin{defn}\label{defn:generalized-one}
$^*$ Suppose that a connected $(p,q)$-graph $G$ admits a gracefully total coloring: (i) If adding a new edge $e$ to $G$ makes a new graph $G+e$ such that the edge-added graph $G+e$ does not admit a gracefully total coloring, we call $G$ a \emph{$C^{+}_{yes}$-gracefully critical graph}, here, it is allowed that $G+e$ is made by adding a new vertex $u$ and joining a vertex $v$ of $G$ with this new vertex to form an edge $e=uv$; (ii) if removing an edge $e$ from $G$ produces a new graph $G-e$ such that this graph $G-e$ (including isolated vertex deleted) does not admit a gracefully total coloring, we call $G$ a \emph{$C^{-}_{yes}$-gracefully critical graph}.

Suppose that another connected $(s,t)$-graph $H$ does not admit a gracefully total coloring: (1) If adding a new edge $e\,'$ to $H$ built a new graph $H+e\,'$ such that the graph $H+e\,'$ admits a gracefully total coloring, we call $H$ a \emph{$C^{+}_{no}$-gracefully critical graph}, notice that it is allowed that $H+e\,'$ is made by adding a new vertex $x$ and joining a vertex $y$ of $H$ with this new vertex to form an edge $e\,'=xy$; (2) if removing an edge $e\,'$ from $H$ yields a new graph $H-e\,'$ such that this graph $H-e\,'$ (including isolated vertex deleted) admits a gracefully total coloring, we call $H$ a \emph{$C^{-}_{no}$-gracefully critical graph}.

An \emph{ex-gracefully total number} $R_{grace}(p,s\mid A,B)$ is extremal and holds: Any red-blue edge-coloring of a complete graph $K_{m}$ of $m$ $(\geq R_{grace}(p,s\mid A,B))$ vertices induces a copy $T_i$ of the $A$-gracefully critical graph $G$ of $p$ vertices, or a copy $L_j$ of the $B$-gracefully critical graph $H$ of $s$ vertices, where each edge of $T_i$ is red and each edge of $L_j$ is blue, and $A,B\in \{C^{+}_{yes},C^{-}_{yes},C^{+}_{no},C^{-}_{no}\}$. The above case does not occur to $K_r$ if $r\leq R_{grace}(p,s\mid A,B)-1$. \qqed
\end{defn}

\begin{defn}\label{defn:generalized-two-red-blue-complete}
$^*$ Let $G$ be a connected $(p,q)$-graph admitting a $W$-type coloring: (a-1) If the edge-added graph $G+e$ does not admit a $W$-type coloring, then $G$ is called a \emph{$C^{+}_{yes}$-$W$-type critical graph}, notice that $G+e$ may be made by adding a new vertex $u$ and joining a vertex $v$ of $G$ with this new vertex to form an edge $e=uv$; (a-2) if the deleting-edge graph $G-e$ (including isolated vertex deleted) does not admit a $W$-type coloring, we call $G$ a \emph{$C^{-}_{yes}$-$W$-type critical graph}.

Suppose that another connected $(s,t)$-graph $H$ does not admit a $W$-type coloring: (b-1) If the edge-added graph $H+e$ admits a $W$-type coloring, we call $H$ a \emph{$C^{+}_{no}$-$W$-type critical graph}, here, $H+e$ may be made by adding a new vertex $x$ and joining a vertex $y$ of $H$ with this new vertex to form an edge $e\,'=xy$; (b-2) if the deleting-edge graph $H-e$ (including isolated vertex deleted) admits a $W$-type coloring, then $H$ is called a \emph{$C^{-}_{no}$-$W$-type critical graph}.

\textbf{The $(p,s\mid I,J)$-2-coloring}. A red-blue edge-coloring of a complete graph $K_{n}$ of $n$ vertices induces a copy $T_i$ of an $I$-type critical graph $G$ of $p$ vertices, or a copy $L_j$ of a $J$-type critical graph $H$ of $s$ vertices for $I,J\in \{C^{+}_{yes}-W$, $C^{-}_{yes}-W,C^{+}_{no}-W,C^{-}_{no}-W\}$, such that each edge of $T_i$ is red, or each edge of $L_j$ is blue, we say that $K_{n}$ admits a \emph{$(p,s\mid I,J)$-2-coloring}.

An \emph{ex-$W$-type number} $m=R_{grace}(p,s\mid I,J)$ is a smallest number such that each complete graph $K_{n}$ of $n~(\geq m)$ vertices admits a \emph{$(p,s\mid I,J)$-2-coloring}, and each complete graph $K_{t}$ with $t~\leq m-1$ vertices does not admit any $(p,s\mid I,J)$-2-coloring.\qqed
\end{defn}

\begin{defn}\label{defn:generalized-three}
$^*$ For $I,J\in \{C^{+}_{yes}-W$, $C^{-}_{yes}-W,C^{+}_{no}-W,C^{-}_{no}-W\}$, if a connected graph $G$ contains a copy $T_i$ of an $I$-type critical graph of $p$ vertices and a copy $L_j$ of a $J$-type critical graph of $s$ vertices, and both critical graphs $T_i$ and $L_j$ are edge-disjoint in $G$, so that $E(G)=E(T_i)\cup E(L_j)$ with $E(T_i)\cap E(L_j)=\emptyset $, and $V(G)=V(T_i)\cup V(L_j)$ with $|V(T_i)\cap V(L_j)|\geq 1$, then we call $G$ a \emph{$(p,s)$-$W$-type $(I,J)$-critical authentication}, and $(T_i,L_j)$ a \emph{$(p,s)$-$W$-type $(I,J)$-graph-critical matching}. A $(p,s)$-$W$-type $(I,J)$-critical authentication $F^*$ is \emph{optimal} if $|V(F^*)|\leq |V(G)|$ for any $(p,s)$-$W$-type $(I,J)$-critical authentication $G$.\qqed
\end{defn}

There are many unsolved problems in graph theory, which can persuade people to believe that Topsnut-gpws can resist cipher's attackers equipped quantum computer and AI technique, such a famous example is: ``\emph{If a graph $R$ with the maximum $R(s,k)-1$ vertices has no a complete graph $K_s$ of $s$ vertices and an independent set of $k$ vertices, then we call $R$ a \emph{Ramsey graph} and $R(s,k)-1$ a \emph{Ramsey number}. As known, it is a terrible job for computer to find Ramsey number $R(5,5)$, although we have known $46\leq R(5,5) \leq49$}''. Joel Spencer said:``\emph{Erd\"{o}s asks us to imagine an alien force, vastly more powerful than us, landing on Earth and demanding the value of $R(5,5)$ or they will destroy our planet. In that case, he claims, we should marshal all our computers and all our mathematicians and attempt to find the value. But suppose, instead, that they ask for $R(6,6)$. In that case, he believes, we should attempt to destroy the aliens}''.

\subsection{Tcn-pure total colorings, Labeling-type restrictive total colorings}

Coloring each of vertices and edges of a graph $G$ with a number in $[1,k]$ makes no two adjacent vertices (resp. edges) or incident edge (resp. vertex) having the same color, we call this coloring a \emph{proper total coloring}, the minimum number of $k$ for which $G$ admits a proper total $k$-colorings is denoted as $\chi ''(G)$.
\begin{defn}\label{defn:edge-magic-tcn-pure-total-coloring}
\cite{Yao-Sun-Zhang-Mu-Sun-Wang-Su-Zhang-Yang-Yang-2018arXiv} Let $f:V(G)\cup E(G)\rightarrow [1,\chi ''(G)]$ be a proper total coloring of $G$ with $\chi ''(G)=|\{f(x):x\in V(G)\cup E(G)\}|$, and we call this coloring $f$ \emph{total chromatic number pure coloring} (tcn-pure coloring). Let $d_f(uv)=f(u)+f(uv)+f(v)$ for each edge $uv\in E(G)$, and set
$$B_{tol}(G,f)=\max_{uv \in E(G)}\{d_f(uv)\}-\min_{uv \in E(G)}\{d_f(uv)\}.$$
Determine a new parameter $\min_fB_{tol}(G,f)$ over all tcn-pure colorings of $G$, and call $\min_fB_{tol}(G,f)$ \emph{pan-bandwidth edge-magic total chromatic number}. Especially, a coloring $h$ is called an \emph{edge-magic tcn-pure coloring} if $B_{tol}(G,h)=0$, or an \emph{equitably edge-magic tcn-pure coloring} if $B_{tol}(G,h)=1$.\qqed
\end{defn}

\begin{figure}[h]
\centering
\includegraphics[width=12cm]{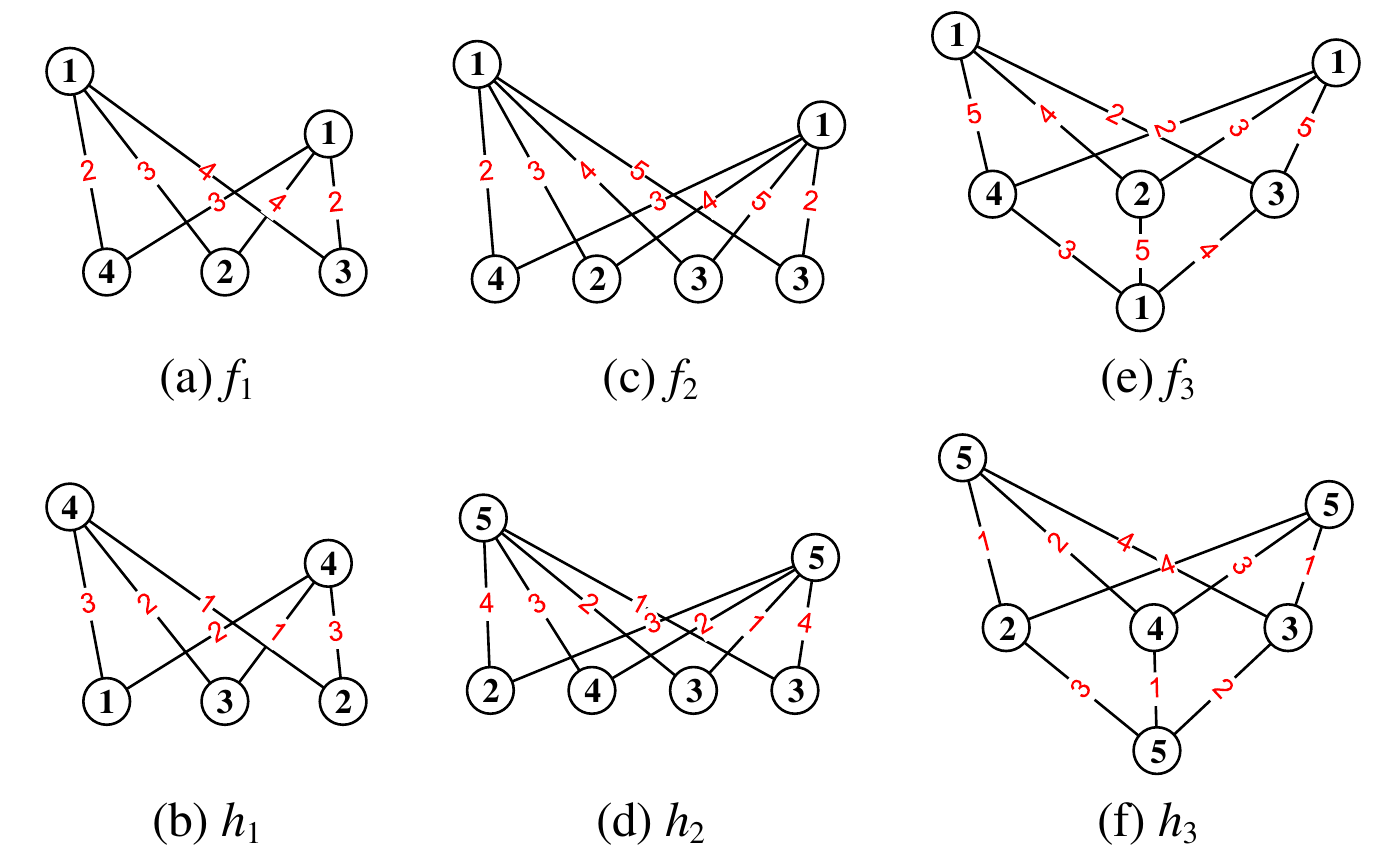}\\
\caption{\label{fig:new-total-colorings}{\small Six tcn-pure colorings of three complete bipartite graphs, in which $h_i$ is the matching total coloring of the tcn-pure coloring $f_i$ with $i=1,2,3$, cited from \cite{Yao-Sun-Zhang-Mu-Sun-Wang-Su-Zhang-Yang-Yang-2018arXiv}.}}
\end{figure}

In Fig.\ref{fig:new-total-colorings}, three complete bipartite graphs $K_{2,3}$, $K_{2,4}$ and $K_{3,3}$ admit six tcn-pure total colorings: each $h_i$ is the \emph{dually total coloring} (also, \emph{matching total coloring}) of $f_i$ with $i\in [1,3]$. Moreover, $B_{tol}(K_{2,3},f_1)=2$, $B_{tol}(K_{2,4},f_2)=3$ and $B_{tol}(K_{3,3},f_3)=4$; $B_{tol}(K_{2,3},h_1)=2$, $B_{tol}(K_{2,4},h_2)=3$, $B_{tol}(K_{3,3},h_3)=3$.

\begin{thm} \label{thm:add-leaves-edge-magic-total}
\cite{Mingjun-Zhang-Bing-Yao-Acta-2020}Adding a new leaf $u$ to a tree $T$ admitting an edge-magic tcn-pure total coloring produces another tree $T+uv$ with $v\in V(T)$ holding $\min_fB_{tol}(T,f)=\min_gB_{tol}(T+uv,g)$ true.
\end{thm}

\begin{thm} \label{thm:666666}
\cite{Mingjun-Zhang-Bing-Yao-Acta-2020} Each tree admits an edge-magic tcn-pure coloring, or an equitably edge-magic tcn-pure coloring.
\end{thm}

\begin{thm} \label{thm:666666}
\cite{Mingjun-Zhang-Bing-Yao-Acta-2020} Any connected graph $G$ admitting an edge-magic tcn-pure coloring or an equitably edge-magic tcn-pure coloring is the result of doing vertex-coinciding operation to some tree $T$ admitting an edge-magic tcn-pure coloring or an equitably edge-magic tcn-pure coloring, that is, $T$ is \emph{graph homomorphism} to $G$, that is, $T\rightarrow G$.
\end{thm}

\begin{defn} \label{defn:combinatoric-definition-total-coloring}
\cite{Bing-Yao-2020arXiv} For a proper total coloring $f:V(G)\cup E(G)\rightarrow [1,M]$ of a graph $G$, we define an \emph{edge-function} $c_f(uv)$ for each edge $uv\in E(G)$, and then have a parameter
\begin{equation}\label{eqa:edge-difference-total-coloring}
B^*_{\alpha}(G,f,M)=\max_{uv \in E(G)}\{c_f(uv)\}-\min_{xy \in E(G)}\{c_f(xy)\}.
\end{equation}
If $B^*_{\alpha}(G,f,M)=0$, we call $f$ an \emph{$\alpha$-proper total coloring} of $G$, the smallest number
\begin{equation}\label{eqa:minimum}
\chi\,''_{\alpha}(G) =\min_f \{M:~B^*_{\alpha}(G,f,M)=0\}
\end{equation}
over all $\alpha$-proper total colorings of $G$ is called \emph{$\alpha$-proper total chromatic number}, and $f$ is called a \emph{perfect $\alpha$-proper total coloring} if $\chi\,''_{\alpha}(G)=\chi\,''(G)$. \textbf{Moreover}
\begin{asparaenum}[\textrm{Tcoloring}-1. ]
\item We call $f$ a \emph{(resp. perfect) \textbf{edge-magic proper total coloring}} of $G$ if $c_f(uv)=f(u)+f(v)+f(uv)$, rewrite $B^*_{\alpha}(G,f,M)=B^*_{emt}(G,f$, $M)$, and $\chi\,''_{\alpha}(G)=\chi\,''_{emt}(G)$ is called \emph{edge-magic total chromatic number} of $G$.
\item We call $f$ a \emph{(resp. perfect) \textbf{edge-difference proper total coloring}} of $G$ if $c_f(uv)=f(uv)+|f(u)-f(v)|$, rewrite $B^*_{\alpha}(G,f,M)=B^*_{edt}(G,f$, $M)$, and $\chi\,''_{\alpha}(G)=\chi\,''_{edt}(G)$ is called \emph{edge-difference total chromatic number} of $G$.
\item We call $f$ a \emph{(resp. perfect) \textbf{felicitous-difference proper total coloring}} of $G$ if $c_f(uv)=|f(u)+f(v)-f(uv)|$, rewrite $B^*_{\alpha}(G,f,M)=B^*_{fdt}(G,f,M)$, and $\chi\,''_{\alpha}(G)=\chi\,''_{fdt}(G)$ is is called \emph{ felicitous-difference total chromatic number} of $G$.
\item We refer to $f$ a \emph{(resp. perfect) \textbf{graceful-difference proper total coloring}} of $G$ if $c_f(uv)=\big ||f(u)-f(v)|-f(uv)\big |$, rewrite $B^*_{\alpha}(G,f,M)=B^*_{gdt}(G,f,M)$, and $\chi\,''_{\alpha}(G)=\chi\,''_{gdt}(G)$ is called \emph{graceful-difference total chromatic number} of $G$.\qqed
\end{asparaenum}
\end{defn}

\begin{figure}[h]
\centering
\includegraphics[width=16.4cm]{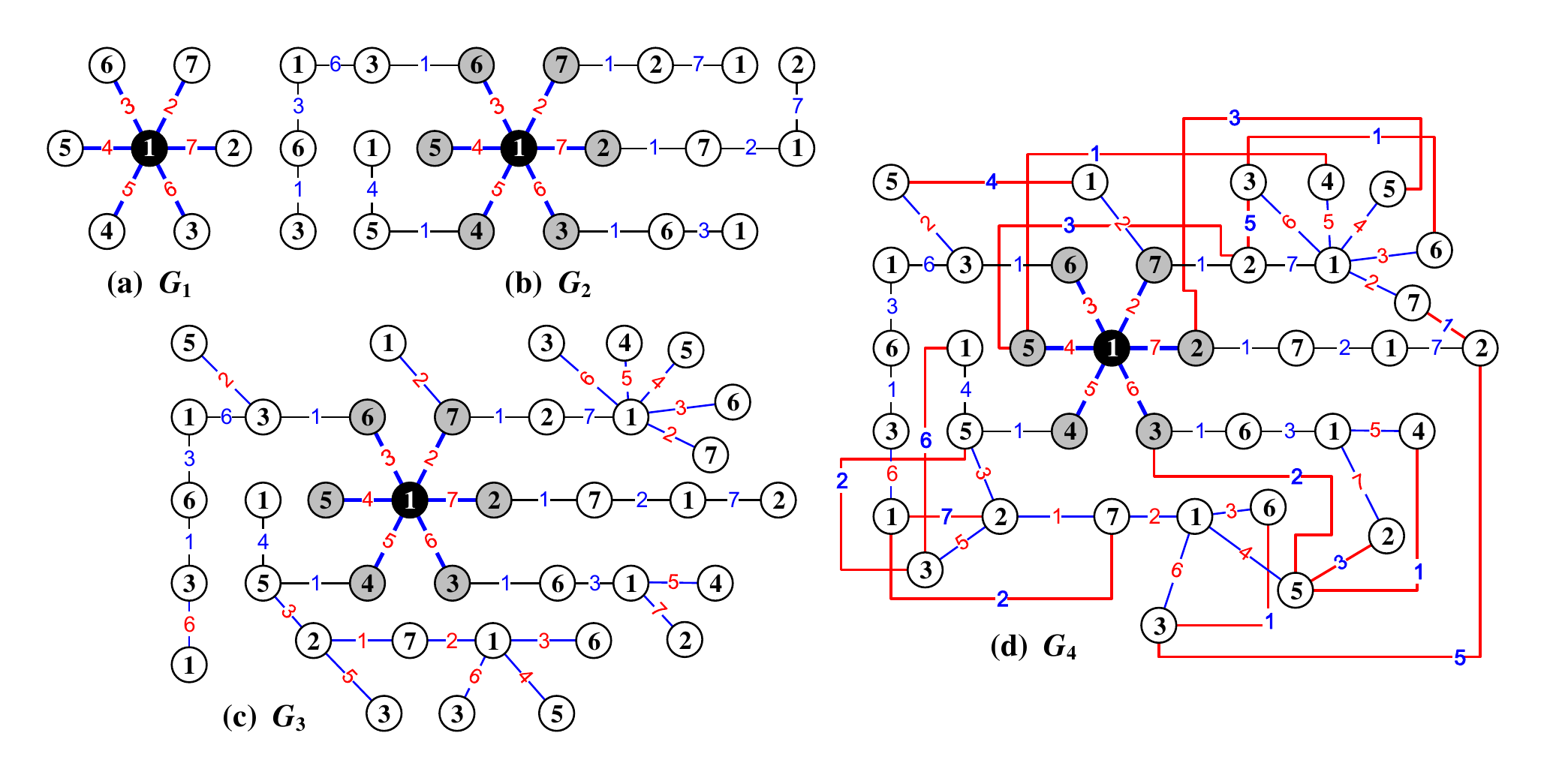}\\
\caption{\label{fig:spider-edge-magic}{\small Four graphs admitting the edge-magic proper total colorings holding $f(u)+f(uv)+f(v)=10$, where (a) is $K_{1,6}$, called a \emph{root}; (b) is a spider $S_{1,3,3,4,3,5}$, called a \emph{rooted Topsnut-gpw}; (c) is a \emph{rooted Topsnut-gpw}, cited from \cite{Bing-Yao-2020arXiv}.}}
\end{figure}

As removing ``proper'' from Definition \ref{defn:combinatoric-definition-total-coloring}, we get

\begin{defn} \label{defn:4-magic-total-colorings}
\cite{Yao-Ma-Wang-ITAIC2020} A randomly growing network model $N(t)$ admits a $W$-type total coloring $f_t:V(N(t))\cup E(N(t))\rightarrow [1,M(t)]$ with $t\in [a,b]^r$ such that there is a fixed constant $k\geq 0$, and a constraint conditional function $F(f_t(x)$, $f_t(xy)$, $f_t(y))=k$ for each edge $xy\in E(N(t))$ is one of the following equations:
\begin{asparaenum}[(\textrm{E}-1) ]
\item \label{mag:edge-magic-total} $f_t(x)+f_t(xy)+f_t(y)=k>0$;
\item \label{mag:graceful-difference-total} $\big ||f_t(x)-f_t(y)|-f_t(xy)\big |=k\geq 0$;
\item \label{mag:edge-difference-total} $f_t(xy)+|f_t(x)-f_t(y)|=k> 0$; and
\item \label{mag:felicitous-difference-total} $\big |f_t(x)+f_t(y)-f_t(xy)\big |=k\geq 0$.
\end{asparaenum}
\textbf{We call $f_t$}:
\begin{asparaenum}[C-1.]
\item An \emph{edge-magic total coloring} if (E-\ref{mag:edge-magic-total}) holds true.
\item A \emph{graceful-difference total coloring} if (E-\ref{mag:graceful-difference-total}) holds true.
\item An \emph{edge-difference total coloring} if (E-\ref{mag:edge-difference-total}) holds true.
\item A \emph{felicitous-difference total coloring} if (E-\ref{mag:felicitous-difference-total}) holds true.\qqed
\end{asparaenum}
\end{defn}

We can add another constraint requirement, such that each edge color $f_t(xy)$ is odd in Definition \ref{defn:4-magic-total-colorings}, so we get new colorings: \emph{odd-edge-magic total coloring}, \emph{odd-graceful-difference total coloring}, \emph{odd-edge-difference total coloring}, \emph{odd-felicitous-difference total coloring}, respectively. Next, we add three parameters to Definition \ref{defn:combinatoric-definition-total-coloring} if $G$ is bipartite, and get another group of particular total colorings as follows:

\begin{defn} \label{defn:combinatoric-definition-total-coloring-abc}
\cite{Bing-Yao-2020arXiv} Suppose that a bipartite graph $G$ admits a proper total coloring $f:V(G)\cup E(G)\rightarrow [1,M]$. We define an \emph{edge-function} $c_f(uv)(a,b,c)$ with three non-negative integers $a,b,c$ for each edge $uv\in E(G)$, and define a parameter
\begin{equation}\label{eqa:edge-difference-total-coloring}
B^*_{\alpha}(G,f,M)(a,b,c)=\max_{uv \in E(G)}\{c_f(uv)(a,b,c)\}-\min_{xy \in E(G)}\{c_f(xy)(a,b,c)\}.
\end{equation}
If $B^*_{\alpha}(G,f,M)(a,b,c)=0$, we call $f$ a \emph{parameterized $\alpha$-proper total coloring} of $G$, the smallest number
\begin{equation}\label{eqa:minimum}
\chi\,''_{\alpha}(G)(a,b,c) =\min_f \{M:~B^*_{\alpha}(G,f,M)(a,b,c)=0\}
\end{equation}
over all parameterized $\alpha$-proper total colorings of $G$ is called \emph{parameterized $\alpha$-proper total chromatic number}, and $f$ is called a \emph{perfect $\alpha$-proper total coloring} if $\chi\,''_{\alpha}(G)(a,b,c)=\chi\,''(G)$.
\begin{asparaenum}[\textrm{TCol}-1. ]
\item We call $f$ a \emph{(resp. perfect) \textbf{parameterized edge-magic proper total coloring}} of $G$ if $c_f(uv)=af(u)+bf(v)+cf(uv)$, rewrite $B^*_{\alpha}(G,f,M)(a,b,c)=B^*_{emt}(G,f$, $M)(a,b,c)$, and $\chi\,''_{\alpha}(G)(a,b,c)=\chi\,''_{emt}(G)(a,b,c)$ is called \emph{parameterized edge-magic total chromatic number} of $G$.
\item We call $f$ a \emph{(resp. perfect) \textbf{parameterized edge-difference proper total coloring}} of $G$ if $c_f(uv)=cf(uv)+|af(u)-bf(v)|$, rewrite $B^*_{\alpha}(G,f,M)(a,b,c)=B^*_{edt}(G,f$, $M)(a,b,c)$, and $\chi\,''_{\alpha}(G)(a,b,c)=\chi\,''_{edt}(G)(a,b,c)$ is called \emph{parameterized edge-difference total chromatic number} of $G$.
\item We call $f$ a \emph{(resp. perfect) \textbf{parameterized felicitous-difference proper total coloring}} of $G$ if $c_f(uv)=|af(u)+bf(v)-cf(uv)|$, rewrite $B^*_{\alpha}(G,f,M)(a,b,c)=B^*_{fdt}(G,f,M)(a,b,c)$, and $\chi\,''_{\alpha}(G)(a,b,c)=\chi\,''_{fdt}(G)(a,b,c)$ is called \emph{parameterized felicitous-difference total chromatic number} of $G$.
\item We refer to $f$ as a \emph{(resp. perfect) \textbf{parameterized graceful-difference proper total coloring}} of $G$ if $c_f(uv)=\big ||af(u)-bf(v)|-cf(uv)\big |$, rewrite
$$B^*_{\alpha}(G,f,M)(a,b,c)=B^*_{gdt}(G,f,M)(a,b,c),
$$ and $\chi\,''_{\alpha}(G)(a,b,c)=\chi\,''_{gdt}(G)(a,b,c)$ is called \emph{parameterized graceful-difference total chromatic number} of $G$.\qqed
\end{asparaenum}
\end{defn}

\begin{rem}\label{rem:333333}
We can put forward various requirements for $(a,b,c)$ in Definition \ref{defn:combinatoric-definition-total-coloring-abc} to increase the difficulty of attacking topological coding, since the ABC-conjecture (also, Oesterl\'{e}-Masser conjecture, 1985) involves the equation $a+b=c$ and the relationship between prime numbers. Proving or disproving the ABC-conjecture could impact: Diophantine (polynomial) math problems including Tijdeman's theorem, Vojta's conjecture, Erd\"{o}s-Woods conjecture, Fermat's last theorem, Wieferich prime and Roth's theorem, and so on \cite{Cami-Rosso2017Abc-conjecture}.\paralled
\end{rem}

\begin{defn} \label{defn:4-dual-total-coloring}
\cite{Bing-Yao-2020arXiv} The \emph{dual colorings} of the colorings defined in Definition \ref{defn:combinatoric-definition-total-coloring-abc} as $(a,b,c)=(1,1,1)$ are in the following:
\begin{asparaenum}[\textrm{\textbf{Dual}}-1. ]
\item For an \emph{edge-magic proper total coloring} $f_{em}$ of a graph $G$, so there exists a constant $k$ such that $f_{em}(u)+f_{em}(uv)+f_{em}(v)=k$ for each edge $uv\in E(G)$. Let $\max f_{em}=\max \{f_{em}(w):w\in V(G)\cup E(G)\}$ and $\min f_{em}=\min \{f_{em}(w):w\in V(G)\cup E(G)\}$. We have the \emph{dual coloring} $g_{em}$ of $f_{em}$ defined as: $g_{em}(w)=(\max f_{em}+\min f_{em})-f_{em}(w)$ for $w\in V(G)\cup E(G)$, and then
\begin{equation}\label{eqa:f-em}
{
\begin{split}
g_{em}(u)+g_{em}(uv)+g_{em}(v)&=3(\max f_{em}+\min f_{em})-[f_{em}(u)+f_{em}(uv)+f_{em}(v)]\\
&=3(\max f_{em}+\min f_{em})-k=k\,'
\end{split}}
\end{equation} for each edge $uv\in E(G)$.
\item For an \emph{edge-difference proper total coloring} $f_{ed}$ of a graph $G$, we have a constant $k$ holding $f_{ed}(uv)+|f_{ed}(u)-f_{ed}(v)|=k$ for each edge $uv\in E(G)$. Let $\max f_{ed}=\max \{f_{ed}(w):w\in V(G)\cup E(G)\}$ and $\min f_{ed}=\min \{f_{ed}(w):w\in V(G)\cup E(G)\}$. We have the \emph{dual coloring} $g_{ed}$ of $f_{ed}$ defined by setting $g_{ed}(x)=(\max f_{ed}+\min f_{ed})-f_{ed}(x)$ for $x\in V(G)$ and $g_{ed}(uv)=f_{ed}(uv)$ for $uv\in E(G)$, and then
\begin{equation}\label{eqa:f-ed}
g_{ed}(uv)+|g_{ed}(u)-g_{ed}(v)|=f_{ed}(uv)+|f_{ed}(u)-f_{ed}(v)|=k
\end{equation} for every edge $uv\in E(G)$.
\item For a \emph{graceful-difference proper total coloring} $f_{gd}$ of a graph $G$, there exists a constant $k$ such that $\big ||f_{gd}(u)-f_{gd}(v)|-f_{gd}(uv)\big |=k$ for each edge $uv\in E(G)$. Let $\max f_{gd}=\max \{f_{gd}(w):w\in V(G)\cup E(G)\}$ and $\min f_{gd}=\min \{f_{gd}(w):w\in V(G)\cup E(G)\}$. We have the \emph{dual coloring} $g_{gd}$ of $f_{gd}$ defined in the way: $g_{gd}(x)=(\max f_{gd}+\min f_{gd})-f_{gd}(x)$ for $x\in V(G)$ and $g_{gd}(uv)=f_{gd}(uv)$ for each edge $uv\in E(G)$, and then
\begin{equation}\label{eqa:f-md}
{
\begin{split}
\big ||g_{gd}(u)-g_{gd}(v)|-g_{gd}(uv)\big |=\big ||f_{gd}(u)-f_{gd}(v)|-f_{gd}(uv)\big |=k
\end{split}}
\end{equation} for each edge $uv\in E(G)$.
\item For a \emph{felicitous-difference proper total coloring} $f_{fd}$ of a graph $G$, we have a constant $k$ such that $|f_{fd}(u)+f_{fd}(v)-f_{fd}(uv)|=k$ for each edge $uv\in E(G)$. Let $\max f_{fd}=\max \{f_{fd}(w):w\in V(G)\cup E(G)\}$ and $\min f_{fd}=\min \{f_{fd}(w):w\in V(G)\cup E(G)\}$. We have the \emph{dual coloring} $g_{fd}$ of $f_{fd}$ defined as: $g_{fd}(w)=(\max f_{fd}+\min f_{fd})-f_{fd}(w)$ for $w\in V(G)\cup E(G)$, and then
\begin{equation}\label{eqa:f-tg}
{
\begin{split}
|g_{fd}(u)+g_{fd}(v)-g_{fd}(uv)|&=|(\max f_{fd}+\min f_{fd})+f_{fd}(u)+f_{fd}(v)-f_{fd}(uv)|\\
&=(\max f_{fd}+\min f_{fd})\pm k
\end{split}}
\end{equation} for each edge $uv\in E(G)$. Here, if $f_{fd}$ is edge-ordered, that is, $f_{fd}(x)+f_{fd}(y)\geq f_{fd}(xy)$ for each edge $xy\in E(G)$, then $$|g_{fd}(u)+g_{fd}(v)-g_{fd}(uv)|=(\max f_{fd}+\min f_{fd})+k=k\,'.$$ We have $$|g_{fd}(u)+g_{fd}(v)-g_{fd}(uv)|=(\max f_{fd}+\min f_{fd})-k=k\,',$$
if $f_{fd}(x)+f_{fd}(y)< f_{fd}(xy)$ for each edge $xy\in E(G)$. \qqed
\end{asparaenum}
\end{defn}

\begin{figure}[h]
\centering
\includegraphics[width=16.4cm]{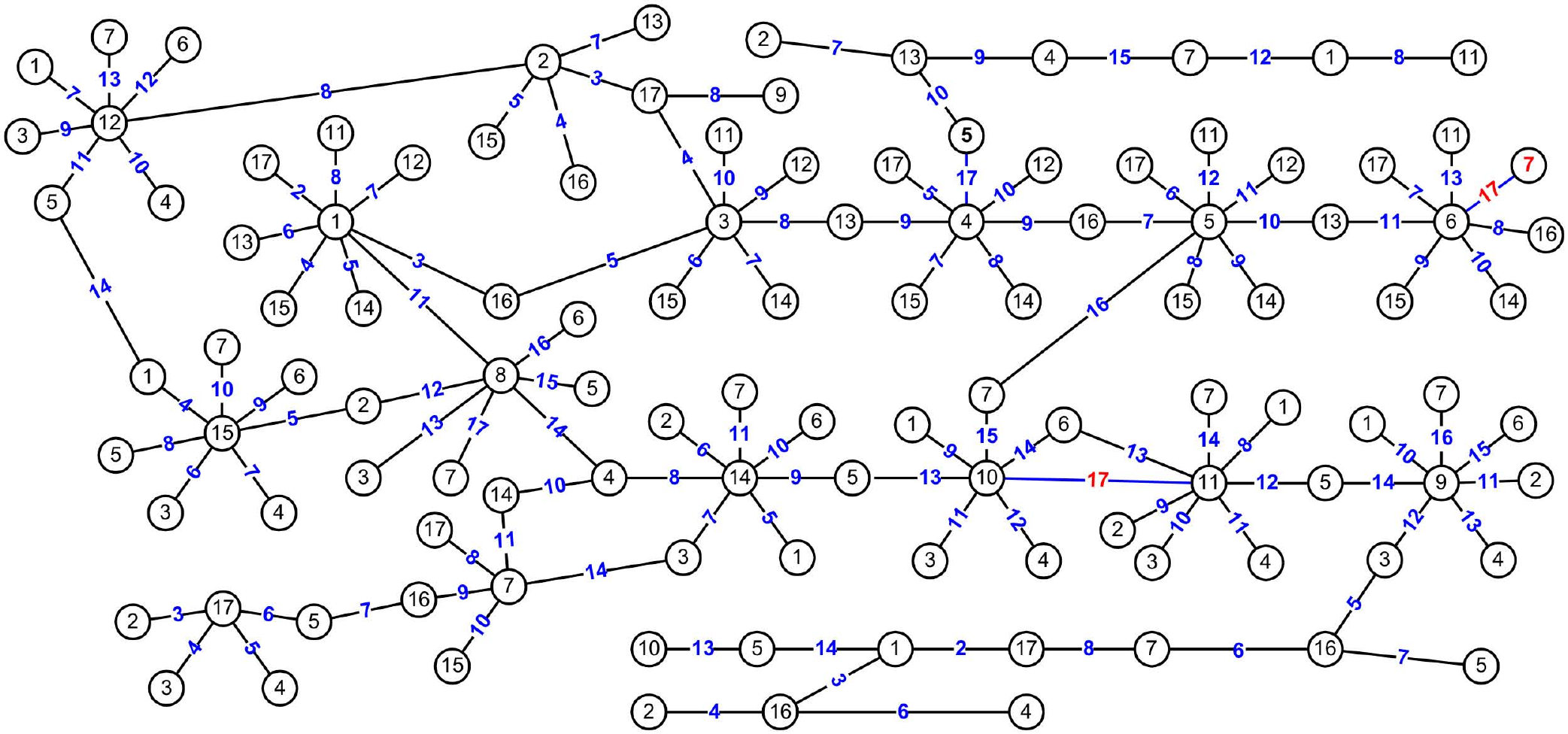}\\
\caption{\label{fig:random-growing-edge-difference}{\small A randomly growing graph admits an edge-difference proper total coloring holding $f(uv)+|f(u)-f(v)|=18$, cited from \cite{Bing-Yao-2020arXiv}.}}
\end{figure}

\begin{rem}\label{rem:333333}
Removing ``proper'' from Definition \ref{defn:combinatoric-definition-total-coloring} produces: A simple graph $G$ admits a total coloring $f:V(G)\cup E(G)\rightarrow [1,M]$ such that $f(u)\neq f(v)$ for each edge $uv\in E(G)$, and $f(xy)\neq f(xw)$ for two adjacent edges $xy,xw\in E(G)$. So, this particular total coloring allows $f(u)=f(uv)$ for some edge $uv\in E(G)$, and is weak than that in Definition \ref{defn:combinatoric-definition-total-coloring}. Similarly, removing ``proper'' from Definition \ref{defn:combinatoric-definition-total-coloring-abc} produces four parameterized $\alpha$-proper total colorings that are weak than that defined in Definition \ref{defn:combinatoric-definition-total-coloring-abc}.\paralled
\end{rem}

\begin{defn} \label{defn:rainbow-proper-total-coloring}
\cite{Bing-Yao-2020arXiv} A \emph{rainbow proper total coloring} $f$ of a connected graph $G$ holds: For any path $x_1x_2x_3x_4x_5$ of $G$, edge colors $f(x_{i}x_{i+1})\neq f(x_{j}x_{j+1})$ with $i,j\in [1,4]$, and each $f(x_{j}x_{j+1})$ is one of $f(x_{j}x_{j+1})=f(x_i)+f(x_ix_{i+1})+f(x_{i+1})$, $f(x_{j}x_{j+1})=f(x_ix_{i+1})+|f(x_i)-f(x_{i+1})|$, $f(x_{j}x_{j+1})=|f(x_i)+f(x_{i+1})-f(x_ix_{i+1})|$ and $f(x_{j}x_{j+1})=\big ||f(x_i)-f(x_{i+1})|-f(x_ix_{i+1})\big |$.\qqed
\end{defn}

\begin{defn} \label{defn:4-ice-flower-total-coloring}
\cite{Bing-Yao-2020arXiv} Let $f:V(G)\cup E(G)\rightarrow [1,M]$ be a proper total coloring of a simple graph $G$. If $E(G)=\bigcup^4_{i=1}E_i$ with $E_i\cap E_j=\emptyset$ and $E_i\neq \emptyset$ for $1\leq i,j\leq 4$, such that $f(u)+f(uv)+f(v)=k_1$ for each edge $uv\in E_1$, $f(xy)+|f(x)-f(y)|=k_2$ for each edge $xy\in E_2$, $|f(s)+f(t)-f(st)|=k_3$ for each edge $st\in E_3$, and $\big ||f(a)-f(b)|-f(ab)\big |=k_4$ for each edge $ab\in E_4$. We call $f$ a \emph{4-ice-flower proper total coloring} of $G$, and the smallest number of $\max\{f(w):w\in V(G)\cup E(G)\}$ for which $G$ admits a 4-ice-flower proper total coloring is denoted as $\chi\,''_{\textrm{4ice}}(G)$, called \emph{4-ice-flower total chromatic number}.\qqed
\end{defn}

\begin{defn} \label{defn:isomorphic-coloring-equals}
\cite{Bing-Yao-2020arXiv} Let $G$ be a totally colored graph admitting a $W_G$-type proper total coloring $f_G$, and let $H$ be another totally colored graph with a $W_H$-type proper total coloring $g_H$. We say $G=H$ if there is a bijection $\varphi: V (G)\rightarrow V (H)$ such that

(i) $G\wedge u\cong H\wedge \varphi(u)$ for each vertex $u \in V (G)$ with $\textrm{deg}_G(u)\geq 2$ (refer to Theorem \ref{thm:2-vertex-split-graphs-isomorphic}); and

(ii) for each element $w \in V(G)\cup E(G)$, there exists an element $w\,' \in V(H)\cup E(H)$ holding $g_H(w\,')=f_G(w)$ when $w\,'=\varphi(w)$.\qqed
\end{defn}

\begin{defn}\label{defn:L-multiple-type-coloring-labeling}
\cite{Bing-Yao-2020arXiv} Suppose that a connected graph $G$ admits a coloring $f$. If there is a spanning subgraph $T$ of $G$, such that $f$ is just a $W$-type coloring of $T$, we call $f$ an \emph{inner $W$-type coloring} of $T$, and say $G$ admits a \emph{coloring including a $W$-type coloring}. Moreover, if there are $L$ spanning subgraphs $H_i,H_2,\dots ,H_L$ with $E(H_i)\neq E(H_j)$ for distinct $i,j\in [1,L]$ and $L\geq 2$ such that $f_i~(=f)$ is an \emph{inner $W_i$-type coloring} of $H_i$ with $i\in [1,L]$, we call $f$ a \emph{coloring including $L$-multiple colorings} of $G$, and furthermore $f$ is called an \emph{$(W_i)^L_1$-type coloring} of $G$ if $E(G)=\bigcup^L_{i=1} E(H_i)$.\qqed
\end{defn}

\begin{defn}\label{defn:5C-5C-labeling}
\cite{Bing-Yao-2020arXiv} If a proper total coloring $f:V(G)\cup E(G)\rightarrow [1,M]$ for a bipartite $(p,q)$-graph $G$ holds:

(i) (e-magic) $f(uv)+|f(u)-f(v)|=k$;

(ii) (ee-difference) each edge $uv$ matches with another edge $xy$ holding $f(uv)=|f(x)-f(y)|$ (or $f(uv)=2(p+q)-|f(x)-f(y)|$);

(iii) (ee-balanced) let $s(uv)=|f(u)-f(v)|-f(uv)$ for $uv\in E(G)$, then there exists a constant $k\,'$ such that each edge $uv$ matches with another edge $u\,'v\,'$ holding $s(uv)+s(u\,'v\,')=k\,'$ (or $2(p+q)+s(uv)+s(u\,'v\,')=k\,'$) true;

(iv) (set-ordered) $\max f(X)<\min f(Y)$ (or $\min f(X)>\max f(Y)$) for the bipartition $(X,Y)$ of $V(G)$.

(v) (edge-fulfilled) $f(E(T))=[1,q]$.

We call $f$ a \emph{$5$C-total coloring} of $G$.\qqed
\end{defn}

\subsection{New parameters on proper vertex colorings}

\begin{figure}[h]
\centering
\includegraphics[width=14cm]{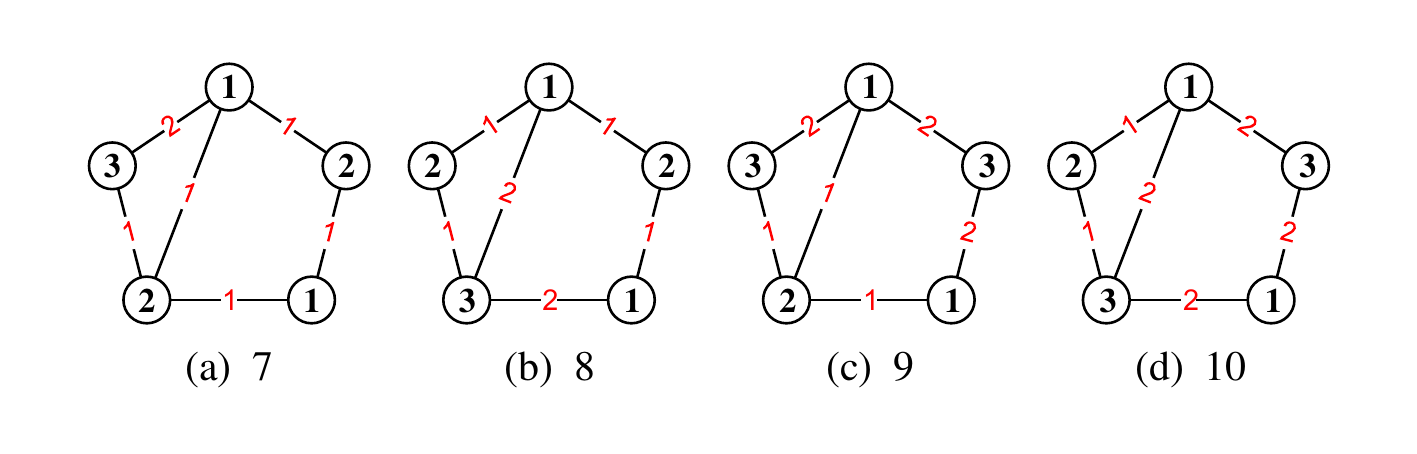}\\
\caption{\label{fig:new-topic-v-coloring-1}{\small A graph $C_5+uv$ admits a group of \emph{consecutive difference proper vertex colorings}, which forms a Topsnut-matching chain, cited from \cite{Yao-Sun-Zhang-Mu-Sun-Wang-Su-Zhang-Yang-Yang-2018arXiv}.}}
\end{figure}

\begin{defn} \label{defn:new-parameters-1}
\cite{Yao-Sun-Zhang-Mu-Sun-Wang-Su-Zhang-Yang-Yang-2018arXiv} Let $f:V(G)\rightarrow [1,k]$ be a proper vertex coloring of a graph $G$ with $k=\chi (G)$. We define a parameter
\begin{equation}\label{eqa:555555}
B_{sub}(G,f)=\sum _{xy\in E(G)}|f(x)-f(y)|,
\end{equation}
and try to determine $\min_fB_{sub}(G,f)$ and $\max_fB_{sub}(G,f)$. Clearly, if $G$ is a bipartite graph, then $$\min_fB_{sub}(G,f)=\max_fB_{sub}(G,f)=|E(G)|.$$
If there exists a proper vertex coloring $f_M$ of a connected graph $G$ such that $B_{sub}(G,f_M)=M$ for each $M$ holding
\begin{equation}\label{eqa:555555}
\min_fB_{sub}(G,f)<M<\max_fB_{sub}(G,f),
\end{equation}we say that $G$ admits \emph{a group of consecutive difference proper vertex colorings} (see an example shown in Fig.\ref{fig:new-topic-v-coloring-1}).\qqed
\end{defn}

\begin{defn} \label{defn:new-parameters-2}
\cite{Yao-Sun-Zhang-Mu-Sun-Wang-Su-Zhang-Yang-Yang-2018arXiv} Let $g:V(G)\rightarrow [1,k]$ be a proper vertex coloring of a graph $G$ with $k=\chi(G)$. We define another parameter
\begin{equation}\label{eqa:555555}
B_{sum}(G,g)=\sum _{xy\in E(G)}[g(x)+g(y)]
\end{equation}
and try to compute two extremal values $\min_gB_{sum}(G,g)$ and $\max_gB_{sum}(G,g)$. If $G$ admits a proper vertex coloring $g_Q$ for each $Q$ satisfying
\begin{equation}\label{eqa:555555}
\min_gB_{sum}(G,g)<Q<\max_hB_{sum}(G,h)
\end{equation}
such that $B_{sum}(G,g_Q)=Q$, we say that $G$ admits \emph{a group of consecutive sum proper vertex colorings}. \qqed
\end{defn}

\subsection{Weak gracefully total colorings}

\begin{defn} \label{defn:weak-graceful-total-colorings}
\cite{Bing-Yao-2020arXiv} If a connected $(p,q)$-graph $G$ admits a total coloring $f:V(G)\cup E(G)\rightarrow [1,M]$, such that $f(uv)=|f(u)-f(v)|$ and $f(u)\neq f(v)$ for each edge $uv\in E(G)$, and $f(E(G))=[1,q]$, as well as $f(V(G))\subseteq [1,q+1]$, we call $f$ a \emph{weak-gracefully total coloring}. It may happen $f(u)=2f(v)$ or $f(v)=2f(u)$ in a weak gracefully total coloring. Moreover, if $G$ is bipartite, and $\max f(X)<\min f(Y)$ for the bipartition $(X,Y)$ of vertex set of $G$, we call $f$ a \emph{set-ordered weak gracefully total coloring.}\qqed
\end{defn}

\begin{defn} \label{defn:weak-graceful-total-colorings}
\cite{Bing-Yao-2020arXiv, Yao-Mu-Sun-Zhang-Yang-Wang-Wang-Su-Ma-Sun-2019} A connected $(p,q)$-graph $G$ admits a total coloring $f:V(G)\cup E(G)\rightarrow [1,M]$, and $f(x)=f(y)$ for some pair of vertices $x,y\in V(G)$. There are the following constraint conditions:
\begin{asparaenum}[(i)]
\item \label{33weak-gracefully-00} each edge $uv\in E(G)$ holds $f(uv)=|f(u)-f(v)|$ true;
\item \label{33weak-gracefully-11} edge color set $f(E(G))=[1,q]$;
\item \label{33weak-proper} vertex color set $f(V(G))\subseteq [1,q+1]$;
\item \label{33weak-odd-gracefully-11} edge color set $f(E(G))=[1,2q-1]^o$;
\item \label{33weak-odd-proper} vertex color set $f(V(G))\subseteq [1,2q]$; and
\item \label{33weak-set-ordered} $G$ is bipartite and its own vertex set $V(G)=X\cup Y$ with $X\cap Y=\emptyset$, such that each edge $uv\in E(G)$ holds $u\in X$ and $v\in Y$, the total coloring $f$ holds $\max f(X)<\min f(Y)$ true.
\end{asparaenum}
\textbf{We have:}
\begin{asparaenum}[\textrm{Weak}-1.]
\item If (\ref{33weak-gracefully-00}) and (\ref{33weak-gracefully-11}) hold true, we call $f$ a \emph{prospective weak gracefully total coloring}.
\item If (\ref{33weak-gracefully-00}), (\ref{33weak-gracefully-11}) and (\ref{33weak-proper}) hold true, so we call $f$ a \emph{weak gracefully total coloring}.
\item If (\ref{33weak-gracefully-00}), (\ref{33weak-gracefully-11}) and (\ref{33weak-set-ordered}) hold true, then we call $f$ a \emph{prospective set-ordered weak gracefully total coloring}.
\item If (\ref{33weak-gracefully-00}), (\ref{33weak-gracefully-11}), (\ref{33weak-proper}) and (\ref{33weak-set-ordered}) hold true, then we call $f$ a \emph{set-ordered weak gracefully total coloring}.
\item If (\ref{33weak-gracefully-00}) and (\ref{33weak-odd-gracefully-11}) hold true, then we call $f$ a \emph{prospective weak odd-gracefully total coloring}.
\item If (\ref{33weak-gracefully-00}), (\ref{33weak-odd-gracefully-11}) and (\ref{33weak-odd-proper}) hold true, so we call $f$ a \emph{weak odd-gracefully total coloring}.
\item If (\ref{33weak-gracefully-00}), (\ref{33weak-odd-gracefully-11}) and (\ref{33weak-set-ordered}) hold true, we call $f$ a \emph{prospective set-ordered weak odd-gracefully total coloring}.
\item If (\ref{33weak-gracefully-00}), (\ref{33weak-odd-gracefully-11}), (\ref{33weak-odd-proper}) and (\ref{33weak-set-ordered}) hold true, then we call $f$ a \emph{set-ordered weak odd-gracefully total coloring}.\qqed
\end{asparaenum}
\end{defn}

\begin{rem} \label{remc:graceful-total-colorings}
(1) Each total coloring defined in Definition \ref{defn:weak-graceful-total-colorings} is a mixed thing of graceful labeling and a total coloring.

(2) Definition \ref{defn:weak-graceful-total-colorings} induces a parameter: A connected $(p,q)$-graph $G$ admits a weak graceful (resp. odd-graceful) total coloring $h$, if $|h(V(G))|\leq |f(V(G))|$ holds true for any \emph{weak graceful (resp. odd-graceful) total coloring} $f$ of $G$, we write $v_{gtc}(G)=|h(V(T))|=\min _f\{|f(V(T))|\}$ over all weak graceful (resp. odd-graceful) total colorings of $G$, called a \emph{weak graceful (resp. odd-graceful) total chromatic number}. It is not easy to compute the exact value of $v_{gtc}(G)$, since it is related with Total Coloring Conjecture.

(3) If a total coloring $f$ defined in Definition \ref{defn:weak-graceful-total-colorings} is \emph{proper}, so we have $f(v)\neq f(uv)=|f(u)-f(v)|$ and $2f(v)\neq f(u)$, or $f(u)\neq f(uv)=|f(u)-f(v)|$ and $2f(u)\neq f(v)$. In general, a total coloring $f$ defined in Definition \ref{defn:weak-graceful-total-colorings} and its dual total coloring may be not \emph{proper}. If a total coloring defined in Definition \ref{defn:weak-graceful-total-colorings} is \emph{proper}, so we call it a graceful (resp. odd-graceful) proper total coloring, or a weak graceful (resp. odd-graceful) proper total coloring, or a set-ordered weak graceful (resp. odd-graceful) total coloring, or a prospective weak graceful (resp. odd-graceful) proper total coloring, or a prospective set-ordered weak graceful (resp. odd-graceful) proper total coloring.\paralled
\end{rem}

\subsection{Parameterized total colorings}

Let integers $a,k,m\geq 0$, $d\geq 1$ and $q\geq 1$ in this subsection. We have two parameterized sets $S_{m,k,a,d}=\{k+ad,k+(a+1)d,\dots ,k+(a+m)d\}$ and $O_{2q-1,k,d}=\{k+d,k+3d,\dots ,k+(2q-1)d\}$. The \emph{cardinality} of a set $S$ is denoted as $|S|$, so $|S_{m,k,a,d}|=m+1$ and $|O_{2q-1,k,d}|=q$. Motivated from the parameterized labelings: the $(k,d)$-arithmetic labeling, the $(k,d)$-graceful labeling, the $(k,d)$-harmonious labeling and the new $(k,d)$-type labelings introduced in \cite{Sun-Zhang-Yao-ICMITE-2017}. Based on the well-defined parameterized labelings, we present the following parameterized total colorings:

\begin{defn} \label{defn:kd-w-type-colorings}
\cite{Yao-Su-Wang-Hui-Sun-ITAIC2020} Let $G$ be a bipartite and connected $(p,q)$-graph, so its vertex set $V(G)=X\cup Y$ with $X\cap Y=\emptyset$ such that each edge $uv\in E(G)$ holds $u\in X$ and $v\in Y$. If there is a coloring $f:X\rightarrow S_{m,0,0,d}=\{0,d,\dots ,md\}$ and $f:Y\cup E(G)\rightarrow S_{n,k,0,d}=\{k,k+d,\dots ,k+nd\}$, here it is allowed $f(x)=f(y)$ for some distinct vertices $x,y\in V(G)$. Let $c$ be a non-negative integer. \textbf{We have}:
\begin{asparaenum}[\textrm{Ptol}-1. ]
\item If $f(uv)=|f(u)-f(v)|$ for $uv\in E(G)$, $f(E(G))=S_{q-1,k,0,d}$ and $f(V(G)\cup E(G))\subseteq S_{m,0,0,d}\cup S_{q-1,k,0,d}$, then $f$ is called a \emph{$(k,d)$-gracefully total coloring}; and moreover $f$ is called a \emph{$(k,d)$-strongly gracefully total coloring} if $f(x)+f(y)=k+(q-1)d$ for each matching edge $xy$ of a matching $M$ of $G$.
\item If $f(uv)=|f(u)-f(v)|$ for $uv\in E(G)$, $f(E(G))=O_{2q-1,k,d}$ and $f(V(G)\cup E(G))\subseteq S_{m,0,0,d}\cup S_{2q-1,k,0,d}$, then $f$ is called a \emph{$(k,d)$-odd-gracefully total coloring}; and moreover $f$ is called a \emph{$(k,d)$-strongly odd-gracefully total coloring} if $f(x)+f(y)=k+(2q-1)d$ for each matching edge $xy$ of a matching $M$ of $G$.
\item If the color set $$\{f(u)+f(uv)+f(v):uv\in E(G)\}=\{2k+2ad,2k+2(a+1)d,\dots ,2k+2(a+q-1)d\}$$ with $a\geq 0$ and $f(V(G)\cup E(G))\subseteq S_{m,0,0,d}\cup S_{2(a+q-1),k,a,d}$, then $f$ is called a \emph{$(k,d)$-edge antimagic total coloring}.
\item If $f(uv)=f(u)+f(v)~(\bmod^*qd)$ defined by $f(uv)-k=[f(u)+f(v)-k](\bmod ~qd)$ for $uv\in E(G)$, and $f(E(G))=S_{q-1,k,0,d}$, then $f$ is called a \emph{$(k,d)$-harmonious total coloring}.
\item If $f(uv)=f(u)+f(v)~(\bmod^*qd)$ defined by $f(uv)-k=[f(u)+f(v)-k](\bmod ~qd)$ for $uv\in E(G)$, and $f(E(G))=O_{2q-1,k,d}$, then $f$ is called a \emph{$(k,d)$-odd-elegant total coloring}.
\item If $f(u)+f(uv)+f(v)=c$ for each edge $uv\in E(G)$, $f(E(G))=S_{q-1,k,0,d}$ and $f(V(G)\cup E(G))\subseteq S_{m,0,0,d}\cup S_{q-1,k,0,d}$, then $f$ is called a \emph{strongly $(k,d)$-edge-magic total coloring}; and moreover $f$ is called an \emph{edge-magic $(k,d)$-total coloring} if $|f(E(G))|\leq q$ and $f(u)+f(uv)+f(v)=c$ for $uv\in E(G)$.
\item If $f(uv)+|f(u)-f(v)|=c$ for each edge $uv\in E(G)$ and $f(E(G))=S_{q-1,k,0,d}$, then $f$ is called a \emph{strongly edge-difference $(k,d)$-total coloring}; and moreover $f$ is called an \emph{edge-difference $(k,d)$-total coloring} if $|f(E(G))|\leq q$ and $f(uv)+|f(u)-f(v)|=c$ for $uv\in E(G)$.
\item If $|f(u)+f(v)-f(uv)|=c$ for each edge $uv\in E(G)$ and $f(E(G))=S_{q-1,k,0,d}$, then $f$ is called a \emph{strongly felicitous-difference $(k,d)$-total coloring}; and moreover $f$ is called a \emph{felicitous-difference $(k,d)$-total coloring} if $|f(E(G))|\leq q$ and $|f(u)+f(v)-f(uv)|=c$ for $uv\in E(G)$.
\item If $\big ||f(u)-f(v)|-f(uv)\big |=c$ for each edge $uv\in E(G)$ and $f(E(G))=S_{q-1,k,0,d}$, then $f$ is called a \emph{strongly graceful-difference $(k,d)$-total coloring}; and $f$ is called a \emph{graceful-difference $(k,d)$-total coloring} if $|f(E(G))|\leq q$ and $\big ||f(u)-f(v)|-f(uv)\big |=c$ for $uv\in E(G)$.\qqed
\end{asparaenum}
\end{defn}

\begin{defn} \label{defn:properk-d-total-colorings}
$^*$ A total coloring $f$ defined in Definition \ref{defn:kd-w-type-colorings} is \emph{proper} if $f(u)\neq f(v)$ for each edge $uv\in E(G)$, and $f(uv)\neq f(uw)$ for any two adjacent edges $uv,uw\in E(G)$, so we call $f$ a \emph{$W$-type proper $(k,d)$-total coloring} of $G$.\qqed
\end{defn}

\begin{rem} \label{rem:kd-w-tupe-colorings-definition}
In general, we call $f$ a \emph{$W$-type $(k,d)$-total coloring} if the constraint function $F(f(u)$, $f(uv)$, $f(v))=0$ and the edge color set $f(E(G))$ hold one of the well-defined total colorings defined in Definition \ref{defn:kd-w-type-colorings}.
\begin{asparaenum}[(1) ]
\item We have some new parameters of graphs based on Definition \ref{defn:kd-w-type-colorings}. For a graph $G$, we have two $W$-type $(k,d)$-total colorings $g_{\min}$ and $g_{\max}$ of $G$ such that
 $$|g_{\min}(V(G))|\leq |g(V(G))|\leq |g_{\max}(V(G))|$$ for each $W$-type $(k,d)$-total coloring $g$ of $G$. As this $W$-type $(k,d)$-total coloring is the graceful $(k,d)$-total coloring, then a graceful $(1,1)$-total coloring $g_{\max}$ of a tree $T$ means that $|g_{\max}(V(T))|$ is the approximate value of a graceful labeling of $T$. The graceful tree conjecture says: $|g_{\max}(V(T))|=|V(T)|$.
\item In application, we may reduce the restrictions of some $W$-type $(k,d)$-total colorings of graphs, such as a $W$-type $(k,d)$-total coloring $g$ of a bipartite graph $G$ with bipartition $(X,Y)$ holds $g:X\rightarrow \{-md,-(m-1)d,\dots ,-2d,-d\}\cup \{0,d,\dots ,md\}$ and $g:Y\cup E(G)\rightarrow \{-k,-k-d,\dots ,-k-nd\}\cup \{-k+d,-k+2d,\dots ,-k+nd\}\cup \{k-d,k-2d,\dots ,k-nd\}\cup \{k,k+d,\dots ,k+nd\}$ with $m,n,k,d>0$.\paralled
\end{asparaenum}
\end{rem}

\subsection{Multiple dimension total colorings}

We put three number-based strings $0112304$, $4500123$ and $0023111$ into one number-based string
$$011230445001230023111, \textrm{or }002311101123044500123, \textrm{or }450012301123040023111$$
and we call them \emph{ordered permutations}, and each one of $0112304$, $4500123$ and $0023111$ is called a \emph{coordinate}. In general, a $n$-dimension number-based string is defined by an ordered permutation $a_1a_2\cdots a_n$ and each coordinate $a_i=\beta_{i,1}\beta_{i,2}\cdots \beta_{i,b_i}$ with $\beta_{i,j}\in [0,9]$ for $j\in [1,b_i]$, and moreover the number $||a_i||=\sum ^{b_i}_{r=1}\beta_{i,r}$ is called the \emph{norm} of the coordinate $a_i$. If $\theta_{i,r}=0$ for $r\in [1,s]$ and $\theta_{i,s+1}\geq 1$ with $s+1\leq b_i$ in some coordinate $a_i=\theta_{i,1}\theta_{i,2}\cdots \theta_{i,b_i}$ of a $n$-dimension number-based string $a_1a_2\cdots a_n$, as a substitution, we use $\theta_{i,s+1}\theta_{i,s+2}\cdots \theta_{i,b_i}\in Z^0$ to participate in addition operation, subtraction operation, multiplication operation and and other mathematical operations. Let $S(n,Z^0)$ be the set of all $n$-dimension number-based strings. We present new colorings of graphs, which are like parameterized colorings in \cite{Yao-Wang-Su-arameterized-2020}.

\begin{defn} \label{defn:n-dimension-total-colorings}
\cite{Yao-Wang-Su-arameterized-2020} A $(p,q)$-graph $G$ admits a coloring $f:S\subseteq V(G)\cup E(G)\rightarrow S(n,Z^0)$. Let $f(u)=a_1a_2\cdots a_n$, $f(v)=b_1b_2\cdots b_n$, $f(uv)=c_1c_2\cdots c_n$ for each edge $uv\in E(G)$, and $\gamma$ be a non-negative integer. There are the following constraint conditions:
\begin{asparaenum}[\textrm{Res}-1. ]
\item \label{dimen:v} $S=V(G)$;
\item \label{dimen:e} $S=E(G)$;
\item \label{dimen:ve} $S=V(G)\cup E(G)$;
\item \label{dimen:adjacent-v} $f(u)\neq f(v)$ for each edge $uv\in E(G)$;
\item \label{dimen:adjacent-e} $f(uv)\neq f(uw)$ for any pair of two adjacent edges $uv,uw\in E(G)$;
\item \label{dimen:incident-v-e} $f(u)\neq f(uv)$ and $f(v)\neq f(uv)$ for each edge $uv\in E(G)$;
\item \label{dimen:e-graceful} $c_j=|a_j-b_j|$ with $j\in [1,n]$;
\item \label{dimen:each-odd} $c_j=|a_j-b_j|$, and each $c_j$ is odd with $j\in [1,n]$;
\item \label{dimen:felicitous} $c_j=a_j+b_j~(\bmod~q)$ with $j\in [1,n]$;
\item \label{dimen:edge-magic} $a_j+b_j+c_j=\gamma$ with $j\in [1,n]$;
\item \label{dimen:edge-difference} $c_j+|a_j-b_j|=\gamma$ with $j\in [1,n]$;
\item \label{dimen:graceful-difference} $\big |c_j-|a_j-b_j|\big |=\gamma$ with $j\in [1,n]$;
\item \label{dimen:felicitous-difference} $|a_j+b_j-c_j|=\gamma$ with $j\in [1,n]$;
\item \label{dimen:each-common-factor} $c_j=\textrm{gcd}(a_j,b_j)$ for each $j\in [1,n]$;
\item \label{dimen:common-factor} Some $j\in [1,n]$ holds $c_j=\textrm{gcd}(a_j,b_j)$;
\item \label{dimen:pairwise-distinct} $c_i\neq c_j$ for any pair of $c_i$ and $c_j$;
\item \label{dimen:part-graceful} $f(E(G))=\{f(e_k)=c_{k,1}c_{k,2}\cdots c_{k,n}:k\in[1$, $q]\}$ such that each $k\in[1,q]$ holds $c_{k,j}=k$ for some $c_{k,j}$ of $f(e_k)$;
\item \label{dimen:part-odd-graceful} $f(E(G))=\{f(e_j)=t_{j,1}t_{j,2}\cdots t_{j,n}:j\in[1$, $2q-1]^o\}$ such that each $j\in[1,2q-1]^o$ holds $t_{j,r}=j$ for some $t_{j,r}$ of $f(e_j)$;
\item \label{dimen:uniform-graceful} $f(E(G))=\{f(e_k)=c_{k,1}c_{k,2}\cdots c_{k,n}:k\in[1$, $q]\}$ such that $c_{k,r}=k\in[1,q]$ for each $r\in [1,n]$; and
\item \label{dimen:uniform-odd-graceful} $f(E(G))=\{f(e_j)=t_{j,1}t_{j,2}\cdots t_{j,n}:j\in[1$, $2q-1]^o\}$ such that $t_{j,r}=j\in[1,2q-1]^o$ for each $r\in [1,n]$.
\end{asparaenum}

\noindent \textbf{We call $f$}:

\noindent ------ \emph{traditional types}

\begin{asparaenum}[\textrm{Dtc}-1. ]
\item A \emph{$n$-dimension proper vertex coloring} if Res-\ref{dimen:v} and Res-\ref{dimen:adjacent-v} hold true.
\item A \emph{$n$-dimension proper edge coloring} if Res-\ref{dimen:e} and Res-\ref{dimen:adjacent-e} hold true.
\item A \emph{$n$-dimension proper total coloring} if Res-\ref{dimen:ve}, Res-\ref{dimen:adjacent-v}, Res-\ref{dimen:adjacent-e} and Res-\ref{dimen:incident-v-e} hold true.

\noindent ------ \emph{other proper types}

\item A \emph{subtraction $n$-dimension proper total coloring} if Res-\ref{dimen:ve}, Res-\ref{dimen:adjacent-v}, Res-\ref{dimen:adjacent-e}, Res-\ref{dimen:incident-v-e} and Res-\ref{dimen:e-graceful} hold true.
\item A \emph{factor $n$-dimension proper total coloring} if Res-\ref{dimen:ve}, Res-\ref{dimen:adjacent-v}, Res-\ref{dimen:adjacent-e}, Res-\ref{dimen:incident-v-e} and Res-\ref{dimen:common-factor} hold true.
\item A \emph{felicitous $n$-dimension proper total coloring} if Res-\ref{dimen:ve}, Res-\ref{dimen:adjacent-v}, Res-\ref{dimen:adjacent-e}, Res-\ref{dimen:incident-v-e} and Res-\ref{dimen:felicitous} hold true.

\item A \emph{graceful $n$-dimension proper total coloring} if Res-\ref{dimen:ve}, Res-\ref{dimen:adjacent-v}, Res-\ref{dimen:adjacent-e}, Res-\ref{dimen:incident-v-e}, Res-\ref{dimen:e-graceful} and Res-\ref{dimen:part-graceful} hold true.
\item An \emph{odd-graceful $n$-dimension proper total coloring} if Res-\ref{dimen:ve}, Res-\ref{dimen:adjacent-v}, Res-\ref{dimen:adjacent-e}, Res-\ref{dimen:incident-v-e}, Res-\ref{dimen:e-graceful} and Res-\ref{dimen:part-odd-graceful} hold true.

\item An \emph{e-magic $n$-dimension proper total coloring} if Res-\ref{dimen:ve}, Res-\ref{dimen:adjacent-v}, Res-\ref{dimen:adjacent-e}, Res-\ref{dimen:incident-v-e}, Res-\ref{dimen:e-graceful} and Res-\ref{dimen:edge-magic} hold true.
\item An \emph{e-difference $n$-dimension proper total coloring} if Res-\ref{dimen:ve}, Res-\ref{dimen:adjacent-v}, Res-\ref{dimen:adjacent-e}, Res-\ref{dimen:incident-v-e}, Res-\ref{dimen:e-graceful} and Res-\ref{dimen:edge-difference} hold true.
\item A \emph{graceful-difference $n$-dimension proper total coloring} if Res-\ref{dimen:ve}, Res-\ref{dimen:adjacent-v}, Res-\ref{dimen:adjacent-e}, Res-\ref{dimen:incident-v-e}, Res-\ref{dimen:e-graceful} and Res-\ref{dimen:graceful-difference} hold true.
\item A \emph{felicitous-difference $n$-dimension proper total coloring} if Res-\ref{dimen:ve}, Res-\ref{dimen:adjacent-v}, Res-\ref{dimen:adjacent-e}, Res-\ref{dimen:incident-v-e}, Res-\ref{dimen:e-graceful} and Res-\ref{dimen:felicitous-difference} hold true.
\item An \emph{anti-equitable $n$-dimension proper total coloring} if Res-\ref{dimen:ve}, Res-\ref{dimen:adjacent-v}, Res-\ref{dimen:adjacent-e}, Res-\ref{dimen:incident-v-e}, Res-\ref{dimen:e-graceful} and Res-\ref{dimen:pairwise-distinct} hold true.

\noindent ------ \emph{sub-proper types}

\item A \emph{subtraction $n$-dimension sub-proper total coloring} if Res-\ref{dimen:ve}, Res-\ref{dimen:adjacent-v}, Res-\ref{dimen:adjacent-e} and Res-\ref{dimen:e-graceful} hold true.
\item A \emph{factor $n$-dimension sub-proper total coloring} if Res-\ref{dimen:ve}, Res-\ref{dimen:adjacent-v}, Res-\ref{dimen:adjacent-e} and Res-\ref{dimen:common-factor} hold true.
\item A \emph{felicitous $n$-dimension sub-proper total coloring} if Res-\ref{dimen:ve}, Res-\ref{dimen:adjacent-v}, Res-\ref{dimen:adjacent-e} and Res-\ref{dimen:felicitous} hold true.

\item A \emph{graceful $n$-dimension sub-proper total coloring} if Res-\ref{dimen:ve}, Res-\ref{dimen:adjacent-v}, Res-\ref{dimen:adjacent-e}, Res-\ref{dimen:e-graceful} and Res-\ref{dimen:part-graceful} hold true.
\item An \emph{odd-graceful $n$-dimension sub-proper total coloring} if Res-\ref{dimen:ve}, Res-\ref{dimen:adjacent-v}, Res-\ref{dimen:adjacent-e}, Res-\ref{dimen:e-graceful} and Res-\ref{dimen:part-odd-graceful} hold true.

\item A \emph{twin-graceful $n$-dimension sub-proper total coloring} if Res-\ref{dimen:ve}, Res-\ref{dimen:adjacent-v}, Res-\ref{dimen:adjacent-e}, Res-\ref{dimen:e-graceful}, Res-\ref{dimen:part-graceful} and Res-\ref{dimen:part-odd-graceful} hold true.

\item A \emph{uniformly graceful $n$-dimension sub-proper total coloring} if Res-\ref{dimen:ve}, Res-\ref{dimen:adjacent-v}, Res-\ref{dimen:adjacent-e}, Res-\ref{dimen:e-graceful} and Res-\ref{dimen:uniform-graceful} hold true.
\item A \emph{uniformly odd-graceful $n$-dimension sub-proper total coloring} if Res-\ref{dimen:ve}, Res-\ref{dimen:adjacent-v}, Res-\ref{dimen:adjacent-e}, Res-\ref{dimen:e-graceful} and Res-\ref{dimen:uniform-odd-graceful} hold true.

\item A \emph{uniformly factor $n$-dimension sub-proper total coloring} if Res-\ref{dimen:ve}, Res-\ref{dimen:adjacent-v}, Res-\ref{dimen:adjacent-e}, Res-\ref{dimen:e-graceful} and Res-\ref{dimen:each-common-factor} hold true.

\item An \emph{anti-equitable $n$-dimension sub-proper total coloring} if Res-\ref{dimen:ve}, Res-\ref{dimen:adjacent-v}, Res-\ref{dimen:adjacent-e}, Res-\ref{dimen:e-graceful} and Res-\ref{dimen:pairwise-distinct} hold true.

\item An \emph{e-magic $n$-dimension sub-proper total coloring} if Res-\ref{dimen:ve}, Res-\ref{dimen:adjacent-v}, Res-\ref{dimen:adjacent-e}, Res-\ref{dimen:e-graceful} and Res-\ref{dimen:edge-magic} hold true.
\item An \emph{e-difference $n$-dimension sub-proper total coloring} if Res-\ref{dimen:ve}, Res-\ref{dimen:adjacent-v}, Res-\ref{dimen:adjacent-e}, Res-\ref{dimen:e-graceful} and Res-\ref{dimen:edge-difference} hold true.
\item A \emph{graceful-difference $n$-dimension sub-proper total coloring} if Res-\ref{dimen:ve}, Res-\ref{dimen:adjacent-v}, Res-\ref{dimen:adjacent-e}, Res-\ref{dimen:e-graceful} and Res-\ref{dimen:graceful-difference} hold true.
\item A \emph{felicitous-difference $n$-dimension sub-proper total coloring} if Res-\ref{dimen:ve}, Res-\ref{dimen:adjacent-v}, Res-\ref{dimen:adjacent-e}, Res-\ref{dimen:e-graceful} and Res-\ref{dimen:felicitous-difference} hold true.\qqed
\end{asparaenum}
\end{defn}

\begin{figure}[h]
\centering
\includegraphics[width=16cm]{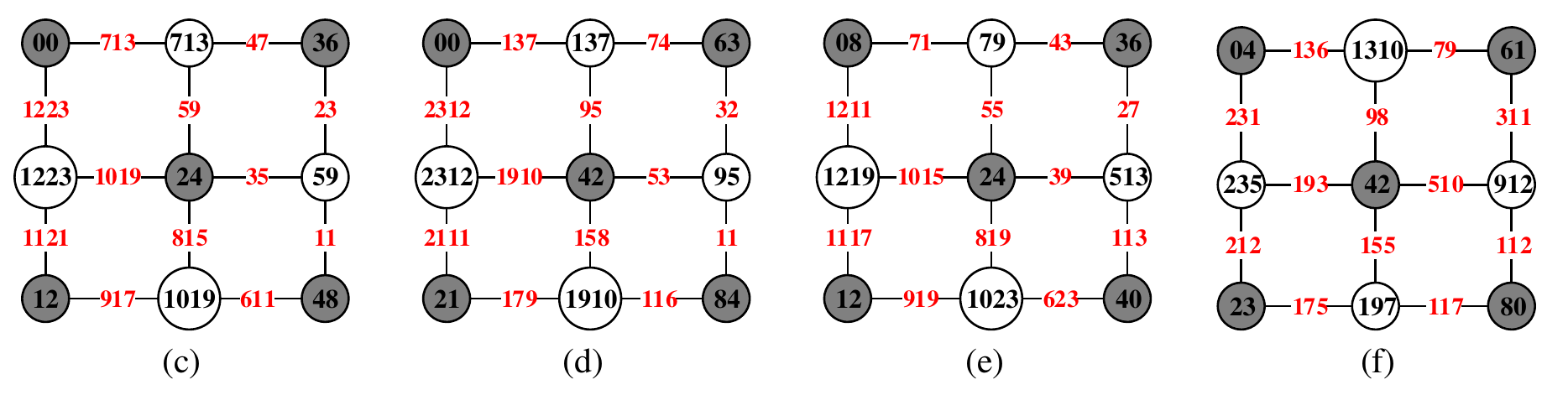}
\caption{\label{fig:n-dimension-11}{\small Four Topsnut-gpws, where both (c) and (d) admit two twin $2$-dimension sub-proper total colorings; both (e) and (f) admit two twin $2$-dimension proper total colorings, cited from \cite{Yao-Wang-Su-arameterized-2020}.}}
\end{figure}

\begin{figure}[h]
\centering
\includegraphics[width=14cm]{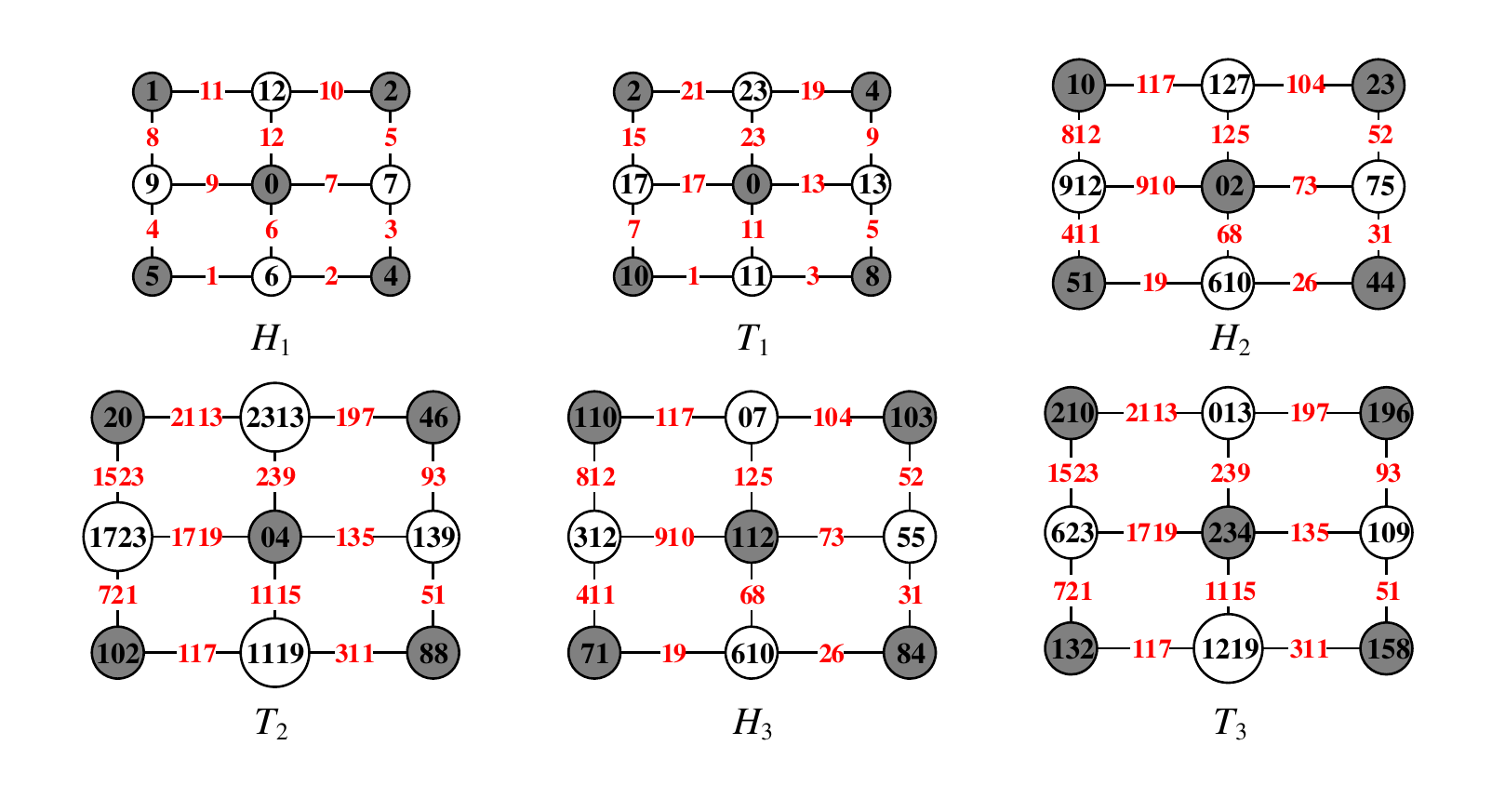}
\caption{\label{fig:n-dimension-22}{\small Six Topsnut-gpws: $H_1$ admits a graceful labeling, $T_1$ admits an odd-graceful labeling; both $H_2$ and $H_3$ admit two graceful $2$-dimension proper total colorings; both $T_2$ and $T_3$ admit two odd-graceful $2$-dimension proper total colorings, cited from \cite{Yao-Wang-Su-arameterized-2020}.}}
\end{figure}

\begin{rem}\label{rem:333333}
We, also, call $f$ defined in Definition \ref{defn:n-dimension-total-colorings} a \emph{$W$-type $n$-dimension total coloring}, since $f$ holds a group of restrictive conditions denoted as $f(uv)=W(f(u),f(v))$ for each edge $uv\in E(G)$. By the way, we can add the restrictive conditions Re-\ref{dimen:adjacent-v}, Re-\ref{dimen:adjacent-e} and Re-\ref{dimen:incident-v-e} to a $W$-type $n$-dimension total coloring for getting a \emph{$W$-type $n$-dimension proper total coloring}. See examples shown in Fig.\ref{fig:n-dimension-11} and Fig.\ref{fig:n-dimension-22}.\paralled
\end{rem}

\begin{thm} \label{thm:graph-graceful-n-dimension}
\cite{Yao-Wang-Su-arameterized-2020} Every connected simple graph admits a \emph{graceful $n$-dimension sub-proper total coloring} for some $n\geq 2$.
\end{thm}
\begin{thm} \label{thm:trees-graceful-2-dimension}
\cite{Yao-Wang-Su-arameterized-2020} Every tree admits a \emph{graceful $2$-dimension proper total coloring}.
\end{thm}
\begin{thm} \label{thm:twin-graceful-2-dimension-tree}
\cite{Yao-Wang-Su-arameterized-2020} Every tree admits a \emph{twin-graceful $2$-dimension sub-proper total coloring}.
\end{thm}

We introduce the following compounded-type multiple dimension total colorings:

\begin{defn} \label{defn:33multiple-dimension-kd-w-type-colorings}
$^*$ By Definition \ref{defn:kd-w-type-colorings}, let $G$ be a bipartite and connected $(p,q)$-graph, so its vertex set $V(G)=X\cup Y$ with $X\cap Y=\emptyset$ such that each edge $uv\in E(G)$ holds $u\in X$ and $v\in Y$. There are a group of colorings $f_s:X\rightarrow S_{m,0,0,d}=\{0,d,\dots ,md\}$ and $f_s:Y\cup E(G)\rightarrow S_{n,k,0,d}=\{k,k+d,\dots ,k+nd\}$ (here it is allowed $f_s(u)=f_s(w)$ for some distinct vertices $u,w\in V(G)$) for $s\in [1,B]$ with integer $B\geq 2$, such that $f_s\in \{$ $(k_s,d_s)$-strongly gracefully total coloring, $(k_s,d_s)$-strongly odd-gracefully total coloring, $(k_s,d_s)$-edge antimagic total coloring, $(k_s,d_s)$-harmonious total coloring, $(k_s,d_s)$-odd-elegant total coloring, strongly $(k_s,d_s)$-edge-magic total coloring, edge-magic $(k_s,d_s)$-total coloring, strongly edge-difference $(k_s,d_s)$-total coloring, edge-difference $(k_s,d_s)$-total coloring, strongly felicitous-difference $(k_s,d_s)$-total coloring, felicitous-difference $(k_s,d_s)$-total coloring, strongly graceful-difference $(k_s,d_s)$-total coloring, graceful-difference $(k_s,d_s)$-total coloring $\}$ with $s\in [1,B]$. Then $G$ admits a \emph{parameterized compounded $B$-dimension total coloring} $F$ holding $F(u)=f_1(u)f_2(u)\cdots f_B(u)$, $F(uv)=f_1(uv)f_2(uv)\cdots f_B(uv)$ and $F(v)=f_1(v)f_2(v)\cdots f_B(v)$ for each edge $uv\in E(G)$.\qqed
\end{defn}

\begin{defn} \label{defn:33multiple-dimension-total-colorings}
$^*$ A $(p,q)$-graph $G$ admits a coloring $g_j:S\subseteq V(G)\cup E(G)\rightarrow S(n,Z^0)$ for $i\in [1,B]$, and $g_j$ is one of \emph{Dtc-$s$ total colorings} with $j\in [1,27]$ defined in Definition \ref{defn:n-dimension-total-colorings}. Then $G$ admits a \emph{compounded $B$-dimension total coloring} $F$ holding $F(u)=g_1(u)g_2(u)\cdots g_B(u)$, $F(uv)=g_1(uv)g_2(uv)\cdots g_B(uv)$ and $F(v)=g_1(v)g_2(v)\cdots g_B(v)$ for each edge $uv\in E(G)$.\qqed
\end{defn}

\begin{rem}\label{rem:333333}
In Definition \ref{defn:33multiple-dimension-kd-w-type-colorings}, each sequence $\{(k_s,d_s)\}^B_1$ will induce random number-based strings generated from the Topcode-matrix of $G$ admitting a parameterized compounded $B$-dimension total coloring. Similarly, a compounded $B$-dimension total coloring $F$ of $G$ defined in Definition \ref{defn:33multiple-dimension-total-colorings} is more complex than any one total coloring defined in Definition \ref{defn:n-dimension-total-colorings}.\paralled
\end{rem}

\subsection{Colorings based on graphic groups and string groups}

Using every-zero graphic groups and edge-every-zero graphic groups to encrypting networks are similar with graph colorings.

\begin{defn} \label{defn:graphic-group-string-coloring}
\cite{Yao-Mu-Sun-Sun-Zhang-Wang-Su-Zhang-Yang-Zhao-Wang-Ma-Yao-Yang-Xie2019} Let $H$ be a $(p,q)$-graph, and $\{F_f(G),\oplus\}$ be an every-zero graphic group containing Topsnut-gpws $G_i$ with $i\in [1,M_G]$. Suppose that $H$ admits a mapping $\theta: S\rightarrow F_f(G)$ with $S\subseteq V(H)\cup E(H)$. Write $\theta(S)=\{\theta(w):w\in S\}$, $\theta(N_{ei}(u))=\{\theta(v):v\in N_{ei}(u)\}$ and $\theta(N\,'_{ei}(u))=\{\theta(uv):v\in N_{ei}(u)\}$. There are the following so-called \emph{graphic group constraints}:
\begin{asparaenum}[\textrm{Gg}-1. ]
\item \label{prob:only-vertex} $S=V(H)$;
\item \label{prob:only-edge} $S=E(H)$;
\item \label{prob:total} $S=V(H)\cup E(H)$;
\item \label{prob:vertex-not} $\theta(u)\neq \theta(v)$ for $v\in N_{ei}(u)$;
\item \label{prob:edge-not} $\theta(uv)\neq \theta(uw)$ for $v,w\in N_{ei}(u)$;
\item \label{prob:vertex-distinghuishing} $\theta(N_{ei}(x))\neq \theta(N_{ei}(y))$ for any pair of vertices $x,y\in V(H)$;
\item \label{prob:adjacent-vertex-distinghuishing} $\theta(N_{ei}(u))\neq \theta(N_{ei}(v))$ for each edge $uv\in E(H)$;
\item \label{prob:vertex-edge-distinghuishing} $\theta(N\,'_{ei}(x))\neq \theta(N\,'_{ei}(y))$ for any pair of vertices $x,y\in V(H)$;
\item \label{prob:adjacent-vertex-edge-distinghuishing} $\theta(N\,'_{ei}(u))\neq \theta(N\,'_{ei}(v))$ for each edge $uv\in E(H)$;
\item \label{prob:adjacent-vertex-equitable} $\big ||\theta(N_{ei}(u))|-|\theta(N_{ei}(v))|\big |\leq 1$ for each edge $uv\in E(H)$;
\item \label{prob:adjacent-edge-equitable} $\big ||\theta(N\,'_{ei}(u))|-|\theta(N\,'_{ei}(v))|\big |\leq 1$ for each edge $uv\in E(H)$;
\item \label{prob:adjacent-total-distinghuishing} $\{\theta(u)\}\cup \theta(N_{ei}(u))\cup \theta(N\,'_{ei}(u))\neq \{\theta(v)\}\cup \theta(N_{ei}(v))\cup \theta(N\,'_{ei}(v))$ for each edge $uv\in E(H)$;
\item \label{prob:induced-edge} There exists an index $k_{uv}$ such that $\theta(uv)=\theta(u)\oplus _{k_{uv}}\theta(v)\in F_f(G)$ for each edge $uv\in E(H)$; and
\item \label{prob:induced-vertex} Each vertex $w$ is assigned with a set $\{\theta(ww_i)\oplus _{k_{ij}}\theta(ww_j): w_i,w_j\in N_{ei}(w)\}$.
\end{asparaenum}
\noindent \textbf{Then we call $\theta$}:
\begin{asparaenum}[1-\textrm{ggc}. ]
\item A \emph{proper gg-coloring} if Gg-\ref{prob:only-vertex} and Gg-\ref{prob:vertex-not} hold true.
\item A \emph{proper edge-gg-coloring} if Gg-\ref{prob:only-edge} and Gg-\ref{prob:edge-not} hold true.
\item A \emph{proper total gg-coloring} if Gg-\ref{prob:total}, Gg-\ref{prob:vertex-not} and Gg-\ref{prob:edge-not} hold true.
\item A \emph{vertex distinguishing proper gg-coloring} if Gg-\ref{prob:only-vertex}, Gg-\ref{prob:vertex-not} and Gg-\ref{prob:vertex-distinghuishing} hold true.
\item An \emph{adjacent vertex distinguishing proper gg-coloring} if Gg-\ref{prob:only-vertex}, Gg-\ref{prob:vertex-not} and Gg-\ref{prob:adjacent-vertex-distinghuishing} hold true.
\item An \emph{edge distinguishing proper gg-coloring} if Gg-\ref{prob:only-edge}, Gg-\ref{prob:edge-not} and Gg-\ref{prob:vertex-edge-distinghuishing} hold true.
\item An \emph{adjacent edge distinguishing proper gg-coloring} if Gg-\ref{prob:only-edge}, Gg-\ref{prob:edge-not} and Gg-\ref{prob:adjacent-vertex-edge-distinghuishing} hold true.
\item An \emph{equitable adjacent-v proper gg-coloring} if Gg-\ref{prob:only-vertex}, Gg-\ref{prob:vertex-not} and Gg-\ref{prob:adjacent-vertex-equitable} hold true.
\item An \emph{equitable adjacent-e proper gg-coloring} if Gg-\ref{prob:only-edge}, Gg-\ref{prob:edge-not} and Gg-\ref{prob:adjacent-edge-equitable} hold true.
\item An \emph{adjacent total distinguishing proper gg-coloring} if Gg-\ref{prob:total}, Gg-\ref{prob:vertex-not}, Gg-\ref{prob:edge-not} and Gg-\ref{prob:adjacent-total-distinghuishing} hold true.
\item A \emph{v-induced total proper gg-coloring} if Gg-\ref{prob:only-vertex}, Gg-\ref{prob:vertex-not} and Gg-\ref{prob:induced-edge} hold true.
\item An \emph{induced e-proper v-set gg-coloring} if Gg-\ref{prob:only-edge}, Gg-\ref{prob:edge-not} and Gg-\ref{prob:induced-vertex} hold true.\qqed
\end{asparaenum}
\end{defn}

\begin{rem}\label{rem:ABC-conjecture}
Based on Definition \ref{defn:graphic-group-string-coloring}, we have new parameters:
\begin{asparaenum}[New-1. ]
\item $\chi _{gg}(H)$ is the minimum number of $M_G$ Topsnut-gpws $G_i$ in some every-zero graphic group $\{F_f(G),\oplus\}$ for which $H$ admits a \emph{proper gg-coloring}. Bruce Reed in 1998 conjectured that $\chi(H)\leq \lceil \frac{1}{2}[\Delta(H)+1+K(H)]\rceil$ (Ref. \cite{Bondy-2008}), where $\chi _{gg}(H)=\chi(H)$.
\item $\chi\,' _{gg}(H)$ is the minimum number of $M_G$ Topsnut-gpws $G_i$ in some every-zero graphic group $\{F_f(G),\oplus\}$ for which $H$ admits a \emph{proper edge-gg-coloring}. We have $\Delta(H)\leq \chi\,'(H)\leq \Delta(H)+1$ (Vadim G. Vizing, 1964; Guppta, 1966 \cite{Bondy-2008}), where $\chi\,' _{gg}(H)=\chi\,'(H)$.

\item $\chi\,'' _{gg}(H)$ is the minimum number of $M_G$ Topsnut-gpws $G_i$ in some every-zero graphic group $\{F_f(G),\oplus\}$ for which $H$ admits a \emph{proper total gg-coloring}. It was conjectured $\Delta(H)+1\leq \chi\,''(H)\leq \Delta(H)+2$ (Behzad, 1965; Vadim G. Vizing, 1964 \cite{Bondy-2008}), where $\chi\,'' _{gg}(H)=\chi\,''(H)$.

\item $\chi _{ggs}(H)$ is the minimum number of $M_G$ Topsnut-gpws $G_i$ in some every-zero graphic group $\{F_f(G),\oplus\}$ for which $H$ admits a \emph{vertex distinguishing proper gg-coloring}.

\item $\chi _{ggas}(H)$ is the minimum number of $M_G$ Topsnut-gpws $G_i$ in some every-zero graphic group $\{F_f(G),\oplus\}$ for which $H$ admits an \emph{adjacent vertex distinguishing proper gg-coloring}.

\item $\chi\,' _{ggs}(H)$ is the minimum number of $M_G$ Topsnut-gpws $G_i$ in some every-zero graphic group $\{F_f(G),\oplus\}$ for which $H$ admits an \emph{edge distinguishing proper gg-coloring}.

\item $\chi\,' _{ggas}(H)$ is the minimum number of $M_G$ Topsnut-gpws $G_i$ in some every-zero graphic group $\{F_f(G),\oplus\}$ for which $H$ admits an \emph{adjacent edge distinguishing proper gg-coloring}. We have a conjecture $\chi\,' _{as}(H)\leq \Delta(H)+2$ by Zhang Zhongfu, Liu Linzhong, Wang Jianfang, 2002 \cite{Zhang-Liu-Wang-2002}, where $\chi\,' _{ggas}(H)=\chi\,' _{as}(H)$.
\item $\chi _{ggeq}(H)$ is the minimum number of $M_G$ Topsnut-gpws $G_i$ in some every-zero graphic group $\{F_f(G),\oplus\}$ for which $H$ admits an \emph{equitable adjacent-v proper gg-coloring}.
\item $\chi\,' _{ggeq}(H)$ is the minimum number of $M_G$ Topsnut-gpws $G_i$ in some every-zero graphic group $\{F_f(G),\oplus\}$ for which $H$ admits an \emph{equitable adjacent-e proper gg-coloring}.
\item $\chi\,'' _{ggas}(H)$ is the minimum number of $M_G$ Topsnut-gpws $G_i$ in some every-zero graphic group $\{F_f(G),\oplus\}$ for which $H$ admits an \emph{adjacent total distinguishing proper gg-coloring}.\paralled
\end{asparaenum}
\end{rem}

\subsection{Directed total colorings with labeling restrictions}

\begin{figure}[h]
\centering
\includegraphics[width=16.4cm]{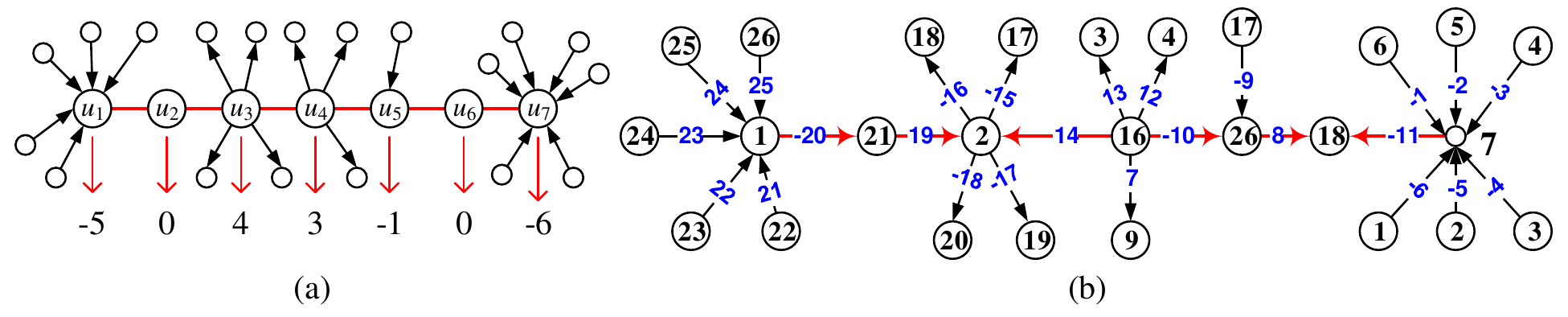}
\caption{\label{fig:directed-caterpillar-coloring}{\small (a) A half-directed caterpillar $T^*$ with its topological vector $V_{ec}(T^*)=(-5,0,4,3,-1,0,-6)$; (b) a proper directed gracefully total coloring of another directed caterpillar, cited from \cite{Bing-Yao-2020arXiv}.}}
\end{figure}

\begin{defn}\label{defn:directed-graceful-coloring}
\cite{Bing-Yao-2020arXiv} Let $\overrightarrow{G}$ be a directed connected graph with $p$ vertices and $Q$ arcs. If $G$ admits a proper total coloring $f:V(\overrightarrow{G})\cup A(\overrightarrow{G})\rightarrow [1,M]$ such that $f(\overrightarrow{uv})=f(u)-f(v)$ for each arc $\overrightarrow{uv}\in A(\overrightarrow{G})$ and $\{|f(\overrightarrow{uv})|:\overrightarrow{uv}\in A(\overrightarrow{G})\}=[1,Q]$, then we call $f$ a \emph{directed gracefully total coloring} of $\overrightarrow{G}$, and moreover $f$ is a \emph{proper directed gracefully total coloring} if $M=Q+1$. (see an example shown in Fig.\ref{fig:directed-caterpillar-coloring}(b))\qqed
\end{defn}

\begin{figure}[h]
\centering
\includegraphics[width=15cm]{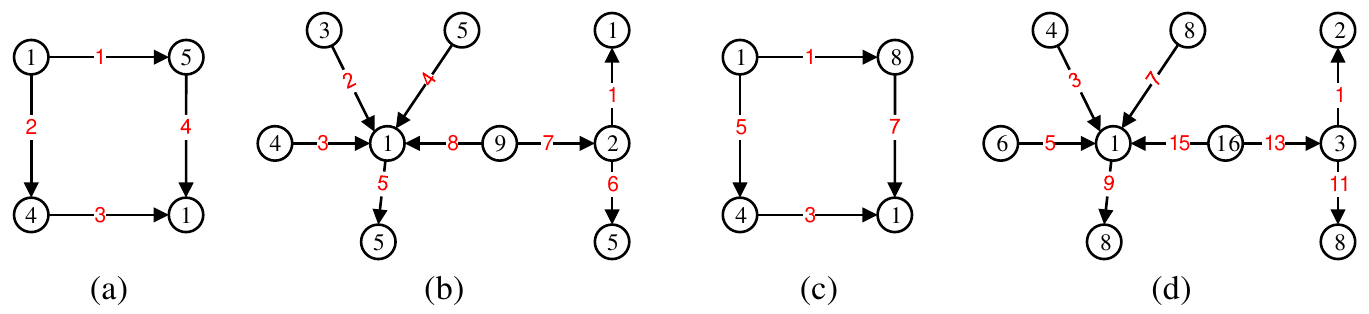}
\caption{\label{fig:digraceful-colorings}{\small (a) and (b) are two digraceful colorings; (c) and (d) are two odd-digraceful colorings.}}
\end{figure}

\begin{figure}[h]
\centering
\includegraphics[width=16.4cm]{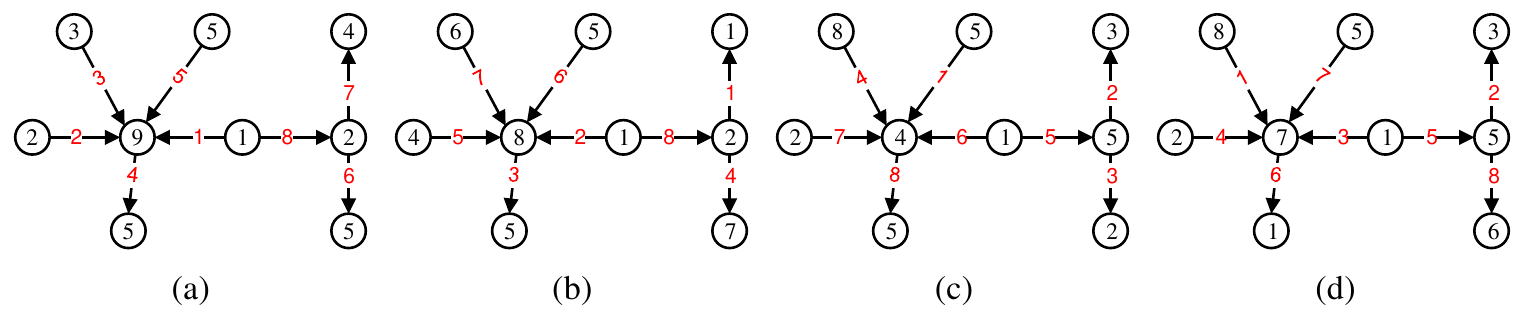}
\caption{\label{fig:digraceful-colorings-11}{\small Four digraceful colorings of a ditree.}}
\end{figure}

\begin{defn}\label{defn:digraceful-labeling-digraph}
$^*$ A digraph $D$ admits a coloring $\theta: V(D)\rightarrow [1, |A(D)|+1]$ (resp. $[1, 2|A(D)|]$), where $A(D)$ is the arc set of $D$, such that each arc $\overrightarrow{uv} \in A(D)$ is colored as $\theta(\overrightarrow{uv}) = \theta(u)-\theta(v)$ if $\theta(u) > \theta(v)$, and $\theta(\overrightarrow{uv}) =|A(D)|+1+ \theta(u)-\theta(v)$ (resp. $\theta(\overrightarrow{uv}) =2|A(D)|+ \theta(u)-\theta(v)$) if $\theta(u) < \theta(v)$. We call $\theta$ a \emph{digraceful labeling} (resp. an \emph{odd-digraceful labeling}) of $D$ if the \emph{arc color set} $\theta(A(D)) =\{\theta(\overrightarrow{uv}) : \overrightarrow{uv}\in A(D)\}= [1, |A(D)|]$ (resp. $[1, 2|A(D)|-1]^o$). Therefore, $D$ is called a \emph{digraceful digraph} (resp. an \emph{odd-digraceful digraph}) (see Fig.\ref{fig:digraceful-colorings} and Fig.\ref{fig:digraceful-colorings-11}). \qqed
\end{defn}

\begin{figure}[h]
\centering
\includegraphics[width=16.4cm]{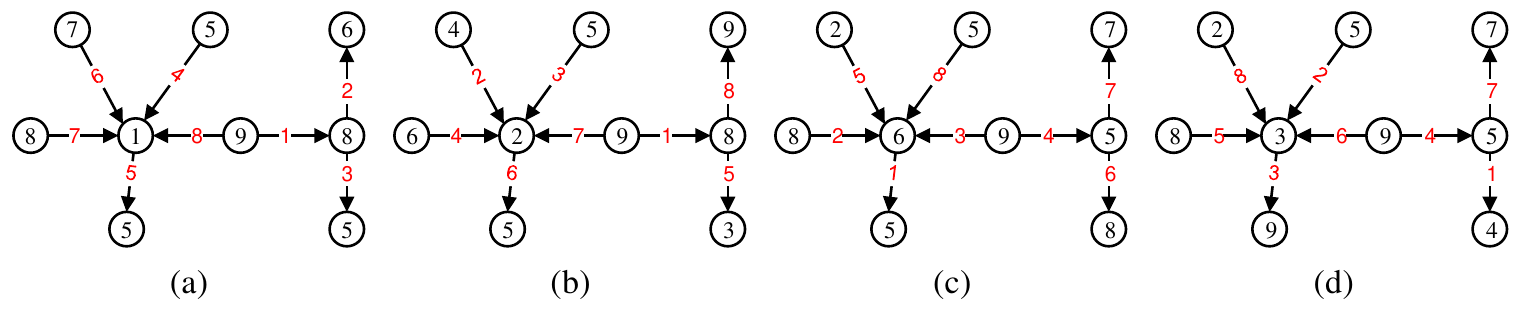}
\caption{\label{fig:digraceful-colorings-22}{\small The dual digraceful colorings of four digraceful colorings shown in Fig.\ref{fig:digraceful-colorings-11}.}}
\end{figure}

\begin{rem}\label{rem:333333}
Suppose that a $(p,q)$-digraph $D$ admits a digraceful labeling (resp. an odd-digraceful labeling) $f$, then the \emph{dual digraceful labeling} $h$ of the digraceful labeling $f$ is defined as:

$h(x)=\max f(V(D))+\min f(V(D))-f(x)$ for each vertex $x\in V(D)$, and

$h(\overrightarrow{uv})=\max f(A(D))+\min f(A(D))-f(\overrightarrow{uv})$ for each arc $\overrightarrow{uv}\in A(D)$.\\
where $f(V(D))=\{f(x):x\in V(D)\}$ and $f(A(D))=\{f(\overrightarrow{uv}):\overrightarrow{uv}\in A(D)\}$ (see examples shown in Fig.\ref{fig:digraceful-colorings-11} and Fig.\ref{fig:digraceful-colorings-22}).\paralled
\end{rem}

\section{Sequence labelings, sequence colorings}

\subsection{Sequence labelings}

\begin{defn}\label{defn:sequence-labelingss}
\cite{Yao-Mu-Sun-Zhang-Wang-Su-2018} Let $G$ be a $(p,q)$-graph, and let $A_M=\{a_1,a_2,\dots, a_M\}=\{a_i\}_1^M$ and $B_q=\{b_1,b_2,\dots,b_q\}=\{b_j\}_1^q$ be two integer sequences with $0\leq a_i<a_{i+1}$, $0\leq b_j<b_{j+1}$ and $m\geq p$, and $k$ be a non-negative integer. There are the following labelings and constraint conditions:
\begin{asparaenum}[Seq-1. ]
\item \label{condi:vertex-mapping} A vertex labeling $f:V(G)\rightarrow A_M$ such that $f(u)\neq f(v)$ for any pair of distinct vertices $u,v\in V(G)$;
\item \label{condi:edge-mapping} an edge labeling $h:E(G)\rightarrow B_q$ such that $h(uv)\neq h(uw)$ for distinct edges $uv,uw\in E(G)$;
\item \label{condi:total-mapping} a total labeling $g:V(G)\cup E(G) \rightarrow A_M\cup B_q$ such that $g(x)\neq g(y)$ for any pair of adjacent or incident elements $x,y\in V(G)\cup E(G)$;
\item \label{condi:half-full} $f(V(G))\subseteq A_M$ and induced edge color set $f(E(G))=B_q$;
\item \label{condi:only-vertex-set} $f(V(G))=A_M$;
\item \label{condi:all-full} $f(V(G))=A_M$ and induced edge color set $f(E(G))=B_q$;
\item \label{condi:not-full-2} $g(V(G)\cup E(G))\subseteq A_M\cup B_q$;
\item \label{condi:not-full-3} $g(V(G))\subseteq A_M$ and $g(E(G))=B_q$;
\item \label{condi:not-full-4} $g(V(G)\cup E(G))=A_M\cup B_q$;
\item \label{condi:not-full-1} $h(E(G))=B_q$;


\item \label{condi:total-perfect-matching} a \emph{matching equation} $\alpha(x)+\alpha(y)=s(p)$ for each matching edge $xy$ of a perfect matching $M$ of $G$, where $s(p)$ is a function of variable $p$, and $\alpha$ is one of the above labelings $f$ and $g$;
\item \label{condi:total-set-ordered} $G$ is a bipartite graph with vertex bipartition $(X,Y)$, such that $\max\beta(X)<\min \beta(Y)$, where $\beta$ is one of the above labelings $f$ and $g$;


\item \label{condi:induced-edge-label} an \emph{induced function} $f(uv)=O(f(u),f(v))$ for each edge $uv\in E(G)$;
\item \label{condi:total-equation} an \emph{$F$-equation} $F(g(u),g(uv),g(v))=k$ for each edge $uv\in E(G)$; and
\item \label{condi:total-equation-edge} an \emph{$M$-equation} $M(f(u),h(uv),f(v))=0$ for each edge $uv\in E(G)$.
\end{asparaenum}
\noindent \textbf{We call $f$}:
\begin{asparaenum}[(1) ]
\item An \emph{induced sequence labeling} if Seq-\ref{condi:vertex-mapping} and Seq-\ref{condi:induced-edge-label} hold true.
\item A \emph{set-ordered induced sequence labeling} if Seq-\ref{condi:vertex-mapping}, Seq-\ref{condi:induced-edge-label} and Seq-\ref{condi:total-set-ordered} hold true.

\item A \emph{strongly induced sequence labeling} if Seq-\ref{condi:vertex-mapping}, Seq-\ref{condi:induced-edge-label} and Seq-\ref{condi:total-perfect-matching} hold true.

\item An \emph{edge sequence-graceful labeling} if Seq-\ref{condi:edge-mapping} and Seq-\ref{condi:not-full-1} hold true.


\item A \emph{generalized edge-magic sequence labeling} if Seq-\ref{condi:total-mapping}, Seq-\ref{condi:not-full-2} and Seq-\ref{condi:total-equation} hold true, where the $F$-equation is $g(u)+g(uv)+g(v)=k$.
\item An \emph{edge-magic total sequence labeling} if Seq-\ref{condi:total-mapping}, Seq-\ref{condi:not-full-4} and Seq-\ref{condi:total-equation} hold true, where the $F$-equation is $g(u)+g(uv)+g(v)=k$.
\item An \emph{$F$-magic sequence-graceful labeling} if Seq-\ref{condi:total-mapping}, Seq-\ref{condi:not-full-3} and Seq-\ref{condi:total-equation} hold true.


\item A \emph{sequence-graceful labeling} if Seq-\ref{condi:vertex-mapping}, Seq-\ref{condi:all-full} and Seq-\ref{condi:induced-edge-label} hold true, where the induced function is $f(uv)=|f(u)-f(v)|$.
\item A \emph{set-ordered sequence-graceful labeling} if Seq-\ref{condi:vertex-mapping}, Seq-\ref{condi:all-full}, Seq-\ref{condi:induced-edge-label} and Seq-\ref{condi:total-set-ordered} hold true, where the induced function is $f(uv)=|f(u)-f(v)|$.
\item A \emph{strongly sequence-graceful labeling} if Seq-\ref{condi:vertex-mapping}, Seq-\ref{condi:all-full}, Seq-\ref{condi:induced-edge-label} and Seq-\ref{condi:total-perfect-matching} hold true, where the induced function is $f(uv)=|f(u)-f(v)|$.


\item A \emph{pan-sequence-graceful labeling} if Seq-\ref{condi:vertex-mapping}, Seq-\ref{condi:half-full} and Seq-\ref{condi:induced-edge-label} hold true, where the induced function is $f(uv)=|f(u)-f(v)|$.
\item A \emph{set-ordered pan-sequence-graceful labeling} if Seq-\ref{condi:vertex-mapping}, Seq-\ref{condi:half-full}, Seq-\ref{condi:induced-edge-label} and Seq-\ref{condi:total-set-ordered} hold true, where the induced function is $f(uv)=|f(u)-f(v)|$.
\item A \emph{strongly pan-sequence-graceful labeling} if Seq-\ref{condi:vertex-mapping}, Seq-\ref{condi:half-full}, Seq-\ref{condi:induced-edge-label} and Seq-\ref{condi:total-perfect-matching} hold true, where the induced function is $f(uv)=|f(u)-f(v)|$.


\item An \emph{$M$-magic sequence-graceful labeling} if Seq-\ref{condi:vertex-mapping}, Seq-\ref{condi:edge-mapping}, Seq-\ref{condi:half-full}, Seq-\ref{condi:only-vertex-set}, Seq-\ref{condi:not-full-1} and Seq-\ref{condi:total-equation-edge} hold true.\qqed
\end{asparaenum}
\end{defn}

\begin{rem}\label{rem:sequence-labelingss}
As two sequences to be consecutive sets, namely, $A_m=[a_1,a_m]$ and $B_q=[b_1,b_q]$, the sequence-type of labellins in Definition \ref{defn:sequence-labelingss} are popular labelings. In fact, Definition \ref{defn:sequence-labelingss} is a pan-definition, there are more labeling-type of constraint conditions as follows:
\begin{asparaenum}[\textbf{\textrm{Cot}}-1.]
\item The induced function $f(uv)=O(f(u),f(v))$ for each edge $uv\in E(G)$ including $f(uv)=|f(u)-f(v)|$, $f(uv)=f(u)+f(v)~(\bmod~\epsilon(q))$, $f(uv)=f(u)+f(v)+r$ with $r\geq 0$, $f(uv)=np-[f(u)+f(v)]$, $f(uv)=\left\lceil \frac{f(u)+f(v)}{2}\right\rceil$, $f(uv)=\textrm{gcd}(f(u),f(v))$; odd or even $f(uv)=|f(u)-f(v)|$, odd or even $f(uv)=f(u)+f(v)~(\bmod~\epsilon(q))$, odd or even $f(uv)=f(u)+f(v)+r$ with $r\geq 0$.

\quad Other functions are as: (1) Linear functions: $f(xy)=|2f(y)-1-2f(x)|$ for each edge $xy\in E(G)$ if $V(G)=X\cup Y$ and $x\in X$ and $y\in Y$ for a bipartite graph $G$, or $f(xy)=|af(y)+b-[cf(x)+d]|$ subject to integers $a,c\geq 1$ and $b,d\geq 0$.

\quad (2) Non-linear functions: $f(xy)=\big \lceil |\sin f(y)-\cos f(x)|\big \rceil $ for each edge $xy\in E(G)$, \emph{etc.}

\item The $F$-equation $F(g(u),g(uv),g(v))=k$ for each edge $uv\in E(G)$ including

\qquad (i) Popular-type: edge-magic $g(u)+g(uv)+g(v)=k$, edge-difference $g(uv)+|g(u)-g(v)|=k$, graceful-difference $\big |g(uv)-|g(u)-g(v)|\big |=k$, felicitous-difference $|g(u)+g(v)-g(uv)|=k$.

\qquad (ii) Generalized-type: parameterized edge-magic $ag(u)+cg(uv)+bg(v)=k$, parameterized edge-difference $cg(uv)+|bg(u)-bg(v)|=k$, parameterized graceful-difference $|cg(uv)-|ag(u)-bg(v)|\big |=k$, parameterized felicitous-difference $|ag(u)+bg(v)-cg(uv)|=k$.

\item An $M$-equation $M(f(u),h(uv),f(v))=0$ for each edge $uv\in E(G)$ is under a composition of a vertex labeling $f$ and an edge labeling $h$, in general, but it differs from a total labeling. Very often, $M$-equations may be composed by constraint conditions of vertex labelings and constraint conditions of edge labelings.

\item Topological structures are as constraint conditions, such as \emph{path}, \emph{cycle}, \emph{clique}, \emph{star tree $K_{1,s}$}, \emph{face}, \emph{bipartite graph}, \emph{extremum graph}, or other \emph{particular subgraphs}. For example, \emph{face labelings} of planar graphs, \emph{subgraph labelings} of graphs.

\quad In Definition \ref{defn:sequence-labelingss}, the vertex labeling $f$ is belong to the topological constraint condition of $K_{1,1}$, both labelings $h$ and $g$ are restricted by the topological constraint condition of star tree $K_{1,d(u)}$, where $d(u)$ is the degree of vertex $u\in V(G)$. \paralled
\end{asparaenum}
\end{rem}

\begin{defn} \label{defn:face-labeling-cycle-labeling}
$^*$ Let $S$ be a set of colors.

(1) Let $H$ be a maximal planar graph admitting a vertex labeling $f:V(H)\rightarrow S$, the \emph{induced face-color function} is $f(\Delta_{xyz})=f(x)+f(y)+f(z)$ for each triangle face $\Delta_{xyz}$ with vertices $x,y,z$ in $H$, and we call $f$ a \emph{$W$-type face labeling} if the face-color set $\{f(\Delta_{xyz}):~\Delta_{xyz}\subset H\}$ holds a $W$-condition true.

(2) A $(p,q)$-graph $G$ admits a vertex labeling $g:V(G)\rightarrow S$, the \emph{induced cycle-color function} is $g(C)=\sum ^n_{i=1}g(x_i)$ for each cycle $C=x_1x_2\cdots x_nx_1$ of $G$, and we call $g$ a \emph{$\{W_i\}^m_1$-cycle labeling} if the cycle-color set $\{g(C):~C=x_1x_2\cdots x_nx_1\subset G\}$ holds a a group of $W_i$-conditions with $i\in [1,m]$ true.

(3) A connected graph $G$ admits a total labeling $h:V(G)\cup E(G)\rightarrow S$, the spanning tree labeling function is $h(T)=F(h(V(T)), h(E(T)))$ for each spanning tree $T$ of $G$, we call $h$ a \emph{$\{W_i\}^m_1$-spanning tree labeling} if the spanning tree-color set $\{h(T):T~\textrm{is a spanning tree of}~G\}$ over all spanning trees of $G$ holds a group of $W_i$-conditions with $i\in [1,m]$ true.\qqed
\end{defn}

\begin{rem}\label{rem:ABC-conjecture}
Replacing ``labeling'' by ``coloring'' in Definition \ref{defn:face-labeling-cycle-labeling} will result: \emph{$W$-type face coloring}, \emph{$\{W_i\}^m_1$-cycle coloring} and \emph{$\{W_i\}^m_1$-spanning tree coloring}.\paralled
\end{rem}

\begin{defn}\label{defn:Yao-Zhang-Sun-Mu-Wang-Zhang2018}
\cite{Yao-Zhang-Sun-Mu-Wang-Zhang2018} A \emph{sequence $(\{a_i\}^M_1,\{b_j\}^q_1)$-harmonious labeling} of a $(p,q)$-graph $G$, where $\{a_i\}^M_1$ and $\{b_j\}^q_1$ are two monotonic decreasing sequences of non-negative numbers with $M\geq p$, is defined by $f:V(G)\rightarrow \{a_i\}^M_1$ such that $f(u)\neq f(v)$ for distinct $u,v\in V(G)$, and the induced edge color is $f(uv)=f(u)+f(v)~(\bmod~\epsilon(q))$ such that $f(E(G))=\{b_j\}^q_1$, where $\epsilon(q)$ is a function of variable $q$.\qqed
\end{defn}

\subsection{Sequence colorings}

Some \emph{sequence-type labelings} of graphs were introduced in \cite{Yao-Mu-Sun-Zhang-Wang-Su-2018}, so we present sequence-type of graph colorings as follows.

\begin{defn}\label{defn:sequence-coloring}
\cite{Yao-Su-Wang-Hui-Sun-ITAIC2020} Let $G$ be a $(p, q)$-graph, and let a sequence $A_M=\{a_i\}_1^M$ hold $0\leq a_i< a_{i+1}$ for $i\in [1, M-1]$ and $p\leq M$, and let another sequence $B_q=\{b_j\}_1^q$ hold $0\leq b_j< b_{j+1}$ for $j\in [1, q-1]$, and let $k$ be a constant. $G$ admits a coloring $f:S\rightarrow C$ with $f(S)=\{f(x):x\in S\}$ and $S\subseteq V(G)\cup E(G)$. There are the following constraint conditions:
\begin{asparaenum}[Rec-1. ]
\item \label{rescondi:vertex-mapping} $S=V(G)$ and $C=A_M$;
\item \label{rescondi:total-mapping} $S=V(G)\cup E(G)$ and $C=A_M\cup B_q$;
\item \label{rescondi:adjacent-different} $f(u)\neq f(v)$ for any edge $uv\in E(G)$;
\item \label{rescondi:incident-different} $f(u)\neq f(uv)$ and $f(v)\neq f(uv)$ for each edge $uv\in E(G)$;
\item \label{rescondi:adjacent-edge-different} $f(uv)\neq f(uw)$ for any two vertices $v, w\in N(u)$;
\item \label{rescondi:not-full-1} $f(E(G))\subseteq B_q$;
\item \label{rescondi:all-full} $f(V(G))\subseteq A_M$ and $p=|V(G)|$;
\item \label{rescondi:edge-set-full} $f(E(G))=B_q$;
\item \label{rescondi:induced-edge-label} an induced function $f(uv)=O(f(u), f(v))$ for each edge $uv\in E(G)$;
\item \label{rescondi:magic-equation} an $F$-equation $F(f(u), f(uv), f(v))=k$ for each edge $uv\in E(G)$; and
\item \label{rescondi:set-ordered} $G$ is a bipartite graph with vertex bipartition $(X, Y)$ such that $\min f(X)<\max f(Y)$.
\end{asparaenum}

\noindent \textbf{We refer to $f$ as}:

\noindent ------ \emph{traditional-type}

\begin{asparaenum}[\textrm{Coloring}-1.]
\item An \emph{edge-induced sequence coloring} of $G$ if Rec-\ref{rescondi:vertex-mapping} and Rec-\ref{rescondi:induced-edge-label} hold true.
\item A \emph{gracefully edge-induced sequence coloring} of $G$ if Rec-\ref{rescondi:vertex-mapping}, Rec-\ref{rescondi:edge-set-full} and Rec-\ref{rescondi:induced-edge-label} hold true.
\item A \emph{set-ordered edge-induced sequence coloring} of $G$ if Rec-\ref{rescondi:vertex-mapping}, Rec-\ref{rescondi:induced-edge-label} and Rec-\ref{rescondi:set-ordered} hold true.
\item A \emph{set-ordered gracefully edge-induced sequence coloring} of $G$ if Rec-\ref{rescondi:vertex-mapping}, Rec-\ref{rescondi:edge-set-full}, Rec-\ref{rescondi:induced-edge-label} and Rec-\ref{rescondi:set-ordered} hold true.

\item A \emph{graceful sequence coloring} of $G$ if Rec-\ref{rescondi:vertex-mapping}, Rec-\ref{rescondi:edge-set-full} and Rec-\ref{rescondi:induced-edge-label} hold true, where $f(uv)=O(f(u), f(v))=|f(u)-f(v)|$.
\item A \emph{fully graceful sequence labeling} of $G$ if Rec-\ref{rescondi:vertex-mapping}, Rec-\ref{rescondi:edge-set-full}, Rec-\ref{rescondi:all-full} and Rec-\ref{rescondi:induced-edge-label} hold true, where $f(uv)=O(f(u), f(v))=|f(u)-f(v)|$.
\item A \emph{fully set-ordered graceful sequence coloring} of $G$ if Rec-\ref{rescondi:vertex-mapping}, Rec-\ref{rescondi:edge-set-full}, Rec-\ref{rescondi:set-ordered} and Rec-\ref{rescondi:induced-edge-label} hold true, where $f(uv)=O(f(u), f(v))=|f(u)-f(v)|$.
\item A \emph{total sequence coloring} of $G$ if Rec-\ref{rescondi:total-mapping} and Rec-\ref{rescondi:adjacent-different} hold true.
\item A \emph{proper total sequence coloring} of $G$ if Rec-\ref{rescondi:total-mapping}, Rec-\ref{rescondi:adjacent-different}, Rec-\ref{rescondi:incident-different} and Rec-\ref{rescondi:adjacent-edge-different} hold true.

------ \emph{graceful-type}

\item A \emph{gracefully total sequence coloring} of $G$ if Rec-\ref{rescondi:total-mapping}, Rec-\ref{rescondi:adjacent-different}, Rec-\ref{rescondi:edge-set-full} and Rec-\ref{rescondi:induced-edge-label} hold true, where $f(uv)=O(f(u), f(v))=|f(u)-f(v)|$.
\item A \emph{gracefully proper total sequence coloring} of $G$ if Rec-\ref{rescondi:total-mapping}, Rec-\ref{rescondi:adjacent-different}, Rec-\ref{rescondi:incident-different}, Rec-\ref{rescondi:adjacent-edge-different}, Rec-\ref{rescondi:edge-set-full} and Rec-\ref{rescondi:induced-edge-label} hold true, where $f(uv)=O(f(u), f(v))=|f(u)-f(v)|$.

\item A \emph{proper graceful-total sequence coloring} of $G$ if Rec-\ref{rescondi:total-mapping}, Rec-\ref{rescondi:adjacent-different}, Rec-\ref{rescondi:incident-different}, Rec-\ref{rescondi:adjacent-edge-different}, Rec-\ref{rescondi:edge-set-full}, Rec-\ref{rescondi:all-full} and Rec-\ref{rescondi:induced-edge-label} hold true, where $f(uv)=O(f(u), f(v))=|f(u)-f(v)|$.

------ \emph{felicitous-type}

\item A \emph{felicitous total sequence coloring} of $G$ if Rec-\ref{rescondi:total-mapping}, Rec-\ref{rescondi:adjacent-different}, Rec-\ref{rescondi:not-full-1} and Rec-\ref{rescondi:induced-edge-label} hold true, where $f(uv)=O(f(u), f(v))=f(u)+f(v)$~($\bmod~\epsilon(q)$).
\item A \emph{felicitous proper total sequence coloring} of $G$ if Rec-\ref{rescondi:total-mapping}, Rec-\ref{rescondi:adjacent-different}, Rec-\ref{rescondi:incident-different}, Rec-\ref{rescondi:adjacent-edge-different}, Rec-\ref{rescondi:not-full-1} and Rec-\ref{rescondi:induced-edge-label} hold true, where $f(uv)=O(f(u), f(v))=f(u)+f(v)$~($\bmod~\epsilon(q)$).

\item A \emph{set-ordered felicitous total sequence coloring} of $G$ if Rec-\ref{rescondi:total-mapping}, Rec-\ref{rescondi:adjacent-different}, Rec-\ref{rescondi:set-ordered}, Rec-\ref{rescondi:not-full-1} and Rec-\ref{rescondi:induced-edge-label} hold true, where $f(uv)=O(f(u), f(v))=f(u)+f(v)$~($\bmod~\epsilon(q)$).
\item A \emph{set-ordered felicitous proper total sequence coloring} of $G$ if Rec-\ref{rescondi:total-mapping}, Rec-\ref{rescondi:adjacent-different}, Rec-\ref{rescondi:incident-different}, Rec-\ref{rescondi:adjacent-edge-different}, Rec-\ref{rescondi:set-ordered}, Rec-\ref{rescondi:not-full-1} and Rec-\ref{rescondi:induced-edge-label} hold true, where $f(uv)=O(f(u), f(v))=f(u)+f(v)$~($\bmod~\epsilon(q)$).

------ \emph{magic-type}

\item An \emph{edge-magic total sequence coloring} of $G$ if Rec-\ref{rescondi:total-mapping}, Rec-\ref{rescondi:adjacent-different}, Rec-\ref{rescondi:adjacent-edge-different} and Rec-\ref{rescondi:magic-equation} hold true, where $F(f(u), f(uv), f(v))=f(u)+f(uv)+f(v)=k$.
\item A \emph{proper edge-magic total sequence coloring} of $G$ if Rec-\ref{rescondi:total-mapping}, Rec-\ref{rescondi:adjacent-different}, Rec-\ref{rescondi:incident-different}, Rec-\ref{rescondi:adjacent-edge-different} and Rec-\ref{rescondi:magic-equation} hold true, where $F(f(u), f(uv), f(v))=f(u)+f(uv)+f(v)=k$.

\item A \emph{set-ordered edge-magic total sequence coloring} of $G$ if Rec-\ref{rescondi:total-mapping}, Rec-\ref{rescondi:adjacent-different}, Rec-\ref{rescondi:adjacent-edge-different}, Rec-\ref{rescondi:set-ordered} and Rec-\ref{rescondi:magic-equation} hold true, where $F(f(u), f(uv), f(v))=f(u)+f(uv)+f(v)=k$.
\item A \emph{set-ordered proper edge-magic total sequence coloring} of $G$ if Rec-\ref{rescondi:total-mapping}, Rec-\ref{rescondi:adjacent-different}, Rec-\ref{rescondi:incident-different}, Rec-\ref{rescondi:adjacent-edge-different}, Rec-\ref{rescondi:set-ordered} and Rec-\ref{rescondi:magic-equation} hold true, where $F(f(u), f(uv), f(v))=f(u)+f(uv)+f(v)=k$.

\item An \emph{edge-difference total sequence coloring} of $G$ if Rec-\ref{rescondi:total-mapping}, Rec-\ref{rescondi:adjacent-different}, Rec-\ref{rescondi:adjacent-edge-different} and Rec-\ref{rescondi:magic-equation} hold true, where $F(f(u), f(uv), f(v))=f(uv)+|f(u)-f(v)|=k$.

\item A \emph{proper edge-difference total sequence coloring} of $G$ if Rec-\ref{rescondi:total-mapping}, Rec-\ref{rescondi:adjacent-different}, Rec-\ref{rescondi:incident-different}, Rec-\ref{rescondi:adjacent-edge-different} and Rec-\ref{rescondi:magic-equation} hold true, where $F(f(u), f(uv), f(v))=f(uv)+|f(u)-f(v)|=k$.

\item A \emph{set-ordered edge-difference total sequence coloring} of $G$ if Rec-\ref{rescondi:total-mapping}, Rec-\ref{rescondi:adjacent-different}, Rec-\ref{rescondi:adjacent-edge-different}, Rec-\ref{rescondi:set-ordered} and Rec-\ref{rescondi:magic-equation} hold true, where $F(f(u), f(uv), f(v))=f(uv)+|f(u)-f(v)|=k$.
\item A \emph{set-ordered proper edge-difference total sequence coloring} of $G$ if Rec-\ref{rescondi:total-mapping}, Rec-\ref{rescondi:adjacent-different}, Rec-\ref{rescondi:incident-different}, Rec-\ref{rescondi:adjacent-edge-different}, Rec-\ref{rescondi:set-ordered} and Rec-\ref{rescondi:magic-equation} hold true, where $F(f(u), f(uv), f(v))=f(uv)+|f(u)-f(v)|=k$.

\item A \emph{graceful-difference total sequence coloring} of $G$ if Rec-\ref{rescondi:total-mapping}, Rec-\ref{rescondi:adjacent-different}, Rec-\ref{rescondi:adjacent-edge-different} and Rec-\ref{rescondi:magic-equation} hold true, where $F(f(u), f(uv), f(v))=\big |f(uv)-|f(u)-f(v)|\big |=k$.

\item A \emph{proper graceful-difference total sequence coloring} of $G$ if Rec-\ref{rescondi:total-mapping}, Rec-\ref{rescondi:adjacent-different}, Rec-\ref{rescondi:incident-different}, Rec-\ref{rescondi:adjacent-edge-different} and Rec-\ref{rescondi:magic-equation} hold true, where $F(f(u), f(uv), f(v))=\big |f(uv)-|f(u)-f(v)|\big |=k$.

\item A \emph{set-ordered graceful-difference total sequence coloring} of $G$ if Rec-\ref{rescondi:total-mapping}, Rec-\ref{rescondi:adjacent-different}, Rec-\ref{rescondi:adjacent-edge-different}, Rec-\ref{rescondi:set-ordered} and Rec-\ref{rescondi:magic-equation} hold true, where $F(f(u), f(uv), f(v))=\big |f(uv)-|f(u)-f(v)|\big |=k$.
\item A \emph{set-ordered proper graceful-difference total coloring} of $G$ if Rec-\ref{rescondi:total-mapping}, Rec-\ref{rescondi:adjacent-different}, Rec-\ref{rescondi:incident-different}, Rec-\ref{rescondi:adjacent-edge-different}, Rec-\ref{rescondi:set-ordered} and Rec-\ref{rescondi:magic-equation} hold true, where $F(f(u), f(uv), f(v))=\big |f(uv)-|f(u)-f(v)|\big |=k$.

\item A \emph{gracefully-total sequence coloring} of $G$ if Rec-\ref{rescondi:total-mapping}, Rec-\ref{rescondi:adjacent-different}, Rec-\ref{rescondi:adjacent-edge-different} and Rec-\ref{rescondi:magic-equation} hold true, where $F(f(u), f(uv), f(v))=|f(u)+f(v)-f(uv)|=k$.

\item A \emph{sequence proper gracefully-total coloring} of $G$ if Rec-\ref{rescondi:total-mapping}, Rec-\ref{rescondi:adjacent-different}, Rec-\ref{rescondi:incident-different}, Rec-\ref{rescondi:adjacent-edge-different} and Rec-\ref{rescondi:magic-equation} hold true, where $F(f(u), f(uv), f(v))=|f(u)+f(v)-f(uv)|=k$.

\item A \emph{set-ordered gracefully-total sequence coloring} of $G$ if Rec-\ref{rescondi:total-mapping}, Rec-\ref{rescondi:adjacent-different}, Rec-\ref{rescondi:adjacent-edge-different}, Rec-\ref{rescondi:set-ordered} and Rec-\ref{rescondi:magic-equation} hold true, where $F(f(u), f(uv), f(v))=|f(u)+f(v)-f(uv)|=k$.
\item A \emph{set-ordered proper gracefully-total sequence coloring} of $G$ if Rec-\ref{rescondi:total-mapping}, Rec-\ref{rescondi:adjacent-different}, Rec-\ref{rescondi:incident-different}, Rec-\ref{rescondi:adjacent-edge-different}, Rec-\ref{rescondi:set-ordered} and Rec-\ref{rescondi:magic-equation} hold true, where $F(f(u), f(uv), f(v))=|f(u)+f(v)-f(uv)|=k$.

------ \emph{gcd-type}

\item An \emph{maxi-common-factor total sequence coloring} of $G$ if Rec-\ref{rescondi:total-mapping}, Rec-\ref{rescondi:adjacent-different}, Rec-\ref{rescondi:adjacent-edge-different} and Rec-\ref{rescondi:induced-edge-label} hold true, where $f(uv)=O(f(u), f(v))=\textrm{gcd}(f(u), f(v))$.
\item A \emph{gracefully maxi-common-factor total sequence coloring} of $G$ if Rec-\ref{rescondi:total-mapping}, Rec-\ref{rescondi:adjacent-different}, Rec-\ref{rescondi:adjacent-edge-different}, Rec-\ref{rescondi:edge-set-full} and Rec-\ref{rescondi:induced-edge-label} hold true, where $f(uv)=O(f(u), f(v))=\textrm{gcd}(f(u), f(v))$.\qqed
\end{asparaenum}
\end{defn}

We say a sentence ``a \emph{$W$-type sequence coloring}'' to represent one of colorings with no ``set-ordered'' defined in Definition \ref{defn:sequence-coloring}, and employ another sentence ``a \emph{set-ordered $W$-type sequence coloring}'' to stand for one of colorings having ``set-ordered'' defined in Definition \ref{defn:sequence-coloring}.

\begin{lem}\label{thm:LEAVES-added-algorithm}
\cite{Yao-Su-Wang-Hui-Sun-ITAIC2020} Let $G$ be a bipartite and connected $(p, q)$-graph with its vertex bipartition $V(G)=X\cup Y$ such that $X\cap Y=\emptyset$, $X=\{x_i:i\in [1, s]\}$ and $Y=\{y_j:j\in [1, t]\}$ holding $s+t=p$. Suppose that $G$ admits a $(k,d)$-gracefully total coloring, then any leaf-added graph $G+L_{\textrm{eaf}}$ admits a \emph{$(k,d)$-gracefully total coloring} too.
\end{lem}

\begin{thm}\label{thm:tree-graceful-total-coloringss}
\cite{Yao-Su-Wang-Hui-Sun-ITAIC2020} Each tree admits a \emph{$(k,d)$-gracefully total coloring} defined in Definition \ref{defn:kd-w-type-colorings}, also, a \emph{set-ordered gracefully total coloring} as $(k, d)=(1, 1)$.
\end{thm}

\begin{thm}\label{thm:graceful-total-sequence-coloring}
\cite{Yao-Su-Wang-Hui-Sun-ITAIC2020} Every tree $T$ with diameter $D(T)\geq 3$ and $s+1=\left \lceil \frac{D(T)}{2}\right \rceil $ admits at least $2^{s}$ different \emph{gracefully total sequence colorings} if two sequences $A_M, B_q$ holding $0<b_j-a_i\in B_q$ for $a_i\in A_M$ and $b_j\in B_q$.
\end{thm}

\begin{lem}\label{thm:adding-leaves-keep-sequence-colorings}
\cite{Yao-Su-Wang-Hui-Sun-ITAIC2020} Suppose that a bipartite and connected graph $G$ admits a gracefully total sequence coloring based on two sequences $A_M, B_q$ holding $0<b_j-a_i\in B_q$ for $a_i\in A_M$ and $b_j\in B_q$, then a new bipartite and connected graph obtained by adding randomly leaves to $G$ admits a \emph{gracefully total sequence coloring} based on two sequences $A\,'_M, B\,'_q$ holding $0<b\,'_j-a\,'_i\in B\,'_q$ for $a\,'_i\in A\,'_M$ and $b\,'_j\in B\,'_q$.
\end{lem}

\noindent \textbf{The $(k,d)$-graceful preparation}. Let $G$ be a bipartite and connected $(p, q)$-graph with its vertex set $V(G)=X\cup Y$ such that $X\cap Y=\emptyset$, $X=\{x_i:i\in [1, s]\}$ and $Y=\{y_j:j\in [1, t]\}$ holding $s+t=p$, and each edge is $x_iy_j$ with $x_i\in X$ and $y_j\in Y$. Suppose that $G$ admits a $(k,d)$-gracefully total coloring $f$, so we have $f:X\rightarrow S_{m, 0, d}$ and $f:Y\cup E(G)\rightarrow S_{q-1, k, d}$ with $k\geq 1$, without loss of generality, there are $0=f(x_1)\leq f(x_i)\leq f(x_{i+1})$ for $i\in [1, s-1]$ and $f(x_{s})\leq f(y_j)\leq f(y_{j+1})\leq f(y_{t})=k+(q-1)d$ for $j\in [1, t-1]$, as well as $f(E(G))=S_{q-1, k, d}$.

\vskip 0.2cm

\noindent \textbf{LEAF-adding algorithm} \cite{Yao-Su-Wang-Hui-Sun-ITAIC2020} By The $(k,d)$-graceful preparation we have

\textbf{Step 1.} Color each vertex $x_i\in X\subset X\,'$ with $g(x_i)=f(x_i)$ for $i\in [1, s]$, and color each vertex $y_j\in Y\subset Y'$ with $g(y_j)=f(y_j)+[A(s)+B(t)]d$ for $j\in [1, t]$, and color edges $x_iy_j\in E(G)\subset E(G+L_{\textrm{eaf}})$ by
$${
\begin{split}
g(x_iy_j)=g(y_j)-g(x_i)=f(y_j)+[A(s)+B(t)]d-f(x_i)=f(x_iy_j)+[A(s)+B(t)]d.
\end{split}}$$ So, we get an edge color set
$${
\begin{split}
g(E(G))=\{k+[A(s)+B(t)]d, k+[A(s)+B(t)+1]d, \dots , k+[A(s)+B(t)+q-1]d\}.
\end{split}}
$$

\textbf{Step 2.} Let edges $e_{i_1} e_{i_2} \dots e_{i_{A(s)+B(t)}}$ be a permutation of edges $x_ix_{i, r}$ for $r\in [1, a_i]$ and $i\in [1, s]$ and edges $y_jy_{j, r}$ for $r\in [1, b_{j}]$ and $j\in [1, t]$, that is, $e_{i_1} e_{i_2} \dots e_{i_{A(s)+B(t)}}$ is a permutation of $x_1x_{1, 1}$ $x_1x_{1, 2}$ $\cdots$ $x_1x_{1, a_1}$ $x_2x_{2, 1}$ $x_2x_{2, 2}$ $\cdots$ $x_2x_{2, a_2}$ $\cdots$ $x_ix_{i, 1}$ $x_ix_{i, 2}$ $\cdots$ $x_ix_{i, a_i}$ $\cdots$ $x_sx_{s, 1}$ $x_sx_{s, 2}$ $\cdots$ $x_sx_{s, a_s}$ $y_1y_{1, 1}$ $y_1y_{1, 2}$ $\cdots$ $y_1y_{1, b_1}$ $y_2y_{2, 1}$ $y_2y_{2, 2}$ $\cdots$ $y_2y_{2, b_2}$ $\cdots$ $y_jy_{j, 1}$ $y_jy_{j, 2}$ $\cdots$ $y_jy_{j, b_j}$ $\cdots$ $y_ty_{t, 1}$ $y_ty_{t, 2}$ $\cdots$ $y_ty_{t, b_t}$. We color each edge $e_{i_r}$ with $g(e_{i_r})=k+(r-1)d$ with $r\in [1, A(s)+B(t)]$, and color each vertex $x_{i, r}$ with $g(x_{i, r})=f(x_i)+g(e_{i_r})$ if vertex $x_{i, r}$ is an end of the edge $e_{i_r}$, and color each vertex $y_{j, r}$ with $g(y_{j, r})=g(y_j)-g(e_{i_r})$ if vertex $y_{j, r}$ is an end of the edge $e_{i_r}$.

\vskip 0.2cm

\begin{figure}[h]
\centering
\includegraphics[width=16.4cm]{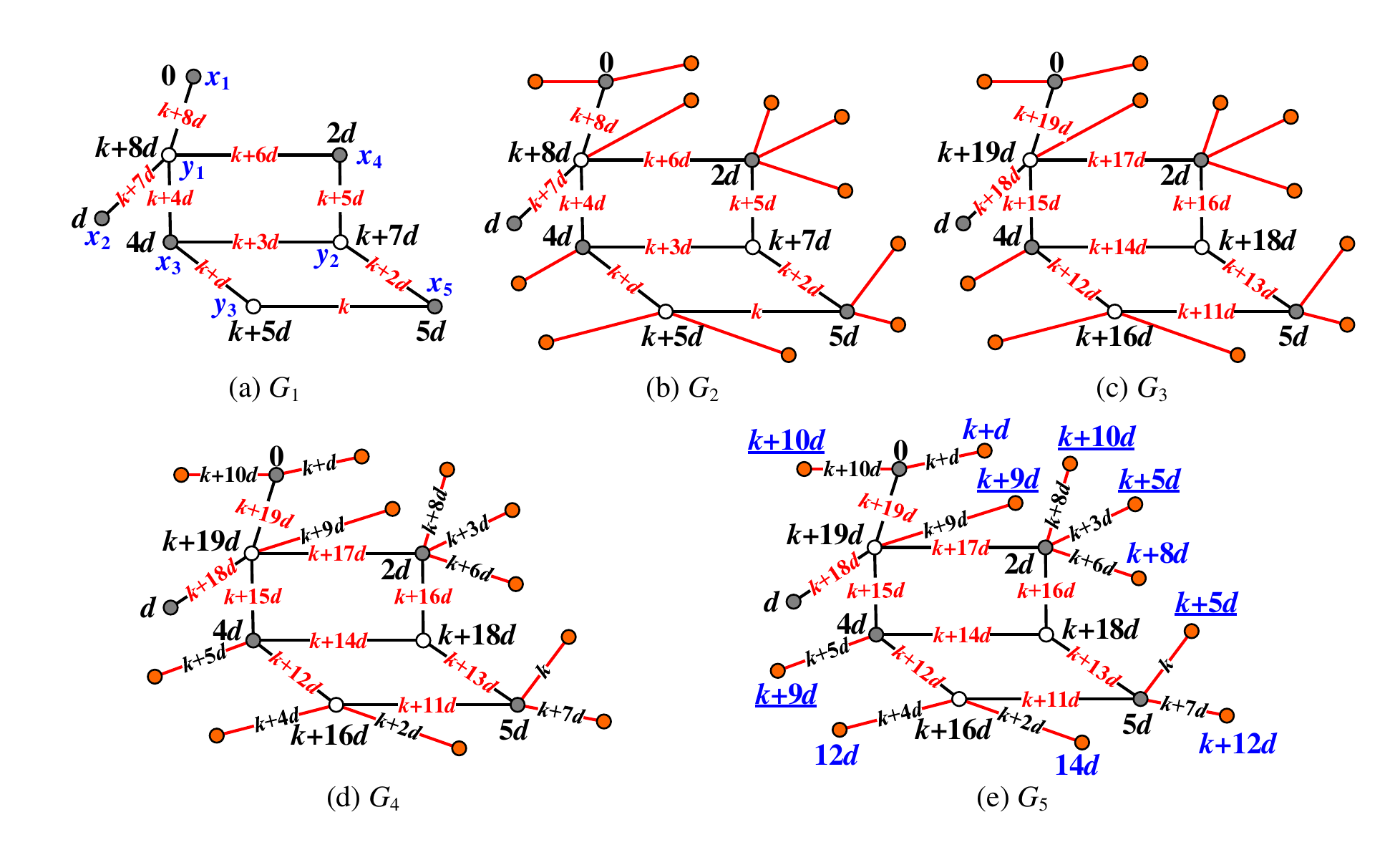}\\
{\small \caption{\label{fig:graceful-sequence-total-coloring} A process for understanding the LEAF-adding algorithm, cited from \cite{Yao-Su-Wang-Hui-Sun-ITAIC2020}.}}
\end{figure}

\begin{figure}[h]
\centering
\includegraphics[width=16.4cm]{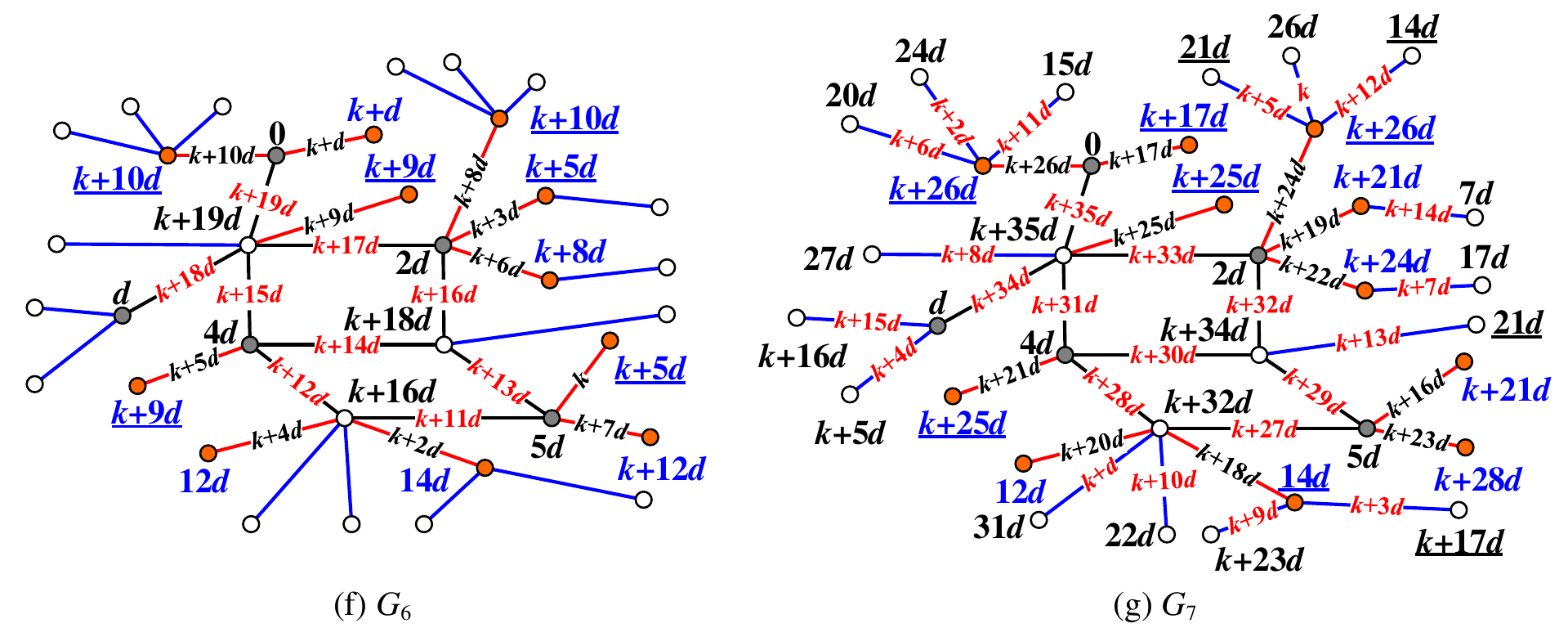}\\
{\small \caption{\label{fig:graceful-sequence-total-coloring11} For understanding the LEAF-adding algorithm, cited from \cite{Yao-Su-Wang-Hui-Sun-ITAIC2020}.}}
\end{figure}

See Fig.\ref{fig:graceful-sequence-total-coloring} and Fig.\ref{fig:graceful-sequence-total-coloring11} for understanding the $(k,d)$-graceful preparation and the LEAF-adding algorithm, and by them and Lemma \ref{thm:adding-leaves-keep-sequence-colorings} we get:

\begin{thm}\label{thm:each-tree-sequence-coloring}
\cite{Yao-Su-Wang-Hui-Sun-ITAIC2020} Each tree on $q$ edges admits a proper graceful total sequence coloring based on two sequences $A_M$ and $B_q$ holding $0<b_j-a_i\in B_q$ for $a_i\in A_M$ and $b_j\in B_q$.
\end{thm}

\begin{rem}\label{rem:Fibonacci-Lucas-sequence-sequences}
We can make number-based strings with long bytes by the help of \emph{Fibonacci-Lucas sequence-sequences} $\{F[w_i, z_i]_n\}^N_{i=1}$, where each $F[w_i, z_i]_n=\{w_i$, $z_i$, $c_{i, 3}$, $c_{i, 4}$, $\dots , c_{i, n-2}\}$ is a \emph{Fibonacci-Lucas sequence} with integers $w_i\geq 1$ and $z_i\geq 1$, $c_{i, 3}=w_i+z_i$, $c_{i, 4}=c_{i, 3}+z_i$, and $c_{i, s+1}=c_{i, s}+c_{i, s-1}$ for $s\in [4, n-2]$ and $i\in [1,N]$. Particularly, a \emph{Fibonacci sequence} is $F[1, 1]_n$, and a \emph{Lucas sequence} is $F[1, 3]_n$. Moreover, as $w_i-w_j\geq 1$ and $z_i-z_j\geq 1$, we have
$$F[w_i, z_i]_n+F[w_j, z_j]_n=F[w_i+w_j, z_i+z_j]_n,~F[w_i, z_i]_n-F[w_j, z_j]_n=F[w_i-w_j, z_i-z_j]_n$$

Thereby, we get $(n\cdot N)!$ permutations $FL_j=m_{j_1}$ $m_{j_2}$ $\cdots $ $m_{j_{nN}}$ of $w_1$ $z_1$ $c_{1, 3}$ $c_{1, 4}$ $\cdots c_{1, n-2}$ $w_2$ $z_2$ $c_{2, 3}$ $c_{2, 4}$ $\cdots c_{2, n-2}$ $\cdots$ $w_i$ $z_i$ $c_{i, 3}$ $c_{i, 4}$ $\cdots c_{i, n-2}$ $\cdots$ $w_N$ $z_N$ $c_{N, 3}$ $c_{N, 4}$ $\cdots c_{N, n-2}$ from a Fibonacci-Lucas sequence-sequence $\{F[w_i, z_i]_n\}^N_{i=1}$. Clearly, each permutation $FL_j$ is just a number-based string that can be used to encrypt digital files. Obviously, it is not easy to partition a number-based string $FL_j$ into a Fibonacci-Lucas sequence-sequence $\{F[w_i, z_i]_n\}^N_{i=1}$.\paralled
\end{rem}

\subsection{Colorings based on abstract sequences}

\begin{defn}\label{defn:colorings-abstract-sequences}
\cite{Yao-Su-Wang-Hui-Sun-ITAIC2020} Suppose that a $(p,q)$-graph $G$ admits a \emph{graceful coloring} $f:V(G)\rightarrow [0,M]$ such that $f(E(G))=\{f(wz)=|f(w)-f(z)|:wz\in E(G)\}=[1,q]$. Let $C_M=(c_i)^M_{i=0}$ and $D_q=(d_j)^q_{j=1}$ be two \emph{abstract sequences}. We define a new coloring $f^*$ of $G$ by letting $f^*(w)=c_i$ if $f(w)=i$ for vertex $w\in V(G)$, and $f^*(wz)=d_j$ if $f(wz)=j$ for edge $wz\in E(G)$. Then we call $f^*$ a \emph{graceful abstract-sequence coloring} of $G$ if $f^*(E(G))=D_q$. Here an abstract sequence $C_M=(c_i)^M_{i=0}$ or $D_q=(d_j)^q_{j=1}$ is consisted of any things in the world. Thereby, $f^*$ is an \emph{abstract substitution} of $f$, conversely, $f$ is a \emph{mapping homomorphism} of $f^*$. Let $E(G)=\{e_i=x_iy_i:~i\in [1,q]\}$. Then this $(p,q)$-graph $G$ has its another Topcode-matrix $T^*_{code}(G)$ defined as
\begin{equation}\label{eqa:topcode-matrix-22}
{
\begin{split}
T^*_{code}(G)= \left(
\begin{array}{cccccccccc}
f^*(x_1) & f^*(x_2) & \cdots & f^*(x_q)\\
f^*(e_1) & f^*(e_2) & \cdots & f^*(e_q)\\
f^*(y_1) & f^*(y_2) & \cdots & f^*(y_q)
\end{array}
\right)_{3\times q}=(f^*(X),f^*(E),f^*(Y))^T
\end{split}}
\end{equation} where three \emph{abstract vectors} $f^*(X)=(f^*(x_1),f^*(x_2), \cdots ,f^*(x_q))$, $f^*(E)=(f^*(e_1),f^*(e_2)$, $ \cdots $, $f^*(e_q))$ and $f^*(Y)=(f^*(y_1)$, $f^*(y_2)$, $ \cdots $, $f^*(y_q))$.\qqed
\end{defn}

In particular case of a Fibonacci-Lucas sequence (refer to Remark \ref{rem:Fibonacci-Lucas-sequence-sequences}), we have $f^*(w)=c_i=F[w_i, z_i]_n$ if $f(w)=i$ for vertex $w\in V(G)$ and $f^*(wz)=d_j=F[w_j, z_j]_n$ if $f(wz)=j$ for edge $wz\in E(G)$. Theorem \ref{thm:tree-graceful-total-coloringss} tells that each tree $T$ admits a \emph{$(k,d)$-gracefully total coloring}, also, a \emph{set-ordered gracefully total coloring} as $(k,d)=(1,1)$ and a \emph{set-ordered odd-gracefully total coloring} as $(k,d)=(1,2)$, by the way, these colorings can be used to provide examples for Ringel-Kotzig Decomposition Conjecture \cite{Ringel-G}. Thereby, we have

\begin{thm}\label{thm:each-tree-graceful-abstract-coloring}
\cite{Yao-Su-Wang-Hui-Sun-ITAIC2020} Each tree admits a \emph{set-ordered graceful abstract sequence coloring}.
\end{thm}


\section{Set-type (vector-type) of colorings and labelings}

\subsection{Graceful set-colorings}

\begin{defn} \label{defn:55-v-set-e-proper-more-labelings}
$^*$ A \emph{v-set e-proper $\varepsilon$-labeling} (resp. $\varepsilon$-\emph{coloring}) of a $(p,q)$-graph $G$ is a mapping $f:V(G)\cup E(G)\rightarrow \Omega$, where $\Omega$ consists of numbers and sets, such that $f(u)$ is a set for each vertex $u\in V(G)$, $f(xy)$ is a number for each edge $xy\in E(G)$, and the edge color set $f(E(G))$ satisfies a $\varepsilon$-condition.\qqed
\end{defn}

\begin{figure}[h]
\centering
\includegraphics[width=13.4cm]{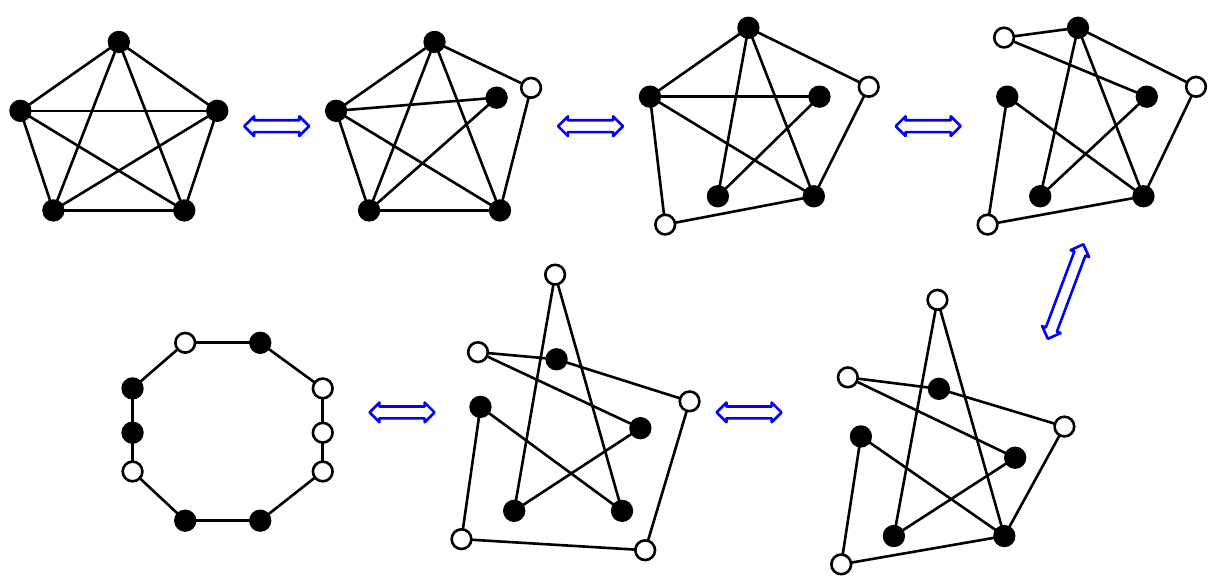}\\
\caption{\label{fig:0-Euler-Hamilton}{\small A procedure of connecting an Euler's graph and a Hamilton cycle of length 10 by the \emph{non-adjacent vertex-coinciding operation} and the \emph{2-edge-connected 2-degree-vertex vertex-splitting operation} introduced in \cite{Sun-Zhang-Yao-IAEAC-2017}.}}
\end{figure}

As $f(E(G))=[1,q]$ in Definition \ref{defn:55-v-set-e-proper-more-labelings}, we get a \emph{v-set e-proper graceful labeling}, and moreover a \emph{v-set e-proper odd-graceful labeling} is $f(E(G))=[1,2q-1]^o$. Euler's graphs and Hamilton cycles are popular researching objects in graph theory. Sun \emph{et al.} \cite{Sun-Zhang-Yao-IAEAC-2017} show a connection between Euler's graphs and Hamilton cycles by an operation, called \emph{non-adjacent identifying operation and the 2-edge-connected 2-degree-vertex splitting operation}. An example is shown in Fig.\ref{fig:0-Euler-Hamilton}.

\begin{figure}[h]
\centering
\includegraphics[width=12.4cm]{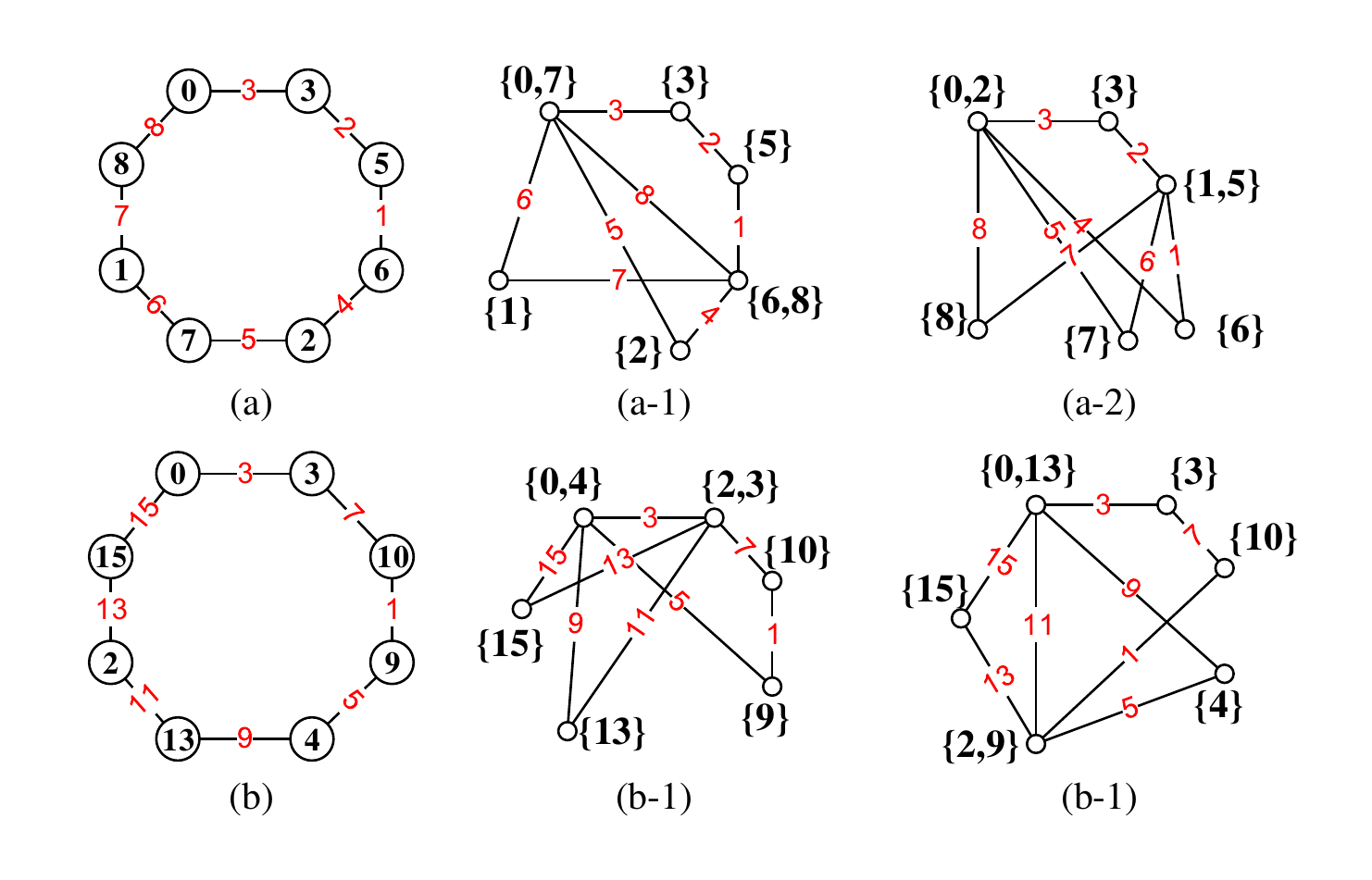}\\
\caption{\label{fig:v-set-e-proper}{\small (a) A cycle admitting a graceful labeling; (a-1) and (a-2) are two Euler's graphs admitting two v-set e-proper graceful labelings; (b) a cycle admitting an odd-graceful labeling; (b-1) and (b-2) are two Euler's graphs admitting two v-set e-proper odd-graceful labelings, cited from \cite{Yao-Zhang-Sun-Mu-Sun-Wang-Wang-Ma-Su-Yang-Yang-Zhang-2018arXiv}.}}
\end{figure}

Sun \emph{et al.} \cite{Hui-Sun-Bing-Yao2018} introduced some v-set e-proper $\varepsilon$-labelings on Euler's graphs, where $\varepsilon \in \{$graceful, odd-graceful, harmonious, $k$-graceful, odd sequential, elegant, odd-elegant, felicitous, odd-harmonious, edge-magic total$\}$, see examples displayed in Fig.\ref{fig:v-set-e-proper}.

\begin{rem}\label{rem:333333}
It is not hard to generate vv-type/vev-type TB-paws from Topsnut-gpws made by Hamilton cycles. We use an example to introduce the Euler-Hamilton-method in the following: We make a vev-type TB-paw $D_{vev}(C_8)=03325164257617880$ form Fig.\ref{fig:v-set-e-proper}(a), thus, $D_{vev}(C_8)$ can be obtained from Fig.\ref{fig:v-set-e-proper}(a-1) too, and vice versa. Obviously, it is not relaxed to pick up $D_{vev}(C_8)$ from Fig.\ref{fig:v-set-e-proper}(a-1) if Topsnut-gpws have large numbers of vertices and edges. Thereby, a vev-type TB-paw (as \emph{a public-key}) made by a colored Hamilton cycle induces directly a vev-type TB-paw (as \emph{a private-key}) generated from a colored Euler's graph. But, such vv-type/vev-type TB-paws can be attacked since colored Hamilton cycles are easy to be found by computer attack.

We can use the Euler-Hamilton-method to produce complex TB-paws in the following way: As known, a graph $G$ is an Euler's graph if and only if there are $m$ edge-disjoint cycles $C_1,C_2,\dots, C_m$ such that $E(G)=\bigcup ^m_{i=1}E(C_i)$ (Ref. \cite{Bondy-2008}). So, we can get $m$ vev-type TB-paws $D_{vev}(C_i)$ with $i\in [1,m]$. Let $i_1,i_2,\dots ,i_m$ be a permutation of $1,2,\dots ,m$. Thereby, we have many vev-type TB-paws like
$$D_{vev}(G)=D_{vev}(C_{i_1})\uplus D_{vev}(C_{i_2})\uplus \cdots \uplus D_{vev}(C_{i_m}),$$
or
$$D'_{vev}(G)=D(C_{i_1})\uplus D(C_{i_2})\uplus \cdots \uplus D(C_{i_j}),~j<m.$$
Clearly, it is an \emph{irreversible} procedure of generating $D'_{vev}(G)$ from Euler's graphs. .\paralled
\end{rem}

We give a character of a non-Euler's graph as follows:
\begin{thm}\label{thm:666666}
\cite{Hui-Sun-Bing-Yao2018} Each non-Euler's graph $G$ corresponds $m$ edge-disjoint paths $P_1,P_2,\dots ,P_m$ such that $|E(G)|=\sum ^m_{i=1}|E(P_i)|$.
\end{thm}

\begin{figure}[h]
\centering
\includegraphics[width=15cm]{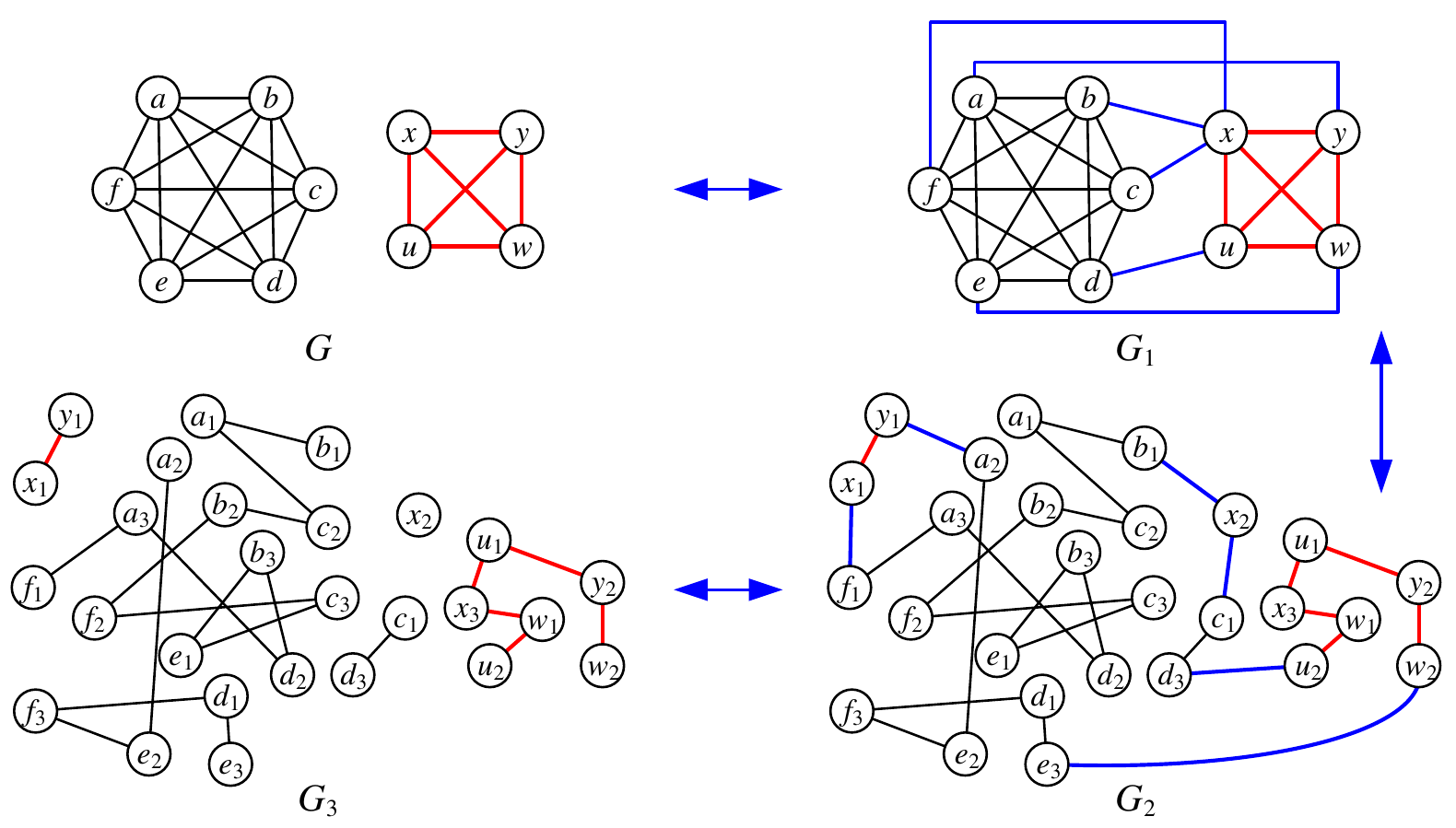}\\
\caption{\label{fig:0-non-Euler-graph}{\small $G$ is a non-Euler's graph; $G_1$ is an Euler's graph obtained by adding six new edges to $G$; $G_2$ is a Hamilton cycle obtained by implementing the non-adjacent vertex-coinciding operation and the 2-edge-connected 2-degree-vertex vertex-splitting operation to $G_1$ in \cite{Sun-Zhang-Yao-IAEAC-2017}; $G_3$ is the desired union of paths after deleting six edges in blue.}}
\end{figure}

In fact, any non-Euler's graph $G$ can be add a set $E^*$ of $m$ new edges such that the resultant graph $G+E^*$ is just an Euler's graph, and $G+E^*$ corresponds a Hamilton cycle $C_q$ with $q=|E(G+E^*)|$ (Ref. \cite{Sun-Zhang-Yao-IAEAC-2017}). Now, we delete all edges of $E^*$ from $C_q$, so $C_q-E^*$ is just a graph consisted of $m$ edge-disjoint paths $P_1,P_2,\dots ,P_m$. The above deduction tells us a way for producing vv-type/vev-type TB-paws from colored edge-disjoint paths $P_1,P_2,\dots ,P_m$ matching with the non-Euler's graph $G$ (see Fig.\ref{fig:0-non-Euler-graph}).

By Definition \ref{defn:topcode-matrix-definition}, an Euler's graph (\textrm{a-2}) admitting two v-set e-proper graceful labeling shown in Fig.\ref{fig:v-set-e-proper} corresponds a Topcode-matrix $T_{code}(\textrm{a-2})$ in (\ref{eqa:T-code-graceful-set-labeling}), and another Euler's graph (\textrm{b-1}) admitting two v-set e-proper graceful labeling shown in Fig.\ref{fig:v-set-e-proper} corresponds a Topcode-matrix $T_{code}(\textrm{b-1})$ in (\ref{eqa:T-code-odd-graceful-set-labeling}). The Topcode-matrix $T_{code}(\textrm{a-2})$ induces a TB-paw
$$D(\textrm{a-2})=15150202021515021224567863367788.
$$ Obviously, finding the Topcode-matrix $T_{code}(\textrm{a-2})$ from the number-based string $D(\textrm{a-2})$ is not easy, even very difficult.

\begin{equation}\label{eqa:T-code-graceful-set-labeling}
\centering
T_{code}(\textrm{a-2})= \left(
\begin{array}{cccccccc}
\{1,5\} & \{1,5\} & \{0,2\} & \{0,2\} & \{0,2\} & \{1,5\} & \{1,5\} & \{0,2\} \\
1 & 2 & 2 & 4 & 5 & 6 & 7 & 8\\
\{6\} & \{3\} & \{3\} & \{6\} & \{7\} & \{7\} & \{8\} & \{8\}
\end{array}
\right)
\end{equation}

\begin{equation}\label{eqa:T-code-odd-graceful-set-labeling}
\centering
T_{code}(\textrm{b-1})= \left(
\begin{array}{cccccccc}
\{2,9\} & \{0,13\} & \{2,9\} & \{3\} & \{0,13\} & \{0,13\} & \{2,9\} & \{0,13\} \\
1 & 3 & 5 & 7 & 9 & 11 & 13 & 15\\
\{10\} & \{3\} & \{10\} & \{10\} & \{4\} & \{2,9\} & \{15\} & \{15\}
\end{array}
\right)
\end{equation}

\begin{defn}\label{defn:graceful-intersection}
\cite{Yao-Zhang-Sun-Mu-Sun-Wang-Wang-Ma-Su-Yang-Yang-Zhang-2018arXiv} A $(p,q)$-graph $G$ admits a vertex set-labeling $f:V(G)\rightarrow [1,q]^2$~(resp. $[1,2q-1]^2)$, and induces an edge set-color $f\,'(uv)=f(u)\cap f(v)$ for $uv \in E(G)$. If we can select a \emph{representative} $a_{uv}\in f\,'(uv)$ for each edge color set $f\,'(uv)$ such that $\{a_{uv}:~uv\in E(G)\}=[1,q]$ (resp. $[1,2q-1]^o$), then $f$ is called a \emph{graceful-intersection (resp. an odd-graceful-intersection) total set-labeling} of $G$.\qqed
\end{defn}

\begin{figure}[h]
\centering
\includegraphics[width=16cm]{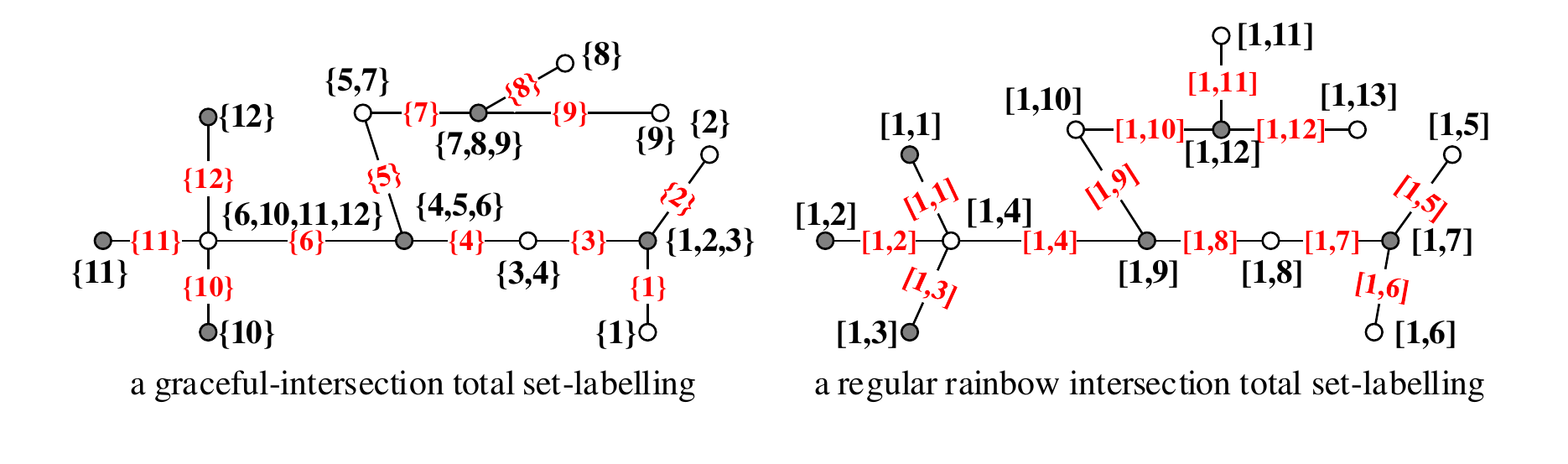}\\
\caption{\label{fig:0-graceful-intersection}{\small Left tree admitting a graceful-intersection total set-labeling for illustrating Theorem \ref{thm:total-set-labelings}; Right tree admitting a regular rainbow intersection total set-labeling for illustrating Theorem \ref{thm:rainbow-total-set-labelings}, cited from \cite{Yao-Zhang-Sun-Mu-Sun-Wang-Wang-Ma-Su-Yang-Yang-Zhang-2018arXiv}.}}
\end{figure}

\begin{thm}\label{thm:total-set-labelings}
\cite{Yao-Zhang-Sun-Mu-Sun-Wang-Wang-Ma-Su-Yang-Yang-Zhang-2018arXiv} Each tree $T$ admits a \emph{graceful-intersection (resp. an odd-graceful-intersection) total set-labeling} (see Fig.\ref{fig:0-graceful-intersection}).
\end{thm}

\begin{thm}\label{thm:rainbow-total-set-labelings}
\cite{Yao-Zhang-Sun-Mu-Sun-Wang-Wang-Ma-Su-Yang-Yang-Zhang-2018arXiv} Each tree $T$ of $q$ edges admits a \emph{regular rainbow intersection total set-labeling} based on a regular rainbow set-sequence $\{R_k\}^{q}_1$ (see Fig.\ref{fig:0-graceful-intersection}).
\end{thm}

\subsection{Set-colorings and set-labelings}

\begin{defn}\label{defn:55-set-labeling}
\cite{Yao-Sun-Zhang-Mu-Sun-Wang-Su-Zhang-Yang-Yang-2018arXiv} Let $G$ be a $(p,q)$-graph.

(i) A set mapping $F: V(G)\cup E(G)\rightarrow [0, p+q]^2$ is called a \emph{total set-labeling} of $G$ if two sets $F(x)\neq F(y)$ for distinct elements $x,y\in V(G)\cup E(G)$.

(ii) A vertex set mapping $F: V(G) \rightarrow [0, p+q]^2$ is called a \emph{vertex set-labeling} of $G$ if two sets $F(x)\neq F(y)$ for distinct vertices $x,y\in V(G)$.

(iii) An edge set mapping $F: E(G) \rightarrow [0, p+q]^2$ is called an \emph{edge set-labeling} of $G$ if two sets $F(uv)\neq F(xy)$ for distinct edges $uv, xy\in E(G)$.

(iv) A vertex set mapping $F: V(G) \rightarrow [0, p+q]^2$ and a proper edge mapping $g: E(G) \rightarrow [a, b]$ are called a \emph{v-set e-proper labeling $(F,g)$} of $G$ if two sets $F(x)\neq F(y)$ for distinct vertices $x,y\in V(G)$ and two edge colors $g(uv)\neq g(wz)$ for distinct edges $uv, wz\in E(G)$.

(v) An edge set mapping $F: E(G) \rightarrow [0, p+q]^2$ and a proper vertex mapping $f: V(G) \rightarrow [a,b]$ are called an \emph{e-set v-proper labeling $(F,f)$} of $G$ if two sets $F(uv)\neq F(wz)$ for distinct edges $uv, wz\in E(G)$ and two vertex colors $f(x)\neq f(y)$ for distinct vertices $x,y\in V(G)$.\qqed
\end{defn}

\begin{defn} \label{defn:set-coloring-definitions}
\cite{Yao-Sun-Zhang-Li-Yan-Zhang-Wang-ITOEC-2017} Let a $(p,q)$-graph $G$ with integers $q\geq p-1\geq 2$ admit a mapping $F: X\rightarrow S$, where $X$ is a subset of $V(G)\cup E(G)$, $S$ is a subset of power set $[0,pq]^2$ of the set $[0,pq]$, and let $Rs(m)=\{c_1,c_2,\dots, c_m\}$ be a constraint set. There are the following constraint conditions:
\begin{asparaenum}[(a)]
\item \label{vertex-set} $X=V(G)$;
\item \label{edge-set} $X=E(G)$;
\item \label{total-set} $X=V(G)\cup E(G)$;
\item \label{adjacent-vertex-labeling} $F(u)\not =F(v)$ if $uv\in E(G)$ (it may happen $F(u)\cap F(v)\neq \emptyset$);
\item \label{adjacent-edge-labeling} $F(uv)\not =F(uw)$ for any pair of adjacent edges $uv$ and $uw$ of $G$ (it may happen $F(uv)\cap F(uw)\neq \emptyset$);
\item \label{vertex-labeling} $|F(V(G))|=p$, also, $F(x)\not =F(y)$ for any pair of vertices $x$ and $y$ of $G$;
\item \label{edge-labeling} $|F(E(G))|=q$, so $F(xy)\not =F(uv)$ for distinct edges $uv$ and $xy$ of $G$;
\item \label{edge-induced} A mapping $F\,': E(G)\rightarrow S$ is induced by $F$ subject to $Rs(m)$, that is, each edge $uv\in E(G)$ is colored by the set $F\,'(uv)$ such that each $c\in F\,'(uv)$ is generated by some $a\in F(u)$, $b\in F(v)$ and one constraint or more constraints of $Rs(m)$; and
\item \label{induced-edge-labeling} $|F\,'(E(G))|=q$.
\end{asparaenum}

\noindent \textbf{We call}:
\begin{asparaenum}[(1)]
\item $F$ a \emph{strong vertex set-labeling} of $G$ if both (\ref{vertex-set}) and (\ref{vertex-labeling}) hold true.
\item $F$ a \emph{strong edge-set-labeling} of $G$ if both (\ref{edge-set}) and (\ref{edge-labeling}) hold true.
\item $F\,'$ an \emph{strongly induced edge-set-labeling} of $G$ if both (\ref{edge-labeling}) and (\ref{edge-induced}) hold true.
\item $F$ a \emph{strongly total set-coloring} of $G$ if (\ref{total-set}), (\ref{vertex-labeling}) and (\ref{edge-labeling}) hold true.
\item $(F,F\,')$ a \emph{strong set-coloring} subject to $Rs(m)$ if (\ref{vertex-set}), (\ref{vertex-labeling}), (\ref{edge-induced}) and (\ref{induced-edge-labeling}) hold true.
\end{asparaenum}
\textbf{We call}:
\begin{asparaenum}[(1')]
\item $F$ a \emph{set-labeling} of $G$ if it satisfies (\ref{vertex-set}) and (\ref{adjacent-vertex-labeling}) simultaneously.
\item $F$ an \emph{edge-set-labeling} of $G$ if it satisfies (\ref{edge-set}) and (\ref{adjacent-edge-labeling}) simultaneously.
\item $F$ a \emph{total set-coloring} of $G$ if it satisfies (\ref{total-set}), (\ref{adjacent-vertex-labeling}) and (\ref{adjacent-edge-labeling}) simultaneously.
\item $(F,F\,')$ a \emph{set-coloring} subject to $Rs(m)$ if (\ref{vertex-set}), (\ref{adjacent-vertex-labeling}), (\ref{adjacent-edge-labeling}) and (\ref{edge-induced}) are true simultaneously.
\end{asparaenum}
\textbf{We call:}
\begin{asparaenum}[(1'')]
\item $F$ a \emph{pseudo-vertex set-labeling} of $G$ if it holds (\ref{vertex-set}), but not (\ref{adjacent-vertex-labeling}).
\item $F$ a \emph{pseudo-edge set-labeling} of $G$ if it holds (\ref{edge-set}), but not (\ref{adjacent-edge-labeling}).
\item $F$ a \emph{pseudo-total set-coloring} of $G$ if it holds (\ref{total-set}), but not (\ref{adjacent-vertex-labeling}), or but not (\ref{adjacent-edge-labeling}), or not both (\ref{adjacent-vertex-labeling}) and (\ref{adjacent-edge-labeling}).\qqed
\end{asparaenum}
\end{defn}

\begin{rem}\label{rem:333333}
Let $v_s=\min \{|F(x)|:x\in V(G)\}$ and $v_l=\max \{|F(y)|:y\in V(G)\}$ in Definition \ref{defn:set-coloring-definitions}. $F$ is called an \emph{$\alpha$-uniformly vertex set-labeling} of $G$ if $v_s=v_l=\alpha$. Similarly, we have $e_s=\min \{|F\,'(uv)|:~uv\in E(G)\}$ and $e_l=\max \{|F\,'(xy)|:~xy\in E(G)\}$, and $F\,'$ is called a \emph{$\beta$-uniformly edge set-labeling} of $G$ if $e_s=e_l=\beta$. As $\alpha=\beta=1$, $(F,F\,')$ is a popular labeling of graph theory (Ref. \cite{Gallian2020}). There is another group of two parameters are $t_s=\min \{|\psi(x)|:x\in V(G)\cup E(G)\}$ and $t_l=\max \{|\psi(y)|:~y\in V(G)\cup E(G)\}$ in Definition \ref{defn:set-coloring-definitions}, and we call $\psi$ a \emph{$k$-uniformly total set-coloring} if $k=t_s=t_l$. Hereafter, we say ``a set-coloring $(F,F\,')$ s.t.$Rs(m)$'' defined in Definition \ref{defn:set-coloring-definitions}, and say ``a total set-coloring $\psi$ s.t.$Rs(m)$'' defined in Definition \ref{defn:set-coloring-definitions}.\paralled
\end{rem}

\begin{figure}[h]
\centering
\includegraphics[width=16cm]{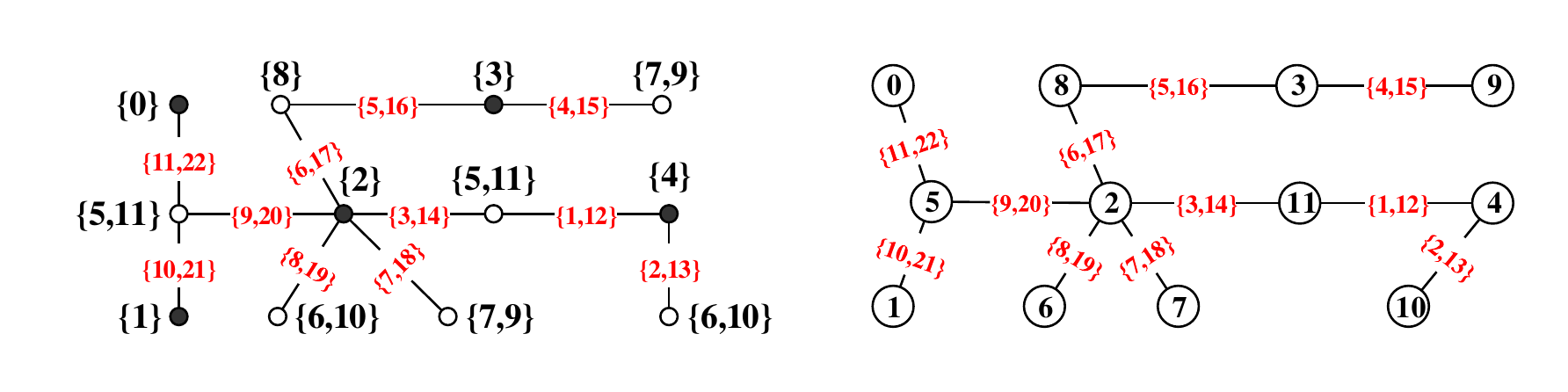}\\
{\small \caption{\label{fig:tu-mixed-set-coloring} Left is a set-coloring $(F,F\,')$, Right is a mixed-vertex set-labeling.}}
\end{figure}

In Figure \ref{fig:tu-mixed-set-coloring}, the left tree $T$ admits a set-coloring $(F,F\,')$ s.t.$Rs(m)=\{c_1,c_2\}$, where $a\in F(u)$, $b\in F(v)$ and $c\in F\,'(uv)$ hold one of $c_1:a+b+c=27$ and $c_2:|a-b|=c$; the right tree $T$ admits a mixed-vertex set-labeling consisted of a vertex labeling $f$ and an edge set-labeling $F\,'$ such that any number $c\in F\,'(uv)$ of each edge $uv\in E(T)$ holds one of $f(u)+c+f(v)=27$ and $f(u)+c+f(v)=16$.

Some set-labelings and set-colorings can be optimal in the following way: $S$ is the power set of an integer set $\{0,1,\dots ,\chi_\epsilon (G)\}=[0,\chi_\epsilon (G)]$ such that $G$ admits a set-labeling or a set-coloring defined In Definition \ref{defn:set-coloring-definitions}, and but $S$ is not the power set of any integer set $[0,M]$ if $M<\chi_\epsilon (G)$, where $\epsilon$ is a combinatoric of some conditions defined in Definition \ref{defn:set-coloring-definitions}. For example, $\epsilon=\{$(\ref{vertex-set}), (\ref{adjacent-vertex-labeling})$\}$ if only about the \emph{set-labeling} of $G$.

\begin{figure}[h]
\centering
\includegraphics[width=16.4cm]{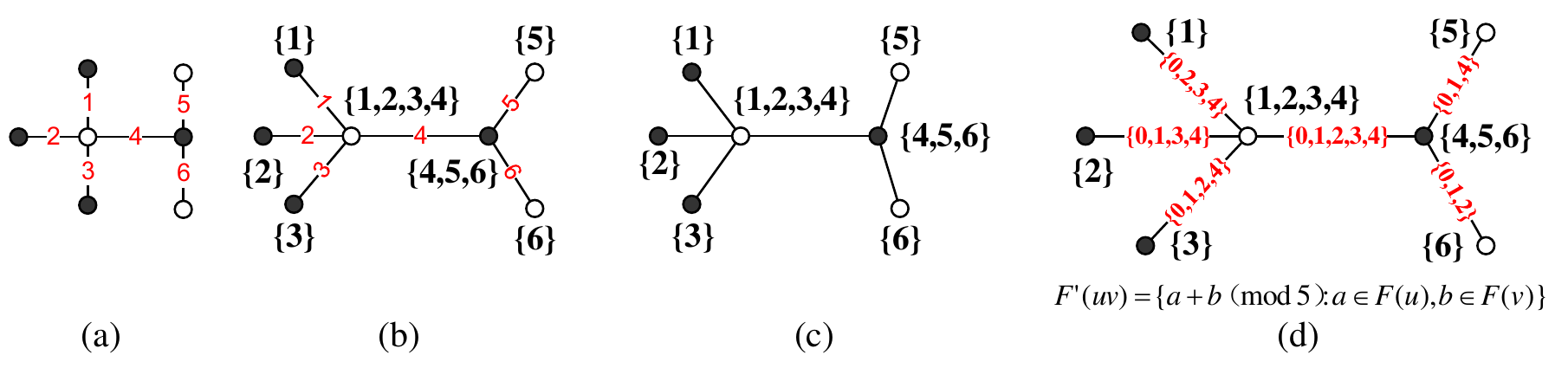}\\
{\small \caption{\label{fig:a-procedure} A procedure of a strong set-coloring generated by a vertex distinguishing edge coloring of a tree shown in (a).}}
\end{figure}

\begin{rem}\label{rem:333333}
A \emph{vertex distinguishing edge coloring} $f$ of a tree $T$ is shown in Fig.\ref{fig:a-procedure}(a). We define a \emph{strong set-labeling} $F$ based on the vertex distinguishing edge coloring $f$ shown in Fig.\ref{fig:a-procedure}(b) and (c); next we define a \emph{strong edge-set-labeling} $F\,'$ by $F$ subject to a unique constraint $c_1:a+b=c~(\bmod~5)$ such that both $a\in F(u)$ and $b\in F(v)$ induce $c\in F\,'(uv)$ (or $F\,'(uv)=\{a+b=c~(\bmod~5):a\in F(u),b\in F(v)\}$ in mathematical notation) for each edge $uv$ of $T$ shown in Fig.\ref{fig:a-procedure}(d).

For a (strongly) total set-coloring $\psi$ s.t.$Rs(m)$ defined in Definition \ref{defn:set-coloring-definitions}, we point out that three numbers $a\in \psi(u)$, $b\in \psi(v)$ and $c\in \psi(uv)$ satisfy a constraint $c_i\in Rs(m)$, by graph colorings (resp. labelings), so $c_i$ holds one of the following constraint conditions:
\begin{asparaenum}[(a)]
\item \label{odd-graceful} $|a-b|=c$ can induce \emph{graceful labeling, or odd-graceful labeling, or odd-elegant labeling, or vertex (distinguishing) coloring if $c\neq 0$}.

\item \label{edge-magic-total} $a+b+c=k$ can induce \emph{edge-magic total labeling} for a constant $k$.

\item \label{filicitous} $a+b=c$ ($\bmod~\eta$) can induce \emph{felicitous labeling, or harmonious labeling}.

\item \label{edge-magic-graceful} $|a+b-c|=k$ can induce \emph{edge-magic graceful labeling}.

\item \label{parameters-edge-magic-graceful} $|a+b-\lambda c|=k$ can induce \emph{$(k,\lambda)$-edge magic graceful labeling}, or \emph{$(k,\lambda)$-odd-magic graceful labeling}.

\item \label{parameters-edge-magic-total} $a+b=k+\lambda c$ can induce \emph{$(k,\lambda)$-magic total labeling}, or \emph{$(k,\lambda)$-odd-magic total labeling}.

\item \label{total-coloring} $a\neq b$, $b\neq c$ and $c\neq a$ can induce \emph{total coloring, vertex distinguishing total coloring, list-coloring}.

\item \label{edge-coloring} $c\in \psi(uv)$ and $c\,'\in \psi(uv')$ hold $c\neq c\,'$ can induce \emph{edge coloring}.

\item \label{couple-edge-magic-total} $a+b+c=k^+$ or $|a+b-c|=k^-$ can induce \emph{$(k^+,k^-)$-couple edge-magic total labeling}.

\item \label{two-magics-magic-graceful} $|a+b-c|=k_1$ or $|a+b-c|=k_2$ can induce \emph{$(k_1,k_2)$-edge magic graceful labeling}.\paralled
\end{asparaenum}
\end{rem}

\begin{defn}\label{defn:111111}
\cite{Yao-Sun-Zhang-Li-Yan-Zhang-Wang-ITOEC-2017} Let $f:V(G)\cup E(G)\rightarrow [1,M]$ be a proper total coloring of a $(p,q)$-graph $G$. A \emph{(pseudo-)total set-labeling} $F$ is defined by $F(x)=\{f(u)\}\cup \{f(xy):y\in N(x)\}$ for each vertex $x\in V(G)$ and $F(uv)=\{f(uv)\}$ for each edge $uv\in E(G)$.\qqed
\end{defn}

Clearly, there may be $F(w)=F(z)$ for some edge $wz\in E(G)$. But, we can obtain \emph{strong set-labelings} on vertex sets of graphs from some particular edge-colorings of graphs.

\begin{rem}\label{rem:333333}
The set-labelings and set-colorings defined in in Definition \ref{defn:set-coloring-definitions} can be optimal in this way: $S$ is the power set of an integer set $[0,\chi _{\epsilon}(G)]$ such that $G$ admits a set-labeling or a set-coloring defined in Definition \ref{defn:set-coloring-definitions}, and but $S$ is not the power set of any integer set $[0,M]$ if $M <\chi _{\epsilon}(G)$, where $\epsilon $ is a combinatoric of some conditions (a)-(g) stated in Definition \ref{defn:set-coloring-definitions}, and $\chi _{\epsilon}(G)$ is called an \emph{$\epsilon $-chromatic number} of $G$. For example, $\epsilon =\{(a), (d)\}$ if only about a set-labeling of $G$. So, we can determine the $\epsilon $-\emph{chromatic number}
$\chi _{\epsilon}(G)$ for a fixed $\epsilon $. As known, there are many long-standing conjectures on graph coloring and graph labelings, so we believe there are new open problems on the set-colorings and set-labelings defined in Definition \ref{defn:set-coloring-definitions}.\paralled
\end{rem}

\begin{lem} \label{components-cycles}
\cite{Yao-Sun-Zhang-Li-Yan-Zhang-Wang-ITOEC-2017} Each vertex distinguishing edge coloring of a $(p,q)$-graph $G$ induces a strong set-labeling $F$ with \begin{equation}\label{eqa:degrees}
\Delta(G)=\max\{|F(x)|:~x\in V(G)\},\quad \delta(G)=\min\{|F(x)|:~x\in V(G)\}.
\end{equation}
\end{lem}

\begin{thm} \label{thm:theorem11}
\cite{Yao-Sun-Zhang-Li-Yan-Zhang-Wang-ITOEC-2017} If a set-labeling $F$ of a graph $G$ holds: $|F(x)|\geq \textrm{deg}_G(x)$ for each vertex $x\in V(G)$, and $|F(u)\cap F(v)|=1$ for each edge $uv\in E(G)$, and $F(u)\cap F(v)\neq F(u)\cap F(w)$ for any adjacent edges $uv,uw\in E(G)$, then $F$ induces a \emph{proper edge-coloring} of $G$.
\end{thm}

\begin{thm} \label{thm:theorem22}
\cite{Yao-Sun-Zhang-Li-Yan-Zhang-Wang-ITOEC-2017} A set-labeling $F$ of a graph $G$ holds: $|F(x)|\geq \textrm{deg}_G(x)$ for each vertex $x\in V(G)$, and $|F(u)\cap F(v)|=1$ for each edge $uv\in E(G)$, and $F(u)\cap F(v)\neq F(u)\cap F(w)$ for any pair of adjacent edges $uv,uw\in E(G)$, then $F$ induces an \emph{adjacent $1$-common edge-coloring} of $G$.
\end{thm}

\begin{thm} \label{thm:box}
\cite{Yao-Sun-Zhang-Li-Yan-Zhang-Wang-ITOEC-2017} Suppose that a graph $G_1$ admits a set-labeling $F_1: V(G_1)\rightarrow S$ where $S$ is a subset of power set of $[1,M]$, and $F_1(u)\cap F_1(v)\neq \emptyset$ for each edge $uv\in E(G_1)$. And there are graphs $G_k=G_{k-1}-u_{k-1}$ for $u_{k-1}\in V(G_{k-1})$ with $k\geq 2$, and for each graph $G_k$ with $k\in [2,|G_1|-2]$ there exists a set-labeling $F_k(x)=F_{k-1}(x)$ if $x\not \in N(u_{k-1})$, and $F_k(y)=F_{k-1}(y)\setminus [F_{k-1}(y)\cap F_{k-1}(u_{k-1})]$ if $y\in N(u_{k-1})$, as well as $F_k(u)\cap F_k(v)\neq \emptyset$ for each edge $uv\in E(G_k)$. If two \emph{systems of distinct representatives} for each edge $uv\in E(G_k)$ holds $|R_{ep}(u)\cap R_{ep}(v)|=1$ with $k\in [2,|G_1|-2]$, then $F_1$ induces a \emph{proper edge-coloring} of $G_1$.
\end{thm}

\begin{defn}\label{defn:set-matrices-from-set-labeling}
\cite{Yao-Sun-Zhang-Li-Yan-Zhang-Wang-ITOEC-2017} A \emph{set-matrix} for a $(p,q)$-graph $G$ admitting an \emph{edge-set-labeling} $F\,'$ is defined by $S_e(G) = (A_{ij})_{q\times q}$ such that $A_{ij}=F\,'(u_iu_j)$ for $u_iu_j\in E(G)$, and $A_{ij}=\emptyset $ otherwise. Suppose that a $(p,q)$-graph $G$ admits a set-labeling $F$ defined on vertex set $V (G)$, we define an operation ``$(\bullet )$'' on two sets, and define a set-matrix $S_v(G) = (B_{ij})_{q\times q}$ of $G$ based on the set-labeling $F$ by $B_{ij}=F(u_i)(\bullet )F(u_j)$ for $u_iu_j\in E(G)$, and $B_{ij}=\emptyset $ otherwise, where the result of $F(u_i)(\bullet )F(u_j)$ is still a set.\qqed
\end{defn}

\begin{defn}\label{defn:edge-graceful-mixed-labeling}
\cite{Yao-Sun-Zhang-Li-Zhao2017} Suppose that a $(p,q)$-graph $G$ admits a \emph{vertex-set labeling} $F:V(G)\rightarrow [0,q]^2$, where $F$ is the set of all subsets of $[0,q]$. If each edge $uv\in E(G)$ can be colored with $f(uv)=|a-b|$ for some $a\in F(u)$ and $b\in F(v)$, and $f(E(G))=[1,q]$, we call $(F,f)$ a \emph{vertex-set edge-graceful mixed labeling} of $G$.\qqed
\end{defn}

\begin{figure}[h]
\centering
\includegraphics[width=15cm]{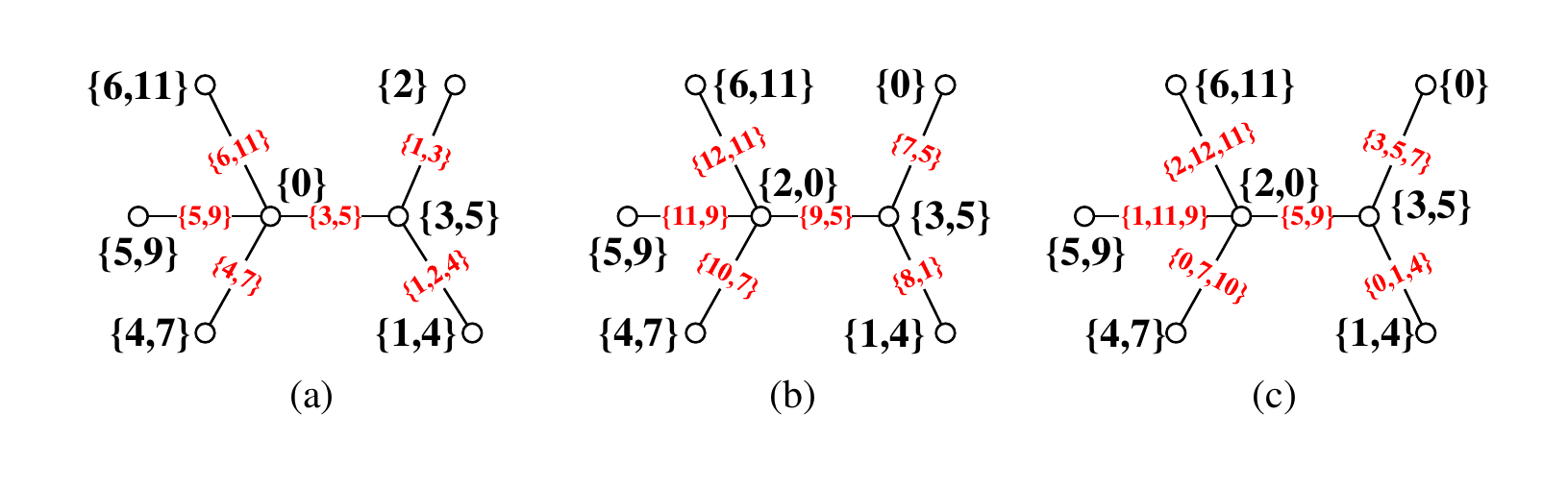}\\
{\small \caption{\label{fig:set-password-22} (a) A strong set-coloring $(F,F\,')$; (b) a strongly total set-coloring $\psi$; (c) another strongly total set-coloring $\theta$, cited from \cite{Yao-Zhang-Sun-Mu-Sun-Wang-Wang-Ma-Su-Yang-Yang-Zhang-2018arXiv}.}}
\end{figure}

The first example is about a \emph{strong set-coloring} $(F,F\,')$ in which the graphical structure is shown in Fig.\ref{fig:set-password-22}(a). We color each vertex $u$ with a set $F(u)$ such that $F(x)\not =F(y)$ for any pair of vertices $x,y$; there are some $a\in F(u)$ and $b\in F(v)$ to hold the unique constraint $|a-b|=c$ ($|Rs(m)|=1$) that induces the edge set $F\,'(uv)$ with $c\in F\,'(uv)$ such that $F\,'(uv)\not =F\,'(xy)$ for any pair of edges $uv$ and $xy$. The desired graphical password is shown in Fig.\ref{fig:set-password-22}(a).

A \emph{strongly total set-coloring} $\psi$, as the second example, is shown in Fig.\ref{fig:set-password-22}(b) with $\psi(x)\not =\psi(y)$ for any two elements $x,y\in V(G)\cup E(G)$, and for an edge $uv$, each $c\in \psi(uv)$ corresponds some $a\in \psi(u)$ and $b\in \psi(v)$ such that at least one of two constraints $|a-b|=c$ and $|a+b-c|=4$ holds true.

The third example on a \emph{strongly total set-coloring} $\theta$ s.t.$R^*s(3)=\{c_1,c_2,c_3\}$ is shown in Fig.\ref{fig:set-password-22}(c), where

$c_1:~|a-b|=c$ for $c\in \theta(uv)$, $a\in \theta(u)$ and $b\in \theta(v)$;

$c_2:~|a\,'+b\,'-c\,'|=4$ for $c\,'\in \theta(uv)$, $a\,'\in \theta(u)$ and $b\,'\in \theta(v)$; and

$c_3:~a\,''+b\,''=c\,''~(\bmod~6)$ for $c\,''\in \theta(uv)$, $a\,''\in \theta(u)$ and $b\,''\in \theta(v)$.

Thereby, each $c\in \theta(uv)$ corresponds some $a\in \theta(u)$ and $b\in \theta(v)$ such that they hold at least one of three constraints of $R^*s(3)$.

\begin{figure}[h]
\centering
\includegraphics[width=8.8cm]{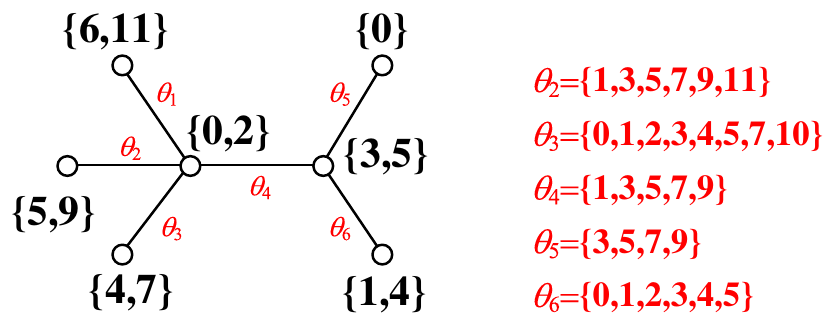}\\
{\small \caption{\label{fig:set-password-33} A strongly set-coloring $(\phi,\phi\,')$ s.t.$R^*s(3)$, cited from \cite{Yao-Zhang-Sun-Mu-Sun-Wang-Wang-Ma-Su-Yang-Yang-Zhang-2018arXiv}.}}
\end{figure}

A \emph{strongly total set-coloring} $(\phi,\phi\,')$ s.t.$R^*s(3)$ shown in Fig.\ref{fig:set-password-33} holds: $\theta(x)=\phi (x)$ for any $x\in V(G)$ and $\theta(uv)\subseteq \phi\,'(uv)$ for each edge $uv\in E(G)$, where $\theta$ is defined in Fig.\ref{fig:set-password-22}(c). We say $(\phi,\phi\,')$ to be the \emph{maximal strongly set-coloring} s.t.$R^*s(3)$.

\begin{thm} \label{them:each-graph-set-labelings}
\cite{Yao-Sun-Zhang-Li-Zhao2017} Each simple and connected $(p,q)$-graph $G$ admits a \emph{strongly total set-labeling} $F$ such that
\begin{equation}\label{eqa:theorem-11}
\max\{|F(x)|:~x\in V(G)\cup E(G)\}=\Delta(G)+1
\end{equation}
and
\begin{equation}\label{eqa:theorem-22}
\max\{c:~c\in F(x),x\in V(G)\cup E(G)\}\in [1,p+q].
\end{equation}
\end{thm}

\begin{thm} \label{them:tree-edge-magic-total-labeling}
\cite{Yao-Sun-Zhang-Li-Zhao2017} If a tree admits a \emph{super edge-magic total labeling} defined in Definition \ref{defn:11-old-labelings-Gallian}, then it admits a \emph{$2$-uniform strongly total set-coloring}.
\end{thm}

Theorem \ref{thm:tree-graceful-total-coloringss} in \cite{Yao-Su-Wang-Hui-Sun-ITAIC2020} says: \emph{Each tree admits a set-ordered gracefully total coloring}, and moreover each simple and connected $(p,q)$-graph $G$ can be vertex-split into a tree of $q+1$ vertices, then we have a result as follows:
\begin{thm} \label{them:graph-edge-graceful-set-labelings}
\cite{Yao-Sun-Zhang-Li-Zhao2017} Each simple and connected $(p,q)$-graph $G$ admits a \emph{v-set e-proper graceful coloring} $f: V(G)\rightarrow [0,q]^2$ defined in Definition \ref{defn:55-v-set-e-proper-more-labelings} and Definition \ref{defn:55-set-labeling}, such that each edge $uv$ is colored with a number $f(uv)=|a_u-b_v|$ for some $a_u\in f(u)$ and $b_v\in f(v)$, and the edge color set $f(E(G))=\{f(uv):uv\in E(G)\}=[1,q]$.
\end{thm}

\begin{defn}\label{defn:new-set-colorings}
$^*$ Let $G$ be a $(p,q)$-graph, and let ``$W$-type'' be one of the existing colorings and the existing labelings.

(i) A \emph{$W$-type ve-set-coloring} $F$ of $G$ holds $F: V(G)\cup E(G)\rightarrow [0, p+q]^2$ such that two sets $F(x)\neq F(y)$ for two adjacent or incident elements $x,y\in V(G)\cup E(G)$.

(ii) A \emph{$W$-type v-set-coloring} $F$ of $G$ holds $F: V(G) \rightarrow [0, p+q]^2$ such that two sets $F(x)\neq F(y)$ for each edge $xy\in E(G)$.

(iii) A \emph{$W$-type e-set-coloring} $F$ of $G$ holds $F: E(G) \rightarrow [0, p+q]^2$ such that two sets $F(uv)\neq F(uw)$ for two adjacent edges $uv, uw\in E(G)$.

(iv) An \emph{e-proper $W$-type v-set-coloring} $(F,g)$ of $G$ is consisted of a vertex set mapping $F: V(G) \rightarrow [0, p+q]^2$ and a proper edge mapping $g: E(G) \rightarrow [a, b]$ such that two sets $F(x)\neq F(y)$ for each edge $xy\in E(G)$ and two adjacent edge colors $g(uv)\neq g(uw)$ for two adjacent edges $uv, uw\in E(G)$.

(v) A \emph{v-proper $W$-type e-set-coloring} $(F,f)$ of $G$ is consisted of an edge set mapping $F: E(G) \rightarrow [0, p+q]^2$ and a proper vertex mapping $f: V(G) \rightarrow [a,b]$ such that two sets $F(uv)\neq F(uw)$ for two adjacent edges $uv, uw\in E(G)$ and two vertex colors $f(x)\neq f(y)$ for each edge $xy\in E(G)$.\qqed
\end{defn}

\begin{thm} \label{them:v-set-e-proper-W-type-colorings}
\cite{Wang-Wang-Yao2019-Euler-Split} Each simple and connected $(p,q)$-graph $G$ can be vertex-split into a tree $T$ of $q+1$ vertices by the vertex-splitting operation, and admits a \emph{v-set e-proper $W$-type coloring} (refer to Definition \ref{defn:55-v-set-e-proper-more-labelings} and Definition \ref{defn:55-set-labeling}) if $T$ admits a \emph{$W$-type coloring}.
\end{thm}

\begin{thm}\label{thm:Euler-v-set-e-proper-graceful-labeling}
\cite{Wang-Wang-Yao2019-Euler-Split} Every connected Euler's graph of $n$ edges with $n\equiv 0,3$ ($\bmod~4$) admits a \emph{v-set e-proper graceful labeling} defined in Definition \ref{defn:55-v-set-e-proper-more-labelings} and Definition \ref{defn:55-set-labeling}.
\end{thm}

\begin{thm}\label{thm:Euler-v-set-e-proper-harmonious-labeling}
\cite{Wang-Wang-Yao2019-Euler-Split} Every connected Euler's graph with odd $n$ edges admits a \emph{v-set e-proper harmonious labeling} defined in Definition \ref{defn:55-v-set-e-proper-more-labelings} and Definition \ref{defn:55-set-labeling}.
\end{thm}

\begin{defn}\label{defn:4C-k-labeling}
\cite{Yao-Mu-Sun-Sun-Zhang-Wang-Su-Zhang-Yang-Zhao-Wang-Ma-Yao-Yang-Xie2019} A $(p,q)$-graph $G$ admits a vertex labeling $f:V(G)\rightarrow \{0,1,\dots ,q\}$, and an edge labeling $g:E(G)\rightarrow \{1,2,\dots \}$, such that

(1) (e-magic) each edge $xy\in E(G)$ holds $g(xy)+|f(x)-f(y)|=k$, a constant;

(2) (ee-balanced) let $s(uv)=|f(u)-f(v)|-g(uv)$ for an edge, and each edge $uv$ matches another edge $xy$ holding $s(uv)+s(xy)=k\,'$ true;

(3) (EV-ordered) $\max f(V(G))< \min g(E(G))$ (or $\min f(V(G))> \max g(E(G))$);

(4) (set-ordered) $\max f(X)<\min f(Y)$ for the bipartition $(X,Y)$ of $V(G)$.

We call the labeling pair $(f,g)$ as a \emph{$(4C,k,k\,')$-labeling}.\qqed
\end{defn}

\begin{thm}\label{thm:3-equivalent-labelings}
\cite{Wang-Wang-Yao2019-Euler-Split} If an Euler's graph has $4m$ edges for $m\geq 2$, then it admits a \emph{v-set e-proper graceful labeling}, a \emph{v-set e-proper odd-graceful labeling}, a \emph{v-set e-proper edge-magic total labeling} (refer to Definition \ref{defn:55-v-set-e-proper-more-labelings} and Definition \ref{defn:55-set-labeling}), and a \emph{v-set e-proper $(4C,k,k\,')$-labeling} defined in Definition \ref{defn:4C-k-labeling}.
\end{thm}

\begin{cor}\label{thm:edge-deleted-graphs}
\cite{Wang-Wang-Yao2019-Euler-Split} If an Euler's graph of $n$ edges holds $n\equiv 1,2~(\bmod~4)$ true, then it admits a \emph{v-set e-proper $\varepsilon$-labeling} defined in Definition \ref{defn:55-v-set-e-proper-more-labelings}, where $\varepsilon$-labeling $\in \{$odd-graceful labeling, edge-magic total labeling, $(4C,k,k\,')$-labeling$\}$.
\end{cor}

\begin{cor}\label{thm:general-graphs}
\cite{Wang-Wang-Yao2019-Euler-Split} Any non-Euler $(p,q)$-graph $G$ of $n$ edges corresponds an Euler's graph $H=G+E^*$ admitting a \emph{v-set e-proper $\varepsilon$-labeling} $f$ for $\varepsilon$-labeling $\in \{$odd-graceful labeling, edge-magic total labeling, $(4C,k,k\,')$-labeling$\}$, such that $G$ admits a \emph{v-set e-proper labeling} $g$ induced by $f$ and $g(E(G))=\{1,2,\dots ,|E^*|+q\}\setminus \{f(xy):xy\in E^*\}$.
\end{cor}

\begin{thm}\label{thm:more-Euler-graphs}
\cite{Wang-Wang-Yao2019-Euler-Split} There are connected and vertex-disjoint Euler's graphs $H_1,H_2,\dots,H_m$, such that another connected Euler's graph $G$ is obtained by vertex-coinciding $H_i$ with some $H_j$ for $i\neq j$. Then $G$ admits a \emph{v-set e-proper $\varepsilon$-labeling} for $\varepsilon$-labeling $\in \{$odd-graceful labeling, edge-magic total labeling, $(4C,k,k\,')$-labeling$\}$ if $\sum^m_{i=1} |E(H_i)|\equiv 0~(\bmod~4)$.
\end{thm}

\begin{thm}\label{thm:v-pseudo-e-proper-graceful}
\cite{Yao-Mu-Sun-Sun-Zhang-Wang-Su-Zhang-Yang-Zhao-Wang-Ma-Yao-Yang-Xie2019} A connected Euler's graph $G$ of $n$ edges admits a graceful labeling $f$ if and only if a cycle $C_n$ obtained by vertex-splitting some vertices of $G$ admits this labeling $f$ as a its \emph{v-pseudo e-proper graceful labeling} defined in Definition \ref{defn:55-set-labeling}.
\end{thm}

\begin{thm}\label{thm:pseudo-v-set-e-proper-graceful}
\cite{Yao-Mu-Sun-Sun-Zhang-Wang-Su-Zhang-Yang-Zhao-Wang-Ma-Yao-Yang-Xie2019} Every connected graph admits a \emph{v-pseudo e-proper graceful labeling} defined in Definition \ref{defn:55-set-labeling}.
\end{thm}

\subsection{Vector-colorings}

Let $V_{ec}(n)$ be the set of $n$-dimension vectors $\textbf{\textrm{d}}_i=(a_{i,1}, a_{i,2}, \dots, a_{i,n})$ with $a_{i,j}\in R_{set}$ being a real number set, and let $V^{I}_{ec}(n)$ be the set of $n$-dimension vectors with each component to be an integer, $V^{+I}_{ec}(n)$ be the set of $n$-dimension vectors with each component to be a non-negative integer. We call $V^{I}_{ec}(n)$ an \emph{integer-vector set}, and $V^{+I}_{ec}(n)$ an \emph{integer$^+$-vector set}. Clearly,
\begin{equation}\label{eqa:555555}
V^{+I}_{ec}(n)\subset V^{I}_{ec}(n)\subset V_{ec}(n).
\end{equation}

\begin{defn} \label{defn:vector-colorings-graphs}
$^*$ Let $R_{es}(m)=(r_1,r_2,\dots ,r_m)$ be a set of mathematical restrictive functions $r_{s}(\alpha,\beta,\gamma)$ of three variables $\alpha,\beta$ and $\gamma$. A mapping $F:V(G)\cup E(G)\rightarrow V_{ec}(n)$ is called a \emph{vector coloring} of a $(p,q)$-graph $G$ if three vectors $F(i)=\textbf{\textrm{d}}_i$, $F(j)=\textbf{\textrm{d}}_j$ and $F(ij)=\textbf{\textrm{d}}_{ij}=(a_{ij,1}, a_{ij,2}, \dots, a_{ij,n})$ for each edge $ij\in E(G)$ hold $r_{ij}(a_{i,k},a_{ij,k},a_{j,k})=0$ and $|a_{i,k}|+|a_{j,k}|\neq 0$ with $k\in [1,n]$ and function $r_{ij}\in R_{es}(m)$. Write $F(V(G)\cup E(G))=\{F(j):~j\in V(G)\cup E(G)\}$. \qqed
\end{defn}

\begin{rem}\label{rem:vector-colorings-graphs-remarks}
(1) A vector coloring differs from a set-coloring, since the vectors $F(i)$, $F(j)$ and $F(ij)$ of each edge $ij\in E(G)$ have the same dimension, and the components of a vector are ordered.

(2) Since new colorings and labelings are coming everyday, determining the maximal $n$ in $F:V(G)\cup E(G)\rightarrow V_{ec}(n)$ defined in Definition \ref{defn:vector-colorings-graphs} is impossible.

(3) A mathematical restrictive function set $R_{es}(m)=(r_1,r_2,\dots ,r_m)$ shown in Definition \ref{defn:vector-colorings-graphs} is coloring-type, or labeling-type, or mixed-type in graph theory. \paralled
\end{rem}

\begin{defn} \label{defn:various-vector-colorings}
$^*$ Let $G$ be a $(p,q)$-graph.
\begin{asparaenum}[\textrm{Vec}-1 ]
\item A vector coloring $F$ defined in Definition \ref{defn:vector-colorings-graphs} is called an \emph{integer-vector coloring} of $G$ if $F:V(G)\cup E(G)\rightarrow V^{I}_{ec}(n)$.

\item A vector coloring $F$ defined in Definition \ref{defn:vector-colorings-graphs} is called a \emph{proper-vector coloring} of $G$ if $F:V(G)\cup E(G)\rightarrow V^{+I}_{ec}(n)$ and each $r_{s}\in R_{es}(m)$ is a coloring-type restrictive function.

\item A vector coloring $F$ defined in Definition \ref{defn:vector-colorings-graphs} is called a \emph{proper-vector labeling} of $G$ if $F:V(G)\cup E(G)\rightarrow V^{+I}_{ec}(n)$ and each $r_{s}\in R_{es}(m)$ is a labeling-type restrictive function.
\item A vector coloring $F$ defined in Definition \ref{defn:vector-colorings-graphs} is called a \emph{mixed proper-vector coloring} of $G$ if $F:V(G)\cup E(G)\rightarrow V^{+I}_{ec}(n)$ and some $r_{s}\in R_{es}(m)$ is a coloring-type restrictive function, and some $r_{t}\in R_{es}(m)$ is a labeling-type restrictive function.\qqed
\end{asparaenum}
\end{defn}

\begin{thm}\label{thm:admits-a-proper-vector-coloring}
$^*$ Since each graph $G$ admits a vertex coloring, an edge coloring and a total coloring in graph theory, then $G$ admits a \emph{proper-vector coloring} $F:V(G)\cup E(G)\rightarrow V^{+I}_{ec}(n_0)$ with $n_0\geq 3$.
\end{thm}

\begin{thm}\label{thm:admits-a-proper-vector-labeling}
$^*$ If a $(p,q)$-graph $G$ admits colorings $f_i:V(G)\cup E(G)\rightarrow [a_i,b_i]$ for $i\in [1,A]$ subject to a mathematical restrictive function $r_{i}$, then $G$ admits a \emph{proper-vector labeling} $F:V(G)\cup E(G)\rightarrow V^{+I}_{ec}(A)$ with $F(u)=(f_1(u), f_2(u), \dots, f_A(u))$ for each vertex $u\in V(G)$ subject to a mathematical restrictive set $R_{es}(m)=(r_1,r_2,\dots ,r_A)$, and $F(xy)=(f_1(xy), f_2(xy), \dots, f_A(xy))$ for each edge $xy\in E(G)$, where $V^{+I}_{ec}(A)=\{F(w):w\in V(G)\cup E(G)\}$.
\end{thm}

\begin{defn} \label{defn:ve-vector-vector-colorings-v-vector-e-number}
$^*$ Let $G$ be a $(p,q)$-graph, and let $R_{set}$ be the set of real numbers.

(1) A \emph{v-vector e-number coloring} is defined as $F:V(G)\rightarrow V_{ec}(n)$ and $F:E(G)\rightarrow R_{set}$ such that the \emph{induced color} $F(ij)=F(i)\bullet F(j)$ for each edge $ij\in E(G)$, where ``$\bullet$'' is the \emph{dot-product operation} of vectors. If $F(E(G))=\{a,a+1,\dots, a+q-1\}$, we call $F$ an \emph{arithmetic v-vector e-number coloring} of the $(p,q)$-graph $G$, and moreover $F$ is called a \emph{graceful v-vector e-number coloring} of $G$ if $F(E(G))=\{1,2,\dots, q\}$, and $F$ is called an \emph{odd-graceful v-vector e-number coloring} of $G$ if $F(E(G))=\{1,3,\dots, 2q-1\}$.

(2) A \emph{ve-vector coloring} is defined as $F:V(G)\rightarrow V_{ec}(n)$ such that the \emph{induced edge vector} $F(ij)=F(i)\times F(j)$ for each edge $ij\in E(G)$, where ``$\times$'' is the \emph{vector-product operation} of vectors.\qqed
\end{defn}

\subsection{Topological expression of integer$^+$-vector sets}

\begin{defn} \label{defn:graph-full-proper-integer-vector-set}
$^*$ Let $I^{+}(n)$ be a finite subset of an integer$^+$-vector set $V^{+I}_{ec}(n)$. If a graph $H$ admitting an \emph{integer$^+$-vector coloring} $F:V(H)\cup E(H)\rightarrow I^{+}(n)$ subject to a mathematical restrictive set $R_{es}(m)=(r_1,r_2,\dots ,r_m)$, such that each vector $\textbf{\textrm{d}}\in I^{+}(n)$ is $F(u)=\textbf{\textrm{d}}$ for some vertex $u\in V(H)$, or $F(xy)=\textbf{\textrm{d}}$ for some edge $xy\in E(H)$, then we say that the set $I^{+}(n)$ to be \emph{graph-full}, call $I^{+}(n)$ a \emph{graph-full integer$^+$-vector set}, and the graph $H$ a \emph{graphic expression} of the \emph{graph-full integer$^+$-vector set} $I^{+}(n)$. \qqed
\end{defn}

An example of vector colorings is shown in Fig.\ref{fig:a-labeling-many-meanings} from (b) to (f), a lobster $T$ admits a vertex coloring defined as: $F(a)=$(0, 0, 0, 0, 0), $F(e)=$(8, 8, 8, 8, 8), $F(w)=$(3, 3, 3, 3, 3), $F(x)=$(9, 9, 9, 9, 9), $F(y)=$(5, 5, 5, 5, 5), $F(d)=$(2, 2, 2, 2, 2), $F(t)=$(11, 11, 11, 11, 1), $F(u)=$(4, 4, 4, 4, 4), $F(v)=$(10, 10, 10, 10, 10), $F(s)=$(7, 7, 7, 7, 7), $F(r)=$(6, 6, 6, 6, 6), $F(c)=$(1, 1, 1, 1, 1), $F(ay)=$(11, 22, 5, 1, 21), $F(yd)=$(9, 20, 7, 3, 17), $F(ed)=$(6, 17, 10, 6, 11), $F(ew)=$(5, 16, 0, 7, 9), $F(wx)=$(4, 15, 1, 8, 7), $F(dt)=$(3, 14, 2, 9, 5), $F(tu)=$(1, 12, 4, 11, 1), $F(uv)=$(2, 13, 3, 10, 3), $F(ds)=$(7, 18, 9, 5, 13), $F(dr)=$(8, 19, 8, 4, 15), $F(yc)=$(10, 21, 6, 2, 19). The mathematical restrictive function set $R_{es}(m)=(r_1,r_2,r_3,r_4,r_5)$ is defined as follows:

$r_1(a_{1,\alpha},a_{1,\alpha\beta},a_{1,\beta})=a_{1,\alpha}+a_{1,\alpha\beta}+a_{1,\beta}=16$ for each edge $\alpha\beta\in E(\textrm{(b)})$

$r_2(a_{2,\alpha},a_{2,\alpha\beta},a_{2,\beta})=a_{2,\alpha}+a_{2,\alpha\beta}+a_{2,\beta}=27$ for each edge $\alpha\beta\in E(\textrm{(c)})$

$a_{3,\alpha\beta}=r_3(a_{3,\alpha},a_{3,\beta})=a_{3,\alpha}+a_{3,\beta}~(\bmod~11)$ for each edge $\alpha\beta\in E(\textrm{(d)})$

$a_{4,\alpha\beta}=r_4(a_{4,\alpha},a_{4,\beta})=a_{4,\alpha}+a_{4,\beta}-4$ for each edge $\alpha\beta\in E(\textrm{(e)})$

$r_5$ is $\{a_{5,\alpha}+a_{5,\alpha\beta}+a_{5,\beta}:\alpha\beta\in E(\textrm{(f)})\}=[16,26]$.

Thereby, $I^{+}(5)=F(V(T)\cup E(T))$ is a set of $5$-dimension vectors, and is a graph-full integer$^+$-vector set too; and moreover $T$ is a \emph{graphic expression} of the graph-full integer$^+$-vector set $I^{+}(5)=F(V(T)\cup E(T))$.

\begin{rem}\label{rem:ABC-conjecture}
As known, trees admitting set-ordered graceful labelings, also, admit many pairwise equivalent labelings (Ref. \cite{SH-ZXH-YB-2018-5, Su-Wang-Yao-Image-labelings-2021-MDPI, Yao-Liu-Yao-2017, Yao-Yao-Hui-Cheng-2012-Malaysian}), so we can construct graph-full integer$^+$-vector sets with finite elements, and these trees are the \emph{graphic expressions} of the graph-full integer$^+$-vector sets, very often, these trees are the \emph{tree-graphic expressions}.\paralled
\end{rem}

\begin{defn} \label{defn:graph-graphic-expression-55}
$^*$ Let $I^{+}(n)$ be a finite subset of an integer$^+$-vector set $V^{+I}_{ec}(n)$. If $I^{+}(n)$ is a graph-full integer$^+$-vector set, then $I^{+}(n)$ is companied with a set of graphs as follows
\begin{equation}\label{eqa:555555}
G_{raph} (I^{+}(n))=\{H: \textrm{$H$ admits a integer$^+$-vector coloring $F:V(H)\cup E(H)\rightarrow I^{+}(n)$}\}.
\end{equation}
A graph $S_{G}(I^{+}(n))$ has its own vertex set $V(S_{G}(I^{+}(n)))=G_{raph} (I^{+}(n))$ and, for two graphs $H_r,H_s\in G_{raph} (I^{+}(n))$ admitting integer$^+$-vector colorings $F_r:V(H_r)\cup E(H_r)\rightarrow I^{+}(n)$ and $F_s:V(H_s)\cup E(H_s)\rightarrow I^{+}(n)$, if $F_r(V(H_r)\cup E(H_r))\cap F_s(V(H_s)\cup E(H_s))\neq \emptyset $, then $S_{G}(I^{+}(n))$ has an edge $H_rH_s$, and moreover $S_{G}(I^{+}(n))$'s edge set is
\begin{equation}\label{eqa:555555}
{
\begin{split}
E(S_{G}(I^{+}(n)))=&\{H_rH_s:~F_r(V(H_r)\cup E(H_r))\cap F_s(V(H_s)\cup E(H_s))\neq \emptyset , \\
&H_r,H_s\in G_{raph} (I^{+}(n))\}.
\end{split}}
\end{equation}
The graph $S_{G}(I^{+}(n))$ is called a \emph{graph-graphic expression} of the graph-full integer$^+$-vector set $I^{+}(n)$. \qqed
\end{defn}


\section{Miscellaneous colorings and labelings}

\subsection{Flawed labelings and colorings}

\begin{defn}\label{defn:flawed-labeling}
\cite{Yao-Mu-Sun-Sun-Zhang-Wang-Su-Zhang-Yang-Zhao-Wang-Ma-Yao-Yang-Xie2019} Suppose that $G=\bigcup^m_{i=1}G_i$ is a disconnected graph, where $G_1,G_2,\dots $, $G_m$ are connected graphs and vertex-disjoint from each other. We have a connected graph $G+E^*$ obtained by adding the edges of an edge set $E^*$ to $G$. If $G+E^*$ admits a $W$-type coloring $f:S\subseteq V(G)\cup E(G)\cup E^*\rightarrow [a,b]$, then we say that the disconnected graph $G$ admits a \emph{flawed $W$-type coloring} $f: S\setminus E^* \rightarrow [a,b]$.\qqed
\end{defn}

\begin{defn}\label{defn:flawed-odd-graceful-labeling}
\cite{Yao-Mu-Sun-Sun-Zhang-Wang-Su-Zhang-Yang-Zhao-Wang-Ma-Yao-Yang-Xie2019} Let $H=E^*+G$ be a connected graph, where $E^*$ is a non-empty set of edges and $G=\bigcup^m_{i=1}G_i$ is a disconnected graph, where $G_1,G_2,\dots, G_m$ are connected graphs and vertex-disjoint from each other. If $H$ admits a (set-ordered) graceful labeling (resp. a (set-ordered) odd-graceful labeling) $f$, then we call $f$ a \emph{flawed (set-ordered) graceful labeling} (resp. a \emph{flawed (set-ordered) odd-graceful labeling}) of $G$.\qqed
\end{defn}

\begin{defn}\label{defn:flawed-directed-graceful}
\cite{Yao-Mu-Sun-Sun-Zhang-Wang-Su-Zhang-Yang-Zhao-Wang-Ma-Yao-Yang-Xie2019} Suppose that the underlying graph of a $(p,q)$-digraph $\overrightarrow{G}$ is disconnected, and $\overrightarrow{G}+E^*$ is a connected directed $(p,q+q\,')$-graph, where $E^*$ is a set of arcs with $q\,'=|E^*|$. Let $f:V(\overrightarrow{G}+E^*) \rightarrow [0,q+q\,']$ (resp. $[0,2(q+q\,')-1]$) be a directed graceful labeling (resp. a directed odd-graceful labeling) $f$ of $\overrightarrow{G}+E^*$, then $f$ is called a \emph{flawed directed graceful labeling} (resp. \emph{flawed directed odd-graceful labeling}) of the $(p,q)$-digraph $\overrightarrow{G}$.\qqed
\end{defn}

\begin{thm} \label{thm:connection-flawed-labelings}
\cite{Yao-Mu-Sun-Sun-Zhang-Wang-Su-Zhang-Yang-Zhao-Wang-Ma-Yao-Yang-Xie2019} Suppose that $T=\bigcup ^m_{i=1}T_i$ is a forest having vertex-disjoint trees $T_1,T_2,\dots ,T_m$, and $(X,Y)$ be the vertex
bipartition of $T$. For integers $k\geq 1$ and $d\geq 1$, the following assertions are mutually equivalent:
\begin{asparaenum}[F-1. ]
\item $T$ admits a \emph{flawed set-ordered graceful labeling} $f$ with $\max f(X)<\min f(Y)$.
\item $T$ admits a \emph{flawed set-ordered odd-graceful labeling} $f$ with $\max f(X)<\min f(Y)$.

\item $T$ admits a f\emph{lawed set-ordered elegant labeling} $f$ with $\max f(X)<\min f(Y)$.

\item $T$ admits a \emph{flawed odd-elegant labeling} $\eta$ with $\eta(x)+\eta(y)\leq 2p-3$ for every edge $xy\in E(T)$.

\item $T$ admits a \emph{super flawed felicitous labeling} $\alpha$ with $\max \alpha(X)<\min \alpha(Y)$.

\item $T$ admits a \emph{super flawed edge-magic total labeling} $\gamma$ with $\max \gamma(X)<\min \gamma(Y)$ and a magic constant $|X|+2p+1$.
\item $T$ admits a \emph{super flawed $(|X|+p+3,2)$-edge antimagic total labeling} $\theta$ with $\max \theta(X)<\min \theta(Y)$.

\item $T$ admits a \emph{flawed harmonious labeling} $\varphi$ with $\max \varphi(X)<\min \varphi(Y\setminus \{y_0\})$ and $\varphi(y_0)=0$.
\end{asparaenum}
\end{thm}

\begin{thm} \label{thm:flawed-(k,d)-labelings}
Let $T=\bigcup ^m_{i=1}T_i$ be a forest having vertex-disjoint trees $T_1,T_2,\dots ,T_m$, and $V(T)=X\cup Y$. For all values of two integers $k\geq 1$ and $d\geq 1$, the following assertions are mutually equivalent:
\begin{asparaenum}[P-1. ]
\item $T$ admits a \emph{flawed set-ordered graceful labeling} $f$ with $\max f(X)<\min f(Y)$.

\item $T$ admits a \emph{flawed $(k,d)$-graceful labeling} $\beta$ with $\max \beta(x)<\min \beta(y)-k+d$ for all $x\in X$ and $y\in Y$.

\item $T$ admits a \emph{flawed $(k,d)$-arithmetic labeling} $\psi$ with $\max \psi(x)<\min \psi(y)-k+d\cdot |X|$ for all $x\in X$ and $y\in Y$.
\item $T$ admits a \emph{flawed $(k,d)$-harmonious labeling} $\varphi$ with $\max \varphi(X)<\min \varphi(Y\setminus \{y_0\})$ and $\varphi(y_0)=0$.
\end{asparaenum}
\end{thm}

\begin{defn}\label{defn:flawed-w-type-labeling}
\cite{Bing-Yao-2020arXiv} Suppose that graphs $H_1,H_2,\dots, H_m$ and $T$ are vertex-disjoint from each other, and $H=\bigcup^m_{i=1}H_i$. A $W$-type coloring means: a $W$-type coloring, or a $W$-type labeling.

(1) If there exists a graph operation ``$(\diamond)$'' on $T$ and $H$ such that the resultant graph $T(\diamond)H$ is a connected graph admitting a $W$-type coloring $f$, then $f$ is called a \emph{flawed $W$-type coloring} of $H$, and $f$ is called a \emph{$W$-type joining coloring} of $T$.

(2) If there is a graph operation ``$(\ast)$'' on $H$ such that the resultant graph $(\ast)H$ is a connected graph admitting a $W$-type coloring $g$, then we call $g$ a \emph{flawed $W$-type coloring} of $H$.\qqed
\end{defn}

\begin{rem}\label{rem:ABC-conjecture}
In topological authentication, the graph $T$ in Definition \ref{defn:flawed-w-type-labeling} is as a \emph{public-key}, and $H_1,H_2,\dots , H_m$ are a group of \emph{private-keys}. The $W$-type joining coloring $f$ of $T$ induces a Topcode-matrix $T_{code}(T)$ defined in Definition \ref{defn:topcode-matrix-definition}, and $H=\bigcup^m_{i=1}H_i$ has its own Topcode-matrix $T_{code}(H)$, such that a number-based string generated from $T_{code}(T)$ encrypts a digital file, and this file will be decrypted by another number-based string made by $T_{code}(H)$. Thereby, the Topcode-matrix $T_{code}(T)$ can be considered a \emph{public-key}, and the Topcode-matrix $T_{code}(H)$ is a \emph{private-key}.\paralled
\end{rem}

\subsection{Vertex distinguishing arc colorings}

Replacing every edge $uv$ of a simple graph $G$ by one of arcs $\overrightarrow{uv}$ and $\overrightarrow{vu}$ products a digraph, denoted as $\overrightarrow{G}$, called the \emph{oriented graph} of $G$.

\begin{defn} \label{defn:arc-chromatic-index-digraphs}
\cite{Bang-Jensen-Gutin-digraphs-2007, D-B-West1996} A mapping $f:A(D)\rightarrow [1,k]$ is called a proper arc $k$-coloring of a digraph $D$ if two arcs
$\overrightarrow{uv}$ and $\overrightarrow{uw}$ (resp. $\overrightarrow{vu}$ and $\overrightarrow{wu}$) are colored with different colors. The smallest number of $k$ colors required for which $D$ admits a proper arc $k$-coloring is called the \emph{di-chromatic index} of $D$, denoted as $\overrightarrow{\chi}'_2(D)$.\qqed
\end{defn}

\begin{rem}\label{rem:333333}
From Definition \ref{defn:arc-chromatic-index-digraphs}, $\left \lfloor \frac{\Delta(UG(D))+1}{2}\right \rfloor\leq \overrightarrow{\chi}'_2(D)\leq \chi\,'(UG(D))\leq \Delta(UG(D))+1$, where $\chi\,'(G)$ is the \emph{chromatic index} of a graph $G$, and $UG(D)$ is the underlying graph of $D$. A \emph{tournament} is a digraph that has an arc between any two vertices. There are tournaments $T$ on $2m+1$ or $2m$ vertices such that $\overrightarrow{\chi}'_2(T)=m$, since $K_{2m+1}$ is the union of $m$ edge-disjoint Hamilton cycles, and $K_{2m}$ is the union of $m$ edge-disjoint Hamilton paths.

Let $D$ be a digraph. A vertex $u$ has its own \emph{out-neighborhood} $N^+(u)=\{w:\overrightarrow{uw}\in A(D)\}$, and \emph{in-neighborhood} $N^-(u)=\{w:\overrightarrow{wu}\in A(D)\}$. So, the out-degree $\textrm{deg}^+ _D(u)$ of $u$ is $|N^+(u)|$, and the in-degree $\textrm{deg}^-_D(u)$ of $u$ $|N^-(u)|$. Let $\Delta^0(D)=\max\{\textrm{deg}^+ _D(u)+\textrm{deg}^- _D(u):u\in V(D)\}$, and let $f$ be an arc-coloring from $A(D)$ to $[1,k]$. For a vertex $u\in V(D)$ we will use a \emph{color set} $C^\pm(u,f)=C^+(u,f)\cup C^-(u,f)$, where $C^+(u,f)=\{f(\overrightarrow{uw}):w\in N^+(u)\}$ and $C^-(u,f)=\{f(\overrightarrow{wu}):w\in N^-(u)\}$. Notice that
$|C^+(u,f)\cup C^-(u,f)|\leq \textrm{deg}^+ _D(u)+\textrm{deg}^- _D(u)$, in general.\paralled
\end{rem}

\begin{defn} \label{defn:distinguishing-arc-digraphs}
\cite{Jing-Su-Bing-Yao-arc-colorings-2019} A proper arc $k$-coloring $f$ of a digraph $D$ without isolated arcs is called a \emph{vertex distinguishing arc (di-vde) $k$-coloring} $f$ if $C^\pm(u,f)\neq C^\pm(v,f)$ for distinct $u,v\in V(D)$. This coloring $f$ is called \emph{adjacent vertex distinguishing arc (di-avde) $k$-coloring} if $C^\pm(u,f)\neq C^\pm(v,f)$ for any arc $\overrightarrow{uv}\in A(D)$. The notation $\overrightarrow{\chi}'_{2s}(D)$ (resp. $\overrightarrow{\chi}'_{2as}(D)$) is the smallest number of $k$ colors required for which $D$ admits a di-vde (resp. a di-avde) $k$-coloring. \qqed
\end{defn}

\begin{figure}[h]
\centering
\includegraphics[width=16cm]{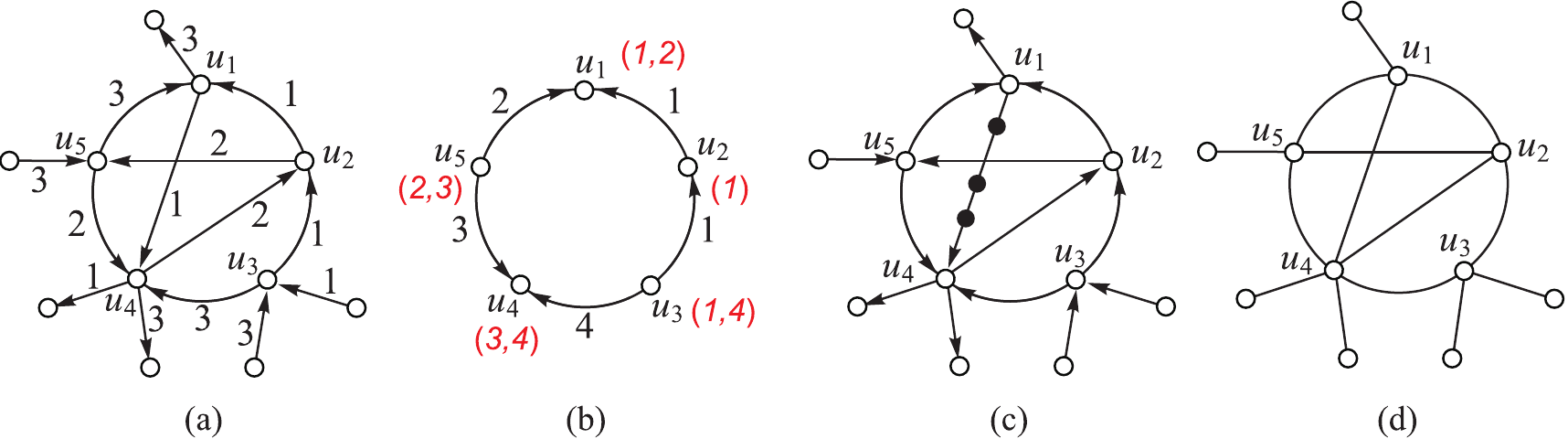}
\caption{\label{fig:Dis-arc-colring001}{\small (a) A digraph $D$ admits a di-avde $3$-coloring; (b) a digraph $H$ admits a di-avde $4$-coloring, also, a di-vde $4$-coloring; (c) a subdivision of an arc $u_1u_4$ of $D$; (d) the underlying graph $UG(D)$ of $D$ shown in (a), cited from \cite{Jing-Su-Bing-Yao-arc-colorings-2019}.}}
\end{figure}

\subsection{Graph-labeling}

\begin{defn}\label{defn:Edge-magic-total-graph-labeling}
\cite{Yao-Zhang-Sun-Mu-Sun-Wang-Wang-Ma-Su-Yang-Yang-Zhang-2018arXiv} Let $M_{pg}(p,q)$ be the set of maximal planar graphs $H_i$ of $i+3$ vertices with $i\in [1,p+q]$, where each face of a planar graph $H_i$ is a triangle. We use a \emph{total labeling} $f$ to color the vertices and edges of a $(p,q)$-graph $G$ with the elements of $M_{pg}(p,q)$, such that $i+ij+j=k$ (a constant), where $f(u_i)=H_i$, $f(u_iv_j)=H_{ij}$ and $f(v_j)=H_j$ for each edges $u_iv_j\in E(G)$. We say $f$ an \emph{edge-magic total graph-labeling} of $G$ based on the graph set $M_{pg}(p,q)$.\qqed
\end{defn}

\begin{figure}[h]
\centering
\includegraphics[width=14cm]{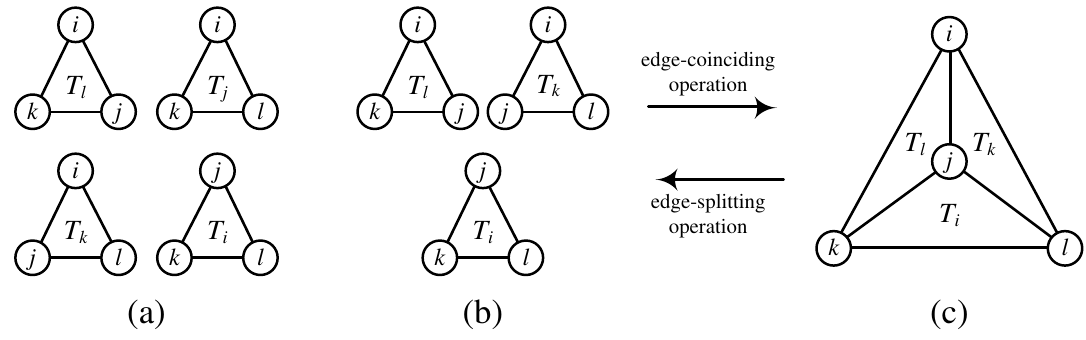}\\
\caption{\label{fig:4-coloring-system}{\small (a) 4-coloring system; (b) and (c) are the edge-splitting operation and the edge-coinciding operation.}}
\end{figure}

\textbf{Four-coloring triangularly edge-coinciding graph-labeling.} In \cite{YAO-SUN-WANG-SU-XU2018arXiv}, the authors introduce the triangularly edge-coinciding operation and triangular edge-splitting operation. Let $F_{\textrm{TPG}}$ be the set of planar graphs such that each one of $F_{\textrm{TPG}}$ has its outer face to be triangle and admits a proper 4-coloring. In Fig.\ref{fig:4-coloring-system}(b) and (c), a triangle $\Delta(T_l,T_k,T_i)$ shown in Fig.\ref{fig:4-coloring-system}(c) admits a 4-coloring obtained by three 4-colorings $f_l$, $f_k$ and $f_i$, so $\Delta(T_l,T_k,T_i)\in F_{\textrm{TPG}}$. The procedure of building up $\Delta(T_l,T_k,T_i)$ is called a \emph{triangularly edge-coinciding operation}. The process of edge-splitting $\Delta(T_l,T_k,T_i)$ into $T_l,T_k$ and $T_i$ is called a \emph{triangular edge-splitting operation}, conversely, the process of obtaining $\Delta(T_l,T_k,T_i)$ from $T_l,T_k$ and $T_i$ is called a \emph{triangular edge-coinciding operation}.

\begin{defn}\label{defn:4-coloring-triangularly-edge-coinciding-graph-labeling}
\cite{Yao-Zhang-Sun-Mu-Sun-Wang-Wang-Ma-Su-Yang-Yang-Zhang-2018arXiv} A $(p,q)$-graph $G$ admits a total labeling $h:V(G)\cup E(G)\rightarrow F_{\textrm{TPG}}$, such that each edge $u_iv_j\in E(G)$ holds that $f(u_i)=T_i$, $f(u_iv_j)=T_{ij}$ and $f(v_j)=T_j$ induce a triangle $\Delta(T_i,T_{ij},T_j)$ admitting a 4-coloring, and then we say $G$ admits a \emph{4-coloring triangularly edge-coinciding graph-labeling}.\qqed
\end{defn}

\begin{thm}\label{thm:4-coloring-triangularly-edge-coinciding-graph-labellin}
Any tree admits a \emph{4-coloring triangularly edge-coinciding graph-labeling}.
\end{thm}

\begin{defn} \label{defn:Topsnut-gpw-sequences-graph-labelings}
\cite{Yao-Zhang-Sun-Mu-Sun-Wang-Wang-Ma-Su-Yang-Yang-Zhang-2018arXiv} Let $\{(k_i,d_i)\}^m_1$ be a sequence with integers $k_i\geq 0$ and $d_i\geq 1$, and $G$ be a $(p,q)$-graph with $p\geq 2$ and $q\geq 1$. A \emph{Topsnut-gpw sequence} $\{G_{(k_i,d_i)}\}^m_1$ is made by an integer sequence $\{(k_i,d_i)\}^m_1$ and each Topsnut-gpw $G_{(k_i,d_i)}\cong G$ for $i\in [1,m]$. We define a labeling $F:V(G)\rightarrow \{G_{(k_i,d_i)}\}^m_1$, and $F(u_iv_j)=(|k_i-k_j|, ~d_i+d_j~(\textrm{mod}~M))$ with $F(u_i)=G_{(k_i,d_i)}$ and $F(v_j)=G_{(k_j,d_j)}$ for each edge $u_iv_j\in E(G)$. Then

(1) If $\{|k_i-k_j|:~u_iv_j\in E(G)\}=[1,2q-1]^o$ and $\{d_i+d_j~(\textrm{mod}~M)):~u_iv_j\in E(G)\}=[0,2q-3]^o$, we call $F$ a \emph{twin odd-type graph-labeling} of $G$.

(2) If $\{|k_i-k_j|:~u_iv_j\in E(G)\}=[1,q]$ and $\{d_i+d_j~(\textrm{mod}~M)):~u_iv_j\in E(G)\}=[0,2q-3]^o$, we call $F$ a \emph{graceful odd-elegant graph-labeling} of $G$.

(3) If $\{|k_i-k_j|:~u_iv_j\in E(G)\}$ and $\{d_i+d_j~(\textrm{mod}~M)):~u_iv_j\in E(G)\}$ are generalized Fibonacci sequences, we call $F$ a \emph{twin Fibonacci-type graph-labeling} of $G$.\qqed
\end{defn}

\begin{defn} \label{defn:total-graph-set-labeling}
\cite{Yao-Sun-Zhang-Mu-Sun-Wang-Su-Zhang-Yang-Yang-2018arXiv} Define a mapping $f:V(H)\cup E(H)\rightarrow \{G_i\}^m_1$ to a graph $H$, such that $f(u)=G_i$, $f(v)=G_j$ and $f(uv)=G_k$ for each edge $uv\in E(H)$.
\begin{asparaenum}[Mg-1. ]
\item If there exists a constant $k^*$, such that $i+k+j=k^*$ for each edge $uv\in E(H)$, we call $f$ an \emph{edge-magic total graph-set labeling} (edge-magic total gs-labeling), a graph $G=\langle \{G_i\}^m_1;H(f)\rangle$ is obtained by joining $G_i$ with $G_k$ and joining $G_k$ with $G_j$ for each edge $uv\in E(H)$ is called an \emph{edge-magic total gs-compound}.
\item If there exists a constant $k^*$ such that $|i-k+j|=k^*$ for each edge $uv\in E(H)$, we call $f$ an \emph{edge-magic total graceful graph-set labeling} (edge-magic total graceful gs-labeling), a graph $G=\langle \{G_i\}^m_1;H(f)\rangle$ obtained by joining $G_i$ with $G_k$ and joining $G_k$ with $G_j$ for each edge $uv\in E(H)$ is called an \emph{edge-magic total graceful gs-compound}.
\item If there exists a constant $k^*$ such that $k+|i-j|=k^*$ for each edge $uv\in E(H)$, we call $f$ an \emph{edge-magic graceful total graph-set labeling} (edge-magic graceful total gs-labeling), a graph $G=\langle \{G_i\}^m_1;H(f)\rangle$ obtained by joining $G_i$ with $G_k$ and joining $G_k$ with $G_j$ for each edge $uv\in E(H)$ is called an \emph{edge-magic graceful total gs-compound}.\qqed
\end{asparaenum}
\end{defn}

\begin{defn} \label{defn:graph-labeling-general}
\cite{Yao-Sun-Zhang-Mu-Sun-Wang-Su-Zhang-Yang-Yang-2018arXiv} Let $H_{ag}$ be a set of graphs. A $(p,q)$-graph $G$ admits a \emph{total graph-coloring} (resp. total graph-labeling) $F:V(G)\cup E(G)\rightarrow H_{ag}$, and each edge color $F(uv)=F(u)(\ast)F(v)$ is just a graph admitting a \emph{$k_{uv}$-matching} based on a graph operation ``$(\ast)$''. Here, a \emph{$k_{uv}$-matching} may be one of a perfect matching of $k_{uv}$ vertices, $k_{uv}$-cycle, $k_{uv}$-connected, $k_{uv}$-edge-connected, $k_{uv}$-colorable, $k_{uv}$-edge-colorable, total $k_{uv}$-colorable, $k_{uv}$-regular, $k_{uv}$-girth, $k_{uv}$-maximum degree, $k_{uv}$-clique, $\{a,b\}_{uv}$-factor, v-split $k_{uv}$-connected, e-split $k_{uv}$-connected, a twin odd-graceful matching $\odot \langle F(u), F(v)\rangle$, and so on. A graph $G(\ast) H_{ag}$ obtained by joining $F(u)$ with $F(uv)$ and joining $F(uv)$ with $F(v)$ for each edge $uv\in E(G)$ under the operation $(\ast)$ is called a \emph{$k_{uv}$-matching graph}.\qqed
\end{defn}

\begin{defn}\label{defn:pan-matching-graphs}
\cite{Yao-Sun-Zhang-Mu-Sun-Wang-Su-Zhang-Yang-Yang-2018arXiv} Let $P_{ag}$ be a set of graphs. A $(p,q)$-graph $G$ admits a \emph{graph-labeling} $F:V(G)\rightarrow P_{ag}$, and each induced edge color $F(uv)=F(u)(\bullet)F(v)$ for $uv\in E(G)$ is just a graph admitting a \emph{pan-matching}, where ``$(\bullet)$'' is a graph operation. Here, a \emph{pan-matching} may be one of a perfect matching of $k_{uv}$ vertices, $k_{uv}$-cycle, $k_{uv}$-connected, $k_{uv}$-edge-connected, $k_{uv}$-colorable, $k_{uv}$-edge-colorable, total $k_{uv}$-colorable, $k_{uv}$-regular, $k_{uv}$-girth, $k_{uv}$-maximum degree, $k_{uv}$-clique, $\{a,b\}_{uv}$-factor, vertex-split $k_{uv}$-connected, edge-split $k_{uv}$-connected. We call the graph $ G(\bullet) P_{ag}$ obtained by joining $F(u)$ with $F(uv)$ and joining $F(uv)$ with $F(v)$ for each edge $uv\in E(G)$ a \emph{pan-matching graph}.\qqed
\end{defn}

\begin{rem}\label{rem:333333}
If each $H_i$ in a \emph{graph set} $P_{ag}$ of graphs admits a labeling $f_i$, we can color a $(p,q)$-graph $G$ in the way: $F:V(G)\rightarrow P_{ag}$, such that each edge $u_iv_j\in E(G)$ is colored by $F(u_iv_j)=F(u_i)\bullet F(v_j)=H_i\bullet H_j=\alpha_{i,j}(f_i,f_j)$, where $\alpha_{i,j}(f_i,f_j)$ is a \emph{reversible function} with $f_j=\alpha_{i,j}(f_i)$, and $f_i=\alpha^{-1}_{i,j}(f_j)$.\paralled
\end{rem}

\subsection{Text-based strings, number-based strings}

A \emph{number-based string} $D=c_1c_2\cdots c_m$ with $c_i\in [0,9]$ has its own \emph{reciprocal number-based string} defined by $D_{-1}=c_mc_{m-1}\cdots c_2c_1$, also, we say both $D$ and $D_{-1}$ match from each other. We consider that $D$ is a \emph{public-key}, and $D_{-1}$ is a \emph{private-key} in topological authentication. For a fixed Topsnut-matrix $T_{code}$ and its reciprocal $T^{-1}_{code}=(X^{-1},E^{-1},T^{-1})^T$ with $X^{-1}=(x_q,x_{q-1},\dots,x_2,x_1)$, $E^{-1}=(e_q,e_{q-1},\dots,e_2,e_1)$ and $Y^{-1}=(y_q,y_{q-1},\dots,y_2,y_1)$, we have the following basic methods for generating vv-type/vev-type TB-paws:

\vskip 0.2cm

\textbf{Met-1.} \textbf{I-route.} $D_1(G)=x_1x_2\cdots x_qe_q e_{q-1} \cdots e_2e_1y_1 y_2\cdots y_q$
with its reciprocal $D_1(G)_{-1}$.

\textbf{Met-2.} $D_2(G)=x_q x_{q-1} \cdots x_1e_1 e_2\cdots e_{q-1}e_qy_q y_{q-1} \cdots y_1$ with its reciprocal $D_2(G)_{-1}$.

\textbf{Met-3.} \textbf{II-route.} $D_3(G)=x_1e_1y_1y_2e_2x_2x_3e_3y_3\cdots x_qe_qy_q$ with its reciprocal $D_3(G)_{-1}$.

\textbf{Met-4.} $D_4(G)=x_q e_q y_q y_{q-1}e_{q-1}x_{q-1} \cdots y_2e_2x_2x_1e_1y_1$ with its reciprocal $D_4(G)_{-1}$.

\textbf{Met-5.} \textbf{III-route.} We set $$D_5(G)=y_2y_1e_1x_1e_2y_3y_4e_3x_2\cdots x_{q-2}e_{q-1}y_qe_{q}x_{q}x_{q-1}$$ with its reciprocal $D_5(G)_{-1}$.

\textbf{Met-6.} We take $D_6(G)= y_{q-1}y_qe_{q}x_{q}e_{q-1}y_{q-2}y_{q-3}e_{q-2}x_{q-1}\cdots x_{2}e_{2}y_1e_{1}x_{1}x_{2}$ with its reciprocal $D_6(G)_{-1}$.

\textbf{Met-7.} Let a mapping $g: \{x_i,e_i,y_i:~x_i\in X,~e_i\in W,~y_i\in Y\}\rightarrow \{a_i:i\in [1,3q]\}$ be a bijection on the Topsnut-matrix $T_{code}(G)$ of $G$, so it induces a vv-type/vev-type TB-paw
\begin{equation}\label{eqa:Topsnut-matrix-vs-TB-paws}
g(G)=g^{-1}(a_{i_1})g^{-1}(a_{i_2})\cdots g^{-1}(a_{i_{3q}})
\end{equation} with its reciprocal $g(G)_{-1}$, where $a_{i_1}a_{i_2}\dots a_{i_{3q}}$ is a permutation of $a_1a_2\dots a_{3q}$. So, there are $(3q)!$ vv-type/vev-type TB-paws by (\ref{eqa:Topsnut-matrix-vs-TB-paws}), in general. Clearly, there are many random routes for inducing vv-type/vev-type TB-paws from Topsnut-matrices.

\begin{rem}\label{rem:333333}
A Topsnut-matrix $T_{code}(G)$ of a $(p,q)$-graph $G$ may has $N_{fL}(m)$ groups of disjoint fold-lines $L_{j,1},L_{j,2},\dots, L_{j,m}$~($=\{L_{j,i}\}^m_1$) for $j\in [1,N_{fL}(m)]$ and $m\in [1,M]$, where each fold-line $L_{j,i}$ has own initial point $(a_{j,i},b_{j,i})$ and terminal point $(c_{j,i},d_{j,i})$ in $xoy$-plan, and $L_{j,i}$ is internally disjoint, such that each element of $T_{code}(G)$ is on one and only one of the disjoint fold-lines $=\{L_{j,i}\}^m_1$ after we put the elements of $T_{code}(G)$ into $xOy$-plane. Notice that each fold-line $L_{j,i}$ has its initial and terminus points, so $2m\leq 3q$, and then $M=\lfloor 3q/2\rfloor $. Each group of disjoint fold-lines $\{L_{j,i}\}^m_1$ can distributes us $m!$ vv-type/vev-type TB-paws, so we have at least $N_{fL}(m)\cdot m!$ vv-type/vev-type TB-paws with $m\in [1,M]$. Thereby, the graph $G$ gives us the number $N^*(G)$ of vv-type/vev-type TB-paws in total as follows
\begin{equation}\label{eqa:random-routes}
{
\begin{split}
N^*(G)=q!\cdot 2^q\sum^{M}_{m=1}N_{fL}(m)\cdot m!
\end{split}}
\end{equation}

Each tree admits a regular odd-rainbow intersection total set-labeling based on a \emph{regular odd-rainbow set-sequence} $\{R_k\}^{q}_1$ defined as: $R_k=[1,2k-1]$ with $k\in [1,q]$, where $[1,1]=\{1\}$. Moreover, we can define a \emph{regular Fibonacci-rainbow set-sequence} $\{R_k\}^{q}_1$ by $R_1=[1,1]$, $R_2=[1,1]$, and $R_{k+1}=R_{k-1}\cup R_{k}$ with $k\in [2,q]$; or a $\tau$-term Fibonacci-rainbow set-sequence $\{\tau,R_i\}^{q}_1$ holds: $R_i=[1,a_i]$ with $a_i>1$ and $i\in [1,q]$, and $R_k=\sum ^{k-1}_{i=k-\tau}R_i$ with $k>\tau$ \cite{Ma-Wang-Wang-Yao-Theoretical-Computer-Science-2018}.\paralled
\end{rem}

\subsection{Algorithms for generating number-based strings from Topcode-matrices}

\begin{figure}[h]
\centering
\includegraphics[width=16.4cm]{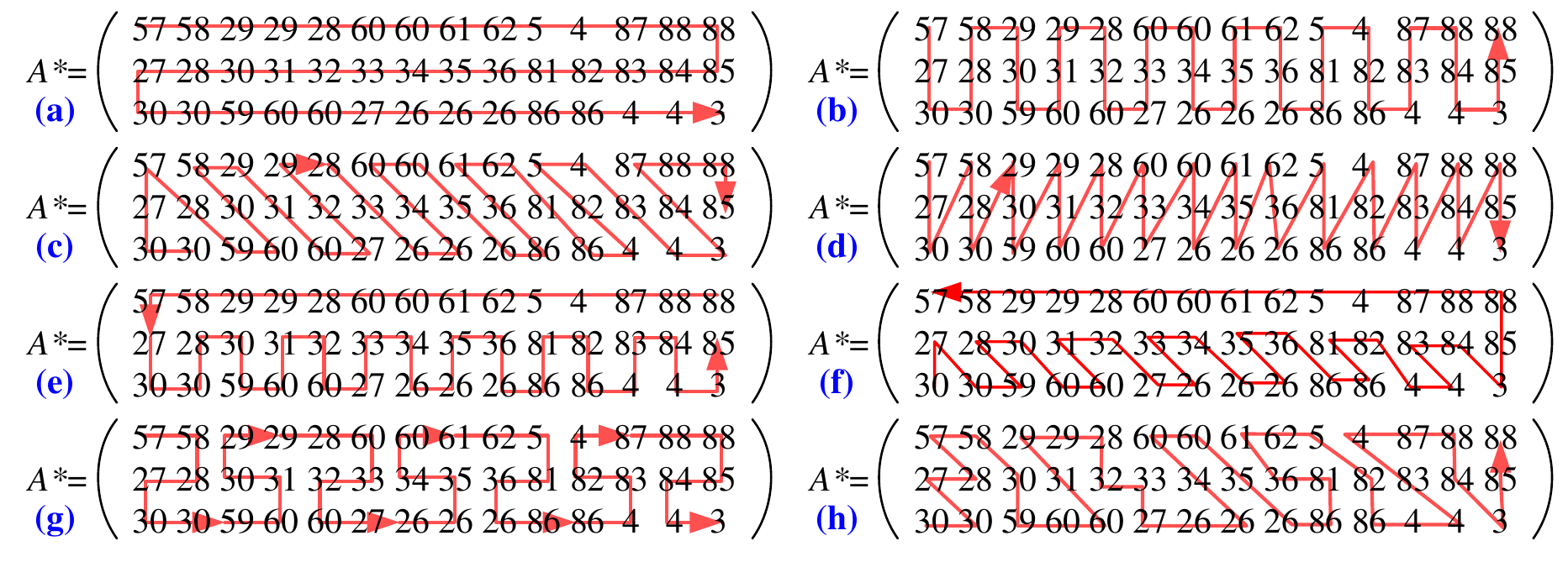}\\
\caption{\label{fig:six-lines-strings}{\small The lines in (a), (b), (c) and (d) are the basic rules for producing number-based strings (TB-paws); others are examples for showing there exist many $1$-line number-based strings, cited from \cite{Yao-Mu-Sun-Sun-Zhang-Wang-Su-Zhang-Yang-Zhao-Wang-Ma-Yao-Yang-Xie2019}.}}
\end{figure}

By the algorithms shown in Fig.\ref{fig:six-lines-strings}, we have the following number-based strings:
{\small
$${
\begin{split}
T^{(a)}_b=&5758292928606061625487888827283031323334353681828384853030596060272626268686443\\
T^{(b)}_b=&5727303028582930596031292832602733606034262635616236268681548286483878884438588\\
T^{(c)}_b=&3030275728596030582931602732292833262634606035268638616281864825483438487888885\\
T^{(d)}_b=&5727305828302930592931602832606033276034266135266236265818648386878348884488853\\
T^{(e)}_b=&8888874562616060282929585727303028305960313260273334262635362686818286483844385\\
T^{(f)}_b=&3027305928306060313227263334262635368686818244838438588888745626160602829295857\\
T^{(g)}_b=&5758282730305960313029292860333260272626353460616258136268686483824878888858443\\
T^{(h)}_b=&3030272857583059606031292928323327262634606035268681366162582864483487888438588
\end{split}}$$
}

\begin{figure}[h]
\centering
\includegraphics[width=16.4cm]{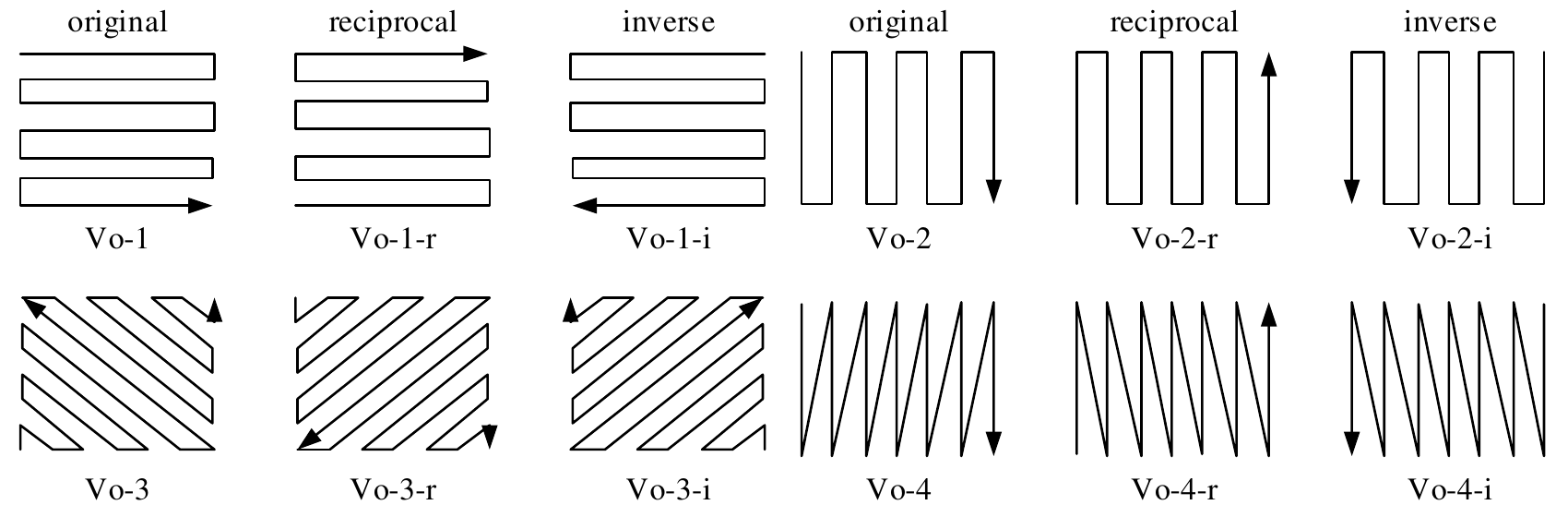}\\
\caption{\label{fig:TB-strings-from-basic-4-curves}{\small Four basic $1$-line O-$k$ with $k\in [1,4]$ used in general matrices, cited from \cite{Yao-Mu-Sun-Sun-Zhang-Wang-Su-Zhang-Yang-Zhao-Wang-Ma-Yao-Yang-Xie2019}.}}
\end{figure}

In Fig.\ref{fig:TB-strings-from-basic-4-curves}, we give each basic $1$-line O-$k$ and its reciprocal $1$-line O-$k$-r and inverse $1$-line O-$k$-i with $k\in [1,4]$. Moreover, we introduce the following efficient algorithms for writing number-based strings from $1$-line O-$k$-r and inverse $1$-line O-$k$-i ($k\in [1,4]$), cited from \cite{Yao-Mu-Sun-Sun-Zhang-Wang-Su-Zhang-Yang-Zhao-Wang-Ma-Yao-Yang-Xie2019}.

\vskip 0.2cm

\textbf{ALGORITHM-I ($1$-line-O-$1$)}

\textbf{Input:} A Topcode-matrix $A_{vev}(G)=(X,~E,~Y)^{T}_{3\times q}$

\textbf{Output:} TB-paws: $1$-line O-$1$ TB-paw $T^{O-1}_b$; $1$-line O-$1$-$r$ TB-paw $T^{O-1-r}_b$; and $1$-line O-$1$-$i$ TB-paw $T^{O-1-i}_b$ as follows:
$${
\begin{split}
T^{O-1}_b=&x_1x_2\cdots x_qe_qe_{q-1}\cdots e_2e_1y_1 y_2\cdots y_q\\
T^{O-1-r}_b=&y_1 y_2\cdots y_qe_qe_{q-1}\cdots e_2e_1x_1x_2\cdots x_q\\
T^{O-1-i}_b=&x_qx_{q-1}\cdots x_2x_1 e_1 e_2\cdots e_qy_qy_{q-1}\cdots y_2y_1
\end{split}}$$

\vskip 0.2cm

\textbf{ALGORITHM-II ($1$-line-O-$2$)}

\textbf{Input:} $A_{vev}(G)=(X,~E,~Y)^{T}_{3\times q}$

\textbf{Output:} TB-paws: $1$-line O-$2$ TB-paw $T^{O-2}_b$; $1$-line O-$2$-$r$ TB-paw $T^{O-2-r}_b$; and $1$-line O-$2$-$i$ TB-paw $T^{O-2-i}_b$ as follows:
$${
\begin{split}
T^{O-2}_b=&x_1e_1y_1y_2e_2x_2x_3e_3y_3y_4\dots x_{q-1}x_qe_qy_q~(\textrm{odd }q)\\
T^{O-2}_b=&x_1e_1y_1y_2e_2x_2x_3e_3y_3y_4\dots y_{q-1}y_qe_qx_q~(\textrm{even }q)\\
T^{O-2-r}_b=&y_1e_1x_1x_2e_2y_2y_3\dots y_{q-1}y_qe_qx_q~(\textrm{odd }q)\\
T^{O-2-r}_b=&y_1e_1x_1x_2e_2y_2y_3\dots x_{q-1}x_qe_qy_q~(\textrm{even }q)\\
T^{O-2-i}_b=&x_qe_qy_qy_{q-1}e_{q-1}x_{q-1}x_{q-2}\dots x_1e_1y_1~(\textrm{odd }q)\\
T^{O-2-i}_b=&x_qe_qy_qy_{q-1}e_{q-1}x_{q-1}x_{q-2}\dots y_1e_1x_1~(\textrm{even }q)
\end{split}}$$

\vskip 0.2cm

\textbf{ALGORITHM-III ($1$-line-O-$3$)}

\textbf{Input:} $A_{vev}(G)=(X,~E,~Y)^{T}_{3\times q}$

\textbf{Output:} TB-paws: $1$-line O-$3$ TB-paw $T^{O-3}_b$; $1$-line O-$3$-$r$ TB-paw $T^{O-3-r}_b$; and $1$-line O-$3$-$i$ TB-paw $T^{O-3-i}_b$ as follows:
$${
\begin{split}
T^{O-3}_b=&y_2y_1e_1x_1e_2y_3y_4e_3x_2x_3e_4y_5y_6\dots y_qe_{q-1}x_{q-2}x_{q-1}x_qe_q\\
T^{O-3-r}_b=&x_2x_1e_1y_1e_2x_3x_4e_3y_2y_3e_4x_5x_6\dots x_qe_{q-1}y_{q-2}y_{q-1}y_qe_q\\
T^{O-3-i}_b=&y_{q-1}y_qe_qx_qe_{q-1}y_{q-2}y_{q-3}e_{q-2}x_{q-1}x_{q-2}\dots y_1e_2x_2x_1e_1
\end{split}}$$

\vskip 0.2cm

\textbf{ALGORITHM-IV ($1$-line-O-$4$)}

\textbf{Input:} $A_{vev}(G)=(X,~E,~Y)^{T}_{3\times q}$

\textbf{Output:} TB-paws: $1$-line O-$4$ TB-paw $T^{O-4}_b$; $1$-line O-$4$-$r$ TB-paw $T^{O-4-r}_b$; and $1$-line O-$4$-$i$ TB-paw $T^{O-4-i}_b$ as follows:
$${
\begin{split}
T^{O-4}_b=&x_1e_1y_1x_2e_2y_2\dots y_{q-1}x_{q}e_qy_q\\
T^{O-4-r}_b=&y_1e_1x_1y_2e_2x_2\dots x_{q-1}y_{q}e_qx_q \\
T^{O-4-i}_b=&x_q e_q y_{q}x_{q-1}e_{q-1}y_{q-1}x_{q-2}\dots x_2e_2y_2x_1e_1 y_1
\end{split}}$$

\subsection{Real-valued colorings}

Suppose that a connected graph $G$ admits a proper total coloring $g$. We select a pair of non-negative functions $\alpha(\varepsilon)$ and $\beta(\varepsilon)$ with $\alpha(\varepsilon)\rightarrow 0$ and $\beta(\varepsilon)\rightarrow 1$ as $\varepsilon\rightarrow 0$. Thereby, we define a real-valued total coloring as
\begin{equation}\label{eqa:0-1-real-valued-total-coloring}
f_{\varepsilon}(w)=\alpha(\varepsilon)+\beta(\varepsilon)g(w),\quad w\in V(G)\cup E(G)
\end{equation}
and call $f_{\varepsilon}(w)$ defined in (\ref{eqa:0-1-real-valued-total-coloring}) a \emph{real-valued $(0,1)$-total coloring} of $G$. As we take $\beta(\varepsilon)=1-\alpha(\varepsilon)$, the equation (\ref{eqa:0-1-real-valued-total-coloring}) is changed as
\begin{equation}\label{eqa:0-1-real-valued-total-coloring-dual}
f_{\varepsilon}(w)=\alpha(\varepsilon)+[1-\alpha(\varepsilon)]g(w)=g(w)+[1-g(w)]\alpha(\varepsilon).
\end{equation}

\begin{defn}\label{defn:real-valued-condition-colorings}
\cite{Yao-Wang-Ma-Su-Wang-Sun-2020ITNEC} Let $R_S$ be a set of non-negative real numbers, and let $G$ be a $(p,q)$-graph. There are the following constraint conditions:
\begin{asparaenum}[Co-1. ]
\item \label{colo:vertex-coloring} A vertex mapping $g:V(G)\rightarrow R_S$;
\item \label{colo:edge-coloring} an edge mapping $g:E(G)\rightarrow R_S$;
\item \label{colo:total-coloring} a total mapping $g:V(G)\cup E(G)\rightarrow R_S$;
\item \label{colo:adjacent-not} $g(u)\neq g(v)$ for any edge $uv\in E(G)$;
\item \label{colo:adjacent-edge-not} $g(uv)\neq g(uw)$ for any pair of adjacent edges $uv,uw\in E(G)$; and
\item \label{colo:total-edge-vertex} $g(uv)\neq g(u)$ and $g(uv)\neq g(v)$ for any edge $uv\in E(G)$.
\end{asparaenum}
\textbf{Then $g$ is}:
\begin{asparaenum}[\textrm{Real}-1. ]
\item A \emph{real-valued coloring} if Co-\ref{colo:vertex-coloring} and Co-\ref{colo:adjacent-not} hold true.
\item A \emph{real-valued edge coloring} if Co-\ref{colo:edge-coloring} and Co-\ref{colo:adjacent-edge-not} hold true.
\item A \emph{real-valued total coloring} if Co-\ref{colo:total-coloring}, Co-\ref{colo:adjacent-not}, Co-\ref{colo:adjacent-edge-not} and Co-\ref{colo:total-edge-vertex} hold true.
\item \emph{Set-ordered} if $\max\{g(x):x\in X\}<\min \{g(y):y\in Y\}$, when $G$ is a bipartite graph with vertex set bipartition $(X,Y)$.\qqed
\end{asparaenum}
\end{defn}

\begin{defn}\label{defn:real-valued-total-cs}
\cite{Yao-Wang-Ma-Su-Wang-Sun-2020ITNEC} Let $R_S$ be a set of non-negative real numbers, and let $G$ be a $(p,q)$-graph. If there exists a real number $k>0$, for any positive real number $\varepsilon$, $G$ admits a real-valued total coloring $f_{\varepsilon}$ on $R_S$, which induces an \emph{edge-function} $C(f_{\varepsilon},uv)$ for each edge $uv\in E(G)$, such that
\begin{equation}\label{eqa:converges-edge-magic-00}
|C(f_{\varepsilon},uv)-k|<\varepsilon
\end{equation}
then we say that $G$ converges to a \emph{$C$-edge-magic constant} $k$, each $f_{\varepsilon}$ is a \emph{$C$-edge-magic real-valued total coloring}, and write this fact as
\begin{equation}\label{eqa:converges-edge-magic-11}
\lim_{\varepsilon\rightarrow 0}C(f_{\varepsilon},uv)_{uv\in E(G)}=k.
\end{equation}
\noindent \textbf{We have}:
\begin{asparaenum}[(i) ]
\item If the edge-function $C(f_{\varepsilon},uv)=f_{\varepsilon}(u)+f_{\varepsilon}(uv)+f_{\varepsilon}(v)$, then $f_{\varepsilon}$ is called an \emph{edge-magic real-valued total coloring}, and $k$ an \emph{edge-magic real constant}.

\item If the edge-function $C(f_{\varepsilon},uv)=f_{\varepsilon}(uv)+|f_{\varepsilon}(u)-f_{\varepsilon}(v)|$, then $f_{\varepsilon}$ is called a \emph{difference edge-magic real-valued total coloring}, and $k$ a \emph{difference edge-magic real constant}.

\item If the edge-function $C(f_{\varepsilon},uv)=\big ||f_{\varepsilon}(u)-f_{\varepsilon}(v)|-f_{\varepsilon}(uv)\big |$, then $f_{\varepsilon}$ is called a \emph{mixed-edge-magic real-valued total coloring}, and $k$ a \emph{mixed-edge-magic real constant}.
\item If the edge-function $C(f_{\varepsilon},uv)=|f_{\varepsilon}(u)+f_{\varepsilon}(v)-f_{\varepsilon}(uv)|$, then $f_{\varepsilon}$ is called a \emph{graceful edge-magic real-valued total coloring}, and $k$ a \emph{graceful edge-magic real constant}.\qqed
\end{asparaenum}
\end{defn}

\begin{defn}\label{defn:real-valued-vertex-coloring}
\cite{Yao-Wang-Ma-Su-Wang-Sun-2020ITNEC} Let $R_S$ be a non-negative real number set, and let $G$ be a $(p,q)$-graph with edge set $E(G)=\{u_kv_k:~k\in [1,q]\}$. If a real-valued vertex coloring $f: V(G) \rightarrow R_S$ induces $f(u_kv_k)=|f(u_k)-f(v_k)|$ for each edge $u_kv_k\in E(G)$, such that $|f(u_kv_k)-k|<\varepsilon<1/2$ (resp. $|f(u_kv_k)-(2k-1)|<\varepsilon<1/2$) for each $k\in [1,q]$, we call $f$ a \emph{graceful real-valued vertex coloring} (resp. an \emph{odd-graceful real-valued vertex coloring}).\qqed
\end{defn}

\begin{thm} \label{thm:real-induced-total-coloring-2020ITNEC}
\cite{Yao-Wang-Ma-Su-Wang-Sun-2020ITNEC} Suppose that a connected graph admits a graceful labeling (resp. an odd-graceful labeling), then it admits a \emph{graceful real-valued total coloring} (resp. an odd-graceful real-valued total coloring).\qqed
\end{thm}

We present another type of real-valued total colorings as follows:
\begin{defn}\label{defn:real-valued-vertex-coloring}
\cite{Yao-Wang-Ma-Su-Wang-Sun-2020ITNEC} A simple $(p,q)$-graph $G$ admits a real-valued vertex coloring $\alpha:V(G)\rightarrow R_S$, and $|\alpha(u)-\alpha(v)|\neq |\alpha(u)-\alpha(w)|$ for any pair of adjacent edges $uv,uw$ for each vertex $u\in V(G)$, then $\alpha$ induces a real-valued total coloring $\beta$ defined by $\beta(x)=\alpha(x)$ for $x\in V(G)$, $\beta(xy)=|\alpha(x)-\alpha(y)|$ for $xy\in E(G)$. We call $\beta$ a \emph{real-valued total coloring} of $G$. If each edge color $\beta(u_kv_k)=|\beta(u_k)-\beta(v_k)|$ for each edge $u_kv_k\in E(G)$ holds $|\beta(u_kv_k)-k|<\varepsilon<1/2$ (resp. $|\beta(u_kv_k)-(2k-1)|<\varepsilon<1/2$) for each $k\in [1,q]$, we call $\beta$ a \emph{graceful real-valued total coloring} (resp. an \emph{odd-graceful real-valued total coloring}).\qqed
\end{defn}

\begin{rem}\label{rem:333333}
Let $[a,b]^r$ be a \emph{real interval}. A real mapping $f: V(G) \rightarrow [0,q+1]^r$ for a $(p,q)$-graph $G$ such that each edge $uv$ of $G$ is colored by $f(uv)=|f(u)-f(v)|$. If the edge set $E(G)=\{u_kv_k:~k\in [1,q]\}$ holds $|f(u_kv_k)-k|<\epsilon<1/2$ for $k\in [1,q]$, we call $f$ an \emph{$\epsilon$-real-valued graceful labeling} of $G$.

We can get $\epsilon$-real-valued graceful labelings if graphs admitting graceful labelings. Suppose that a $(p,q)$-graph $G$ admits a graceful labeling $h: V(G) \rightarrow [0,q]$. We use a random number generator $g(x)$ such that $g(k)$, for each non-negative integer $k$, generates a random number falling into $k-\epsilon <g(k)<k+\epsilon$ with $0<\epsilon <1/2$. So, $G$ admits an $\epsilon$-real-valued graceful labeling $I$ defined by $I(u)=g(h(u))$ for each $u\in V(G)$. Furthermore, we define the \emph{dual $\epsilon$-real-valued graceful labeling} $I^*$ of $I$ by $I^*(u)=2f(u)-I(u)$ for each $u\in V(G)$. Such doing can be realized on many graph labelings introduced in \cite{Gallian2020}.

By the way, we define $f: V(G) \rightarrow [1,p+1]^r$ (or $f: E(G) \rightarrow [1,p+1]^r$, or $f: V(G)\cup E(G) \rightarrow [1,p+1]^r$) for a $(p,q)$-graph $G$. If there is a positive integer $m$ such that each $x\in V(G)$ holds $|f(x)-k_x|<\epsilon <1/2$ for some integer $k_x\in [1,m]$, and $k_x\neq k_y$ for any pair of vertices $x,y$ of $G$, we call $f$ to be an \emph{$(m,\epsilon)$-real-valued (edge-, total-)coloring}. Define the smallest integer $\min \{m:~f\textrm{ is a $(m,\epsilon)$-real-valued (edge-, total-)coloring}\}$ as $\lambda(G)$, and call it the \emph{real chromatic number} of $G$. Moreover, the \emph{dual $(m,\epsilon)$-real-valued (edge-, total-)coloring} $f^*$ of $f$ is defined by $f^*(u)=2k_x-f(x)$ for each $x\in V(G)$.

Clearly, it is impossible to estimate the spaces made by the above real-valued labelings/colorings, since these labelings/colorings are related with real numbers. Notice that some real-valued labelings have been investigated. For example, Vietri \cite{A-Vietri-Ars2011} generalizes the notion of a graceful labeling by allowing the vertex colors of a graph with $q$ edges to be real numbers in the interval $[0,q]^r$. For a simple graph $G(V,E)$ he defines an injective map $\gamma$ from $V$ to $[0,q]$ to be a real-graceful labeling of $G$ provided that
\begin{equation}\label{eqa:Vietri-formula}
\sum_{uv\in E(G)} \big [2^{\gamma(u)-\gamma(v)}+2^{\gamma(v)-\gamma(u)}\big ]=2^{q+1}-2^{-q}-1,
\end{equation}
where the sum is taken over all edges $uv$ of $G$. In the case that the colors are integers, he shows that a \emph{real-graceful labeling} is equivalent to a graceful labeling.\paralled
\end{rem}

\subsection{Algebraic operation of real-valued Topcode-matrices}

An algebraic operation on Topcode-matrices is intoduced in \cite{Yao-Wang-Ma-Su-Wang-Sun-2020ITNEC}. In the Topcode-matrix $I_{code}=(X, E, Y)^{T}$ with $x_i=1$, $e_i=1$ and $y_i=1$ for $i\in [1,q]$ is called the \emph{identity Topcode-matrix}. For two Topcode-matrices $T^j_{code}=(X^j, E^j, Y^j)^{T}$ with $j=1,2$, where
$$X^j=(x^j_1, x^j_2, \dots , x^j_q),~E^j=(e^j_1, e^j_2, \dots , e^j_q),~Y^j=(y^j_1, y^j_2, \dots , y^j_q),$$ the \emph{coefficient multiplication} of a function $f(x)$ and a Topcode-matrix $T^j_{code}$ is defined by
$${
\begin{split}
f(x)\cdot T^j_{code}=f(x)\cdot(X^j, E^j, Y^j)^{T}=(f(x)\cdot X^j, f(x)\cdot E^j, f(x)\cdot Y^j)^{T}
\end{split}}
$$
where $f(x)\cdot X^j=(f(x)\cdot x^j_1, f(x)\cdot x^j_2, \dots , f(x)\cdot x^j_q)$, $f(x)\cdot E^j=(f(x)\cdot e^j_1, f(x)\cdot e^j_2, \dots , f(x)\cdot e^j_q)$ and $f(x)\cdot Y^j=(f(x)\cdot y^j_1, f(x)\cdot y^j_2, \dots , f(x)\cdot y^j_q)$. And the \emph{addition} between two Topcode-matrices $T^1_{code}$ and $T^2_{code}$ is denoted as
$$T^1_{code}+T^2_{code}=(X^1+X^2, E^1+E^2, Y^1+Y^2)^{T},$$
where $X^1+X^2=(x^1_1+x^2_1, x^1_2+x^2_2, \cdots , x^1_q+x^2_q)$, $E^1+E^2=(e^1_1+e^2_1, e^1_2+e^2_2, \cdots , e^1_q+e^2_q)$ and $Y^1+Y^2=(y^1_1+y^2_1, y^1_2+y^2_2, \cdots , y^1_q+y^2_q)$. We have a \emph{real-valued Topcode-matrix} $R_{code}$ defined as:
$R_{code}=\alpha(\varepsilon)T^1_{code}+\beta(\varepsilon)T^2_{code}$
and another real-valued Topcode-matrix
\begin{equation}\label{eqa:real-valued-topcode-matrix}
R_{code}(f_\varepsilon,G)=\alpha(\varepsilon)I_{code}+\beta(\varepsilon)T_{code}(G)
\end{equation}corresponding (\ref{eqa:0-1-real-valued-total-coloring}), where $I_{code}$ is the identity Topcode-matrix, and $T_{code}(G)$ is a Topcode-matrix of $G$.

Clearly, the number-based strings induced by the real-valued Topcode-matrix $R_{code}(f_\varepsilon,G)$ are completive than that induced by a Topcode-matrix of $G$ admitting a $W$-type coloring or a $W$-type labeling, and have huge numbers, since two functions $\alpha(\varepsilon)$ and $\beta(\varepsilon)$ are real and various. By (\ref{eqa:integer-topcode-matrix}), we get the following number-based strings

\begin{equation}\label{eqa:text-based-1}
{
\begin{split}
&D_1=10700220131197531111057111313,~D_2=10111103705577029111311201313\\
&D_3=11110531075711700913211131320
\end{split}}
\end{equation}

The Topcode-matrix $T_{code}(G)$ shown in (\ref{eqa:integer-topcode-matrix}) can provide us with a total of $(21)!$ number-based strings like $D_1,D_2,D_3$ shown in (\ref{eqa:text-based-1}).

\begin{equation}\label{eqa:integer-topcode-matrix}
\centering
{
\begin{split}
T_{code}(G)&= \left(
\begin{array}{ccccccc}
10 & 7 & 0 & 0&2&2&0\\
1 & 3 & 5 & 7&9&11&13\\
11 &10 & 5&7&11&13&13
\end{array}
\right)
\end{split}}
\end{equation}

For example, after taking $\alpha(\varepsilon)=0.32$ and $\beta(\varepsilon)=1-0.32=0.68$, we get a real-valued Topcode-matrix $$R_{code}(f_\varepsilon,G)=0.32\cdot I_{code}+0.68\cdot T_{code}(G)$$ about the Topsnut-matrix $T_{code}(G)$ shown in (\ref{eqa:integer-topcode-matrix}) and its own Topcode-matrix $T_{code}(G)$, where $R_{code}(f_\varepsilon,G)$ is
\begin{equation}\label{eqa:real-valued-Topcode-matrix}
\centering
{
\begin{split}
R_{code}(f_\varepsilon,G)=\left(
\begin{array}{ccccccc}
7.12 & 5.08 & 0.32 & 0.32&1.68&1.68&0.32\\
1 & 2.36 & 3.72 & 5.08&6.44&7.8&9.16\\
7.8 &7.12 &3.72&5.08&7.8&9.16&9.16
\end{array}
\right)
\end{split}}
\end{equation}

We apply $R_{code}(f_\varepsilon,G)$ to induce the following number-based strings:
$${
\begin{split}
D^r_1=7125080320321681680329167864450837223617871237250878916916\\
D^r_2=7817123722367125083725087850803203264491691678168168916032.
\end{split}}$$
Obviously, $D^r_1$ and $D^r_1$ are complex than $D_1,D_2,D_3$ shown in (\ref{eqa:text-based-1}). We have the following relationships between a Topcode-matrix $T_{code}(G)$ and a real-valued Topcode-matrix $R_{code}(f_\varepsilon,G)$:

(1) $e_i=|x_i-y_i|$ in a Topcode-matrix $T_{code}(G)$ of a $(p,q)$-graph $G$ corresponds $\alpha(\varepsilon)+\beta(\varepsilon)|x_i-y_i|$ of the real-valued Topcode-matrix $R_{code}(f_\varepsilon,G)$;

(2) $x_i+e_i+y_i=k$ in $T_{code}(G)$ corresponds $3\alpha(\varepsilon)+\beta(\varepsilon)\cdot k$ of $R_{code}(f_\varepsilon,G)$;

(3) $e_i+|x_i-y_i|=k$ in $T_{code}(G)$ corresponds $\alpha(\varepsilon)+\beta(\varepsilon)\cdot k$ of $R_{code}(f_\varepsilon,G)$;

(4) $|x_i+y_i-e_i|=k$ in $T_{code}(G)$ corresponds $\alpha(\varepsilon)+\beta(\varepsilon)\cdot k$ of $R_{code}(f_\varepsilon,G)$ if $e_i-(x_i+y_i)\geq 0$, otherwise $|x_i+y_i-e_i|=k$ corresponds $|\beta(\varepsilon)\cdot k-\alpha(\varepsilon)|$;

(5) $\big ||x_i-y_i|-e_i\big |=k$ in $T_{code}(G)$ corresponds $\alpha(\varepsilon)+\beta(\varepsilon)\cdot k$ of $R_{code}(f_\varepsilon,G)$ if $|x_i+y_i|-e_i< 0$, otherwise $\big ||x_i-y_i|-e_i\big |=k$ corresponds $|\beta(\varepsilon)\cdot k-\alpha(\varepsilon)|$.

\subsection{Colorings and labelings in graph homomorphisms}

\subsubsection{Colorings and labelings based on graph homomorphisms}

\begin{defn}\label{defn:definition-graph-homomorphism}
\cite{Bondy-2008} A \emph{graph homomorphism} $G\rightarrow H$ from a graph $G$ into another graph $H$ is a mapping $f: V(G) \rightarrow V(H)$ such that $f(u)f(v)\in E(H)$ for each edge $uv \in E(G)$.\qqed
\end{defn}

\begin{defn}\label{defn:faithful}
\cite{Gena-Hahn-Claude-Tardif-1997} A graph homomorphism $\varphi: G \rightarrow H$ is called \emph{faithful} if $\varphi(G)$ is an induced subgraph of $H$, and called \emph{full} if $uv\in E(G)$ if and only if $\varphi(u)\varphi(v) \in E(H)$.\qqed
\end{defn}

\begin{thm}\label{thm:bijective-graph-homomorphism}
\cite{Gena-Hahn-Claude-Tardif-1997} A faithful bijective graph homomorphism is an isomorphism, that is $G \cong H$.
\end{thm}

\begin{thm}\label{thm:sequence-graph-homomorphisms}
\cite{Bing-Yao-Hongyu-Wang-arXiv-2020-homomorphisms} There are infinite graphs $G^*_n$ forming a sequence $\{G^*_n\}^{\infty}_{n=1}$, such that $G^*_n\rightarrow G^*_{n-1}$ is really a graph homomorphism for $n\geq 1$.
\end{thm}

\begin{defn}\label{defn:gracefully-graph-homomorphism}
\cite{Bing-Yao-Hongyu-Wang-arXiv-2020-homomorphisms, Bing-Yao-Hongyu-Wang-graph-homomorphisms-2020} Let $G\rightarrow H$ be a graph homomorphism from a $(p,q)$-graph $G$ to another $(p\,',q\,')$-graph $H$ based on a mapping $\alpha: V(G) \rightarrow V(H)$ such that $\alpha(u)\alpha(v)\in E(H)$ for each edge $uv \in E(G)$. The graph $G$ admits a total coloring $f$, the graph $H$ admits a total coloring $g$, so $G\rightarrow H$ is a \emph{totally-colored graph homomorphism}. Write $f(E(G))=\{f(uv):uv \in E(G)\}$, $g(E(H))=\{g(\alpha(u)\alpha(v)):\alpha(u)\alpha(v)\in E(H)\}$. There are constraints as follows:
\begin{asparaenum}[(\textrm{C}-1) ]
\item \label{bipartite} $V(G)=X\cup Y$, each edge $uv \in E(G)$ holds $u\in X$ and $v\in Y$ true. $V(H)=W\cup Z$, each edge $\alpha(u)\alpha(v)\in E(G)$ holds $\alpha(u)\in W$ and $\alpha(v)\in Z$ true;
\item \label{edge-difference} $f(uv)=|f(u)-f(v)|$ for each $uv \in E(G)$, and $g(\alpha(u)\alpha(v))=|g(\alpha(u))-g(\alpha(v))|$ for each $\alpha(u)\alpha(v)\in E(H)$;
\item \label{edge-homomorphism} $f(uv)=g(\alpha(u)\alpha(v))$ for each $uv \in E(G)$;
\item \label{vertex-color-set} $f(x)\in [1,q+1]$ for $x\in V(G)$ and $g(y)\in [1,q\,'+1]$ with $y\in V(H)$;
\item \label{odd-vertex-color-set} $f(x)\in [1,2q+2]$ for $x\in V(G)$ and $g(y)\in [1,2q\,'+2]$ with $y\in V(H)$;
\item \label{grace-color-set} $[1,q]=f(E(G))=g(E(H))=[1,q\,']$;
\item \label{odd-grace-color-set} $[1,2q-1]^o=f(E(G))=g(E(H))=[1,2q\,'-1]^o$; and
\item \label{set-ordered} $\max f(X)<\min f(Y)$ and $\max g(W)<\min g(Z)$.
\end{asparaenum}
\noindent \textbf{We say $G\rightarrow H$ to be}:
\begin{asparaenum}[(i) ]
\item A \emph{bipartite graph homomorphism} if (C-\ref{bipartite}) holds true.
\item A \emph{graceful graph homomorphism} if (C-\ref{edge-difference}), (C-\ref{edge-homomorphism}), (C-\ref{vertex-color-set}) and (C-\ref{grace-color-set}) hold true.
\item A \emph{set-ordered graceful graph homomorphism} if (C-\ref{bipartite}), (C-\ref{edge-difference}), (C-\ref{edge-homomorphism}), (C-\ref{grace-color-set}), (C-\ref{vertex-color-set}) and (C-\ref{set-ordered}) hold true.
\item An \emph{odd-graceful graph homomorphism} if (C-\ref{bipartite}), (C-\ref{edge-difference}), (C-\ref{edge-homomorphism}), (C-\ref{odd-vertex-color-set}) and (C-\ref{odd-grace-color-set}) hold true.
\item A \emph{set-ordered odd-graceful graph homomorphism} if (C-\ref{bipartite}), (C-\ref{edge-difference}), (C-\ref{edge-homomorphism}), (C-\ref{odd-vertex-color-set}), (C-\ref{odd-grace-color-set}) and (C-\ref{set-ordered}) hold true.\qqed
\end{asparaenum}
\end{defn}

\begin{rem} \label{rem:gracefully-graph-homomorphism}
\cite{Bing-Yao-Hongyu-Wang-arXiv-2020-homomorphisms}
\begin{asparaenum}[(i) ]
\item A graceful graph homomorphism $G\rightarrow H$ holds $|f(E(G))|=|g(E(H))|$ with $q=q\,'$ and $|f(V(G))|\geq |g(V(H))|$, in general. In Definition \ref{defn:gracefully-graph-homomorphism}, $G$ admits a \emph{set-ordered gracefully total coloring} $f$ and $H$ admits a \emph{set-ordered gracefully total coloring} $g$ in topological coding.
\item We call the totally-colored graph homomorphisms defined in Definition \ref{defn:gracefully-graph-homomorphism} as \emph{$W$-type totally-colored graph homomorphisms}, where a ``$W$-type totally-colored graph homomorphism'' is one of the totally-colored graph homomorphisms defined in Definition \ref{defn:gracefully-graph-homomorphism}.
\item We say that the graph $G$ admits a $W$-type totally-colored graph homomorphism to $H$ in a $W$-type totally-colored graph homomorphism $G\rightarrow H$ of Definition \ref{defn:gracefully-graph-homomorphism}.
\item If $T\rightarrow H$ is a $W$-type totally-colored graph homomorphism, and so is $H\rightarrow T$, we say $T$ and $H$ are homomorphically equivalent from each other, denoted as $T\leftrightarrow H$.
\item If two graphs $G$ admitting a coloring $f$ and $H$ admitting a coloring $g$ hold $f(x)=g(\varphi(x))$ and $G\cong H$ for a bijection $\varphi:V(G)\rightarrow V(H)$, we write this case by $G=H$.\paralled
\end{asparaenum}
\end{rem}

\begin{thm}\label{thm:parti-gracecular-bijective}
\cite{Bing-Yao-Hongyu-Wang-arXiv-2020-homomorphisms} If a (set-ordered) graceful graph homomorphism $\varphi: G \rightarrow H$ defined in Definition \ref{defn:gracefully-graph-homomorphism} holds $f(V(G))=g(V(H))$ and $f(E(G))=g(E(H))$ true, then $G=H$.
\end{thm}

\begin{thm}\label{thm:set-ordered-gracefully-bijective}
\cite{Bing-Yao-Hongyu-Wang-arXiv-2020-homomorphisms} If a (set-ordered) graceful graph homomorphism $\varphi: G \rightarrow H$ is faithful bijective, then $G=H$.
\end{thm}

\subsubsection{Colored graph homomorphisms}

\begin{defn}\label{defn:v-colored-graph-homomorphism}
\cite{Bing-Yao-Hongyu-Wang-arXiv-2020-homomorphisms} Suppose that a $(p,q)$-graph $G$ admits a $W$-type coloring (resp. labeling) $f$, and a $(p\,',q\,')$-graph $H$ admits a $W$-type coloring (resp. labeling) $g$. There exists a mapping $\varphi:V(G)\rightarrow V(H)$ such that each edge $uv\in E(G)$ holds $\varphi(u)\varphi(v)\in E(H)$ true, and there are two positive constants $k_{\textrm{vertex}}$ and $k_{\textrm{edge}}$, we have:

(i) If $f(x)+g(\varphi(x))=k_{\textrm{vertex}}$ for each vertex $x\in V(G)$ and $\varphi(x)\in V(H)$, we say that $G$ admits a \emph{$W$-type v-colored-graph homomorphism} to $H$, denoted as $G_f\rightarrow ^c_v H_g$.

(ii) If $f(uv)+g(\varphi(u)\varphi(v))=k_{\textrm{edge}}$ for each edge $uv\in E(G)$ and $\varphi(u)\varphi(v)\in E(H)$, we say that $G$ admits a \emph{$W$-type e-colored-graph homomorphism} to $H$, denoted as $G_f\rightarrow ^c_e H_g$.

(iii) If $f(x)+g(\varphi(x))=k_{\textrm{vertex}}$ for each vertex $x\in V(G)$ and $\varphi(x)\in V(H)$ and $f(uv)+g(\varphi(u)\varphi(v))=k_{\textrm{edge}}$ for each edge $uv\in E(G)$ and $\varphi(u)\varphi(v)\in E(H)$, we say that $G$ admits a \emph{$W$-type totally colored-graph homomorphism} to $H$, denoted as $G_f\rightarrow ^c_t H_g$.

Moreover, we call $(f,g)$ a \emph{matching of $W$-type colorings (resp. labelings)}.\qqed
\end{defn}

\begin{rem}\label{rem:333333}
If $G\cong H$ and $f\neq g$ in Definition \ref{defn:v-colored-graph-homomorphism}, $(f,g)$ is called a \emph{matching of $W$-type image-colorings} defined in \cite{Jing-Su-Hongyu-Wang-Bing-Yao-2020-image-labelings}. Since each graph $G$ admits a total coloring $h$, then we have the dual total coloring $h\,'$ of the total coloring $h$ defined as: $h\,'(w)=(\max h+\min h)-h(w)$ for $w\in V(G)\cup E(G)$, where $\max h=\max \{h(w):w\in V(G)\cup E(G)\}$ and $\min h=\min \{h(w):w\in V(G)\cup E(G)\}$. Clearly, $h(w)+h\,'(w)=\max h+\min h$ to be a constant. So $G$ admits a \emph{$W$-type totally colored-graph homomorphism} to itself, $G_h\rightarrow ^c_t G_{h\,'}$.\paralled
\end{rem}

\begin{figure}[h]
\centering
\includegraphics[width=16cm]{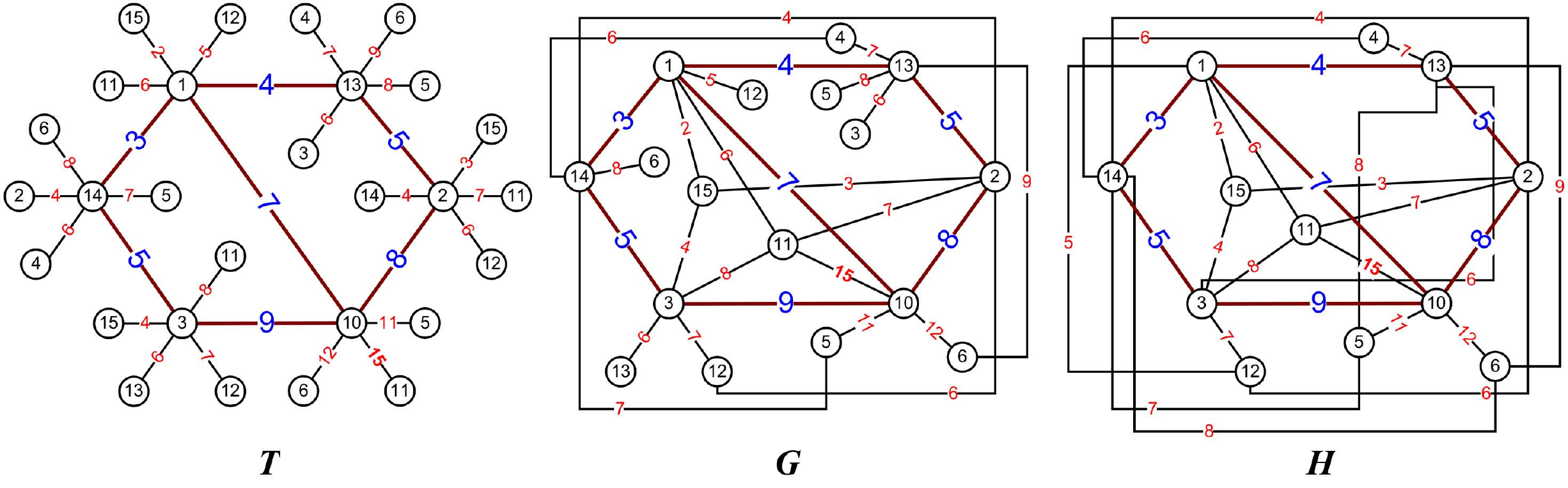}\\
\caption{\label{fig:colored-graph-homomorphism}{\small $T$ admits an edge-difference totally colored-graph homomorphism to $G$, and $G$ admits an edge-difference totally colored-graph homomorphism to $H$, cited from \cite{Bing-Yao-2020arXiv}.}}
\end{figure}

In Definition \ref{defn:v-colored-graph-homomorphism} and Fig.\ref{fig:colored-graph-homomorphism}, we can see two edge-difference totally colored-graph homomorphisms $T_h\rightarrow ^c_t G_f$ and $G_f\rightarrow ^c_t H_g$, where $T$, $G$ and $H$ admit three edge-difference proper total colorings $h$, $f$ and $g$ respectively, and moreover $h(uv)+|h(u)-h(v)|=16$ for each edge $uv\in E(T)$, $f(xy)+|f(x)-f(y)|=16$ for each edge $xy\in E(G)$, and $g(wz)+|g(w)-g(z)|=16$ for each edge $wz\in E(H)$. Notices that three numbers of vertices of three colored-graphs $T,G$ and $H$ satisfy $|V(T)|>|V(G)|>|V(H)|$, and their edge numbers hold $|E(T)|=|E(G)|=|E(H)|$ true.

\subsection{Set-ordered proper total colorings}

\begin{defn} \label{defn:66-optimal-e-v-set-ordered-total-colorings}
$^*$ Let $f:V(G)\cup E(G)\rightarrow [1,M]$ be a proper total coloring of a $(p,q)$-graph $G$.

(1) If $f(V(G))=[1,s]$ and $f(E(G))\subseteq [1,M]$, we call $f$ a \emph{v-set-ordered proper total coloring}, and moreover if $s$ is just the vertex chromatic number $\chi(G)$ of $G$, namely, $s=\chi(G)$ (resp. $|f(V(G))|=\chi(G)$), we call $f$ an \emph{optimal v-set-ordered proper total coloring}.

(2) If $f(E(G))=[1,b]$ and $f(V(G))\subseteq [1,M]$, we say that $f$ an \emph{e-set-ordered proper total coloring}, and moreover if $b$ is just the chromatic index $\chi\,'(G)$ of $G$ (resp. $|f(E(G))|=\chi\,'(G)$), also, $b=\chi\,'(G)$, we call $f$ an \emph{optimal e-set-ordered proper total coloring}.\qqed
\end{defn}

\begin{figure}[h]
\centering
\includegraphics[width=16.4cm]{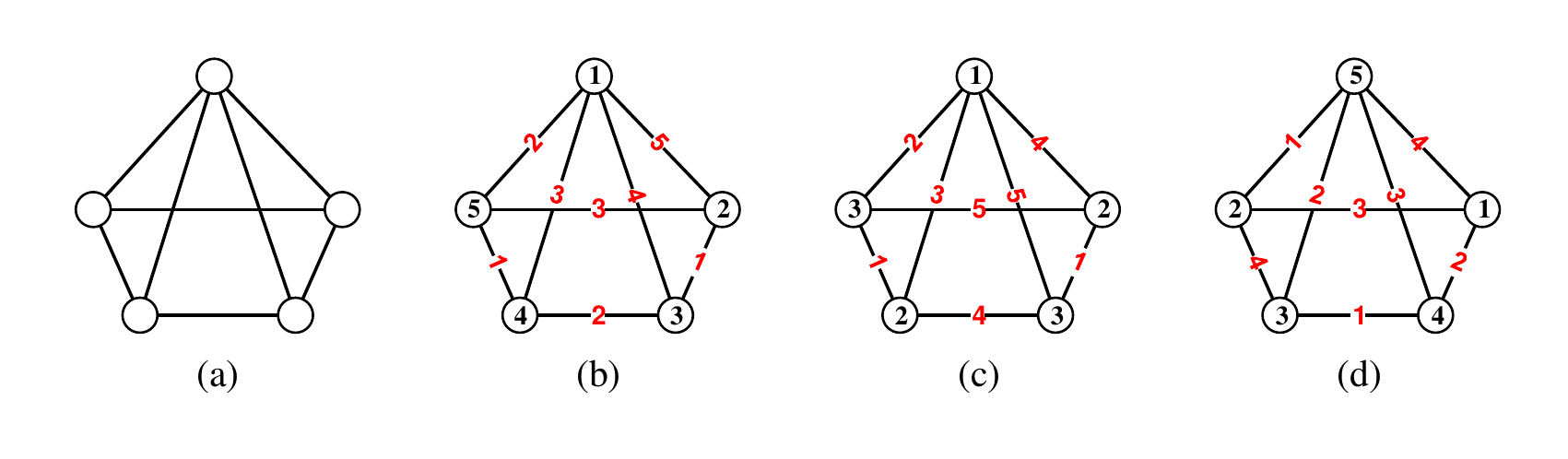}
\caption{\label{fig:optimal-total-colorings}{\small (a) A $(5,8)$-graph $G$; (b) a proper total coloring $f$ of $G$ holding $\chi\,''(G)=5$; (c) an optimal v-set-ordered proper total coloring $g$ of $G$ holding $g(V(G))=[1,3]$ and $\chi(G)=3$; (d) an optimal e-set-ordered proper total coloring $h$ of $G$ holding $h(E(G))=[1,4]$ and $\chi\,'(G)=4$ true.}}
\end{figure}

\begin{thm}\label{thm:66-optimal-e-v-set-ordered-total-colorings}
$^*$ Each $(p,q)$-graph admits an \emph{optimal v-set-ordered proper total coloring} and an \emph{optimal e-set-ordered proper total coloring}.
\end{thm}

\subsection{Directed labelings}

We call a one-way directed edge as an \emph{arc} $\overrightarrow{uv}$, where $v$ with a narrow is the \emph{head} of the arc $\overrightarrow{uv}$ and $u$ is the \emph{tail} of the arc $\overrightarrow{uv}$. A $(p,q)$-graph having some arcs and the remainder being proper edges is called a \emph{half-directed $(p,q)$-graph}, denoted as $G^{\rightarrow}$; a $(p,q)$-graph having only arcs is called a \emph{directed $(p,q)$-graph} or a \emph{$(p,q)$-digraph}, and denoted as $\overrightarrow{G}$. Particularly, the set of all arcs of a digraph $\overrightarrow{G}$ is denoted as $A(\overrightarrow{G})$.

Here, replacing arcs by proper edges in a directed graph, or a half-directed graph produces a graph, we call this graph the \emph{underlying graph} of the directed graph, or the half-directed graph.

\begin{defn}\label{defn:half-directed-graceful}
\cite{Yao-Mu-Sun-Sun-Zhang-Wang-Su-Zhang-Yang-Zhao-Wang-Ma-Yao-Yang-Xie2019} Let $f:V(G^{\rightarrow}) \rightarrow [0,q]$ (resp. $[0,2q-1]$) be a labeling of a half-directed $(p,q)$-graph $G^{\rightarrow}$ with its connected underlying graph, we define $f(xy)=|f(x)-f(y)|$ for proper edge $xy$ of $G^{\rightarrow}$ and $f(\overrightarrow{uv})=f(u)-f(v)$ for each arc $\overrightarrow{uv}$ of $G^{\rightarrow}$. If $\{f(xy),f(\overrightarrow{uv}):~uv, \overrightarrow{uv}\in E(G)\cup A(G)\}=[1,q]$ (resp. $[1,2q-1]^o$), we call $f$ a \emph{half-directed graceful labeling} (resp. \emph{half-directed odd-graceful labeling}) of $G^{\rightarrow}$.\qqed
\end{defn}

\begin{defn}\label{defn:directed-graceful}
\cite{Yao-Mu-Sun-Sun-Zhang-Wang-Su-Zhang-Yang-Zhao-Wang-Ma-Yao-Yang-Xie2019} Let $g:V(\overrightarrow{G}) \rightarrow [0,q]$ (resp. $[0,2q-1]$) be a labeling of a $(p,q)$-digraph $\overrightarrow{G}$ with its connected underlying graph, and define $g(\overrightarrow{uv})=g(u)-g(v)$ for each arc $\overrightarrow{uv}$ of the digraph $\overrightarrow{G}$. If $\{|g(\overrightarrow{uv})|:~\overrightarrow{uv}\in A(\overrightarrow{G})\}=[1,q]$ (resp. $[1,2q-1]^o$), we call $g$ a \emph{directed graceful labeling} (resp. \emph{directed odd-graceful labeling}) of the digraph $\overrightarrow{G}$. Moreover, if $g(\overrightarrow{uv})>0$ (resp. $g(\overrightarrow{uv})<0$) for each arc $\overrightarrow{uv}\in A(\overrightarrow{G})$, we say $g$ to be \emph{uniform}.\qqed
\end{defn}

\begin{equation}\label{eqa:directed-1}
\centering
A(G^{\rightarrow})=\left(
\begin{array}{rrrrrrrrr}
6 & 6 & 7 & 7 & 2 & 9 & 10 & 10 & 10\\
1 & 2 & 4 & 5 & -6 & 7 & 8 & 9 & 10\\
5 & 4 & 3 & 2 & 8 & 2 & 2 & 1 & 0
\end{array}
\right)^{+}_{-}
\end{equation}
\begin{equation}\label{eqa:directed-2}
\centering
A(\overrightarrow{G})=\left(
\begin{array}{rrrrrrrrr}
6 & 4 & 7 & 2 & 2 & 9 & 10 & 1 & 10\\
1 & -2 & 4 & -5 & -6 & 7 & 8 & -9 & 10\\
5 & 6 & 3 & 7 & 8 & 2 & 2 & 10 & 0
\end{array}
\right)^{+}_{-}
\end{equation}

A half-directed Topcode-matrix $A(G^{\rightarrow})$ is shown in (\ref{eqa:directed-1}), and a directed Topcode-matrix $A(\overrightarrow{G})$ is shown in (\ref{eqa:directed-2}) (refer to defined in Definition \ref{defn:directed-Topcode-matrix}). We can derive TB-paws from half-directed Topcode-matrices and directed Topcode-matrices by replacing minus sign ``$-$'' with a letter ``$x$'' (resp. other letters and signs). For example, the half-directed Topcode-matrix $A(G^{\rightarrow})$ shown in (\ref{eqa:directed-1}) distributes us a TB-paw
$$D(A(G^{\rightarrow}))=66772910101010987x65421543282.$$

\begin{rem}\label{rem:ABC-conjecture}
There is no equivalent relation between uniformly directed graceful labelings and set-ordered graceful labelings. In general, directed labelings (resp. colorings) of directed graphs are complex than labelings (resp. colorings) of undirected graphs.\paralled
\end{rem}

\begin{prop}\label{thm:GTC-conjecture}
\cite{Yao-Mu-Sun-Sun-Zhang-Wang-Su-Zhang-Yang-Zhao-Wang-Ma-Yao-Yang-Xie2019} Let $\overrightarrow{O}(T)$ be a set of directed trees having the same underlying tree $T$. If Graceful Tree Conjecture (resp. Odd-graceful Tree Conjecture) holds true, then $\overrightarrow{O}(T)$ contains at least two directed trees admitting \emph{uniformly directed (odd-)graceful labelings}, and each directed tree of $\overrightarrow{O}(T)$ admits a \emph{directed (odd-)graceful labeling}.
\end{prop}

\begin{rem}\label{rem:GTC-conjecture}
Since there are $2^{p-1}$ directed trees in $\overrightarrow{O}(T)$ with the same underlying tree $T$ of $p$ vertices, so we can make $M$ number-based strings by directed Topcode-matrices on $\overrightarrow{O}(T)$, where $M=2^{p-1}\cdot (3p-3)!$.

If the underlying graph of a half-directed $(p,q)$-graph $G^{\rightarrow}$, or a $(p,q)$-digraph $\overrightarrow{G}$ is disconnected, we can add a set $E^*$ of arcs or edges to join some vertices of $G^{\rightarrow}$ or $\overrightarrow{G}$ such that the resultant half-directed $G^{\rightarrow}+E^*$ or the resultant digraph $\overrightarrow{G}+E^*$ has a connected underlying graph, we say $G^{\rightarrow}+E^*$ to be a \emph{connected half-directed graph}, or $\overrightarrow{G}+E^*$ a \emph{connected digraph}, not based on directed paths. \paralled
\end{rem}

We define a flawed half-directed graceful labeling and a flawed directed graceful labeling on half-directed graphs and digraphs in the following Definition \ref{defn:flawed-half-directed-graceful} and Definition \ref{defn:flawed-directed-graceful}, respectively.

\begin{defn}\label{defn:flawed-half-directed-graceful}
\cite{Yao-Mu-Sun-Sun-Zhang-Wang-Su-Zhang-Yang-Zhao-Wang-Ma-Yao-Yang-Xie2019} Suppose that a half-directed $(p,q)$-graph $G^{\rightarrow}$ has its disconnected underlying graph, and $G^{\rightarrow}+E^*$ is a connected half-directed $(p,q+q\,')$-graph, where $q\,'=|E^*|$. Let $f:V(G^{\rightarrow}+E^*) \rightarrow [0,q+q\,']$ (resp. $[0,2(q+q\,')-1]$) be a half-directed graceful labeling (resp. a half-directed odd-graceful labeling) $f$ of $G^{\rightarrow}+E^*$, then $f$ is called a \emph{flawed half-directed graceful labeling} (resp. \emph{flawed half-directed odd-graceful labeling}) of the half-directed $(p,q)$-graph $G^{\rightarrow}$.\qqed
\end{defn}

\begin{defn}\label{defn:flawed-directed-graceful}
\cite{Yao-Mu-Sun-Sun-Zhang-Wang-Su-Zhang-Yang-Zhao-Wang-Ma-Yao-Yang-Xie2019} Suppose that the underlying graph of a $(p,q)$-digraph $\overrightarrow{G}$ is disconnected, and $\overrightarrow{G}+E^*$ is a connected directed $(p,q+q\,')$-graph, where $q\,'=|E^*|$. Let $f:V(\overrightarrow{G}+E^*) \rightarrow [0,q+q\,']$ (resp. $[0,2(q+q\,')-1]$) be a directed graceful labeling (resp. a directed odd-graceful labeling) $f$ of $\overrightarrow{G}+E^*$, then $f$ is called a \emph{flawed directed graceful labeling} (resp. \emph{flawed directed odd-graceful labeling}) of the $(p,q)$-digraph $\overrightarrow{G}$.\qqed
\end{defn}

\begin{rem}\label{rem:flawed-directed-graceful}
We divide the arc color set $f(A(\overrightarrow{G}))$ into $f^+(A(\overrightarrow{G}))=\{f(\overrightarrow{uv})>0:~\overrightarrow{uv}\in A(\overrightarrow{G})\}$ and $f^-(A(\overrightarrow{G}))=\{f(\overrightarrow{uv})<0:~\overrightarrow{uv}\in A(\overrightarrow{G})\}$, namely, $f(A(\overrightarrow{G}))=f^+(A(\overrightarrow{G}))\cup f^-(A(\overrightarrow{G}))$. We say $f$ to be a \emph{uniformly (flawed) directed (odd-)graceful labeling} if one of $f^+(A(\overrightarrow{G}))=\emptyset $ and $f^-(A(\overrightarrow{G}))=\emptyset $ holds true. Determining the existence of directed (odd-)graceful labelings of digraphs are not sight, even for directed trees \cite{Yao-Yao-Cheng-2012}.

It is not hard to define other \emph{flawed directed labelings} or \emph{flawed half-directed labelings} of digraphs and half-directed graphs by the ways used in Definition \ref{defn:flawed-half-directed-graceful} and Definition \ref{defn:flawed-directed-graceful}.\paralled
\end{rem}

\begin{defn}\label{defn:directed-6C-labeling}
\cite{Yao-Mu-Sun-Sun-Zhang-Wang-Su-Zhang-Yang-Zhao-Wang-Ma-Yao-Yang-Xie2019} A total labeling $f:V(\overrightarrow{G})\cup A(\overrightarrow{G})\rightarrow [1,p+q]$ for a bipartite and connected $(p,q)$-digraph $\overrightarrow{G}$ is a bijection and holds:

(i) (e-magic) $f(\overrightarrow{uv})+|f(u)-f(v)|=k$, a constant;

(ii) (ee-difference) each arc $\overrightarrow{uv}$ matches with another arc $\overrightarrow{xy}$ holding $f(\overrightarrow{uv})=|f(x)-f(y)|$ (resp. $f(\overrightarrow{uv})=2(p+q)-|f(x)-f(y)|$);

(iii) (ee-balanced) let $s(\overrightarrow{uv})=|f(u)-f(v)|-f(\overrightarrow{uv})$ for $\overrightarrow{uv}\in A(\overrightarrow{G})$, then there exists a constant $k\,'$ such that each arc $\overrightarrow{uv}$ matches with another arc $\overrightarrow{u\,'v\,'}$ holding $s(\overrightarrow{uv})+s(\overrightarrow{u\,'v\,'})=k\,'$ (resp. $2(p+q)+s(\overrightarrow{uv})+s(\overrightarrow{u\,'v\,'})=k\,'$) true;

(iv) (EV-ordered) $f_{\min}(V(\overrightarrow{G}))>f_{\max}(A(\overrightarrow{G}))$ (resp. $\max f(V(\overrightarrow{G}))<\min f(A(\overrightarrow{G}))$, or $f(V(\overrightarrow{G}))\subseteq f(A(\overrightarrow{G}))$, or $f(A(\overrightarrow{G}))\subseteq f(V(\overrightarrow{G}))$, or $f(V(\overrightarrow{G}))$ is an odd-set and $f(A(\overrightarrow{G}))$ is an even-set);

(v) (ve-matching) there exists a constant $k\,''$ such thateach arc $\overrightarrow{uv}$ matches with one vertex $w$ such that $f(\overrightarrow{uv})+f(w)=k\,''$, and each vertex $z$ matches with one arc $\overrightarrow{xy}$ such that $f(z)+f(\overrightarrow{xy})=k\,''$, except the \emph{singularity} $f(x_0)=\lfloor \frac{p+q+1}{2}\rfloor $;

(vi) (set-ordered) $\max f(X)<\min f(Y)$ (resp. $\min f(X)>\max f(Y)$) for the bipartition $(X,Y)$ of $V(\overrightarrow{G})$.

We call $f$ a \emph{directed 6C-labeling}.\qqed
\end{defn}

\begin{figure}[h]
\centering
\includegraphics[width=16.4cm]{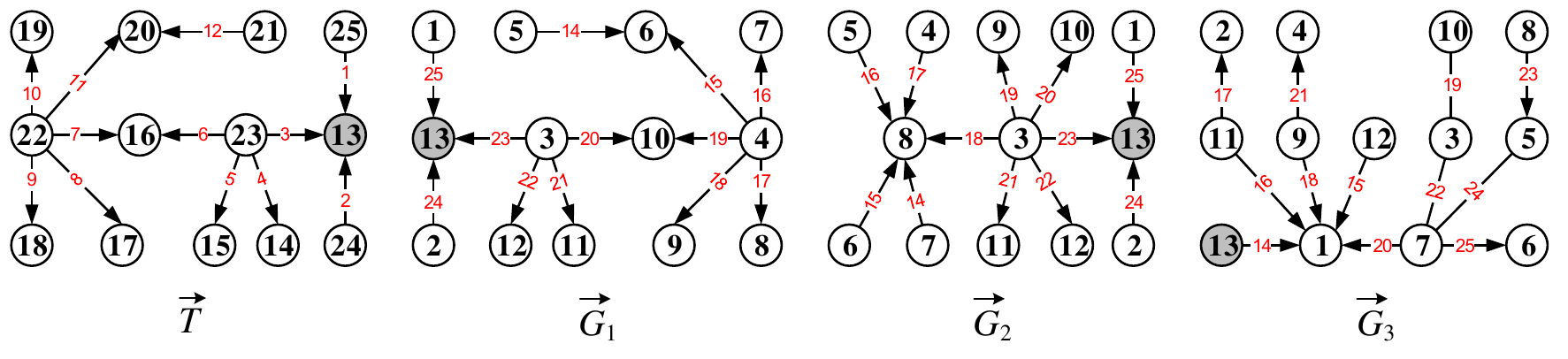}\\
\caption{\label{fig:66-directed-6c-labelings}{\small Four bipartite and connected digraphs admit directed \emph{6C-labelings} defined in Definition \ref{defn:directed-6C-labeling}.}}
\end{figure}

In Fig.\ref{fig:66-directed-6c-labelings}, a directed tree $\overrightarrow{T}$ can be regarded as a \emph{public-key}, and each directed tree $\overrightarrow{G}_i$ is a \emph{private-key} for $i=1,2,3$. So, each vertex-coincided digraph $\odot \langle \overrightarrow{T},\overrightarrow{G}_i\rangle $ obtained by coinciding the vertex $13$ of the directed tree $\overrightarrow{T}$ with the vertex $13$ of each directed tree $\overrightarrow{G}_i$ into one vertex forms a topological authentication based on the directed 6C-labelings.

\subsection{A general definition for matching-type of colorings and labelings}

\begin{defn}\label{defn:general-defi-matching-type-colorings}
$^*$ Suppose that each connected graph $G_i$ admits a $W_i$-type coloring (resp. labeling) $f_i$ for $i\in [1,m]$ with $m\geq 2$, and each edge color set $f(E(G_i))\neq \emptyset$, and $f(E(G_i))\neq f(E(G_j))$ if $i\neq j$. The graph $G^*=(\ast)^m_{i=1}G_i$ is obtained by a graph operation ``$(\ast)$''. We have the following restrictive conditions:
\begin{asparaenum}[\textrm{Ma}-1.]
\item \label{matching:vertex-small-big-inyerval} $\bigcup^m_{i=1}f(V(G_i))\subseteq [a,b]$;
\item \label{matching:edge-small-big-inyerval} $\bigcup^m_{i=1}f(E(G_i))\subseteq [c,d]$;
\item \label{matching:vertex-full-big-inyerval} $\bigcup^m_{i=1}f(V(G_i))=[a,b]$;
\item \label{matching:edge-full-big-inyerval} $\bigcup^m_{i=1}f(E(G_i))=[c,d]$; and
\item \label{matching:ve-mixed-small-big-inyerval} $(W_i)^m_{1}$-type includes some $W_i$-type coloring and $W_j$-type labeling at the same time.
\end{asparaenum}
\noindent \textbf{Then}

(i) the graph $G^*$ admits a \emph{2-whole $(W_i)^m_{1}$-type compound coloring} (resp. \emph{labeling}) $F$ compounded by $W_1,W_2,\dots ,W_m$ if Ma-\ref{matching:vertex-small-big-inyerval} and Ma-\ref{matching:edge-small-big-inyerval} hold true.

(ii) the graph $G^*$ admits a \emph{2-full-whole $(W_i)^m_{1}$-type compound coloring} (resp. \emph{labeling}) $F$ compounded by $W_1,W_2,\dots ,W_m$ if Ma-\ref{matching:vertex-full-big-inyerval} and Ma-\ref{matching:edge-full-big-inyerval} hold true.

(iii) the graph $G^*$ admits a \emph{2-whole mixed-$(W_i)^m_{1}$-type compound coloring} $F$ compounded by $W_1,W_2,\dots ,W_m$ if Ma-\ref{matching:vertex-small-big-inyerval}, Ma-\ref{matching:edge-small-big-inyerval} and Ma-\ref{matching:ve-mixed-small-big-inyerval} hold true.

(iv) the graph $G^*$ admits a \emph{2-full-whole mixed-$(W_i)^m_{1}$-type compound coloring} $F$ compounded by $W_1,W_2,\dots ,W_m$ if Ma-\ref{matching:vertex-full-big-inyerval}, Ma-\ref{matching:edge-full-big-inyerval} and Ma-\ref{matching:ve-mixed-small-big-inyerval} hold true.\qqed
\end{defn}

\begin{rem}\label{rem:ABC-conjecture}
We have defined several 2-whole $(W_f,W_g)$-type compound colorings (resp. labelings) defined in Definition \ref{defn:general-defi-matching-type-colorings}, for example, the 2-odd graceful-elegant labeling defined in Definition \ref{defn:2odd2-matching-labeling}, the twin odd-graceful labeling defined in Definition \ref{defn:22-twin-odd-graceful-labeling}, the twin odd-elegant labeling defined in Definition \ref{defn:twin-odd-elegant-graph}, the ve-exchanged matching labeling defined in Definition \ref{defn:ve-exchanged-labeling}, the twin ($W_f,W_g$)-type coloring defined in Definition \ref{defn:twin-colorings-general}, the ve-inverse edge-magic graceful labelings defined in Definition \ref{defn:ve-inverse-edge-magic-graceful-labeling}, the odd-edge-magic matching labeling defined in Definition \ref{defn:Oemt-labeling}, the ee-difference odd-edge-magic matching labeling defined in Definition \ref{defn:relaxed-Oemt-labeling}, the ee-difference graceful-magic matching labeling defined in Definition \ref{defn:Dgemm-labeling}, the (resp. even)odd-harmonious $s$-matching pair defined in Definition \ref{defn:even-odd-harmonious-matching-pair}, the $k$-matching odd-graceful labeling defined in Definition \ref{defn:matching-odd-graceful-labeling}, the twin edge module-$k$ odd-graceful graph defined in Definition \ref{defn:module-k-twin-odd-graceful-graph}, the duality-image $(W_f,W_g)$-type labelings (resp. colorings) defined in Definition \ref{defn:new-image-labeling-colorings}, and so on.\paralled
\end{rem}

\subsection{A general definition for Topsnut-gpws}

We use the function concept to present a general definition of Topsnut-gpws as follows:
\begin{defn} \label{defn:Topsnut-gpw-function}
\cite{YAO-SUN-WANG-SU-XU2018arXiv} Let $l$ stand up a lock (authentication), $k$ be a key (password), and $h$ be the password rule (a procedure of authentication). The function $l=h(k)$ represents the state of ``a key $k$ open a lock $l$ through the password rule $h$. If $l=h(k)$ is consisted of topological structure and mathematical rule, then $l=h(k)$ is called a \emph{Topsnut-gpw}.\qqed
\end{defn}

\begin{rem}\label{rem:333333}
Let $D(h)$ and $R(h)$ be the domain and the range of the function $h$ in Definition \ref{defn:Topsnut-gpw-function}, respectively. The complex of the Topsnut-gpw $l=h(k)$ is determined by the password rule $h$, the domain $D(h)$ and the range $R(h)$. If one of cardinalities of $D(h)$ and $R(h)$ is the exponential form, then the complex of the Topsnut-gpw $l=h(k)$ is not polynomial; if the password rule $h$ is NP-hard, thereby, so is the Topsnut-gpw $l=h(k)$.

In visualization, a Topsnut-gpw $l=h(k)$ is a colored graph $G$ with a $W$-type coloring (resp. $W$-type labeling) $f$ belonging to a particular class $F$. So, $D_{dif}(G)=(p,q,r,s)$ is defined as the basic difficulty of the Topsnut-gpw $G$, where parameters $p=|V(G)|$, $q=|E(G)|$, $r$ graph properties and $s$ colorings (resp. labelings) belonging to $F\setminus \{f\}$. All topological structures used in Topsnut-gpws are storage in computer by various graph matrices. Thereby, these graph matrices enable us to judge different Topsnut-gpws.\paralled
\end{rem}

\begin{figure}[h]
\centering
\includegraphics[width=16.4cm]{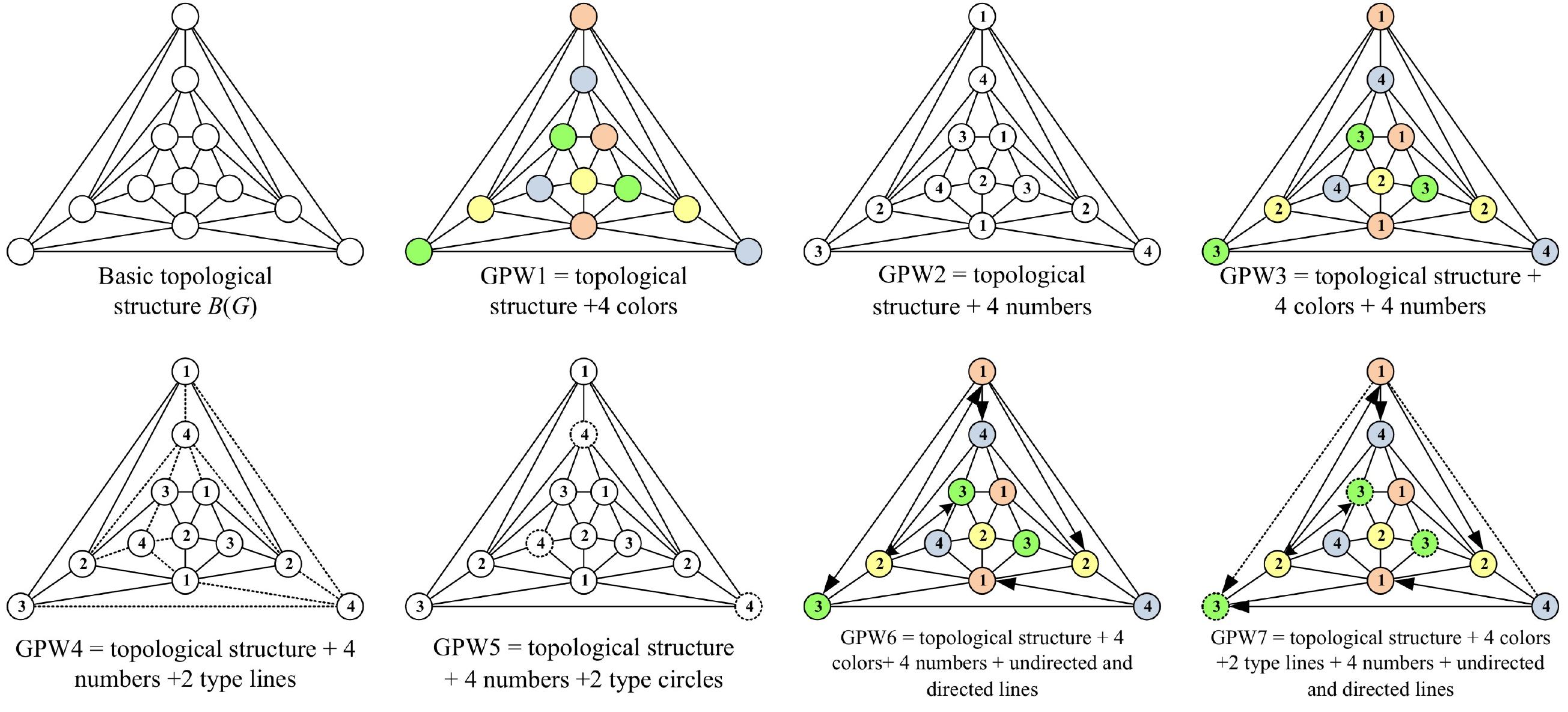}\\
\caption{\label{fig:various-Topsnut-gpws}{\small Various Topsnut-gpws.}}
\end{figure}

\begin{figure}[h]
\centering
\includegraphics[width=11cm]{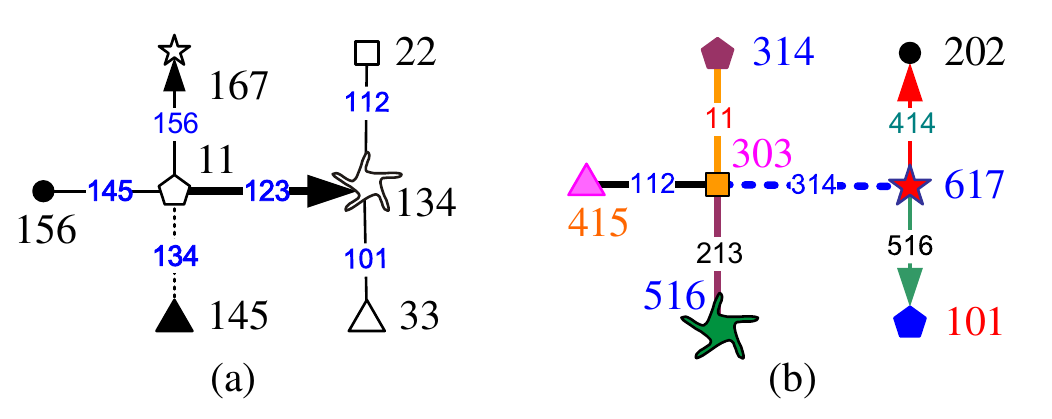}\\
\caption{\label{fig:Pan-various-Topsnut-gpws}{\small Two Pan-Topsnut-gpws.}}
\end{figure}

All of human activities can be expressed in ``\emph{language}'' we have thought, where ``\emph{language}'' is a combination of pictures, sounds, videos, scientific languages, scientific symbols, scientific knowledge, biological techniques, \emph{etc.} So, Topsnut-gpw is a particular ``\emph{language}'', which can be considered as a \emph{platform} for designing various Topsnut-gpws. The study of the ``\emph{language}'' has the significance of graph theory and practical application. We have the following advantages of ``\emph{language}'':
\begin{asparaenum}[C-1.]
\item \textbf{Not pictures}. Topsnutgpws run fast in communication networks because they are saved in computer by popular matrices rather than pictures. (A-1) Topsnut-gpws $\rightarrow$ (A-2) text-based (number-based) strings $\rightarrow$ (A-3) encrypt files; conversely, (B-1) public-keys $\rightarrow$ (B-2) text-based (resp. number-based) strings $\rightarrow$ (B-3) Topsnut-gpws $\rightarrow$ (B-4) decrypt files, where the procedure of (B-2) $\rightarrow$ (B-3) is impossible.
\item \textbf{Incalculable amount of graphs, colorings and labelings.} There are enormous numbers of graph colorings and labelings in graph theory (Ref. \cite{Gallian2020}). And new graph colorings (resp. labelings) come into being everyday. The number of one $W$-type different labelings (resp. colorings) of a graph maybe large, and no method is reported to find out all of such labelings (resp. colorings) for a graph.
\item \textbf{Based on modern mathematics.} For easy memory, some simpler operations like addition, subtraction, absolution and finite modular operations are applied. Number theory, algebra and graph theory are the strong support to Topsnut-gpws of topological coding. It is noticeable, a topological authentication consists of the following two aspects: (1) complex topological structures (vertices are different geometry figures, edges are undirected lines, or directed lines, or solid lines, or dotted line, \emph{ect.}) dyed by hundreds of colors and, (2) various mathematical constraints (see Fig.\ref{fig:various-Topsnut-gpws} and Fig.\ref{fig:Pan-various-Topsnut-gpws}).
\item \textbf{Diversity of asymmetric ciphers.} One key corresponds to more locks, or more keys corresponds to one lock only. Topsnut-gpws realize the coexistence of two or more labelings and colorings on a graph, which leads to the problem of multi-labeling (resp. multi-coloring) decomposition of graphs, and brings new research objects and new problems to mathematics.
\item \textbf{Connection}. There are connections between Topsnut-gpws and other type of passwords. For example, small circles in the Topsnut-gpws can be equipped with fingerprints and other biological information requirements, and users' pictures can be embedded in small circles, greatly reflects personalization.
\item \textbf{Simplicity and convenience.} Topsnut-gpws are suitable for people who need not learn new rules and are allowed to use their private knowledge in making Topsnut-gpws for the sake of remembering easily. For example, Chinese characters are naturally topological structures (also graphs in Topsnut-gpws) to produce Hanzi-graphs for topological coding. Chinese people can generate Topsnut-gpws simply by speaking and writing Chinese directly.
\item \textbf{Ease created.} Many labelings of trees are convertible from each other \cite{Yao-Liu-Yao-2017}. Tree structures can adapt to a large number of labelings and colorings, in addition to graceful labeling, no other labelings reported that were established in those tree structures having smaller vertex numbers by computer, almost no computer proof. Because construction methods are complex, this means that using computers to break down Topsnut-gpws will be difficult greatly.
\item \textbf{Irreversibility.} Topsnut-gpws can generate quickly text-based (resp. number-based) strings with bytes as long as desired, but these strings can not reconstruct the original Topsnut-gpws. The confusion of text-based (resp. number-based) strings and Topsnut-gpws can't be erased, as our knowledge.
\item \textbf{Computational security.} There are many non-polynomial algorithms in making Topsnut-gpws. For example, drawing non-isomorphic graphs is very difficult and non-polynomial; for a given graph, finding out all possible colorings (resp. labelings) are impossible, since these colorings (resp. labelings) are massive data, and many graph problems have been proven to be NP-complete.
\item \textbf{Provable security.} There are many mathematical conjectures (also, \emph{open problems}) in graph labelings and colorings, such as the famous graceful tree conjecture, odd-graceful tree conjecture, total coloring conjecture \emph{etc.}
\end{asparaenum}

\subsection{Distinguishing colorings}

\begin{defn} \label{defn:public-colorings-general-framework}
\cite{Yao-Yang-Yao-2020-distinguishing} Let $Q$ be a \emph{\textbf{graph property}} consisted of one or more \emph{\textbf{constraints}} on graphs. A $k$-mapping $\pi$ of a graph $G$ is defined by partitioning a subset $S\subseteq V(G)\cup E(G)$ into disjoint non-empty subsets $S_1,S_2,\dots ,S_k$. We say each $i$ a color, and write $\pi(x)=i$ for $x\in S_i$ and $\pi:S \rightarrow [1,k]$. Let $\chi _Q(G)$ be the minimum number of colors required for which $G$ admits a $k$-mapping $\pi$ with $k\geq \chi _Q(G)$ such that the graph property $Q$ holds true.\qqed
\end{defn}

We restate the existing distinguishing-type colorings and define new distinguishing-type colorings and $V_d$-type colorings including: (adjacent-)vertex distinguishing (vertex, edge, total) colorings in Definition \ref{defn:combinator-distinguishing-colorings}. Suppose that a graph $G$ admits a coloring $\theta:S\rightarrow [1,k]$, where $S\subseteq V(G)\cup E(G)$. There are the following neighbor color sets:
\begin{equation}\label{eqa:66-first-group}
{
\begin{split}
C_v(x,\theta)=&\{\theta(y):y\in N(x)\}\textrm{ (local v-color set)};\\
C_v[x,\theta]=&\{\theta(x)\}\cup C_v(x,\theta)\textrm{ (closed local v-color set)};\\
C_e(x,\theta)=&\{\theta(xy):y\in N(x)\}\textrm{ (local e-color set)};\\
C_e[x,\theta]=&\{\theta(x)\}\cup C_e(x,\theta)\textrm{ (closed local e-color set)};\\
C_{ve}(x,\theta)=&C_v(x,\theta)\cup C_e(x,\theta)\textrm{ (local ve-color set)};\\
C_{ve}[x,\theta]=&\{\theta(x)\}\cup C_v(x,\theta)\cup C_e(x,\theta)\textrm{ (closed local ve-color set)};\\
C_{ve}\{w,\theta\}=&\{C_e(w,\theta), C_e[w,\theta], C_v[w,\theta],C_{ve}[w,\theta]\}\textrm{~for $w\in V(G)$ (closed local (4)-color set)}.
\end{split}}
\end{equation}

\begin{defn}\label{defn:combinator-distinguishing-colorings}
\cite{Yao-Yang-Yao-2020-distinguishing} Suppose that a graph $G$ admits a coloring $\theta:S\rightarrow [1,k]$, where $S\subseteq V(G)\cup E(G)$, such that $S=\bigcup ^k_{i=1}S_i$, and no two elements of $S_i$ with $i\in [1,k]$ are adjacent or
incident in $G$, each $S_i$ is called an independent (stable) set. By the color sets in (\ref{eqa:66-first-group}), there are the following constraint conditions:
\begin{asparaenum}[(\textrm{R}-1) ]
\item \label{asp:vertex-only} $S=V(G)$ (vertex-set);
\item \label{asp:edge-only} $S=E(G)$ (edge-set);
\item \label{asp:total} $S=V(G)\cup E(G)$ (ve-set);
\item \label{asp:adjacent-vertices} $\theta(u)\neq \theta(v)$ for $uv\in E(G)$ (adjacent-vertices);
\item \label{asp:adjacent-edges} $\theta(xy)\neq \theta(xw)$ for $y,w\in N(x)$ (adjacent-edges);
\item \label{asp:adjacent-vertices-edge} $\theta(u)\neq \theta(uv)$ and $\theta(v)\neq \theta(uv)$ for $uv\in E(G)$ (incident vertices and edges);
\item \label{asp:universal-vertices} $\theta(u)\neq \theta(x)$ for $ux\not \in E(G)$ (no-adjacent-vertices);
\item \label{asp:universal-edges} $\theta(xy)\neq \theta(uv)$ for $xy,uv\in E(G)$ with $x\neq u$, $x\neq v$, $y\neq u$ and $y\neq v$ (no-adjacent-edges);
\item \label{asp:neighbor-distinguishing} $C_v(x,\theta)\neq C_v(y,\theta)$ with $y\in N(x)$ for each vertex $x\in V(G)$ (local vertex distinguishing);
\item \label{asp:closed-neighbor-distinguishing} $C_v[x,\theta]\neq C_v[y,\theta]$ with $y\in N(x)$ for each vertex $x\in V(G)$ (closed local vertex distinguishing);
\item \label{asp:neighbor-e-distinguishing} $C_e(x,\theta)\neq C_e(y,\theta)$ for $y\in N(x)$ (local edge distinguishing);
\item \label{asp:closed-neighbor-e-distinguishing}$C_e[x,\theta]\neq C_e[y,\theta]$ with $y\in N(x)$ for each vertex $x\in V(G)$ (closed local ve-distinguishing);
\item \label{asp:neighbor-ve-distinguishing} $C_{ve}(x,\theta)\neq C_{ve}(y,\theta)$ for $y\in N(x)$ (local ve-distinguishing);
\item \label{asp:closed-neighbor-ve-distinguishing} $C_{ve}[x,\theta]\neq C_{ve}[y,\theta]$ for $y\in N(x)$ (closed local ve-distinguishing);

\item \label{asp:universal-v-distinguishing} $C_v(x,\theta)\neq C_v(w,\theta)$ for any pair of distinct vertices $x,w\in V(G)$ (universal vertex distinguishing);
\item \label{asp:closed-universal-v-distinguishing} $C_v[x,\theta]\neq C_v[w,\theta]$ for any pair of distinct vertices $x,w\in V(G)$ (closed universal vertex distinguishing);
\item \label{asp:universal-e-distinguishing} $C_e(x,\theta)\neq C_e(w,\theta)$ for any pair of distinct vertices $x,w\in V(G)$ (universal edge distinguishing);
\item \label{asp:closed-universal-e-distinguishing} $C_e[x,\theta]\neq C_e[w,\theta]$ for any pair of distinct vertices $x,w\in V(G)$ (universal edge distinguishing);
\item \label{asp:universal-ve-distinguishing} $C_{ve}(x,\theta)\neq C_{ve}(w,\theta)$ for any pair of distinct vertices $x,w\in V(G)$ (universal ve-distinguishing);
\item \label{asp:closed-universal-ve-distinguishing} $C_{ve}[x,\theta]\neq C_{ve}[w,\theta]$ for any pair of distinct vertices $x,w\in V(G)$ (closed universal ve-distinguishing);
\item \label{asp:local-totally-ve-distinguishing} $C_{ve}\{x,\theta\}\neq C_{ve}\{y,\theta\}$ for $y\in N(x)$ (local totally ve-distinguishing);

------ \emph{distance}

\item \label{asp:distance-v-distinguishing} $C_v(u,\theta)\neq C_v(v,\theta)$ for any pair of vertices $u$ and $v$ with distance $d(u,v)\leq \beta$ ($\beta$-distance vertex distinguishing);
\item \label{asp:distance-closed-v-distinguishing} $C_v[u,\theta]\neq C_v[v,\theta]$ for any pair of vertices $u$ and $v$ with
distance $d (u,v)\leq \beta$ ($\beta$-distance closed vertex distinguishing);
\item \label{asp:distance-e-distinguishing} $C_e(u,\theta)\neq C_e(v,\theta)$ for any pair of vertices $u$ and $v$ with distance $d(u,v)\leq \beta$ ($\beta$-distance edge distinguishing);
\item \label{asp:distance-closed-e-distinguishing} $C_e[u,\theta]\neq C_e[v,\theta]$ for any pair of vertices $u$ and $v$ with
distance $d (u,v)\leq \beta$ ($\beta$-distance closed edge distinguishing);
\item \label{asp:distance-ve-distinguishing} $C_{ve}(u,\theta)\neq C_{ve}(v,\theta)$ for any pair of vertices $u$ and $v$ with distance $d(u,v)\leq \beta$ ($\beta$-distance total distinguishing);
\item \label{asp:distance-closed-ve-distinguishing} $C_{ve}[u,\theta]\neq C_{ve}[v,\theta]$ for any pair of vertices $u$ and $v$ with distance $d (u,v)\leq \beta$ ($\beta$-distance closed total distinguishing);

------ \emph{equitable, acyclic}

\item \label{asp:equitable-constraint} $\big||S_i|-|S_j|\big |\leq 1$ with $i,j\in [1,k]$ (equitable sets); and
\item \label{asp:acyclic-constraint-3} The induced subgraph by $S_i\cup S_j$ with $i\neq j$ contains no cycle.
\end{asparaenum}

\noindent \textbf{We call $\theta$}:

\noindent ------ \emph{proper, traditional, non-distinguishing}

\begin{asparaenum}[\textrm{Color}-1. ]
\item A \emph{proper vertex coloring} if (R-\ref{asp:vertex-only}) and (R-\ref{asp:adjacent-vertices}) hold true, the $V_d$-type chromatic number $\chi(G)$.
\item A \emph{proper edge coloring} if (R-\ref{asp:edge-only}) and (R-\ref{asp:adjacent-edges}) hold true, the $V_d$-type chromatic number $\chi\,'(G)$.
\item A \emph{proper total coloring} if (R-\ref{asp:total}), (R-\ref{asp:adjacent-vertices}), (R-\ref{asp:adjacent-edges}) and (R-\ref{asp:adjacent-vertices-edge}) hold true, the $V_d$-type chromatic number $\chi\,''(G)$.
\item A \emph{labeling} if (R-\ref{asp:vertex-only}), (R-\ref{asp:adjacent-vertices}) and (R-\ref{asp:universal-vertices}) hold true.

------ \emph{improper}

\item A \emph{vertex coloring} if (R-\ref{asp:vertex-only}) holds true.
\item An \emph{edge coloring} if (R-\ref{asp:edge-only}) holds true.
\item A \emph{total coloring} if (R-\ref{asp:total}) holds true.
\item An \emph{adjacent-vertex distinguishing v-coloring} if (R-\ref{asp:vertex-only}) and (R-\ref{asp:neighbor-distinguishing}) hold true.
\item An \emph{adjacent-vertex distinguishing closed v-coloring} if (R-\ref{asp:vertex-only}) and (R-\ref{asp:closed-neighbor-distinguishing}) hold true.
\item An \emph{adjacent-vertex distinguishing e-coloring} if (R-\ref{asp:edge-only}) and (R-\ref{asp:neighbor-e-distinguishing}) hold true.
\item An \emph{adjacent-vertex distinguishing closed e-coloring} if (R-\ref{asp:edge-only}) and (R-\ref{asp:closed-neighbor-e-distinguishing}) hold true.
\item An \emph{adjacent-vertex distinguishing total coloring} if (R-\ref{asp:total}) and (R-\ref{asp:neighbor-ve-distinguishing}) hold true.
\item An \emph{adjacent-vertex distinguishing closed total coloring} if (R-\ref{asp:total}) and (R-\ref{asp:closed-neighbor-ve-distinguishing}) hold true.

------ \emph{local proper}

\item An \emph{adjacent-vertex distinguishing proper v-coloring} if (R-\ref{asp:vertex-only}), (R-\ref{asp:adjacent-vertices}) and (R-\ref{asp:neighbor-distinguishing}) hold true.
\item An \emph{adjacent-vertex distinguishing closed proper v-coloring} if (R-\ref{asp:vertex-only}), (R-\ref{asp:adjacent-vertices}) and (R-\ref{asp:closed-neighbor-distinguishing}) hold true.

\item \cite{Zhang-Liu-Wang-2002} an \emph{adjacent-vertex distinguishing proper e-coloring} if (R-\ref{asp:edge-only}), (R-\ref{asp:adjacent-edges}) and (R-\ref{asp:neighbor-e-distinguishing}) hold true, the \emph{adjacent-vertex distinguishing chromatic index} $\chi\,'_{as}(G)$.
\item \cite{Zhang-Liu-Wang-2002} an \emph{adjacent-vertex distinguishing closed proper e-coloring} if (R-\ref{asp:edge-only}), (R-\ref{asp:adjacent-edges}) and (R-\ref{asp:closed-neighbor-e-distinguishing}) hold true, the \emph{adjacent-vertex distinguishing total chromatic number} $\chi\,'_{as}(G)$.

\item \cite{Zhang-Chen-Li-Yao-Lu-Wang-China-2005} an \emph{adjacent-vertex distinguishing proper total coloring} if (R-\ref{asp:total}), (R-\ref{asp:adjacent-vertices}), (R-\ref{asp:adjacent-edges}), (R-\ref{asp:adjacent-vertices-edge}) and (R-\ref{asp:neighbor-ve-distinguishing}) hold true, the $V_d$-type chromatic number $\chi\,''_{as}(G)$.
\item \cite{Zhang-Cheng-Yao-Li-Chen-Xu2008} an \emph{adjacent-vertex distinguishing closed proper total coloring} if (R-\ref{asp:total}), (R-\ref{asp:adjacent-vertices}), (R-\ref{asp:adjacent-edges}), (R-\ref{asp:adjacent-vertices-edge}) and (R-\ref{asp:closed-neighbor-ve-distinguishing}) hold true, the $V_d$-type chromatic number $\chi\,''_{ast}(G)$.

------ \emph{distance}

\item A \emph{$\beta$-distance vertex distinguishing proper v-coloring} if (R-\ref{asp:vertex-only}), (R-\ref{asp:adjacent-vertices}) and (R-\ref{asp:distance-v-distinguishing}) hold true.
\item A \emph{$\beta$-distance vertex distinguishing closed proper v-coloring} if (R-\ref{asp:vertex-only}), (R-\ref{asp:adjacent-vertices}) and (R-\ref{asp:distance-closed-v-distinguishing}).

\item \label{Distance-distinguishing1} a \emph{$\beta$-distance vertex distinguishing proper e-coloring} if (R-\ref{asp:edge-only}), (R-\ref{asp:adjacent-edges}) and (R-\ref{asp:distance-e-distinguishing}) hold true, the \emph{$\beta$-distance vertex distinguishing chromatic index} $\chi\,'_{\beta}(G)$, where $\chi\,'_{1}(G)=\chi\,'_{as}(G)$, $\chi\,'_{D}(G)=\chi\,'_{s}(G)$.

\item A \emph{$\beta$-distance vertex distinguishing closed proper e-coloring} if (R-\ref{asp:edge-only}), (R-\ref{asp:adjacent-edges}) and (R-\ref{asp:distance-closed-e-distinguishing}).

\item A \emph{$\beta$-distance vertex distinguishing proper total coloring} if (R-\ref{asp:total}), (R-\ref{asp:adjacent-vertices}), (R-\ref{asp:adjacent-edges}), (R-\ref{asp:adjacent-vertices-edge}) and (R-\ref{asp:distance-ve-distinguishing}) hold true, the \emph{$\beta$-distance vertex distinguishing total chromatic number} $\chi\,''_{\beta}(G)$, where $\chi\,''_{1}(G)=\chi\,''_{as}(G)$, $\chi\,''_{D}(G)=\chi\,''_{s}(G)$.

\item A \emph{$\beta$-distance vertex distinguishing closed proper total coloring} if (R-\ref{asp:total}), (R-\ref{asp:adjacent-vertices}), (R-\ref{asp:adjacent-edges}), (R-\ref{asp:adjacent-vertices-edge}) and (R-\ref{asp:distance-closed-ve-distinguishing}), the \emph{$\beta$-distance vertex distinguishing total chromatic number} $\chi\,''_{\beta\textrm{-s}}(G)$.

------ \emph{$(4)$-adjacent}

\item \cite{Yang-Ren-Yao-Ars-2016} a \emph{$(4)$-adjacent-vertex distinguishing closed proper total coloring} if (R-\ref{asp:total}), (R-\ref{asp:adjacent-vertices}), (R-\ref{asp:adjacent-edges}), (R-\ref{asp:adjacent-vertices-edge}) and (R-\ref{asp:local-totally-ve-distinguishing}) hold true, the $V_d$-type chromatic number $\chi\,''_{(4)as}(G)$.

------ \emph{universal proper}

\item A \emph{vertex distinguishing proper v-coloring} if (R-\ref{asp:vertex-only}), (R-\ref{asp:adjacent-vertices}) and (R-\ref{asp:universal-v-distinguishing}) hold true.
\item A \emph{vertex distinguishing closed proper v-coloring} if (R-\ref{asp:vertex-only}), (R-\ref{asp:adjacent-vertices}) and (R-\ref{asp:closed-universal-v-distinguishing}) hold true.
\item \cite{Burris-Schelp} a \emph{vertex distinguishing proper e-coloring} if (R-\ref{asp:edge-only}), (R-\ref{asp:adjacent-edges}) and (R-\ref{asp:universal-e-distinguishing}) hold true, the $V_d$-type chromatic number $\chi\,'_s(G)$.
\item A \emph{vertex distinguishing closed proper e-coloring} if (R-\ref{asp:edge-only}), (R-\ref{asp:adjacent-edges}) and (R-\ref{asp:closed-universal-e-distinguishing}) hold true.

\item A \emph{vertex distinguishing proper total coloring} if (R-\ref{asp:total}), (R-\ref{asp:adjacent-vertices}), (R-\ref{asp:adjacent-edges}), (R-\ref{asp:adjacent-vertices-edge}) and (R-\ref{asp:universal-ve-distinguishing}) hold true, the $V_d$-type chromatic number $\chi\,''_s(G)$.
\item A \emph{vertex distinguishing closed proper total coloring} if (R-\ref{asp:total}), (R-\ref{asp:adjacent-vertices}), (R-\ref{asp:adjacent-edges}), (R-\ref{asp:adjacent-vertices-edge}) and (R-\ref{asp:closed-universal-ve-distinguishing}) hold true.

------ \emph{equitably}


\item An \emph{equitably adjacent-vertex distinguishing proper e-coloring} if (R-\ref{asp:edge-only}), (R-\ref{asp:adjacent-edges}), (R-\ref{asp:neighbor-e-distinguishing}) and (R-\ref{asp:equitable-constraint}) hold true, the \emph{adjacent-vertex distinguishing chromatic index} $\chi\,'_{eas}(G)$.
\item An \emph{equitably adjacent-vertex distinguishing closed proper e-coloring} if (R-\ref{asp:edge-only}), (R-\ref{asp:adjacent-edges}), (R-\ref{asp:closed-neighbor-e-distinguishing}) and (R-\ref{asp:equitable-constraint}) hold true.
\item An \emph{equitably adjacent-vertex distinguishing proper total coloring} if (R-\ref{asp:total}), (R-\ref{asp:adjacent-vertices}), (R-\ref{asp:adjacent-edges}), (R-\ref{asp:adjacent-vertices-edge}), (R-\ref{asp:neighbor-ve-distinguishing}) and (R-\ref{asp:equitable-constraint}) hold true, the \emph{equitably adjacent-vertex distinguishing total chromatic number} $\chi\,''_{eas}(G)$.


\item An \emph{equitably vertex distinguishing proper e-coloring} if (R-\ref{asp:edge-only}), (R-\ref{asp:adjacent-edges}), (R-\ref{asp:universal-e-distinguishing}) and (R-\ref{asp:equitable-constraint}) hold true, the \emph{equitably vertex distinguishing chromatic index} $\chi\,'_{es}(G)$.
\item An \emph{equitably vertex distinguishing proper total coloring} if (R-\ref{asp:total}), (R-\ref{asp:adjacent-vertices}), (R-\ref{asp:adjacent-edges}), (R-\ref{asp:adjacent-vertices-edge}), (R-\ref{asp:closed-universal-ve-distinguishing}) and (R-\ref{asp:equitable-constraint}) hold true, the \emph{equitably vertex distinguishing total chromatic number} $\chi\,''_{es}(G)$.

------ \emph{acyclic}

\item An \emph{acyclic adjacent-vertex distinguishing proper e-coloring} if (R-\ref{asp:edge-only}), (R-\ref{asp:adjacent-edges}), (R-\ref{asp:neighbor-e-distinguishing}) and (R-\ref{asp:acyclic-constraint-3}) hold true, the \emph{acyclic adjacent-vertex distinguishing chromatic index} $\chi\,'_{aas}(G)$.
\item \label{color:acyclic-adjacent-vertex-distinguishing-total} an \emph{acyclic adjacent-vertex distinguishing proper total coloring} if (R-\ref{asp:total}), (R-\ref{asp:adjacent-vertices}), (R-\ref{asp:adjacent-edges}), (R-\ref{asp:adjacent-vertices-edge}), (R-\ref{asp:neighbor-ve-distinguishing}) and (R-\ref{asp:acyclic-constraint-3}) hold true, the \emph{acyclic adjacent-vertex distinguishing total chromatic number} $\chi\,''_{aas}(G)$.

\item An \emph{acyclic vertex distinguishing proper total coloring} if (R-\ref{asp:total}), (R-\ref{asp:adjacent-vertices}), (R-\ref{asp:adjacent-edges}), (R-\ref{asp:adjacent-vertices-edge}), (R-\ref{asp:closed-neighbor-ve-distinguishing}) and (R-\ref{asp:acyclic-constraint-3}) hold true, the \emph{acyclic adjacent-vertex distinguishing total chromatic number} $\chi\,''_{aas}(G)$.\qqed
\end{asparaenum}
\end{defn}

\begin{rem}\label{rem:333333}
A graph $G$ admits a $V_d$-type coloring defined in Definition \ref{defn:combinator-distinguishing-colorings}, then the smallest number of $k$ colors required for which $G$ admits a $V_d$-type coloring is called \emph{$V_d$-type chromatic number (index)} and denoted as $\chi^{\varepsilon}_{vd}(G)$, where a \emph{$V_d$-type chromatic number} $\chi_{vd}$ is for vertex coloring (also, v-coloring), a \emph{$V_d$-type chromatic index} $\chi\,'_{vd}$ is for edge coloring (also, e-coloring) and a \emph{$V_d$-type total chromatic number} $\chi\,''_{vd}$ is for total coloring.\paralled
\end{rem}

About any vertex $u$ of a graph $G$ admitting a total coloring $\varphi:S\rightarrow [1,k]$, we have the following weighted sums:
\begin{equation}\label{eqa:66-second-group}
{
\begin{split}
W_v(u,\varphi)=&\sum_{y\in N(u)}\varphi(y)\textrm{ (a weighted sum of neighbors of $u$)};\\
W_e(u,\varphi)=&\sum_{y\in N(u)}\varphi(uy)\textrm{ (an edge weighted sum of}\textrm{edges incident with $u$)};\\
W_v[u,\varphi]=&\varphi(u)+W_v(u,\varphi)\textrm{ (a closed vertex weighted sum)};\\
W_e[u,\varphi]=&\varphi(u)+W_e(u,\varphi)\textrm{ (a closed edge weighted sum)};\\
W_{ve}(u,\varphi)=&W_v(u,\varphi)+W_e(u,\varphi)\textrm{ (a total weighted sum)};\\
W_{ve}[u,\varphi]=&\varphi(u)+W_{ve}(u,\varphi)\textrm{ (a closed total weighted sum)}.
\end{split}}
\end{equation}We introduce the following weighted distinguishing colorings.

\begin{defn}\label{defn:combinator-weighted-sum-distinguishing}
\cite{Yao-Yang-Yao-2020-distinguishing} Suppose that a graph $G$ admits a total coloring $\varphi:S\rightarrow [1,k]$, where $S\subseteq V(G)\cup E(G)$. By the weighted sums defined in (\ref{eqa:66-second-group}) and the restrictions defined in Definition \ref{defn:combinator-distinguishing-colorings}, there are the following distinguishing-type weighted sums:

\noindent ------ \emph{local weight}

\begin{asparaenum}[(\textrm{W}1) ]
\item \label{weight:v-distinguishing} $W_v(u,\varphi)\neq W_v(v,\varphi)$ for edge $uv\in E(G)$;
\item \label{weight:closed-v-distinguishing} $W_v[u,\varphi]\neq W_v[v,\varphi]$ for edge $uv\in E(G)$;
\item \label{weight:e-distinguishing} $W_e(u,\varphi)\neq W_e(v,\varphi)$ for edge $uv\in E(G)$;
\item \label{weight:closed-e-distinguishing} $W_e[u,\varphi]\neq W_e[v,\varphi]$ for edge $uv\in E(G)$;
\item \label{weight:ve-distinguishing} $W_{ve}(u,\varphi)\neq W_{ve}(v,\varphi)$ for edge $uv\in E(G)$;
\item \label{weight:closed-ve-distinguishing} $W_{ve}[u,\varphi]\neq W_{ve}[v,\varphi]$ for edge $uv\in E(G)$;

------ \emph{universal weight}

\item \label{weight:universal-v-distinguishing} $W_v(x,\varphi)\neq W_v(y,\varphi)$ for distinct $x,y\in V(G)$;
\item \label{weight:universal-closed-v-distinguishing} $W_v[x,\varphi]\neq W_v[y,\varphi]$ for distinct $x,y\in V(G)$;
\item \label{weight:universal-e-distinguishing} $W_e(x,\varphi)\neq W_e(y,\varphi)$ for distinct $x,y\in V(G)$;
\item \label{weight:universal-closed-e-distinguishing} $W_e[x,\varphi]\neq W_e[y,\varphi]$ for distinct $x,y\in V(G)$;
\item\label{weight:universal-ve-distinguishing} $W_{ve}(x,\varphi)\neq W_{ve}(y,\varphi)$ for distinct $x,y\in V(G)$; and
\item \label{weight:universal-closed-ve-distinguishing} $W_{ve}[x,\varphi]\neq W_{ve}[y,\varphi]$ for distinct $x,y\in V(G)$.
\end{asparaenum}

\noindent \textbf{We call $\varphi$:}

\noindent ------ \emph{local improper}

\begin{asparaenum}[(\textrm{Wcolor}-1) ]
\item An \emph{adjacent-vertex weighted distinguishing v-coloring} if (R-\ref{asp:vertex-only}) and (R-\ref{weight:v-distinguishing}) hold true.
\item An \emph{adjacent-vertex weighted distinguishing closed v-coloring} if (R-\ref{asp:vertex-only}) and (R-\ref{weight:closed-v-distinguishing}) hold true.

\item An \emph{adjacent-vertex weighted distinguishing e-coloring} if (R-\ref{asp:edge-only}) and (R-\ref{weight:e-distinguishing}) hold true.
\item An \emph{adjacent-vertex weighted distinguishing closed e-coloring} if (R-\ref{asp:edge-only}) and (R-\ref{weight:closed-e-distinguishing}) hold true.

\item An \emph{adjacent-vertex weighted distinguishing total coloring} if (R-\ref{asp:total}) and (R-\ref{weight:ve-distinguishing}) hold true.
\item An \emph{adjacent-vertex weighted distinguishing closed total coloring} if (R-\ref{asp:total}) and (R-\ref{weight:closed-ve-distinguishing}) hold true.

------ \emph{local proper}

\item An \emph{adjacent-vertex weighted distinguishing proper v-coloring} if (R-\ref{asp:vertex-only}), (R-\ref{asp:adjacent-vertices}) and (R-\ref{weight:v-distinguishing}) hold true.
\item An \emph{adjacent-vertex weighted distinguishing closed proper v-coloring} if (R-\ref{asp:vertex-only}), (R-\ref{asp:adjacent-vertices}) and (R-\ref{weight:closed-v-distinguishing}) hold true.

\item An \emph{adjacent-vertex weighted distinguishing proper e-coloring} if (R-\ref{asp:edge-only}), (R-\ref{asp:adjacent-edges}) and (R-\ref{weight:e-distinguishing}) hold true.
\item An \emph{adjacent-vertex weighted distinguishing closed proper e-coloring} if (R-\ref{asp:edge-only}), (R-\ref{asp:adjacent-edges}) and (R-\ref{weight:closed-e-distinguishing}) hold true.

\item An \emph{adjacent-vertex weighted distinguishing proper total coloring} if (R-\ref{asp:total}), (R-\ref{asp:adjacent-vertices}), (R-\ref{asp:adjacent-edges}), (R-\ref{asp:adjacent-vertices-edge}) and (R-\ref{weight:ve-distinguishing}) hold true.
\item An \emph{adjacent-vertex weighted distinguishing closed proper total coloring} if (R-\ref{asp:total}), (R-\ref{asp:adjacent-vertices}), (R-\ref{asp:adjacent-edges}), (R-\ref{asp:adjacent-vertices-edge}) and (R-\ref{weight:closed-ve-distinguishing}) hold true.

------ \emph{universal improper}

\item An \emph{adjacent-vertex weighted distinguishing v-coloring} if (R-\ref{asp:vertex-only}) and (R-\ref{weight:universal-v-distinguishing}) hold true.
\item An \emph{adjacent-vertex weighted distinguishing closed v-coloring} if (R-\ref{asp:vertex-only}) and (R-\ref{weight:universal-closed-v-distinguishing}) hold true.

\item An \emph{adjacent-vertex weighted distinguishing e-coloring} if (R-\ref{asp:edge-only}) and (R-\ref{weight:universal-e-distinguishing}) hold true.
\item An \emph{adjacent-vertex weighted distinguishing closed e-coloring} if (R-\ref{asp:edge-only}) and (R-\ref{weight:universal-closed-e-distinguishing}) hold true.

\item An \emph{adjacent-vertex weighted distinguishing total coloring} if (R-\ref{asp:total}) and (R-\ref{weight:universal-ve-distinguishing}) hold true.
\item An \emph{adjacent-vertex weighted distinguishing closed total coloring} if (R-\ref{asp:total}) and (R-\ref{weight:universal-closed-ve-distinguishing}) hold true.

------ \emph{universal proper}

\item An \emph{adjacent-vertex weighted distinguishing proper v-coloring} if (R-\ref{asp:vertex-only}), (R-\ref{asp:adjacent-vertices}) and (R-\ref{weight:universal-v-distinguishing}) hold true.
\item An \emph{adjacent-vertex weighted distinguishing closed proper v-coloring} if (R-\ref{asp:vertex-only}), (R-\ref{asp:adjacent-vertices}) and (R-\ref{weight:universal-closed-v-distinguishing}) hold true.

\item An \emph{adjacent-vertex weighted distinguishing proper e-coloring} if (R-\ref{asp:edge-only}), (R-\ref{asp:adjacent-edges}) and (R-\ref{weight:universal-e-distinguishing}) hold true.
\item An \emph{adjacent-vertex weighted distinguishing closed proper e-coloring} if (R-\ref{asp:edge-only}), (R-\ref{asp:adjacent-edges}) and (R-\ref{weight:universal-closed-e-distinguishing}) hold true.

\item An \emph{adjacent-vertex weighted distinguishing proper total coloring} if (R-\ref{asp:total}), (R-\ref{asp:adjacent-vertices}), (R-\ref{asp:adjacent-edges}), (R-\ref{asp:adjacent-vertices-edge}) and (R-\ref{weight:universal-ve-distinguishing}) hold true.
\item An \emph{adjacent-vertex weighted distinguishing closed proper total coloring} if (R-\ref{asp:total}), (R-\ref{asp:adjacent-vertices}), (R-\ref{asp:adjacent-edges}), (R-\ref{asp:adjacent-vertices-edge}) and (R-\ref{weight:universal-closed-ve-distinguishing}) hold true.\qqed
\end{asparaenum}
\end{defn}

\begin{defn}\label{defn:distinguishing-ice-flower-systems}
\cite{Yao-Yang-Yao-2020-distinguishing} Suppose that each star $K^c_{1,m_j}$ of a colored ice-flower system $\textbf{\textrm{K}}^c=(K^c_{1,m_j})^n_{j=1}$ admits a $W$-type coloring $f_i$, using the notations defined in (\ref{eqa:66-first-group}), then we call $\textbf{\textrm{K}}^c$:
\begin{asparaenum}[\textrm{Ice}-1. ]
\item A \emph{vertex distinguishing v-ice-flower system} if two local v-color sets $C_v(x_i,f_i)\neq C_v(x_j,f_j)$ for two centers $x_i\in V(K^c_{1,m_i})$ and $x_j\in V(K^c_{1,m_j})$.
\item A \emph{vertex distinguishing closed v-ice-flower system} if two closed local v-color sets $C_v[x_i,f_i]\neq C_v[x_j,f_j]$ for two centers $x_i\in V(K^c_{1,m_i})$ and $x_j\in V(K^c_{1,m_j})$.
\item A \emph{vertex distinguishing e-ice-flower system} if two local e-color sets $C_e(x_i,f_i)\neq C_e(x_j,f_j)$ for two centers $x_i\in V(K^c_{1,m_i})$ and $x_j\in V(K^c_{1,m_j})$.
\item A \emph{vertex distinguishing closed e-ice-flower system} if two closed local e-color sets $C_e[x_i,f_i]\neq C_e[x_j,f_j]$ for two centers $x_i\in V(K^c_{1,m_i})$ and $x_j\in V(K^c_{1,m_j})$.
\item A \emph{vertex distinguishing total ice-flower system} if two local ve-color sets $C_{ve}(x_i,f_i)\neq C_{ve}(x_j,f_j)$ for two centers $x_i\in V(K^c_{1,m_i})$ and $x_j\in V(K^c_{1,m_j})$.
\item A \emph{vertex distinguishing closed total ice-flower system} if two closed local ve-color sets $C_{ve}[x_i,f_i]\neq C_{ve}[x_j,f_j]$ for two centers $x_i\in V(K^c_{1,m_i})$ and $x_j\in V(K^c_{1,m_j})$. and
\item A \emph{vertex distinguishing closed (4)-ice-flower system} if two closed local (4)-color sets $C_{ve}\{x_i,f_i\}\neq C_{ve}\{x_j,f_j\}$ for two centers $x_i\in V(K^c_{1,m_i})$ and $x_j\in V(K^c_{1,m_j})$.\qqed
\end{asparaenum}
\end{defn}

\begin{defn} \label{defn:adjacent-vertex-equipotential}
\cite{Yao-Yang-Yao-2020-distinguishing} Let $f:V(G)\rightarrow [1,k]$ be a proper vertex coloring of a graph $G$, we call $f$ an \emph{adjacent-vertex equipotential proper v-coloring} if $C_v[u,f]=C_v[v,f]$ for each edge $uv\in E(G)$.\qqed
\end{defn}

A $4$-colorable maximal planar graph $G$ holds $C_v[u,f]=C_v[v,f]$ for each edge $uv\in E(G)$, where $f$ is an arbitrary $4$-coloring of $G$.

\begin{rem}\label{rem:333333}
For making Topsnut-gpws and Topsnut-matchings more complex, we can use distinguishing edge-colorings and distinguishing total colorings, since they match with the following open problems:

(i) In \cite{Zhang-Liu-Wang-2002} Zhang \emph{et al.} show a famous conjecture: For every graph $G$ with no $K_2$ or $C_5$ component, then the adjacent strong edge chromatic number $\chi\,'_{as}(G) \leq \Delta(G)+2$.

(ii) In \cite{Zhang-Chen-Li-Yao-Lu-Wang-China-2005} Zhang \emph{et al.} introduced a concept of adjacent vertex distinguishing total coloring (AVDTC), and show a conjecture: Let $G$ be a simple graph with order $n \geq 2$; then $G$ has its AVDTC chromatic number $\chi\,''_{as}(G) \leq \Delta(G)+2$.

(iii) In \cite{Yang-Ren-Yao-Ars-2016} and \cite{Yang-Yao-Ren-Information-2016}, Yang \emph{et al.} proposed the following conjectures (\cite{Yang-Ren-Yao-Ars-2016, Yang-Yao-Ren-Information-2016}): Every simple graph $G $ having at least a $4$-avdtc holds $\chi\,''_{4as}(G) \leq \Delta(G)+4$.\paralled
\end{rem}


\section{Matrices in topological coding }

In \cite{Yao-1909-01587-2019} the authors introduce topological coding matrices (abbreviated as Topcode-matrices) and topological matrices. Topcode-matrices are matrices of order $3\times q$ and differ from popular matrices applied in linear algebra and computer science. Topcode-matrices can use numbers, letters, Chinese characters, sets, graphs, algebraic groups \emph{etc.} as their elements. One important thing is that Topcode-matrices of numbers can derive easily number strings, since number strings are text-based passwords used in information security. Topcode-matrices can be used to describe topological graphic passwords (Topsnut-gpws) used for solving some problems coming from the investigation of Graph Networks and Graph Neural Networks proposed by GoogleBrain and DeepMind \cite{Battaglia-27-authors-arXiv1806-01261v2}.

As known, both \emph{adjacency matrix} and \emph{incidence matrix} are traditional matrices of graph theory. There are other type of matrices: Topcode-matrices, set-matrices, Hanzi-matrices, adjacent e-value matrices, Cds-matrices and adjacent ve-value matrices, and so on in topological coding.

\begin{figure}[h]
\centering
\includegraphics[width=16cm]{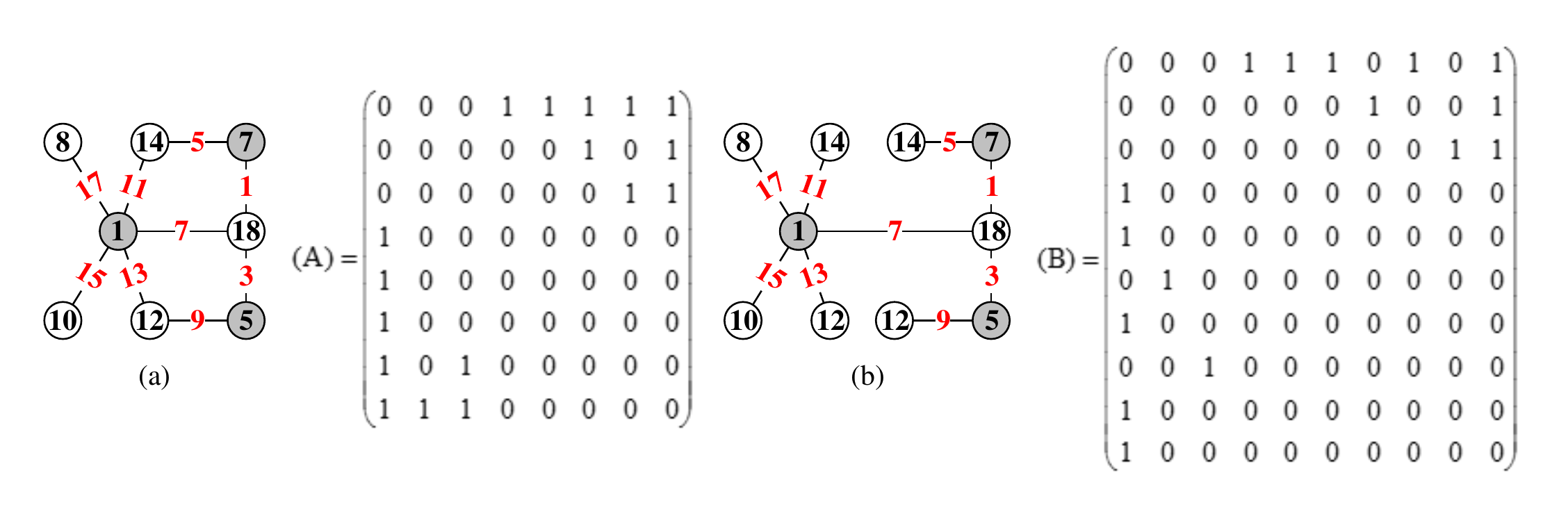}\\
\caption{\label{fig:adjacency-edge-value-matrix}{\small Two Topsnut-gpws (a) and (b) and their adjacent matrices (A) and (B).}}
\end{figure}

\begin{equation}\label{eqa:Six-Topsnut-gpws-vs-one-matrix}
\centering
{
\begin{split}
A(6)= \left(
\begin{array}{ccccccccc}
7 & 5 & 7 &1 & 5 & 1 &1 & 1& 1\\
1 & 3 & 5 &7 & 9 & 11 &13& 15& 17\\
18& 18 & 14 &18& 12 & 14 &12& 10& 8
\end{array}
\right)
\end{split}}
\end{equation}

\begin{figure}[h]
\centering
\includegraphics[width=16.2cm]{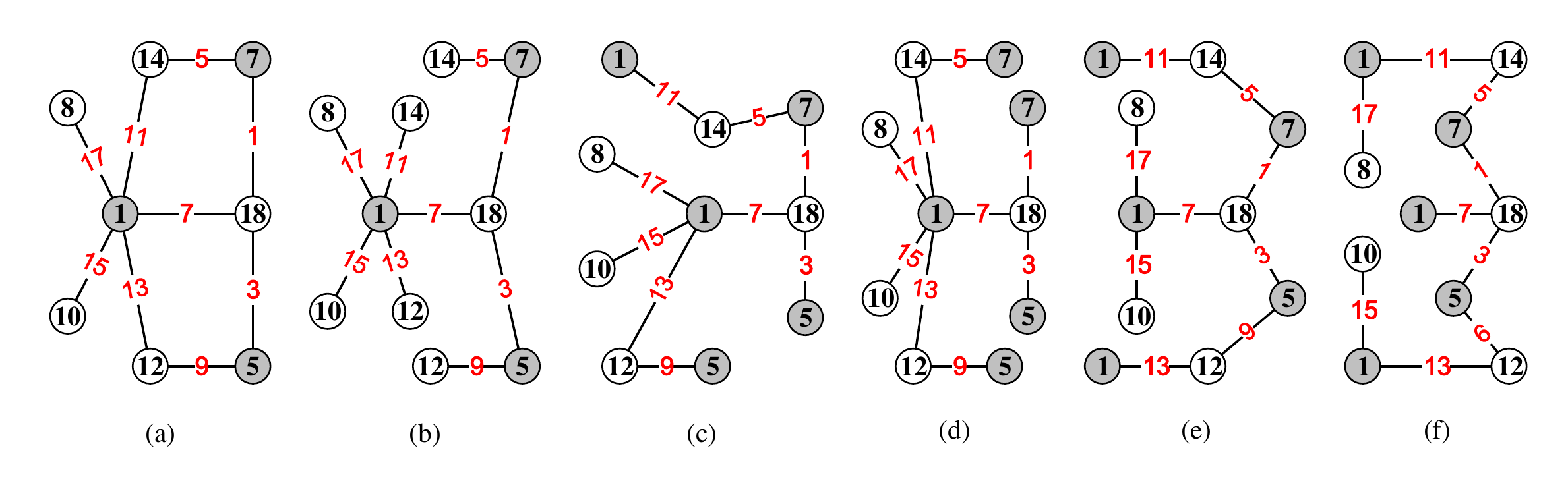}\\
\caption{\label{fig:1-matrix-vs-more-gpws}{\small Six Topsnut-gpws correspond a common Topcode-matrix $A(6)$ shown in (\ref{eqa:Six-Topsnut-gpws-vs-one-matrix}).}}
\end{figure}

A common Topcode-matrix $A(6)$ shown in (\ref{eqa:Six-Topsnut-gpws-vs-one-matrix}) corresponds six Topsnut-gpws shown in Fig.\ref{fig:1-matrix-vs-more-gpws}, and these six Topsnut-gpws are not isomorphic from each other.

\begin{rem}\label{rem:333333}
Graphs have their own incident matrices similarly with Topcode-matrices defined in Definition \ref{defn:topcode-matrix-definition}. A similar concept, called \emph{Topsnut-matrix}, was introduced in \cite{Yao-Sun-Zhao-Li-Yan-ITNEC-2017, Yao-Zhang-Sun-Mu-Sun-Wang-Wang-Ma-Su-Yang-Yang-Zhang-2018arXiv} for studying Topsnut-gpws and producing number-based strings. Pr\"{u}fer Code is a part of Topcode-matrices and can be used to prove Cayley's formula $\tau(K_n) = n^{n-2}$, where $\tau(K_n)$ is the number of spanning trees in a complete graph $K_n$ \cite{Bondy-2008}.\paralled
\end{rem}

\subsection{Traditional Topcode-matrices}

In Definition \ref{defn:topcode-matrix-definition}, we collect all different elements of two v-vectors $X$ and $Y$ into a set $(XY)^*$, and all of different elements of the e-vector $E$ into a set $E^*$, as well as $x_i\neq y_i$ with $i\in [1,q]$.
\begin{defn}\label{defn:particular-topcode-matrices}
\cite{Yao-1909-01587-2019} Let a Topcode-matrix $T_{code}$ be defined in Definition \ref{defn:topcode-matrix-definition} and let $k$ be a non-negative integer, the following constraint conditions:
\begin{asparaenum}[\textrm{Cond}-1. ]
\item \label{asparae:no-same} $(XY)^*=[0,p-1]$ with $p\leq q+1$;
\item \label{asparae:full} $|(XY)^*|=p\leq q+1$;
\item \label{asparae:graceful-no-same} $(XY)^*\subset [0,q]$;
\item \label{asparae:odd-no-same} $(XY)^*\subset [0,2q-1]$;
\item \label{asparae:odd-odd-magic} $(XY)^*\subset [0,2q]$;
\item \label{asparae:edge-magic-total-1} $(XY)^*\cup E^*=[1,p+q]$;
\item \label{asparae:only-E-graceful} $E^*=[1,q]$;
\item \label{asparae:only-E-odd} $E^*=[1,2q-1]^o$;
\item \label{asparae:super-edge-magic-total} $(XY)^*=[1,p]$ and $E^*=[p+1,p+q]$;
\item \label{asparae:graceful} $e_i=|x_i-y_i|$ for $i\in [1,q]$, and $E^*=[1,q]$;
\item \label{asparae:odd-graceful} $e_i=|x_i-y_i|$ for $i\in [1,q]$, and $E^*=[1,2q-1]^o$;
\item \label{asparae:odd-elegant} $e_i=x_i+y_i~(\bmod~2q)$ for $i\in [1,q]$, and $E^*=[1,2q-1]^o$;
\item \label{asparae:elegant} $e_i=x_i+y_i~(\bmod~q)$, $i\in [1,q]$, and $E^*=[0,q-1]$;
\item \label{asparae:edge-magic-total-2} $x_i+e_i+y_i=k$ for $i\in [1,q]$;
\item \label{asparae:edge-vs-difference} $e_i+|x_i-y_i|=k$ for $i\in [1,q]$;
\item \label{asparae:felicitous-differences} $|x_i+y_i-e_i|=k$ for $i\in [1,q]$;
\item \label{asparae:graceful-differences} $\big |e_i-|x_i-y_i|\big |=k$ for $i\in [1,q]$; and
\item \label{asparae:set-ordered} $\max \{x_i:$ $i\in [1,q]\}<\min \{y_j:~j\in [1,q]\}$.
\end{asparaenum}
\noindent \textbf{There are particular Topcode-matrices as follows}:
\begin{asparaenum}[\textrm{Topmatrix}-1. ]
\item $T_{code}$ is a \emph{set-ordered Topcode-matrix} if Cond-\ref{asparae:set-ordered} holds true.
\item $T_{code}$ is a \emph{graceful Topcode-matrix} if Cond-\ref{asparae:graceful-no-same}, Cond-\ref{asparae:only-E-graceful} and Cond-\ref{asparae:graceful} hold true.
\item $T_{code}$ is a \emph{set-ordered graceful Topcode-matrix} if Cond-\ref{asparae:graceful-no-same}, Cond-\ref{asparae:only-E-graceful}, Cond-\ref{asparae:set-ordered} and Cond-\ref{asparae:graceful} hold true.
\item $T_{code}$ is an \emph{odd-graceful Topcode-matrix} if Cond-\ref{asparae:odd-no-same}, Cond-\ref{asparae:only-E-odd} and Cond-\ref{asparae:odd-graceful} hold true.
\item $T_{code}$ is a \emph{set-ordered odd-graceful Topcode-matrix} if Cond-\ref{asparae:odd-no-same}, Cond-\ref{asparae:only-E-odd}, Cond-\ref{asparae:odd-graceful} and Cond-\ref{asparae:set-ordered} hold true.
\item $T_{code}$ is an \emph{elegant Topcode-matrix} if Cond-\ref{asparae:no-same} and Cond-\ref{asparae:elegant} hold true.
\item $T_{code}$ is an \emph{et-ordered elegant Topcode-matrix} if Cond-\ref{asparae:no-same}, Cond-\ref{asparae:elegant} and Cond-\ref{asparae:set-ordered} hold true.
\item $T_{code}$ is an \emph{edge-magic total Topcode-matrix} if Cond-\ref{asparae:edge-magic-total-1} and Cond-\ref{asparae:edge-magic-total-2} hold true.
\item $T_{code}$ is a \emph{super edge-magic total Topcode-matrix} if Cond-\ref{asparae:super-edge-magic-total} and Cond-\ref{asparae:edge-magic-total-2} hold true.
\item $T_{code}$ is an \emph{edge-difference Topcode-matrix} if Cond-\ref{asparae:edge-magic-total-1} and Cond-\ref{asparae:edge-vs-difference} hold true.
\item $T_{code}$ is a \emph{felicitous-difference Topcode-matrix} if Cond-\ref{asparae:only-E-graceful} and Cond-\ref{asparae:felicitous-differences} hold true.
\item $T_{code}$ is a \emph{graceful-difference Topcode-matrix} if Cond-\ref{asparae:only-E-graceful} and Cond-\ref{asparae:graceful-differences} hold true.
\item $T_{code}$ is an \emph{odd-edge-magic total Topcode-matrix} if Cond-\ref{asparae:full}, Cond-\ref{asparae:odd-no-same} and Cond-\ref{asparae:odd-graceful} hold true.
\item $T_{code}$ is an \emph{odd-elegant Topcode-matrix} if Cond-\ref{asparae:odd-no-same} and Cond-\ref{asparae:odd-elegant} hold true.
\item $T_{code}$ is a \emph{harmonious Topcode-matrix} if Cond-\ref{asparae:graceful-no-same} and Cond-\ref{asparae:elegant} hold true.
\item $T_{code}$ is a \emph{perfect odd-graceful Topcode-matrix} if Cond-\ref{asparae:odd-no-same}, Cond-\ref{asparae:only-E-odd} and $\{|a-b|:~a,b\in (XY)^*\}=[1,|(XY)^*|]$ hold true.\qqed
\end{asparaenum}
\end{defn}

There are four parameterized Topcode-matrices shown in (\ref{eqa:Four-parameter-Topcode-matrices-11}), (\ref{eqa:Four-parameter-Topcode-matrices-22}), (\ref{eqa:Four-parameter-Topcode-matrices-33}) and (\ref{eqa:Four-parameter-Topcode-matrices-44}), cited from \cite{Bing-Yao-2020arXiv}.

\begin{equation}\label{eqa:Four-parameter-Topcode-matrices-11}
\centering
{
\begin{split}
T^1_{code}= \left(
\begin{array}{ccccccccc}
2d & 2d & d & 2d & d & 0 & d & 0 & 0\\
k & k+d & k+2d & k+3d & k+4d & k+5d & k+6d & k+7d & k+8d\\
k+2d & k+3d & k+3d & k+5d & k+5d & k+5d & k+7d & k+7d & k+8d
\end{array}
\right)
\end{split}}
\end{equation}

\begin{equation}\label{eqa:Four-parameter-Topcode-matrices-22}
\centering
{
\begin{split}
T^2_{code}= \left(
\begin{array}{ccccccccc}
d & 2d & 2d & 0 & 0 & d & 0 & d & 2d\\
k & k+d & k+2d & k+3d & k+4d & k+5d & k+6d & k+7d & k+8d\\
k+8d & k+8d & k+9d & k+3d & k+4d & k+4d & k+6d & k+6d & k+6d
\end{array}
\right)
\end{split}}
\end{equation}

\begin{equation}\label{eqa:Four-parameter-Topcode-matrices-33}
\centering
{
\begin{split}
T^3_{code}= \left(
\begin{array}{ccccccccc}
2d & 2d & d & 2d & d & 0 & d & 0 & 0\\
k & k+d & k+2d & k+3d & k+4d & k+5d & k+6d & k+7d & k+8d\\
k+8d & k+7d & k+7d & k+5d & k+5d & k+5d & k+3d & k+3d & k+2d
\end{array}
\right)
\end{split}}
\end{equation}

\begin{equation}\label{eqa:Four-parameter-Topcode-matrices-44}
\centering
{
\begin{split}
T^4_{code}= \left(
\begin{array}{ccccccccc}
0 & 0 & d & 0 & d & 2d & d & 2d & 2d\\
k & k+d & k+2d & k+3d & k+4d & k+5d & k+6d & k+7d & k+8d\\
k+2d & k+3d & k+3d & k+5d & k+5d & k+5d & k+7d & k+7d & k+8d
\end{array}
\right)
\end{split}}
\end{equation}

\subsection{Particular sub-Topcode-matrices}

Let $T_{code}$ be a Topcode-matrix defined in Definition \ref{defn:topcode-matrix-definition} in the following discussion \cite{Yao-1909-01587-2019}:

(1) \emph{Perfect matching.} If we have a subset $E_S=\{e_{i_1},e_{i_2},\dots ,e_{i_m}\}\subset E^*$ in $T_{code}$, and there exists $w\in (XY)^*$ to be not a common end of any two $e_{i_j},e_{i_t}$ of $E_S$, and the set of all ends of $E_S$ is just equal to $(XY)^*$, then we call $E_S$ a \emph{perfect matching} of $T_{code}$, and $(X_S,E_S,Y_S)^{T}$ a \emph{perfect matching sub-Topcode-matrix} of $T_{code}$.

2) \emph{Complete Topcode-matrices.} If any $x_i\in (XY)^*$ matches with each $y_j\in (XY)^*\setminus \{x_i\}$ such that $x_i,y_j$ are the ends of some $e_{ij}\in E^*$ in $T_{code}$, and $q=\frac{p(p-1)}{2}$ with $p=|(XY)^*|$, we call $T_{code}$ a \emph{complete Topcode-matrix}.

(3) \emph{Clique Topcode-matrices.} Let $T\,'_{code}$ be a proper sub-Topcode-matrix of $T_{code}$. If $T\,'_{code}$ is a complete Topcode-matrix, we call $T\,'_{code}$ a \emph{clique Topcode-matrix}.

(4) \emph{Paths and cycles in Topcode-matrices.} Notice that $x_i$ and $y_i$ are the \emph{ends} of $e_i$ in $T_{code}$. We define two particular phenomenons in Topcode-matrices: If there are $e_{i_1},e_{i_2},\dots ,e_{i_m}\in E^*$ in $T_{code}$, such that $e_{i_j}$ and $e_{i_{j+1}}$ with $j\in [1,m-1]$ have a common end $w_{i_j}$ holding $w_{i_s}\neq w_{i_t}$ for $s\neq t$ and $1\leq s,t\leq m$. Then $T_{code}$ has a \emph{path}, denote this path as
$$P(x\,'_{i_1}\rightarrow y\,'_{i_m})=x\,'_{i_1}e_{i_1}w_{i_1}e_{i_2}w_{i_2}\cdots e_{i_m-1}w_{i_m-1}e_{i_m}y\,'_{i_m}$$
where $x\,'_{i_1}$ is an end of $e_{i_1}$, and $y\,'_{i_m}$ is an end of $e_{i_m}$ with $x\,'_{i_1}\neq y\,'_{i_m}$, $x\,'_{i_1}\neq w_{i_j}$ and $y\,'_{i_m}\neq w_{i_j}$ with $j\in[1,m-1]$; and moreover if $x\,'_{i_1}= y\,'_{i_m}=ww_{i_m}$ in $P(x\,'_{i_1}\rightarrow y\,'_ {i_m})$, we say that $T_{code}$ has a \emph{cycle} $C$, denoted as $$C=w_{i_m}e_{i_1}w_{i_1}e_{i_2}w_{i_2}\cdots e_{i_m-1}w_{i_m-1}e_{i_m}w_{i_m}.$$
We call $T_{code}$ to be \emph{connected} if any pair of distinct numbers $w_i,w_j\in (XY)^*$ is connected by a path $P(w_{i}\rightarrow w_{j})$.

(5) \emph{Spanning sub-Topcode-matrices.} Let $T\,'_{code}=(X_L$,$E_L$,$Y_L)^{T}$ be a sub-Topcode-matrix of $T_{code}$, where $E_L=\{e_{i_1},e_{i_2},\dots ,e_{i_m}\}$ with $e_{i_j}\in E^*$. If $(X_LY_L)^*=$ $(XY)^*$, we call $T\,'_{code}$ a \emph{spanning sub-Topcode-matrix} of $T_{code}$. Moreover, if this spanning sub-Topcode-matrix $T\,'_{code}$ corresponds a tree, we call it a \emph{spanning tree Topcode-matrix}.

(6) \emph{Euler's Topcode-matrices and Hamilton Topcode-matrices.} If the number of times for each $w\in (XY)^*$ appeared in $T_{code}$ is even, then we call $T_{code}$ an \emph{Euler's Topcode-matrix}. Moreover, if $T_{code}$ has a cycle with $q$ edges $e_{i_k}$, then we call $T_{code}$ a \emph{Hamilton Topcode-matrix}.

(7) \emph{Neighbor sets of Topcode-matrices.} For each $(x_i$, $e_i$, $y_i)^{T}$ $\subset T_{code}$, we call set $N_{ei}(x_i)=\{w_j:$ $(x_i,e_{s}$, $w_j)^{T}\subset T_{code}\}$ as \emph{v-neighbor set} of $x_i$, $N_{ei}(y_i)=\{z_t:~(z_t$, $e_{r},y_i)^{T}\subset T_{code}\}$ as \emph{v-neighbor set} of $y_i$ and $N'_{ei}(x_i)=\{e_{s}:$ $(x_i,e_{s},w_j)^{T}\subset T_{code}\}$ as \emph{e-neighbor set} of $e_i$.

\subsection{Topcode-matrices with more restrictions}

\begin{defn}\label{defn:twin-(k,d)-Topcode-matrices}
\cite{Yao-1909-01587-2019} Let $T_{code}$ be a Topcode-matrix defined in Definition \ref{defn:topcode-matrix-definition}. There are the following restrictions on Topcode-matrices:
\begin{asparaenum}[Comp-1. ]
\item \label{asparae:strongly-matching} There exists a subset $M(T_{code})=\{e_{k_i}:i\in[1,m],e_{k_i}\in E^*\}\subset E^*$ with $q=2m$ such that each $x_i$ in $X$ is an end of some $e_{k_i}\in M(T_{code})$, and each $y_j$ in $Y$ is an end of some $e_{k_j}\in M(T_{code})$;
\item \label{asparae:strongly-constant} there exists a constant $k$ such that two ends $x_{k_i},y_{k_i}$ of each $e_{k_i}\in M(T_{code})$ hold $x_{k_i}+y_{k_i}=k$;
\item \label{asparae:(k,d)-11} each $e_i\in E^*$ is valuated as $e_i=|x_i-y_i|$;
\item \label{asparae:relaxed} $e_i=|x_j-y_j|$ or $e_i=2M-|x_j-y_j|$ for each $i\in [1,q]$ and some $j\in [1,q]$, and a constant $M=\theta(q,|(XY)^*|)$;
\item \label{asparae:ee-difference-00} (ee-difference) each $e_i$ with ends $x_{i}$ and $y_{i}$ matches with another $e_j$ with ends $x_{j}$ and $y_{j}$ holding $e_i=2q+|x_j-y_j|$, or $e_i=2q-|x_j-y_j|$, $e_i+e_j=2q$ for each $i\in [1,q]$ and some $j\in [1,q]$;
\item \label{asparae:(k,d)-22} each $e_i\in E^*$ is valuated as $e_i=x_i+y_i~(\bmod~q)$;
\item \label{asparae:edge-magic-graceful} There exists a constant $k$ such that $|x_i+y_i-e_i|=k$ for each $i\in [1,q]$;
\item \label{asparae:total-graceful} $|(XY)^*|=p$, $|E^*|=q$ and $(XY)^*\cup E^*=[1,p+q]$;
\item \label{asparae:total-graceful-k-d} $|(XY)^*|=p$, $|E^*|=q$ and $(XY)^*\cup E^*\subseteq [0,k+(q-1)d]$;
\item \label{asparae:edge-plus-difference} (e-magic) there exists a constant $k$ such that $e_i+|x_i-y_i|=k$ for $i\in [1,q]$;
\item \label{asparae:edge-magic-total-graceful} (total magic) There exists a constant $k$ such that $x_i+e_i+y_i=k$ with $i\in [1,q]$;
\item \label{asparae:EV-ordered} (EV-ordered) $\min (XY)^*>\max E^*$, or $\max (XY)^*$ $<\min E^*$, or $(XY)^*\subseteq E^*$, or $E^*\subseteq(XY)^*$, or $(XY)^*$ is an odd-set and $E^*$ is an even-set;
\item \label{asparae:ee-balanced} (ee-balanced) Let $s(e_i)=|x_i-y_i|-e_i$ for $i\in [1,q]$. There exists a constant $k\,'$ such that each $e_i$ with $i\in [1,q]$ matches with another $e_j$ holding $s(e_i)+s(e_j)=k\,'$ (or $2(q+|(XY)^*|)+s(e_i)+s(e_j)=k\,'$, or $(|(XY)^*|+q+1)+s(e_i)+s(e_j)=k\,'$) true;

\item \label{asparae:ve-matching} (ve-matching) there exists a constant $k\,''$ such that each $e_i\in E^*$ matches with $w\in (XY)^*$, $e_i+w=k\,''$, and each vertex $z\in (XY)^*$ matches with $e_t\in E^*$ such that $z+e_t=k\,''$, except the \emph{singularity} $w=\lfloor \frac{|(XY)^*|+q+1}{2}\rfloor $;
\item \label{asparae:(k,d)-felicitous} $e_i+k=[x_i+y_i-k]~(\bmod~qd)$ for $i\in [1,q]$;
\item \label{asparae:(k,d)-00} $E^*=S_{q-1,k,d}$;
\item \label{asparae:(k,d)-33} $x_i+e_i+y_i\in S_{q-1,2k,2d}$ for $i\in [1,q]$;
\item \label{asparae:(k,d)-arithmetic} $e_i=x_i+y_i\in S_{q-1,k,d}$ for $i\in [1,q]$;
\item \label{asparae:v-general-e-odd} $(XY)^*\subset [0,q-1]$, and $E^*=[1,2q-1]^o$;
\item \label{asparae:odd-6C} Each $e_i$ is odd, $(XY)^*\cup E^*\subset [1,4q-1]$; and
\item \label{asparae:total-consecutive} $\{x_i+e_i+y_i:~i\in [1,q]\}=[a,b]$, where $q=b-a+1$.
\end{asparaenum}

\noindent \textbf{The particular Topcode-matrices are defined as follows}:
\begin{asparaenum}[\textrm{ComMatrix}-1. ]
\item A graceful Topcode-matrix $T_{code}$ is called a \emph{strongly graceful Topcode-matrix} if Comp-\ref{asparae:strongly-matching} and Comp-\ref{asparae:strongly-constant} hold true.
\item An odd-graceful Topcode-matrix $T_{code}$ is called a \emph{strongly odd-graceful Topcode-matrix} if Comp-\ref{asparae:strongly-matching} and Comp-\ref{asparae:strongly-constant} hold true.
\item A Topcode-matrix $T_{code}$ is called a \emph{$(k,d)$-graceful Topcode-matrix} if Comp-\ref{asparae:(k,d)-11} and Comp-\ref{asparae:(k,d)-00} hold true.
\item A Topcode-matrix $T_{code}$ is called a \emph{$(k,d)$-felicitous Topcode-matrix} if Comp-\ref{asparae:(k,d)-felicitous} and Comp-\ref{asparae:(k,d)-00} hold true.
\item A Topcode-matrix $T_{code}$ is called a \emph{$(k,d)$-edge-magic total Topcode-matrix} if Comp-\ref{asparae:edge-magic-total-graceful}, Comp-\ref{asparae:ve-matching} and Comp-\ref{asparae:(k,d)-00} hold true.

\item A Topcode-matrix $T_{code}$ is called a \emph{$(k,d)$-edge antimagic total Topcode-matrix} if Comp-\ref{asparae:total-graceful-k-d}, Comp-\ref{asparae:(k,d)-00} and Comp-\ref{asparae:(k,d)-33} hold true.

\item A Topcode-matrix $T_{code}$ is called a \emph{total graceful Topcode-matrix} if Comp-\ref{asparae:(k,d)-11} and Comp-\ref{asparae:total-graceful} hold true.
\item A Topcode-matrix $T_{code}$ is called a \emph{ve-magic total graceful Topcode-matrix} if Comp-\ref{asparae:total-graceful} and Comp-\ref{asparae:edge-plus-difference} hold true.
\item A Topcode-matrix $T_{code}$ is called a \emph{relaxed edge-magic total Topcode-matrix} if Comp-\ref{asparae:total-graceful}, Comp-\ref{asparae:edge-magic-total-graceful} and Comp-\ref{asparae:relaxed} hold true.

\item A Topcode-matrix $T_{code}$ is called an \emph{edge-magic graceful Topcode-matrix} if Comp-\ref{asparae:edge-magic-graceful} and Comp-\ref{asparae:total-graceful} hold true.

\item A Topcode-matrix $T_{code}$ is called a \emph{6C-Topcode-matrix} if Comp-\ref{asparae:relaxed}, Comp-\ref{asparae:total-graceful}, Cond-\ref{asparae:set-ordered}, Comp-\ref{asparae:edge-plus-difference}, Comp-\ref{asparae:EV-ordered}, Comp-\ref{asparae:ee-balanced} and Comp-\ref{asparae:ve-matching} hold true.

\item A Topcode-matrix $T_{code}$ is called an \emph{odd-6C Topcode-matrix} if it holds $\{|a-b|:a,b\in (XY)^*\}=[1,2q-1]^o$, Comp-\ref{asparae:ee-difference-00}, Cond-\ref{asparae:set-ordered}, Comp-\ref{asparae:edge-plus-difference}, Comp-\ref{asparae:EV-ordered}, Comp-\ref{asparae:odd-6C}, Comp-\ref{asparae:ee-balanced} and there are two constants $k_1,k_2$ such that each $e_i$ matches with $w\in (XY)^*$ such that $e_i+w=k_1~(\textrm{or }k_2)$ true.

\item A Topcode-matrix $T_{code}$ is called an \emph{ee-difference odd-edge-magic matching Topcode-matrix} if Comp-\ref{asparae:relaxed}, Comp-\ref{asparae:edge-plus-difference}, Comp-\ref{asparae:ee-balanced} and Comp-\ref{asparae:v-general-e-odd} hold true.
\item A Topcode-matrix $T_{code}$ is called an \emph{odd-edge-magic matching Topcode-matrix} if Comp-\ref{asparae:edge-magic-total-graceful} and Comp-\ref{asparae:v-general-e-odd} hold true.
\item A Topcode-matrix $T_{code}$ is called an \emph{edge-odd-graceful total Topcode-matrix} if Comp-\ref{asparae:v-general-e-odd} and Comp-\ref{asparae:total-consecutive} hold true.
\item A Topcode-matrix $T_{code}$ is called an \emph{multiple edge-meaning vertex Topcode-matrix} if $(XY)^*=[0,p-1]$ with $p=|(XY)^*|$, and there are three constants $k,k\,'$ and $k\,''$ such that

\quad (1) $E^*=[1,q]$ and $x_i+e_i+y_i=k$.

\quad (2) $E^*=[p,p+q-1]$ and $x_i+e_i+y_i=k\,'$.

\quad (3) $E^*=[0,q-1]$ and $e_i=x_i+y_i~(\bmod~q)$.

\quad (4) $E^*=[1,q]$ and $|x_i+y_i-e_i|=k\,''$.

\quad (5) $E^*=[1,2q-1]^o$, and $\{x_i+e_i+y_i:~i\in [1,q]\}=[a,b]$ with $b-a+1=q$.\qqed
\end{asparaenum}
\end{defn}

\subsection{Topcode$^+$-matrix groups and Topcode$^-$-matrix groups}

\begin{defn}\label{defn:77-two-graphic-set-groups}
Let $F_m(T^i_{code})=\{T^1_{code},T^2_{code},\dots, T^m_{code}\}$ be a set of Topcode-matrices with each Topcode-matrix $T^i_{code}=(X_i,E_i,Y_i)^{T}$, where v-vector $X_i=(x_{i,1},~x_{i,2},~\cdots,~x_{i,q})$, e-vector $E_i=(e_{i,1},~e_{i,2},~\cdots,~e_{i,q})$ and v-vector $Y_i=(y_{i,1},~y_{i,2},~\cdots,~y_{i,q})$ and there are functions $f_i$ holding $e_{i,r}=f_i(x_{i,r},y_{i,r})$ with $i\in [1,m]$ and $r\in [1,q]$.

(i) For a fixed positive integer $k$, if there exists a constant $M$, such that
\begin{equation}\label{eqa:77-group-operation-1}
(x_{i,r}+x_{j,r}-x_{k,r})~(\bmod~ M)=x_{\lambda,r}\in X_{\lambda}
\end{equation}
and
\begin{equation}\label{eqa:77-group-operation-2}
(y_{i,r}+y_{j,r}-y_{k,r})~(\bmod~ M)=y_{\lambda,r}\in Y_{\lambda}
\end{equation}
where $\lambda=i+j-k~(\bmod~ M)\in [1,m]$ and, $T^{\lambda}_{code}=(X_{\lambda}$, $E_{\lambda}$, $Y_{\lambda})^{T}$ is a Topcode-matrix with $e_{\lambda,r}=f(x_{\lambda,r},y_{\lambda,r})$ for $r\in [1,q]$. Then we say (\ref{eqa:77-group-operation-1}) and (\ref{eqa:77-group-operation-2}) to be an \emph{additive v-operation} on the set $F_mF_m(T^i_{code})$, and we write this operation by ``$\oplus$'', so $T^i_{code}\oplus T^j_{code}=T^{\lambda}_{code}$ under any \emph{preappointed zero} $T^k_{code}$. Then we call $F_m(T^i_{code})$ an \emph{every-zero Topcode$^+$-matrix group}, denoted as $\{F_m(T^i_{code});\oplus\}$.

(ii) For a fixed Topcode-matrix $T^k_{code}\in F_m$, if there exists a constant $M$, such that
\begin{equation}\label{eqa:77-subtraction-group-operation-1}
(x_{i,r}-x_{j,r}+x_{k,r})~(\bmod~ M)=x_{\mu,r}\in X_{\mu}
\end{equation}
and
\begin{equation}\label{eqa:77-subtraction-group-operation-2}
(y_{i,r}-y_{j,r}+y_{k,r})~(\bmod~ M)=y_{\lambda,r}\in Y_{\mu}
\end{equation}
where $\mu=i-j+k~(\bmod~ M)\in [1,m]$ and, $T^{\mu}_{code}=(X_{\mu},E_{\mu},Y_{\mu})^{T}\in F_m$, where $e_{\mu,s}=f(x_{\mu,s},y_{\mu,s})$ for $s\in [1,q]$. Then two equations (\ref{eqa:77-subtraction-group-operation-1}) and (\ref{eqa:77-subtraction-group-operation-2}) define a new operation, called the \emph{subtractive v-operation}, denoted as $T^i_{code}\ominus_k T^j_{code}=T^{\lambda}_{code}$ under any \emph{preappointed zero} $T^k_{code}$. Thereby, we call $F_m(T^i_{code})$ an \emph{every-zero Topcode$^-$-matrix group}, denoted as $\{F_m(T^i_{code});\ominus\}$.\qqed
\end{defn}
\begin{rem}\label{rem:77-two-graphic-set-groups}
\textbf{Two graphic-set groups.} Since each Topcode-matrix $T^i_{code}$ is accompanied by a graph set $G_{rap}(T^i_{code})$ such that every graph $G\in G_{rap}(T^i_{code})$ has its own Topcode-matrix $T_{code}(G)=T^i_{code}$, so the set $F_m(G_{rap}(T^i_{code}))$ of graph sets $G_{rap}(T^1_{code}),G_{rap}(T^2_{code}),\dots ,G_{rap}(T^m_{code})$, also, forms an \emph{every-zero graphic-set group}, denoted as $\{F_m(G_{rap}(T^i_{code}));\oplus\}$ based on the operation ``$\oplus$'' defined in (\ref{eqa:77-group-operation-1}) and (\ref{eqa:77-group-operation-2}), that is,
\begin{equation}\label{eqa:555555}
G_{rap}(T^i_{code})\oplus G_{rap}(T^j_{code})=G_{rap}(T^{\lambda}_{code})
\end{equation}
with $\lambda=i+j-k~(\bmod~M)$ for any \emph{preappointed zero} $G_{rap}(T^k_{code})$. And moreover we have another every-zero graphic-set group $\{F_m(G_{rap}(T^i_{code}));\ominus\}$ based on the operation ``$\ominus$'' defined in (\ref{eqa:77-subtraction-group-operation-1}) and (\ref{eqa:77-subtraction-group-operation-2}), so
\begin{equation}\label{eqa:555555}
G_{rap}(T^i_{code})\ominus G_{rap}(T^j_{code})=G_{rap}(T^{\mu}_{code})
\end{equation}
with $\mu=i-j+k~(\bmod~M)$ for any \emph{preappointed zero} $G_{rap}(T^k_{code})$.\paralled
\end{rem}

\subsection{Matching type of Topcode-matrices}

We show some connections between Topcode-matrices here. Let $X_0=\{0,d,2d, \dots ,(q-1)d\}$. The set $S^*$ collects all different elements of a vector (or collection) $S$, and Topcode-matrices are defined in Definition \ref{defn:topcode-matrix-definition} hereafter.
\begin{defn}\label{defn:twin-(k,d)-Topcode-matrices}
\cite{Yao-1909-01587-2019} The following Topcode-matrices are defined in Definition \ref{defn:topcode-matrix-definition}.
\begin{asparaenum}[\textrm{Matching}-1 ]
\item If two Topcode-matrices $T_{code}=(X,E,Y)^{T}$ and $\overline{T}_{code}=(U,W,V)^{T}$ hold $(X_0\cup S_{q-1,k,d})\setminus (XY)^*\cup E^*=(UV)^*\cup W^*$ true, then $\overline{T}_{code}$ is called a \emph{complementary $(k,d)$-Topcode-matrix} of $T_{code}$, and $(T_{code},\overline{T}_{code})$ is called a \emph{twin $(k,d)$-Topcode-matrices}.
\item Two Topcode-matrices $T^t_{code}=(X^t$, $E^t$, $Y^t)^{T}$ with $t=1,2$ correspond a $(p,q)$-graph $G$, and satisfy that $(X^tY^t)^*\subset X_0\cup S_{q-1,k,d}$, and $e^t_i-k=[x^t_i+y^t_i-k~(\textrm{mod}~qd)]$ with $t=1,2$ and $i\in[1,q]$. If $e^1_i+e^2_i=2k+(q-1)d$ with $i\in[1,q]$, then $(T^1_{code},T^2_{code})$ is called a \emph{matching of $(k,d)$-harmonious image Topcode-matrices}.
\item Two Topcode-matrices $T^t_{code}=(X^t$, $E^t$, $Y^t)^{T}$ with $t=1,2$ correspond a $(p,q)$-graph $G$, and $e^t_i=|x^t_i-y^t_i|$ with $t=1,2$ and $i\in[1,q]$. If there exists a positive constant $k$ such that $e^1_i+e^2_i=k$ with $i\in[1,q]$, then $(T^1_{code},T^2_{code})$ is called a \emph{matching of image Topcode-matrices}, and $T^t_{code}$ is the \emph{mirror-image} of $T^{3-t}_{code}$ with $t=1,2$.
\item Suppose that a Topcode-matrix $T_{code}=(X,E,Y)^{T}=\uplus^m_{k=1} T^k_{code}$ with $m\geq 2$, where $T^k_{code}=(X^k,E^k,Y^k)^{T}$. If $(XY)^*=[0, |(XY)^*|-1]$, and each $T^k_{code}$ is a $W$-type Topcode-matrix, then $T_{code}$ is called a \emph{multiple matching $W$-type Topcode-matrix}.\qqed
\end{asparaenum}
\end{defn}

\subsection{Pan-Topcode-matrices with elements are not numbers}

Pan-Topcode-matrices are some particular Topcode-matrices having elements being sets, Hanzis (Chinese characters), graphs, graphic groups and things around us.

\begin{defn}\label{defn:pan-Topcode-matrices}
\cite{Yao-1909-01587-2019} Let $T_{pcode}=(P_X,P_E,P_Y)^{T}_{3\times q}$ be a \emph{pan-Topcode-matrix}, where three vectors $P_X=(\alpha_1,\alpha_2,\dots ,\alpha_q)$, $P_E=(\gamma_1,\gamma_2,\dots ,\gamma_q)$ and $P_Y=(\beta_1,\beta_2,\dots ,\beta_q)$. The notation $(P_XP_Y)^*$ is the set of different elements of the union set $P_X\cup P_Y$, and the symbol $P^*_E$ is the set of different elements of the vector $P_E$. If there exists an operation ``$(\ast)$'' such that $\gamma_i=(\ast)\langle \alpha_i, \beta_i\rangle $ for $i\in [1,q]$, we say $T_{pcode}$ to be \emph{operation-valued}.\qqed
\end{defn}

\begin{defn}\label{defn:set-odd-graceful-Topcode-matrix}
\cite{Yao-1909-01587-2019} Using notation and terminology defined in Definition \ref{defn:pan-Topcode-matrices}. Topcode-matrices related with sets are defined as follows:
\begin{asparaenum}[\textrm{Set}-1. ]
\item A pan-Topcode-matrix $T_{pcode}$ has $(P_XP_Y)^*\subseteq [1,q]^2$ and $P^*_E\subseteq [1,q]^2$, and each set $\gamma_i=\alpha_i\cap \beta_i$. If we can select a \emph{representative} $a_{i}\in \gamma_i$ such that the representative set $\{a_{i}:~\gamma_i\in P^*_E\}=[1,q]$, then $T_{pcode}$ is called a \emph{graceful intersection total set-Topcode-matrix}.
\item A pan-Topcode-matrix $T_{pcode}$ holds $(P_XP_Y)^*\subseteq [1,2q]^2$ and $P^*_E\subseteq [1,2q-1]^2$, and each set $\gamma_i=\alpha_i\cap \beta_i$. If we can select a \emph{representative} $a_{i}\in \gamma_i$ such that the representative set $\{a_{i}:~\gamma_i\in P^*_E\}=[1,2q-1]^o$, then $T_{pcode}$ is called an \emph{odd-graceful intersection total set-Topcode-matrix}.
\item A pan-Topcode-matrix $T_{pcode}$ has $(P_XP_Y)^*\subseteq [1,q]^2$ (resp. $[1,2q]^2$) and $P^*_E\subseteq [1,q]^2$ (resp. $[1,2q-1]^2$), and each set $\gamma_i=\alpha_i\cap \beta_i$, $|\gamma_i|=|\{e_i\}|=1$ and $\bigcup ^q_{i=1}\gamma_i=[1,q]$ (resp. $[1,2q-1]^o$), we call $T_{pcode}$ a \emph{set-graceful pan-Topcode-matrix} (resp. a \emph{set-odd-graceful Topcode-matrix}).
\item A pan-Topcode-matrix $T_{pcode}$ has $(P_XP_Y)^*\subseteq [1,q]^2$ and $P^*_E\subseteq [1,q]^2$, and each set $\gamma_i=\alpha_i\setminus \beta_i$ or $\gamma_i=\beta_i\setminus \alpha_i$, $\gamma_i\cap \gamma_j=\emptyset$ for $i\neq j$ and $\bigcup ^q_{i=1}\gamma_i=[a,b]$, we call $T_{pcode}$ a \emph{set-subtraction graceful pan-Topcode-matrix}.

\item A \emph{set-intersecting rainbow Topcode-matrix} $C$ corresponds a tree $H$ with a \emph{set coloring} $h: V(H)\rightarrow S$, where $S=\{[1,1],[1,2],\dots ,[1,n]\}$, such that each set $h(uv)=h(u)\cap h(v)$ for each edge $uv\in E(H)$, and $h(E(H))=\{[1,1],[1,2],\dots ,[1,n-1]\}$. Similarly, we can have a set-union rainbow Topcode-matrix by defining each set $h(uv)=h(u)\cup h(v)$.
\item We call a Topcode-matrix $T_{code}$ defined in Definition \ref{defn:topcode-matrix-definition} a \emph{v-distinguishing Topcode-matrix} if $N_{ei}(x_i)\neq N_{ei}(w_j)$ for any pair of distinct $x_i,w_j\in (XY)^*$. $T_{code}$ is called an \emph{adjacent v-distinguishing Topcode-matrix} if $N_{ei}(x_i)\neq N_{ei}(y_i)$ for each $(x_i$, $e_i$, $y_i)^{T}$ $\in T_{code}$; and $T_{code}$ is called an \emph{adjacent e-distinguishing Topcode-matrix} if $N\,'_{ei}(x_i)\neq N\,'_{ei}(y_i)$ for each $(x_i$, $e_i$, $y_i)^{T}$ $\in T_{code}$,
 $$N_{ei}(s_i)=\{t_j:~\textrm{$s_i$ and $t_j$ are the ends of $e_j$}\},~N\,'_{ei}(s_i)=\{e_j:~\textrm{$s_i$ is an end of $e_j$}\}.$$
 $T_{code}$ is called an \emph{adjacent total-distinguishing Topcode-matrix} if $N_{ei}(x_i)\cup N\,'_{ei}(x_i)\neq N_{ei}(y_i)\cup N\,'_{ei}(y_i)$ for each $e_i$ with $i\in [1,q]$.\qqed
\item Suppose that $(P_XP_Y)^*$ and $P^*_E$ are subsets of an abstract set $A_{bstract}$, and $\gamma_i=\alpha_i\cap \beta_i$ for $\gamma_i\in P_E$ and $\alpha_i, \beta_i\in (P_XP_Y)^*$. If $|\gamma_i|=1$ for $i\in [1,q]$, we call $T_{pcode}$ an \emph{e-graceful pan-Topcode-matrix}; and if $|\gamma_i|=2i-1$ for $i\in [1,q]$, we call $T_{pcode}$ an \emph{e-odd-graceful pan-Topcode-matrix}.
\end{asparaenum}
\end{defn}

\begin{defn}\label{defn:group-odd-graceful-Topcode-matrices}
\cite{Yao-1909-01587-2019} Suppose that $\{F_{n}(H);\oplus\}$ is an \emph{every-zero graphic group} based on a graphic set $F_{n}(H)=\{H_i:i\in [1,n]\}$ with $n\geq q$ ($n\geq 2q-1$) and the additive operation ``$\oplus$'', and moreover $\{F_{n}(H);\ominus\}$ is an \emph{every-zero graphic group} based on the subtractive operation ``$\ominus$''. Using notation and terminology defined in Definition \ref{defn:pan-Topcode-matrices}, there are the following constraint conditions:
\begin{asparaenum}[\textrm{Gcond}-1. ]
\item \label{asparae:vertex-restriction} $(P_XP_Y)^*\subseteq F_{n}(H)$;
\item \label{asparae:edge-restriction} $P^*_E\subseteq F_{n}(H)$;
\item \label{asparae:edge-restriction-grace} $P^*_E=\{H_1,H_2,\dots ,H_q\}$;
\item \label{asparae:edge-restriction-odd-grace} $P^*_E=\{H_1,H_3,\dots ,H_{2q-1}\}$;
\item \label{asparae:additive-operation} $\gamma_i=\alpha_i\oplus \beta_i$;
\item \label{asparae:subtractive-operation} $\gamma_i=\alpha_i\ominus \beta_i$;
\item \label{asparae:vertex-grace} $(P_XP_Y)^*=\{H_1,H_2,\dots ,H_q\}$;
\item \label{asparae:vertex-odd-grace} $(P_XP_Y)^*=\{H_1,H_3,\dots ,H_{2q-1}\}$; and
\item \label{asparae:vertex-set-ordered} $\max\{i:\alpha_i=H_i\}<\min\{j:\beta_j=H_j\}$.
\end{asparaenum}

\noindent \textbf{We have}:
\begin{asparaenum}[\textrm{Matrix} 1. ]
\item A pan-Topcode-matrix $T_{pcode}$ is called an \emph{e-graceful group Topcode$^+$-matrix} if Gcond-\ref{asparae:vertex-restriction}, Gcond-\ref{asparae:edge-restriction-grace} and Gcond-\ref{asparae:additive-operation} hold true.
\item A pan-Topcode-matrix $T_{pcode}$ is called an \emph{e-odd-graceful group Topcode$^+$-matrix} if Gcond-\ref{asparae:vertex-restriction}, Gcond-\ref{asparae:edge-restriction-odd-grace} and Gcond-\ref{asparae:additive-operation} hold true.

\item A pan-Topcode-matrix $T_{pcode}$ is called an \emph{e-graceful group Topcode$^-$-matrix} if Gcond-\ref{asparae:vertex-restriction}, Gcond-\ref{asparae:edge-restriction-grace} and Gcond-\ref{asparae:subtractive-operation} hold true.
\item A pan-Topcode-matrix $T_{pcode}$ is called an \emph{e-odd-graceful group Topcode$^-$-matrix} if Gcond-\ref{asparae:vertex-restriction}, Gcond-\ref{asparae:edge-restriction-odd-grace} and Gcond-\ref{asparae:subtractive-operation} hold true.

\item An e-graceful group Topcode$^+$-matrix $T_{pcode}$ is called a \emph{ve-graceful group Topcode$^+$-matrix} if Gcond-\ref{asparae:vertex-grace} holds true.
\item An e-odd-graceful group Topcode$^+$-matrix $T_{pcode}$ is called a \emph{ve-odd-graceful group Topcode$^+$-matrix} if Gcond-\ref{asparae:vertex-odd-grace} holds true.

\item An e-graceful group Topcode$^-$-matrix $T_{pcode}$ is called a \emph{ve-graceful group Topcode$^-$-matrix} if Gcond-\ref{asparae:vertex-grace} holds true.
\item An e-odd-graceful group Topcode$^-$-matrix $T_{pcode}$ is called a \emph{ve-odd-graceful group Topcode$^-$-matrix} if Gcond-\ref{asparae:vertex-odd-grace} holds true.

\item A Topcode-matrix $T_{pcode}$ is called a \emph{set-ordered group Topcode-matrix} if Gcond-\ref{asparae:vertex-restriction}, Gcond-\ref{asparae:edge-restriction} and Gcond-\ref{asparae:vertex-set-ordered} hold true.

\item An $\epsilon$-matrix $T_{pcode}$ is called a \emph{set-ordered $\varepsilon$-matrix} if Gcond-\ref{asparae:vertex-set-ordered} holds true, where $\varepsilon\in \{$e-graceful group Topcode$^+$, e-odd-graceful group Topcode$^+$, ve-graceful group Topcode$^+$, ve-odd-graceful group Topcode$^+$, e-graceful group Topcode$^-$, e-odd-graceful group Topcode$^-$, ve-graceful group Topcode$^-$, ve-odd-graceful group Topcode$^-$$\}$.\qqed
\end{asparaenum}
\end{defn}

\subsection{Adjacent e-value matrices and adjacent ve-value matrices}

Each colored graph $G$ corresponds an \emph{adjacent e-value matrix} and an \emph{adjacent ve-value matrix} defined in the following:

\begin{defn}\label{defn:ve-value-matrix}
\cite{Yao-Mu-Sun-Sun-Zhang-Wang-Su-Zhang-Yang-Zhao-Wang-Ma-Yao-Yang-Xie2019} An \emph{adjacent ve-value matrix} $VE_{color}(G)=(a_{i,j})_{(p+1)\times (p+1)}$ is defined on a $(p,q)$-graph $G$ admitting a total coloring $g: V(G)\cup E(G)\rightarrow [a,b]$ for $V(G)=\{x_1,x_2,\dots ,x_p\}$ in the way:

(i) $a_{1,1}=0$, $a_{1,j+1}=g(x_j)$ with $j\in[1,p]$, and $a_{k+1,1}=g(x_k)$ with $k\in[1,p]$;

(ii) $a_{i+1,i+1}=0$ with $i\in[1,p]$;

(iii) For an edge $x_{ij}=x_ix_j\in E(G)$ with $i,j\in[1,p]$, then $a_{i+1,j+1}=g(x_{ij})$, and $a_{i+1,j+1}=0$ otherwise.\qqed
\end{defn}

\begin{defn}\label{defn:e-value-matrix}
$^*$ An \emph{adjacent e-value matrix} $E_{color}(G)=(c_{i,j})_{p\times p}$ is defined on a $(p,q)$-graph $G$ admitting a total coloring $f: V(G)\cup E(G)\rightarrow [a,b]$ for $V(G)=\{x_1,x_2,\dots ,x_p\}$ in the way:

(i) $a_{i,j}=f(x_ix_j)$ with $i\in[1,p]$, $j\in[1,p]$ and $i\neq j$;

(ii) $a_{i,i}=0$ with $i\in[1,p]$.\qqed
\end{defn}

\begin{figure}[h]
\centering
\includegraphics[width=10.6cm]{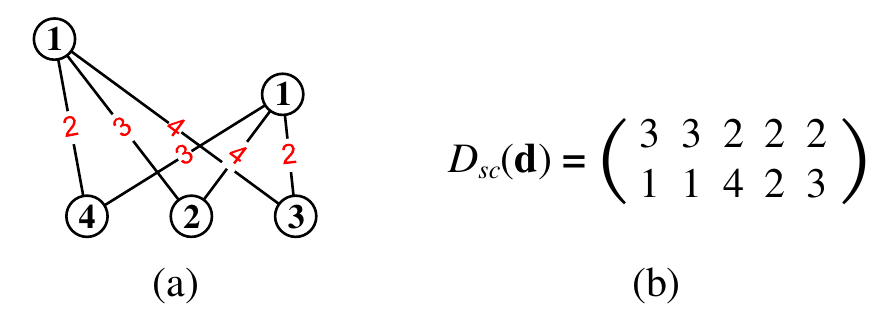}\\
\caption{\label{fig:for-explain-various-matrices}{\small (a) The colored complete graph $K_{2,3}$; (b) the colored degree-sequence matrix of the colored complete graph $K_{2,3}$.}}
\end{figure}

As an example, a matrix $A_{djacent}(K_{2,3})$ is for illustrating the adjacent matrix of a graph $K_{2,3}$ shown in Fig.\ref{fig:for-explain-various-matrices}(a), so $K_{2,3}$ has its own adjacent matrix $A(K_{2,3})$ shown in (\ref{eqa:matrix-t-111}). For generating number-based strings in real application, we have designed the adjacent e-value matrix $E_{color}(K_{2,3})$ and adjacent ve-value matrix $VE_{color}(K_{2,3})$ shown in (\ref{eqa:matrix-t-222}), in which we can construct more number-based strings with longer bytes.

\begin{equation}\label{eqa:matrix-t-111}
\centering
A_{djacent}(K_{2,3})= \left(
\begin{array}{cccccc}
\textrm{vertices} & x_1 & x_2 & x_3 & x_4 & x_5 \\
x_1 & 0 & 0 & 1 & 1 & 1 \\
x_2 & 0 & 0 & 1 & 1 & 1 \\
x_3 & 1 & 1 & 0 & 0 & 0 \\
x_4 & 1 & 1 & 0 & 0 & 0 \\
x_5 & 1 & 1 & 0 & 0 & 0
\end{array}
\right),~A(K_{2,3})= \left(
\begin{array}{ccccc}
0 & 0 & 1 & 1 & 1 \\
0 & 0 & 1 & 1 & 1 \\
1 & 1 & 0 & 0 & 0 \\
1 & 1 & 0 & 0 & 0 \\
1 & 1 & 0 & 0 & 0
\end{array}
\right)
\end{equation}

\begin{equation}\label{eqa:matrix-t-222}
\centering
E_{color}(K_{2,3})= \left(
\begin{array}{cccccc}
0 & 0 & 2 & 3 & 4 \\
 0 & 0 & 3 & 4 & 2 \\
 2 & 3 & 0 & 0 & 0 \\
 3 & 4 & 0 & 0 & 0 \\
 4 & 2 & 0 & 0 & 0
\end{array}
\right),~VE_{color}(K_{2,3})= \left(
\begin{array}{cccccc}
0 & 1 & 1 & 4 & 2 & 3 \\
1 & 0 & 0 & 2 & 3 & 4 \\
1 & 0 & 0 & 3 & 4 & 2 \\
4 & 2 & 3 & 0 & 0 & 0 \\
2 & 3 & 4 & 0 & 0 & 0 \\
3 & 4 & 2 & 0 & 0 & 0
\end{array}
\right)
\end{equation}

\subsection{Cds-matrices based on degree sequences and colorings}

\begin{defn}\label{defn:colored-degree-sequence-matrix}
$^*$ Let $\textbf{\textrm{d}}=(a_1,a_2,\dots ,a_p)$ with $a_j\in Z^0\setminus \{0\}$ be a $p$-rank degree-sequence of a $(p,q)$-graph. There exists a coloring $f: \textbf{\textrm{d}}\rightarrow \{b_1,b_2,\dots ,b_p\}$ with $b_j\in Z^0$, and the following matrix
\begin{equation}\label{eqa:degree-sequence-matrix}
\centering
{
\begin{split}
D_{sc}(\textbf{\textrm{d}})= \left(
\begin{array}{ccccc}
a_1 & a_2 & \cdots & a_p\\
f(a_1) & f(a_2) & \cdots & f(a_p)\\
\end{array}
\right)_{2\times p}=(\textbf{\textrm{d}},~~f(\textbf{\textrm{d}}))^T
\end{split}}
\end{equation} is called a \emph{colored degree-sequence matrix} (Cds-matrix), where $f(\textbf{\textrm{d}})=(f(a_1), f(a_2), \dots , f(a_p))$ is a \emph{colored vector} of the degree sequence $\textbf{\textrm{d}}$.\qqed
\end{defn}

\begin{figure}[h]
\centering
\includegraphics[width=16.4cm]{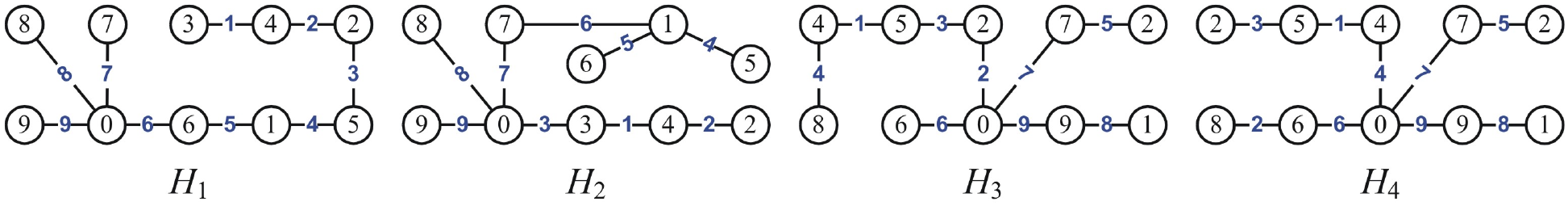}\\
\caption{\label{fig:degree-sequence-colorings}{\small A degree-sequence $\textrm{\textbf{d}}=$(4, 2, 2, 2, 2, 2, 1, 1, 1, 1) is colored by different colorings and labelings, where each graph $H_i$ has its own degree-sequence $\textrm{Deg}(H_i)=\textrm{\textbf{d}}$ with $i\in [1,4]$.}}
\end{figure}

\textbf{Particular degree-sequences.} Let $\textrm{\textbf{d}}=(a_1, a_2, \dots, a_n)$ be a $n$-rank degree-sequence, there are particular degree-sequences as follows \cite{Yao-Zhang-Wang-Su-Integer-Decomposing-2021}:

\begin{asparaenum}[\textrm{\textbf{Par}}-1. ]
\item $\overline{\textbf{\textrm{d}}}=(n-1-a_1$, $n-1-a_2, \dots , n-1-a_n)$ is the \emph{complementary degree-sequence} of the degree-sequence $\textbf{\textrm{d}}$.
\item Partition integer sequence $(a_1, a_2, \dots, a_n)$ into $k$ groups of disjoint integer sequence $(a_{i,1},a_{i,1}, \dots, a_{i,n_i})$ with $i\in [1, k]$ with $k\geq 2$, such that $\{a_{i,1},a_{i,1}, \dots, a_{i,n_i}\}\cap \{a_{j,1},a_{j,1}, \dots, a_{j,n_j}\}=\emptyset $ for $i\neq j$, and $\{a_1, a_2, \dots, a_n\}=\bigcup^k_{i=1} \{a_{i,1},a_{i,1}, \dots, a_{i,n_i}\}$. We call these disjoint integer sequences $(a_{i,1},a_{i,1}, \dots, a_{i,n_i})$ with $i\in [1, k]$ to be a \emph{$k$-partition} of the degree-sequence $\textrm{\textbf{d}}$.
\item A \emph{perfect degree-sequence} $\textrm{\textbf{d}}=(a_1, a_2, \dots, a_n)$ holds: Select $a_{j_1}, a_{j_2}, \dots, a_{j_r}$ from $\textrm{\textbf{d}}$ with $r\geq 2$, such that $a_{j_i}\geq a_{j_{i+1}}$ and $a_{j_1}\leq r-1$, and the new integer sequence $(a_{j_1}, a_{j_2}, \dots, a_{j_r})$ is just a degree-sequence. Some examples of perfect degree-sequences are:

\quad (i) $a_1=n-1$, $a_2=\dots=a_n=k$;

\quad (ii) $\textrm{deg}(K_{m,n})$;

\quad (iii) $a_1=a_2=n-2$, $a_3=\dots=a_n=k$.
\item A \emph{unique graph degree-sequence} corresponds one graph only, for example, (5, 2, 2, 1, 1, 1).
\item A\emph{ prime degree-sequence} with prime number $a_i$ for $i\in [1,n]$, for instance, (5, 3, 3, 3, 3, 1, 1, 1, 1).
\item An \emph{Euler degree-sequence} with each $a_i$ is even for $i\in [1,n]$.
\end{asparaenum}

\begin{prop}\label{thm:unique-graph-degree-sequence}
If a degree-sequence $\emph{\textbf{\textrm{d}}}=(a_1, a_2, \dots, a_n)$ holds $a_1=n-1$ true, and let $\emph{\textbf{\textrm{d}}}\,'=(a_2-1, a_3-1, \dots, a_n-1)$. Then $\emph{\textbf{\textrm{d}}}$ is a unique graph degree-sequence if and only if $\emph{\textbf{\textrm{d}}}\,'$ is a unique graph degree-sequence too.
\end{prop}

\textbf{Operations on degree-sequences. }Let $\textrm{\textbf{d}}=(a_1, a_2, \dots, a_n)$ be a degree-sequence, and let $\textrm{\textbf{d}}\,'=(a\,'_1, a\,'_2, \dots, a\,'_m)$ be another degree-sequence. Some operations shown in the following have been introduced in \cite{Yao-Wang-Ma-Wang-Degree-sequences-2021}.

\begin{asparaenum}[\textrm{\textbf{Doper}}-1. ]
\item \textbf{Increasing (decreasing) degree component operation}: Increasing a new degree component $k$ to a $n$-rank degree-sequence $\textbf{\textrm{d}}_n=(a_1$, $a_2, \dots , a_n)$ produces a $(n+1)$-rank degree-sequence $\textbf{\textrm{d}}_{n+1}=(a\,'_1$, $a\,'_2, \dots , a\,'_n, k)$, where $a\,'_{j_i}=a_{j_i}+1$ for $i\in [1,k]$, others $a\,'_{j_r}=a_{j_r}$ if $r\not \in [1,k]$, denote $\textbf{\textrm{d}}_{n+1}=\textbf{\textrm{d}}_n\uplus (k)$. For example, (3, 2, 2, 1)$\uplus (4)=$(4, 3, 3, 2, 4), (3, 3, 3, 3)$\uplus (1)=$(4, 3, 3, 3, 1), (3, 3, 3, 3)$\uplus (2)=$(4, 4, 3, 3, 2), (3, 3, 3, 3)$\uplus (3)=$(4, 4, 4, 3, 3), and (3, 3, 3, 3)$\uplus (4)=$(4, 4, 4, 4, 4).

\quad The \emph{inverse} of the increasing degree component operation ``$\uplus$'' is denoted as ``$\uplus^{-1}$'', so $\textbf{\textrm{d}}_n=\textbf{\textrm{d}}_{n+1}\uplus^{-1}(a_i)$ defined by removing a degree component $a_i$ from $\textbf{\textrm{d}}_{n+1}$ and arbitrarily select $a_i$ degree components from the remainder to subtract one from each of them. So, ``$\uplus^{-1}$'' is the \textbf{decreasing degree component operation}.

\quad The increasing degree component operation can produce an algorithm for constructing large-scale degree-sequences from small-scale degree-sequences.

\item \textbf{Degree-sequence union (subtraction) operation}:
$$\textbf{\textrm{d}}\cup \textbf{\textrm{d}}\,'=(a_1, a_2, \dots, a_n)\cup (a\,'_1, a\,'_2, \dots, a\,'_m)=(b_1,b_2, \dots, b_{n+m}),
$$ such that $a_i,a\,'_j\in \{b_1, b_2, \dots, b_{n+m}\}$ and $|\textbf{\textrm{d}}\cup \textbf{\textrm{d}}\,'|=|\textrm{\textbf{d}}|+|\textrm{\textbf{d}}\,'|$. Thereby, $\textbf{\textrm{d}}$ and $\textbf{\textrm{d}}\,'$ are two \emph{subdegree-sequences} of $\textbf{\textrm{d}}\cup \textbf{\textrm{d}}\,'$.

\quad The degree-sequence subtraction operation ``$-$'' is the inverse of the degree-sequence union operation: $\textbf{\textrm{d}}-\textbf{\textrm{d}}\,'=(c_1,c_2,\dots, c_t)$ such that $c_i\in \{a_1, a_2, \dots, a_n\}$ and $c_i\not \in \{a_1$, $a_2, \dots, a_n\}\cap \{a\,'_1, a\,'_2, \dots, a\,'_m\}$.
\item \textbf{Component-coinciding (component-splitting) operation}:
$$
\textbf{\textrm{d}}\odot_s \textbf{\textrm{d}}\,'=(b_1, b_2, \dots, b_{n+m-s})
$$ with $\min\{m,n\}\geq s\geq 1$, where $b_{i_j}=a_{r_j}+a\,'_{k_j}$ for $j\in [1,s]$, and others $b_{t_j}\in \{a_1,a_2,\dots$, $a_n$, $a\,'_1$, $a\,'_2$, $\dots $, $a\,'_m\}\setminus $ $\{a_{r_j}, a\,'_{k_j}:j\in [1,s]\}$. For example, (\textbf{\textcolor[rgb]{0.00,0.00,1.00}{4}}, \textbf{\textcolor[rgb]{0.00,0.00,1.00}{3}}, 2, 2, 1)$\odot_2$(\textbf{\textcolor[rgb]{0.00,0.00,1.00}{3}}, \textbf{\textcolor[rgb]{0.00,0.00,1.00}{3}}, 2, 2, 2)=(\textbf{\textcolor[rgb]{0.00,0.00,1.00}{7}}, \textbf{\textcolor[rgb]{0.00,0.00,1.00}{6}}, 2, 2, 2, 2, 2, 1), and (4, 3, 2, \textbf{\textcolor[rgb]{0.00,0.00,1.00}{2}}, \textbf{\textcolor[rgb]{0.00,0.00,1.00}{1}})$\odot_2$(3, 3, \textbf{\textcolor[rgb]{0.00,0.00,1.00}{2}}, \textbf{\textcolor[rgb]{0.00,0.00,1.00}{2}}, 2)=(4, \textbf{\textcolor[rgb]{0.00,0.00,1.00}{4}}, \textbf{\textcolor[rgb]{0.00,0.00,1.00}{3}}, 3, 3, 3, 2, 2).

\quad The degree component-splitting operation ``$\odot^{-1}_s$'' is the inverse of the degree component-coinciding operation ``$\odot$'': $\odot^{-1}_s(\textbf{\textrm{d}})=(\textbf{\textrm{d}}_1, \textbf{\textrm{d}}_2)$, such that $\textbf{\textrm{d}}=\textbf{\textrm{d}}_1\odot_s \textbf{\textrm{d}}_2$. For example, $\odot^{-1}_2$(\textbf{\textcolor[rgb]{0.00,0.00,1.00}{7}}, \textbf{\textcolor[rgb]{0.00,0.00,1.00}{6}}, 2, 2, 2, 2, 2, 1)$=$((\textbf{\textcolor[rgb]{0.00,0.00,1.00}{4}}, \textbf{\textcolor[rgb]{0.00,0.00,1.00}{3}}, 2, 2, 1),(\textbf{\textcolor[rgb]{0.00,0.00,1.00}{3}}, \textbf{\textcolor[rgb]{0.00,0.00,1.00}{3}}, 2, 2, 2)).

\item \textbf{Component decomposition operation} ``$\wedge$'': Decompose some components $a_{i_j}$ into $b_{i_j,1},b_{i_j,2},\dots ,b_{i_j,n_j}$ with $j\in [1,s]$, $a_{i_j}=$ $b_{i_j,1}+b_{i_j,2}+\cdots +b_{i_j,n_j}$, and write $D_{co}=$ $\{a_1, a_2$, $\dots, a_n\}\setminus \{a_{i_1},a_{i_2},\dots ,a_{i_s}\}$, the new degree sequence is denoted as $$\wedge(\textbf{\textrm{d}})=(c_1,c_2,\dots,c_{n-s},b_{i_1,1}, b_{i_1,2},\dots,b_{i_1,n_1} , b_{i_2,1},b_{i_2,2},\dots ,b_{i_2,n_2},\dots ,b_{i_s,1},b_{i_s,2},\dots ,b_{i_s,n_s})
 $$ with $c_i\in $ $D_{co}$. For example, $d_1=$(4, 2, 2, 2, 2, 2, 1, 1, 1, 1), $d_2=$(5, 3, 2, 2, 2, 2, 2) and $d_3=$(5, 4, 3, 2, 2, 2), we have $\wedge(d_{i+1})=d_{i}$ for $i=1,2$.

\quad \textbf{Component compound operation} ``$\vee$'' is the inverse operation of the component-degree decomposition operation ``$\wedge$'': $\vee(\textbf{\textrm{d}})=(e_1,e_2,\dots, e_s)$, where $e_{t_j}=a_{t_j,1}+a_{t_j,2}+\cdots +a_{t_j,n_j}$ with $j\in [1,r]$, and $e_i\in \{a_1, a_2, \dots, a_n\}\setminus \{a_{t_j,1},a_{t_j,2},\dots ,a_{t_j,n_j}:j\in [1,r]\}$ for $i\neq t_j$. For example, we have $\vee(d_{i})=d_{i+1}$ for $i=1,2$.

\item \textbf{Degree-sequence direct-sum operation}: For $m\leq n$, our direct-sum operation $\textbf{\textrm{d}}+ \textbf{\textrm{d}}\,'=\textbf{\textrm{d}}\odot_m \textbf{\textrm{d}}\,'$. Especially, $\textbf{\textrm{d}}+ \overline{\textbf{\textrm{d}}}=(n-1,n-1, \dots , n-1)$, where $\overline{\textbf{\textrm{d}}}=(n-1-a_1, n-1-a_2, \dots, n-1-a_n)$ is the \emph{complementary degree-sequence} of $\textbf{\textrm{d}}$.
\item \textbf{Degree-sequence linear-sum operation ``$\Sigma $''}: Let $\textbf{\textrm{d}}^*=(d^1_n,d^2_n,\dots,d^m_n)$ be a \emph{degree-sequence base}, where $d^k_n=(b_{k,j})^n_{j=1}$ with $k\in [1,m]$ is a degree-sequence. A new integer sequence $(y_{j})^n_{j=1}=\sum^m_{k=1} \lambda_kd^k_n$ with $\lambda_k\in Z^0$, where each $y_{j}=\sum^m_{k=1}\lambda_kb_{k,j}$ with $j\in [1,n]$. We claim that the integer sequence $(y_{j})^n_{j=1}$ is just a degree-sequence by the vertex-coinciding operation on the graphs having degree-sequences $d^k_n$ with $k\in [1,m]$.
\end{asparaenum}

\subsection{Dynamic Topcode-matrices from dynamic networks}
Let $\mathcal {N}(t)$ be a dynamic network at time step $t\in [a,b]$. Thereby, $\mathcal {N}(t)$ has an its own \emph{dynamic Topcode-matrix} $T_{code}(\mathcal {N}(t))$ at each time step $t$, where $T_{code}(\mathcal {N}(t))=(X(t),E(t),Y(t))^{T}$ with three dynamic vectors $X(t)=(x^t_1$, $x^t_2$, $\dots $, $x^t_{q(t)})$, $E(t)=(e^t_1,e^t_2,\dots ,e^t_{q(t)})$ and $Y(t)=(y^t_1,y^t_2,\dots ,y^t_{q(t)})$. Sometimes, there is a function $F$, such that $e^t_k=F(x^t_k,y^t_k)$ for $k\in [1,q(t)]$. Conversely, $T_{code}(\mathcal {N}(t))$ may corresponds two or more dynamic networks at time step $t$, and these dynamic networks match from each other \cite{Li-Gu-Dullien-Vinyals-Kohli-arXiv2019}. Obviously, there are interesting researching topics on dynamic Topcode-matrices.

\subsection{Directed Topcode-matrices}

Directed Topcode-matrices are related with directed colorings (resp. labelings) of graphs.

\begin{defn}\label{defn:directed-Topcode-matrix}
\cite{Yao-1909-01587-2019} A \emph{directed Topcode-matrix} is defined as
\begin{equation}\label{eqa:Topcode-dimatrix}
\centering
{
\begin{split}
\overrightarrow{T}_{code}= \left(
\begin{array}{ccccc}
x_{1} & x_{2} & \cdots & x_{q}\\
e_{1} & e_{2} & \cdots & e_{q}\\
y_{1} & y_{2} & \cdots & y_{q}
\end{array}
\right)^{+}_{-}=
\left(\begin{array}{c}
X\\
\overrightarrow{E}\\
Y
\end{array} \right)^{+}_{-}=[(X,\overrightarrow{E},Y)^{+}_{-}]^{T}
\end{split}}
\end{equation}\\
where \emph{v-vector} $X=(x_1 , x_2 , \dots ,x_q)$, \emph{v-vector} $Y=(y_1 $, $y_2$, $\dots $, $y_q)$ and \emph{directed-e-vector} $\overrightarrow{E}=(e_1$, $e_2 $, $ \dots $, $e_q)$, such that each arc $e_i$ has its own \emph{head} $x_i$ and \emph{tail} $y_i$ with $i\in [1,q]$, and $q$ is the \emph{size} of $\overrightarrow{T}_{code}$. Moreover, the directed Topcode-matrix $\overrightarrow{T}_{code}$ is \emph{evaluated} if there exists a function $\varphi$ such that $e_i=\varphi(x_i,y_i)$ for $i\in [1,q]$.\qqed
\end{defn}


\section{Graphic groups made by colorings and labelings}

In encrypting networks, we employee graphic groups to make passwords for encrypting large-scale networks in short time. Our every-zero graphic groups can ba used to encrypt every network/graph, namely ``graph-to-graph'', and such graphic groups run in encrypting network/graphs are supported well by coloring theory of graph theory, such as traditional colorings, distinguishing colorings etc. It is noticeable, graphic groups can be used in machine encryption, or AI encryption for networks. We hope algebraic methods may solve some open problems of graph theory, for example, Graceful Tree Conjecture, and try to use algebraic methods for discovering possible regularity among labelings all in confusion \cite{Yao-Mu-Sun-Sun-Zhang-Wang-Su-Zhang-Yang-Zhao-Wang-Ma-Yao-Yang-Xie2019}.

\subsection{Abelian additive graphic groups}

\begin{defn}\label{defn:88-odd-graceful-Topsnut-group}
\cite{Yao-Sun-Zhao-Li-Yan-ITNEC-2017} Let $f_1:V(G_1)\rightarrow [0,q]$ be an odd-graceful labeling such that $f_1(E(G_1))=[1,2q-1]^o$ for a $(p,q)$-graph $G_1$. Define a \emph{modular labeling} $f_k(x)=f_1(x)+(k-1)~(\bmod~2q)$ with $x\in V(G)$ and $k\in [1,2q]$, the resultant graph is denoted as $F(G_k)$. Define a 2-element operation ``$F(G_i)\oplus_k F(G_j)$'' on the set $\{F(G_k):~k\in [1,2q]\}$ by $[f_i(x)+f_j(x)-f_k(x)]~(\bmod~2q)$ with $x\in V(G)=V(G_k)$ for $k\in [1,2q]$, where $[f_i(x)+f_j(x)-f_k(x)]~(\bmod~2q)$ means that
\begin{equation}\label{eqa:odd-modular-labeling11}
[f_i(x)+f_j(x)-f_k(x)]~(\bmod~2q)=2q-[f_i(x)+f_j(x)-f_k(x)]
\end{equation}
if one of $f_i(x)+f_j(x)-f_k(x)<0$ and $f_i(x)+f_j(x)-f_k(x)\geq 2q$ holds true, otherwise,
\begin{equation}\label{eqa:odd-modular-labeling22}
[f_i(x)+f_j(x)-f_k(x)]~(\bmod~2q)=f_i(x)+f_j(x)-f_k(x).
\end{equation}
We write the set of the graphs $F(G_{1}),F(G_{1}),\dots F(G_{2q})$ as $\{F_{2q}(\textbf{\textrm{G}});\oplus\}$, and call it an \emph{every-zero graphic group}.\qqed
\end{defn}

\begin{rem}\label{rem:333333}
For showing $\{F_{2q}(\textbf{\textrm{G}});\oplus\}$ to be an every-zero graphic group, first of all, the equation
\begin{equation}\label{eqa:basice-formula}
F(G_i)\oplus_k F(G_j)=F(G_{\lambda})
\end{equation} with $\lambda=i+j-k~(\bmod~2q)\in [1,2q]$ follows the equation $[f_i(x)+f_j(x)-f_k(x)]~(\bmod~2q)$ defined in (\ref{eqa:odd-modular-labeling11}) and (\ref{eqa:odd-modular-labeling22}) of Definition \ref{defn:88-odd-graceful-Topsnut-group} for any \emph{preappointed zero} $F(G_{k})$. We have

(1) \emph{Zero element}. Each element $F(G_k)$ of $\{F_{2q}(\textbf{\textrm{G}});\oplus\}$ is as a \emph{preappointed zero} in the equation (\ref{eqa:basice-formula}).

(2) \emph{Inverse}. The inverse of each element $F(G_i)\in \{F_{2q}(\textbf{\textrm{G}});\oplus\}$ is $F(G_{2k-i})\in \{F_{2q}(\textbf{\textrm{G}});\oplus\}$. In fact,
$$F(G_i)\oplus_k F(G_{2k-i})=F(G_k).$$

(3) \emph{Uniqueness and Closure}. $F(G_i)\oplus_k F(G_j)\in \{F_{2q}(\textbf{\textrm{G}});\oplus\}$ follows the equation (\ref{eqa:basice-formula}).

(4) \emph{Associative law}. $$[F(G_i)\oplus_k F(G_j)]\oplus_k F(G_l)=F(G_i)\oplus_k [F(G_j)\oplus_k F(G_l)].$$ Since
$${
\begin{split}
[F(G_i)\oplus_k F(G_j)]\oplus_k F(G_l)&=F(G_{i+j-1})\oplus_k F(G_l)=F(G_{i+j+l-2}),
\end{split}}
$$
and
$${
\begin{split}
F(G_i)\oplus_k [F(G_j)\oplus_k F(G_l)]&=F(G_i)\oplus_k F(G_{j+l-1})=F(G_{i+j+l-2}),
\end{split}}
$$ so the associative law holds true.

We conclude that $\{F_{2q}(\textbf{\textrm{G}});\oplus\}$ is an odd-graceful every-zero graphic group. \paralled
\end{rem}

\begin{defn}\label{defn:graph-set-set-group}
$^*$ The Topcode-matrices $T_{code}(F(G_{1})),T_{code}(F(G_{2})),\dots,T_{code}(F(G_{2q}))$ of graphs $F(G_{1}),F(G_{1}),\dots ,F(G_{2q})$ in an every-zero graphic group $\{F_{2q}(\textbf{\textrm{G}});\oplus\}$, also, form an \emph{every-zero Topcode-matrix group} $\{F_{2q}(T_{code}(F(\textbf{\textrm{G}})))$; $\oplus\}$. In an every-zero Topcode-matrix group $\{F_{n}(\textbf{\textrm{T}}_{code})$; $\oplus\}$ based on a set $F_{n}(\textbf{\textrm{T}}_{code})=\{T\,^1_{code},T\,^2_{code},\dots,T\,^n_{code}\}$, each Topcode-matrix $T\,^r_{code}$ with $r\in [1,n]$ corresponds a graph-set $G_{raph}(T\,^r_{code})$ of graphs having the same Topcode-matrix $T\,^r_{code}$. So, the graph-set set $\{G_{raph}(T\,^r_{code})\}^n_{r=1}$ also produces an \emph{every-zero graph-set set group} based on the operation ``$\oplus$'', we write this group as $\{\{G_{raph}(T\,^r_{code})\}^n_{r=1};\oplus\}$.\qqed
\end{defn}

\begin{rem}\label{rem:333333}
Select arbitrarily a graph $H_{s,k}\in G_{raph}(T\,^s_{code})$ as a \emph{preappointed zero}, do the operation ``$\oplus$'' to graph-sets
\begin{equation}\label{eqa:graph-set-set-operation}
G_{raph}(T\,^i_{code}) \oplus G_{raph}(T\,^{j}_{code})=G_{raph}(T\,^{\lambda}_{code})
\end{equation} with $\lambda=i+j-k~(\bmod~M)$ for some integer $M\geq 1$, which means that $H_{s,i}\oplus H_{s,j}=H_{s,\lambda}$ based on any \emph{preappointed zero} $H_{s,k}$ for $H_{s,i}\in G_{raph}(T\,^i_{code})$ and $H_{s,j}\in G_{raph}(T\,^j_{code})$.

In fact, the operation shown in (\ref{eqa:graph-set-set-operation}) is based on the operation of the an every-zero Topcode-matrix group $\{F_{n}(\textbf{\textrm{T}}_{code});\oplus\}$ as follows
\begin{equation}\label{eqa:topcode-matrix-group-operation}
T\,^i_{code}\oplus T\,^{j}_{code}=T\,^{\lambda}_{code}
\end{equation} with $\lambda=i+j-k~(\bmod~M)$ under any \emph{preappointed zero} $T\,^{k}_{code}\in \{F_{n}(\textbf{\textrm{T}}_{code});\oplus\}$.\paralled
\end{rem}

\begin{defn}\label{defn:every-zero-total-graphic-group}
$^*$ Let $G_1,G_2,\dots, G_{p+q}$ be the copies of a $(p,q)$-graph $G$ admitting a magic total labeling $h:V(G)\cup E(G)\rightarrow [1,p+q]$ such that $h(u)+h(v)=k+h(uv)$ for each edge $uv\in E(G)$. Each $G_i$ with $i\in [1,p+q]$ admits a magic total labeling $h_i:V(G_i)\cup E(G_i)\rightarrow [1,p+q]$ holding
\begin{equation}\label{eqa:555555}
h_i(x)+h_i(y)=k_i+h_i(xy)(\textrm{mod}~p+q)
\end{equation} true for each edge $xy\in E(G_i)$, and $\{k_1,k_2,\dots ,k_{p+q}\}=[1,p+q]$, write $F_{p+q}(G,h)=\{G_1,G_2,\dots, G_{p+q}\}$. For any \emph{preappointed zero} $G_k\in F_{p+q}(G,h)$, we have
\begin{equation}\label{eqa:graphic-group-1111}
h_i(w)+h_i(w)-h_k(w)=h_{\lambda}(w)
\end{equation}
with $\lambda=i+j-k~(\bmod~p+q)$ for each $w\in V(G)\cup E(G)$, and we call $F_{p+q}(G,h)$ an \emph{every-zero total graphic group}, and rewrite it as $\{F_{p+q}(G,h);\oplus\}$.\qqed
\end{defn}

\begin{defn}\label{defn:general-Topsnut-group}
\cite{Yao-Sun-Zhao-Li-Yan-ITNEC-2017} Let $f_{0,0}$ be an $\epsilon$-labeling of a $(p,q)$-graph $G_{0,0}$. For $i\in [1,p]$ and $j\in [1,q]$, we define a Topsnut-gpw $G_{i,j}$ admitting an $\epsilon$-labeling $f_{i,j}$ defined by $f_{i,j}(x)=f_{0,0}(x)+i~(\bmod~p\,')$ for each vertex $x\in V(G_{i,j})$, $f_{i,j}(uv)=f_{0,0}(uv)+j~(\bmod~q\,')$ for each edge $uv\in E(G_{i,j})$, and $G_{i,j}\cong G_{0,0}$ for $i\in [1,p]$ and $j\in [1,q]$. We call the set $\{G_{i,j}:~i\in [1,p],j\in [1,q]\}$ an \emph{$\epsilon$-Topsnut-group}, denoted by $\epsilon$-$L(G_{0,0})$.\qqed
\end{defn}

\begin{defn}\label{defn:graphic-group-definition}
\cite{Yao-Zhang-Sun-Mu-Sun-Wang-Wang-Ma-Su-Yang-Yang-Zhang-2018arXiv} An \emph{every-zero graphic group} $\{F_{n}(H);\oplus\}$ made by a Topsnut-gpw $H$ admitting an $\varepsilon$-labeling $h$ contains its own elements $H_i\in F_{n}(H)=\{H_i:i\in [1,n]\}$ with $n\geq q$ ($n\geq 2q-1$) holding $H\cong H_i$ and admitting an $\varepsilon$-labeling $h_i$ induced by $h$ with $i\in [1,n]$ and hold an additive operation $H_i\oplus H_j$ defined as
\begin{equation}\label{eqa:graphic-group-definition}
h_i(x)+h_j(x)-h_k(x)=h_{\lambda}(x)
\end{equation}
with $\lambda=i+j-k\,(\bmod\,n)$ for each element $x\in V(H)$ under any \emph{preappointed zero} $H_k$.\qqed
\end{defn}

\begin{defn}\label{defn:graph-graceful-group-labeling}
\cite{Yao-Zhang-Sun-Mu-Sun-Wang-Wang-Ma-Su-Yang-Yang-Zhang-2018arXiv} Let $\{F_{n}(H);\oplus\}$ be an every-zero graphic group based on a graphic set $F_{n}(H)=\{H_i:i\in [1,n]\}$ with $n\geq q$ ($n\geq 2q-1$) and the additive operation ``$\oplus$''. A $(p,q)$-graph $G$ admits a \emph{graceful group-labeling} (resp. an \emph{odd-graceful group-labeling}) $F:V(G)\rightarrow \{F_{n}(H);\oplus\}$ such that each edge $uv$ is colored by $F(uv)=F(u)\oplus F(v)$ under any \emph{preappointed zero} $H_k$, and the edge color set $F(E(G))=\{F(uv):uv \in E(G)\}=\{H_1,H_2,\dots ,H_q\}$ (resp. $F(E(G))=\{F(uv):uv \in E(G)\}=\{H_1,H_3,\dots ,H_{2q-1}\}$).\qqed
\end{defn}

\begin{rem}\label{rem:333333}
In Definition \ref{defn:graph-graceful-group-labeling}, we have:

(i) If $n=q$ (or $n=2q-1$), then $F$ is called a \emph{pure graceful group-labeling} (resp. a pure odd-graceful group-labeling).

(ii) If $F(x)=F(y)$ for some vertices $x,y\in V(G)$, we call $F$ a \emph{graceful group-coloring} (resp. an \emph{odd-graceful group-coloring}).\paralled
\end{rem}

\begin{figure}[h]
\centering
\includegraphics[width=15.4cm]{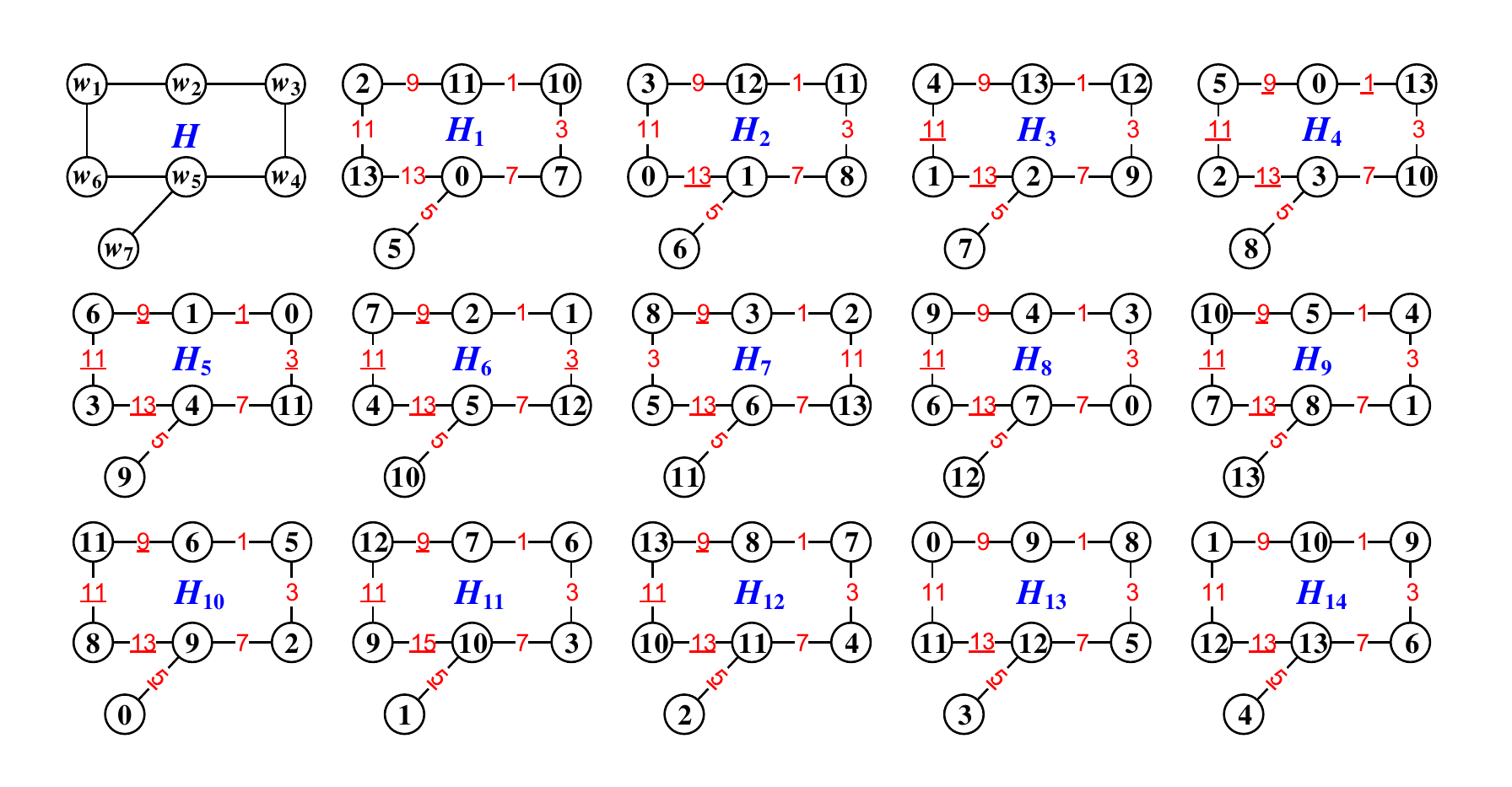}\\
\caption{\label{fig:group-in-group}{\small An odd-graceful every-zero graphic group, cited from \cite{Yao-Zhang-Sun-Mu-Sun-Wang-Wang-Ma-Su-Yang-Yang-Zhang-2018arXiv}.}}
\end{figure}

\begin{figure}[h]
\centering
\includegraphics[width=14.4cm]{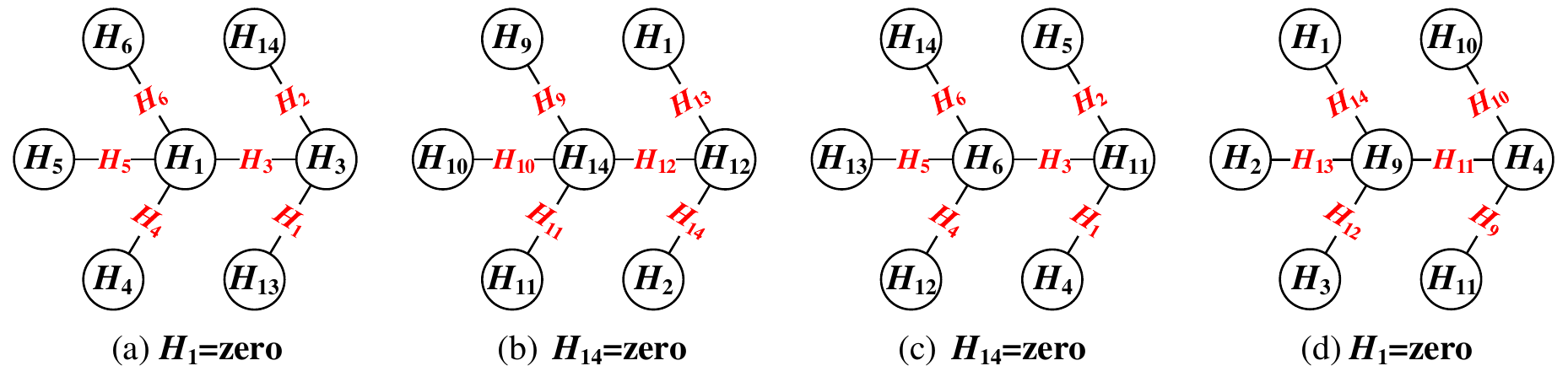}\\
\caption{\label{fig:encrypting-network}{\small Four networks encrypted by an odd-graceful graphic group shown in Fig.\ref{fig:group-in-group}, cited from \cite{Yao-Zhang-Sun-Mu-Sun-Wang-Wang-Ma-Su-Yang-Yang-Zhang-2018arXiv}.}}
\end{figure}

\begin{figure}[h]
\centering
\includegraphics[width=16.4cm]{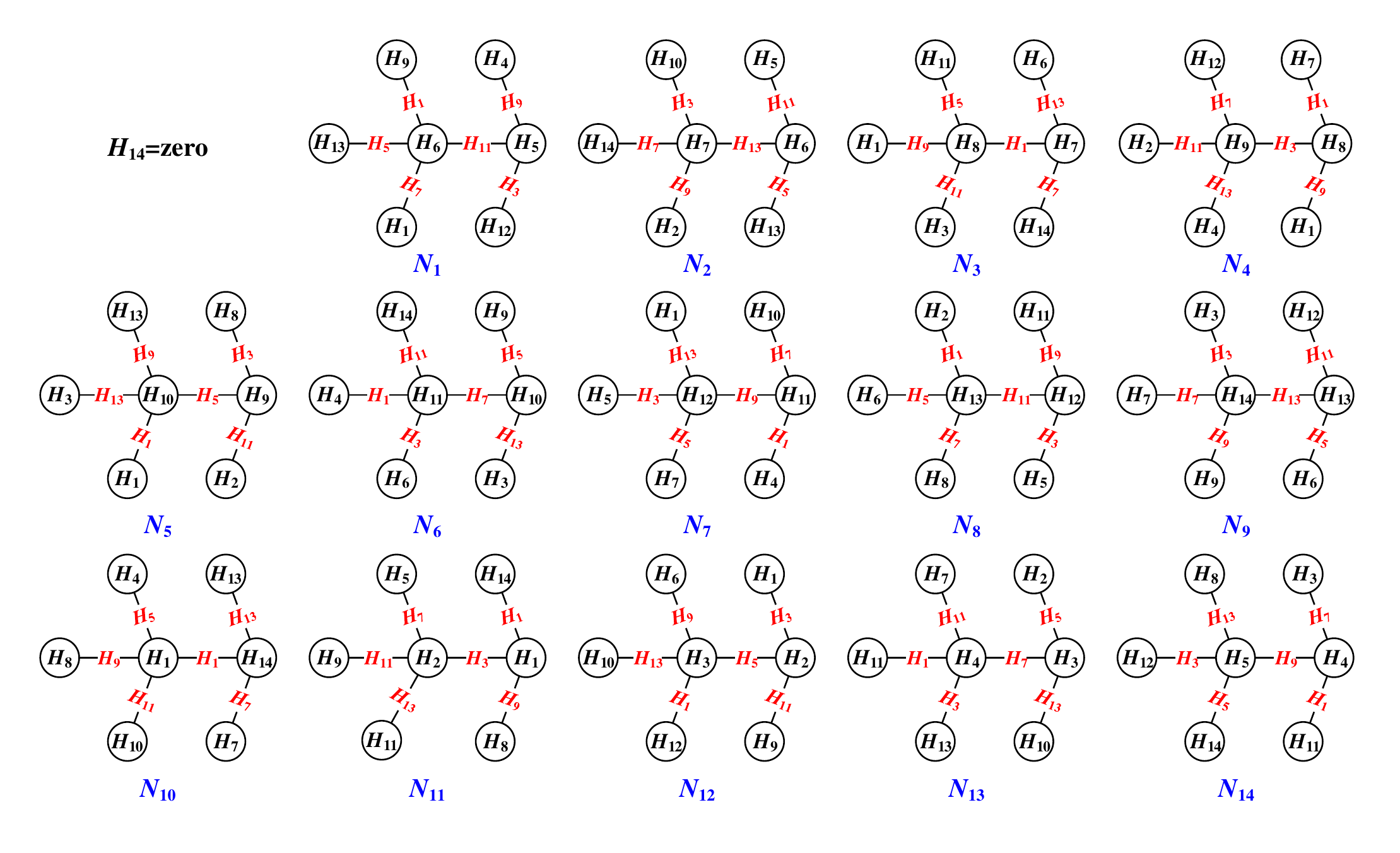}\\
\caption{\label{fig:group-in-group-more}{\small An odd-graceful graphic-group group made by the odd-graceful graphic group shown in Fig.\ref{fig:group-in-group}.}}
\end{figure}

\begin{defn}\label{defn:graphic-group-group-group-labeling}
\cite{Yao-Zhang-Sun-Mu-Sun-Wang-Wang-Ma-Su-Yang-Yang-Zhang-2018arXiv} Let $\{F_{n}(H);\oplus\}$ be an every-zero graphic group based on a graphic set $F_{n}(H)=\{H_i:i\in [1,n]\}$ with $n\geq q$ ($n\geq 2q-1$), and $\{H_{i_j}\}^q_1$ be a subset of $\{F_{n}(H);\oplus\}$. Suppose that a $(p,q)$-graph $G$ admits a mapping $F:V(G)\rightarrow \{F_{n}(H);\oplus\}$ such that each edge $uv$ is colored by $F(uv)=F(u)\oplus F(v)$ under any \emph{preappointed zero} $H_k$. If $F(x)\neq F(y)$ for any pair of vertices $x,y$, and the edge color set $F(E(G))=\{H_{i_j}\}^q_1$, we call $F$ an \emph{$\{H_{i_j}\}^q_1$-sequence group-labeling}; if $F(w)=F(z)$ for some two distinct vertices $w,z$, and the edge color set $F(E(G))=\{H_{i_j}\}^q_1$, we call $F$ an \emph{$\{H_{i_j}\}^q_1$-sequence group-coloring}. The colored graph made by joining $F(u)$ with $F(uv)$ and joining $F(uv)$ with $F(v)$ for each edge $uv\in E(G)$ is denoted as $\{N_{et}(G; F_{n}(H));\oplus\}$, see Fig.\ref{fig:group-in-group-more}.\qqed
\end{defn}

\begin{thm} \label{them:trees-sequence-group-coloring}
\cite{Yao-Zhang-Sun-Mu-Sun-Wang-Wang-Ma-Su-Yang-Yang-Zhang-2018arXiv} For any sequence $\{H_{i_j}\}^q_1$ of an every-zero graphic group $\{F_{n}(H);\oplus\}$, any tree of $q$ edges admits an \emph{$\{H_{i_j}\}^q_1$-sequence group-coloring}, or an \emph{$\{H_{i_j}\}^q_1$-sequence group-labeling}.
\end{thm}

By the induction proof on Theorem \ref{them:trees-sequence-group-coloring}, we can set randomly the elements of the set $\{H_1,H_2,\dots $, $H_q\}=\{H_{i}\}^q_1$ on the edges of any tree $T$ having $q$ edges, where $H_i\neq H_j$ for $i\neq j$, and then color the vertices of $T$ with the elements of an every-zero graphic group $\{F_{n}(H);\oplus\}$. We provide a sequence group-coloring $F$ of $T$ through the following algorithm.

\vskip 0.2cm

\textbf{TREE-GROUP-COLORING algorithm.}

\textbf{Input:} A tree $T$ of $q$ edges, and $\{H_{i}\}^q_1\subseteq \{F_{n}(H);\oplus\}$.

\textbf{Output:} An $\{H_{i}\}^q_1$-sequence group-coloring (or group-labeling) of $T$.

\textbf{Step 1.} Select an initial vertex $u_1\in V(T)$, its neighbor set $N(u_1)=\{v_{1,1},v_{1,2},\dots ,v_{1,d_1}\}$, where $d_1$ is the degree of the vertex $u_1$; next, select $H_1$ as any \emph{preappointed zero}, and color $u_1$ with $F(u_1)=H_1$ and $F(u_1v_{1,j})=H_{1,j}$ with $j\in [1,d_1]$. From $h_1(w)+h_z(w)-h_1(w)=h_{i_j~(\bmod ~n)}(w)$, where $h_z$ is the labeling of $H_{1,j}$, immediately, we get solutions $z={i_j}$, that is $F(v_{1,j})=H_{1,j}$ with $j\in [1,d_1]$. Let $V_1\leftarrow V(T)\setminus \{u_1,v_{1,1},v_{1,2},\dots ,v_{1,d_1}\}$, $L_1\leftarrow \{F(u_1v_{1,j})\}^{d_1}_1$.

\textbf{Step 2.} If $V_{k-1}=\emptyset$ (resp. $L_{k-1}=\{H_{i}\}^q_1$), go to Step 4.

\textbf{Step 3.} If $V_{k-1}\neq \emptyset$ (resp. $L_{k-1}\neq \{H_{i}\}^q_1$), select $u_k\in V_{k-1}$ such that $N(u_k)=\{v_{k,1},v_{k,2},\dots $, $v_{k,d_k}\}$ contains the unique vertex $v_{k,1}$ being colored with $F(v_{k,1})=H_{\alpha}$. Label $F(u_kv_{k,j})=H_{k,j}\in \{H_{i}\}^q_1\setminus L_{k-1}$ with $j\in [1,d_k]$. Assume $F(u_k)=H_{k}$, solve $h_k(w)+h_{\alpha}(w)-h_1(w)=h_{k,1~(\bmod ~n)}(w)$, then $k+\alpha-1=(k,1)~(\bmod ~n)$, thus, $k=1-\alpha+(k,1)~(\bmod ~n)$. Next, solve $h_k(w)+h_z(w)-h_1(w)=h_{k，j~(\bmod ~n)}(w)$, where $h_z$ is the labeling of $F(v_{k,j})$ with $j\geq 2$. Then $k+z-1=(k,j)~(\bmod ~n)$, so $z=1-k+(k,j)~(\bmod ~n)$, also, $F(v_{k,j})=H_{1-k+(k,j)~(\bmod ~n)}$ with $j\in [2,d_k]$. Let $V_{k}\leftarrow V(T)\setminus \big (V_{k-1}\cup \{u_k,v_{k,2},\dots ,v_{k,d_k}\}\big )$, and $L_k\leftarrow L_{k-1}\cup \{F(u_kv_{k,j}):j\in [1,d_k]\}$, and then go to Step 2.

\textbf{Step 4.} Return an $\{H_{i}\}^q_1$-sequence group-coloring $F$ of $T$.

\vskip 0.2cm

The TREE-GROUP-COLORING algorithm is polynomial and efficient, and it can quickly set Topsnut-gpws to a tree-like network.

\subsection{Connections between Topcode-matrices and graphic groups}

\begin{defn}\label{defn:graphic-group-matrix}
\cite{Yao-Zhang-Sun-Mu-Sun-Wang-Wang-Ma-Su-Yang-Yang-Zhang-2018arXiv} For an every-zero graphic group $\{F_n(H);\oplus\}$ based on a graph set $F_n(H)=\{H_i,H\,'_i,H\,''_i\}^n_1$ with $n\geq q>0$, a \emph{graphic group-matrix} $P_{vev}(G)$ of a $(p,q)$-graph $G$ is defined as $P_{vev}(G)=(X_P,~W_P,~Y_P)^{T}$ with
\begin{equation}\label{eqa:two-vectors}
{
\begin{split}
X_P=(H_1, ~ H_2, ~ \cdots ,~H_q), W_P=(H\,'_1, ~ H\,'_2, ~ \cdots ,~H\,'_q), Y_P=(H\,''_1, ~ H\,''_2, ~\cdots ,~ H\,''_q),
\end{split}}
\end{equation}
where each edge $u_iv_i$ of $G$ with $i\in [1,q]$ is colored with a graph $H\,'_i$, and its own two ends $u_i$ and $v_i$ are colored with graphs $H_i$ and $H\,''_i$ respectively, where $X_P,Y_P$ are called two \emph{pan-v-vectors}, $W_P$ is called a \emph{pan-e-vector}; and $G$ corresponds another \emph{graphic group-matrix} $P_{vv}(G)$ defined as $P_{vv}(G)=(X_P,~Y_P)^{T}$.\qqed
\end{defn}

\begin{rem}\label{rem:ABC-conjecture}
Color a $(p,q)$-graph $H$ with the elements of an every-zero graphic group $\{F_n(G);\oplus\}=\{G_{x,i}$, $G_{e,j}$, $G_{y,k}:1\leq i,j,k\leq q\}$, in other view, the graph $H$ is made by the graphic group $\{F_n(G);\oplus\}$. Thereby, we get $H$'s Topcode-matrix as follows
\begin{equation}\label{eqa:Topcode-matrix}
\centering
{
\begin{split}
M_{atrix}(H)= \left(
\begin{array}{ccccc}
G_{x,1} & G_{x,2} & \cdots & G_{x,q}\\
G_{e,1} & G_{e,2} & \cdots & G_{e,q}\\
G_{y,1} & G_{y,2} & \cdots & G_{y,q}
\end{array}
\right)=
\left(\begin{array}{c}
G_X\\
G_E\\
G_Y
\end{array} \right)=(G_X,G_E,G_Y)^{T}
\end{split}}
\end{equation}where three \emph{graph-graphic vectors} $G_X=(G_{x,1},G_{x,2},\dots ,G_{x,q})$, $G_E=(G_{e,1}$, $G_{e,2}$, $\dots ,G_{e,q})$ and $G_Y=(G_{y,1},G_{y,2},\dots ,G_{y,q})$, also, the matrix $M_{atrix}(H)$ is \emph{graphicable}.\paralled
\end{rem}

\subsection{Mixed graphic groups}

\textbf{MIXED Graphic-group Algorithm}. Let $f: V(G)\cup E(G)\rightarrow [1,M]$ be a $W$-type proper total coloring of a graph $G$ such that two color sets $f(V(G))=\{f(x):x\in V(G)\}$ and $f(E(G))=\{f(uv):uv\in E(G)\}$ hold a collection of restrictions true. We define a $W$-type proper total coloring $g_{s,k}$ by setting $g_{s,k}(x)=f(x)+s~(\bmod~p)$ for every vertex $x\in V(G)$, and $g_{s,k}(uv)=f(uv)+k~(\bmod~q)$ for each edge $uv\in E(G)$. Let $F_f(G)$ be the set of graphs $G_{s,k}$ admitting $W$-type proper total colorings $g_{s,k}$ defined above, and each graph $G_{s,k}\cong G$ in topological structure. We define an additive operation ``$\oplus$'' on the elements of $F_f(G)$ in the following way: Take arbitrarily an element $G_{a,b}\in F_f(G)$ as \emph{zero}, and $G_{s,k}\oplus G_{i,j}$ is defined by the following computation
\begin{equation}\label{eqa:mixed-graphic-group}
[g_{s,k}(w)+g_{i,j}(w)-g_{a,b}(w)]~(\bmod~\varepsilon)=g_{\lambda,\mu}(w)
\end{equation}
for each element $w\in V(G)\cup E(G)$, where $\lambda=s+i-a~(\bmod~p)$ and $\mu=k+j-b~(\bmod~q)$. As $w=x\in V(G)$, the form (\ref{eqa:mixed-graphic-group}) is just equal to
\begin{equation}\label{eqa:mixed-graphic-group11}
[g_{s,k}(x)+g_{i,j}(x)-g_{a,b}(x)]~(\bmod~p)=g_{\lambda,\mu}(x)
 \end{equation}and as $w=uv\in E(G)$, the form (\ref{eqa:mixed-graphic-group}) is defined as follows:
 \begin{equation}\label{eqa:mixed-graphic-group22}
[g_{s,k}(uv)+g_{i,j}(uv)-g_{a,b}(uv)]~(\bmod~q)=g_{\lambda,\mu}(uv).
\end{equation}
Especially, as $s=i=a=\alpha$, we have $\bmod~\varepsilon=\bmod~q$ in (\ref{eqa:mixed-graphic-group}), and
\begin{equation}\label{eqa:mixed-graphic-group-edge}
[g_{\alpha,k}(uv)+g_{\alpha,j}(uv)-g_{\alpha,b}(uv)]~(\bmod~q)=g_{\alpha,\mu}(uv)
\end{equation} for $uv\in E(G)$; and when $k=j=b=\beta$, so $\bmod~\varepsilon=\bmod~p$ in (\ref{eqa:mixed-graphic-group}), we have
\begin{equation}\label{eqa:mixed-graphic-group-vertex}
[g_{s,\beta}(x)+g_{i,\beta}(x)-g_{a,\beta}(x)]~(\bmod~p)=g_{\lambda,\beta}(x),~x\in V(G).
\end{equation}

\begin{rem}\label{rem:333333}
Sice the graph set $F_f(G)$ made by the MIXED Graphic-group Algorithm holds:
\begin{asparaenum}[(1) ]
\item \emph{Zero.} Each graph $G_{a,b}\in F_f(G)$ can be determined as a \emph{preappointed zero} such that $G_{s,k}\oplus G_{a,b}=G_{s,k}$.

\item \emph{Uniqueness.} For $G_{s,k}\oplus G_{i,j}=G_{c,d}\in F_f(G)$ and $G_{s,k}\oplus G_{i,j}=G_{r,t}\in F_f(G)$, then $c=s+i-a~(\bmod~p)=r$ and $c=k+j-b~(\bmod~q)=t$ under any \emph{preappointed zero} $G_{a,b}$.

\item \emph{Inverse.} Each graph $G_{s,k}\in F_f(G)$ has its own \emph{inverse} $G_{s',k'}\in F_f(G)$ such that $G_{s,k}\oplus G_{s',k'}=G_{a,b}$ determined by $[g_{s,k}(w)+g_{i,j}(w)]~(\bmod~\varepsilon)=2g_{a,b}(w)$ for each element $w\in V(G)\cup E(G)$.
\item \emph{Associative law.} Each triple $G_{s,k},G_{i,j},G_{c,d}\in F_f(G)$ holds
$$G_{s,k}\oplus [G_{i,j}\oplus G_{c,d}]=[G_{s,k}\oplus G_{i,j}]\oplus G_{c,d}.$$
\item \emph{Commutative law.} $G_{s,k}\oplus G_{i,j}=G_{i,j}\oplus G_{s,k}$.
\end{asparaenum}

Thereby, we call $F_f(G)=\{G_{s,k}:s\in [0,p],k\in [0,q]\}$ an \emph{every-zero mixed graphic group} based on the additive operation ``$\oplus$'' defined in (\ref{eqa:mixed-graphic-group}), and write this group by $\textbf{\textrm{G}}=\{F_f(G);\oplus\}$.

There are $pq$ graphs in the every-zero mixed graphic group $\textbf{\textrm{G}}$. There are two particular \emph{every-zero graphic subgroups} $\{F_{v}(G);\oplus\}\subset \textbf{\textrm{G}}$ and $\{F_{e}(G);\oplus\}\subset \textbf{\textrm{G}}$, where $F_{v}(G)=\{G_{s,0}:s\in [0,p]\}$ and $F_{e}(G)=\{G_{0,k}:k\in [0,q]\}$. In fact, $\textbf{\textrm{G}}$ contains at least $(p+q)$ different every-zero graphic subgroups.\paralled
\end{rem}

\begin{defn}\label{defn:111111}
\cite{Bing-Yao-Hongyu-Wang-graph-homomorphisms-2020} For two every-zero graphic groups $\{F_f(G);\oplus\}$ based on a graph set $F_f(G)=\{G_i\}^m_1$ and $\{F_h(H);\oplus\}$ based on a graph set $F_h(H)=\{H_i\}^m_1$, suppose that there are graph homomorphisms $G_i\rightarrow H_i$ defined by $\theta_i:V(G_i)\rightarrow V(H_i)$ with $i\in [1,m]$. We define $\theta=\bigcup^m_{i=1}\theta_i$, and have an \emph{every-zero graphic group homomorphism} $\{F_f(G);\oplus\}\rightarrow \{F_h(H);\oplus\}$ from a graph set $F_f(G)$ to another graph set $F_h(H)$.\qqed
\end{defn}

\subsection{Infinite graphic-sequence groups}

\begin{figure}[h]
\centering
\includegraphics[width=16cm]{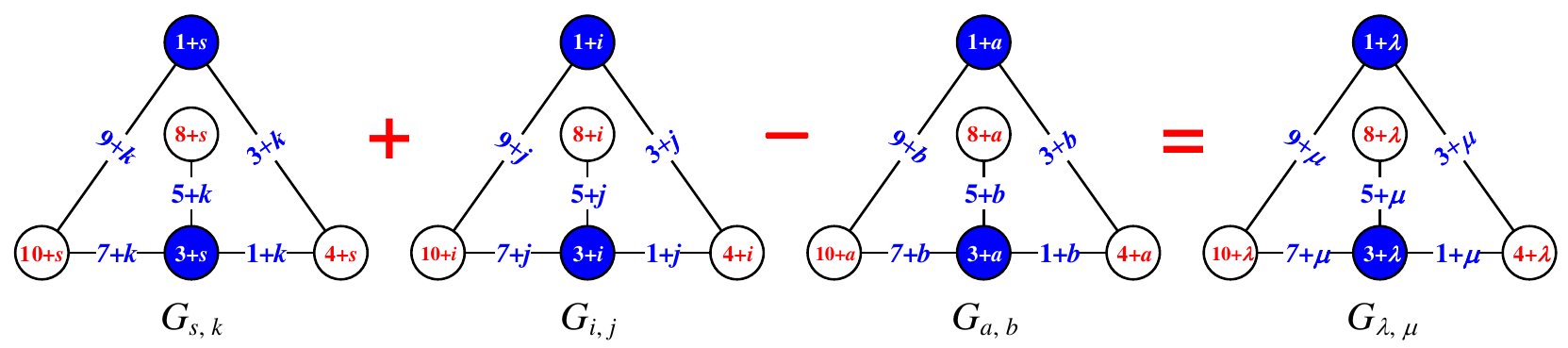}\\
\caption{\label{fig:infinite-group}{\small The additive operation on every-zero infinite graphic-sequence groups, cited from \cite{Bing-Yao-Hongyu-Wang-graph-homomorphisms-2020}.}}
\end{figure}

Infinite graphic-sequence groups have been introduced in \cite{yao-sun-su-wang-matching-groups-zhao-2020}. Suppose that a connected $(p,q)$-graph $G$ admits a $W$-type total coloring $f$, we define $W$-type total colorings $f_{s,k}$ by setting $f_{s,k}(x)=f(x)+s$ for every vertex $x\in V(G)$, and $f_{s,k}(uv)=f(uv)+k$ for each edge $uv\in E(G)$ as two integers $s,k$ belong to the set $Z$ of integers. So, we have each connected $(p,q)$-graph $G_{s,k}\cong G$ admits a $W$-type total coloring $f_{s,k}$ defined above, immediately, we get an infinite graphic-sequence $\{\{G_{s,k}\}^{+\infty}_{-\infty}\}^{+\infty}_{-\infty}$. We take $G_{a,b}\in \{\{G_{s,k}\}^{+\infty}_{-\infty}\}^{+\infty}_{-\infty}$ as a \emph{preappointed zero}, and any two $G_{s,k}$ and $G_{i,j}$ in $\{\{G_{s,k}\}^{+\infty}_{-\infty}\}^{+\infty}_{-\infty}$ to do the additive computation ``$G_{s,k}\oplus G_{i,j}$'' in the following way : For $uv\in E(G)$,
\begin{equation}\label{eqa:edge-graphic-group}
[f_{s,k}(uv)+f_{i,j}(uv)-f_{a,b}(uv)]~(\bmod~ q_W)=f_{\lambda,\mu}(uv).
\end{equation} with $\mu=k+j-b~(\bmod~ q_W)$; and for $x\in V(G)$,
\begin{equation}\label{eqa:vertex-graphic-group}
[f_{s,k}(x)+f_{i,j}(x)-f_{a,b}(x)]~(\bmod~ p_W)=f_{\lambda,\mu}(x)
\end{equation} with $\lambda=s+i-a~(\bmod~ p_W)$. See Fig.\ref{fig:infinite-group} for understanding $G_{s,k}\oplus G_{i,j}=G_{\lambda,\mu}$.

Here, $p_W=|V(G)|$ and $q_W=|E(G)|$ if the $W$-type total coloring $f$ is a gracefully total coloring; and $p_W=q_W=2|E(G)|$ if the $W$-type total coloring $f$ is an odd-gracefully total coloring.

Especially, for an edge subsequence $G_{s,k}, G_{s,k+1},\dots ,G_{s,k+q_W}$, and a vertex subsequence $G_{s,k}$, $G_{s+1,k}$, $\dots $, $G_{s+p_W,k}$, we have two sets $F(\{G_{s,k+j}\}^{q_W}_{j=1};\oplus;(G,f))$ and $F(\{G_{s+i,k}\}^{p_W}_{i=1};\oplus;(G,f))$. By the operation ``$\oplus$'' defined in (\ref{eqa:edge-graphic-group}), we claim that $F(\{G_{s,k+j}\}^{q_W}_{j=1};\oplus;(G,f))$ is an \emph{every-zero edge-graphic group}, since there are the following facts:

(i) \emph{Zero}. Every graph $G_{s,k+j}$ of $F(\{G_{s,k+j}\}^q_{j=1};\oplus)$ is as a \emph{preappointed zero} such that $G_{s,k+r}\oplus G_{s,k+j}=G_{s,k+r}$ for any graph $G_{s,k+r}\in F(\{G_{s,k+j}\}^q_{j=1};\oplus)$.

(ii) \emph{Closure law}. For $r=i+j-j_0~(\bmod~ q_W)$, $G_{s,k+i}\oplus G_{s,k+j}=G_{s,k+r}\in F(\{G_{s,k+j}\}^q_{j=1};\oplus)$ under any \emph{preappointed zero} $G_{s,k+j_0}$.

(iii) \emph{Inverse.} For $i+j=2j_0~(\bmod~ q_W)$, $G_{s,k+i}\oplus G_{s,k+j}=G_{s,k+j_0}$ under any \emph{preappointed zero} $G_{s,k+j_0}$.

(iv) \emph{Associative law}. $G_{s,k+i}\oplus (G_{s,k+j}\oplus G_{s,k+r})=(G_{s,k+i}\oplus G_{s,k+j})\oplus G_{s,k+r}$.

(v) \emph{Commutative law}. $G_{s,k+i}\oplus G_{s,k+j}=G_{s,k+j}\oplus G_{s,k+i}$.

Similarly, $F(\{G_{s+i,k}\}^{p_W}_{i=1};\oplus;(G,f))$ is an \emph{every-zero vertex-graphic group} by the operation ``$\oplus$'' defined in (\ref{eqa:vertex-graphic-group}).

As considering some graphs arbitrarily selected from the sequence $\{\{G_{s,k}\}^{+\infty}_{-\infty}\}^{+\infty}_{-\infty}$, we have
\begin{equation}\label{eqa:mixed-infinite-graphic-group}
[f_{s,k}(w)+f_{i,j}(w)-f_{a,b}(w)]~(\bmod~ p_W, q_W)=f_{\lambda,\mu}(w).
\end{equation} with $\lambda=s+i-a~(\bmod~ p_W)$ and $\mu=k+j-b~(\bmod~ q_W)$ for each element $w\in V(G)\cup E(G)$.

Thereby, we call the set $F(\{\{G_{s,k}\}^{+\infty}_{-\infty}\}^{+\infty}_{-\infty};\oplus;(G,f))$ an \emph{every-zero infinite graphic-sequence group} under the additive operation ``$\oplus$'' based on two modules $p_W$ and $q_W$ and a connected $(p,q)$-graph $G$ admitting a $W$-type total coloring, since it possesses the properties of Zero, Closure law, Inverse, Associative law and Commutative law.

\begin{rem}\label{rem:infinite-graphic-sequence-group}
Let $F^*(G,f)=F(\{\{G_{s,k}\}^{+\infty}_{-\infty}\}^{+\infty}_{-\infty};\oplus;(G,f))$ be an every-zero infinite graphic-sequence group. The elements of an every-zero infinite graphic-sequence group $F^*(G,f)$ can tile fully each point $(x,y)$ of $xOy$-plane. And moreover, $F^*(G,f)$ contains infinite every-zero graphic groups having finite elements, such as $F(\{G_{s+i,k}\}^{p_W}_{i=1}$; $\oplus;(G,f))$ and $F(\{G_{s,k+j}\}^{q_W}_{i=1};\oplus;(G,f))$. Also, $F^*(G,f)$ contains infinite every-zero graphic groups of infinite elements.

Clearly, particular every-zero graphic groups having infinite elements or finite elements can be used easily to encrypt randomly networks. Suppose that the coloring $f$ of $G$ in $F^*(G,f)$ is equivalent to another $W_g$-type total coloring $g$ of $G$. Then we get another every-zero infinite graphic-sequence group $F^*(G,g)=F(\{\{G_{i,j}\}^{+\infty}_{-\infty}\}^{+\infty}_{-\infty};\oplus;(G,g))$ with $G\cong G_{i,j}$. Thereby, the every-zero infinite graphic-sequence group $F^*(G,f)$ is a \emph{public-key graphic-sequence group}, the every-zero infinite graphic-sequence group $F^*(G,g)$ is a \emph{private-key graphic-sequence group} accordingly.

Since there exists a mapping $\varphi: V(G)\cup E(G)\rightarrow V(G)\cup E(G)$ such that $g(w)=\varphi(f(w))$ for $w\in V(G)\cup E(G)$, we claim that $F^*(G,f)$ admits an every-zero graphic-sequence homomorphism to $F^*(G,g)$, and moreover
\begin{equation}\label{eqa:555555}
F^*(G,f)\leftrightarrow F^*(G,g),
\end{equation} is a pair of \emph{homomorphically equivalent every-zero graphic-sequence homomorphisms}.\paralled
\end{rem}

\begin{defn}\label{rem:infinite-graphic-sequence-group}
\cite{yao-sun-su-wang-matching-groups-zhao-2020} Let a connected graph $G$ admit a \emph{graph homomorphism} to another connected graph $H$ under a mapping $\varphi:V(G)\rightarrow V(H)$, where $H$ admits a $W\,'$-type total coloring $g$. By the construction of an infinite graphic-sequence $\{\{G_{s,k}\}^{+\infty}_{-\infty}\}^{+\infty}_{-\infty}$, we get another infinite graphic-sequence $\{\{H_{a,b}\}^{+\infty}_{-\infty}\}^{+\infty}_{-\infty}$, and moreover, we have an \emph{infinite graphic-sequence homomorphism} as follows:
\begin{equation}\label{eqa:555555}
\{\{G_{s,k}\}^{+\infty}_{-\infty}\}^{+\infty}_{-\infty}\rightarrow \{\{H_{a,b}\}^{+\infty}_{-\infty}\}^{+\infty}_{-\infty},
\end{equation}
where each $H_{a,b}$ is a copy of $H$ and admit a $W\,'$-type total colorings $g_{a,b}$ defined by $g_{a,b}(x)=g(x)+a$ for every vertex $x\in V(H)$, and $g_{a,b}(xy)=g(xy)+b$ for each edge $xy\in E(H)$, and as two integers $a,b\in Z$.\qqed
\end{defn}

\subsection{Twin-type of graphic group, Matching graphic groups}

Wang \emph{et al.} in \cite{Wang-Xu-Yao-2017-Twin} introduced the \emph{twin odd-graceful labeling} as: For two connected $(p_i,q)$-graphs $G_i$ with $i=1,2$, and let $p=p_1+p_2-2$, if a $(p,q)$-graph $G=\odot \langle G_1, G_2\rangle$ admits a vertex labeling $f$: $V(G)\rightarrow [0, q]$ such that (i) $f$ is just an odd-graceful labeling of of $G_1$, so $f(E(G_1))=\{f(uv)=|f(u)-f(v)|: uv\in E(G_1)\}=[1, 2q-1]^o$; (ii) $f(E(G_2))=\{f(uv)=|f(u)-f(v)|: uv\in E(G_2)\}=[1,2q-1]^o$; and (iii) $|f(V(G_1))\cap f(V(G_2))|=k$ and $f(V(G_1))\cup f(V(G_2))\subseteq [0, 2q-1]$. Then $f$ is called a \emph{twin odd-graceful labeling}of $G$. Thereby, we get two \emph{twin odd-graceful graphic groups} $\{F_f(G);\oplus\}$ and $\{F_g(H);\oplus\}$ based on a twin odd-graceful labeling $(f,g)$. Notice that $G\not \cong H$, in general.

\begin{defn}\label{defn:twin-type-every-zero-graphic-groups}
$^*$ Let $\{F_{n}(G);\oplus\}$ be an every-zero graphic group based on a graph set $F_{n}(G)=\{G_i:i\in [1,n]\}$ and the additive operation ``$\oplus$'' defined in Definition \ref{defn:88-odd-graceful-Topsnut-group}, where each $G_i$ is a copy of $G$ and admits a $W$-type labeling $f_i$ induced by the $W$-type labeling $f$ of $G$; and let $\{F_{n}(H);\oplus\}$ be another every-zero graphic group based on the graph set $F_{n}(H)=\{H_i:i\in [1,n]\}$ and the additive operation ``$\oplus$'', where each $H_i$ is a copy of $H$ and admits a $W$-type labeling $g_i$ induced by the $W$-type labeling $g$ of $H$. We have the following twin-type every-zero graphic groups:
\begin{asparaenum}[(i)]
\item If both $W$-type labelings $f$ and $g$ are a twin $(k,d)$-labeling, and $g$ is a \emph{complementary $(k,d)$-labeling} of $f$. Then, we call both every-zero graphic groups $\{F_{n}(G);\oplus\}$ and $\{F_{n}(H);\oplus\}$ a \emph{twin $(k,d)$-labeling graphic group}.
\item If both $W$-type labelings $f$ and $g$ are a twin odd-graceful labeling. Then, we call both every-zero graphic groups $\{F_{n}(G);\oplus\}$ and $\{F_{n}(H);\oplus\}$ a \emph{twin odd-graceful graphic group}.
\item If both $W$-type labelings $f$ and $g$ are a twin odd-elegant labeling. Then, we call both every-zero graphic groups $\{F_{n}(G);\oplus\}$ and $\{F_{n}(H);\oplus\}$ a \emph{twin odd-elegant graphic group}.
\item By Definition \ref{defn:twin-colorings-general}, if \emph{each vertex-coincided graph} $\odot \langle G_i,H_i\rangle$ admits a twin ($W_f$, $W_g$)-type coloring (resp. ($W_f$, $W_g$)-type labeling), then both every-zero graphic groups $\{F_{n}(G);\oplus\}$ and $\{F_{n}(H);\oplus\}$ are called a \emph{twin ($W_f$, $W_g$)-type coloring graphic group} (resp. \emph{($W_f$, $W_g$)-type labeling graphic group}).
\end{asparaenum}

Also, $\{F_{n}(H);\oplus\}$ is called an \emph{accompanied graphic group} of the graphic group $\{F_{n}(G);\oplus\}$, and vice versa.\qqed
\end{defn}

\begin{defn}\label{defn:multiple-tree-graphic-groups}
$^*$ Suppose that each spanning tree $T_k$ of a connected $(p,q)$-graph $G=\oplus_F\langle T_i\rangle ^m_1$ with $k\in [1,m]$ admits a proper labeling $f_k$ induced by an e-set v-proper labeling $(F,f)$ of $G$, so we have an every-zero graphic group $\{G;F_{n}(T_k);\oplus\}$ based on the set $F_{n}(T_k)=\{T_{k,i}:i\in [1,n]\}$ with $k\in [1,m]$, such that $E(G)=\bigcup^m_{i=1}E(T_{k,i})$, and $E(T_{k,i})\cap E(T_{k,j})=\emptyset$ for $i\neq j$. We call $\{G;F_{n}(T_1);\oplus\}$, $\{G;F_{n}(T_2);\oplus\}$, $\dots$, $\{G;F_{n}(T_m);\oplus\}$ a matching of \emph{multiple-tree graphic groups}.\qqed
\end{defn}

\begin{defn}\label{defn:multiple-tree-graphic-groups}
$^*$ Suppose that a $(p,q)$-graph $G$ admits a $W$-type coloring $f$. Let $\max f=\max\{f(w): w\in S\subseteq V(G)\cup E(G)\}$ and $\min f=\min\{f(w): w\in S\subseteq V(G)\cup E(G)\}$. We call $g(w)=\max f+\min f-f(w)$ for each element $w\in S\subseteq V(G)\cup E(G)$ the \emph{dual $W$-type coloring} of the coloring $f$. Then, $\{F_g(G);\oplus\}$ is called the \emph{dual graphic group} of the graphic group $\{F_f(G);\oplus\}$ based on a pair of mutually dual $W$-type colorings $f$ and $g$. \qqed
\end{defn}

\begin{rem}\label{rem:ABC-conjecture}
Notice that these two graphic groups defined in Definition \ref{defn:multiple-tree-graphic-groups} were built up on the same graph $G$. If a graph $G$ is bipartite and admits a \emph{set-ordered graceful labeling} $f$, then there are a collection of labelings $g_i$ being equivalent with $f$ (Ref. \cite{Yao-1909-01587-2019, Yao-Mu-Sun-Sun-Zhang-Wang-Su-Zhang-Yang-Zhao-Wang-Ma-Yao-Yang-Xie2019, Yao-Zhang-Sun-Mu-Sun-Wang-Wang-Ma-Su-Yang-Yang-Zhang-2018arXiv, Yao-Sun-Zhang-Mu-Sun-Wang-Su-Zhang-Yang-Yang-2018arXiv}), so we get a collection of \emph{matching-labeling graphic groups} $\{F_f(G);\oplus\}$ and $\{F_{g_i}(H_i);\oplus\}$ with $G\cong H_i$ for $i\in [1,m]$ and $m\geq 2$. For example, these labelings $g_i$ are odd-graceful labeling, odd-elegant labeling, edge-magic total labeling, image-labeling, 6C-labeling, odd-6C-labeling, even-odd separated 6C-labeling, and so on (Ref. \cite{Yao-Sun-Zhang-Mu-Sun-Wang-Su-Zhang-Yang-Yang-2018arXiv}). Here, we refer to $\{F_f(G);\oplus\}$ as a \emph{public-key}, and each $\{F_{g_i}(H_i);\oplus\}$ as a \emph{private-key} in encrypting networks. Let $G^c$ be the complementary graph of $G$, that is, $V(G)=V(G^c)=V(K_n)$, $E(G)\cup E(G^c)=E(K_n)$ and $E(G)\cap E(G^c)=\emptyset$. So, we have $\{F_f(G);\oplus\}$ and $\{F_g(G^c);\oplus\}$ as a pair of\emph{ matching graphic groups}.\paralled
\end{rem}

\subsection{Graphic group sequences, graphic group homomorphisms}

\begin{defn}\label{defn:graphic-group-sequences}
\cite{Bing-Yao-2020arXiv} Let $G^{(1)}_{ro}(H)=\{F_f(G^{(1)})$; $\oplus\}$ be an every-zero graphic group. We get an encrypted graph $G^{(2)}(H)=H\triangleleft G^{(1)}_{ro}(H)$ to be one graph of the set
$$\left \{H\triangleleft_{a,b} |^{(p,q)}_{(s,k)}a^{(2)}_{s,k}G^{(2)}_{s,k}: a^{(2)}_{s,k}\in Z^0,~G^{(2)}_{s,k}\in F_f(G^{(1)})\right \}
$$ with $\sum a^{(2)}_{s,k}\geq 1$ after encrypting a graph $H$ with the elements of the every-zero graphic group $G^{(1)}_{ro}(H)$. Immediately, we get an every-zero graphic group $G^{(2)}_{ro}(H)=\{F_f(G^{(2)});\oplus\}$ made by the graph $G^{(2)}(H)$, the operation ``$\oplus$'' and the MIXED Graphic-group Algorithm. Go on in this way, we get an \emph{every-zero $H$-graphic group sequence} $\{G^{(t)}_{ro}(H)\}$ based on the initial every-zero graphic group $G^{(1)}_{ro}(H)=\{F_f(G^{(1)});\oplus\}$ and the graph $H$, where $G^{(t)}_{ro}(H)=H\triangleleft G^{(t-1)}_{ro}(H)$. \qqed
\end{defn}
Clearly, every-zero graphic group each $G^{(t)}_{ro}(H)$ is a network at time step $t$.

\begin{defn}\label{defn:graphic-group-homomorphisms}
\cite{Bing-Yao-2020arXiv} Suppose that a graph sequence $\{L_k\}^m_{k=1}$ holds graph homomorphisms $L_k\rightarrow L_{k+1}$ for $k\in [1,m-1]$, and each graph $L_j$ admits a $W_j$-type coloring $f_j$ for $j\in [1,m]$. We have every-zero graphic groups $\{F_{f_j}(L_j);\oplus\}$ for $j\in [1,m]$. If each graph $G\in \{F_{f_k}(L_k);\oplus\}$ corresponds a graph $H\in \{F_{f_{k+1}}(L_{k+1});\oplus\}$ to form $G\rightarrow H$, and vice versa for $k\in [1,m-1]$, then we define an every-zero graphic group $\{F_{f_k}(L_k);\oplus\}$ is \emph{graphic group homomorphism} an every-zero graphic group $\{F_{f_{k+1}}(L_{k+1});\oplus\}$, i.e.
\begin{equation}\label{eqa:555555}
\{F_{f_k}(L_k);\oplus\}\rightarrow \{F_{f_{k+1}}(L_{k+1});\oplus\}
\end{equation} with $k\in [1,m-1]$.\qqed
\end{defn}


\section{Graphic lattices with colorings and labelings}

\subsection{Graphic lattices based on gracefully total colorings}

\begin{defn}\label{defn:111111}
\cite{Wang-Yao-2021-SCI} A \emph{graphic base} $\textbf{\textrm{H}}=(H_k)^n_{k=1}$ is consisted of $n$ vertex-disjoint connected bipartite graphs $H_1, H_2,\dots, H_n$ with $n\geq 2$, and

(i) if one edge of a new edge set $E^*$ joins a vertex of $H_i$ with a vertex of $H_j$ together, the resultant graph is called an \emph{edge-joining graph}, and denoted as $E^*\oplus^n_{k=1}H_k$;

(ii) suppose that a graph $F$ has $n$ vertices $x_1, x_2, \dots, x_n$, we vertex-coincide a vertex $x_i$ of $F$ with a vertex $u_i$ of $H_i$ into one vertex $x_i\odot u_i$, the resultant graph is called a \emph{$F$-graph}, written as $F\odot |^n_{k=1}H_k$.\qqed
\end{defn}

\begin{thm} \label{thm:set-graceful-total-coloring}
\cite{Wang-Yao-2021-SCI} Suppose that a graphic base $\textbf{\textrm{H}}=(H_k)^n_{k=1}$ is consisted of $n$ vertex-disjoint connected bipartite graphs $H_1, H_2,\dots, H_n$ with $n\geq 2$, and $e_k=|E(H_k)|$ with $e_1\geq e_2\geq \cdots \geq e_n$. If each connected bipartite graph $H_k$ is connected and admits a \emph{set-ordered graceful total coloring}, then there exists an edge set $E^*$, such that the edge-joining graph $E^*\oplus^n_{k=1}H_k$ admits a \emph{set-ordered graceful total coloring} too.
\end{thm}

Theorem \ref{thm:set-graceful-total-coloring} enables us to construct a \emph{graphic lattice} as
\begin{equation}\label{eqa:lattice-edge-sets}
\textbf{\textrm{L}}(\textbf{\textrm{E}}\oplus \textbf{\textrm{H}})=\left \{E^*\oplus ^n_{k=1}a_kH_k:~ a_k\in Z^0, H_{k}\in \textbf{\textrm{H}}, E^*\in \textbf{\textrm{E}}\right \},
\end{equation} where $\textbf{\textrm{E}}$ is a set of isolated edge sets $E^*$. Some graphs of $\textbf{\textrm{L}}(\textbf{\textrm{E}}\oplus \textbf{\textrm{H}})$ admit set-ordered graceful total colorings by Theorem \ref{thm:set-graceful-total-coloring}.

\begin{thm} \label{thm:coincide-graceful-total-coloring}
\cite{Wang-Yao-2021-SCI} If each connected bipartite graph $H_k$ of a graphic base $\textbf{\textrm{H}}=(H_k)^n_{k=1}$ admits a set-ordered graceful total coloring, and a connected bipartite graph $F$ with $n$ vertices admits a set-ordered graceful total coloring, then the $F$-graph $F\odot |^n_{k=1}H_k$ is a connected bipartite graph and admits a \emph{set-ordered graceful total coloring}.
\end{thm}

Theorem \ref{thm:coincide-graceful-total-coloring} induces a \emph{finite $F$-graphic lattice}
\begin{equation}\label{eqa:lattice-graph-coincide-operation}
\textbf{\textrm{L}}(\textbf{\textrm{F}}^*(n)\odot \textbf{\textrm{H}})=\left \{F\odot |^n_{k=1}H_k:~ H_{k}\in \textbf{\textrm{H}}, F\in \textbf{\textrm{F}}^*(n)\right \},
\end{equation} where $\textbf{\textrm{F}}^*(n)$ is a set of connected bipartite graphs of $n$ vertices, and many graphs of $\textbf{\textrm{F}}^*(n)$ admit set-ordered graceful total colorings by Theorem \ref{thm:coincide-graceful-total-coloring}. In general, we have a \emph{proper $F$-graphic lattice} as:
\begin{equation}\label{eqa:large-lattice-graph}
\textbf{\textrm{L}}(\textbf{\textrm{F}}^*\odot \textbf{\textrm{H}})=\left \{G\odot |^n_{k=1}a_kH_k:~ a_k\in Z^0, H_{k}\in \textbf{\textrm{H}}, G\in \textbf{\textrm{F}}^*\right \},
\end{equation} where $\textbf{\textrm{F}}^*$ is a set of connected bipartite graphs admitting set-ordered graceful total colorings, and $|V(G)|=\sum ^n_{k=1}a_k$.

\begin{rem}\label{rem:333333}
These graphic lattices $\textbf{\textrm{L}}(\textbf{\textrm{F}}^*(n)\odot \textbf{\textrm{H}})$ and $\textbf{\textrm{L}}(\textbf{\textrm{F}}^*\odot \textbf{\textrm{H}})$ are called to be closed to the set-ordered graceful total coloring if each graph in them admits a set-ordered graceful total coloring. Thereby, each graph of $\textbf{\textrm{L}}(\textbf{\textrm{F}}^*(n)\odot \textbf{\textrm{H}})$ and $\textbf{\textrm{L}}(\textbf{\textrm{F}}^*\odot \textbf{\textrm{H}})$ is a Topsnut-gpw, and can be applied to actual information encryption.\paralled
\end{rem}

We have a connection between graphic lattices and traditional lattices as follows: Suppose that each connected bipartite graph $H_k$ of the connected bipartite graph $F\odot |^n_{k=1}H_k$ of $\textbf{\textrm{L}}(\textbf{\textrm{F}}^*(n)\odot \textbf{\textrm{H}})$ has just $m$ vertices, and $\textbf{\textrm{d}}_k=(d_{k,1},d_{k,2},\dots ,d_{k,m})=(d_{k,i})^m_{i=1}$ with $d_{k,i}\geq d_{k,i+1}$ is the degree-sequence of $H_k$, after vertex-coincide a vertex $x_k$ of $F$ with a vertex $w_{k,j}$ of $H_k$ into one vertex $x_k\odot w_{k,j}$, we get a new degree-sequence $\textbf{\textrm{d}}'_k=(d\,'_{k,1},d\,'_{k,2},\dots ,d\,'_{k,m})=(d\,'_{k,j})^m_{j=1}$, where only one $d\,'_{k,j}=d_{k,j}+\textrm{deg}_F(x_k)$. Thereby, the graph $L=F\odot |^n_{k=1}H_k$ has its own degree-sequence
$$
\textbf{\textrm{d}}_L=\sum^n_{k=1} \textbf{\textrm{d}}'_k=\sum^n_{k=1} (d\,'_{k,j})^m_{j=1}=\left (\sum^n_{k=1} d\,'_{k,1},\sum^n_{k=1} d\,'_{k,2},\dots ,\sum^n_{k=1} d\,'_{k,m}\right )=\left (\sum^n_{k=1} d\,'_{k,j}\right )^m_{j=1},
$$ and these degree-sequences build up a traditional lattice
\begin{equation}\label{eqa:555555}
\textbf{\textrm{L}}(\textbf{\textrm{d}})=\left \{\sum^n_{k=1} b_k\textbf{\textrm{d}}'_k:~ b_k\in Z^0, \textbf{\textrm{d}}'_k=(d\,'_{k,j})^m_{j=1}, H_{k}\in \textbf{\textrm{H}}, F\in \textbf{\textrm{F}}^*(n)\right \}.
\end{equation}

\subsection{Graphic lattices based on ice-follower systems}

There are many results based on the ice-flower systems can be found in \cite{Bing-Yao-2020arXiv}.

\textbf{Ice-flower systems.} A star $K_{1,n}$ is a tree with diameter two, and it has its own vertex set $V(K_{1,n})=\{x,y_i:i\in [1,n]\}$ and edge set $E(K_{1,n})=\{xy_j:j\in [1,n]\}$, and each one-degree vertex $y_i$ is called a \emph{leaf} and each edge $xy_j$ is called a \emph{leaf-edge} of the star $K_{1,n}$, we call vertex $x$ the \emph{center} of the star $K_{1,n}$. Ice-flower systems based on stars have been introduced in \cite{Bing-Yao-2020arXiv}.

Doing the leaf-coinciding operation to a \emph{star base} $\textbf{\textrm{K}}=(K_{1,m_{1}},K_{1,m_{2}},\dots ,K_{1,m_{n}})$ in the following way: We leaf-coincide a leaf-edge $x_iy_{i,r}$ of a star $K_{1,m_{i}}$ with a leaf-edge $x_{j}y_{j,s}$ of another star $K_{1,m_{j}}$ into one edge $x_ix_{j}=x_iy_{i,r}\overline{\ominus} x_{j}y_{j,s}$ for $i\neq j$, which joins $K_{1,m_{i}}$ and $K_{1,m_{j}}$ together, the resultant graph is denoted as $K_{1,m_{i}}\overline{\ominus} K_{1,m_{j}}$, such that any two stars $K_{1,m_{a}}$ and $K_{1,m_{b}}$ with $a,b\in [1,n]$ and $a\neq b$ participate the leaf-coinciding operation at most once. If each star has participated the leaf-coinciding operation, and the resultant graph $F$ has not any leaf of each star of the star base $\textbf{\textrm{K}}$, then we write this graph as $F=\overline{\ominus} ^n_{i=1}K_{1,m_{i}}$, and call $\textbf{\textrm{K}}$ an \emph{ice-flower system}. In general, $\overline{\ominus} ^n_{i=1}K_{1,m_{i}}$ produces a set of graphs with the same degree-sequence $(m_1,m_2, \dots , m_n)$ and vertex set $\{x_1,x_2,\dots ,x_n\}$, where each vertex $x_i$ is the center of the star $K_{1,m_{i}}$ of $\textbf{\textrm{K}}$ for $i\in [1,n]$.

\begin{figure}[h]
\centering
\includegraphics[width=16cm]{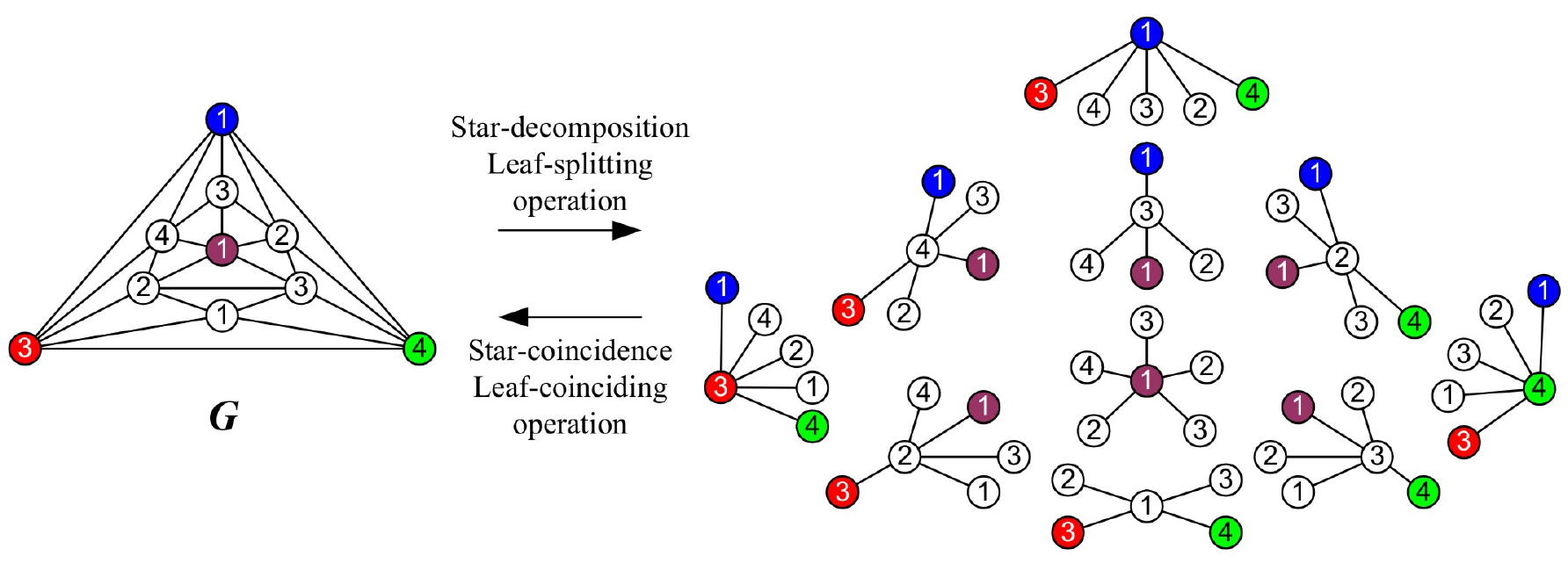}\\
\caption{\label{fig:star-decompose}{\small A star decomposition.}}
\end{figure}

\subsection{Lattices and homomorphisms based on degree-sequences}

\subsubsection{Complex graphs}

\begin{defn}\label{defn:imaginary-graph}
\cite{Wang-Yao-Su-Wanjia-Zhang-2021-IMCEC} A complex graph $H\overline{\ominus} iG$ has its own vertex set $=V(H)$, and \emph{imaginary vertex set} $V(iG)$, and edge set $E(H\overline{\ominus} iG)=E(H)\cup E(iG)\cup E_{1/2}$, where $E_{1/2}=\{xy:x\in V(H), y\in V(iG)\}$ is \emph{half-half edge set}, $E(H)$ is \emph{popular edge set}, and $E(iG)$ is \emph{imaginary edge set}. Moreover, each vertex $x\in V(H)$ is called a \emph{vertex} again, each vertex $y\in V(iG)$ is called an \emph{imaginary vertex}; each edge of $E(H)$ is called an \emph{edge} again, each edge of $E(iG)$ is called an \emph{imaginary edge}, and each edge of the half-half edge set is called a \emph{half-half edge}.\qqed
\end{defn}

A complex graph $H\overline{\ominus} iG$ has its own \emph{complex degree-sequence}
\begin{equation}\label{eqa:555555}
\textrm{Cdeg}(H\overline{\ominus} iG)=(a_1,a_2, \dots ,a_{p_r}, -b_1,-b_2, \dots ,-b_{p_i}),
\end{equation} where $p_r=|V(H)|$ and $p_i=|V(iG)|$, and \emph{complex degree} $\textrm{Cdeg}(x_j)=a_j\geq 0$ for each vertex $x_j\in V(H)$ with $j\in [1$, $p_r]$, and complex degree $\textrm{Cdeg}(y_s)=b_s\geq 0$ for each imaginary vertex $y_s\in V(iG)$ with $s\in [1,p_i]$, and moreover the size of the complex degree-sequence $\textrm{Cdeg}(H\overline{\ominus} iG)$ is $|\textrm{Cdeg}(H\overline{\ominus} iG)|=\sum^{p_r}_{i=1}a_{i}+\sum^{p_i}_{j=1}b_{j}$. In general, the complex degree-sequence of a complex graph $H\overline{\ominus} iG$ is written as $\textrm{Cdeg}(H\overline{\ominus} iG)=(c_1,c_2, \dots , c_{n})$ with $n=p_r+p_i$, where $c_1c_2 \cdots c_{n}$ is a permutation of $a_1a_2 \cdots a_{p_r} (-b_1)(-b_2) \cdots (-b_{p_i})$. In the view of vectors, a complex degree-sequence $\textrm{Cdeg}(H\overline{\ominus} iG)$ is just a \emph{popular vector}, so we call $\textrm{Cdeg}(H\overline{\ominus} iG)$ a \emph{graphic vector} . The \emph{proper degree-sequence} $\textrm{\textbf{d}}=(c\,'_{j_1},c\,'_{j_2}, \dots , c\,'_{j_n})$ with $c\,'_{j_s}=|c_{j_s}|$ for $c_{j_s}\in \textrm{Cdeg}(H\overline{\ominus} iG)$ and $c\,'_{j_s}\geq c\,'_{j_{s+1}}$ for $s\in $ $ [1,n]$ holds Erd\"{o}s-Gallai's theorem shown in (\ref{eqa:Erdos-Gallai}).

\begin{figure}[h]
\centering
\includegraphics[width=14cm]{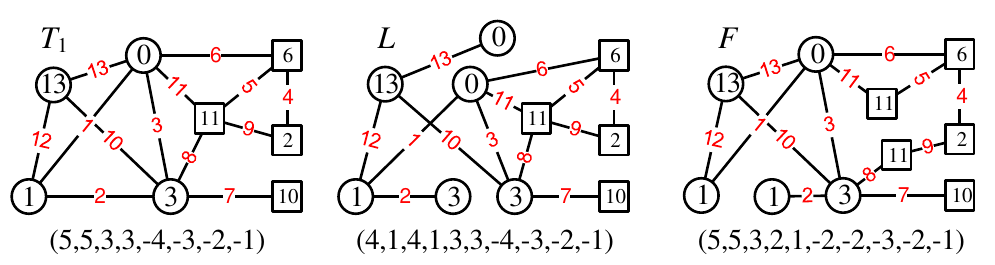}\\
\caption{\label{fig:Imaginary-graphs}{\small Three colored complex graphs have the same Topcode-matrix $A$ shown in (\ref{eqa:Three-imaginary-graphs-same-Topcode-matrix}) and different degree-sequences and, they admit gracefully total colorings, cited from \cite{Wang-Yao-Su-Wanjia-Zhang-2021-IMCEC}.}}
\end{figure}

\begin{equation}\label{eqa:Three-imaginary-graphs-same-Topcode-matrix}
\centering
{
\begin{split}
A=\left(
\begin{array}{ccccccccccccc}
0 & 1 & 0 & 2 & 6 & 0 & 3& 3 & 2 & 3 & 0 & 1 & 0\\
1 & 2 & 3 & 4 & 5 & 6 & 7& 8 & 9 & 10 & 11 & 12 & 13\\
1 & 3 & 3 & 6 & 11 &6 & 10 & 11 & 11 & 13 & 11 & 13 & 13
\end{array}
\right)
\end{split}}
\end{equation}

An integer sequence $\textbf{\textrm{d}}=(c_1$, $ c_2, \dots , c_p)$ with $|c_{i}|\geq |c_{i+1}|>0$ for $i\in [1,p-1]$ is the \emph{degree-sequence} of a certain graph $G$ of $p$ vertices and $q$ edges if and only if $\sum^n_{i=1}|c_{i}|=2q$ and
\begin{equation}\label{eqa:Erdos-Gallai}
\sum^k_{i=1}|c_{i}|\leq k(k-1)+\sum ^p_{i=k+1}\min\{k,|c_{i}|\}
\end{equation}
distributed by Erd\"{o}s and Gallai in 1960 \cite{Bondy-2008}, and we write $\textrm{deg}(G)=\textbf{\textrm{d}}$, and call $\textbf{\textrm{d}}$ \emph{proper degree-sequence}, $p=L_{ength}(\textbf{\textrm{d}})$ the \emph{length} of $\textbf{\textrm{d}}$ and $|\textrm{deg}(G)|=\sum^n_{i=1}|c_i|=2q$ \emph{size} of $\textrm{deg}(G)$.

\subsubsection{Complex degree-sequence lattices}

\begin{defn}\label{defn:99-complex-degree-sequence-lattice}
\cite{Wang-Yao-Su-Wanjia-Zhang-2021-IMCEC} For a \emph{complex degree-sequence base} $\textbf{\textrm{C}}^*=(C^1_n,C^2_n,\dots$, $C^m_n)$, where each $C^k_n=(I_{k,1},I_{k,2},\dots ,I_{k,n})=(I_{k,j})^n_{j=1}$ with $k\in [1,m]$ is a $n$-rank complex degree-sequence, then we get a new complex degree-sequence $(b_{j})^n_{j=1}=\sum^m_{k=1} a_kC^k_n$ with $a_k\in Z$, where each component $b_{j}=\sum^m_{k=1}a_kI_{k,j}$ for $j\in [1,n]$. In \cite{Wang-Yao-Su-Wanjia-Zhang-2021-IMCEC}, the following set
\begin{equation}\label{eqa:complex-degree-sequence-lattice}
\textbf{\textrm{L}}(\Sigma \textbf{\textrm{C}}^*)=\left \{\sum^m_{k=1} a_kC^k_n:a_k\in Z,C^k_n\in \textbf{\textrm{C}}^*\right \}
\end{equation} is called a \emph{complex degree-sequence lattice}, and define the following graph set
\begin{equation}\label{eqa:555555}
\textbf{\textrm{C}}_{omp}(\textbf{\textrm{L}}(\Sigma \textbf{\textrm{C}}^*))=\{H:~\textrm{Cdeg}(H)\in \textbf{\textrm{L}}(\Sigma \textbf{\textrm{C}}^*)\}
\end{equation}
the \emph{accompany graphic lattice} of $\textbf{\textrm{L}}(\Sigma \textbf{\textrm{C}}^*)$. \qqed
\end{defn}

\begin{thm}\label{thm:linear-independent-degree-sequences}
\cite{Wang-Yao-Su-Wanjia-Zhang-2021-IMCEC} An im-graph $H\overline{\ominus} iG$ with $n$ vertices has at least $n!$ groups of complex degree-sequences $d\,^1_t,d\,^2_t,d\,^3_t,\dots, d\,^n_t$ with $t\in [1,n!]$, in which the complex degree-sequences in each group are linear independent from each other.
\end{thm}

Clearly, any two graphs $L_i,L_k\in C_{omp}(\textbf{\textrm{L}}(\Sigma \textbf{\textrm{C}}^*))$ admit a \emph{graph base} $\textbf{\textrm{H}}$-\emph{similarity}, where $\textbf{\textrm{H}}=(H_1,H_2,\dots ,H_n)$ and each complex degree-sequence $\textrm{Cdeg}(H_i)=C^i_n$ with $i\in [1,n]$. We have a result like that in \cite{Yao-Wang-Ma-Wang-Degree-sequences-2021} as follows:

\begin{thm}\label{thm:integer-lattice}
\cite{Wang-Yao-Su-Wanjia-Zhang-2021-IMCEC} A complex degree-sequence lattice $\textbf{\textrm{L}}(\Sigma \textbf{\textrm{C}}^*)$ is equivalent to an integer lattice $\textrm{\textbf{L}}(\textbf{ZB})$.
\end{thm}

Without loss of generality, let $\big \| C^1_n\big \|=\min \{\big \| C^k_n\big \|:k\in [1,m]\}$ and $\big \| C^m_n\big \|=\max \{\big \| C^k_n\big \|:k\in [1,m]\}$, so we have
$$M\big \| C^1_n\big \|\leq \| \textbf{\textrm{y}}^*- \textbf{\textrm{w}}^*\|\leq M\big \| C^m_n\big \|,$$
where $M=\sqrt{\sum^m_{k=1}\big (t_k-r_k\big )^2}$, and moreover we take $t_k=r_k$ for $k\in [2,m]$ and $t_k=r_k+1$, so $\| \textbf{\textrm{y}}^*- \textbf{\textrm{w}}^*\|=\big \| C^1_n\big \|$ is smallest.

A complex ice-flower system is $\textbf{\textrm{CK}}=(K_{1,r_{1}},K_{1,r_{2}},\dots $, $K_{1,r_{m}}, iK_{1,s_{1}},iK_{1,s_{2}},\dots ,iK_{1,s_{n}})$, where each $iK_{1,s_{j}}$ for $j\in [1,n]$ is a \emph{complex star}. Each graph $G$ made by this complex ice-flower system, after running the leaf-coinciding operation, is denoted as $G=\overline{\ominus} ^{(m,n)}_{(r_{j},s_{i})}\langle K_{1,r_{j}}, iK_{1,s_{i}}\rangle$, such that $G$ has $|V(G)|=m+n$ vertices and $|E(G)|=\frac{1}{2}\big (\sum ^{m}_{j=1}r_{j}+\sum ^{n}_{i=1}s_{i}\big )$ edges. We make graphs $T=\overline{\ominus} ^{(m,n)}_{(r_{j},s_{i})}\langle a_{r_{j}}K_{1,r_{j}},b_{s_{i}}iK_{1,s_{i}}\rangle$ for $a_{r_{j}}, b_{s_{i}}\in Z^0$ by the leaf-coinciding operation, such that $T$ has no any leaf of ant star of $\textbf{\textrm{CK}}$, and call the following graph set
\begin{equation}\label{eqa:complex-graphic-lattices}
\textbf{\textrm{L}}(\textbf{\textrm{CK}})=\{\overline{\ominus} ^{(m,n)}_{(r_{j},s_{i})}\langle a_{r_{j}}K_{1,r_{j}},b_{s_{i}}iK_{1,s_{i}}\rangle:a_{r_{j}}, b_{s_{i}}\in Z^0\}
\end{equation}
a \emph{complex graphic lattices} based on the complex ice-flower system $\textbf{\textrm{CK}}$ with $\sum^m_{j=1}a_{r_{j}}\geq 1$ and $\sum^n_{i=1}b_{s_{i}}\geq 1$.

\subsubsection{Degree-sequence homomorphisms}

\begin{defn}\label{defn:degree-sequence-homomorphism}
\cite{Yao-Zhang-Wang-Su-Integer-Decomposing-2021} The degree-sequence homomorphism ``$\rightarrow $'' is defined as: For two graphs $G$ and $H$ with $G\not \cong H$, if there exists a mapping $\varphi :V(G)\rightarrow V(H)$ such that each edge $uv\in E(G)$ holds $\varphi(u)\varphi(v)\in E(H)$, where the degree-sequence $\textrm{deg}(G)=\textbf{\textrm{d}}$ and the degree-sequence $\textrm{deg}(H)=\textbf{\textrm{d}}\,'$, then we say that $\textbf{\textrm{d}}$ is \emph{degree-sequence homomorphism} to $\textbf{\textrm{d}}\,'$, denoted as $\textbf{\textrm{d}}\rightarrow \textbf{\textrm{d}}\,'$, and $L_{ength}(\textbf{\textrm{d}})>L_{ength}(\textbf{\textrm{d}}\,')$ if $\textbf{\textrm{d}}\neq \textbf{\textrm{d}}\,'$.\qqed
\end{defn}

\begin{rem}\label{rem:ABC-conjecture}
A \emph{graph homomorphism} $G\rightarrow H$ from a graph $G$ into another graph $H$ is a mapping $f: V(G) \rightarrow V(H)$ such that $f(u)f(v)\in E(H)$ for each edge $uv \in E(G)$ \cite{Bondy-2008}. We define a graph-splitting homomorphism as $H\rightarrow_{split} T$ by the vertex-splitting operation, in other words, a \emph{graph-split homomorphism} is the \emph{inverse} of a graph homomorphism. Similarly, $\textbf{\textrm{d}}\rightarrow \textbf{\textrm{d}}\,'$ is accompanied by $\textbf{\textrm{d}}\,'\rightarrow_{split} \textbf{\textrm{d}}$, called \emph{degree-sequence splitting homomorphism}.\paralled
\end{rem}

\begin{figure}[h]
\centering
\includegraphics[width=16.4cm]{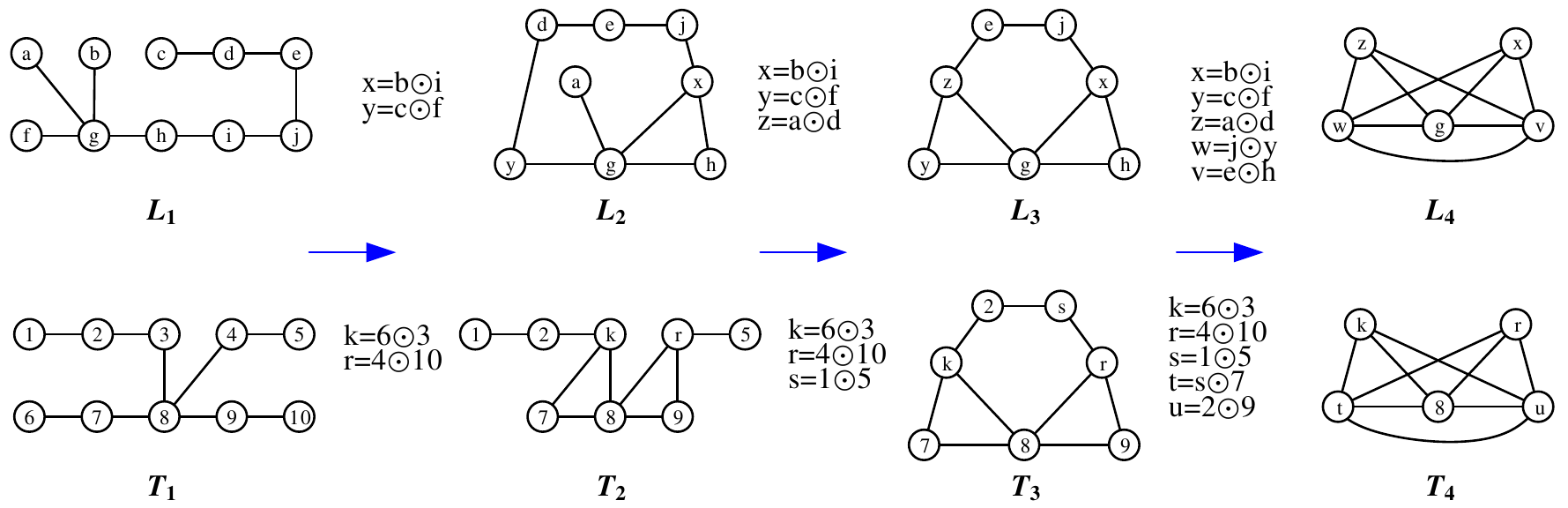}\\
\caption{\label{fig:degree-colored-homomorphism}{\small For understanding the degree-sequence homomorphism.}}
\end{figure}

\begin{figure}[h]
\centering
\includegraphics[width=16.4cm]{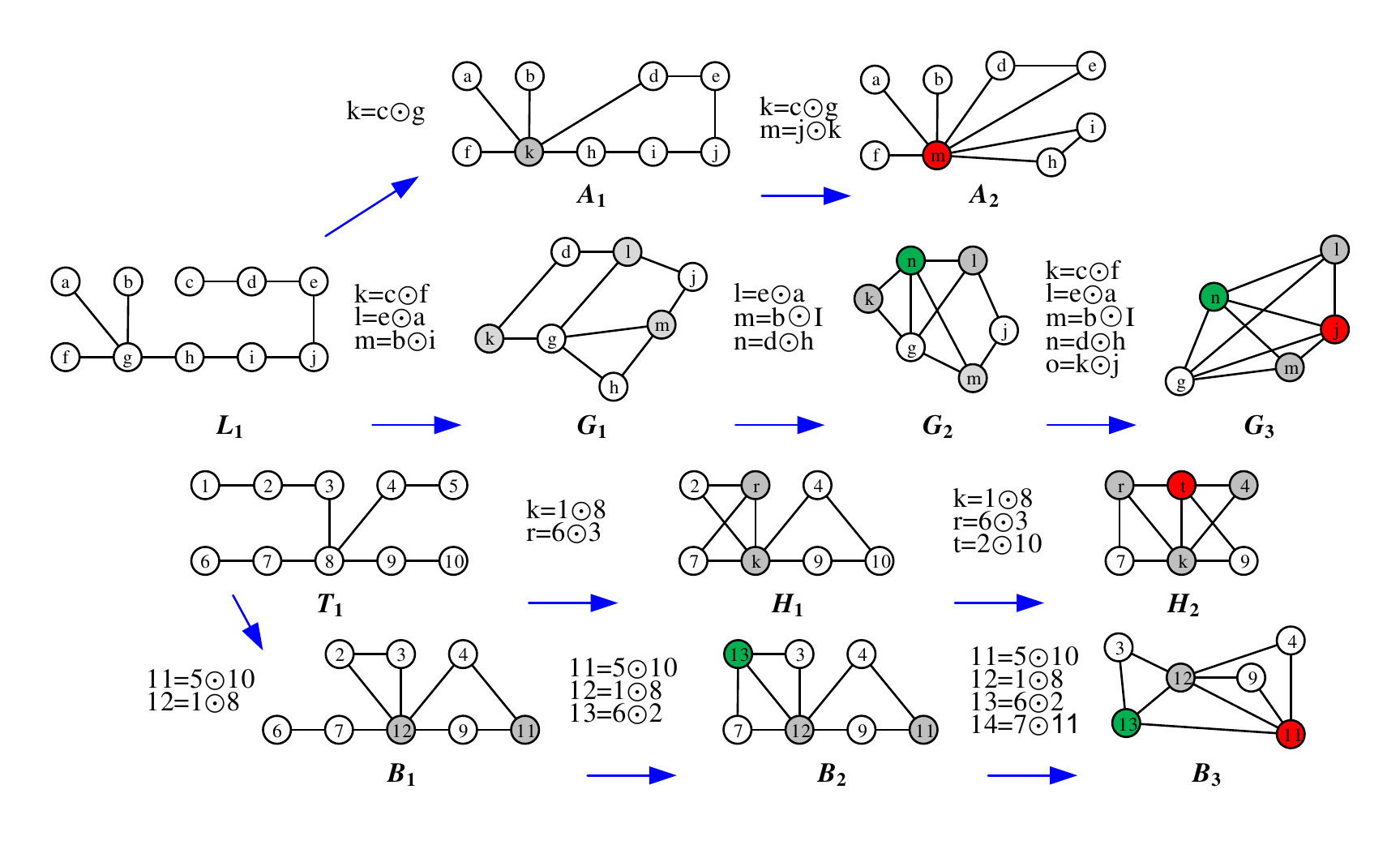}\\
\caption{\label{fig:degree-colored-homomorphism-2}{\small More degree-sequence homomorphisms and graph homomorphisms.}}
\end{figure}

Two examples about degree-sequence homomorphism are shown in Fig.\ref{fig:degree-colored-homomorphism}, we have \emph{degree-sequence homomorphisms} $\textrm{deg}(L_i)\rightarrow \textrm{deg}(L_{i+1})$ and $\textrm{deg}(T_i)\rightarrow \textrm{deg}(T_{i+1})$ with $i=1,2,3$. Let $\textbf{\textrm{d}}=$(4, 2, 2, 2, 2, 2, 1, 1, 1, 1), since $\textrm{deg}(L_1)=\textrm{deg}(T_1)=\textbf{\textrm{d}}$, we have a group of \emph{degree-sequence homomorphisms}
\begin{equation}\label{eqa:555555}
\textbf{\textrm{d}}=\textrm{deg}(L_1)\rightarrow \textrm{deg}(L_2)\rightarrow \textrm{deg}(L_3)\rightarrow \textrm{deg}(L_4)
\end{equation}
where $\textrm{deg}(L_2)=$(4, 3, 2, 2, 2, 2, 2, 1), $\textrm{deg}(L_3)=$(4, 3, 3, 2, 2, 2, 2) and $\textrm{deg}(L_4)=$(4, 4, 4, 3, 3). We have another group of degree-sequence homomorphisms as follows
\begin{equation}\label{eqa:555555}
\textbf{\textrm{d}}= \textrm{deg}(T_1)\rightarrow \textrm{deg}(T_2)\rightarrow \textrm{deg}(T_3)\rightarrow \textrm{deg}(T_4)
\end{equation}
where $\textrm{deg}(T_2)=$(4, 3, 3, 2, 2, 2, 1, 1), and moreover $\textrm{deg}(L_k)=\textrm{deg}(T_k)$ and $L_k\cong T_k$ for $j=3,4$. Moreover, $\textrm{deg}(L_k)=\textrm{deg}(T_k)$ and $L_k\cong T_k$ for $k=2,3$. Fig.\ref{fig:degree-colored-homomorphism-2} shows us more examples of degree-sequence homomorphisms:
$${
\begin{split}
&\textbf{\textrm{d}}=\textrm{deg}(L_1)\rightarrow \textrm{deg}(A_1)\rightarrow \textrm{deg}(A_2),\\
&\textbf{\textrm{d}}=\textrm{deg}(T_1)\rightarrow \textrm{deg}(H_1)\rightarrow \textrm{deg}(H_2),\\
&\textbf{\textrm{d}}=\textrm{deg}(L_1)\rightarrow \textrm{deg}(G_1)\rightarrow \textrm{deg}(G_2)\rightarrow \textrm{deg}(G_3),\\
&\textbf{\textrm{d}}=\textrm{deg}(L_1)\rightarrow \textrm{deg}(B_1)\rightarrow \textrm{deg}(B_2)\rightarrow \textrm{deg}(B_3).
\end{split}}$$

\begin{lem}\label{thm:degree-sequence-homomorphism}
\cite{Yao-Zhang-Wang-Su-Integer-Decomposing-2021} If a simple graph $G$ holds $N(u)\cap N(v)\neq \emptyset$ for any edge $uv\not \in E(G)$ and $u,v\in V(G)$, then its degree-sequence $\textrm{deg}(G)$ is not \emph{degree-sequence homomorphism} to any degree-sequence, except $\textrm{deg}(G)$ itself.
\end{lem}

By the vertex-splitting operation and the vertex-coinciding operation, we can prove the following result:

\begin{thm}\label{thm:keep-degree-sequence-homomorphism}
\cite{Yao-Zhang-Wang-Su-Integer-Decomposing-2021} Suppose that $\textrm{deg}(G)=\textbf{\textrm{d}}$ and $\textrm{deg}(H)=\textbf{\textrm{d}}\,'$ for two graphs $G$ and $H$, where two degree-sequences $\textbf{\textrm{d}}=(a_1,a_2, \dots , a_p)$ and $\textbf{\textrm{d}}\,'=(b_1,b_2, \dots , b_{p-1})$ with $a_1+a_2=b_1$ and $a_{i}=b_{i-1}$ for $i\in [3,p]$. We have a degree-sequence homomorphism $\textbf{\textrm{d}}\rightarrow \textbf{\textrm{d}}\,'$ if and only if we have a graph homomorphism $G\rightarrow H$.
\end{thm}

\begin{figure}[h]
\centering
\includegraphics[width=16.4cm]{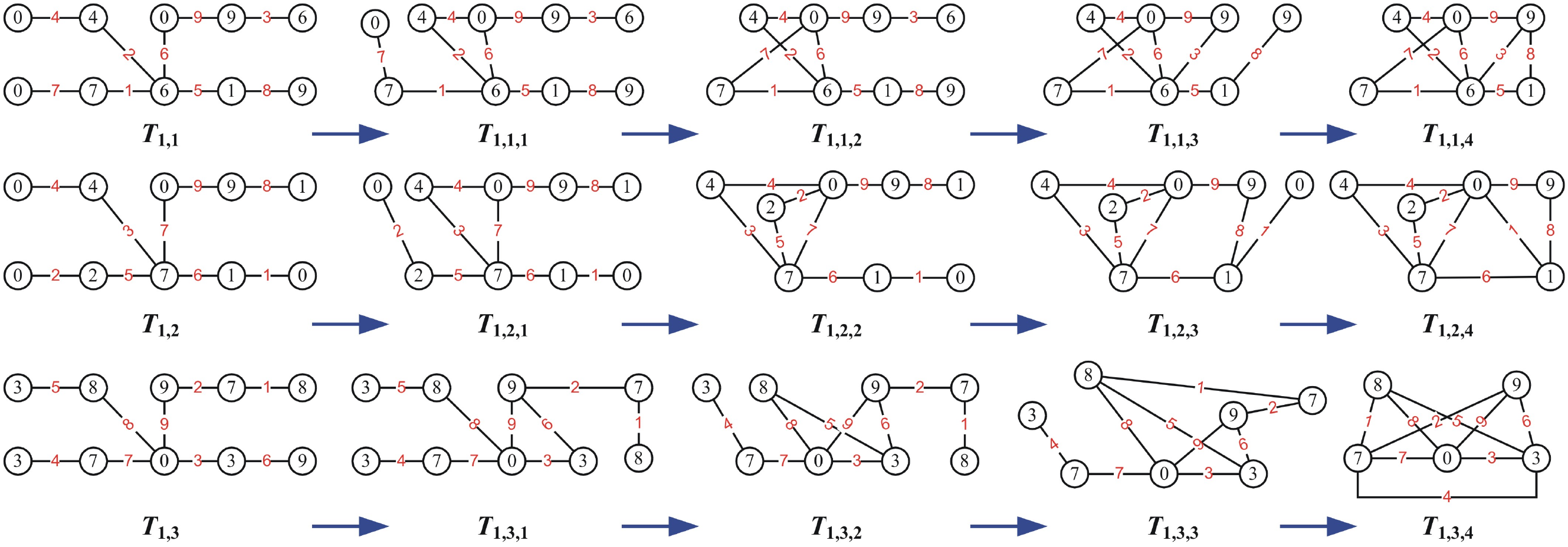}\\
\caption{\label{fig:colored-ds-homomo}{\small Examples on colored degree-sequence homomorphisms.}}
\end{figure}

\begin{rem}\label{rem:ABC-conjecture}
A degree-sequence $\textbf{\textrm{d}}(t)=(a_1(t),a_2(t), \dots , a_m(t))$ for $t\in [a,b]$ is called a \emph{dynamic degree-sequence}. In general, the length $s(t)=L_{ength}(\textbf{\textrm{d}}(t))$ of a dynamic degree-sequence $\textbf{\textrm{d}}(t)$ is a function of time $t$, so we write $\textbf{\textrm{d}}(t)=(d_1(s(t),t),d_2(s(t),t), \dots , d_{s(t)}(s(t),t))$ with $t\in [a,b]$. It may be interesting to characterize dynamic degree-sequences by means of scale-free degree-sequences of scale-free graphs \cite{Li-Alderson-Doyle-Willinger-2006scale-free-graphs}.\paralled
\end{rem}

\subsubsection{Degree-sequence lattices}

\begin{defn}\label{defn:self-contraction-operation}
\cite{Yao-Wang-Su-Jianmin-Xie-2021-Conference} Suppose that two sequences $\textbf{\textrm{d}}=(a_1,a_2,\dots ,a_p)$ and $\textbf{\textrm{d}}\,'=(b_1,b_2, \dots , b_{p-1})$ hold $a_1+a_2=b_1$ and $a_{i}=b_{i-1}$ for $i\in [3,p]$ true, we write $\textbf{\textrm{d}}\,'=\odot_1(\textbf{\textrm{d}})$, and call this operation a \emph{self-contraction operation}, and moreover $\textbf{\textrm{d}}=\wedge_1(\textbf{\textrm{d}}\,')$ is called a \emph{self-splitting degree-sequence} under the \emph{self-splitting operation}.\qqed
\end{defn}

Doing the one-order self-contraction operation, we get $\textbf{\textrm{d}}_1=\odot_{1}(\textbf{\textrm{d}})$, and $\textbf{\textrm{d}}_{i+1}=\odot_{1}(\textbf{\textrm{d}}_{i})$ for $i\in [1,k-1]$, very often, we write $\textbf{\textrm{d}}_k=\odot_{k}(\textbf{\textrm{d}})$, so $\textbf{\textrm{d}}_{i}\rightarrow \textbf{\textrm{d}}_{i+1}$, and $\textbf{\textrm{d}}\rightarrow \textbf{\textrm{d}}_{k}$. If $\textbf{\textrm{d}}_k$ is a degree-sequence, but any $\odot_{1}(\textbf{\textrm{d}}_k)$ is no longer a degree-sequence, we get a set $e_{nd}(\textbf{\textrm{d}})$ of degree-sequences such that the self-contraction degree-sequence $\odot_{1}(\textbf{\textrm{d}}^*)$ of any degree-sequence $\textbf{\textrm{d}}^*\in e_{nd}(\textbf{\textrm{d}})$ is no longer a degree-sequence.

\begin{defn}\label{defn:degree-coinciding-operation}
\cite{Yao-Wang-Su-Jianmin-Xie-2021-Conference} For two sequences $\textbf{\textrm{d}}=(a_1,a_2, \dots , a_m)$ and $\textbf{\textrm{d}}\,'=(c_1,c_2, \dots , c_{n})$, make a new sequence by adding a degree component $a_i\in \textbf{\textrm{d}}$ with another degree component $c_j\in \textbf{\textrm{d}}\,'$ together, and put other degree components of two sequences $\textbf{\textrm{d}}$ and $\textbf{\textrm{d}}\,'$ into the new sequence as follows
\begin{equation}\label{eqa:degree-coinciding-operation}
{
\begin{split}
\odot\langle \textbf{\textrm{d}},\textbf{\textrm{d}}\,'\rangle=(a_1,a_2, \dots , a_{i-1},a_{i+1}, \dots ,a_m,c_1,c_2, \dots , c_{j-1},c_{j+1}, \dots ,c_{n},a_i+c_j)
\end{split}}
\end{equation} we call the procedure of obtaining $\odot\langle \textbf{\textrm{d}},\textbf{\textrm{d}}\,'\rangle$ a \emph{degree-coinciding operation}. The following degree-sequence
\begin{equation}\label{eqa:degree-joining-operation}
{
\begin{split}
\ominus\langle \textbf{\textrm{d}},\textbf{\textrm{d}}\,'\rangle=(a_1,a_2, \dots , a_{i-1},1+a_i,a_{i+1}, \dots ,a_m,c_1,c_2, \dots , c_{j-1},1+c_j,c_{j+1}, \dots ,c_{n})
\end{split}}
\end{equation} is the result of a \emph{degree-joining operation}.\qqed
\end{defn}

\begin{thm}\label{thm:degree-coinciding-joining-operation}
\cite{Yao-Wang-Su-Jianmin-Xie-2021-Conference} For two sequences $\textbf{\textrm{d}}=(a_1,a_2, \dots , a_m)$ and $\textbf{\textrm{d}}\,'=(c_1,c_2, \dots , c_{n})$, the sequence $\odot\langle \textbf{\textrm{d}},\textbf{\textrm{d}}\,'\rangle$ (resp. $\ominus\langle \textbf{\textrm{d}}_1,\textbf{\textrm{d}}_2\rangle$) is a degree-sequence if and only if both sequences $\textbf{\textrm{d}}$ and $\textbf{\textrm{d}}\,'$ are degree-sequences.
\end{thm}

\begin{thm}\label{thm:degree-coinciding-joining-operation}
\cite{Yao-Wang-Su-Jianmin-Xie-2021-Conference} If there are four degree-sequences $\textbf{\textrm{d}},\textbf{\textrm{d}}_1,\textbf{\textrm{d}}\,',\textbf{\textrm{d}}_2$ holding $\textbf{\textrm{d}}\rightarrow \textbf{\textrm{d}}_1$ and $\textbf{\textrm{d}}\,'\rightarrow \textbf{\textrm{d}}_2$ true, then we have $\odot\langle \textbf{\textrm{d}},\textbf{\textrm{d}}\,'\rangle\rightarrow \odot\langle \textbf{\textrm{d}}_1,\textbf{\textrm{d}}_2\rangle$, and $\ominus\langle \textbf{\textrm{d}},\textbf{\textrm{d}}\,'\rangle\rightarrow \ominus\langle \textbf{\textrm{d}}_1,\textbf{\textrm{d}}_2\rangle$
\end{thm}

\begin{defn}\label{defn:Ds-homomorphism-sequence-graph-set}
\cite{Yao-Wang-Su-Jianmin-Xie-2021-Conference} Suppose that $\{\textbf{\textrm{d}}_1,\textbf{\textrm{d}}_2,\dots, \textbf{\textrm{d}}_m\}$ is a degree-sequence set, and there are degree-sequence homomorphisms $\textbf{\textrm{d}}_{k}\rightarrow \textbf{\textrm{d}}_{k+1}$ with $L_{ength}(\textbf{\textrm{d}}_{k})>L_{ength}(\textbf{\textrm{d}}_{k+1})$ for $k\in [1,m-1]$ (refer to Definition \ref{defn:degree-sequence-homomorphism}), we denote them as $H_{omo}\{\textbf{\textrm{d}}_k\}^m_{k=1}$, and call $H_{omo}\{\textbf{\textrm{d}}_k\}^m_{k=1}$ a \emph{degree-sequence homomorphism sequence} (Ds-homomorphism sequence). Since each $\textbf{\textrm{d}}_{k}\in H_{omo}\{\textbf{\textrm{d}}_k\}^m_{k=1}$ corresponds a set $G_{raph}(\textbf{\textrm{d}}_{k})$ of graphs having their degree-sequences to be the same $\textbf{\textrm{d}}_{k}$, so we define a \emph{graph-set homomorphism} $G_{raph}(\textbf{\textrm{d}}_{k})\rightarrow G_{raph}(\textbf{\textrm{d}}_{k+1})$ if each $H_k\in G_{raph}(\textbf{\textrm{d}}_{k})$ with $\textrm{deg}(H_k)=\textbf{\textrm{d}}_{k}$ is graph homomorphism to a graph $H_{k+1}\in G_{raph}(\textbf{\textrm{d}}_{k+1})$ with $\textrm{deg}(H_{k+1})=\textbf{\textrm{d}}_{k+1}$ for $k\in [1,m-1]$.\qqed
\end{defn}

In the following statement, we will apply Definition \ref{defn:colored-degree-sequence-matrix}: A coloring $f: \textbf{\textrm{d}}\rightarrow \{b_1,b_2,\dots ,b_p\}$ with $b_j\in Z^0$ for a $p$-rank degree-sequence $\textbf{\textrm{d}}=(a_1,a_2,\dots ,a_p)$ of a $(p,q)$-graph with $a_j\in Z^0\setminus \{0\}$, and a \emph{colored degree-sequence matrix} (Cds-matrix)
\begin{equation}\label{eqa:xxxxxxxxxxxxxxx}
D_{sc}(\textbf{\textrm{d}})= \left(
\begin{array}{ccccc}
a_1 & a_2 & \cdots & a_p\\
f(a_1) & f(a_2) & \cdots & f(a_p)\\
\end{array}
\right)=(\textbf{\textrm{d}},f(\textbf{\textrm{d}}))^T,
\end{equation} where \emph{color vector} $f(\textbf{\textrm{d}})=(f(a_1), f(a_2), \dots , f(a_p))$.

\begin{defn}\label{defn:Cds-matrix-homomorphism}
\cite{Yao-Wang-Su-Jianmin-Xie-2021-Conference} For a $p$-rank degree-sequence $D_{sc}(\textbf{\textrm{d}})$ based on $\textbf{\textrm{d}}=(a_1,a_2,\dots ,a_p)$, if there is another colored degree-sequence matrix $D_{sc}(\textbf{\textrm{d}}\,')=(\textbf{\textrm{d}}\,',g(\textbf{\textrm{d}}\,'))^T$ based on a $(p-1)$-rank degree-sequence $\textbf{\textrm{d}}\,'=(b_1,b_2,\dots ,b_{p-1})$ of a $(p-1,q)$-graph with $b_j\in Z^0\setminus \{0\}$, such that $b_1=a_1+a_2$ and $g(b_1)=f(a_1)=f(a_2)$, as well as $b_{i-1}=a_i$ and $g(b_{i-1})=f(a_i)$ for $i\in [3,p]$. We say that the colored degree-sequence matrix $D_{sc}(\textbf{\textrm{d}})$ is \emph{degree-sequence matrix homomorphism} to the colored degree-sequence matrix $D_{sc}(\textbf{\textrm{d}}\,')$, and write this fact as $D_{sc}(\textbf{\textrm{d}})\rightarrow^{cdsm} D_{sc}(\textbf{\textrm{d}}\,')$ (resp. the inverse $D_{sc}(\textbf{\textrm{d}}\,')\rightarrow^{cdsm}_{split} D_{sc}(\textbf{\textrm{d}})$).

Each $D_{sc}(\textbf{\textrm{d}})$ corresponds a set $G_{raph}(D_{sc}(\textbf{\textrm{d}}))$ of colored graphs having their colored degree-sequence matrices to be the same $D_{sc}(\textbf{\textrm{d}})$. So we have a \emph{graph-set homomorphism} $G_{raph}(D_{sc}(\textbf{\textrm{d}}))\rightarrow G_{raph}(D_{sc}(\textbf{\textrm{d}}\,'))$ since $D_{sc}(\textbf{\textrm{d}})\rightarrow^{cdsm} D_{sc}(\textbf{\textrm{d}}\,')$.\qqed
\end{defn}

\begin{thm}\label{thm:xxxxxx}
\cite{Yao-Wang-Su-Jianmin-Xie-2021-Conference} In a degree-sequence matrix homomorphism $D_{sc}(\textbf{\textrm{d}})\rightarrow^{cdsm} D_{sc}(\textbf{\textrm{d}}\,')$, if $D_{sc}(\textbf{\textrm{d}})=(\textbf{\textrm{d}},f(\textbf{\textrm{d}}))^T$ holds that $f$ induces a graceful (or odd-graceful) coloring of a simple graph $G$ with $\textrm{deg}(G)=\textbf{\textrm{d}}$, then $D_{sc}(\textbf{\textrm{d}}\,')=(\textbf{\textrm{d}}\,',g(\textbf{\textrm{d}}\,'))^T$ holds that $g$ induces a graceful (resp. odd-graceful) coloring of a simple graph $H$ with $\textrm{deg}(H)=\textbf{\textrm{d}}\,'$, such that $G\rightarrow H$.
\end{thm}

By the graph $T_{1,1}$ shown in Fig.\ref{fig:colored-ds-homomo}, we have a Cds-matrix as
\begin{equation}\label{eqa:cds-matrix-11}
{
\begin{split}
D_{sc}(\textrm{deg}(T_{1,1}))= \left(
\begin{array}{cccccccccc}
4& 2& 2& 2& 2& 2& 1& 1& 1& 1\\
6& 7& 4& 0& 9& 1& 0& 6& 0& 9
\end{array}
\right)=(\textrm{deg}(T_{1,1}),f_{1,1}(\textrm{deg}(T_{1,1}))^T
\end{split}}
\end{equation}
and a Topcode-matrix as
\begin{equation}\label{eqa:topcode-matrix-11}
{
\begin{split}
T_{code}(T_{1,1})= \left(
\begin{array}{cccccccccc}
7& 4& 9& 4& 6& 6& 7& 1& 9\\
1& 2& 3& 4& 5& 6& 7& 8& 9\\
6& 6& 6& 0& 1& 0& 0& 9& 0
\end{array}
\right)
\end{split}}
\end{equation}
By the graph $T_{1,1}$ shown in Fig.\ref{fig:colored-ds-homomo}, there are two matrices as follows
\begin{equation}\label{eqa:cds-matrix-11}
{
\begin{split}
D_{sc}(\textrm{deg}(T_{1,1,4}))= \left(
\begin{array}{cccccc}
5& 4& 3& 2& 2& 2\\
6& 0& 9& 4& 7& 1
\end{array}
\right)=(\textrm{deg}(T_{1,1,4}),f_{1,1}(\textrm{deg}(T_{1,1,4}))^T
\end{split}}
\end{equation}
and $T_{code}(T_{1,1,4})=T_{code}(T_{1,1})$, since there is a \emph{graph homomorphic chain} $ T_{1,1}\rightarrow T_{1,1,1}\rightarrow T_{1,1,2}\rightarrow T_{1,1,3}\rightarrow T_{1,1,4}$.

\begin{figure}[h]
\centering
\includegraphics[width=16.4cm]{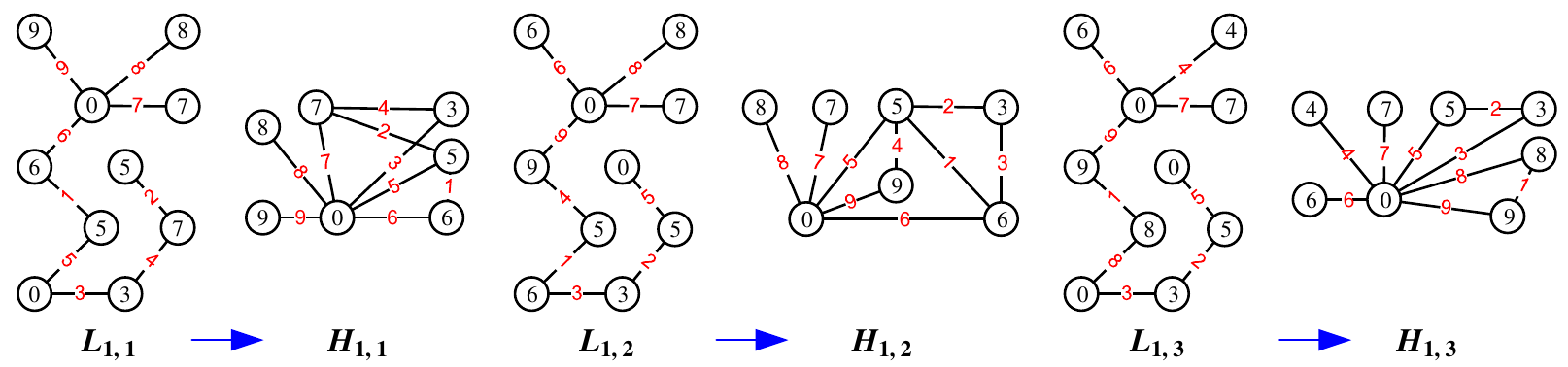}\\
\caption{\label{fig:colored-ds-homomo-11}{\small Three colored degree-sequence homomorphisms.}}
\end{figure}

\begin{defn} \label{defn:Cds-matrix-group-definition}
\cite{Yao-Wang-Su-Jianmin-Xie-2021-Conference} Let $D^1_{sc}(\textbf{\textrm{d}})=(\textbf{\textrm{d}},f_1(\textbf{\textrm{d}}))^T$ be a Cds-matrix defined in Definition \ref{defn:colored-degree-sequence-matrix}, and $M=\max\{a_i\in \textbf{\textrm{d}}:i\in [1,p]\}$ be a positive integer. By $f_j(a_s)=f_1(a_s)+j~(\bmod~M)$ for each degree component $a_s\in \textbf{\textrm{d}}$ with $s\in [1,p]$, we define $f_{j}(\textbf{\textrm{d}})=f_1(\textbf{\textrm{d}})+j$ $(\bmod~M)$ for each Cds-matrix $D^j_{sc}(\textbf{\textrm{d}})=(\textbf{\textrm{d}},f_j(\textbf{\textrm{d}}))^T$. A set of Cds-matrices $D^j_{sc}(\textbf{\textrm{d}})$ is denoted by $\big \{D^{(M)}_{sc}(\textbf{\textrm{d}}),\oplus \big \}$. For a fixed $D^k_{sc}(\textbf{\textrm{d}})\in \big \{D^{(M)}_{sc}(\textbf{\textrm{d}}),\oplus \big \}$ being consisted of a graph set $\{D^k_{sc}(\textbf{\textrm{d}})\}^M_{k=1}$ and an algebraic operation ``$\oplus$'', we have
\begin{equation}\label{eqa:Cds-matrix-group-operation}
f_i(\textbf{\textrm{d}})+f_j(\textbf{\textrm{d}})-f_k(\textbf{\textrm{d}})=f_{\lambda}(\textbf{\textrm{d}})
\end{equation} with $\lambda=i+j-k~(\bmod~M)$ and a \emph{preappointed zero} $f_{k}(\textbf{\textrm{d}})\in \{D^{(M)}_{sc}(\textbf{\textrm{d}}),\oplus \}$, where the formula (\ref{eqa:Cds-matrix-group-operation}) is defined by
\begin{equation}\label{eqa:Cds-matrix-group-operation11}
f_i(a_s)+f_j(a_s)-f_k(a_s)=f_{\lambda}(a_s)
\end{equation} for each degree component $a_s\in \textbf{\textrm{d}}$ with $s\in [1,p]$. We call the set $\big \{D^{(M)}_{sc}(\textbf{\textrm{d}}),\oplus \big \}$ an \emph{every-zero Cds-matrix group}.\qqed
\end{defn}

The proof of every-zero Cds-matrix group is similar with that in \cite{Yao-Sun-Zhang-Mu-Sun-Wang-Su-Zhang-Yang-Yang-2018arXiv}.

\begin{defn} \label{defn:degree-coinciding-ds-lattices}
\cite{Yao-Wang-Su-Jianmin-Xie-2021-Conference} Let $\textbf{\textrm{I}}=(\textbf{\textrm{d}}_1,\textbf{\textrm{d}}_2,\dots ,\textbf{\textrm{d}}_m)$ be a \emph{degree-sequence vector}, where degree-sequences $\textbf{\textrm{d}}_1,\textbf{\textrm{d}}_2,\dots ,\textbf{\textrm{d}}_m$ are independent from each other under the degree-coinciding operation ``$\odot$''. For $\sum^m_{k=1} a_k\geq 1$ with $a_k\in Z^0$, we have a set $\{a_1\textbf{\textrm{d}}_1,a_2\textbf{\textrm{d}}_2,\dots ,a_m\textbf{\textrm{d}}_m\}= \{a_k\textbf{\textrm{d}}_k\}^m_{k=1}$, and do the degree-coinciding operation ``$\odot$'' defined in definition \ref{defn:degree-coinciding-operation} to the elements of the set $\{a_k\textbf{\textrm{d}}_k\}^m_{k=1}$ such that each of $a_k$ degree-sequences $\textbf{\textrm{d}}_k$ appears in some $\odot\langle \textbf{\textrm{d}}_k,\textbf{\textrm{d}}_j\rangle$ if $a_k\neq 0$ and $a_j\neq 0$. The resulting degree-sequences are collected into a set $\odot\langle a_k\textbf{\textrm{d}}_k\rangle^m_{k=1}$. We call the following set
\begin{equation}\label{eqa:555555}
\textbf{\textrm{L}}(Z^0\odot \langle \textbf{\textrm{I}}\rangle )=\{\odot\langle a_k\textbf{\textrm{d}}_k\rangle^m_{k=1}:~a_k\in Z^0,\textbf{\textrm{d}}_k\in \textbf{\textrm{I}}\}
\end{equation} a \emph{degree-coinciding degree-sequence lattice}.\qqed
\end{defn}

\begin{defn} \label{defn:degree-joining-ds-lattices}
\cite{Yao-Wang-Su-Jianmin-Xie-2021-Conference} Let $\textbf{\textrm{I}}=(\textbf{\textrm{d}}_1,\textbf{\textrm{d}}_2,\dots ,\textbf{\textrm{d}}_m)$ be a \emph{degree-sequence vector}, where degree-sequences $\textbf{\textrm{d}}_1,\textbf{\textrm{d}}_2,\dots ,\textbf{\textrm{d}}_m$ are independent from each other under the degree-coinciding operation ``$\ominus$''. For $\sum^m_{k=1} a_k\geq 1$ with $a_k\in Z^0$, we have a set $\{a_1\textbf{\textrm{d}}_1,a_2\textbf{\textrm{d}}_2,\dots ,a_m\textbf{\textrm{d}}_m\}= \{a_k\textbf{\textrm{d}}_k\}^m_{k=1}$, and do the degree-coinciding operation ``$\ominus$'' defined in definition \ref{defn:degree-coinciding-operation} to the elements of the set $\{a_k\textbf{\textrm{d}}_k\}^m_{k=1}$ such that each of $a_s$ degree-sequences $\textbf{\textrm{d}}_s$ appears in some $\ominus\langle \textbf{\textrm{d}}_s,\textbf{\textrm{d}}_i\rangle$ if $a_s\neq 0$ and $a_i\neq 0$. The resulting degree-sequences are collected into a set $\ominus\langle a_k\textbf{\textrm{d}}_k\rangle^m_{k=1}$. We call the following set
\begin{equation}\label{eqa:555555}
\textbf{\textrm{L}}(Z^0\ominus \langle \textbf{\textrm{I}}\rangle )=\{\ominus\langle a_k\textbf{\textrm{d}}_k\rangle^m_{k=1}:~a_k\in Z^0,\textbf{\textrm{d}}_k\in \textbf{\textrm{I}}\}
\end{equation} a \emph{degree-joining degree-sequence lattice}.\qqed
\end{defn}

\begin{defn} \label{defn:degree-sequence-lattices-homomorphism}
\cite{Yao-Wang-Su-Jianmin-Xie-2021-Conference} If a degree-sequence vector $\textbf{\textrm{I}}=(\textbf{\textrm{d}}_1,\textbf{\textrm{d}}_2$, $\dots ,\textbf{\textrm{d}}_m)$ in a degree-coinciding degree-sequence lattice $\textbf{\textrm{L}}(Z^0\odot \langle \textbf{\textrm{I}}\rangle )$ (resp. $\textbf{\textrm{L}}(Z^0\ominus \langle \textbf{\textrm{I}}\rangle )$) is \emph{degree-sequence vector homomorphism} to another degree-sequence vector $\textbf{\textrm{I}}\,'=(\textbf{\textrm{d}}\,'_1,\textbf{\textrm{d}}\,'_2,\dots ,\textbf{\textrm{d}}\,'_m)$ in another degree-coinciding degree-sequence lattice $\textbf{\textrm{L}}(Z^0\odot \langle \textbf{\textrm{I}}\,'\rangle )$ (resp. $\textbf{\textrm{L}}(Z^0\ominus \langle \textbf{\textrm{I}}\,'\rangle )$), such that each degree-sequence homomorphism $\textbf{\textrm{d}}_k\rightarrow \textbf{\textrm{d}}\,'_k$ with $k\in [1,m]$ holds true, then $\textbf{\textrm{L}}(Z^0\odot \langle \textbf{\textrm{I}}\rangle )$ is \emph{homomorphism} to $\textbf{\textrm{L}}(Z^0\odot \langle \textbf{\textrm{I}}\,'\rangle )$ (resp. $\textbf{\textrm{L}}(Z^0\ominus \langle \textbf{\textrm{I}}\rangle )$ is \emph{homomorphism} to $\textbf{\textrm{L}}(Z^0\ominus \langle \textbf{\textrm{I}}\,'\rangle )$), we call
\begin{equation}\label{eqa:ds-lattice-homomorphism}
\textbf{\textrm{L}}(Z^0\odot \langle \textbf{\textrm{I}}\rangle )\rightarrow \textbf{\textrm{L}}(Z^0\odot \langle \textbf{\textrm{I}}\,'\rangle )
\end{equation} (resp. $\textbf{\textrm{L}}(Z^0\ominus \langle \textbf{\textrm{I}}\rangle )\rightarrow \textbf{\textrm{L}}(Z^0\ominus \langle \textbf{\textrm{I}}\,'\rangle )$) a \emph{degree-sequence lattice homomorphism}.\qqed
\end{defn}

\begin{rem}\label{rem:ABC-conjecture}
For $(\ast)\in \{\ominus, \odot\}$, each $\textbf{\textrm{L}}(Z^0(\ast) \langle \textbf{\textrm{I}}\rangle )$ corresponds a set $G_{raph}(\textbf{\textrm{L}}(Z^0(\ast) \langle \textbf{\textrm{I}}\rangle ))$ of graphs having their degree-sequences to be in $\textbf{\textrm{L}}(Z^0(\ast) \langle \textbf{\textrm{I}}\rangle )$. So we have a \emph{graph-set homomorphism}
\begin{equation}\label{eqa:555555}
G_{raph}(\textbf{\textrm{L}}(Z^0(\ast) \langle \textbf{\textrm{I}}\rangle ))\rightarrow G_{raph}(\textbf{\textrm{L}}(Z^0(\ast) \langle \textbf{\textrm{I}}\,'\rangle ))
\end{equation} where $G_{raph}(\textbf{\textrm{L}}(Z^0(\ast) \langle \textbf{\textrm{I}}\,'\rangle ))$ is the set of graphs having their degree-sequences to be in $\textbf{\textrm{L}}(Z^0(\ast) \langle \textbf{\textrm{I}}\,'\rangle )$ if $\textbf{\textrm{L}}(Z^0(\ast) \langle \textbf{\textrm{I}}\rangle )\rightarrow \textbf{\textrm{L}}(Z^0(\ast) \langle \textbf{\textrm{I}}\,'\rangle )$ holds true.\paralled
\end{rem}

\subsection{Connections between graphic lattices}

\subsubsection{Systems of graphic equations}

\textbf{1. Coefficient matrices are popular integer matrices}

Let
\begin{equation}\label{eqa:number-coefficient-matrix}
\centering
{
\begin{split}
A_{coe}= \left(
\begin{array}{ccccc}
a_{1,1} & a_{1,2} & \cdots & a_{1,n}\\
a_{2,1} & a_{2,2} & \cdots & a_{2,n}\\
\cdots & \cdots & \cdots & \cdots \\
a_{m,1} & a_{m,2} & \cdots & a_{m,n}\\
\end{array}
\right)_{m\times n}
\end{split}}
\end{equation}
be a coefficient matrix with non-negative integer elements $a_{i,j}$ with $i\in [1,m]$ and $j\in [1,n]$, and two \emph{graph vectors} $Y(H)=(H_1$, $H_2$, $\dots $, $H_m)^{T}$ and $X(G)=(G_1$, $G_2$, $\dots $, $G_n)^{T}$, where $G_i$ and $H_j$ are (resp. colored) graphs with $i\in [1,n]$ and $j\in [1,m]$. We have a \emph{system of graphic equations}
\begin{equation}\label{eqa:number-coefficient-graph-equation}
\centering
Y(H)=\bigcup\langle A_{coe},X(G)\rangle
\end{equation}
with $H_i=\bigcup^n_{j=1}a_{i,j}G_j$ for $i\in [1,m]$, where ``$\bigcup$'' is the union operation on graphs of graph theory, and $a_{i,j}G_j$ is a disconnected graph having $a_{i,j}$ components such that each component is isomorphic to $G_i$, that is

\begin{equation}\label{eqa:number-graph-equations-total}
\centering
{
\begin{split}
Y(H)=\left(\begin{array}{c}
H_1\\
H_2\\
\cdots\\
H_m
\end{array} \right)= \left(
\begin{array}{ccccc}
a_{1,1} & a_{1,2} & \cdots & a_{1,n}\\
a_{2,1} & a_{2,2} & \cdots & a_{2,n}\\
\cdots & \cdots & \cdots & \cdots \\
a_{m,1} & a_{m,2} & \cdots & a_{m,n}\\
\end{array}
\right)\bigcup \left(\begin{array}{c}
G_1\\
G_2\\
\cdots\\
G_n
\end{array} \right)=\bigcup\langle A_{coe},X(G)\rangle
\end{split}}
\end{equation}

\begin{rem}\label{rem:coefficient-matrices}
Substituting by a graphic operation ``$(\bullet)$'' the operation ``$\bigcup $'' in (\ref{eqa:number-coefficient-graph-equation}), so we have an equation $Y(H)=(\bullet)\langle A_{coe},X(G)\rangle$ with each graph $H_j$ of $Y(H)$ is constructed by $H_j=(\bullet)^n_{k=1}b_{j,k}G_k$. It is natural to ask for a problem: If the determinant $|A_{coe}|\neq 0$, then we have $X(G)=(\bullet)^{-1}\langle A^{-1}_{coe},Y(H)\rangle$ for the inverse operation ``$(\bullet)^{-1}$'' of the operation ``$(\bullet)$''?\paralled
\end{rem}

\textbf{2. Coefficient matrices with elements to be graphs}

A \emph{graphic matrix} is one with every element to be a graph or a (resp. colored) graph. A \emph{graph coefficient matrix} $G_{\textrm{mat}}$ is defined as
\begin{equation}\label{eqa:graph-coefficient-matrix}
\centering
{
\begin{split}
G_{\textrm{mat}}= \left(
\begin{array}{ccccc}
G_{1,1} & G_{1,2} & \cdots & G_{1,n}\\
G_{2,1} & G_{2,2} & \cdots & G_{2,n}\\
\cdots & \cdots & \cdots & \cdots \\
G_{p,1} & G_{m,2} & \cdots & G_{m,n}\\
\end{array}
\right)_{m\times n}
\end{split}}
\end{equation}
with each element $G_{i,j}$ is a (resp. colored) graph. There are two \emph{graph vectors} $Y(H)=(H_1$, $H_2$, $\dots $, $H_m)^{T}$ and $X(G)=(G_1$, $G_2$, $\dots $, $G_n)^{T}$, where $G_i$ and $H_j$ are (resp. colored) graphs with $i\in [1,n]$ and $j\in [1,m]$. Similarly with (\ref{eqa:number-graph-equations-total}), we have a \emph{system of graphic equations}
\begin{equation}\label{eqa:graph-matrix-equation}
\centering
Y(H)=\bigcup (\ast)\langle G_{\textrm{mat}}, X(G)\rangle
\end{equation}
with $H_i=\bigcup^n_{j=1}[G_{i,j}(\ast)G_j]$ for $i\in [1,m]$, where ``$(\ast)$'' is an operation on (resp. colored) graphs in topological coding (Ref. \cite{Bang-Jensen-Gutin-digraphs-2007, Bondy-2008}). Moreover, we have $Y(H)=(\bullet) (\ast)\langle G_{\textrm{mat}}, X(G)\rangle$ with $H_i=(\bullet)^n_{j=1}[G_{i,j}(\ast)G_j]$ for $i\in [1,m]$.

\subsubsection{Connecting graphic lattices}

Before building connection between graphic lattices, we need to do some preparation works.
\begin{asparaenum}[\textrm{Preparation}-1. ]
\item Suppose that a graph coefficient matrix $G_{\textrm{mat}}$ be a matrix of rank $n\times n$, and let $T$ be a (resp. colored) graph. We define a \emph{mixed identity matrix} $I(T)$ of rank $n\times n$ by
\begin{equation}\label{eqa:identity-graph-coefficient-matrix}
G_{\textrm{mat}}(\ast)G^{-1}_{\textrm{mat}}=I(T)=(T_1,T_2,\cdots ,T_n)^T,
\end{equation} under an operation ``$(\ast)$'', in which each \emph{mixed vector} $T_k=(a_{k,1}, a_{k,2}, \cdots ,a_{k,n})$ with $k\in [1,n]$ holds $a_{k,k}=T$ and $a_{k,j}=0$ if $j\neq k$, especially, as $a_{k,k}=1$, we get an identity matrix $I=(c_{ij})_{n\times n}$ with $c_{ii}=1$ and $c_{ij}=0$ if $i\neq j$. We call $G^{-1}_{\textrm{mat}}$ the \emph{inverse matrix} of the graph coefficient matrix $G_{\textrm{mat}}$.

\item For two degree sequences $\textbf{\textrm{d}}_a=(a_1,a_2,\dots ,a_n)$ and $\textbf{\textrm{d}}_b=(b_1,b_2,\dots ,b_n)$, we regard them as vectors, and define the dot-product on $\textbf{\textrm{d}}_a$ and $\textbf{\textrm{d}}_b$
by $\textbf{\textrm{d}}_a \cdot \textbf{\textrm{d}}_b=\sum^n_{k=1}a_kb_k$ like the dot product of vectors of algebra, the degree-sequence product is a degree sequence $\textbf{\textrm{d}}=\textbf{\textrm{d}}_a \times \textbf{\textrm{d}}_b$ like the vector product in algebra.

\item By Definition \ref{defn:imaginary-graph}, a complex graph $L=H\overline{\ominus} iG$ with degree sequence $\textbf{\textrm{d}}(L)=(c_1,c_2,\dots ,c_n)$ having some integers $c_i<0$, so we can get $\textbf{\textrm{d}}(L)\cdot \textbf{\textrm{d}}(L\,')=1$ for another complex graph $L\,'=H\,'\overline{\ominus} iG\,'$, and a new complex graph $L\,''$ having its degree sequence $\textbf{\textrm{d}}(L\,'')=\textbf{\textrm{d}}(L)\times \textbf{\textrm{d}}(L\,')$ is denoted as $L\,''=L\times L\,'$ directly.
\end{asparaenum}

\vskip 0.4cm

Let $\textbf{\textrm{L}}((\ast) Y(\textbf{\textrm{H}}))=\{(\ast)^n_{k=1}\beta_kH_k:~ \beta_k\in Z^0, H_{k}\in Y(\textbf{\textrm{H}})\}$ be a \emph{graphic lattice} made by a graph operation ``$(\ast)$'' and a \emph{graph base} $Y(\textbf{\textrm{H}})=(H_1,H_2,\dots ,H_n)^T$, and another graphic lattice $\textbf{\textrm{L}}((\ast) X(\textbf{\textrm{G}}))=\{(\ast)^n_{k=1}\alpha_kG_k:~ \alpha_k\in Z^0, G_{k}\in X(\textbf{\textrm{G}})\}$ is defined on the graph operation ``$(\ast)$'' and another graphic lattice base $X(\textbf{\textrm{G}})=(G_1,G_2,\dots ,G_n)^T$.

Suppose that there exists a non-negative integer matrix $A_{coe}=(a_{ij})_{n\times n}$, such that $Y(\textbf{\textrm{H}})=\bigcup \langle A_{coe}, X(\textbf{\textrm{G}})\rangle$ with $H_k=\bigcup^n_{j=1}a_{k,j}G_j$ for $k\in [1,n]$. We get a connection between two graphic lattices $\textbf{\textrm{L}}((\ast) Y(\textbf{\textrm{H}}))$ and $\textbf{\textrm{L}}((\ast) Y(\textbf{\textrm{G}}))$ as follows

\begin{equation}\label{eqa:555555}
(\ast)^n_{k=1}\beta_kH_k=(\ast)^n_{k=1}\beta_k\bigcup\langle A_k,X(\textbf{\textrm{G}})\rangle=(\ast)^n_{k=1}\bigcup\langle \beta_kA_k,X(\textbf{\textrm{G}})\rangle
\end{equation}
since
$$\beta_k\left (\bigcup^n_{j=1}a_{k,j}G_j\right )=\beta_ka_{k,1}G_1\cup \beta_ka_{k,2}G_2\cup \cdots \cup \beta_ka_{k,n}G_n=\beta_k\cup\langle A_k,X(\textbf{\textrm{G}})\rangle,
$$ and $A_{coe}=(A_{1},A_{2},\cdots ,A_{n})$ with $A_k=(a_{k,1},a_{k,2},\cdots ,a_{k,n})$ for $k\in [1,n]$.

\begin{thm} \label{thm:99-connection-graphic-lattices}
$^*$ For a given graphic lattice $\textbf{\textrm{L}}((\ast) Y(\textbf{\textrm{G}}))$, there are infinite graphic lattices $\textbf{\textrm{L}}((\ast) Y(\textbf{\textrm{H}}))$ such that the graphic lattice base $Y(\textbf{\textrm{H}})$ of each $\textbf{\textrm{L}}((\ast) Y(\textbf{\textrm{H}})$ holds a system of graphic equations $Y(H)=(\bullet)\langle A_{coe},X(G)\rangle$ based on the graphic lattice base $Y(\textbf{\textrm{G}})$ of $\textbf{\textrm{L}}((\ast) Y(\textbf{\textrm{G}}))$, and two graph operations ``$(\bullet)$'' and ``$(\ast)$'', and a coefficient matrix $A_{coe}$ with elements to be non-negative integers.
\end{thm}

\begin{rem}\label{rem:ABC-conjecture}
If we consider $\textbf{\textrm{L}}((\ast) Y(\textbf{\textrm{G}}))$ as a \emph{public-key set}, then there are infinite \emph{private-key sets} $\textbf{\textrm{L}}((\ast) Y(\textbf{\textrm{H}}))$ by Theorem \ref{thm:99-connection-graphic-lattices} and various graph operations.\paralled
\end{rem}

\subsection{Graphic lattice made by the cycle-type of operations}

\subsubsection{Cycle-joining operation}
\begin{figure}[h]
\centering
\includegraphics[width=16cm]{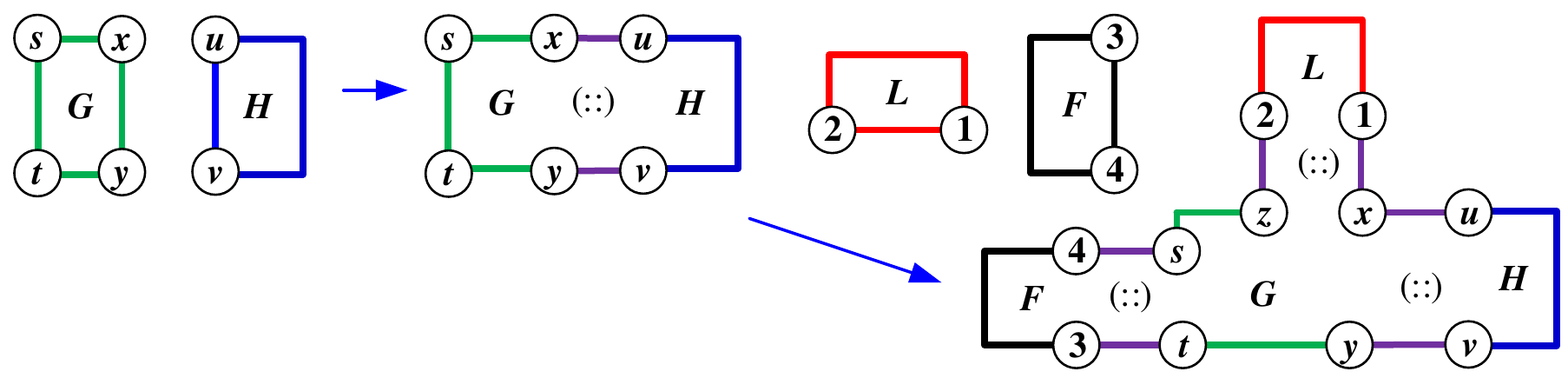}\\
\caption{\label{fig:cycle-join-hamilton}{\small A scheme for illustrating the cycle-joining operation.}}
\end{figure}

In \cite{Yao-Wang-Ma-Wang-Degree-sequences-2021}, the \emph{cycle-joining operation} ``$(\overline{\circ \circ})$'' on hamiltonian graphs is defined as: Let $xy$ be an edge of a Hamilton cycle of a Hamilton graph $G$, $uv$ be an edge of a Hamilton cycle of a Hamilton graph $H$, and $12$ be an edge of a Hamilton cycle of a Hamilton graph $L$ show in Fig.\ref{fig:cycle-join-hamilton}. Remove the edge $xy$ from $G$ and remove the edge $uv$ from $H$, and join the vertex $x$ with the vertex $u$ by a new edge $xu$, next join the vertex $y$ with the vertex $v$ by a new edge $yv$, the resultant graph is denoted as $G(\overline{\circ \circ})H$, so there is a Hamilton cycle in the \emph{Hamilton cycle-joining graph} $G(\overline{\circ \circ})H$ (see Fig.\ref{fig:cycle-join-hamilton}). Moreover, we can do the cycle-joining operation on three Hamilton graphs $L$, $F$ and $G(\overline{\circ \circ})H$, and get a Hamilton cycle-joining graph $\{[G(\overline{\circ \circ})H](\overline{\circ \circ})L\}(\overline{\circ \circ})F$.

Suppose that $G_k$ is a hamiltonian graph satisfied a hamiltonian condition $k$ with $k\in [1, A]$, the \emph{hamiltonian graph base} $\textbf{\textrm{H}}=(G_k)^A_{k=1}$ is linear independent from each other, that is, each hamiltonian graph $G_k$ is not the result of some $G_{i_1},G_{i_2},\dots ,G_{i_j}$ under the cycle-joining operation ``$(\overline{\circ \circ})$''. Each graph of the following set
\begin{equation}\label{eqa:555555}
\textbf{\textrm{L}}(Z^0(\overline{\circ \circ})\textbf{\textrm{H}})=\{(\overline{\circ \circ})^A_{k=1}a_kG_k:a_k\in Z^0,G_k\in \textbf{\textrm{H}}\}
\end{equation} is a hamiltonian graph, so we call $\textbf{\textrm{L}}(Z^0(\overline{\circ \circ})\textbf{\textrm{H}})$ a \emph{hamiltonian graph lattice}. Obviously, each graph $H\in \textbf{\textrm{L}}(Z^0(\overline{\circ \circ})\textbf{\textrm{H}})$ does not obey any one of these $A$ hamiltonian conditions when $\sum ^A_{k=1}a_k\geq 2$. Thereby, we have at least $Z=2^A-1=\sum ^A_{k=1}{A \choose k}$ different hamiltonian graphs $H_i\in \textbf{\textrm{L}}(Z^0(\overline{\circ \circ})\textbf{\textrm{H}})$ with $i\in [1,Z]$, such that $(H_k)^{Z}_{k=1}$ is a new hamiltonian graph base.

\subsubsection{Cycle-coinciding operation}
The \emph{cycle-coinciding operation} ``$(\circ \odot \circ)$'' on hamiltonian graphs is defined in the following way \cite{Yao-Wang-Ma-Wang-Degree-sequences-2021}: Let $xy$ be an edge of a Hamilton cycle of a Hamilton graph $G$, $uv$ be an edge of a Hamilton cycle of a Hamilton graph $H$ show in Fig.\ref{fig:cycle-join-hamilton}. Remove the edge $xy$ from $G$ and remove the edge $uv$ from $H$, respectively, and coincide the vertex $x$ with the vertex $u$ into a new vertex $x\odot u$, next coincide the vertex $y$ with the vertex $v$ into a new vertex $y\odot v$, the resultant graph is denoted as $G(\circ \odot \circ)H$, which has a Hamilton cycle, immediately, we get another \emph{hamiltonian graph lattice} below
\begin{equation}\label{eqa:555555}
\textbf{\textrm{L}}(Z^0(\circ \odot \circ)\textbf{\textrm{H}})=\{(\circ \odot \circ)^A_{k=1}a_kG_k:a_k\in Z^0,G_k\in \textbf{\textrm{H}}\}.
\end{equation}

\begin{rem}\label{rem:ABC-conjecture}
Let $C$ be a Hamilton cycle of a graph $T$ in a hamiltonian graph lattice $\textbf{\textrm{L}}(Z^0(\overline{\circ \circ})\textbf{\textrm{H}})$ (resp. $\textbf{\textrm{L}}(Z^0(\circ \odot \circ)\textbf{\textrm{H}})$). We add some edges $x_iy_i$ to join some pairs of non-adjacent vertices $x_i$ and $y_i$ of $C$ for $i\in [1,s]$, the resultant graph $T+E_{s}$ has its own Hamilton cycle $C$, so that $G_k\in T+E_{s}$ if $a_k\neq 0$.\paralled
\end{rem}

\subsection{Planar graphic lattices}

\begin{figure}[h]
\centering
\includegraphics[width=16.2cm]{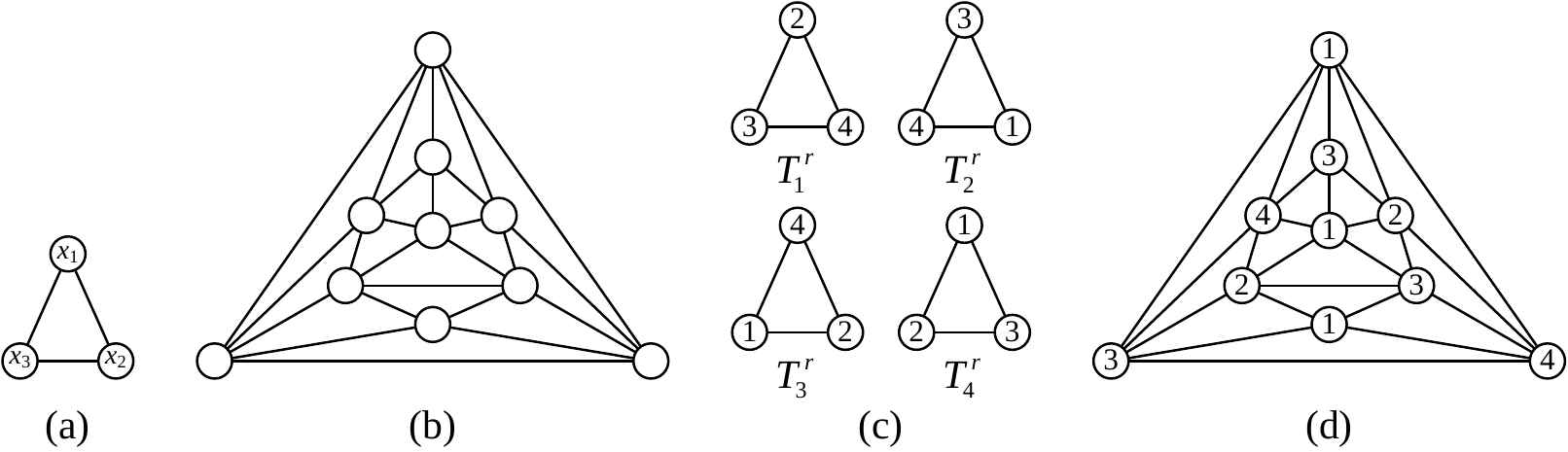}
\caption{\label{fig:4-color-planar-tile-lattice}{\small (a) A triangle; (b) a maximal planar graph $H$; (c) an every-zero graphic group $\{F_{\textrm{inner}\bigtriangleup};\oplus\}$ cited from \cite{YAO-SUN-WANG-SU-XU2018arXiv}; (d) a maximal planar graph $H$ tiled by the every-zero graphic group $\{F_{planar};\oplus\}$ shown in (c).}}
\end{figure}

The authors in \cite{YAO-SUN-WANG-SU-XU2018arXiv} guessed: A maximal planar graph is 4-colorable if and only it can be tiled by the every-zero graphic group $\{F_{\textrm{inner}\bigtriangleup};\oplus\}$ shown in Fig.\ref{fig:4-color-planar-tile-lattice}(c).

\begin{defn}\label{defn:99-4-color-planar-graphic-lattice}
\cite{YAO-SUN-WANG-SU-XU2018arXiv} Let $F_{\textrm{inner}\bigtriangleup}$ be the set of planer graphs having triangular inner faces. If a planar graph $H$ with each inner face to be a triangle can be titled by the every-zero graphic group $\{F_{\textrm{inner}\bigtriangleup};\oplus\}$, write the resultant graph as $H\bigtriangleup ^4_{k=1}a_kT^r_k$ with $a_k\in Z^0$ and $\sum a_k\geq 1$. The following set
\begin{equation}\label{eqa:planar-lattice}
{
\begin{split}
\textbf{\textrm{L}}(\textbf{\textrm{T}}^r\bigtriangleup F_{\textrm{inner}\bigtriangleup})=\{H\bigtriangleup ^4_{k=1}a_kT^r_k:a_k\in Z^0,H\in F_{\textrm{inner}\bigtriangleup}\}
\end{split}}
\end{equation} with $\sum^4_{k=1} a_k\geq 1$ is called a \emph{planar graphic lattice}, where the \emph{planar graphic lattice base} is $\textbf{\textrm{T}}^r=(T^r_1,T^r_2,T^r_3,T^r_4)$. \qqed
\end{defn}

Thereby, each planar graph $G\in \textbf{\textrm{L}}(\textbf{\textrm{T}}^r\bigtriangleup F_{\textrm{inner}\bigtriangleup})$ is a $4$-colorable planar graph with each inner face to be a triangle.

\subsection{Graphic lattices subject to graph operations}

\subsubsection{Graph lattices based on the vertex-coinciding operation}

\begin{defn}\label{defn:99-colored-graphic-lattice}
\cite{Bing-Yao-2020arXiv} Let $\textbf{\textrm{H}}=(H_1,H_2,\dots, H_n)$ be a group of $n$ \emph{linearly independent graphic vectors} (also, a \emph{graphic base}) under a graph operation ``$(\bullet)$'', where each $H_i$ is a colored/uncolored graph, and $F_{p,q}$ is a set of colored/uncolored graphs of $\lambda$ vertices and $\mu$ edges with respect to $\lambda \leq p$, $\mu \leq q$ and $2n-2\leq p$. The result graph obtained by doing a graph operation $(\bullet)$ on $G$ and the base $\textbf{\textrm{H}}$ with $a_i\in Z^0$ is denoted as $\textbf{\textrm{H}}(\bullet)G=G(\bullet)^n_{i=1}a_iH_i$. In general, we call the following graph set
\begin{equation}\label{eqa:graphic-lattice-graph-operation}
\textbf{\textrm{L}}(\textbf{\textrm{H}}(\bullet)F_{p,q})=\{G(\bullet)^n_{i=1}a_iH_i:~a_i\in Z^0,~G\in F_{p,q}\}
\end{equation}
with $\sum^n_{i=1} a_i\geq 1$ a \emph{graphic lattice} (or \emph{colored graphic lattice}), $\textbf{\textrm{H}}$ a \emph{graphic lattice base}, $p$ is the \emph{dimension} and $n$ is the \emph{rank} of $\textbf{\textrm{L}}(\textbf{\textrm{H}}(\bullet)F_{p,q})$.\qqed
\end{defn}

\begin{rem}\label{rem:xxxxxxxxxx}
Moreover, $\textbf{\textrm{L}}(\textbf{\textrm{H}}(\bullet)F_{p,q})$ is called a \emph{linear graphic lattice} if every $G\in F_{p,q}$, each base $H_i$ of $\textbf{\textrm{H}}$ and $G(\bullet)^n_{i=1}a_iH_i$ are forests or trees. An uncolored tree-graph lattice, or a colored tree-graph lattice is \emph{full-rank} $p=n$ in the equation (\ref{eqa:graphic-lattice-graph-operation}). Especially, if each $H\in \textbf{\textrm{H}}$ is a (colored) \emph{Hanzi-graph}, we call $\textbf{\textrm{L}}(\textbf{\textrm{H}}(\bullet)F_{p,q})$ a (colored) \emph{Hanzi-lattice}.\paralled
\end{rem}

\begin{defn}\label{defn:99-colored-directed-graphic-lattice}
\cite{Bing-Yao-2020arXiv} Let $\overrightarrow{F}_{p,q}$ be a set of directed graphs of $p$ vertices and $q$ arcs with $n\leq p$, and let $\overrightarrow{\textbf{\textrm{H}}}=(\overrightarrow{H}_1,\overrightarrow{H}_2,\dots, \overrightarrow{H}_n)$ be a group of $n$ linearly independent directed-graphic vectors, where each $\overrightarrow{H}_i$ is a directed graph. By an operation ``$(\bullet)$'' on directed graphs, we have a \emph{directed-graphic lattice} (or \emph{colored directed-graphic lattice}) as follows
\begin{equation}\label{eqa:directed-graphic-lattice-graph-operation}
\overrightarrow{\textbf{\textrm{L}}}(\overrightarrow{\textbf{\textrm{H}}}(\bullet)\overrightarrow{F}_{p,q})=\left \{\overrightarrow{G}(\bullet)^n_{i=1}a_i\overrightarrow{H}_i:~a_i\in Z^0,~\overrightarrow{G}\in \overrightarrow{F}_{p,q}\right \}
\end{equation}
with $\sum^n_{i=1} a_i\geq 1$.\qqed
\end{defn}

\begin{defn}\label{defn:99-Topcode-matrix-lattice}
\cite{Bing-Yao-2020arXiv} For a given Topcode-matrix $T_{code}$ if a graph $G$ admits a $W$-type coloring $f$, such that each $x_i=f(x)$ for some vertex $x\in V(G)$ and each $y_j=f(y)$ for some vertex $y\in V(G)$ and every $e_i=f(vu)$ for some edge $uv\in E(G)$, then we say $T_{code}$ corresponds the graph $G$, conversely, $G$ has its own Topcode-matrix, denoted as $T_{code}(G)$. Let $T^i_{code}=(X_i,~E_i,~Y_i)^{T}_{3\times q_i}$, where $X_i=(x^i_1, x^i_2, \cdots, x^i_{q_i})$, $E_i=(e^i_1, e^i_2, \cdots, e^i_{q_i})$ and $Y_i=(y^i_1, y^i_2, \cdots, y^i_{q_i})$ with $i=1,2$. The union operation ``$\uplus$'' of two Topcode-matrices $T^1_{code}$ and $T^2_{code}$ is defined by
\begin{equation}\label{eqa:xxx}
T^1_{code}\uplus T^2_{code}=(X_1\cup X_2,~E_1\cup E_2,~Y_1\cup Y_2)^{T}_{3\times (q_1+q_2)}.
\end{equation}
with $X_1\cup X_2=(x^1_1, x^1_2, \cdots, x^1_{q_1},x^2_1, x^2_2, \cdots, x^2_{q_2})$, $Y_1\cup T_2=(y^1_1, y^1_2, \cdots, y^1_{q_1},y^2_1, y^2_2, \cdots, y^2_{q_2})$ and $E_1\cup E_2=(e^1_1, e^1_2, \cdots, e^1_{q_1},e^2_1, e^2_2, \cdots, e^2_{q_2})$. Moreover, each Topcode-matrix $T^i_{code}$ corresponds a graph $G_i$, and each of $X_1\cap X_2\neq \emptyset $ and $Y_1\cap Y_2\neq \emptyset $ holds true, thus the Topcode-matrix of the vertex-coincided graph $G_1\odot G_2$ is just $T_{code}(G_1\odot G_2)=T^1_{code}\uplus T^2_{code}$. So, we vertex-split the vertex-coincided graph $H\odot |^n_{i=1}a_iT_i$ with $a_i\in \{0,1\}$ into vertex-disjoint graphs $T_1,T_2,\dots ,T_n,H$. In the expression of Topcode-matrices, we have the Topcode-matrix of a vertex-coincided graph $H\odot |^n_{i=1}a_iT_i$ as follows
\begin{equation}\label{eqa:xxx}
{
\begin{split}
T_{code}(H\odot |^n_{i=1}T_i)=T_{code}(H)\uplus |^n_{i=1}T^i_{code}=T_{code}(H)\uplus T^1_{code}\uplus T^2_{code}\uplus \cdots \uplus T^n_{code},
\end{split}}
\end{equation}
where $T^i_{code}=(X_i,~E_i,~Y_i)^{T}_{3\times q_i}=T_{code}(T_i)$ with $i\in [1,n]$, such that $T_{code}(H)=(X_H,~E_H,~Y_H)^{T}_{3\times q_H}$ and one of $X_H\cap X_i\neq \emptyset $ and $Y_H\cap Y_i\neq \emptyset $ for $i\in [1,n]$ holds true, and the size of $T_{code}(H\odot |^n_{i=1}T_i)$ is equal to $q_H+\sum^n_{i=1} q_i$. We call the following set
\begin{equation}\label{eqa:Topcode-matrices-lattice}
{
\begin{split}
\textbf{\textrm{L}}(\textbf{\textrm{T}}_{\textrm{code}}\uplus F_{p,q})=\{T_{code}(H)\uplus |^n_{i=1}a_iT^i_{code}:~a_i\in Z^0,~H\in F_{p,q}\}
\end{split}}
\end{equation}
a \emph{Topcode-matrix lattice} with $\sum^n_{i=1}a_i\geq 1$, where $\textbf{\textrm{T}}_{\textrm{code}}=(T^1_{code}$, $T^2_{code}$, $\dots $, $T^n_{code})$ is a group of \emph{linearly independent Topcode-matrix vectors} under the vertex-coinciding operation.\qqed
\end{defn}

\begin{defn}\label{defn:99-graphic-lattice-dot-coninciding}
\cite{Bing-Yao-2020arXiv} Let $\textbf{\textrm{T}}=(T_1,T_2,\dots, T_n)$ be a group of $n$ \emph{linearly independent graphic vectors} under the vertex-coinciding operation, also, a \emph{graphic base}, and let $H\in F_{p,q}$ be a connected graph of vertices $u_{1},u_{2},\dots, u_{m}$, and $F_{p,q}$ is a set of colored/uncolored graphs of $\lambda$ vertices and $\mu$ edges with respect to $\lambda \leq p$, $\mu \leq q$ and $2n-2\leq p$. We write the result graph obtained by vertex-coinciding a vertex $v_i$ of some base $T_i$ with some vertex $u_{i_j}$ of the connected graph $H$ into one vertex $w_i=u_{i_j}\odot v_i$ as $H\odot \textbf{\textrm{T}}=H\odot |^n_{i=1}a_iT_i$ with $a_i\in Z^0$ and $\sum^n_{i=1} a_i\geq 1$. Since there are two or more vertices of the graphic lattice base $T_i$ that can be vertex-coincided with some vertex of the connected graph $H$, so $H\odot \textbf{\textrm{T}}$ is not unique in general, in other word, these graphs $H\odot \textbf{\textrm{T}}$ forms a set. We call the following set
\begin{equation}\label{eqa:graphic-lattice-00}
\textbf{\textrm{L}}(\textbf{\textrm{T}}\odot F_{p,q})=\{H\odot |^n_{i=1}a_iT_i:~a_i\in Z^0,~H\in F_{p,q}\}
\end{equation} with $\sum^n_{i=1} a_i\geq 1$ a \emph{graphic lattice}, and $p$ is the \emph{dimension}, and $n$ is the \emph{rank} of the graphic lattice. Moreover $\textbf{\textrm{L}}(\textbf{\textrm{T}}\odot F_{p,q})$ is called a \emph{linear graphic lattice} if every $H\in F_{p,q}$, each base $T_i$ of the lattice base $\textbf{\textrm{T}}$ and $H\odot \textbf{\textrm{T}}$ are forests or trees.\qqed
\end{defn}

\begin{defn} \label{defn:99-colored-graphic-lattice}
\cite{Bing-Yao-2020arXiv} \textbf{Colored graphic lattices.} Let $F\,^c_{p,q}$ be a set of colored graphs of $\lambda $ vertices and $\mu $ edges with respect to $\lambda \leq p$, $\mu \leq q$ and $2n-2\leq p$, where each graph $H\,^c\in F\,^c_{p,q}$ is colored by a $W$-type coloring $f$, and let $\textbf{\textrm{T}}\,^c=(T\,^c_1,T\,^c_2,\dots, T\,^c_n)$ with $n\leq p$ be a \emph{linearly independent colored graphic base} under the vertex-coinciding operation, we have two particular cases: (i) each graph $T\,^c_i$ of $\textbf{\textrm{T}}\,^c$ admits a $W_i$-type coloring $g_i$; (ii) the union graph $\bigcup ^n_{i=1}T\,^c_i$ admits a flawed $W_i$-type coloring. Vertex-coinciding a vertex $x_i$ of some base $T\,^c_i$ with some vertex $y_{i_j}$ of the colored graph $H\,^c\in F\,^c_{p,q}$ into one vertex $z_i=y_{i_j}\odot x_i$ produces a vertex-coincided graph $H\,^c\odot |^n_{i=1}T\,^c_i$ admitting a coloring induced by $g_1, g_2, \dots ,g_n$ and $f$, here, two vertices $x_i,y_{i_j}$ are colored the same color $\gamma$, then the vertex $z_i=y_{i_j}\odot x_i$ is colored with the color $\gamma$ too. We call the following set
\begin{equation}\label{eqa:graphic-lattice-colored}
\textbf{\textrm{L}}(\textbf{\textrm{T}}\,^c\odot F\,^c_{p,q})=\left \{H\,^c\odot |^n_{i=1}a_iT\,^c_i:~a_i\in Z^0,~H\,^c\in F\,^c_{p,q}\right \}
\end{equation} with $\sum^n_{i=1} a_i\geq 1$ a \emph{colored graphic lattice}, where $p$ is the \emph{dimension}, and $n$ is the \emph{rank} of the colored graphic lattice. We call the colored graphic lattice $\textbf{\textrm{L}}(\textbf{\textrm{T}}\,^c\odot F\,^c_{p,q})$ a \emph{linear colored graphic lattice} if every colored graph $H\,^c\in F\,^c_{p,q}$, each graph $T\,^c_i$ of $\textbf{\textrm{T}}\,^c$ and the graph $H\,^c\odot |^n_{i=1}a_iT\,^c_i$ are colored forests, or colored trees. Clearly, each element of the lattice $\textbf{\textrm{L}}(\textbf{\textrm{T}}\,^c\odot F\,^c_{p,q})$, each graph $T\,^c_i$ and each colored graph $H\,^c\in F\,^c_{p,q}$ may admit the same $W$-type colorings.\qqed
\end{defn}

\subsubsection{Graph lattices based on the edge-coinciding operation}

\begin{defn} \label{defn:99-felicitous-difference-star-graphic-lattice}
\cite{Bing-Yao-2020arXiv} \textbf{Felicitous-difference star-graphic lattices}. Since a colored leaf-coinciding operation $F_{1,n}D_j\overline{\ominus} F_{1,n}D_k$ between two colored stars $F_{1,n}D_k$ and $F_{1,n}D_j$ produces a graph with diameter three, so each group of colored stars $K_{1,n_1}$, $K_{1,n_2}$, $\dots$, $K_{1,n_m}$ is linearly independent under the colored leaf-coinciding operation. By the colored leaf-coinciding operation and the felicitous-difference ice-flower systems $I_{ce}(F_{1,n}D_k)^{2n}_{k=1}$ and $I_{ce}(SF_{1,n}D_k)^{n}_{k=1}$, each graph contained in the following graphic lattice
\begin{equation}\label{eqa:stars-lattice}
\textrm{\textbf{L}}(\overline{\ominus} \textbf{I}_{ce}(FD)) =\left \{\overline{\ominus}^{2n}_{i=1}a_iF_{1,n}D_i: a_i\in Z^0, F_{1,n}D_i\in I_{ce}(F_{1,n}D_k)^{2n}_{k=1}\right \}
\end{equation} is $\Delta$-saturated, where $\sum ^{2n}_{i=1}a_i\geq 1$ and the base is $\textbf{I}_{ce}(FD)=I_{ce}(F_{1,n}D_k)^{2n}_{k=1}$. We call $\textrm{\textbf{L}}(\overline{\ominus} \textbf{I}_{ce}(FD))$ a \emph{felicitous-difference star-graphic lattice}. Similarly, by the smallest felicitous-difference ice-flower system $I_{ce}(SF_{1,n}D_k)^{n}_{k=1}$, we have another \emph{felicitous-difference star-graphic lattice} defined as follows:
\begin{equation}\label{eqa:smallest-stars-lattice}
\textrm{\textbf{L}}(\overline{\ominus} \textbf{I}_{ce}(SFD)) =\left \{\overline{\ominus}^n_{j=1}a_jSF_{1,n}D_j: a_j\in Z^0, SF_{1,n}D_j\in I_{ce}(SF_{1,n}D_k)^{n}_{k=1}\right \}
\end{equation} with $\sum^n_{j=1} a_j\geq 1$ and the base is $\textbf{I}_{ce}(SFD)=I_{ce}(SF_{1,n}D_k)^{n}_{k=1}$.\qqed
\end{defn}

\begin{defn}\label{defn:99-general-star-graphic-lattice}
\cite{Bing-Yao-2020arXiv} Suppose that $F^c_{star\Delta}$ is the set of colored stars $C^K_{1,k}$ ($\cong K_{1,k}$) with $k\leq \Delta$, in general. We define a \emph{general star-graphic lattice}
\begin{equation}\label{eqa:general-stars-lattice}
\textrm{\textbf{L}}(\overline{\ominus} \textbf{F}^c_{star\Delta}) =\left \{\overline{\ominus}^{\Delta}_{k=2}a_k C^K_{1,k}: a_k\in Z^0, C^K_{1,k}\in F^c_{star\Delta}\right \}
\end{equation} with $\sum^{\Delta}_{k=2} a_k\geq 1$, this graphic lattice contains all colored graphs with maximum degrees no more than $ \Delta$. \qqed
\end{defn}

Clearly, by Definition \ref{defn:99-felicitous-difference-star-graphic-lattice}, we have
$$
\textrm{\textbf{L}}(\overline{\ominus} \textbf{I}_{ce}(FD))\subset \textrm{\textbf{L}}(\overline{\ominus} \textbf{F}^c_{star\Delta}),\textrm{\textbf{L}}(\overline{\ominus} \textbf{I}_{ce}(SFD))\subset \textrm{\textbf{L}}(\overline{\ominus} \textbf{F}^c_{star\Delta}), \textrm{\textbf{L}}(\odot \textbf{I}_{ce}(FD))\subset \textrm{\textbf{L}}(\overline{\ominus} \textbf{F}^c_{star\Delta}).
$$

\begin{defn}\label{defn:99-4-color-star-graphic-lattice}
\cite{Bing-Yao-2020arXiv} \textbf{A 4-color ice-flower system.} Each star $K_{1,d}$ with $d\in [2,M]$ admits a proper vertex-coloring $f_i$ with $i\in [1,4]$ defined as $f_i(x_0)=i$, $f_i(x_j)\in [1,4]\setminus\{i\}$, and $f_i(x_s)\neq f_i(x_t)$ for some $s,t\in [1,d]$, where $V(K_{1,d})=\{x_0,x_j:j\in [1,d]\}$. For each pair of $d$ and $i$, $K_{1,d}$ admits $n(d,i)$ proper vertex-colorings like $f_i$ defined above. Such colored star $K_{1,d}$ is denoted as $P_dS_{i,k}$, we have a set $(P_{d}S_{i,k})^{n(a,i)}_{k=1}$ with $i\in [1,4]$ and $d\in [2,M]$, and moreover we obtain a \emph{4-color ice-flower system} $\textbf{\textrm{I}}_{ce}(PS,M)=I_{ce}(P_{d}S_{i,k})^{n(a,i)}_{k=1})^{4}_{i=1})^{M}_{d=2}$, which induces a \emph{$4$-color star-graphic lattice}
\begin{equation}\label{eqa:4-color-star-system-lattices}
\textbf{\textrm{L}}(\Delta\overline{\ominus} \textbf{\textrm{I}}_{ce}(PS,M))=\left \{\Delta\overline{\ominus}^{A}_{(d,i,k)} a_{d,i,k}P_{d}S_{i,j}:~a_{d,i,k}\in Z^0,~P_{d}S_{i,j}\in \textbf{\textrm{I}}_{ce}(PS,M)\right \}
\end{equation} with $\sum ^{A}_{(d,i,k)} a_{d,i,k}\geq 3$, and the base is $\textbf{\textrm{I}}_{ce}(PS,M)=I_{ce}(P_{d}S_{i,k})^{n(a,i)}_{k=1})^{4}_{i=1})^{M}_{d=2}$, where $A=|\textbf{\textrm{I}}_{ce}(PS$, $M)|$, and the operation ``$\Delta\overline{\ominus}$'' is doing a series of leaf-coinciding operations to colored stars $a_{d,i,k}P_{d}S_{i,j}$ such that the resultant graph to be a planar with each inner face being a triangle.\qqed
\end{defn}

\begin{defn} \label{defn:99-spanning-star-graphic-lattice}
\cite{Bing-Yao-2020arXiv} \textbf{Spanning star-graphic lattices.} Aa known, a connected graph $T$ is a tree if and only if $n_1(T)=2+\Sigma_{d\geq 3}(d-2)n_d(T)$ (Ref. \cite{Yao-Zhang-Yao-2007, Yao-Zhang-Wang-Sinica-2010, Yao-Chen-Yao-Jiajuan-Zhang-Jingxia-Guo-2010}). An ice-flower system $\textbf{\textrm{K}}^c$ defined as: Each $K^c_{1,m_j}$ of $\textbf{\textrm{K}}^c$ admits a proper vertex coloring $g_j$ such that $g_j(x)\neq g_j(y)$ for any pair of vertices $x,y$ of $K^c_{1,m_j}$, and each leaf-coinciding graph $T=\overline{\ominus}|^n_{j=1}K^c_{1,m_j}$ is connected based on the leaf-coinciding operation ``$K^c_{1,m_i}\overline{\ominus} K^c_{1,m_j}$'', such that

(1) $n_1(T)=2+\Sigma_{d\geq 3}(d-2)n_d(T)$ holds true; and

(2) $T$ admits a proper vertex coloring $f=\overline{\ominus}|^n_{j=1}g_j$ with $f(u)\neq f(w)$ for any pair of vertices $u,w$ of $T$.

We get a set $\textbf{\textrm{L}}(\overline{\ominus} (m,g)\textbf{\textrm{K}}^c)$ containing the above leaf-coinciding graphs $T=\overline{\ominus}|^n_{j=1}K^c_{1,m_j}$ if $|V(T)|=m$. Since each graph $T\in\textbf{\textrm{L}}(\overline{\ominus} (m,g)\textbf{\textrm{K}}^c)$ is a tree, and Cayley's formula $\tau(K_m)=m^{m-2}$ in graph theory (Ref. \cite{Bondy-2008}) tells us the number of elements of $\textbf{\textrm{L}}(\overline{\ominus} (n)\textbf{\textrm{K}}^c)$ to be equal to $m^{m-2}$. We call this set
$$\textbf{\textrm{L}}(\overline{\ominus} (m,g)\textbf{\textrm{K}}^c)=\{\overline{\ominus}|^n_{j=1}K^c_{1,m_j},~K^c_{1,m_j}\in \textbf{\textrm{K}}^c\}$$ a \emph{spanning star-graphic lattice}.\qqed
\end{defn}

\begin{defn} \label{defn:99twin-odd-graceful-graphic-lattice}
\cite{Bing-Yao-2020arXiv} Let $F_{\textrm{odd}}(G)$ be the set of graphs $G_t$ with $G_t\cong G$ and admitting odd-graceful labelings, and let $\textrm{\textbf{H}}=(H_{i,j})^{M_{\textrm{odd}}}_{i,j}$ be the \emph{base}, where $M_{\textrm{odd}}$ is the number of twin odd-graceful matchings $(G_t,H_{i,j})$ for $G_t\in F_{\textrm{odd}}(G)$. Joining a vertex $x_{t,s}$ of $G_t$ for $s\in [1,|V(G_t)|]$ with a vertex $y^{k}_{i,j}$ of $H_{i,j}$ for $k\in [1,|V(H_{i,j})|]$ by an edge $x_{t,s}y^{k}_{i,j}$ produces a graph, denoted as $L=G_t\ominus |^{M_{\textrm{odd}}}_{k,j}a_{k,j}H_{k,j}$ with $\sum a_{k,j}=|V(G_t)|$, and the graph $L$ admits a labeling $h$ defined by $h(w)=f_t(w)$ for each element $w\in V(G_t)\cup E(G_t)$, $h(w)=g_{i,j}(w)$ for each element $w\in V(H_{i,j})\cup E(H_{i,j})$ and $h(x_{t,s}y^{k}_{i,j})=|f_t(x_{t,s})-g_{i,j}(y^{k}_{i,j})|$. We obtain a \emph{twin odd-graceful graphic lattice}
\begin{equation}\label{eqa:twin-odd-graceful-lattice}
\textrm{\textbf{L}}(\textrm{\textbf{H}}\ominus \textbf{F}_{\textrm{odd}}(G)) =\left \{G_t\ominus |^{M_{\textrm{odd}}}_{k,j}a_{k,j}H_{k,j}: a_k\in Z^0, G_t\in F_{\textrm{odd}}(G)\right \}
\end{equation} with $\sum a_{k,j}=|V(G_t)|$, where each matching $(G_t,H_{k,j})$ is called a \emph{twin odd-graceful matching}.\qqed
\end{defn}

\vskip 0.4cm

\noindent \textbf{A constructional condition for Hamilton graphs.} Let $C=x_1x_2\cdots x_px_1$ be a H-cycle of a connected $(p,q)$-graph $G$, where $|E(G)|\geq p+1$. Suppose that an edge $x_ix_j$ of $G$ for $j-i\geq 2$ is not in the H-cycle $C$, we edge-split this edge $x_ix_j$ into two edges $x\,'_ix\,'_j$ and $x\,''_ix\,''_j$, so the H-cycle $C$, also, is split into $C_1=x_1x_2\cdots x\,'_ix\,'_jx_{j+1}\cdots x_px_1$ and $C_2=x\,''_ix_{i+1}x_{i+2}\cdots x_{j-1}x\,''_jx\,''_i$. Let

$V_1=\{x_1,x_2,\cdots ,x\,'_i,x\,'_j,x_{j+1},\cdots ,x_p\}$ and $V_2=\{x\,''_i,x_{i+1},x_{i+2},\cdots ,x_{j-1},x\,''_j\}$,\\
and we get two graphs $G_r=[V_r]$ induced by vertex set $V_r$ for $r=1,2$. It is noticeable, each graph $G_r$ has a H-cycle $C_r$ for $r=1,2$. Let $V^*_1=V_1\setminus \{x\,'_i,x\,'_j\}$, $V^*_2=V_2\setminus \{x\,''_i,x\,''_j\}$, so

$${
\begin{split}
E(G_1)=&\{x\,'_ix\,'_j\}\cup \{uv\in E(G):u,v\in V^*_1\subset V(G)\}\cup \\
&\cup \{x_iu\in E(G),x_jv\in E(G):u,v\in V^*_1\subset V(G),x\,'_i=x_i, x\,'_j=x_j\}
\end{split}}
$$

$${
\begin{split}
E(G_2)=&\{x\,''_ix\,''_j\}\cup \{wz\in E(G):w,z\in V^*_2\subset V(G)\}\cup \\
&\cup \{x_iw\in E(G),x_jz\in E(G):w,z\in V^*_2\subset V(G),x\,''_i=x_i, x\,''_j=x_j\}
\end{split}}
$$

We get an edge set $E(G_1,G_2)=E\langle V^*_1; V^*_2\rangle$, where $E\langle V^*_1; V^*_2\rangle=\{x_ax_b\in E(G), x_aw\in E(G),x_bz\in E(G):w\in V^*_2, x_a\in V^*_1; z\in V^*_1,x_b\in V^*_2\}$. Thereby, $V(G)=V^*_1\cup V^*_2\cup\{x_i,x_j\}$, and $E(G)=[E(G_1)\setminus \{x\,'_ix\,'_j\}]\cup [E(G_2)\setminus \{x\,''_ix\,''_j\}]\cup \{x_ix_j\}\cup E(G_1,G_2)$.

The procedure of obtaining two graphs $G_1,G_2$ and an edge set $E(G_1,G_2)$ is denoted as $G\|x_ix_j$, conversely, the procedure of doing an edge-coinciding operation to two edges $x\,'_ix\,'_j$ and $x\,''_ix\,''_j$ of $G_1,G_2$ and add the edges of $E(G_1,G_2)$ is denoted as $G=G_1(x\,'_ix\,'_j)\overline{\odot} G_2(x\,''_ix\,''_j)+E(G_1,G_2)$, since $E(G_1,G_2)$ does not contain any one of four edges $x\,'_ix\,''_j, x\,''_ix\,'_j,x\,'_ix\,''_i$ and $x\,'_jx\,''_j$. Clearly, two Hamilton graphs $G_1$ and $G_2$ both are the proper subgraphs of $G$.

\begin{thm}\label{thm:hamiltonian-iff}
$^*$ A connected $(p,q)$-graph $G$ with $p\geq 4$ and $q\geq p+1$ is hamiltonian if and only if there exists an edge $x_ix_j$ of $G$, such that doing an edge-splitting operation to the edge $x_ix_j$ produces two Hamilton graphs $G_1,G_2$ and an edge set $E(G_1,G_2)$ between $G_1$ and $G_2$ in $G\|x_ix_j$.
\end{thm}

\begin{rem}\label{rem:hamiltonian-iff}
The result in Theorem \ref{thm:hamiltonian-iff} is constructional. Some one of $G_1$ and $G_2$ maybe a cycle, so we can use Theorem \ref{thm:hamiltonian-iff} to another one, until, each Hamilton graph is a cycle only, we stop to apply Theorem \ref{thm:hamiltonian-iff} to graphs.\paralled
\end{rem}

\vskip 0.4cm

\noindent \textbf{Algorithms for construction of Hamilton graphs.} Let $N_{ec}(1),N_{ec}(2),\dots ,N_{ec}(m)$ be the existing necessary conditions of Hamilton graphs, and let $S_{uf}(1),S_{uf}(2),\dots ,S_{uf}(n)$ be the existing sufficient conditions of Hamilton graphs. We get graph sets $G_{nec}(i)$ for $i\in [1,m]$ containing all graphs hold the existing necessary conditions of $N_{ec}(i)$ for $i\in [1,m]$ and graph sets $G_{suf}(j)$ for $j\in [1,n]$ containing all graphs hold the existing sufficient conditions of $S_{uf}(j)$ for $j\in [1,n]$. Clearly, $N_{ec}(i)\cap S_{uf}(j)=\emptyset$ and $G_{nec}(i)\cap G_{suf}(j)=\emptyset$, since no report say that a necessary and sufficient condition for Hamilton graphs has been found nowadays.

We define a graph operation ``$(\ast)$'' on graphs $H^{nec}_i\in G_{nec}(i)$ and $H^{suf}_j\in G_{suf}(j)$, the resultant graph $L$ is denoted as $L=(\ast)\langle H^{nec}_i |^{m}_{i=1}, H^{suf}_j |^{n}_{j=1}\rangle$, such that $L$ is a Hamilton graph.
\begin{equation}\label{eqa:555555}
F=\{(\ast)\langle a_iH^{nec}_i |^{m}_{i=1}, b_jH^{suf}_j|^{n}_{j=1}\rangle: a_i,b_j\in Z^0,H^{nec}_i\in G_{nec}(i),H^{suf}_j\in G_{suf}(j)\}
\end{equation} with $\sum^{m}_{i=1}a_i\geq 1$ and $\sum^{n}_{j=1}b_j\geq 1$.

It is noticeable, each Hamilton graph $T=(\ast)\langle a_iH^{nec}_i |^{m}_{i=1},b_jH^{suf}_j|^{n}_{j=1}\rangle\in F$ does not obey two conditions simultaneously: one is in $N_{ec}(i)$ for $i\in [1,m]$, and another one is in $S_{uf}(j)$ for $j\in [1,n]$, since $N_{ec}(i)\cap S_{uf}(j)=\emptyset$.

We define an operation ``$(\ast)=\square $'' on vertex-disjoint Hamilton graphs as: Let $u_{1,i}u_{1,i+1}$ be an edge of a H-cycle $C_1=u_{1,1}u_{1,2}\cdots u_{1,p_1}u_{1,1}$ of a Hamilton $(p_1,q_1)$-graph $G_1$, and let $u_{2,j}u_{2,j+1}$ be an edge of a H-cycle $C_2=u_{2,1}u_{2,2}\cdots u_{2,p_2}u_{2,1}$ of a Hamilton $(p_2,q_2)$-graph $G_2$. Next, we add two new edges $u_{1,i}u_{2,j}$ and $u_{1,i+1}u_{2,j+1}$ to join the vertex $u_{1,i}$ with the vertex $u_{2,j}$ together, and join the vertex $u_{1,i+1}$ with the vertex $u_{2,j+1}$ together, so the resultant graph is a Hamilton graph, denoted as $\square \langle G_1, G_2\rangle$, it contains a H-cycle
$${
\begin{split}
\square \langle C_1-u_{1,i}u_{1,i+1}, C_2-u_{2,j}u_{2,j+1}\rangle=&u_{1,1}u_{1,2}\cdots u_{1,i}u_{2,j}u_{2,j-1} \cdots u_{2,1} u_{2,p_2}\\
& u_{2,p_2-1}u_{2,j+1}u_{1,i+1}u_{1,i+2}\cdots u_{1,p_1}u_{1,1}.
\end{split}}$$ We say the Hamilton graph $\square \langle G_1, G_2\rangle$ is the result of doing a \emph{$2e$-join-$2e$ operation}.

Let $H_s$ be a Hamilton $(p_s,q_s)$-graph with a H-cycle $C_s=v_{s,1}v_{s,2}\cdots v_{s,p_s}v_{s,1}$ for $s\in [1,n]$.

\vskip 0.4cm

\textbf{H-CYCLE-CHAIN algorithm.} We do a series of $2e$-join-$2e$ operations to Hamilton graphs $H_1,H_2,\dots ,H_n$ in the following way: $G_{1,2}=\square \langle H_1, H_2\rangle$, $G_{2,3}=\square \langle G_{1,2}, H_3\rangle$, $\dots $, $G_{n-1,n}=\square \langle G_{n-2,n-1}, H_n\rangle$. For simplicity, we write $G_{n-1,n}=\square _{chain}\langle H_s\rangle^n_{s=1}$.

\vskip 0.2cm

\textbf{H-CYCLE-TREE algorithm.} Let $T$ be a tree of $n$ vertices, and its maximal degree $\Delta(T)\leq \min\{p_1,p_2,\dots ,p_n\}$. We replace each vertex $x_i$ of $T$ with a Hamilton graph $H_i$, and each edge $x_iy_j$ means $\square \langle H_i, H_j\rangle$ under the $2e$-join-$2e$ operation, such that each edge $v_{i,a}v_{i,a+1}$ of the H-cycle $C_i$ of $H_i$ is at most in a unique H-cycle $\square \langle C_i-v_{i,a}v_{i,a+1}, C_j-v_{j,b}v_{j,b+1}\rangle$, but other $\square \langle C_i-v_{i,a}v_{i,a+1}, C_s-v_{s,c}v_{s,c+1}\rangle$ for $s\neq j$. The result Hamilton graph is denoted as $\square _{tree}\langle H_s\rangle^n_{s=1}$. Clearly, $\square _{chain}\langle H_s\rangle^n_{s=1}$ is a particular case of $\square _{tree}\langle H_s\rangle^n_{s=1}$.

\vskip 0.4cm

\begin{rem}\label{rem:xxxxxxxxxxxxx}
In general, we have the following complex Hamilton graph sets by H-CYCLE-CHAIN algorithm and H-CYCLE-TREE algorithm based on the $2e$-join-$2e$ operation:
\begin{equation}\label{eqa:H-cycle-chain}
F_{chain}=\{\square _{chain}\langle a_iH^{nec}_i |^{m}_{i=1}, b_jH^{suf}_j|^{n}_{j=1}\rangle: a_i,b_j\in Z^0,H^{nec}_i\in G_{nec}(i),H^{suf}_j\in G_{suf}(j)\}
\end{equation} with $\sum^{m}_{i=1}a_i\geq 1$ and $\sum^{n}_{j=1}b_j\geq 1$. And
\begin{equation}\label{eqa:H-cycle-chain}
F_{tree}=\{\square _{tree}\langle \alpha_iH^{nec}_i |^{m}_{i=1}, \beta_jH^{suf}_j|^{n}_{j=1}\rangle: \alpha_i,\beta_j\in Z^0,H^{nec}_i\in G_{nec}(i),H^{suf}_j\in G_{suf}(j)\}
\end{equation} with $\sum^{m}_{i=1}\alpha_i\geq 1$ and $\sum^{n}_{j=1}\beta_j\geq 1$.\paralled
\end{rem}

\subsubsection{Graph lattices based on graph anti-homomorphism and non-multiple-edge full-v-coinciding operation}

Considering the \emph{inverse} of the \emph{graph homomorphism} defined in Definition \ref{defn:definition-graph-homomorphism}, we defined a new homomorphism as follows:

\begin{defn}\label{defn:new-graph-anti-homomorphisms}
$^*$ A \emph{graph anti-homomorphism} $L\rightarrow_{anti} T$ from a graph $L$ into another graph $T$ is a mapping $g: V(L) \rightarrow V(T)$ such that $f(u)f(v)\not \in E(T)$ for each edge $uv \in E(L)$, and $xy\not \in E(L)$ for each edge $f(x)f(y)\in E(T)$.\qqed
\end{defn}

\textbf{Non-multiple-edge full-v-coinciding operation ``$[\odot ]$''.} Let $L\rightarrow_{anti} T$ be a graph anti-homomorphism defined in Definition \ref{defn:new-graph-anti-homomorphisms}, we vertex-coincide each pair of vertices $u\in V(L)$ and $g(u)\in V(T)$ into one vertex $w=u\odot g(u)$ such that the \emph{gah-v-coincided graph} (\emph{graph anti-homomorphism vertex-coincided graph}) is rewritten as $G=[\odot ]\langle L,T\rangle$ with vertex number $|V(G)|=|V(T)|$ and edge set $E(G)=E(L)\cup E(T)$, where $V(G)=\{w=u\odot g(u):u\in V(L),g(u)\in V(T)\}$. Moreover, $T=G-E(L)$, and $L=G-E(T)$ when as $|V(L)|=|V(T)|$. We call ``$[\odot ]$'' a \emph{non-multiple-edge full-v-coinciding operation}.

\begin{figure}[h]
\centering
\includegraphics[width=16cm]{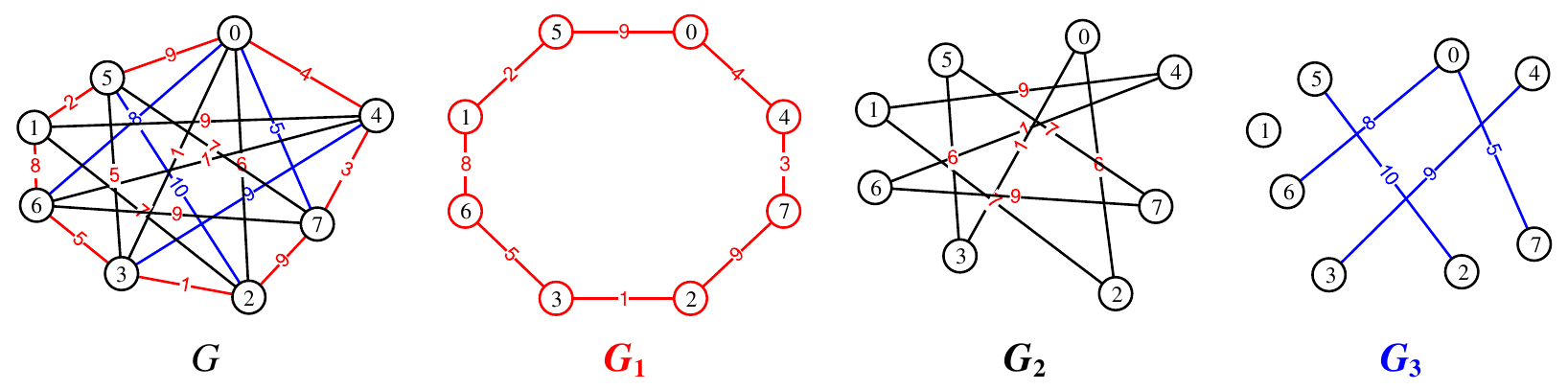}\\
{\small \caption{\label{fig:anti-homomorphism-11} A graph $G$ has two edge-disjoint Hamilton cycles, and $G=G_1[\odot]G_2[\odot]G_3$.}}
\end{figure}

\begin{rem}\label{rem:333333}
See an example $G=G_1[\odot]G_2[\odot]G_3$ shown in Fig.\ref{fig:anti-homomorphism-11} for understanding the \emph{non-multiple-edge full-v-coinciding operation}, in which the graph $G$ has two edge-disjoint Hamilton cycles. If we can cut a Topcode-matrix $T_{code}(G)$ of the graph $G$ into three subTopcode-matrices $T_{code}(G_1)$, $T_{code}(G_2)$ and $T_{code}(G_3)$, where each subTopcode-matrix $T_{code}(G_i)$ is a Topcode-matrix $T_{code}(G_i)$ of the graph $G_i$ for $i\in [1,3]$. See $T_{code}(G_i)=A_i$ for $i\in [1,3]$ shown in (\ref{eqa:cut-Topcode-matrix-into-3-11}), and the Topcode-matrix $T_{code}(G)$ is shown in (\ref{eqa:cut-Topcode-matrix-into-3-00}), such that
\begin{equation}\label{eqa:555555}
T_{code}(G)=T_{code}(G_1)\uplus T_{code}(G_2)\uplus T_{code}(G_3)
\end{equation}
Then we have decomposed $G$ into three graphs $G_1$, $G_2$ and $G_3$ by decomposing the Topcode-matrix $T_{code}(G)$. Thereby, decomposing a Topcode-matrix will produce algorithms for judging Hamilton graphs and graphs contains more edge-disjoint Hamilton cycles.\paralled
\end{rem}

{\footnotesize
\begin{equation}\label{eqa:cut-Topcode-matrix-into-3-11}
\centering
{
\begin{split}
A_1= \left(
\begin{array}{cccccccccccccccccccc}
0 & 4 & 7 & 2 & 3 & 6 & 1 & 5\\
4 & 3 & 9 & 1 & 5 & 8 & 2 & 9\\
4 & 7 & 2 & 3 & 6 & 1 & 0 & 5
\end{array}
\right),A_2= \left(
\begin{array}{cccccccccccccccccccc}
0 & 2 & 1 & 4 & 6 & 7 & 5 & 3\\
6 & 7 & 9 & 1 & 9 & 7 & 6 & 1\\
2 & 1 & 4 & 6 & 7 & 5 & 0 & 3
\end{array}
\right),A_3= \left(
\begin{array}{cccccccccccccccccccc}
0 & 0 & 5 & 3 \\
8 & 5 & 8 & 9 \\
6 & 7 & 2 & 4
\end{array}
\right)
\end{split}}
\end{equation}
}

{\footnotesize
\begin{equation}\label{eqa:cut-Topcode-matrix-into-3-00}
\centering
{
\begin{split}
T_{code}(G)= \left(
\begin{array}{cccccccccccccccccccc}
0 & 4 & 7 & 2 & 3 & 6 & 1 & 5 & 0 & 2 & 1 & 4 & 6 & 7 & 5 & 3 & 0 & 0 & 5 & 3\\
4 & 3 & 9 & 1 & 5 & 8 & 2 & 9 & 6 & 7 & 9 & 1 & 9 & 7 & 6 & 1 & 8 & 5 & 8 & 9\\
4 & 7 & 2 & 3 & 6 & 1 & 0 & 5 & 2 & 1 & 4 & 6 & 7 & 5 & 0 & 3 & 6 & 7 & 2 & 4
\end{array}
\right)
\end{split}}
\end{equation}
}

\textbf{Non-multiple-edge partial-v-coinciding operation ``$[\odot ]$''.} Let $H$ be a subgraph of a connected graph $L$, and let $H\,'$ be a subgraph of another connected graph $T$. If $H$ is graph anti-homomorphism to $H\,'$, $H\rightarrow_{anti} H\,'$, under a mapping $h$, we vertex-coincide each pair of vertices $x\in V(H)$ and $h(x)\in V(H\,')$ into one vertex $y=x\odot h(x)$, the \emph{partial gah-v-coincided graph} is denoted as $G^*=\partial [\odot ] \langle H\subseteq L,H\,'\subseteq T\rangle$ with $|V(G^*)|=|V(L)|-|V(H)|+|V(T)|$, $E(G^*)=E(L)\cup E(T)$, and we call ``$\partial [\odot ]$'' \emph{non-multiple-edge partial-v-coinciding operation}.

If the mapping $g$ defined in Definition \ref{defn:new-graph-anti-homomorphisms} is a \emph{one-one mapping}, that is, $g^{-1}: V(T) \rightarrow V(L)$, and then we have a graph anti-homomorphism $T\rightarrow_{anti} L$, and write this fact as $T\leftrightarrow_{anti} L$.

\vskip 0.4cm

\textbf{Graph anti-homomorphism vertex-coincided graph lattices.} Let $\textbf{\textrm{H}}=(H_1,H_2,\dots ,H_n)$ be a \emph{connected graph base} with each graph $H_i$ bo be connected, and $V(H_i)\cap V(H_j)$ and $H_i\leftrightarrow_{anti} H_j$ if $i\neq j$. Notice that $p=|V(H_i)|=|V(H_j)|$. We do the \emph{non-multiple-edge full-v-coinciding operation} on the connected graphs of the base $\textbf{\textrm{H}}$, and get the gah-v-coincided graphs $G_1=[\odot ] \langle H_1,H_2\rangle$, $G_2=[\odot ] \langle G_1,H_3\rangle$, and $G_s=[\odot ] \langle G_{s-1},H_{s+1}\rangle$ for $s\in [2,n-1]$, and write the last gah-v-coincided graph $G_{n-2}=[\odot ]^n_{k=1} \langle H_{k}\rangle$ with $p$~($=|V(G_{n-2})|=|V(H_j)|$ for $i\in [1,n]$) vertices and $q$~($=|E(G_{n-2})|$) edges with $q\leq p(p-1)/2$, where $E(G_{n-2})=\bigcup ^n_{k=1}E(H_k)$.

For $a_k$ copies of $H_k$ with $a_k\in Z^0$ and $H_k\in \textbf{\textrm{H}}$, we write them as $a_kH_k$, and moreover we do the \emph{non-multiple-edge full-v-coinciding operation} on $a_1H_1,a_2H_2,\dots ,a_nH_n$ by the way: Let $G_{i,1},G_{i,2},\dots ,G_{i,M}$ be the $i$th permutation of graphs $a_1H_1,a_2H_2,\dots ,a_nH_n$, where $M=\sum ^n_{k=1}a_k$, so we get a gah-v-coincided graph $[\odot ]^M_{k=1} \langle G_{i,k}\rangle$, we use $[\odot ]^n_{k=1}a_kH_k$ to represent each gah-v-coincided graph $[\odot ]^M_{k=1} \langle G_{i,k}\rangle$, and call the following set
\begin{equation}\label{eqa:gah-v-coincided-graph-lattices}
\textbf{\textrm{L}}(G_{\textrm{anti-ho}}[\odot ]\textbf{\textrm{H}})=\{[\odot ]^n_{k=1}a_kH_k:~ a_k\in Z^0, H_{k}\in \textbf{\textrm{H}}\}
\end{equation} a \emph{gah-v-coincided graph lattice} (\emph{graph anti-homomorphism vertex-coincided graph lattice}) with $\sum ^n_{k=1}a_k\geq 2$, and each graph $G\in \textbf{\textrm{L}}(G_{\textrm{anti-ho}}[\odot ]\textbf{\textrm{H}})$ is \emph{simple} and holds $|E(G)|\leq |V(G)|(|V(G)|-1)/2$ true.

\begin{rem}\label{rem:333333}
Especially, if each connected graph $H_i$ has a H-cycle, we call $\textbf{\textrm{H}}$ a \emph{hamiltonian graph base}, then each graph $G\in \textbf{\textrm{L}}(G_{\textrm{anti-ho}}[\odot ]\textbf{\textrm{H}})$ defined in (\ref{eqa:gah-v-coincided-graph-lattices}) has $m$ edge-disjoint H-cycles with $m=\sum ^n_{k=1}a_k$, so each edge-added graph $G+E^*\subseteq K_{mp}$ is hamiltonian, where $E^*$ is subset of edge set $V(G^c)$ of the complementary $G^c$ of the graph $G$, and $p=|V(H_i)|=|V(H_j)|$ for $i,j\in [1,n]$. We can claim: \emph{A connected graph $H$ has $m$ edge-disjoint H-cycles with $m\geq 2$ if and only if $H$ is isomorphic to an edge-added graph $G+E^*$~($\subseteq K_{mp}$) with $G\in \textbf{\textrm{L}}(G_{\textrm{anti-ho}}[\odot ]\textbf{\textrm{H}})$ based on a hamiltonian graph base $\textbf{\textrm{H}}$}.\paralled
\end{rem}

\vskip 0.4cm

\textbf{Infinite graph-base graph lattices.} Let $\textbf{\textrm{G}}^{\infty}=(G_1,G_2,\dots ,G_n,\dots )$ be an \emph{infinite graph base} and let $G_{i,1},G_{i,2},\dots ,G_{i,n_i}$ be a group of graphs selected from $\textbf{\textrm{G}}^{\infty}$ for $n_i\in Z^+=Z^0\setminus \{0\}$, and ``$(\ast )$'' be a graph operation. We call the following set
\begin{equation}\label{eqa:infinite-base-graph-lattices}
\textbf{\textrm{L}}(Z^0 (\ast )\textbf{\textrm{G}}^{\infty})=\{(\ast )^{n_i}_{k=1}b_{i,k}G_{i,k}:~ b_{i,k}\in Z^0, n_i\in Z^+, G_{i,k}\in \textbf{\textrm{G}}^{\infty}\}
\end{equation} an \emph{infinite graph-base graph lattice} with $\sum^{n_i}_{k=1}b_{i,k}\geq 1$ such that each graph of $\textbf{\textrm{L}}(Z^0 (\ast )\textbf{\textrm{G}}^{\infty})$ is \emph{simple}.

For example, each graph $G_i\in \textbf{\textrm{G}}^{\infty}$ is a cycle $C_{i+2}$ of $(i+2)$ vertices, the graph operation ``$(\ast )$'' is the vertex-coinciding operation ``$\odot$'', so $\odot |^{n_i}_{k=1}b_{i,k}G_{i,k}$ is an \emph{Euler's graph}. Moreover, suppose that $\textbf{\textrm{E}}^{\infty}_{\textrm{uler}}=(E^u_1,E^u_2,\dots ,E^u_n,\dots )$ is an \emph{infinite graph base} with each graph $E^u_i$ is an Euler's graph. Thereby, the vertex-coinciding operation ``$\odot$'' produces each graph $\odot |^{n_i}_{k=1}c_{i,k}E^u_{i,k}$ with $E^u_{i,k}\in \textbf{\textrm{E}}^{\infty}_{\textrm{uler}}$ to be an Euler's graph, and we get an \emph{infinite-base Euler-graph lattice} as follows
\begin{equation}\label{eqa:infinite-base-Euler-graph-lattices}
\textbf{\textrm{L}}(Z^0 \odot \textbf{\textrm{E}}^{\infty}_{\textrm{uler}})=\{\odot |^{n_i}_{k=1}c_{i,k}E^u_{i,k}:~ c_{i,k}\in Z^0, n_i\in Z^+, E^u_{i,k}\in \textbf{\textrm{E}}^{\infty}_{\textrm{uler}}\}
\end{equation} with $\sum^{n_i}_{k=1}c_{i,k}\geq 1$ such that each graph of $\textbf{\textrm{L}}(Z^0 \odot \textbf{\textrm{E}}^{\infty}_{\textrm{uler}})$ is \emph{simple}.

Let $\textbf{\textrm{H}}_p=(H_{p,1},H_{p,2},\dots ,H_{p,n_p})$ be a \emph{hamiltonian graph base} with $|V(H_{p,i})|\neq |V(H_{p,j})|$ for $i\neq j$, in which each edge-added graph $H_{p,k}=C_p+E_{p,k}$ consists of a cycle $C_p$ of $p$ vertices and a set $E_{p,k}$ of edges, where $E_{p,1}=\emptyset$. Suppose that $\textbf{\textrm{H}}^\infty$ contains any hamiltonian graph base $\textbf{\textrm{H}}_p$, by the help of the \emph{non-multiple-edge full-v-coinciding operation} ``$[\odot ]$'', we get an \emph{infinite H-graph-base graph lattice}
\begin{equation}\label{eqa:infinite-base-Euler-graph-lattices}
\textbf{\textrm{L}}(Z^0 [\odot ]\textbf{\textrm{H}}^\infty)=\{[\odot ]^{n_p}_{k=1}d_{p,k}H_{p,k}:~ d_{p,k}\in Z^0, n_p\in Z^+, H_{p,k}\in \textbf{\textrm{H}}_p\subset \textbf{\textrm{H}}^\infty\}
\end{equation} with $\sum^{n_p}_{k=1}d_{p,k}\geq 1$ such that each graph of $\textbf{\textrm{L}}(Z^0 [\odot ]\textbf{\textrm{H}}^\infty)$ is a Euler's graph.

\subsection{Graph-splitting families}

Doing a series of vertex-splitting operations to a connected graph $G$ produces $n_G$ \emph{edge-disjoint graph-splitting families} $G_k=(G_{k,1},G_{k,2},\dots$, $G_{k,m_k})$ with $m_k\geq 2$ and $|E(G_{k,i})|\geq 1$ for $k\in [1,n_G]$, such that $E(G_{k,i})\cap E(G_{k,j})=\emptyset$ if $i\neq j$, and $E(G)=\bigcup^{n_G}_{k=1}E(G_k)$. Conversely, we can do the vertex-coinciding operation to an edge-disjoint graph-splitting group $G_k$ for getting the connected graph $G$, written this case by $G=\odot |^{m_k}_{i=1}G_{k,i}$ with $k\in [1,n_G]$. In general, $\odot |^{m_k}_{i=1}G_{k,i}$ produces many connected graphs differing from $G$. We call the graph set $\textbf{\textrm{L}}(\wedge G)=\bigcup^{n_G}_{k=1} \{\odot |^{m_k}_{i=1}G_{k,i}\}$ as a \emph{$G$-v-splitting quasi-lattice} (Ref. \cite{Wang-Su-Yao-divided-2020}). We have a result as follows:

\begin{thm}\label{thm:decompose-degree-sequence}
\cite{Yao-Wang-Ma-Wang-Degree-sequences-2021} For a connected graph $G$ and each edge-disjoint graph-splitting group $G_k\in S_{plit}(G)$, we have $\textrm{deg}(G)=\odot |^{m_k}_{j=1} \textrm{deg}(G_{k,j})$ and $\textrm{deg}(G_k)=\bigcup^{m_k}_{j=1} \textrm{deg}(G_{k,j})$, so the degree-sequence $\textrm{deg}(G)$ has been decomposed into degree-sequences $\textrm{deg}(G_k)$ for $k\in [1,n_G]$.
\end{thm}

\begin{thm}\label{thm:degree-sequence-decomposed}
\cite{Yao-Wang-Ma-Wang-Degree-sequences-2021} Let $G_{raph}(\textbf{\textrm{d}})$ be the set of graphs having degree-sequence $\textbf{\textrm{d}}$. Then the degree-sequence $\textbf{\textrm{d}}$ can be decomposed into the degree-sequence $\textrm{deg}(G_k)$ for each edge-disjoint graph-splitting group $G_k\in S_{plit}(G)$, $G\in G_{raph}(\textbf{\textrm{d}})$, that is, $d_{ecom} (\textbf{\textrm{d}})=\textrm{deg}(G_k)$.
\end{thm}

\begin{defn} \label{defn:more-definitions-on-degree-sequence}
\cite{Yao-Wang-Ma-Wang-Degree-sequences-2021} For a connected graph $G$, there is a graph-splitting family $S_{plit}(G)=\{G_k=(G_{k,1},G_{k,2},\dots, G_{k,m_k}):~m_k\geq 2,k\in [1,n_G]\}$, we have the following particular degree-sequences $\textrm{deg}(G_k)$:
\begin{asparaenum}[\textrm{\textbf{Par}}-1. ]
\item If each connected subgraph $G_{k,i}$ ($i\in [1,m_k]$) holding $|V(G_{k,i})|=2$, call $\textrm{deg}(G_k)$ \emph{trivial degree-sequence}.
\item If some connected subgraph $G_{k,i}$ holds $|V(G_{k,i})|\geq 3$ true, call $\textrm{deg}(G_k)$ \emph{normal degree-sequence}.
\item If $|V(G_{k,1})|\geq 3$, others $|V(G_{k,i})|=2$ with $i\in [2,m_k]$, call $\textrm{deg}(G_k)$ \emph{unique non-$K_2$ degree-sequence}.
\item If $|V(G_{k,i})|\geq 3$ for $i\in [1,m_k]$, call $\textrm{deg}(G_k)$ \emph{non-$K_2$ degree-sequence}.
\item If $|V(G_{k,i})|\geq 3$ and $G_{k,i}$ is a path for $i\in [1,m_k]$, call $\textrm{deg}(G_k)$ \emph{pure-path degree-sequence}.
\item If $|V(G_{k,i})|\geq 3$ and $G_{k,i}$ is a cycle for $i\in [1,m_k]$, call $\textrm{deg}(G_k)$ \emph{pure-cycle degree-sequence}.
\item If $|V(G_{k,i})|\geq 3$ and $G_{k,i}$ is a tree for $i\in [1,m_k]$, call $\textrm{deg}(G_k)$ \emph{tree-type degree-sequence}. Moreover, if each diameter $D(G_{k,i})\geq 3$, call $\textrm{deg}(G_k)$ \emph{pure tree-type sequence}.
\item If $|V(G_{k,i})|\geq 3$ and $G_{k,i}$ is a star $K_{v(k,i)}$ with $v(k,i)=|V(G_{k,i})|$ for $i\in [1,m_k]$, and some $v(k,i)\geq 2$, call $\textrm{deg}(G_k)$ \emph{star degree-sequence}.
\item If each connected subgraph $G_{k,i}$ is a Hamilton graph for $i\in [1,m_k]$, call $\textrm{deg}(G_k)$ \emph{hamiltonian degree-sequence}.
\item If each connected subgraph $G_{k,i}$ has no odd-degree for $i\in [1,m_k]$, call $\textrm{deg}(G_k)$ \emph{Euler's degree-sequence}.
\item If $G_{k,i}\not \cong G_{k,j}$ with $i\neq j$ and $i,j\in [1,m_k]$, call $\textrm{deg}(G_k)$ \emph{mutually no-isomorphic degree-sequence}.
\item If $|E(G_{k,j})|>|E(G_{k,j+1})|$ with $j\in [1,m_k-1]$, call $\textrm{deg}(G_k)$ \emph{strictly edge-decreasing degree-sequence}.
\item If $|V(G_{k,j})|>|V(G_{k,j+1})|$ ($j\in [1,m_k-1]$), call $\textrm{deg}(G_k)$ \emph{strictly vertex-decreasing degree-sequence}.
\item Randomly select $a_{j_1}, a_{j_2}, \dots, a_{j_m}$ from a degree-sequence $\textrm{\textbf{d}}=(a_1, a_2, \dots, a_n)$, such that $a_{j_i}\geq a_{j_{i+1}}$ with $i\in [1,m-1]$ and $a_{j_1}\leq m-1$, and the new sequence $(a_{j_1}, a_{j_2}, \dots, a_{j_m})$ to be really a degree-sequence. We call $\textrm{\textbf{d}}$ a \emph{perfect degree-sequence}. For example, (1) $a_1=n-1$, $a_2=\dots=a_n=k$; (2) $a_1=a_2=n-2$, $a_3=\dots=a_n=k$; (3) $\textrm{\textbf{d}}=\textrm{deg}(K_{m,n})$.
\item A base $\textbf{\textrm{d}}^*=\{d^1_n,d^2_n,\dots,d^m_n\}$ with each degree-sequence $d^k_n=(b_{k,j})^m_{j=1}$ derives a new degree-sequence $(a_{j})^n_{j=1}$ with $a_{j}=\sum^m_{k=1}b_{k,j}$ for $j\in [1,n]$. If sequences $(a_{j})^n_{j=1}$ and $(b_{k,j})^m_{k=1}$ with $j\in [1,n]$ are degree-sequences, we call $\textbf{\textrm{d}}^*$ a \emph{right-angled degree-sequence base}.\qqed
\end{asparaenum}
\end{defn}


\section{Recent problems}

Up to today, people do not solve a conjecture ``\emph{Every tree admits a graceful labeling}'' first proposed by Rosa in \cite{A-Rosa-graceful-1967}, and Gallian in \cite{Gallian2020} pointed that most graph labeling methods trace their origin to one introduced by Rosa in 1967. Unfortunately, we do not know what kind of graphs admitting set-ordered graceful labelings, even for trees, to this day. In this section we have collected some problems without thorough inspection and verification, maybe part of them are meaningless, maybe there are errors in them, however, we will be very happy to get criticism and corrections about them.

\subsection{Problems from graph colorings and labelings}

\begin{figure}[h]
\centering
\includegraphics[width=13cm]{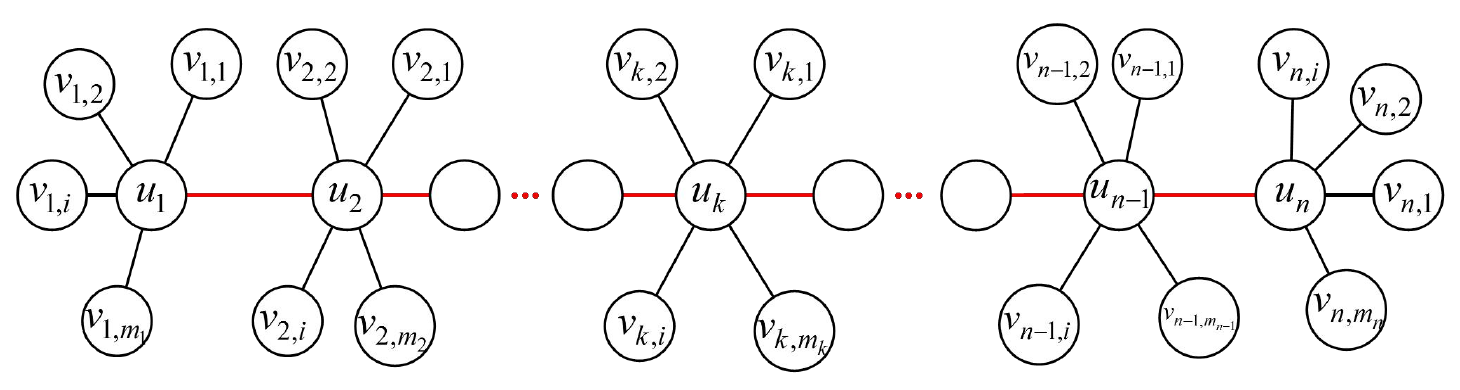}
\caption{\label{fig:1010-caterpillar}{\small A caterpillar.}}
\end{figure}

\begin{problem}\label{problem:xxxxxx}
\textbf{Set-ordered odd-graceful (resp. odd-elegant) labelings for caterpillars.} \textbf{Determine} the number of non-isomorphic caterpillars of $p$ vertices. \textbf{How} many set-ordered odd-graceful/odd-elegant labelings does a caterpillar admit? (see Fig.\ref{fig:1010-caterpillar})
\end{problem}

\begin{problem}\label{problem:xxxxxx}
\textbf{Determine} the value of $G_{race}(H)$ for each simple graph $H$, refer to Definition \ref{defn:sequce-graceful-labeling-k-f}.
\end{problem}

\begin{problem}\label{problem:xxxxxx}
\textbf{Edge-magic proper total colorings and equitably proper total colorings.} \cite{Yao-Sun-Zhang-Mu-Sun-Wang-Su-Zhang-Yang-Yang-2018arXiv} Let $f:V(G)\cup E(G)\rightarrow [1,\chi \,''(G)]$ be a proper total coloring of a graph $G$ with $\chi \,''(G)=|\{f(x):x\in V(G)\cup E(G)\}|$, and we call $f$ a \emph{total chromatic number pure (tcn-pure) coloring}. Let $d_f(uv)=f(u)+f(uv)+f(v)$, and
$$B_{tol}(G,f)=\max_{uv \in E(G)}\{d_f(uv)\}-\min_{uv \in E(G)}\{d_f(uv)\}.$$
Especially, a coloring $h$ is called an \emph{edge-magic proper total coloring} (\emph{edge-magic tcn-pure coloring}) if $B_{tol}(G,h)=0$, or an \emph{equitably edge-magic proper total coloring} (\emph{equitably edge-magic tcn-pure coloring}) if $B_{tol}(G,h)=1$, refer to Definition \ref{defn:edge-magic-tcn-pure-total-coloring}.

(1) \textbf{Determine} the parameter $\min_fB_{tol}(G,f)$ over all tcn-pure colorings of $G$, as well as its edge-magic proper total colorings or equitably edge-magic proper total colorings. If a total coloring $g$ of $G$ arrives at $B_{tol}(G,g)=\min_fB_{tol}(G,f)$, \textbf{is} $\chi \,''(G)=|\{g(x):x\in V(G)\cup E(G)\}|$ true?

(2) \textbf{Find} connected graphs admitting one of the edge-magic proper total coloring and the equitably edge-magic proper total coloring.
\end{problem}

\begin{problem}\label{qeu:perfect-graph-vs-subgraph-labelings}
\cite{Yao-Zhang-Sun-Mu-Sun-Wang-Wang-Ma-Su-Yang-Yang-Zhang-2018arXiv} Suppose that a $\theta$-labeling is one of the existing graph labelings, \textbf{characterize} perfect $\theta$-labeling graphs defined in Definition \ref{defn:perfect-graph-vs-subgraph-labelings}.
\end{problem}

\begin{problem}\label{problem:xxxxxx}
\textbf{Characterize twin odd-graceful chains.} (\cite{Wang-Xu-Yao-2017-Twin, Wang-Xu-Yao-2017}) For each tree $T_i$ admitting an odd-graceful labeling (resp. an odd-elegant labeling) $f_i$ such that $f_i(V(T_i))\subset [0,2q-1]$ and $f_i(E(T_i))=[1,2q-1]^o$ with $i\in [1,m]$ (Ref. Definition \ref{defn:22-twin-odd-graceful-labeling}). If each vertex-coincided graph $\odot_{k_{i,i+1}} \langle T_i,T_{i+1}\rangle$ obtained by doing the vertex-coinciding operation on both $T_{i}$ and $T_{i+1}$ admits a \emph{twin odd-graceful labeling} $f\odot h$ (resp. a \emph{twin odd-elegant labeling}), where $f_i(V(T_i))\cup f_{i+1}(V(T_{i+1}))= [0,2q-1]$ and $|f_i(V(T_i))\cap f_{i+1}(V(T_{i+1}))|= k_{i,i+1}$ with $i\in [1,m-1]$, then we get a \emph{twin odd-graceful chain} (resp. a \emph{twin odd-elegant chain}) $\odot \langle T_i\rangle ^m_1$.
\end{problem}
\begin{problem}\label{problem:xxxxxx}
\textbf{Find 6C-complementary matchings.} For a given $(p,q)$-tree $G$ admitting a 6C-labeling $f$ defined in Definition \ref{defn:6C-labeling}, \textbf{find} a $(p,q)$-tree $H$ admitting a 6C-labeling $g$ such that the vertex-coincided graph $\odot\langle G,H \rangle $ is a 6C-complementary matching defined in Definition \ref{defn:6C-complementary-matching}.
\end{problem}
\begin{problem}\label{problem:xxxxxx}
\cite{Yao-Sun-Zhang-Mu-Sun-Wang-Su-Zhang-Yang-Yang-2018arXiv} \textbf{Find reciprocal-inverse matchings.} Suppose that a $(p,q)$-graph $G$ and a $(q,p)$-graph $H$ admit two edge-magic graceful labelings $f$ and $g$, respectively, and both $f$ and $g$ are reciprocal inverse from each other as if $f(E(G))=g(V(H))\setminus X^*$ and $f(V(G))\setminus X^*=g(E(H))$ for $X^*=f(V(G))\cap g(V(H))$. \textbf{Find} such pairs of graphs $G$ and $H$, and characterize them, refer to Definition \ref{defn:6C-complementary-matching}.
\end{problem}

\begin{problem}\label{qeu:PNBSPP}
\textbf{Parameterized Number-based String Partition Problem (PNBSPP).} Given a number-based string $s(n)=c_1c_2\cdots c_n$ with $c_i\in [0,9]$, partition it into $3q$ segments $c_1c_2\cdots c_n=a_1a_2\cdots a_{3q}$ with $a_j=c_{n_j}c_{n_j+1}\cdots c_{n_{j+1}}$ with $j\in [1,3q-1]$, where $n_1=1$ and $n_{3q}=n$. And use $a_k$ with $k\in [1,3q]$ to reform a Topcode-matrix $T_{code}(G)$ (refer to Definition \ref{defn:topcode-matrix-definition}), and moreover use the Topcode-matrix $T_{code}(G)$ to reconstruct all of Topsnut-gpws of $q$ edges. By the found Topsnut-gpws corresponding the common Topcode-matrix $T_{code}(G)$, \textbf{find} the desired \emph{public Topsnut-gpws} $H_i$ and the desired \emph{private Topsnut-gpws} $G_i$, such that each mapping $\varphi_i:V(G_i)\rightarrow V(H_i)$ forms a \emph{colored graph homomorphism} $G_i\rightarrow H_i$ with $i\in [1,r]$.
\end{problem}

\begin{rem}\label{rem:ABC-conjecture}
About PNBSPP we will face the following difficult problems:
\begin{asparaenum}[\textrm{H}-1. ]
\item \label{hard:matrices} A number-based string $s(n)$ may correspond no matrix, or two or more matrices, or Topcode-matrices, as well as parameterized Topcode-matrices.
\item \label{hard:no-method} No method for cutting a number-based string $s(n)$ into $3q$ segments $c_1c_2\cdots c_n=a_1a_2\cdots a_{3q}$ with $a_j=c_{n_j}c_{n_j+1}\cdots c_{n_{j+1}}$ with $j\in [1,3q-1]$, where $n_1=1$ and $n_{3q}=n$, such that $a_1,a_2,\dots , a_{3q}$ are just the elements of some Topcode-matrix $T_{code}(G)$.
\item \label{hard:coloring-labeling} Since a Topcode-matrix $T_{code}(G)$ with parameterized elements is related with one of thousand and thousand of graph colorings and graph labelings, so determining this Topcode-matrix $T_{code}(G)$ is not a slight job.
\item \label{hard:graph-isomorphic} If a parameterized Topcode-matrix $T_{code}(G)$ corresponds a number-based string $s(n)$ has been determined, finding all non-isomorphic Topsnut-gpws will meet the graph isomorphic problem, as known, it is NP-hard.\paralled
\end{asparaenum}
\end{rem}

\begin{figure}[h]
\centering
\includegraphics[width=16.4cm]{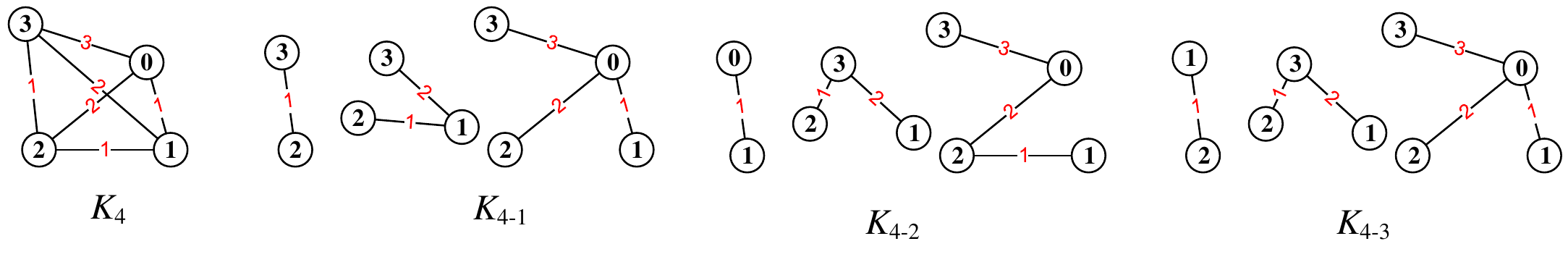}
\caption{\label{fig:Kn-colored-trees-1}{\small A compound of $K_4$ by three groups of colored trees.}}
\end{figure}

\begin{figure}[h]
\centering
\includegraphics[width=16.4cm]{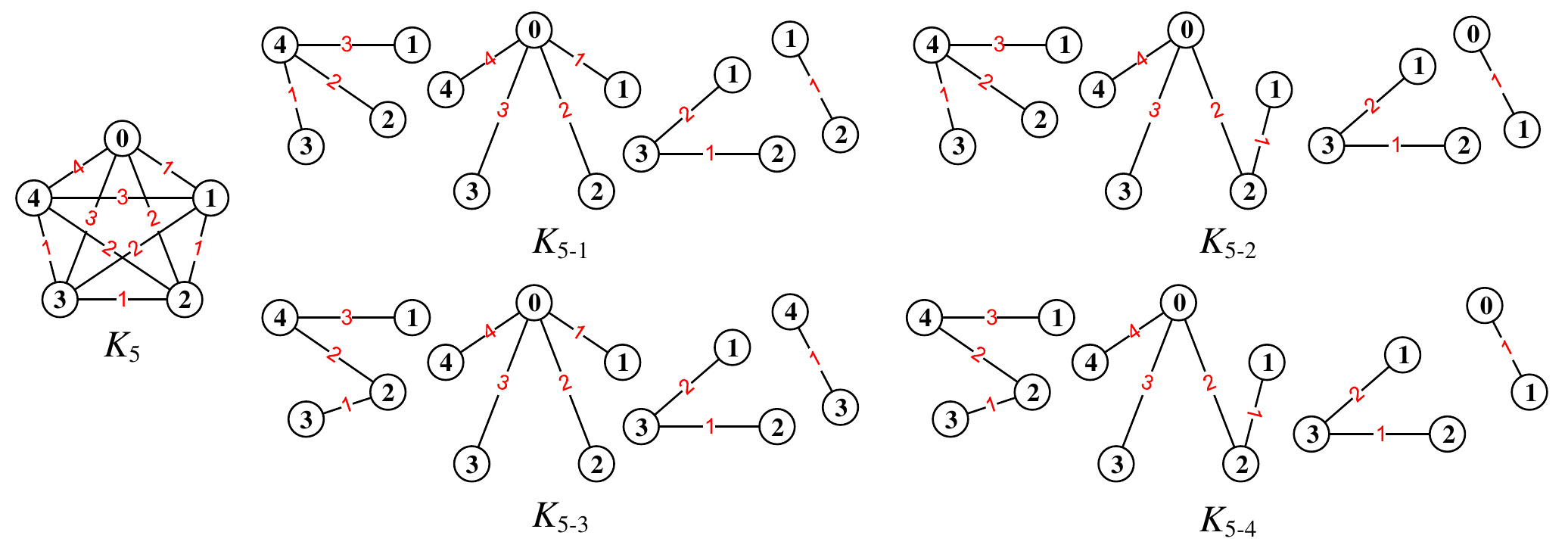}
\caption{\label{fig:Kn-colored-trees-2}{\small A compound of $K_5$ by four groups of colored trees.}}
\end{figure}

\begin{problem}\label{problem:1010-complete-graphs}
\textbf{Colored complete graphs obtained by vertex-coinciding colored trees.} For any group of vertex-disjoint trees $T_2,T_3,\dots,T_n$ with each $T_k$ having $k$ vertices for $k\in [2,n]$, \textbf{can} we find a $W$-type labeling $f_k$ for each tree $T_k$ such that $f_k:V(T_k)\rightarrow [0,n-1]$ and $\{|f_k(u)-f_k(v)|:uv\in E(T_k)\}=[1,|V(T_k)|-1]=[1, k-1]$, and then vertex-coincide the vertices of $\bigcup^n_{k=2}V(T_k)$ having been colored the same color into one, such that the resultant graph is just a complete graph $K_n=\odot |^n_2\langle T_k\rangle$ with $E(K_n)=\bigcup^n_{k=2}E(T_k)$ and $E(T_i)\cap E(T_j)=\emptyset$ for $i\neq j$? See Fig.\ref{fig:Kn-colored-trees-1} and Fig.\ref{fig:Kn-colored-trees-2} for understanding the complex of this problem.
\end{problem}

\begin{problem}\label{problem:1010-colored-regular-graphs}
\textbf{Colored regular graphs obtained by vertex-coinciding colored trees.} What conditions does a group of vertex-disjoint trees $H_1,H_2,\dots,H_m$ hold, such that we can find a $W_i$-type total labeling $g_i:V(H_i)\cup E(H_i)\rightarrow [1,M]$ for each tree $H_i$ with $i\in [1,m]$, and then vertex-coincide the vertices of $\bigcup^m_{i=1}V(H_i)$ having been colored the same color into one, so we get the resultant graph to be just a regular (or multiple-edge) graph $G=\odot |^m_1\langle H_i\rangle$ with $E(G)=\bigcup^m_{i=1}E(H_i)$ and $E(H_i)\cap E(H_j)=\emptyset$ for $i\neq j$? Problem \ref{problem:1010-complete-graphs} provides some solutions for this problem.
\end{problem}

\begin{problem}\label{problem:set-ordered-graceful}
\textbf{Graphs admitting set-ordered graceful labelings.} \textbf{Characterize} graphs admitting set-ordered graceful labelings, especially, \textbf{find} trees admitting set-ordered graceful labelings defined in Definition \ref{defn:basic-W-type-labelings}.
\end{problem}

\begin{problem}\label{problem:especial-total-colorings-3}
\textbf{Color-valued graphic authentication problem (CVGAP).} For a given connected non-tree $(p,q)$-graph $G$, we have two graph sets: A \emph{\textbf{public-key set}} $S_v$ and a \emph{\textbf{private-key set}} $S_e$, such that each graph $H_i$ of $S_v$ admits a proper vertex coloring, and each graph $L_j$ of $S_e$ admits a proper edge coloring, as well as $|E(G)|=|E(H_i)|=|E(L_j)|$. \textbf{Can} we find a graph $H_i\in S_v$ and another graph $L_j\in S_e$, and do the vertex-coinciding operation to $H_i$ and $L_j$ respectively, such that the resultant graphs $H\,'_i$ and $L\,'_j$ hold two graph homomorphisms $H\,'_i\rightarrow G$ and $L\,'_j\rightarrow G$ true, and two colorings of $H\,'_i$ and $L\,'_j$ induce just a proper total coloring of $G$ (as a \emph{topological authentication})? Since we can vertex-split the vertices of a $(p,q)$-graph admitting a $W$-type total coloring into at least $q-p+1$ different connected graphs, so $S_v\neq \emptyset $ and $S_e\neq \emptyset $, and there exists a solution for this problem.
\end{problem}

\begin{rem}\label{rem:333333}
According to Definition \ref{defn:77-two-graphic-set-groups} and Remark \ref{rem:77-two-graphic-set-groups}, we have two every-zero graphic-set groups $Gg^+=\{F_m(G_{rap}(T^i_{code}));\oplus\}$ and $Gg^-=\{F_m(G_{rap}(T^i_{code}));\ominus\}$. Since they used the same set $F_m(T^i_{code})=\{T^1_{code},T^2_{code},\dots, T^m_{code}\}$ of Topcode-matrices with each Topcode-matrix $T^i_{code}=(X_i,E_i,Y_i)^{T}$, where v-vector $X_i=(x_{i,1},~x_{i,2},~\cdots,~x_{i,q})$, e-vector $E_i=(e_{i,1},~e_{i,2},~\cdots,~e_{i,q})$ and v-vector $Y_i=(y_{i,1},~y_{i,2},~\cdots,~y_{i,q})$ and there are functions $f_i$ holding $e_{i,r}=f_i(x_{i,r},y_{i,r})$ with $i\in [1,m]$ and $r\in [1,q]$. We can set the \emph{every-zero graphic-set group} $Gg^+$ as a \emph{public-key group}, and the \emph{every-zero graphic-set group} $Gg^-$ as a \emph{private-key group}.\paralled
\end{rem}

\begin{problem}\label{problem:especial-total-colorings-6}
\textbf{Splitting-coinciding total coloring problem (SCTCP).} Given two connected graphs $W$ and $U$ with $\chi\,''(W)=\chi\,''(U)$ and $|E(W)|=|E(U)|$, \textbf{does} doing a series of vertex-splitting operations and vertex-coinciding operations to $W$ (resp. $U$) produce $U$ (resp. $W$)?
\end{problem}

\begin{problem}\label{qeu:1010-OSOPTCP}
\textbf{Optimal set-ordered proper total coloring problem (OSOPTCP).} By Definition \ref{defn:66-optimal-e-v-set-ordered-total-colorings}, \textbf{determine} \emph{optimal v-set-ordered proper total colorings} $f$ and \emph{optimal e-set-ordered proper total colorings} $g$ for a connected graph $G$ with $f:V(G)\cup E(G)\rightarrow [1,\chi\,''(G)]$ and $g:V(G)\cup E(G)\rightarrow [1,\chi\,''(G)]$.
\end{problem}

\begin{figure}[h]
\centering
\includegraphics[width=16.4cm]{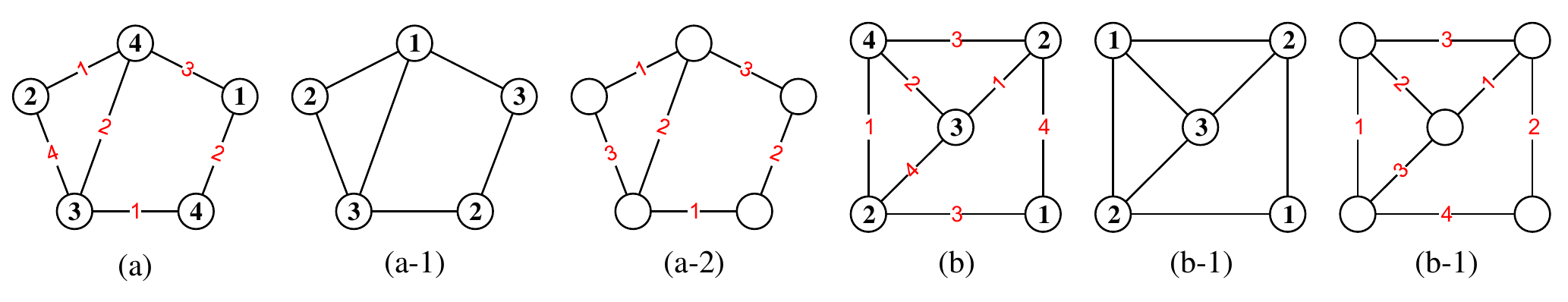}
\caption{\label{fig:OSOPTCP}{\small Examples for understanding Problem \ref{qeu:1010-OSOPTCP}.}}
\end{figure}

Two graphs $G$ and $H$ shown in Fig.\ref{fig:OSOPTCP} do not admit any optimal v-set-ordered proper total coloring, in which (a) a connected $(5,6)$-graph $G$ admits a proper total coloring $f$ with $|f(V(G)\cup E(G))|=4$, $\chi\,''(G)=4=\Delta(G)+1$, but $|f(V(G))|=4$; (a-1) $\chi(G)=3$; (a-2) $\chi\,'(G)=3$. (b) A connected $(5,7)$-graph $H$ admits a proper total coloring $g$ with $|g(V(H)\cup E(H))|=4$, $\chi\,''(H)=4=\Delta(H)+1$, however $|g(V(H))|=4$; (b-1) $\chi(H)=3$; (b-2) $\chi\,'(H)=4$.

\begin{problem}\label{problem:general-ve-inverse-labeling}
$^*$ Find ve-inverse $(W_f,W_g)$-type labelings $(f,g)$, or ve-inverse $(W_f,W_g)$-type colorings $(f,g)$, such that $f(E(G))=g(V(H))$ and $f(V(G))=g(E(H))$ if a graph $G$ admits a $W_f$-type labeling (resp. coloring) $f$ and another graph $H$ admits a $W_g$-type labeling (resp. coloring) $g$ (refer to Definition \ref{defn:general-ve-inverse-labeling}).
\end{problem}

\subsubsection{Multiple dimension total colorings}

\begin{problem}\label{problem:xxxxxxxxx}
\textbf{Does} every connected simple graph admit a graceful $3$-dimension sub-proper total coloring (refer to Definition \ref{defn:n-dimension-total-colorings})?
\end{problem}

\begin{problem}\label{problem:xxxxxxxxx}
By Definition \ref{defn:n-dimension-total-colorings}, \textbf{is} there a strongly graceful $n$-dimension sub-proper total coloring for a tree having a perfect matching?
\end{problem}

\begin{problem}\label{problem:xxxxxxxxx}
By Definition \ref{defn:n-dimension-total-colorings}, a $n$-dimension text-based string $D=\alpha_1\alpha_2\cdots \alpha_n$, where $\alpha_i=\beta_{i,1}\beta_{i,2}\cdots\beta_{i,n_i}$ with $\beta_{i,j}$ is a letter or a number in $[0,9]$. \textbf{Consider} various $W$-type total colorings made by $n$-dimension text-based strings.
\end{problem}

\begin{problem}\label{problem:xxxxxxxxx}
By Definition \ref{defn:n-dimension-total-colorings}, \textbf{what} is the norm of a tree colored with $W$-type total colorings based on $n$-dimension digital-based strings?
\end{problem}
\begin{problem}\label{problem:xxxxxxxxx}
By Definition \ref{defn:n-dimension-total-colorings}, since every tree $T$ admits a graceful $2$-dimension proper total coloring, \textbf{how} far away from Graceful Tree Conjecture?
\end{problem}

\begin{problem}\label{problem:xxxxxxxxx}
By Definition \ref{defn:edge-magic-tcn-pure-total-coloring}: (1) The distance between two maximal-degree vertices of a connected graph admitting an edge-magic tcn-pure coloring is at lest three. (2) For a fixed positive integer $q$, characterize connected $(p,q)$-graphs admitting edge-magic tcn-pure colorings. (3) What is the minimum circle length in a connected graph admitting an edge-magic tcn-pure coloring?
\end{problem}

\subsubsection{Conjectures related with colorings and labelings}

\begin{conj} \label{conj:xxxxxxxxx}
\textbf{A conjecture related to Four Color Conjecture.} \cite{YAO-SUN-WANG-SU-XU2018arXiv} A maximal planar graph $G$ is 4-colorable if and only if four colored triangles shown in Fig.\ref{fig:4-coloring-system}(a) can tile fully the maximal planar graph $G$.
\end{conj}
\begin{conj} \label{conj:xxxxxxxxx}
\textbf{A conjecture on the v-set e-proper labeling.} By Definition \ref{defn:55-set-labeling}, each connected graph with no multiple edges and self-loops admits a v-set e-proper (odd-)graceful labeling defined in Definition \ref{defn:55-v-set-e-proper-more-labelings} and Definition \ref{defn:55-set-labeling}.
\end{conj}

\begin{conj} \label{conj:path-conjecture-magically-total-labelings}
\cite{Yao-Chen-Yao-Cheng2013JCMCC} \textbf{A conjecture on $(k,\lambda _k)$-magically total labellins of paths}. For each $k\in [2,2(n-1)]$, each path $P_n$ of $n$ vertices admits a magic total labeling $f_k:V(P_n)\cup E(P_n)\rightarrow [1,2n-1]$ such that $f_k(u)+f_k(v)=k+f_k(uv)$ for each edge $uv\in E(P_n)$ (refer to Definition \ref{defn:k-lambda-magically-total-labeling}).
\end{conj}

The authors in \cite{Yao-Chen-Yao-Cheng2013JCMCC} have verified Conjecture \ref{conj:path-conjecture-magically-total-labelings} for all trees of order $\leq 9$ and Conjecture \ref{conj:path-conjecture-magically-total-labelings} for all paths on $n\leq 10$ vertices. An example for Conjecture \ref{conj:path-conjecture-magically-total-labelings} is: A path on $n$ vertices is denoted as $P_n=u_1e_1u_2e_2\cdots u_{n-1}e_{n-1}u_n$, and we color each vertex $u_i$ of $P_n$ with a small circle ${\Large \textcircled{\small $m$}}$ colored with a color $m$ and we color each edge $e_i$ of $P_n$ with a number, respectively. For a path $P_7$, $\lambda _k=1$ and each integer $k\in [2,12]$ we have
\begin{multicols}{2}
$k=2$\quad ${\Large \textcircled{\small 1}}2{\Large
\textcircled{\small 3}}10{\Large \textcircled{\small 9}}13{\Large
\textcircled{\small 6}}12{\Large \textcircled{\small 8}}11{\Large
\textcircled{\small 5}}7{\Large \textcircled{\small 4}}$

$k=3$\quad ${\Large \textcircled{\small 5}}3{\Large
\textcircled{\small 1}}2{\Large \textcircled{\small 4}}10{\Large
\textcircled{\small 9}}13{\Large \textcircled{\small 7}}12{\Large
\textcircled{\small 8}}11{\Large \textcircled{\small 6}}$

$k=4$\quad ${\Large \textcircled{\small 7}}12{\Large
\textcircled{\small 9}}13{\Large \textcircled{\small 8}}6{\Large
\textcircled{\small 2}}1{\Large \textcircled{\small 3}}4{\Large
\textcircled{\small 5}}11{\Large \textcircled{\small 10}}$

$k=5$\quad ${\Large \textcircled{\small 1}}5{\Large
\textcircled{\small 9}}11{\Large \textcircled{\small 7}}12{\Large
\textcircled{\small 10}}13{\Large \textcircled{\small 8}}6{\Large
\textcircled{\small 3}}2{\Large \textcircled{\small 4}}$

$k=6$\quad ${\Large \textcircled{\small 3}}2{\Large
\textcircled{\small 5}}7{\Large \textcircled{\small 8}}12{\Large
\textcircled{\small 10}}13{\Large \textcircled{\small 9}}4{\Large
\textcircled{\small 1}}6{\Large \textcircled{\small 11}}$

$k=7$\quad ${\Large \textcircled{\small 3}}6{\Large
\textcircled{\small 10}}12{\Large \textcircled{\small 9}}4{\Large
\textcircled{\small 2}}8{\Large \textcircled{\small 13}}7{\Large
\textcircled{\small 1}}5{\Large \textcircled{\small 11}}$

$k=8$\quad ${\Large \textcircled{\small 4}}8{\Large
\textcircled{\small 12}}7{\Large \textcircled{\small 3}}5{\Large
\textcircled{\small 10}}11{\Large \textcircled{\small 9}}2{\Large
\textcircled{\small 1}}6{\Large \textcircled{\small 13}}$

$k=9$\quad ${\Large \textcircled{\small 6}}7{\Large
\textcircled{\small 10}}9{\Large \textcircled{\small 8}}12{\Large
\textcircled{\small 13}}5{\Large \textcircled{\small 1}}3{\Large
\textcircled{\small 11}}4{\Large \textcircled{\small 2}}$

$k=10$\quad ${\Large \textcircled{\small 2}}5{\Large
\textcircled{\small 13}}4{\Large \textcircled{\small 1}}3{\Large
\textcircled{\small 12}}8{\Large \textcircled{\small 6}}7{\Large
\textcircled{\small 11}}10{\Large \textcircled{\small 9}}$

$k=11$\quad ${\Large \textcircled{\small 4}}5{\Large
\textcircled{\small 12}}2{\Large \textcircled{\small 1}}3{\Large
\textcircled{\small 13}}11{\Large \textcircled{\small 9}}6{\Large
\textcircled{\small 8}}7{\Large \textcircled{\small 10}}$

$k=12$\quad ${\Large \textcircled{\small 4}}3{\Large
\textcircled{\small 11}}12{\Large \textcircled{\small 13}}9{\Large
\textcircled{\small 8}}2{\Large \textcircled{\small 6}}1{\Large
\textcircled{\small 7}}5{\Large \textcircled{\small 10}}$
\end{multicols}

\begin{conj} \label{conj:xxxxxxxxxxxxxxx}
\cite{Yao-Chen-Yao-Cheng2013JCMCC} Every tree admits a super $(k,\lambda)$-magically total labeling for some integers $k$ and $\lambda\neq 0$ defined in Definition \ref{defn:k-lambda-magically-total-labeling}.
\end{conj}

\begin{problem} \label{problem:xxxxxxxxxxxxxxxxxxx}
\cite{Yao-Chen-Yao-Cheng2013JCMCC} If a connected $(p,q)$-graph graph admits a $(k,\lambda)$-magically total labeling defined in Definition \ref{defn:k-lambda-magically-total-labeling}, whether does it have $q=2p-3$ edges at most?
\end{problem}

\begin{conj} \label{conj:xxxxxxxxx}
\cite{Yao-Sun-Zhang-Zhang-Yang-Wang-Zhang-ICISCE2017} Every tree $T$ admits an edge-magic graceful labeling $f$ defined in Definition \ref{defn:AM-odd-elegant-perfect-matchings} such that each edge $uv$ of $T$ holds $f(u)+f(v)-f(uv)=k\geq 0$ for some fixed constant $k$.
\end{conj}

\begin{conj}\label{conj:parameted-8-labelings}
\cite{Sun-Wang-Yao-2020} Trees having perfect matchings admit an arm-2-parameter totally odd/even-graceful labeling, an arm-2-parameter totally odd/even-elegant labeling, an arm-2-parameter arithmetic labeling, an arm-2-parameter super felicitous labeling, an arm-2-parameter super edge-magic total labeling, and an arm-2-parameter super edge-magic graceful labeling defined in Definition \ref{defn:parameted-perfect-matchings}.
\end{conj}

\begin{figure}[h]
\centering
\includegraphics[width=16.4cm]{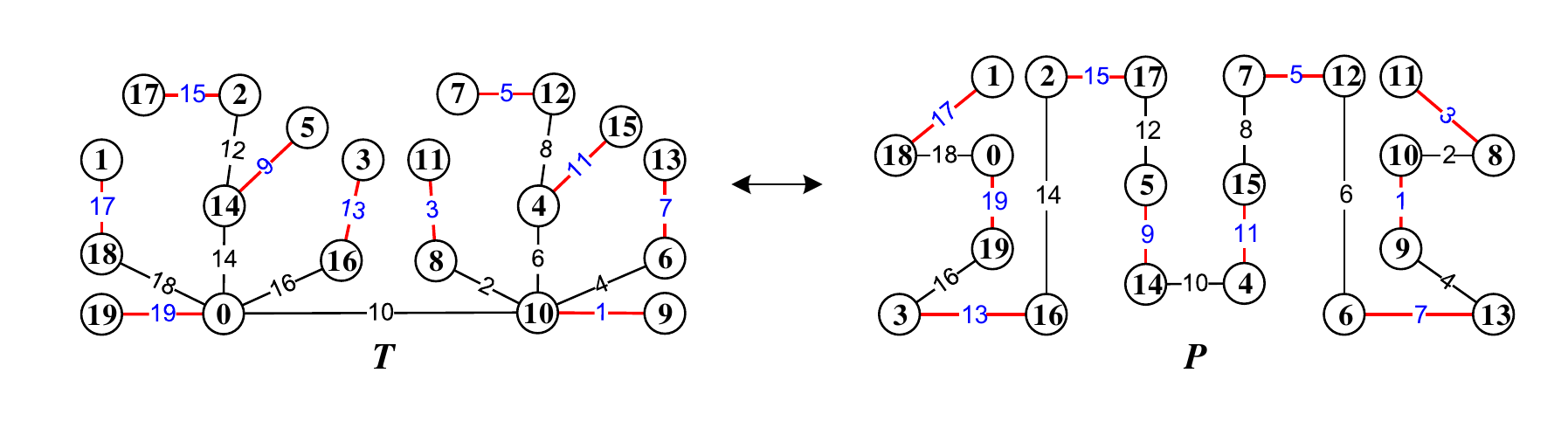}\\
\caption{\label{fig:parameted-perfect-conjec}{\small A strongly graceful tree $T$ having perfect matching is transformed into a strongly graceful path $P$ having perfect matching.}}
\end{figure}

\begin{conj}\label{conj:tree-perfect-matching-path}
\cite{Sun-Wang-Yao-2020} \textbf{Strongly Graceful Graph Homomorphism Conjecture.} A strongly graceful tree having perfect matchings can be transformed into a strongly graceful path having perfect matching (see an example shown in Fig.\ref{fig:parameted-perfect-conjec}).
\end{conj}

\begin{rem}\label{rem:ABC-conjecture}
In Conjecture \ref{conj:tree-perfect-matching-path}, a strongly graceful tree $T$ having perfect matching is \emph{graph homomorphism to} a strongly graceful path $P$ having perfect matching, here, $P$ is as a public-key, and it has many private-key $T$, this graph homomorphism is consisted of topological structures having perfect matchings and strongly graceful labelings.\paralled
\end{rem}

\begin{conj}\label{conj:gcd-labelings}
$^*$ Each tree admits an odd-graceful gcd-labeling (refer to Definition \ref{defn:graceful-gcd-colorings} and Definition \ref{defn:graceful-gcd-colorings}).
\end{conj}

\begin{conj} \label{conj:box}
$^*$ Any tree admits a $k$-graceful labeling some $k>1$ (refer to Definition \ref{defn:11-old-labelings-Gallian}).
\end{conj}

\begin{conj}\label{conj:flawed-abc-total-colorings}
It is natural based on Definition \ref{defn:flawed-odd-graceful-labeling}, the authors in \cite{Yao-Mu-Sun-Sun-Zhang-Wang-Su-Zhang-Yang-Zhao-Wang-Ma-Yao-Yang-Xie2019} conjecture: ``\emph{Each forest $T=\bigcup ^m_{i=1}T_i$ with vertex-disjoint trees $T_1,T_2,\dots ,T_m$ admits a flawed graceful (resp. odd-graceful) labeling}''. \textbf{Determine} positive integers $A_e$ and $B_e$ such that $H=E^*+T$ admits a (set-ordered) graceful (resp. odd-graceful) labeling as $A_e\leq |E^*|\leq B_e$.
\end{conj}

\begin{conj}\label{conj:Max-Min-difference-sum}
\cite{Yao-Mu-Sun-Sun-Zhang-Wang-Su-Zhang-Yang-Zhao-Wang-Ma-Yao-Yang-Xie2019} By Definition \ref{defn:difference-sum-matching}, each tree induces a perfect Max-Min difference-sum matching partition and a perfect Max-Min felicitous-sum matching partition (see Fig.\ref{fig:felicitous-sum-matching} for understanding this conjecture).
\end{conj}

\subsubsection{Parameterized total colorings}

\begin{problem}\label{problem:parameterized-edge-magics}
For a bipartite graph $G$, finding three parameters $a,b,c$ holding $(a,b,c)\neq (1,1,1)$ under a proper total coloring $f:V(G)\cup E(G)\rightarrow [1,M]$ realizes $B^*_{\alpha}(G,f,M)(a,b,c)=0$ holding each one of the parameterized edge-magic proper total coloring, the parameterized edge-difference proper total coloring, the parameterized felicitous-difference proper total coloring and the parameterized graceful-difference proper total coloring defined in Definition \ref{defn:combinatoric-definition-total-coloring-abc}.
\end{problem}

\begin{rem}\label{rem:ABC-conjecture}
In a parameterized edge-magic proper total coloring $f$ defined in Definition \ref{defn:combinatoric-definition-total-coloring-abc}, the number $B^*_{\alpha}(G,f,M)=0$ means that $c_f(uv)=af(u)+bf(v)+cf(uv)=k$ for each edge $uv\in E(G)$. If there are $(a_0,b_0,c_0)\neq (1,1,1)$ holding $c_f(uv)=a_0f(u)+b_0f(v)+c_0f(uv)=k$, then we have $c_f(uv)=\beta a_0f(u)+\beta b_0f(v)+\beta c_0f(uv)=\beta k$ for each edge $uv\in E(G)$ with $\beta > 0$ and $(\beta a_0,\beta b_0,\beta c_0)\neq (\beta ,\beta ,\beta )$. So, there are infinite groups of parameters $a,b,c$ holding $(a,b,c)\neq (1,1,1)$ for the total colorings.\paralled
\end{rem}

\begin{problem}\label{problem:especial-total-colorings-1}
\textbf{Estimate} the bounds of the constant $k_i$ with $i\in [1,4]$ in each especial proper total coloring defined in Definition \ref{defn:combinatoric-definition-total-coloring}, where $k_1=f(u)+f(uv)+f(v)$, $k_2=f(uv)+|f(u)-f(v)|$, $k_3=|f(u)+f(v)-f(uv)|$ and $k_4=\big ||f(u)-f(v)|-f(uv)\big |$.
\end{problem}

\begin{problem}\label{problem:especial-total-colorings-2}
\textbf{Multiple-magic proper total coloring.} For any group of positive integers $k_1,k_2,k_3,k_4$, \textbf{find} a connected graph $G$ admitting a $\{k_i\}^4_1$-magic proper total coloring $h$ defined in Definition \ref{defn:combinatoric-definition-total-coloring}, such that there are edges $u_iv_i\in E(G)$ with $i\in [1,4]$ holding $k_1=h(u_1)+h(u_1v_1)+h(v_1)$, $k_2=h(u_2v_2)+|h(u_2)-h(v_2)|$, $k_3=|h(u_3)+h(v_3)-h(u_3v_3)|$ and $k_4=\big ||h(u_4)-h(v_4)|-h(u_4v_4)\big |$ true, and each edge holds one of the above four equations true.
\end{problem}

\begin{problem}\label{problem:especial-total-colorings-4}
\textbf{Find} a simple and connected graph $G$ admitting a proper total coloring $f:V(G)\cup E(G)\rightarrow [1,M]$ and inducing an edge-function $c_f(uv)$ for each edge $uv\in E(G)$ according to Definition \ref{defn:combinatoric-definition-total-coloring}, and find constants $k_1,k_2,\dots ,k_m$, such that each edge $uv\in E(G)$ corresponds some $k_i$ holding $c_f(uv)=k_i$ true, and each constant $k_j$ corresponds at least one edge $xy$ holding $c_f(xy)=k_j$.
\end{problem}

\begin{problem}\label{problem:especial-total-colorings-5}
For any integer sequence $\{k_i\}^n_1$ with $k_i<k_{i+1}$ for $i\in [1,n-1]$, \textbf{find} a simple and connected graph $G$ such that each $k_i$ corresponds a proper total coloring $f_i:V(G)\cup E(G)\rightarrow [1,M]$ defined in Definition \ref{defn:combinatoric-definition-total-coloring}, and $f_i$ induces an edge-function $c_{f_i}(uv)=k_i$ for each edge $uv\in E(G)$.
\end{problem}

\begin{problem}\label{problem:trees-various-colorings}
Since a colored connected $(p,q)$-graph $G$ can be vertex-split into some colored trees, or be leaf-split into colored trees, we have questions as follows:
\begin{asparaenum}[\textbf{\textrm{Tree}}-1. ]
\item \textbf{Construct} graphs or graphic lattices admitting set-ordered gracefully total colorings defined in Definition \ref{defn:kd-w-type-colorings}, or set-ordered graceful labelings.
\item If a tree $T$ admits a graceful labeling, then \textbf{does} it admit a gracefully total coloring?
\item \textbf{Determine} each $W$-type coloring defined in Definition \ref{defn:new-graceful-strongly-colorings} for trees.
\item \textbf{Characterize} graphs admitting set-ordered gracefully total colorings defined in Definition \ref{defn:kd-w-type-colorings}.
\item About the parameter $v_{W}(G)=\min_f\{|f(V(G))|\}$ over all $W$-type coloring $f$ of the $(p,q)$-graph $G$ for a fixed $W\in[1,27]$ based on Definition \ref{defn:new-graceful-strongly-colorings}, for each integer $m$ subject to $v_{W}(G)<m\leq p-1$, \textbf{does} there exist a $W$-type coloring $g$ holding $|g(V(G))|=m$?
\end{asparaenum}
\end{problem}

\begin{problem}\label{problem:4-parameters-graceful-odd-graceful}
Since each tree $T$ admits set-ordered gracefully total colorings and set-ordered odd-gracefully total colorings defined in Definition \ref{defn:kd-w-type-colorings}, \textbf{determine} the following parameters:
$$m_{\textrm{sogtc}}(T)=\min _f\{|f(V(T))|\}\textrm{ and }M_{\textrm{sogtc}}(T)=\max _f\{|f(V(T))|\}
$$ over all set-ordered gracefully total colorings of $G$, and
$$m^o_{\textrm{sogtc}}(T)=\min _g\{|g(V(T))|\}\textrm{ and }M^o_{\textrm{sogtc}}(T)=\max _g\{|g(V(T))|\}
$$ over all set-ordered odd-gracefully total colorings of $G$. Clearly, $M_{\textrm{sogtc}}(T)=|V(T)|$ and $M^o_{\textrm{sogtc}}(T)=|V(T)|$ means the Graceful Tree Conjecture and the Odd-graceful Tree Conjecture.
\end{problem}

\begin{problem}\label{problem:Multiple-coloring}
We have the following questions about the coloring including $L$-multiple colorings defined in Definition \ref{defn:L-multiple-type-coloring-labeling}:
\begin{asparaenum}[\textbf{\textrm{Mul}}-1. ]
\item It, by Graceful Tree Conjecture, is natural to \textbf{guess}: ``\emph{Every connected graph $G$ admits a coloring including a graceful labeling}''.
\item For a complete bipartite graph $K_{1,n}$ with vertices $x_0,x_i,\dots ,x_n$ and edge set $E(K_{1,n})=\{x_0x_i:i\in [1,n]\}$, we define a graceful labeling $f$ of $K_{1,n}$ as: $f(x_0)=n$, $f(x_j)=j$ with $j\in [1,n]$, so we have $f(x_0x_j)=|f(x_0)-f(x_j)|=n-j$, and $f(E(K_{1,n}))=[1,n]$. However, we can see $f(x_0)+f(x_0x_j)+f(x_j)=2n$, in other word, $f$ is an edge-magic total labeling of $K_{1,n}$ too. \textbf{Is} this going to happen to other graphs ($\neq K_{1,n}$) else?
\end{asparaenum}
\end{problem}

\begin{figure}[h]
\centering
\includegraphics[width=14cm]{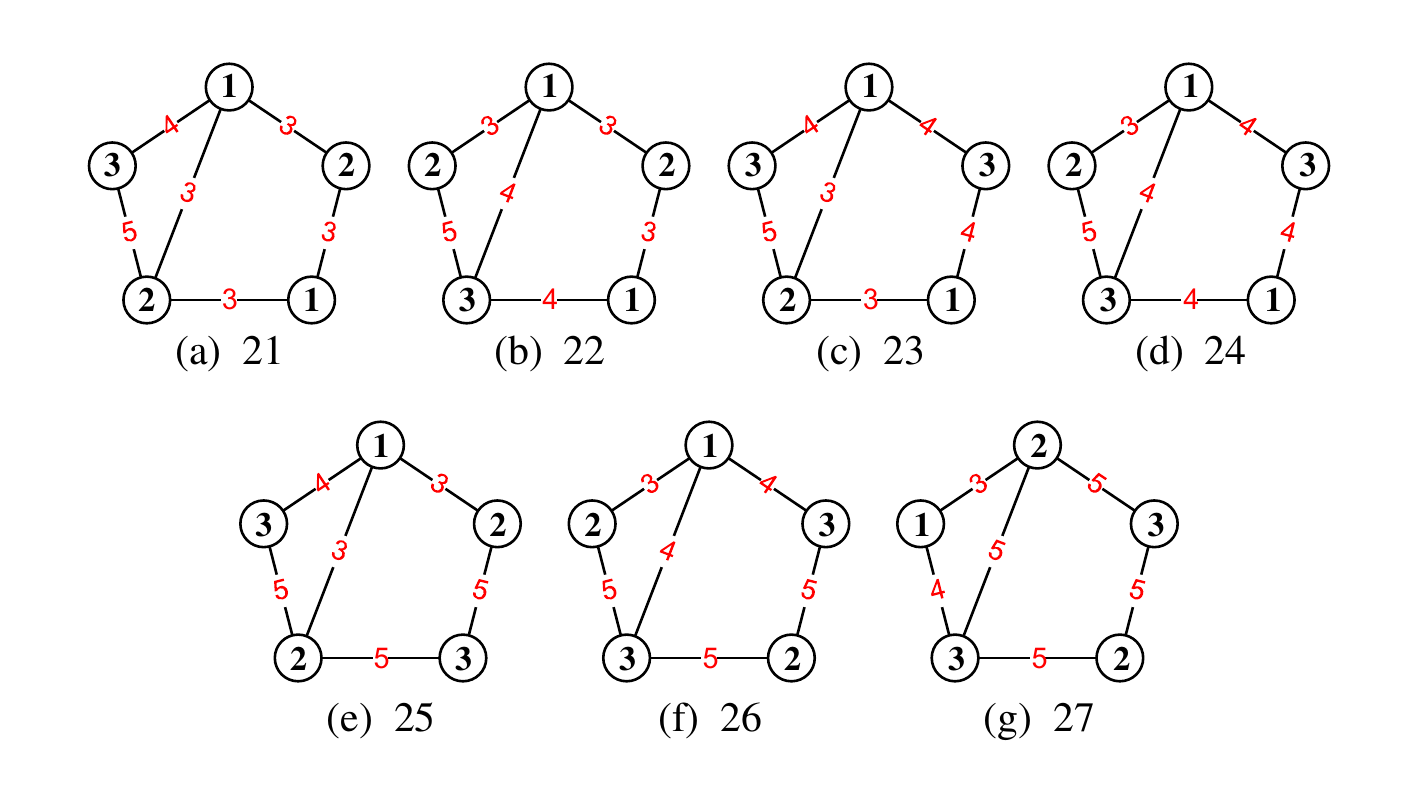}\\
\caption{\label{fig:new-topic-v-coloring-2}{\small The graph $C+uv$ admits a group of \emph{consecutive sum} proper vertex colorings, which forms a Topsnut-matching chain, cited from \cite{Yao-Sun-Zhang-Mu-Sun-Wang-Su-Zhang-Yang-Yang-2018arXiv}.}}
\end{figure}

\begin{problem}\label{problem:xxxxxx}
Observed that the colored graph (a) in Fig.\ref{fig:new-topic-v-coloring-1} is equal to the colored graph (a) in Fig.\ref{fig:new-topic-v-coloring-2}, according to Definition \ref{defn:new-parameters-2}. If a proper vertex coloring $h^*$ holds $B_{sub}(G,h^*)=\min_fB_{sub}(G$, $f)$ true, then \textbf{does} $h^*$ hold $B_{sum}(G,h^*)=\min_fB_{sum}(G,f)$ true too?
\end{problem}

\subsection{Problems with extremal and critical conditions}

\begin{problem}\label{problem:gracefully-total-numbers-graphs}
\cite{Bing-Yao-2020arXiv} The gracefully total number $R_{grace}(p,s)$ has been designed by means of the idea of Ramsey number of graph theory. It seems to be not easy to \textbf{determine} the exact value of a gracefully total number $R_{grace}(p,s)$ defined in Definition \ref{defn:gracefully-critical-graph} and Definition \ref{defn:gracefully-total-authentication}.

A connected graph $G$ containing a $(p,s)$-gracefully total matching $(H^{+},H^{-})$ is just a $(p,s)$-gracefully total authentication when $H^{+}$ is as a \emph{public-key} and $H^{-}$ is as a \emph{private-key}, and $G$ is optimal if $|V(G)|\leq |V(H)|$ and $|E(G)|\leq |E(H)|$ for any $(p,s)$-gracefully total authentication $H$. So, there are many optimal $(p,s)$-gracefully total authentications by Proposition \ref{thm:gracefully-total-numbers-graphs}, \textbf{find} optimal $(p,s)$-gracefully total authentications for integers $p,s\geq 6$.
\end{problem}

\begin{problem}\label{problem:generalized-one}
$^*$ By Definition \ref{defn:gracefully-critical-graph} and Definition \ref{defn:gracefully-total-authentication}, an \emph{ex-gracefully total number} $R_{grace}(p,s\mid A,B)$ is extremum and holds: Any red-blue edge-coloring of a complete graph $K_{m}$ of $m~(\geq R_{grace}(p,s\mid A,B))$ vertices induces a copy $T_i$ of the $A$-gracefully critical graph $G$ of $p$ vertices, or a copy $L_j$ of the $B$-gracefully critical graph $H$ of $s$ vertices, where each edge of $T_i$ is red and each edge of $L_j$ is blue, and $A,B\in \{C^{+}_{yes},C^{-}_{yes},C^{+}_{no},C^{-}_{no}\}$ (refer to Definition \ref{defn:generalized-one}). And, some red-blue edge-coloring of a complete graph $K_{n}$ with $n\leq R_{grace}(p,s)-1$ does not induce any one of copies of the gracefully$^{+}$ critical graph $H^{+}$ of $p$ vertices and copies of the gracefully$^{-}$ critical graph $H^{-}$ of $s$ vertices. \textbf{Determine} ex-gracefully total numbers $R_{grace}(p,s\mid A,B)$ for $A,B\in \{C^{+}_{yes},C^{-}_{yes},C^{+}_{no},C^{-}_{no}\}$ and integers $p,s\geq 6$.
\end{problem}

\begin{problem}\label{problem:generalized-three}
$^*$ By Definition \ref{defn:gracefully-critical-graph} and Definition \ref{defn:gracefully-total-authentication}, \textbf{determine} optimal $(p,s)$-$W$-type $(I,J)$-critical authentications and ex-$W$-type numbers $R_{grace}(p,s\mid I,J)$ for $I,J\in \{C^{+}_{yes}-W$, $C^{-}_{yes}-W,C^{+}_{no}-W,C^{-}_{no}-W\}$ and integers $p,s\geq 6$ defined in Definition \ref{defn:generalized-two-red-blue-complete} and Definition \ref{defn:generalized-three}, where $W$-type is one of the existing colorings and labelings.
\end{problem}

\begin{problem}\label{qeu:1010}
By Definition \ref{defn:graceful-gcd-colorings}, Definition \ref{defn:22-graceful-gcd-labelings}, Remark \ref{rem:graceful-gcd-colorings} and Remark \ref{rem:22edge-prime-labelings}, for a $(p,q)$-graph $G$, \textbf{determine}: graceful gcd-labeling number $G_{gcd}(G)$, odd-graceful gcd-labeling number $OG_{gcd}(G)$, graceful gcd-coloring M-m-number $M_{mgrac}(G)$, odd-graceful gcd-coloring M-m-number $M_{moddgra}(G)$; prime gcd-labeling number $P_{rime}(G)$, edge-prime gcd-labeling number $P\,'_{rime}(G)$; and graceful gcd-labeling M-m-number $M_{mgral}(G)$ and odd-graceful gcd-labeling M-m-number $M_{moddgral}(G)$.
\end{problem}

\subsection{Sum-type of colorings}

\begin{problem}\label{qeu:10-various-sums-total-coloring}
$^*$ For a $(p,q)$-graph $G$ admits a proper total coloring $f:V(G)\cup E(G)\rightarrow [1,M]$.
\begin{asparaenum}[\textbf{\textrm{Variousum}}-1.]
\item $TC_{dsum}(G,f)=\displaystyle \sum_{uv\in E(G)}|f(u)-f(v)|$ is a \emph{total coloring difference-sum number}. \textbf{Find} two extremal numbers $\max_f TC_{dsum}(G,f)$ and $\min_f TC_{dsum}(G,f)$ over all proper total colorings of $G$.
\item $TC_{fsum}(G,g)=\displaystyle \sum_{uv\in E(G)}[g(u)+g(v)]~(\bmod ~q+1)$ is a \emph{total coloring felicitous-sum number}. \textbf{Determine} two extremal numbers $\max_g TC_{fsum}(G,g)$ and $\min_g TC_{fsum}(G,g)$ over all felicitous-sum colorings of $G$.
\item $TC_{edsum}(G,f)=\displaystyle \sum _{uv\in E(G)}\big [f(uv)+|f(u)-f(v)|\big ]$ is a \emph{total coloring edge-difference-sum number}, \textbf{find} two extremal numbers $\max_f TC_{edsum}(G,f)$ and $\min_f TC_{edsum}(G,f)$ over all proper total colorings of $G$.
\item $TC_{gdsum}(G,f)=\displaystyle \sum _{uv\in E(G)}\big |f(u)+f(v)-f(uv)\big |$ is a \emph{total coloring graceful-difference number}, \textbf{determine} two extremal numbers $\max_f TC_{gdsum}(G,f)$ and $\min_f TC_{gdsum}(G,f)$ over all felicitous-sum colorings of $G$.
\item $TC_{magicsum}(G,f)=\displaystyle \sum _{uv\in E(G)}\big [f(u)+f(uv)+f(v)\big ]=kq$ is a \emph{total coloring edge-magic number}, \textbf{compute} two extremal numbers $\max_f TC_{magicsum}(G,f)$ and $\min_f TC_{magicsum}(G,f)$ over all proper total colorings of $G$.
\end{asparaenum}
\end{problem}

\subsection{Problems of decomposition and partition}

\begin{problem}\label{problem:text-based-string-sequences}
\cite{Yao-Su-Wang-Hui-Sun-ITAIC2020} \textbf{Text-based String Partition Problem (TBSPP).} Partition a given text-based string $D$ (as an authentication) or two given text-based strings $D_1,D_2$ into two sequences $A_M$ (as a \emph{public-key}) and $B_q$ (as a \emph{private-key}), and then finding a topological structure (graph, network) $G$ admitting a $W$-type sequence coloring defined on $A_M$ and $B_q$ in Definition \ref{defn:sequence-coloring}, such that the Topcode-matrix $T_{code}(G)$ defined in Definition \ref{defn:topcode-matrix-definition} can derive the text-based string $D$ or two text-based strings $D_1$ and $D_2$.
\end{problem}

\begin{problem}\label{problem:number-based-string}
\cite{Yao-Wang-Su-Sun-ITOEC-2020}~\textbf{Number-based String Partition Problem (NBSPP).} Suppose that a number-based string $s(n)=c_1c_2\cdots c_n$ with $c_j\in [0,9]$ was generated from some Topcode-matrix, \textbf{cut} $s(n)$ into substrings $a_{1}a_{2}\dots a_{3q}$ holding $a_{1}=c_1c_2\cdots c_{j_1}$, $a_{2}=c_{j_1+1}c_{j_1+2}\cdots c_{j_1+j_2}$, $\dots $, $a_{3q}=c_{m+1}c_{m+2}\cdots c_n$, where each $j_k\geq 1$ for $k\in [1,3q]$, $n=m+j_{3q}$ and $m=\sum ^{3q-1}_{k=1}j_k$, such that there exists at least a colored graph $H$ with its own Topcode-matrix $T_{code}(H)$ defined in Definition \ref{defn:topcode-matrix-definition} (resp. directed Topcode-matrix defined in Definition \ref{defn:directed-Topcode-matrix}), which contains each substring $a_{i}$ with $i\in [1,3q]$ as its own elements and can deduces $s(n)$ from $T_{code}(H)$.
\end{problem}

\begin{problem}\label{problem:string-decomposition-matrices}
\cite{Bing-Yao-2020arXiv} \textbf{Number-based Strings and Matrices Problem (NBSMP).} In general, we want to cut a number-based string $D=c_1c_2\cdots c_n$ with $c_j\in [0,9]$ into $n^2$ segments $a_{ij}$ such that these segments $a_{ij}$ are just the elements of the adjacency matrix $A(a_{ij})_{n\times n}(G)$, or the adjacency e-value matrix (refer to Definition \ref{defn:e-value-matrix}), or the adjacency ve-value matrix (refer to Definition \ref{defn:ve-value-matrix}) of a graph $G$ with $n$ vertices.
\end{problem}

\begin{problem}\label{problem:infinite-number-based-string}
\cite{Bing-Yao-2020arXiv} \textbf{Infinite Number-based String Problem (INBSP).} Let $I_{string}=\{c_k\}^{\infty}_{k=1}$ be an infinite number-based string with $c_k\in [0,9]$, and let $T_{code}(G)$ be a Topcode-matrix of a $(p,q)$-graph $G$ admitting a $W$-type total coloring. For each finite number-based string $s(n)$ induced from $T_{code}(G)$, \textbf{is} $s(n)$ a segment of $I_{string}$?
\end{problem}

\begin{problem}\label{problem:IDTAP-CDSP-SDSP-11}
\cite{Yao-Zhang-Wang-Su-Integer-Decomposing-2021} \textbf{Integer-Decomposing Topological Authentication Problem (IDTAP).} Decompose an even integer $m$ into positive integers $m_{k,1},m_{k,2},\dots ,m_{k,n_k}$ with $n_k\geq 2$ and $k\in [1,D_{comp}(m)]$ with $m=m_{k,1}+m_{k,2}+\cdots +m_{k,n_k}$, such that a number-based string $m_{k,1}m_{k,2}\cdots m_{k,n_k}$ is as a \emph{public-key}, and there exists a uncolored graph $G_k$ (as a \emph{private-key}) having its own degree-sequence $\textbf{\textrm{d}}_k=(m_{k,1},m_{k,2},\dots ,m_{k,n_k})$, where $D_{comp}(m)$ is the number of different groups of decompositions of the even integer $m$.
\end{problem}

\begin{problem}\label{problem:IDTAP-CDSP-SDSP-22}
\cite{Yao-Zhang-Wang-Su-Integer-Decomposing-2021} \textbf{Colored Integer-Decomposing Topological Authentication Problem (Colored-IDTAP).} Given an even integer $m$ companied by a number-based string $s(n)$ (as a \emph{combinatorial public-key}), \textbf{do}:

(1) first decompose this even integer $m$ into positive integers $m_{k,1},m_{k,2},\dots ,m_{k,n_k}$ with $n_k\geq 2$ and $k\in [1,D_{comp}(m)]$, such that $m=m_{k,1}+m_{k,2}+\cdots +m_{k,n_k}$ and there exists a colored graph $G$ (as a \emph{private-key}) having its own degree-sequence $\textbf{\textrm{d}}=(m_{k,1},m_{k,2},\dots ,m_{k,n_k})$, where $D_{comp}(m)$ is the number of different groups of decompositions of the even integer $m$, and $G$ is colored by a total coloring $f$;

(2) cut the number-based string $s(n)=c_1c_2\cdots c_n$ into $3q$ segments $s(n)=a_{1}a_{2}\dots a_{3q}$ holding $a_{1}=c_1c_2\cdots c_{j_1}$, $a_{2}=c_{j_1+1}c_{j_1+2}\cdots c_{j_1+j_2}$, $\dots $, $a_{3q}=c_{m+1}c_{m+2}\cdots c_n$, where each $j_k\geq 1$ for $k\in [1,3q]$, $n=m+j_{3q}$ and $m=\sum ^{3q-1}_{k=1}j_k$, which are just the elements of the Topcode-matrix $T_{code}(G)$ under the total coloring $f$, where $q=|E(G)|$.
\end{problem}

\begin{rem}\label{rem:IDTAP-Colored-IDTAP}
It is not easy to answer Problem \ref{problem:IDTAP-CDSP-SDSP-11} and Problem \ref{problem:IDTAP-CDSP-SDSP-22}: A positive integer $m$ is decomposed into a group of $k$ parts $m_1,m_2,\cdots ,m_k$ such that $m=m_1+m_2+\cdots +m_k$ with $m_i>0$. Suppose there are $P(m,k)$ groups of such $k$ parts. There is a recursive formula
\begin{equation}\label{eqa:555555}
A(m,k)=A(m,k-1)+A(m-k,k)
\end{equation}
with $0 \leq k\leq m$. It is not easy to compute the exact value of $A(m,k)$. As we cut $m$ into a group of $k$ parts $m_1,m_2,\cdots ,m_k$, let $m_{i_1},m_{i_2},\cdots ,m_{i_k}$ be a permutation from a group of $k$ parts $m_1,m_2,\cdots ,m_k$, so the number of such permutations is a factorial $k!$, finally, we have $k!$ possible degree-sequences $(m_{i_1},m_{i_2},\cdots ,m_{i_k})$. Thereby, there are $k!\cdot P(m,k)$ degree-sequences of form $(m_{i_1},m_{i_2},\cdots ,m_{i_k})$ in total. \paralled
\end{rem}

\begin{problem}\label{problem:decompose-evaluated-topcode-matrix-submatrices}
\cite{Bing-Yao-2020arXiv} \textbf{Decompose} an evaluated Topcode-matrix $T_{code}$ defined in Definition \ref{defn:topcode-matrix-definition} into submatrices $T^1_{code}$, $T^2_{code}$, $\dots $, $T^m_{code}$, such that $T_{code}=\uplus |^m_{k=1}T^k_{code}$, and each submatrices $T^i_{code}$ is just a Topcode-matrix of a connected graph $H_i$ admitting a $W_i$-type coloring $h_i$ for $i\in [1,m]$.
\end{problem}

\begin{problem}\label{problem:decompose-topcode-matrix-submatrices}
$^*$ \textbf{Decompose} a Topcode-matrix $T_{code}(G)$ of a graph $G$ defined in Definition \ref{defn:topcode-matrix-definition} into submatrices $T^1_{code}$, $T^2_{code}$, $\dots $, $T^n_{code}$, such that $T_{code}(G)=\uplus |^n_{k=1}T^k_{code}$, and each submatrices $T^k_{code}$ is just a Topcode-matrix $T^k_{code}(G_k)$ of a particular graph $G_k$ for $k\in [1,n]$ holding a graph property true.
\end{problem}

\begin{rem}\label{rem:333333}
For understanding Problem \ref{problem:decompose-topcode-matrix-submatrices}, we show an example: A connected graph $H$ of $p$ vertices contains two edge-disjoint Hamilton cycles $C_p$ and $C\,'_p$, we have an edge subset $E^*=E(H)\setminus [E(C_p)\cup E(C\,'_p)]$, so $H^*=C_p [\odot] C\,'_p$ by the \emph{non-multiple-edge full-v-coinciding operation} ``$[\odot]$'' based on the \emph{graph anti-homomorphism} defined in Definition \ref{defn:new-graph-anti-homomorphisms}, and $H=H^*+E^*$. Moreover,
$$T_{code}(H)=T_{code}(H^*)\uplus T_{code}(E^*)=T_{code}(C_p)\uplus T_{code}(C\,'_p)\uplus T_{code}(E^*).$$
This is an application of Topcode-matrices in graph decomposition. It is meaningful to develop algebraic technique or other operations for Topcode-matrices, such as coloring Topcode-matrices directly instead of colored graphs, building connections between matrices and the existing matrices equipped with rich algebraic technique, and so on.\paralled
\end{rem}

\begin{problem}\label{problem:multiple-tree-partition}
\textbf{Find} conditions for a connected graph $G$ to be a multiple-tree matching partition $G=\oplus_F\langle T_i\rangle ^m_1$ with $m\geq 2$ (refer to Definition \ref{defn:multiple-trees-labelings} and \cite{Yao-Sun-Zhang-Mu-Sun-Wang-Su-Zhang-Yang-Yang-2018arXiv}). \textbf{Does} each lobster admit a multiple edge-meaning vertex labeling?
\end{problem}

\begin{problem}\label{problem:string-vs-NBSPAEVEVMP}
$^*$ \textbf{Number-Based String Partition Into Adjacent E-Value And Ve-Value Matrices Problem (NBSPAEVEVMP).} Partition a number-based string $s(n)=c_1c_2\cdots c_n$ with $c_j\in [0,9]$ into $m^2$ segments $a_1,a_2,\dots, a_{m^2}$ with $a_1=c_1c_2\cdots c_{s_1}$, $a_2=c_{s_1+1}c_{s_1+2}\cdots c_{s_2}$, $\dots$, $a_k=c_{s_{k-1}+1}c_{s_{k-1}+2}\cdots c_{s_k}$, $\dots$, $a_{m^2}=c_{s_{m^2-1}+1}c_{s_{m^2-1}+2}\cdots c_{n}$, such that $a_1,a_2,\dots, a_{m^2}$ form just one of an adjacent e-value matrix, or an adjacent ve-value matrix for some graph $G$. See Definition \ref{defn:ve-value-matrix} and Definition \ref{defn:e-value-matrix} for the definitions of adjacent e-value matrix and adjacent ve-value matrix. Here, we say that the number-based string $s(n)$ is \emph{block-cut}.
\end{problem}

\begin{problem}\label{problem:string-vs-no-order-partition}
$^*$ \textbf{Number-Based String No-order Decomposition Problem (NBSNODP).} Decomposing a number-based string $s(n)=c_1c_2\cdots c_n$ with $c_j\in [0,9]$ (as a \emph{public-key}) produces smaller number-based strings $b_1,b_2,\dots, b_{k}$ such that

(1) $b_i=c_{i,1}c_{i,2}\cdots c_{i,m_i}$ with $i\in [1,k]$;

(2) if a number $c_j$ of $s(n)$ appears in $b_j$ for some $j$, then $c_j$ does not appear in any $b_i$ for $i\neq j$.

We want that these number-based strings $b_1,b_2,\dots, b_{k}$ can construct just one of a Topcode-matrix $T_{code}(G)$, a colored degree-sequence matrix $D_{sc}(\textbf{\textrm{d}})$, a degree-sequence $\textrm{deg}(G)$, an adjacent e-value matrix $E_{color}(G)$, and an adjacent ve-value matrix $VE_{color}(G)$ for some colored graph $G$ (as a \emph{private-key}). Read Definition \ref{defn:topcode-matrix-definition}, Definition \ref{defn:colored-degree-sequence-matrix}, Definition \ref{defn:ve-value-matrix} and Definition \ref{defn:e-value-matrix} for the definitions of Topcode-matrix, adjacent e-value matrix and adjacent ve-value matrix.
\end{problem}

\subsection{Sequence-type of colorings and labelings}

\begin{problem}\label{problem:magic-graceful-sequence}
\cite{Yao-Mu-Sun-Zhang-Wang-Su-2018} \textbf{What conditions} do two sets $A_M=\{a_1,a_2,\dots ,a_M\}$ and $B_q=\{b_1,b_2, \dots, b_q\}$ satisfy such that there exists a $(p,q)$-graph $G$ admitting one of full sequence labeling, graceful sequence labeling, edge-magic sequence labeling, $F$-magic graceful sequence graceful labeling and $M$-magic graceful sequence labeling? (refer to Definition \ref{defn:sequence-labelingss})
\end{problem}

\begin{problem}\label{problem:2}
\cite{Yao-Mu-Sun-Zhang-Wang-Su-2018} Given an integer set $N_m=\{a_1,a_2, \dots, a_m\}$ with $a_i\in Z^0$, does there exist a Topsnut-gpw $H$ admitting a $W$-type labeling (resp. coloring) $f$ such that one of $f(V(H))=N_m$, $f(E(H))=N_m$ and $f(V(H)\cup E(H))=N_m$ holds true?
\end{problem}

\begin{rem}\label{rem:ABC-conjecture}
As known, a set-ordered gracefully total coloring of a tree is equivalent to many $W$-type total colorings of the tree, so this tree admits some $W$-type abstract-sequence colorings defined in Definition \ref{defn:colorings-abstract-sequences}. \paralled
\end{rem}

\begin{problem}\label{problem:1}
\cite{Yao-Su-Wang-Hui-Sun-ITAIC2020} \textbf{What conditions} do two sequences $A_M=(a_1$, $a_2$, $ \dots , a_M)$ and $B_q=(b_1, b_2, \dots, b_q)$ satisfy such that there exists a $(p, q)$-graph $G$ admitting a graceful sequence total coloring? or a $W$-type sequence coloring defined on $A_M$ and $B_q$ in Definition \ref{defn:sequence-coloring}?
\end{problem}

\begin{problem}\label{problem:2}
\cite{Yao-Su-Wang-Hui-Sun-ITAIC2020} Given a group $N_m$ of non-negative integers $a_1$, $ a_2$, $ \dots, a_m$ with $a_i<a_{i+1}$ for $i\in [1,m-1]$, does there exist a graph $H$ admitting a sequence total coloring $f$ defined in Definition \ref{defn:sequence-coloring} such that one of $f(V(H))=N_m$, $f(E(H))=N_m$ and $f(V(H)\cup E(H))=N_m$ holds true?
\end{problem}

\begin{problem} \label{prob:new-EMGTLs}
For every $(p,q)$-graph $G$, there exists a bijection $f: V(G)\cup E(G)\rightarrow [1,p+q]$ and a group of integers $0\leq k_1<k_2<\cdots < k_m$ such that each edge $uv\in E(G)$ holds $|f(u)+f(v)-f(uv)|=k_i$ for some $i$, and each $k_j$ corresponds at least an edge $x_jy_j$ holding $|f(x_j)+f(y_j)-f(x_jy_j)|=k_j$. We call $f$ a \emph{$(k_i)^m_1$-edge magic graceful totally labeling} (or $(k_i)^m_1$-EMGTL), \textbf{determine} the following parameters:

(1) $E_{mgtl}(G) = \min\{m : \textrm{each $(k_i)^m_1$-EMGTLs of }~G\}$;

(2) $K_{mgtl}(G) = \min\{k_{m_0}: \textrm{each $(k_i)^{m_0}_1$-EMGTLs of}~ G, m_0=E_{mgtl}(G)\}$.
\end{problem}

\begin{problem}\label{problem:new-set-colorings}
Let $S_{split}(G)$ be the set of all trees of $q+1$ vertices obtained by doing vertex-splitting operations to a simple and connected $(p,q)$-graph $G$, so each tree $H$ of $S_{split}(G)$ is graph homomorphism to $G$, that is, $H\rightarrow G$.

(1) \textbf{Estimate} the value of $|S_{split}(G)|$.

(2) By the colorings defined in Definition \ref{defn:new-set-colorings}, \textbf{compute} parameter
$$\min(W,G)=\min_f\{ |\{f(x):|f(x)|\geq 2,~x\in V(G)\}|:~\textrm{$f$ is a v-set e-proper $W$-type coloring of $G$}\}
$$ obtained from the $W$-type colorings of trees in $S_{split}(G)$. If $\min(W,G)=0$, then $G$ admits a $W$-type coloring, for example, $G$ is a tree admitting a graceful labeling, or $G$ is a connected bipartite graph admitting an odd-graceful labeling. And, the case of $\min(W,G)\neq 0$ means that $G$ does not admit any $W$-type coloring.

(3) Suppose that $G$ is a maximal planar graph, a tree $T$ obtained from by vertex-splitting the vertices of $G$ admits a $W$-type vertex $k$-coloring for $k\geq 2$. So, $T$ is graph homomorphism to $G$, which shows that $G$ admits a $W$-type $k$-color v-set-coloring defined in Definition \ref{defn:new-set-colorings}. If $\min(\textrm{$k$-color},G)=0$ for some $k=2,3,4$, we get a proper $k$-coloring of $G$.
\end{problem}

\begin{problem}\label{problem:xxxxxx}
Let $V_{ec}(n)$ be the set of $n$-dimension vectors $\textbf{\textrm{d}}_i=(a_{i,1}, a_{i,2}, \dots, a_{i,n})$, and let an \emph{integer-vector set} $V^{I}_{ec}(n)$ be the set of $n$-dimension vectors with each component to be an integer, as well as an \emph{integer$^+$-vector set} $V^{+I}_{ec}(n)$ be the set of $n$-dimension vectors with each component to be a non-negative integer. Suppose that $I^{+}(n)$ is a finite subset of an integer$^+$-vector set $V^{+I}_{ec}(n)$. By Definition \ref{defn:vector-colorings-graphs}, Definition \ref{defn:various-vector-colorings}, Definition \ref{defn:ve-vector-vector-colorings-v-vector-e-number}, Definition \ref{defn:graph-full-proper-integer-vector-set} and Definition \ref{defn:graph-graphic-expression-55}, there are the following problems:
\begin{asparaenum}[\textrm{Topexp}-1. ]
\item \textbf{Determine} graceful v-vector e-number colorings (resp. odd-graceful v-vector e-number colorings) of a $(p,q)$-graph $G$.
\item \textbf{Determine} arithmetic v-vector e-number colorings of a $(p,q)$-graph $G$.
\item \textbf{Find} trees $G$ admitting proper-vector labelings $F:V(G)\cup E(G)\rightarrow I^*(n_0)$ for an integer $n_0>0$ as big as possible, where $I^*(n_0)$ is a finite subset of an integer$^+$-vector set $V^{+I}_{ec}(n_0)$.
\item \textbf{Character} an integer$^+$-vector set $I^{+}(n)$ if it has a \emph{graphic expression}.
\item \textbf{Determine} the smallest \emph{graph-graphic expressions} $S_{G}(I^{+}(n))$ for a given graph-full integer$^+$-vector set $I^{+}(n)$.
\end{asparaenum}
\end{problem}

\begin{problem}\label{problem:xxxxxx}
\cite{Yao-Zhang-Sun-Mu-Sun-Wang-Wang-Ma-Su-Yang-Yang-Zhang-2018arXiv} For a subset sequence $\{S_i\}^q_1$ with $S_i\in [1,q+1]^2$, here it is allowed $S_i=S_j$ for some $i\neq j$. \textbf{Find} a vertex labeling $f:V(G)\rightarrow [1,q+1]^2$ for a $(p,q)$-graph $G$, and induced edge color set $f(u_iv_i)=f(u_i)\cap f(v_i)=S_i$, such that:

\quad (1) Consecutive sets $S_i=[a_i,b_i]$, $a_{i+1}=a_i+1$, $b_{i+1}=b_i+1$.

\quad (2) \emph{Fibonacci sequences} $|S_1|=1$, $|S_2|=1$, and $|S_{i+1}|=|S_{i-1}|+|S_{i}|$ for $i\in [2,q-1]$.

\quad (3) Generalized rainbow sequence $S_i=[a,b_i]$ with $b_i<b_{i+1}$, the \emph{regular rainbow sequence} $S_i=[1,i]$ with $i\in [2,q]$.

\quad (4) $|S_i|=i$, where $S_i=\{a_{i,1}, a_{i,2}, \dots ,a_{i,i}\}$.

\quad (5) $f:V(G)\rightarrow [1,q+1]^2$, $f(uv)=f(u)\cup f(v)$, and $f(V(G))\cup f(E(G))=[1,N]^2$ with $N\leq q+1$.
\end{problem}

\begin{problem}\label{problem:xxxxxx}
As known, Topsnut-gpw sequences $\{G_{(k_i,d_i)}\}^m_1$ can be used to encrypt graphs/networks \cite{Yao-Zhang-Sun-Mu-Sun-Wang-Wang-Ma-Su-Yang-Yang-Zhang-2018arXiv}, where $\{G_{(k_i,d_i)}\}^m_1$ is defined in Remark \ref{rem:3-parameter-labelings}. \textbf{Determine} what graphs/networks can be encrypted by what Topsnut-gpw sequences $\{G_{(k_i,d_i)}\}^m_1$.
\end{problem}

\subsection{Miscellaneous problems}

\begin{problem} \label{problem:distinguishing-problem1}
According to the notation in (\ref{eqa:66-first-group}), \textbf{what} connections between $C_{ve}\{w,\lambda\}$ and $C_{ve}\{w,\theta\}$ with $\lambda,\theta\in F_c(G)$ are there, where $F_c(G)$ is the set of all distinguishing-type colorings of a graph $G$?
\end{problem}
\begin{problem} \label{problem:distinguishing-problem2}
\textbf{Determine} each \emph{$V_d$-type chromatic number} (\emph{index}) $\chi^{\varepsilon}_{vd}(G)$ defined in Definition \ref{defn:combinator-distinguishing-colorings}.
\end{problem}
\begin{problem} \label{problem:distinguishing-problem3}
According to the notation in (\ref{eqa:66-first-group}), \textbf{determine} a graph $H$ holds $C_v[u,h]=C_v[v,h]$ for each edge $uv\in E(H)$, where $h$ is a proper coloring $f$ of $H$; or $H$ admits a proper edge coloring $g$ holding $C_e(u,g)=C_e(v,g)$ for each edge $uv\in E(H)$; or $H$ admits a proper total coloring $\varphi$ holding $C_{ve}(u,\varphi)=C_{ve}(v,\varphi)$ for each edge $uv\in E(H)$.
\end{problem}

\begin{problem} \label{problem:distinguishing-problem5}
A graph $G$ admits an $RC(m)$-coloring if a $W$-type coloring of $G$ holds a group of restrict conditions $RC_1,RC_2,\dots ,RC_m$ true, write $RC(m)=\{RC_1,RC_2,\dots ,RC_m\}$. In general, the graph $G$ may admit $R_iC(m)$-colorings with $R_iC(m)=\{R_iC_1,R_iC_2,\dots ,R_iC_m\}$ for $i\in [1,n]$. Analyze the complex of $R_iC(m)$-colorings of $G$ with $i\in [1,n]$ for serving quantum cryptography.
\end{problem}

\subsubsection{Degree sequences}

\begin{problem}\label{problem:ds-sequences-1}
By Definition \ref{defn:Ds-homomorphism-sequence-graph-set}, \textbf{determine} a maximal positive integer $M(\textbf{\textrm{d}})$ for a given degree-sequence $\textbf{\textrm{d}}$, and \textbf{find}:

(1) All of degree-sequence homomorphic chains $H_{omo}\{\textbf{\textrm{d}}_{j,k}\}^{m_j}_{k=1}$ for $j\in [1,M(\textbf{\textrm{d}})]$.

(2) All graph-set homomorphisms $G_{raph}(\textbf{\textrm{d}}_{j,k})\rightarrow G_{raph}(\textbf{\textrm{d}}_{j,k+1})$ for $\textbf{\textrm{d}}_{j,i}\in H_{omo}\{\textbf{\textrm{d}}_{j,k}\}^{m_j}_{k=1}$ and $j\in [1,M(\textbf{\textrm{d}})]$.
\end{problem}

\begin{problem}\label{problem:ds-sequences-2}
\textbf{Write} a degree-sequence $\textbf{\textrm{d}}$ as $\textbf{\textrm{d}}=\odot\langle \textbf{\textrm{d}}_1,\textbf{\textrm{d}}_2\rangle$, or $\textbf{\textrm{d}}=\ominus\langle \textbf{\textrm{d}}_1,\textbf{\textrm{d}}_2\rangle$ defined in Definition \ref{defn:degree-coinciding-operation}, where both $\textbf{\textrm{d}}_1$ and $\textbf{\textrm{d}}_2$ are degree-sequences.
\end{problem}

\begin{problem}\label{problem:ds-sequences-3}
Suppose that two degree-sequences $\textrm{deg}(G)=\textrm{deg}(H)$ and $G\not \cong H$ for two graphs $G$ and $H$. \textbf{Are} there two graphs $G^*$ and $H^*$ with $G^*\cong H^*$, such that two graph homomorphisms $G\rightarrow G^*$ and $H\rightarrow H^*$ hold true?
\end{problem}

\begin{problem}\label{problem:accompany-graphic-lattice}
By Definition \ref{defn:99-complex-degree-sequence-lattice}, \textbf{characterize} a complex degree-sequence accompany graphic lattice $\textbf{\textrm{C}}_{omp}(\textbf{\textrm{L}}(\Sigma \textbf{\textrm{C}}^*))$ of $\textbf{\textrm{L}}(\Sigma \textbf{\textrm{C}}^*)$.
\end{problem}

\begin{problem}\label{problem:unique-graphic}
By Definition \ref{defn:99-complex-degree-sequence-lattice}, a complex degree-sequence $\textbf{\textrm{d}}$ is \emph{unique graphic} if any two graphs $L$ and $T$ having complex degree-sequences $\textrm{Cdeg}(L)=\textbf{\textrm{d}}=\textrm{Cdeg}(T)$ hold $L\cong T$ true. \textbf{Find} conditions for unique graphic complex degree-sequences.
\end{problem}

\begin{problem}\label{problem:xxxxxxxxx}
By Definition \ref{defn:99-complex-degree-sequence-lattice}, for a given complex degree-sequence $\textbf{\textrm{y}}^*=\sum^m_{k=1} t_kC^k_n\in \textbf{\textrm{L}}(\Sigma \textbf{\textrm{C}}^*)$ defined in (\ref{eqa:complex-degree-sequence-lattice}) and $\sum^m_{k=1} |t_k|\neq 1$, find another complex degree-sequence $\textbf{\textrm{w}}^*=\sum^m_{k=1} r_kC^k_n\in \textbf{\textrm{L}}(\Sigma \textbf{\textrm{C}}^*)$, such that the Euclidean norm
\begin{equation}\label{eqa:555555}
\| \textbf{\textrm{y}}^*- \textbf{\textrm{w}}^*\|=\sqrt{\sum^m_{k=1}\big (t_k-r_k\big )^2\big \| C^k_n\big \|^2}
\end{equation} is smallest, called the closest complex degree-sequence problem (CCDSP) of complex degree-sequence lattice. We have the shortest complex degree-sequence problem (SCDSP) of complex degree-sequence lattice: \textbf{Find} a complex degree-sequence $\textbf{\textrm{z}}^*=\sum^m_{k=1} s_kC^k_n\in \textbf{\textrm{L}}(\Sigma \textbf{\textrm{C}}^*)$ and $\sum^m_{k=1} |s_k|$ $\neq 1$, such that the Euclidean norm $\| \textbf{\textrm{z}}^*\|$ is a smallest one in $\textbf{\textrm{L}}(\Sigma \textbf{\textrm{C}}^*)$.
\end{problem}

\begin{problem}\label{qeu:xxxxxx}
\cite{Yao-Wang-Ma-Wang-Degree-sequences-2021} By Definition \ref{defn:more-definitions-on-degree-sequence}, \textbf{Characterize and construct:} right-angled degree-sequences, incompressible degree-sequences, equipotential degree-sequences, perfect degree-sequences, right-angled degree-sequence bases.
\end{problem}

\begin{problem}\label{qeu:topological-authentication-11}
\cite{Yao-Wang-Ma-Wang-Degree-sequences-2021} \textbf{Topological Authentication Problem-I (TAP-I)}~For a given number-based string $s(n)=c_1c_2\cdots c_n$ with $c_i\in [0,9]$, we cut $s(n)$ into $p$ segments $s(n)=a_1a_2\cdots a_p$, where $a_1=c_1c_2\cdots c_{b_1}$, $a_2=c_{1+b_1}c_{2+b_1}\cdots c_{b_2}$, $\dots$, $a_{k+1}=c_{1+b_k}$ $c_{2+b_k}\cdots c_{b_k}$, $\dots$, $a_{p}=c_{1+b_{p-1}}c_{2+b_{p-1}}\cdots c_n$. What condition does this string-based sequence $\textbf{\textrm{A}}=(a_1,a_2,\dots ,a_p)$ satisfy such that it is just a degree-sequence? Thereby, a graph $H$ has its own degree-sequence $\textrm{deg}(H)=\textbf{\textrm{A}}$. Here, the given number-based string $s(n)$ is a \emph{public key}, and the graph $H$ is just a \emph{private key}, so they form a \emph{topological authentication}.
\end{problem}

\begin{rem}\label{rem:xxxxxxxxxxxx}
It is not easy to answer Problem \ref{qeu:topological-authentication-11}, since (i) Judge a number-based string to be made by a degree-sequence or by a Topcode-matrix; (ii) a degree-sequence may correspond two or more no-isomorphic graphs; (iii) a Topcode-matrix may correspond two or more no-isomorphic graphs; (iv) no polynomial algorithm for cutting a number-based string into a degree-sequence, or $(3q)$ number-based strings for reconstructing a Topcode-matrix. The complexity of decomposing a number-based string into a degree-sequence in Problem \ref{qeu:topological-authentication-11} is NP-hard.\paralled
\end{rem}

\begin{problem}\label{qeu:topological-authentication-22}
\cite{Yao-Wang-Ma-Wang-Degree-sequences-2021} \textbf{Topological Authentication Problem-II (TAP-II).}~For a number-based string $s(n)=c_1c_2\cdots c_n$ with $c_i\in [0,9]$, \textbf{how} to rearrange the numbers in $s(n)$ into $p$ number-based strings $b_1,b_2,\dots ,b_p$, where $b_i=c\,'_{i,1}c\,'_{i,2}\cdots c\,'_{i,m_i}$ with $c\,'_{i,j}\in \{c_1,c_2$, $\dots ,c_m\}$ and $m=\sum^p_{i=1}m_i$, such that the sequence $\textbf{\textrm{d}}(s(n))=(b_1,b_2,\dots ,b_p)$ to be a degree-sequence just?
\end{problem}

\begin{problem}\label{qeu:topological-authentication-33}
\cite{Yao-Wang-Ma-Wang-Degree-sequences-2021} \textbf{Topological Authentication Problem-III (TAP-III).}~For a given number-based string $s(n)$, find another number-based string $s^*$ and two graphs $H$ and $H^*$ such that the degree-sequence $\textrm{deg}(H)$ is made by all numbers of $s(n)$ and the degree-sequence $\textrm{deg}(H^*)$ is made by all numbers of $s^*$, as well as $H$ is graph homomorphism to $H^*$, namely, $H\rightarrow H^*$. Here, the given number-based string $s(n)$ is a \emph{public key}, and the founded number-based string $s^*$ is a \emph{private key}, and the graph homomorphism $H\rightarrow H^*$ is just a \emph{topological authentication}.
\end{problem}

\begin{rem}\label{rem:topological-authentication-33}
In Problem \ref{qeu:topological-authentication-33}, we can replace the graph homomorphism $H\rightarrow H^*$ by the following topological authentications:

(i) $H$ admits a $W$-coloring $f$, and $H^*$ admits a $W$-coloring $g$, and $(f,g)$ is a $W$-coloring matching, $H$ is colored graph homomorphism to $H^*$;

(ii) $H$ and $H^*$ admit a graphical similarity; and

(iii) $H$ admits an odd-graceful labeling $f$, and $H^*$ admits a labeling $g$, so $(f,g)$ is a twin odd-graceful labeling matching.\paralled
\end{rem}

\subsubsection{Set, lines}

\begin{problem}\label{problem:xxxxxx}
\cite{Yao-Zhang-Sun-Mu-Sun-Wang-Wang-Ma-Su-Yang-Yang-Zhang-2018arXiv} \textbf{Fold-lines covering all points on lattices.} Let $P_3\times P_q$ be a lattice in $xOy$-plane. There are points $(i,j)$ on the lattice $P_3\times P_q$ with $i\in[1,3]$ and $j\in [1,q]$. If a continuous fold-line $L$ with initial point $(a,b)$ and terminal point $(c,d)$ on $P_3\times P_q$ is internally disjoint and contains all points $(i,j)$ of $P_3\times P_q$, we call $L$ a \emph{total TB-paw line}. \textbf{Find} all possible total TB-paw lines.

In general, let $\{L_i\}^m_1=\{L_1,L_2,\dots, L_m\}$ be a set of $m$ continuous disjoint fold-lines on $P_3\times P_q$, where each $L_i$ has own initial point $(a_i,b_i)$ and terminal point $(c_i,d_i)$. If $\{L_i\}^m_1$ contains all points $(i,j)$ of $P_3\times P_q$, we call $\{L_i\}^m_1$ a \emph{group of TB-paw lines}, here $(a_i,b_i)\neq (c_i,d_i)$ for each $L_i$. \textbf{Find} all possible groups $\{L_i\}^m_1$ of TB-paw lines for $m\in [1,3q]$.
\end{problem}

\begin{problem}\label{problem:xxxxxx}
Given a matrix $A_{3\times q}$ with integer elements, by \textbf{what} condition $A_{3\times q}=(a_{ij})_{3\times q}$ is a Topsnut-matrix $T_{code}(G)$ of some colored graph $G$ in \cite{Yao-Zhang-Sun-Mu-Sun-Wang-Wang-Ma-Su-Yang-Yang-Zhang-2018arXiv}? (refer to Definition \ref{defn:topcode-matrix-definition})
\end{problem}

\begin{problem}\label{problem:xxxxxx}
\textbf{How} to construct a matrix $A_{3\times q}=(a_{ij})_{3\times q}$ by a given number-based string such that $A_{3\times q}$ is just a Topsnut-matrix of some Topsnut-gpw? (refer to Definition \ref{defn:topcode-matrix-definition})
\end{problem}

\begin{problem}\label{problem:xxxxxx}
\textbf{Total colorings with consecutive color sets.} Let $f: V(G)\cup E(G)\rightarrow [1,\chi\,''(G)]$ be a proper total coloring of a graph $G$, and let the set $f^*(E(G))=\{f(u)+f(uv)+f(v):uv \in E(G)\}$. \textbf{Characterize} $G$ if $f^*(E(G))=[m,n]$.
\end{problem}

\subsection{Graphic colorings, graphic labelings}

\begin{problem}\label{problem:xxxxxx}
\textbf{Find} \emph{graph-labelings} of graphs \cite{Yao-Sun-Zhang-Mu-Sun-Wang-Su-Zhang-Yang-Yang-2018arXiv}. By Definition \ref{defn:graph-labeling-general}, let $H_{ag}$ be a set of graphs. A $(p,q)$-graph $G$ admits a graph-labeling $F:V(G)\cup E(G)\rightarrow H_{ag}$, and each edge color $F(uv)=F(u)(\bullet)F(v)$ is just a graph having a \emph{$k_{uv}$-matching} based on an operation $(\bullet)$ on graphs. Here, a \emph{$k_{uv}$-matching} may be one of a perfect matching of $k_{uv}$ vertices, $k_{uv}$-cycle, $k_{uv}$-connected, $k_{uv}$-edge-connected, $k_{uv}$-colorable, $k_{uv}$-edge-colorable, total $k_{uv}$-colorable, $k_{uv}$-regular, $k_{uv}$-girth, $k_{uv}$-maximum degree, $k_{uv}$-clique, $\{a,b\}_{uv}$-factor, v-split $k_{uv}$-connected, e-split $k_{uv}$-connected, a twin odd-graceful matching $\odot \langle F(u), F(v)\rangle$, and so on. The graph $\langle G(\bullet) H_{ag}\rangle$ obtained by joining $F(u)$ with $F(uv)$ and joining $F(uv)$ with $F(v)$ for each edge $uv\in E(G)$ under the operation $(\bullet)$ is called a \emph{$k_{uv}$-matching graph}.
\end{problem}

\begin{problem}\label{problem:xxxxxx}
\textbf{Find} \emph{pan-matching graphs} \cite{Yao-Sun-Zhang-Mu-Sun-Wang-Su-Zhang-Yang-Yang-2018arXiv}. By Definition \ref{defn:pan-matching-graphs}, let $P_{ag}$ be a set of graphs. A $(p,q)$-graph $G$ admits a graph-labeling $F:V(G)\rightarrow H_{ag}$, and induced edge color $F(uv)=F(u)(\bullet)F(v)$ is just a graph having a pan-matching, where $(\bullet)$ is an operation on graphs. Here, a \emph{pan-matching} may be one of a perfect matching of $k_{uv}$ vertices, $k_{uv}$-cycle, $k_{uv}$-connected, $k_{uv}$-edge-connected, $k_{uv}$-colorable, $k_{uv}$-edge-colorable, total $k_{uv}$-colorable, $k_{uv}$-regular, $k_{uv}$-girth, $k_{uv}$-maximum degree, $k_{uv}$-clique, $\{a,b\}_{uv}$-factor, vertex-split $k_{uv}$-connected, edge-split $k_{uv}$-connected. We call the graph $\langle G(\bullet) P_{ag}\rangle$ obtained by joining $F(u)$ with $F(uv)$ and joining $F(uv)$ with $F(v)$ for each edge $uv\in E(G)$ a \emph{pan-matching graph}. If each $H_i$ of a \emph{graph set} $P_{ag}$ of graphs admits a labeling $f_i$, we can color a $(p,q)$-graph $G$ in the way: $F:V(G)\rightarrow P_{ag}$, such that each edge $u_iv_j\in E(G)$ is colored by $F(u_iv_j)=F(u_i)\bullet F(v_j)=H_i\bullet H_j=\alpha_{i,j}(f_i,f_j)$, where $\alpha_{i,j}(f_i,f_j)$ is a \emph{reversible function} with $f_j=\alpha_{i,j}(f_i)$, and $f_i=\alpha^{-1}_{i,j}(f_j)$.
\end{problem}

\begin{problem}\label{problem:xxxxxxxxx}
\textbf{Are} there infinite kinds of colorings on graphs?
\end{problem}

\begin{problem}\label{qeu:isomorphic-mathematical-problems}
There are the following mathematical problems in the topological authentication:
\begin{asparaenum}[\quad \textbf{\textrm{Iso}}-1.]
\item \textbf{Find} a multivariate function $\theta$ of vertex colors and edges colors of each particular subgraph $H$ of a graph $G$ such that $G$ admits a proper total coloring $h$ holding $\theta(h(V(H)\cup E(H)))=$ a constant for each particular subgraph $H$, where

\quad 1-1. $H$ may be an edge, or a face $f_i$ having bound $B(f_i)$, or a cycle $C_n$, or a path $P_n$, and so on.

\quad 1-2. Find more multivariate functions $\theta_i$ of vertex colors and edge colors such that each $\theta_i$ is equal to a constant under a proper total coloring of $G$.

\item Given a set $V_{co}$ of colored vertices and a set $E_{co}$ of colored edges, \textbf{how} to assemble some elements of two sets into a graph $G$ such that $G$ is just colored by a $W$-type total coloring $f$ holding $f(V(G)\cup E(G))\subseteq V_{co}\cup E_{co}$.
\item \textbf{$J$-graphic isomorphic problem.} Let $G$ and $H$ be two graphs of $p$ vertices with $h:V(G)\rightarrow V(H)$, and let $J$ be a particular graph. If $G-V(J)\cong H-h(V(J))$ for each particular graph $J$ of $G$ and $H$, can we claim $G\cong H$? Here, $J$ may be a path of $p$ vertices, or a cycle of $p$ vertices, or a complete graph of $p$ vertices, \emph{etc.} Recall, let $G$ and $H$ be two graphs that have the same number of vertices. If there exists a bijection $\varphi: V (G)\rightarrow V (H)$ such that $uv \in E(G)$ if and only if $\varphi(u)\varphi(v) \in E(H)$, then we say both graphs $G$ and $H$ to be isomorphic to each other, denoted by $G\cong H$ in \cite{Bondy-2008}. A long-standing Kelly-Ulam's Reconstruction Conjecture proposed in 1942: Let both $G$ and $H$ be graphs with $n$ vertices. If there is a bijection $f: V (G)\rightarrow V (H)$ such that $G-u\cong H-f(u)$ for each vertex $u \in V (G)$, then $G\cong H$. This conjecture supports some cryptosystems consisted of graphic isomorphism to be ``Resisting classical computers and quantum computers''.
\item A \emph{topological coloring isomorphism} consists of graph isomorphism and coloring isomorphism. For a colored graph $G$ admitting a $W$-type total coloring $f$ and another colored graph $H$ admitting a $W$-type total coloring $g$, if there is a mapping $\varphi$ such that $w\,'=\varphi(w)$ for each element $w\in V(G)\cup E(G)$ and each element $w\,'\in V(H)\cup E(H)$, then we say they are \emph{isomorphic} to each other, and write this case by $G\cong H$, and moreover if $g(w\,')=f(w)$ for $w\,'=\varphi(w)$, we say they are subject to \emph{coloring isomorphic} to each other, so we denoted $G=H$ for expressing the combination of \emph{topological isomorphism} and \emph{coloring isomorphism}.
\end{asparaenum}
\end{problem}

\begin{problem}\label{qeu:xxxxxxxxxx}
 \cite{YAO-SUN-WANG-SU-XU2018arXiv} Let ``\emph{$W$-type labeling}'' be one of the existing graph labelings, and let a connected graph $G$ admit a $W$-type labeling. If every connected proper subgraph of $G$ also admits a labeling to be a $W$-type labeling, then we call $G$ a \emph{perfect $W$-type labeling graph}. We ask for: \emph{If every connected proper subgraph of a connected graph $G$ admits a $W$-type labeling, then \textbf{does} $G$ admit this $W$-type labeling too}?
\end{problem}

\subsection{Graphic groups}

\begin{problem}\label{problem:xxxxxx}
An \emph{every-zero total graphic group} is defined in Definition \ref{defn:every-zero-total-graphic-group}. \textbf{Find every-zero total graphic groups} for a $(p,q)$-graph admitting magic-type total labelings.
\end{problem}

\begin{problem}\label{problem:graphic-groups-11}
By Definition \ref{defn:graph-graceful-group-labeling}, \textbf{does} any lobster $T$ admit an \emph{odd-graceful graphic group-labeling} $\varphi$ by any \emph{preappointed zero} $G_k\in \{F_f(G);\oplus\}$, such that $\varphi(u)\neq \varphi(v)$ for distinct vertices $u,v\in V(T)$, and the edge index set $\{k:~\varphi(xy)=G_k,xy\in E(T)\}=\{1,3,5$, $\dots $, $2|V(T)|-3\}$?
\end{problem}
\begin{problem}\label{problem:graphic-groups-22}
Let $\{F_f(G);\oplus\}$ be an every-zero graphic group. By Definition \ref{defn:graph-graceful-group-labeling}, if a bipartite graph $T$ admits a set-ordered graceful labeling (\cite{Yao-Liu-Yao-2017}, \cite{Zhou-Yao-Chen-Tao2012}), \textbf{does} $T$ admit a \emph{graceful graphic group-labeling} $\theta$ by any \emph{preappointed zero} $G_k\in \{F_f(G);\oplus\}$, such that $\theta(u)\neq \theta(v)$ for distinct vertices $u,v\in V(T)$, the \emph{edge index set} $\{j:~\theta(xy)=G_j,xy\in E(T)\}=[1,|E(T)|]$?
\end{problem}
\begin{problem}\label{problem:graphic-groups-33}
By Definition \ref{defn:graph-graceful-group-labeling}, \textbf{find} \emph{$W$-type graphic group-labelings}, such as, $W$-type is one of edge-magic total, elegant, felicitous, and so on.
\end{problem}

\begin{conj} \label{conj:xxxxxxxxxxxxxxx}
Motivated from Graceful Tree Conjecture (Alexander Rosa, 1966), we \textbf{guess}: Every tree admits a graceful graphic group-labeling defined in Definition \ref{defn:graph-graceful-group-labeling}.
\end{conj}

\begin{problem}\label{problem:graphic-groups-55}
By Definition \ref{defn:graph-graceful-group-labeling}, \textbf{find} \emph{$W$-type graphic group colorings} $\varphi$ with $\varphi(x)=\varphi(y)$ for some distinct vertices $x,y$ (resp. $\varphi(uv)=\varphi(wz)$ for some non-adjacent edges $uv,wz$), and the number of these vertex pairs $(x,y)$ (resp. the number of edge pairs $(uv,wz)$~) is as less as possible.
\end{problem}
\begin{problem}\label{problem:graphic-groups-66}
\textbf{Consider} various \emph{graphic group proper total colorings} related with the famous Total Coloring Conjecture (Behzad, 1965; Vadim G. Vizing, 1964).
\end{problem}

\begin{problem}\label{problem:xxxxxx}
\cite{Yao-Zhang-Sun-Mu-Sun-Wang-Wang-Ma-Su-Yang-Yang-Zhang-2018arXiv} An \emph{every-zero graphic group} $\{F_{n}(H,h);\oplus\}$ made by a graph $H$ admitting an $\varepsilon$-labeling $h$ contains its elements $H_i$ holding $H\cong H_i$ and admitting an $\varepsilon$-labeling $h_i$ induced by $h$ with $i\in [1,n]$ and hold an additive operation $H_i\oplus H_j$ defined as
\begin{equation}\label{eqa:graphic-group-definition}
h_i(x)+h_j(x)-h_k(x)=h_{\lambda}(x)
\end{equation}
with $\lambda=i+j-k\,(\bmod\,n)$ for each element $x\in V(H)$ under any \emph{preappointed zero} $H_k$.

For a sequence $\{H_{i_j}\}^q_1$ of an every-zero graphic group $\{F_{n}(H,h);\oplus\}$, \textbf{determine}: (1) an $\{H_{i_j}\}^q_1$-sequence group-labeling of a tree having $q$ edges; (2) an $\{H_{i_j}\}^q_1$-sequence group-labeling or an $\{H_{i_j}\}^q_1$-sequence group-coloring of a connected $(p,q)$-graph. \textbf{Does} the graph $H$ in $\{F_{n}(H,f);\oplus\}$ admit an odd-graceful (an $\{H_{i_j}\}^q_1$-sequence) group-coloring under each self-zero $H_i\in F_{n}(H,f)$ with index set $\{i_1,i_2,i_3,\dots ,i_{q}\}=[1,2q-1]^o$?
\end{problem}

\begin{problem}\label{problem:mixed-graphic-group-algorithm}
A graphic group sequence $\left \{G^{(t)}_{ro}(H)\right \}$ is defined in Definition \ref{defn:graphic-group-sequences}.
\begin{asparaenum}[\textrm{Ggseq}-1. ]
\item \textbf{Characterize} the topological structure of $\left \{G^{(t)}_{ro}(H)\right \}$. \textbf{Is} $G^{(t)}_{ro}(H)$ scale-free? \textbf{Is} $G^{(t)}_{ro}(H)$ self-similar?
\item \textbf{Determine} colorings admitted by each element of $\;\left \{G^{(t)}_{ro}(H)\right \}$.
\item \textbf{Estimate} the cardinality of $\left \{G^{(t)}_{ro}(H)\right \}$.
\item \textbf{Study} \emph{every-zero $\textrm{\textbf{H}}$-graphic group sequence} $\left \{G^{(t)}_{ro}(\textrm{\textbf{H}})\right \}$ for a base $\textrm{\textbf{H}}=(H_1,H_2$, $\dots $, $H_m)$.
\end{asparaenum}
\end{problem}

\begin{problem}\label{problem:xxxxxx}
\cite{Yao-Zhang-Sun-Mu-Sun-Wang-Wang-Ma-Su-Yang-Yang-Zhang-2018arXiv} Suppose that two trees $T,H$ of $p$ vertices are 6C-complementary matching from each other, refer to Definition \ref{defn:6C-complementary-matching}. We use two every-zero graphic groups $F_n(T,f)$ and $F_n(H,g)$ to encrypt a connected $(p,q)$-graph $G$ respectively, and then get two encrypted networks $N_{et}(G,F_n(T,f))$ and $N_{et}(G,F_n(H,g))$. \textbf{Does} $F_n(T,f)$ match with $F_n(H,g)$? Moreover, \textbf{does} $N_{et}(G,F_n(T,f))$ match with $N_{et}(G,F_n(H,g))$?
\end{problem}

\subsection{Graphic lattices}

\begin{problem}\label{problem:topcode-matrix-lattices}
If a Topcode-matrix base $\textbf{\textrm{T}}_{\textrm{code}}=\{T^i_{code}\}^n_1$ is an every-zero graphic group, \textbf{show} properties of the Topcode-matrix lattice $\textbf{\textrm{L}}(\textbf{\textrm{T}}_{\textrm{code}}\uplus F_{p,q})$ defined in Definition \ref{defn:99-Topcode-matrix-lattice}.
\end{problem}

\begin{problem}\label{problem:extrems-in-lattices-00}
\cite{Bing-Yao-2020arXiv} \textbf{Characterize} the connection between the graphic lattice base $\textbf{\textrm{H}}$ and the graph set $F_{p,q}$, in other words, the graphic lattice $\textbf{\textrm{L}}(\textbf{\textrm{H}}(\bullet)F_{p,q})$ defined in Definition \ref{defn:99-colored-graphic-lattice} is not empty as $F_{p,q}$ holds what conditions.
\end{problem}

\begin{problem}\label{problem:extrems-in-lattices-11}
\cite{Bing-Yao-2020arXiv} \textbf{Find} a graph $G^*$ of a graphic lattice $\textbf{\textrm{L}}(\textbf{\textrm{H}}(\bullet)F_{p,q})$ defined in Definition \ref{defn:99-colored-graphic-lattice}, such that $G^*$ has the shortest diameter, or $G^*$ is hamiltonian, or $G^*$ contains a spanning tree with the most leaves in $\overrightarrow{\textbf{\textrm{L}}}(\overrightarrow{\textbf{\textrm{H}}}(\bullet)\overrightarrow{F}_{p,q})$ defined in Definition \ref{defn:99-colored-directed-graphic-lattice}, and so on.
\end{problem}

\begin{problem}\label{qeu:topcode-matrix-lattices}
\cite{Bing-Yao-2020arXiv} About Topcode-matrix lattices, we have the following questions:
\begin{asparaenum}[\textbf{\textrm{Pro}}-1. ]
\item If $\{T^i_{code}\}^n_1$ is an every-zero graphic group, show properties of the Topcode-matrix lattice $\textbf{\textrm{L}}(\textbf{\textrm{T}}_{\textrm{code}}\uplus F_{p,q})$ defined in Definition \ref{defn:99-Topcode-matrix-lattice}.

\item \textbf{Find} some connections between two Topcode-matrix lattices $\textbf{\textrm{L}}(\textbf{\textrm{T}}_{\textrm{code}}\uplus F_{p,q})$ and $\textbf{\textrm{L}}(\textbf{\textrm{T}}^*_{\textrm{code}}\uplus F_{p,q})$ defined in Definition \ref{defn:99-Topcode-matrix-lattice}.
\item \textbf{Define} Topcode-matrix lattices for other graphic lattices.
\end{asparaenum}
\end{problem}

\begin{problem}\label{problem:structures-lattice}
\cite{Bing-Yao-2020arXiv} \textbf{Determine} a graph $G^*$ of a graphic lattice $\textbf{\textrm{L}}(\textbf{\textrm{T}}\odot F_{p,q})$ defined in Definition \ref{defn:99-graphic-lattice-dot-coninciding} such that diameters $D(G^*)\leq D(G)$ for any $G\in \textbf{\textrm{L}}(\textbf{\textrm{T}}\odot F_{p,q})$.
\end{problem}

\begin{problem}\label{problem:structures-lattice-00}
\cite{Bing-Yao-2020arXiv} \textbf{Estimate} the \emph{cardinality} of a graphic lattice $\textbf{\textrm{L}}(\textbf{\textrm{T}}\odot F_{p,q})$ defined in Definition \ref{defn:99-graphic-lattice-dot-coninciding}, however this will be related with the graph isomorphic problem, a NP-hard problem.
\end{problem}

\begin{problem}\label{problem:structures-lattice-11}
\cite{Bing-Yao-2020arXiv} By Definition \ref{defn:99-graphic-lattice-dot-coninciding}, \textbf{do} we have $H\,'\odot |^n_{i=1}T_i\cong H\,''\odot |^n_{i=1}T_i$ if $H\,'\cong H\,''$?
\end{problem}

\begin{problem}\label{problem:structures-lattice-22}
\cite{Bing-Yao-2020arXiv} By Definition \ref{defn:99-graphic-lattice-dot-coninciding}, let $\overline{T}_i$ be the \emph{complementary graph} of each graph $T_i$ of a graphic base $\textbf{\textrm{T}}$, and let $\overline{H}$ be the complement of $H\in F_{p,q}$. \textbf{Is} the graph $\overline{H}\odot |^n_{i=1}\overline{T}_i$ the complementary graph of the graph $H\odot |^n_{i=1}T_i$?
\end{problem}

\begin{problem}\label{problem:GTC}
\cite{Bing-Yao-2020arXiv} The $4$-coloring star-graphic lattice helps us to ask for the following questions:
\begin{asparaenum}[\textrm{GTC}-1. ]
\item \textbf{Find} various sublattices of the general star-graphic lattice $\textrm{\textbf{L}}(\overline{\ominus} \textbf{F}^c_{star\Delta})$ by graph colorings (resp. labelings) defined in Definition \ref{defn:99-general-star-graphic-lattice}.
\item Let $\textrm{\textbf{L}}(\overline{\ominus} \textbf{G}_{race}(T))$ be the set of graceful trees of $\textrm{\textbf{L}}(\overline{\ominus} \textbf{F}^c_{star\Delta})$ defined in Definition \ref{defn:99-general-star-graphic-lattice}. \textbf{Does} $\textrm{\textbf{L}}(\overline{\ominus} \textbf{G}_{race}(T))$ contains every tree with maximum degree $\Delta$?
\item A \emph{gracefully total coloring} $f$ of a tree $T$ is a proper total coloring $f:V(T)\cup E(T)\rightarrow [1,M]$ such that such that $f(x)=f(y)$ for some pair of vertices $x,y\in V(T)$, $f(uv)=|f(u)-f(v)|$ for each edge $uv\in E(T)$, and the edge color set $f(E(T))=[1,|V(T)|-1]$. A star-graphic sublattice $\textrm{\textbf{L}}(\overline{\ominus} \textbf{T}_{grace})\subset \textrm{\textbf{L}}(\overline{\ominus} \textbf{F}^c_{star\Delta})$ is formed by all trees admitting gracefully total colorings in $\textrm{\textbf{L}}(\overline{\ominus} \textbf{F}^c_{star\Delta})$ defined in Definition \ref{defn:99-general-star-graphic-lattice}. \textbf{Does} each tree $H$ with maximum degree $\Delta$ correspond a tree $H\,'\in \textrm{\textbf{L}}(\overline{\ominus} \textbf{T}_{grace})$ such that $H\cong H\,'$?
\item \textbf{Is} every planar graph isomorphic to a group of 4-colored planar graphs of the $4$-coloring star-graphic lattice $\textbf{\textrm{L}}(\overline{\ominus} \textbf{\textrm{I}}_{ce}(PS,M))$ defined in Definition \ref{defn:99-4-color-star-graphic-lattice}?
\item \textbf{Find} connection between a planar graphic lattice $\textbf{\textrm{L}}(\textbf{\textrm{T}}\,^r\bigtriangleup F_{\textrm{inner}\bigtriangleup})$ defined in Definition \ref{defn:99-4-color-planar-graphic-lattice} and a $4$-coloring star-graphic lattice $\textbf{\textrm{L}}(\overline{\ominus} \textbf{\textrm{I}}_{ce}(PS,M))$ defined in Definition \ref{defn:99-4-color-star-graphic-lattice}.
\item \textbf{Tree and planar graph authentication.} As known, each tree $T$ admits a proper $k$-coloring $f$ with $k\geq 2$, we do a vertex-coinciding operation to some vertices $x,y$ with $f(x)=f(y)$, such that $w=x\odot y$ and $f(w)=f(x)=f(y)$, and the resultant graph $T^*$ obtained by doing a series of vertex-coinciding operations to those vertices of $T$ colored with the same color is just a $k$-colorable graph with some particular properties. For instance, $T^*$ is a $k$-colorable Euler's graph holding the \emph{chromatic number} $\chi(T^*)=k$, or a $k$-colorable planar graph with each inner face to be a triangle, or a $k$-colorable Hamilton graph, \emph{etc.} \textbf{Characterize} a $4$-colorable tree $T$ (as a \emph{public-key}) such that $T^*$ (as a \emph{private-key}) is a $4$-colorable planar graph, or a $4$-colorable planar graph with each inner face to be a triangle, that is, $T$ admits a $4$-colorable graph homomorphism to $T^*$, namely, $T\rightarrow T^*$.
\item \textbf{Tree topological authentication.} Each connected graph $G$ corresponds a tree $T$ based on the vertex-splitting operation and the leaf-splitting operation, denoted as $(\wedge, \prec )(G)=T$. Conversely, a tree $T$ can produce a connected graph $G$ by means of the vertex-coinciding operation and the leaf-coinciding operation, or a mixed operation of them, for the convenience of statement, we write this process as $(\odot, \overline{\ominus})(T)=G$. Suppose that a tree $T$ (as a public-key) admits a $W$-type coloring $f_T$ and $G$ (as a private-key) admits a $W$-type coloring $h_G$ too, if $(\odot, \overline{\ominus})(f_T)=h_G$ in the process $(\odot, \overline{\ominus})(T)=G$, we say $(\odot, \overline{\ominus})(T)=G$ to be a \emph{topological authentication}. Oppositely, $(\wedge, \prec )(G)=T$ is called a topological authentication if $(\wedge, \prec )(h_G)=f_T$. For a given public tree $T$, \textbf{find} a $W$-type coloring $f_T$ of $T$, and \textbf{determine} a connected graph $G$ admitting a $W$-type coloring $h_G$ such that $(\odot, \overline{\ominus})(T)=G$ and $(\odot, \overline{\ominus})(f_T)=h_G$ hold true.
\end{asparaenum}
\end{problem}

\begin{figure}[h]
\centering
\includegraphics[width=16.4cm]{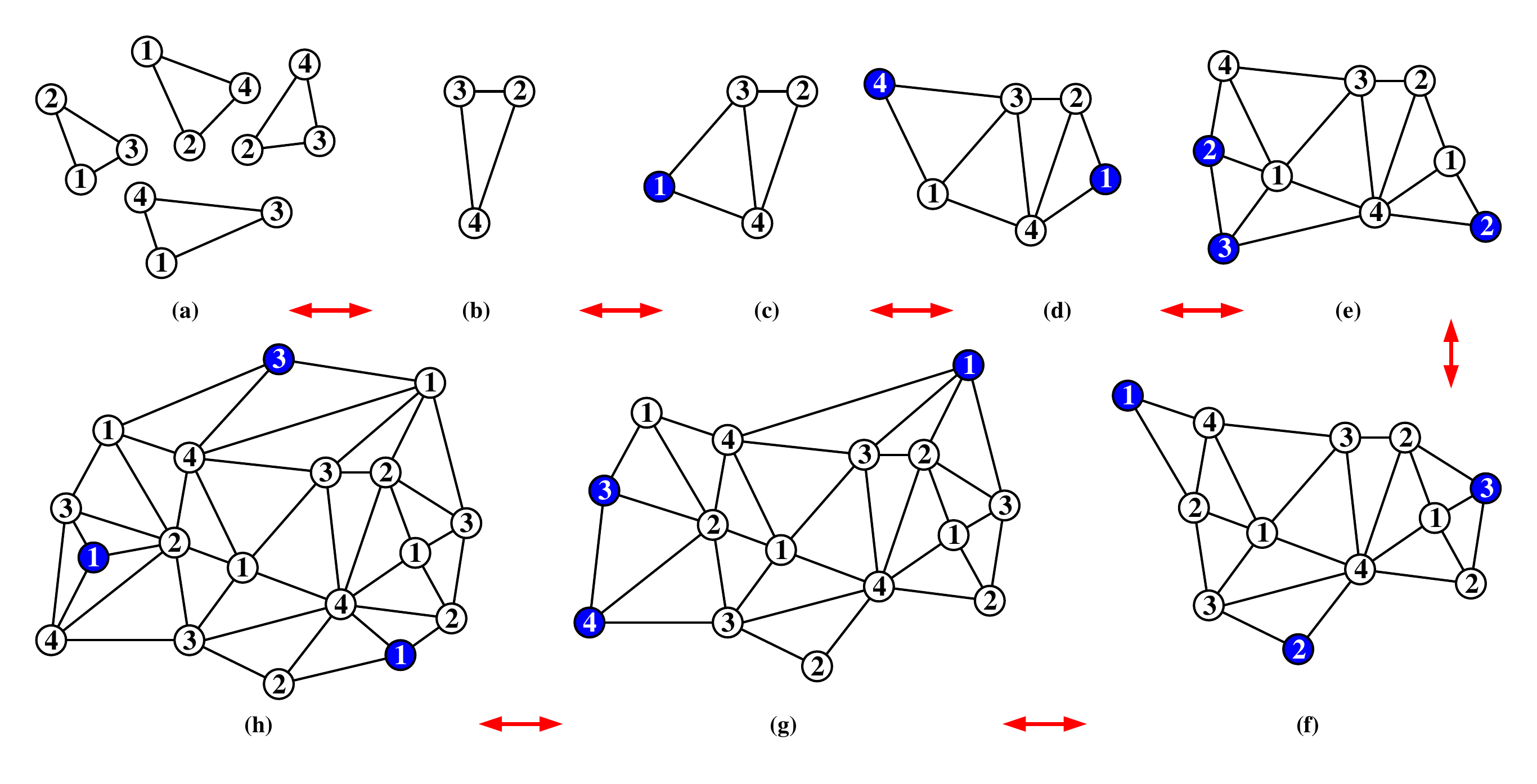}\\
\caption{\label{fig:4-color-plane}{\small $xOy$-plane can be titled with four colored triangles shown in (a).}}
\end{figure}

\begin{problem}\label{problem:99-4-color-planar-graphic-lattice}
\cite{Bing-Yao-2020arXiv} By the notation and terminology introduced in Definition \ref{defn:99-4-color-planar-graphic-lattice}, there are the following problems:
\begin{asparaenum}[\textrm{4C}-1. ]
\item \textbf{Determine} $\{H\bigtriangleup [F(H)-1]\cdot T^r_k: H\in F_{\textrm{inner}\bigtriangleup}\}$, where $F(H)$ is the face number of $H\in F_{\textrm{inner}\bigtriangleup}$ and a fixed $T^r_k\in \{F_{\textrm{inner}\bigtriangleup};\oplus\}$. In other word, each $H\bigtriangleup [F(H)-1]\cdot T^r_k$ is tiled by one $T^r_k$ only, so $H$ is 3-colorable.
\item \textbf{Find} conditions for a planar graph $H\in F_{\textrm{inner}\bigtriangleup}$, such that $H$ must be tiled with all elements of the planar graphic lattice base $\textbf{\textrm{T}}\,^r$ only.
\item \textbf{Estimate} the cardinality of a planar graphic lattice $\textbf{\textrm{L}}(\textbf{\textrm{T}}\,^r\bigtriangleup F_{\textrm{inner}\bigtriangleup})$.
\item For each uncolored planar graph $H\in F_{\textrm{inner}\bigtriangleup}$, \textbf{does} there exist a 4-colorable planar graph $G=T\bigtriangleup ^4_{k=1}a_kT^r_k$ with $a_k\in Z^0$ and $\sum a_k\geq 1$ such that $H\cong G$?
\item Use the elements of the planar graphic lattice base $\textbf{\textrm{T}}\,^r$ defined in Definition \ref{defn:99-4-color-planar-graphic-lattice} to tile completely the entire $xOy$-plane, such that the resultant plane, denoted as $P_{\textrm{4C}}$, is $4$-colorable, and the plane $P_{\textrm{4C}}$ contains infinite triangles $T^r_k$ for each $k\in [1,4]$, we call $P_{\textrm{4C}}$ a \emph{$4$-colorable plane}, see examples shown in Fig.\ref{fig:4-color-plane}. For any given planar graph $G$, \textbf{prove} $G$ in, or not in the plane $P_{\textrm{4C}}$. Similarly, we can consider: Any 3-colorable planar graph is in the plane $P_{\textrm{3C}}$ tiled completely by one element of the planar graphic lattice base $\textbf{\textrm{T}}\,^r$.
\end{asparaenum}
\end{problem}

\begin{problem}\label{qeu:especial-total-colorings}
Definition \ref{defn:99twin-odd-graceful-graphic-lattice} enables us to propose two problems as follows:
\begin{asparaenum}[\textrm{Twin}-1.]
\item \textbf{Find} an algorithm for figuring all graphs $H_{i,j}$ of the base $\textrm{\textbf{H}}$, that is, \textbf{find} all twin odd-graceful matchings $(G_t,H_{i,j})$ for each colored graph $G_t\in F_{\textrm{odd}}(G)$, and determine $M_{\textrm{odd}}$.
\item \textbf{Find} the smallest $\sum h(x_{t,s}y^{k}_{i,j})$ in all graphs $L=G_t\ominus ^3_{k=1}|^{M_{\textrm{odd}}}_{k,j}a_{k,j}H_{k,j}$ of a twin odd-graceful graphic lattice $\textrm{\textbf{L}}(\textrm{\textbf{H}}\ominus \textbf{F}_{\textrm{odd}}(G))$.
\end{asparaenum}
\end{problem}

\begin{problem}\label{problem:standard-vs-graphic-lattices}
For the research of various topological coding lattices and directed topological coding lattices, we present the following questions:
\begin{asparaenum}[\textrm{D}-1. ]
\item \textbf{Determine} the number of trees corresponding a common Topcode-matrix $T_{code}$.

\item Let $IS(n)$ be the set of trees of $n$ vertices. We define a graph $G_{IS}$ with vertex set $V(G_{IS})=IS(n)$, two vertices $T_x$ and $T_y$ of $G_{IS}$ are adjacent from each other if doing a vertex-coinciding operation to two non-adjacent vertices $u$ and $v$, and obtain a unicycle graph $H=T_x(u\odot v)$, and then vertex-split a vertex $w$ of the unicycle graph $H$ into two vertices, such that the resultant graph $H\wedge w$ is just $T_y$, and vice versa. \textbf{Find} the shortest path connecting two vertices $L_x$ (as a \emph{public-key}) and $L_y$ (as a \emph{private-key}) of $G_{IS}$. \textbf{Consider} this question about the graph $G_{IS^c}$ having the vertex set $V(G_{IS^c})=IS^c(n)$, where $IS^c(n)$ is the set of colored trees of $n$ vertices, each tree $T_x$ of $IS^c(n)$ admits a $W$-type total coloring $f$, such that a unicycle graph $H=T_x(u\odot v)$ holds $f(u)=f(v)$.

\item \textbf{Connections between traditional lattices and graphic lattices.} \textbf{Translate} a traditional integer lattice $\textrm{\textbf{L}}(\textbf{ZB})$ into some (colored) graphic lattice.
\item \textbf{Translate} some problems of traditional lattices into graphic lattices, such as: Shortest Vector Problem (SVP, NP-hard), Shortest Independent Vector Problem (IVP), Unique Shortest Vector Problem, Closest Vector Problem (CVP, NP-C), Bounded Distance Decoding (BDD), Shortest Independent Vector Problem (SIVP, NP-hard), and so on.
\item There are many methods to build up topological vectors of graphs, for example, a spider tree $S_{pider}$ with $m$ legs $P_i$ of length $p_i$ for $i\in [1,m]$, so this spider tree $S_{pider}$ has its own topological vector $V_{ec}(S_{pider})=(p_1,p_2,\dots ,p_m)$; directly, a graph $G$ has its own topological vector $V_{ec}(G)=(d_1,d_2,\dots ,d_n)$, where $d_1,d_2,\dots ,d_n$ is the \emph{degree sequence} of $G$. \textbf{Find} other ways for making topological vectors of graphs.

\item By Definition \ref{defn:directed-graceful-coloring}, \textbf{determine} $v_{gra}(\overrightarrow{G})=\min_f\left \{|f(V(\overrightarrow{G}))|\right \}$ over all of directed gracefully total colorings of a connected digraph $\overrightarrow{G}$.
\item \textbf{Define} more directed $W$-type total colorings according to Definition \ref{defn:directed-graceful-coloring}.
\end{asparaenum}
\end{problem}

\begin{problem}\label{problem:weak-gracefully-total-coloring}
Let $v_{\textrm{weakgtc}}(G)=\min_f\{|f(V(G))|\}$ over all weak gracefully total colorings of a connected graph $G$, refer to Definition \ref{defn:weak-graceful-total-colorings}. Then \textbf{try to claim} that any tree $T$ with diameter at least three holds $v_{\textrm{weakgtc}}(T)\geq \frac{1}{2}|V(T)|$ true.
\end{problem}

\begin{problem}\label{problem:felicitous-difference-colorings}
By Definitions \ref{defn:99-felicitous-difference-star-graphic-lattice}, we present the following questions:
\begin{asparaenum}[\textrm{FDQ}-1. ]
\item \textbf{Characterize} the structures of two graphic lattices $\textrm{\textbf{L}}(\overline{\ominus} \textbf{I}_{ce}(FD))$ and $\textrm{\textbf{L}}(\overline{\ominus} \textbf{I}_{ce}(SFD))$.
\item \textbf{Find} all $k$ with $|f_k(x)+f_k(y)-f_k(xy)|=k$ for $\max \{f_k(w):w\in V(F)\cup E(G)\}=\chi\,''_{fdt}(G)$, see examples shown in Fig.\ref{fig:more-magic-numbers-big}.
\begin{figure}[h]
\centering
\includegraphics[width=16.2cm]{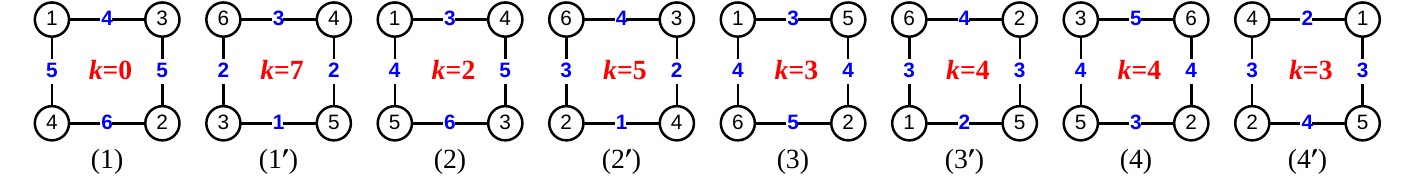}
\caption{\label{fig:more-magic-numbers-big}{\small A cycle $C_4$ admits four pairs of all-dual felicitous-difference proper total colorings ($k$) and ($k\,'$) with $k\in [1,4]$, refer to Definition \ref{defn:combinatoric-definition-total-coloring}.}}
\end{figure}
\item \textbf{Does} every planar graph $H$ belong to the felicitous-difference star-graphic lattice $\textrm{\textbf{L}}(\overline{\ominus} \textbf{I}_{ce}(SFD))$, in other word, $\chi\,''_{fdt}(H)\leq 1+2\Delta(H)$?
\item Since each graph $H\in \textrm{\textbf{L}}(\overline{\ominus} \textbf{I}_{ce}(SFD))$ holding $\chi\,''_{fdt}(H)\leq 1+2\Delta(H)$, so \textbf{find} other subset $S\subset \textrm{\textbf{L}}(\overline{\ominus} \textbf{F}^c_{star\Delta})$ such that each graph $L\in S$ holding $\chi\,''_{fdt}(L)\leq 1+2\Delta(L)$.
\item Since a caterpillar $T$ corresponds a topological vector $V_{ec}(T)$, \textbf{can} we characterize a traditional lattice by some graphic lattices?
\item \textbf{Plant} some results of a traditional lattice $\textrm{\textbf{L}}(\textbf{B})$ to the felicitous-difference star-graphic lattices.
\item \textbf{Optimal felicitous-difference ice-flower system.} Find a $L$-magic felicitous-difference ice-flower system $I_{ce}(LF_{1,n}D_k)^{n_{fdt}}_{k=1}$, such that each graph $G$ is colored well by $G=\overline{\ominus}^{n_{fdt}}_{j=1}a_jLF_{1,n}D_j$ with color set $[1,\chi\,''_{fdt}(G)]$, where $LF_{1,n}D_j\in I_{ce}(LF_{1,n}D_k)^{n_{fdt}}_{k=1}$, $\sum^{n_{fdt}}_{j=1}a_j\geq 1$ and $a_j\in Z^0$. In other word, this felicitous-difference ice-flower system $I_{ce}(LF_{1,n}D_k)^{n_{fdt}}_{k=1}$ is \emph{optimal}.
\end{asparaenum}
\end{problem}

\begin{problem}\label{problem:complex-problems}
For a colored graphic lattice $\textbf{\textrm{L}}(\textbf{\textrm{T}}\,^c\odot F\,^c_{p,q})$ defined in Definition \ref{defn:99-colored-graphic-lattice}, we propose the following complex problems:
\begin{asparaenum}[\textrm{C}-1. ]
\item \textbf{Classify} a colored graphic lattice $\textbf{\textrm{L}}(\textbf{\textrm{T}}\,^c\odot F\,^c_{p,q})$ defined in Definition \ref{defn:99-colored-graphic-lattice} into some particular subsets $L^{sub}_k$ with $k\in [1,m]$, find particular subsets, such as each graph of $L^{sub}_i$ is a tree, or an Euler's graph, or a Hamiltonian graph; if the colored graphic lattice base $\textbf{\textrm{T}}\,^c$ admits a flawed $W$-type coloring, then each graph of $L^{sub}_j$ admits a $W$-type coloring too.
\item \textbf{Find} a graph of $\textbf{\textrm{L}}(\textbf{\textrm{T}}\,^c\odot F\,^c_{p,q})$ with the shortest diameter $D(G^*)$, such that $D(G^*)\leq D(G)$ for any graph $G\in \textbf{\textrm{L}}(\textbf{\textrm{T}}\,^c\odot F\,^c_{p,q})$.
\item \textbf{List} possible $W$-type colorings for constructing a colored graphic lattice $\textbf{\textrm{L}}(\textbf{\textrm{T}}\,^c\odot F\,^c_{p,q})$.
\item \textbf{Do} we have the topological coloring isomorphism $H\,^c\odot |^n_{i=1}T\,^c_i= G^c\odot |^n_{i=1}T\,^c_i$ in a colored graphic lattice $\textbf{\textrm{L}}(\textbf{\textrm{T}}\,^c\odot F\,^c_{p,q})$ when $H\,^c\cong G^c$ or $H\,^c\not \cong G^c$?
\item If $T\,^c_i\cong T\,^c_j$ and $T\,^c_i\cong H\,^c$ for some distinct $i,j\in [1,n]$, \textbf{characterize} $H\,^c\odot |^n_{i=1}T\,^c_i$.
\item Since a tree admits a set-ordered graceful labeling if and only if it admits a set-ordered odd-graceful labeling, we consider: For two colored graphic lattices $\textbf{\textrm{L}}(\textbf{\textrm{T}}\,^c_i\odot F\,^c_{p,q})$ and two bases $\textbf{\textrm{T}}\,^c_i=(T\,^c_{i,1}$, $T\,^c_{i,2}$, $\dots $, $T\,^c_{i,n})$ with $i=1,2$, each $X_i=H\,^c\odot |^n_{j=1}T\,^c_{i,j}$ admits a $W_i$-type coloring, if both $X_i$ and $X_{3-i}$ are equivalent from each other. \textbf{Is} $\textbf{\textrm{L}}(\textbf{\textrm{T}}\,^c_i\odot F\,^c_{p,q})$ equivalent to $\textbf{\textrm{L}}(\textbf{\textrm{T}}\,^c_{3-i}\odot F\,^c_{p,q})$ with $i=1,2$?
\item If the graphic lattice base $\textbf{\textrm{T}}\,^c=(T\,^c_1,T\,^c_2,\dots, T\,^c_n)$ forms an every-zero graphic group based under a $W$-type coloring, \textbf{does} the corresponding colored graphic lattice form a graphic group too?
\item If each graphic base of the graphic lattice base $\textbf{\textrm{T}}\,^c=(T\,^c_1,T\,^c_2,\dots, T\,^c_n)$ admits a $W$-type coloring defined in Definition \ref{defn:combinatoric-definition-total-coloring}, \textbf{determine} a subset $S(\textbf{\textrm{L}})$ of the graphic lattice $\textbf{\textrm{L}}(\textbf{\textrm{T}}\,^c\odot F\,^c_{p,q})$, such that each connected graph of $S(\textbf{\textrm{L}})$ admits a rainbow proper total coloring defined in Definition \ref{defn:rainbow-proper-total-coloring}.
\item \textbf{Find} a graph $G\in \textbf{\textrm{L}}(\textbf{\textrm{T}}\,^c\odot F\,^c_{p,q})$, such that for any $H\in \textbf{\textrm{L}}(\textbf{\textrm{T}}\,^c\odot F\,^c_{p,q})$, we have

\quad (1) the proper total chromatic numbers satisfy $\chi\,''(G)\leq \chi\,''(H)$; or

\quad (2) the edge-magic total chromatic number $\chi\,''_{emt}$, the edge-difference total chromatic number $\chi\,''_{edt}$, the felicitous-difference total chromatic number $\chi\,''_{fdt}$ and the graceful-difference total chromatic number $\chi\,''_{gdt}$ hold $\chi\,''_{\varepsilon}(G)\leq \chi\,''_{\varepsilon}(H)$ for $\varepsilon\in \{$\emph{emt, edt, fdt, gdt}$\}$ defined in Definition \ref{defn:edge-magic-tcn-pure-total-coloring}, Definition \ref{defn:combinatoric-definition-total-coloring} and Definition \ref{defn:combinatoric-definition-total-coloring-abc}; and or

\quad (3) the diameters obey $D_{\textrm{iameter}}(G)\leq D_{\textrm{iameter}}(H)$.
\end{asparaenum}
\end{problem}

\begin{problem}\label{problem:even-odd-harmonious-matching-pair}
\cite{Yao-Zhang-Sun-Mu-Wang-Zhang2018} Suppose that each graph $T_i$ of $q$ edges with $i=1,2$ admits an (resp. even)odd-harmonious labeling $f_i$ defined in Definition \ref{defn:more-harmonious-labelings}. \textbf{\textrm{Find}} the vertex-coinciding graph $\odot_s\langle T_1,T_2\rangle$ made by an (resp. even)odd-harmonious $s$-matching pair of $T_1$ and $T_2$.
\end{problem}

\begin{problem} \label{prob:optimal-00}
A vector $\textbf{\textrm{x}}=(x_1,x_2,\dots ,x_n)$ has its norm $||\textbf{\textrm{x}}||=\sqrt[m]{\sum ^n_{i=1}x^m_i}$. Minimization $\min ||\textbf{\textrm{x}}||=\sqrt[m]{\sum ^n_{i=1}x^m_i}$ is regarded by computer scientist as an NP-hard problem. Definition \ref{defn:set-coloring-definitions} induces some parameters $v_l$ and $e_l$ for a $(p,q)$-graph $G$. So, we can borrow the word `norm' to optimize
\begin{equation}\label{eqa:555555}
\min \max\left \{||F(u)_m||=\sqrt[m]{\sum ^{r_u}_{i=1}u^m_i}\right \}
\end{equation} for $F(u)=\{u_{1},u_{2},\dots ,u_{r_u}\}$ for $u\in V(G)$ s.t. $R_s(m)$, and
\begin{equation}\label{eqa:555555}
\min_F \{v_l+e_l:\textrm{ each total set-coloring $F$ of $G$}\}~s.t. R_s(m).
\end{equation}
\end{problem}

\begin{problem} \label{prob:optimal-11}
Definition \ref{defn:set-coloring-definitions} can derivative so-called set-matrices by adjacent and incident matrices of graphs, in these set-matrices each element is a set only. Find more properties of set-matrices.
\end{problem}

\begin{problem}\label{problem:mixed-Graphic-group-Algorithm}
By Definition \ref{defn:graphic-group-sequences},
\begin{asparaenum}[\textrm{Seq}-1. ]
\item \textbf{Characterize} the topological structure of $\big \{G^{(t)}_{ro}(H)\big \}$. \textbf{Is} $G^{(t)}_{ro}(H)$ scale-free? \textbf{Is} $G^{(t)}_{ro}(H)$ self-similar?
\item \textbf{Determine} colorings admitted by each element of $\big \{G^{(t)}_{ro}(H)\big \}$.
\item \textbf{Estimate} the cardinality of $\big \{G^{(t)}_{ro}(H)\big \}$.
\item For $\textrm{\textbf{H}}=(H_1,H_2,\dots ,H_m)$, \textbf{study} \emph{every-zero $\textrm{\textbf{H}}$-graphic group sequence} $\big \{G^{(t)}_{ro}(\textrm{\textbf{H}})\big \}$.
\end{asparaenum}
\end{problem}

\begin{problem}\label{problem:real-valued-total-coloring}
\cite{Yao-Wang-Ma-Su-Wang-Sun-2020ITNEC} We vertex-split some vertices of a connected graph $G^e_i$ (resp. a connected graph $G^v_i$) to obtain connected graphs admitting real-valued edge colorings (resp. real-valued edge colorings). After doing vertex-splitting operation on $G^e_i$ and $G^v_i$, we get a set $S^e(G,g_i)^N_1$ containing all connected graphs admitting real-valued edge colorings, and another set $S^v(G,g_i)^N_1$ containing all connected graphs admitting real-valued vertex colorings defined in Definition \ref{defn:real-valued-condition-colorings} and Definition \ref{defn:real-valued-total-cs}. We propose:
\begin{asparaenum}[\quad \textbf{\textrm{Que}}-1. ]
\item \textbf{A topological color-valued authentication problem with the vertex-splitting and vertex-coinciding operations.} For any graph $A_e\in S^e(G,g_i)^N_1$, we want to find another graph $A_v\in S^v(G,g_i)^N_1$, and then doing the vertex-coinciding operation to $A_e$ and $A_v$, respectively, produces two graphs $B_e$ and $B_v$, such that $B_e\cong G \cong B_v$, and the real-valued vertex coloring of $B_e$ and the real-valued edge coloring of $B_v$ both induce just a $C$-edge-magic real-valued total coloring of $G$.

\item As an application of real-valued total colorings, we build two graph sets $X(G)$ and $Y(G)$ based on a connected $(p,q)$-graph $G$, such that each graph $H_i\in X(G)$ admits a real-valued vertex coloring and $H_i\cong G$, and each graph $T_j\in Y(G)$ admits a real-valued edge coloring and $T_j\cong G$. As known, $G$ admits a $C$-edge-magic real-valued total coloring $f$ defined in Definition \ref{defn:real-valued-total-cs}. Then we remove the vertex colors of $G$, the resultant graph is denoted as $T^*$, so it admits a real-valued edge coloring induced by $f$; next we remove the edge colors of $G$ to make a graph $H^*$ admitting a real-valued vertex coloring induced by $f$. Clearly, $H^*\in X(G)$, and $T^*\in Y(G)$. For designing \emph{topological color-valued authentications}, we propose the following questions:

\quad (i) \textbf{Does} each graph $H_i\in X(G)$ admitting a real-valued vertex coloring $f_i$ correspond a graph $T_j\in Y(G)$ admitting a real-valued edge coloring $g_j$, such that combing two real-valued colorings $f_i$ and $g_j$ produces just a real-valued total coloring of $G$?

\quad (ii) \textbf{Does} there exist two subsets $X_C(G)\subset X(G)$ and $Y_C(G)\subset Y(G)$ such that each graph $H\,'_i\in X_C(G)$ admitting a real-valued vertex coloring $f\,'_i$ corresponds a graph $T\,'_j\in Y_C(G)$ admitting a real-valued edge coloring $g\,'_j$, and furthermore two real-valued colorings $f\,'_i$ and $g\,'_j$ produces a $C$-edge-magic real-valued total coloring of $G$?

\quad (iii) Suppose that a graph set $X^*$ contains all graphs of $p$ vertices, and each graph of $X^*$ admits a real-valued vertex coloring; and another graph set $Y^*$ contains all graphs of $p$ vertices, and each graph of $Y^*$ admits a real-valued edge coloring. Clearly, $X(G)\subset X^*$ and $Y(G)\subset Y^*$. For a given graph $H$ of $p$ vertices, \textbf{do} there exist two graphs $T\in X^*$ and $L\in Y^*$ such that $T\cong H\cong L$, and the real-valued vertex coloring $f_T$ of $T$ and the real-valued edge coloring $f_L$ of $L$ induce just a ($C$-edge-magic) real-valued total coloring $f_H$ of $H$? Of cause, the existences of $T$ and $L$ are true, we have \emph{difficult points} as: (1) no polynomial algorithm for judging $T\cong H\cong L$; (2) no existing way for determining the real-valued vertex coloring $f_T$ and the real-valued edge coloring $f_L$, such that they compose $f_H$.
\end{asparaenum}
\end{problem}

\subsection{Problems on set partitions}

Since many readers do not learn graph theory and cryptographs, topological coding cryptosystems, it should do translate colorings and labelings into set language, so we will meet new problems and new motivation from set theory. In \cite{Bing-Yao-Yarong-Mu-ITOEC-2018}, the authors introduce the so-called \textbf{Set Partition} that is a natural phenomenon in mathematics. For instance, an integer set $[1,10]$ contains two subsets $[1,10]^o=\{1,3,5,7,9\}$ and $[1,10]^e=\{2,4,6,8,10\}$, conversely, $[1,10]^o\cup [1,10]^e=[1,10]$, so both sets $[1,10]^o$ and $[1,10]^e$ match with each other, and are called a \emph{set partition} of $[1,10]$. We say a $p$-set $S$ if the cardinality of the set $S$ is just $|S|=p$.

\begin{defn} \label{defn:set-even-odd-harmonious-matching-pair}
\cite{Yao-Zhang-Sun-Mu-Wang-Zhang2018} By Definition Definition \ref{defn:more-harmonious-labelings}, if there are two subsets $S,L\subset [0,2q]$ holding $S\cup L=[0,2q]$ and $|S\cap L|=k$ such that there are two induced sets $S^*=\{s_i=x+y~(\bmod~2q):~x,y\in S\}$ and $L^*=\{l_i=x+y~(\bmod~2q):~x,y\in L\}$ holding $S^*\subset [0,2q]$ and $L^*\subset [0,2q]$, then we call two pairs $(S,S^*),(L,L^*)$ a \emph{harmonious $k$-matching set-pair}, denoted as $\odot_k\langle (S,S^*),(L,L^*)\rangle$. If $S^*=\{1,3,5,\dots ,2q-1\}$, and $L^*=S^*$ or $L^*=S^*\setminus \{2q-1\}$, we say $\odot_k\langle (S,S^*),(L,L^*)\rangle$ to be an \emph{odd-harmonious $k$-matching set-pair}; if $S^*=\{2,4,6,\dots ,2q\}$, and $L^*=S^*$ or $L^*=S^*\setminus \{2q\}$, we say $\odot_k\langle (S,S^*),(L,L^*)\rangle$ to be an \emph{even-harmonious $k$-matching set-pair}.\qqed
\end{defn}
\begin{problem}\label{problem:10-set-partition-problems}
For two integers $p\geq 2$ and $q\geq 1$, there are the following \emph{set partition problems} of topological coding:
\begin{asparaenum}[\textbf{Set}-1. ]
\item \textbf{Find} set-ordered graceful partition and strongly graceful partition (\cite{Gallian2020, Zhou-Yao-Chen-Tao2012}). Partition a set $[0,q]$ into two sets $X$ and $Y$ such that $X\cup Y\subseteq [0,q]$ and $X\cap Y=\emptyset$. If $[1,q]\subseteq \{|a-b|:~a\in X,b\in Y\}$, we say the partition $(X,Y)$ is a \emph{graceful partition}, and $(X,Y)$ is called a \emph{set-ordered graceful partition} if $\max X< \min Y$.

\quad A \textbf{\emph{perfect set}} $A(X,Y)=\{(a_i,b_j):~a_i\in X,b_j\in Y\}$ holds no one of $a_i=a_s$ and $b_j=b_t$ for any two $(a_i,b_j),(a_s,b_t)\in A(X,Y)$. As $(X,Y)$ is an odd-graceful partition, and $a_i+b_j=q$ for each $(a_i,b_j)\in A(X,Y)$, we call $(X,Y)$ a \emph{strongly graceful partition}, and moreover $(X,Y)$ is called a \emph{set-ordered strongly graceful partition} if $\max X<\min Y$.

\item \textbf{Find} set-ordered odd-graceful partitions and strongly odd-graceful partitions (\cite{Gallian2020, Zhou-Yao-Chen-Tao2012}). Partition a set $[0,2q-1]$ into two sets $X$ and $Y$ such that $X\cup Y\subseteq [0,2q-1]$ and $X\cap Y=\emptyset$. If $[1,2q-1]^o\subseteq \{|a-b|:~a\in X,b\in Y\}$, we say the partition $(X,Y)$ an \emph{odd-graceful partition}, and $(X,Y)$ is called a \emph{set-ordered odd-graceful partition} if $\max X< \min Y$.

\quad A \textbf{\emph{perfect set}} $A(X,Y)=\{(a_i,b_j):~a_i\in X,b_j\in Y\}$ holds no one of $a_i=a_s$ and $b_j=b_t$ for any two $(a_i,b_j),(a_s,b_t)\in A(X,Y)$. As $(X,Y)$ is an odd-graceful partition, and $a_i+b_j=2q-1$ for each $(a_i,b_j)\in A(X,Y)$, we call $(X,Y)$ a \emph{strongly odd-graceful partition}, and moreover $(X,Y)$ is called a \emph{set-ordered strongly odd-graceful partition} if $\max X<\min Y$.

\item \textbf{Find} perfect odd-graceful partitions and perfect $\varepsilon$-partitions \cite{Yao-Zhang-Sun-Mu-Sun-Wang-Wang-Ma-Su-Yang-Yang-Zhang-2018arXiv}: (1) Suppose that $(X,Y)$ is an odd-graceful partition of $[0,2q-1]$. If $\{|x-y|:~x\in X,y\in Y\}=[1,p]$, we call $(X,Y)$ a \emph{perfect odd-graceful partition} of $[0,2q-1]$.

\quad (2) Partitioning $[0,p+q]$ into $V$ and $E$ such that each $c\in E$ corresponds $a,b\in V$ holding an $\varepsilon$-restriction $c=\varepsilon(a,b)$, we call $(V,E)$ an \emph{$\varepsilon$-partition}, and moreover if $\{|a-b|:~a,b\in V\}=[1,p]$, we call $(V,E)$ a \emph{perfect $\varepsilon$-partition}.

\item \textbf{Find} mirror-images. Partition $[0,q]$ into $(V_i,E_i)$ with $i=1,2$ such that each $c_i\in E_i$ corresponds $a_i,b_i\in V_i$ holding $c_i=|a_i-b_i|$, and $|E_i|=q$. If there is a constant $k$ such that $c_i+c_{3-i}=k$ with $i=1,2$, then we call $(V_i,E_i)$ the \emph{mirror-image} of $(V_{3-i},E_{3-i})$ with $i=1,2$.

\item \textbf{Find} twin odd-graceful and odd-elegant KL-partitions (\cite{Wang-Xu-Yao-2017-Twin, Wang-Xu-Yao-2017}).

\quad (A) Partition $[0,2q]$ into two sets $X_1$ and $X_2$ such that: (1) $X_1\cup X_2=[0,2q]$ and $\{0\}\subseteq X_1\cap X_2\subseteq [0,2q]$; (2) $\{|a-b|:~a\in X_i^{o},b\in X_i^{e}\}=[1,2q-1]^o$, where $X_i^{o}$ is an odd subset and $X_i^{e}$ is an even subset of $X_i$ such that $X_i^{o}\cup X_i^{e}=X_i$ and $X_i^{o}\cap X_i^{e}=\emptyset$ with $i=1,2$. We say $(X_1,X_2)$ to be an \emph{odd-graceful KL-partition}.

\quad (B) Suppose that $Y_1\cup Y_2=[0,2q-1]$ and $\{0,k\}=Y_1\cap Y_2\subseteq [0,2q-1]$ for some $k\in[1,2q-1]$; and the set $\{a+b~(\bmod~2q):~a\in Y_i^{o},b\in Y_i^{e}\}=[1,2q-1]^o$, where $Y_i^{o}$ is an odd subset and $Y_i^{e}$ is an even subset of $Y_i$ holding $Y_i^{o}\cup Y_i^{e}=Y_i$ and $Y_i^{o}\cap Y_i^{e}=\emptyset$ with $i=1,2$. We call $(Y_1,Y_2)$ an \emph{odd-elegant KL-partition}.

\item \cite{Yao-Sun-Zhang-Mu-Sun-Wang-Su-Zhang-Yang-Yang-2018arXiv} Partition $[1, p+q]$ into two disjoint subsets $V$ and $E$ with $V\cup E=[1, p+q]$ such that: (i) each $c\in E$ corresponds to distinct $a,b\in V$ holding $c=|a-b|$; (ii) there exists a constant $k$, each $c\,'\in E$ matches with distinct $a\,',b\,'\in V$ holding $a\,'+c\,'+b\,'=k$ true. We call $(V,E)$ a \emph{relaxed edge-magic total partition} of $[1, p+q]$. \textbf{Find} all possible relaxed edge-magic total partitions $(V,E)$ of $[1, p+q]$.

\item \cite{Yao-Sun-Zhang-Mu-Sun-Wang-Su-Zhang-Yang-Yang-2018arXiv} Select two subsets $V,E\subset [1,2q-1]$ with $E=[1,2p-1]^o$ such that there is a constant $k$, each $c\in E$ corresponds two distinct $a,b\in V$ holding $a+b+c=k$ true. We call $(V,E)$ an \emph{odd-edge-magic partition} of $[1,2q-1]$. \textbf{Find} all possible odd-edge-magic partitions of $[1,2q-1]$.

\item \cite{Yao-Sun-Zhang-Mu-Sun-Wang-Su-Zhang-Yang-Yang-2018arXiv} Selecting two subsets $V,E\subset [1,2q-1]$ with $E=[1,2p-1]^o$ holds: (i) each $c\in E$ corresponds $a,b\in V$ to form an \emph{ev-matching} $(acb)$; (ii) each $c\in E$ corresponds to $z\in E$ with its ev-matching $(xzy)$ such that $c=|x-y|$; (iii) each $c\in E$ with the ev-matching $(acb)$ corresponds to $c\,'\in E$ with its ev-matching $(a\,'c\,'b\,')$ holding $(|a-b|-c)+(|a\,'-b\,'|-c\,')=0$ true. We call $(V,E)$ an \emph{ee-difference odd-edge-magic partition} of $[1,2q-1]$. \textbf{Find} all possible ee-difference odd-edge-magic partitions $(V,E)$ of $[1,2q-1]$.

\item \cite{Yao-Sun-Zhang-Mu-Sun-Wang-Su-Zhang-Yang-Yang-2018arXiv} Partition $[1,p+q]$ into two subsets $V$ and $E$, such that: (i) each $c\in E$ corresponds $a,b\in V$ to form an \emph{ev-matching} $(acb)$; (ii) (e-magic) each $c\in E$ with the ev-matching $(acb)$ hold $c+|a-b|=k$; (iii) (ee-difference) each $c\in E$ corresponds to $z\in E$ with the ev-matching $(xzy)$ such that $c=|x-y|$; (iv) (ee-bandwiden) each $c\in E$ with the ev-matching $(acb)$ corresponds to $c\,'\in E$ with the ev-matching $(a\,'c\,'b\,')$ such that $(|a-b|-c)+(|a\,'-b\,'|-c\,')=0$; (iv) (EV-ordered) $\max V<\min E$ (resp. $\max V>\min E$); (v) (ve-matching) each ev-matching $(acb)$ matches with another ev-matching $(uwv)$ such that $a+w=k\,'$ or $b+w=k\,'$, a constant, except the \emph{singularity} $\lfloor \frac{p+q+1}{2}\rfloor $. We call $(V,E)$ a \emph{6C-partition} of $[1,p+q]$. \textbf{Find} all possible 6C-partitions $(V,E)$ of $[1,p+q]$.

 \qquad For a given 6C-partition $(V,E)$ of $[1,p+q]$, if there exists another 6C-partition $(V\,',E\,')$ of $[1,p+q]$ such that $V\setminus (V\cap V\,')=E\,'$, $E=V\,'\setminus (V\cap V\,')$ for $V\cap V\,'=\{\lfloor \frac{p+q+1}{2}\rfloor \}$, we get a partition $(V\cup E\,', E\cup V\,')$, and call it a \emph{6C-complementary partition} of $[1,p+q]$. \textbf{Find} all possible 6C-complementary partitions $(V\cup E\,', E\cup V\,')$.

\item \cite{Yao-Sun-Zhang-Mu-Sun-Wang-Su-Zhang-Yang-Yang-2018arXiv} Partitioning $[0,p+q-1]$ into two subsets $V$ and $E$ with $E=[1,q]$ and $V\subseteq [0,q-1]$ satisfies: (i) each $c\in E$ corresponds $a,b\in V$ to form an \emph{ev-matching} $(acb)$; (ii) (ee-difference) each $c\in E$ corresponds to $z\in E$ with the ev-matching $(xzy)$ such that $c=|x-y|$; (iii) (ee-bandwidth) each $c\in E$ with the ev-matching $(acb)$ corresponds to $c\,'\in E$ with the ev-matching $(a\,'c\,'b\,')$ such that $(|a-b|-c)+(|a\,'-b\,'|-c\,')=|a-b|+|a\,'-b\,'|-(c+c\,')=0$; (iv) there exists a constant $k$ such that each $c\in E$ with its ev-matching $(acb)$ holds $c+|a-b|=k$; (v) each $c\in E$ corresponds another $c\,'\in E$ with its ev-matching $(a\,'c\,'b\,')$ such that $c+a\,'=p$ or $c+b\,'=p$. We call $(V,E)$ an \emph{ee-difference graceful-magic partition} of $[0,p+q-1]$. \textbf{Find} all possible ee-difference graceful-magic partitions of $[0,p+q-1]$.

\item \cite{Yao-Sun-Zhang-Mu-Sun-Wang-Su-Zhang-Yang-Yang-2018arXiv} Partition $[1, p+q]$ into two disjoint subsets $V$ and $E$ with $V\cup E=[1, p+q]$ such that each $c\in E$ corresponds $a,b\in V$ to form an \emph{ev-matching} $(acb)$ holding $c=|a-b|$, and there exists a constant $k$ satisfying $a+c+b=k$ for each $c\in E$ and its ev-matching. We call $(V,E)$ an \emph{edge-magic graceful partition} of $[1, p+q]$. If $(E, V)$ is another edge-magic graceful partition of $[1, p+q]$, we say $(V,E)$ (resp. $(E, V)$) to be a \emph{ve-exchanged matching partition} of $[1, p+q]$. \textbf{Find} all possible edge-magic graceful partitions and ve-exchanged partitions of $[1, p+q]$.
\item \cite{Yao-Sun-Zhang-Mu-Sun-Wang-Su-Zhang-Yang-Yang-2018arXiv} There are two subsets $V\subseteq [0,q]^2$ (resp. $[0,2q-1]^2$) and $E\subseteq [1,q]$ (resp. $[1,2q-1]$) such that each $c\in E$ with its ev-matching $(acb)$ holds $c=|a-b|$, where $a\in A\subset V$ and $b\in B\subset V$ with $A\cap B=\emptyset$, then we call $(V,E)$ a \emph{v-set e-proper graceful (resp. odd-graceful) partition}. \textbf{Find} all possible v-set e-proper graceful (resp. odd-graceful) partition of $[0,q]^2$.

\item \cite{Yao-Sun-Zhang-Mu-Sun-Wang-Su-Zhang-Yang-Yang-2018arXiv} Partition $[0,2q]$ into two subsets $S_1,S_2$ such that $S_1\subset [0,2q-1]$, $S_2\subset [1,2q]$, $|S_1\cap S_2|\geq 1$ and $S_1\cup S_2=[0,2q]$. If there are $E_1=E_2=[1,2q-1]^o$, such that each $c_i\in E_i$ corresponds two numbers $a_i,b_i\in S_i$ holding $c_i=|a_i-b_i|$ (resp. $c_i=a_i+b_i~(\bmod~2q)$) with $i=1,2$, then we call $(S_1,S_2)$ a \emph{twin odd-graceful (resp. odd-elegant) partition} of $[0,2q]$. \textbf{Characterize} twin odd-graceful (resp. odd-elegant) partitions, and \textbf{find} them.

\item \cite{Yao-Sun-Zhang-Mu-Sun-Wang-Su-Zhang-Yang-Yang-2018arXiv} Select a subset $E\subset [0,p-1]$ such that each $c\in E$ corresponds two distinct $a,b\in V=[0,p-1]$ to hold $c=|a-b|$ (resp. $c=a+b~(\bmod~|E|)$), we call $f_E=(V,E)$ a \emph{graph matching partition}, and call $S_{um}(G,f_E)=\sum_{c\in E}|a-b|$ a \emph{difference-sum number} (resp. $F_{um}(G,f_E)=\sum_{c\in E}(a+b)~(\bmod~|E|)$ is a \emph{felicitous-sum number}). \textbf{Determine} $\max_{f_E} S_{um}(G,f_E)$ and $\min_{f_E} S_{um}(G,f_E)$ over all difference-sum partitions $f_E=(V,E)$ of $[0,p-1]$. For all felicitous-sum partitions $f_E=(V,E)$ of $[0,p-1]$, \textbf{find} $\max_{f_E} F_{um}(G,f_E)$ and $\min_{f_E} F_{um}(G,f_E)$.

\item \cite{Yao-Zhang-Sun-Mu-Sun-Wang-Wang-Ma-Su-Yang-Yang-Zhang-2018arXiv} Let $V$ and $E$ be two subsets of $[1,q]^2$~(resp. $[1,2q-1]^2)$, such that each set $c\in E$ corresponds two subsets $a,b\subset V$ to form an \emph{ev-matching} $(acb)$. Suppose that $c=a\cap b$ and $d_c\in c$ is a \emph{representative} of $c$. If $\{d_c:~c\in E\}=[1,q]$ (resp. $[1,2q-1]^o$), then we call $(V,E)$ a \emph{graceful-intersection (an odd-graceful-intersection) total set-partition} of $[1,q]^2$~(resp. $[1,2q-1]^2)$. \textbf{Find} all graceful-intersection (an odd-graceful-intersection) total set-partition of $[1,q]^2$~(resp. $[1,2q-1]^2)$.
\item \cite{Yao-Zhang-Sun-Mu-Sun-Wang-Wang-Ma-Su-Yang-Yang-Zhang-2018arXiv} Suppose that $E$ and $V=[0,q-1]$ are two subsets of $[0,2q-1]$, such that each $c\in E$ corresponds $a,b\in V$ to form an \emph{ev-matching} $(acb)$. We call $(V,E)$ a \emph{multiple-meaning vertex partition} of $[0,2q-1]$, for each $c\in E$ and its ev-matching $(acb)$, if:

\quad (1) $a+c+b=$a constant $k$, and $E=[1,q]$;

\quad (2) $a+c+b=$a constant $k\,'$, and $E=[p,p+q-1]$;

\quad (3) $c=a+b~(\bmod~q)$, and $E=[0,q-1]$;

\quad (4) $|a+b-c|=$a constant $k\,''$, and $E=[1,q]$;

\quad (5) $c=$an odd number for each $c\in E$ holding $E=[1,2q-1]^o$, and $\{a+c+b:c\in E\textrm{ and its ev-matching }(acb)\}=[\alpha,\beta]$ with $\beta-\alpha+1=q$.

\quad \textbf{Find} all multiple-meaning vertex partitions $(V,E)$ of $[0,2q-1]$.
\end{asparaenum}
\end{problem}

\begin{thm}\label{thm:set-partitions-1}
For a $W$-type partition $(X,Y)$ (resp. $(V,E)$) defined in Set-$i$ for $i\in [1,14]$ in Problem \ref{problem:10-set-partition-problems}, there exists a graph $H$ admitting a $W$-type labeling (resp. $W$-type coloring) $f$ holds $f(V(H))=X\cup Y$ (resp. $f(V(H))=X$ and $f(E(H))=E$) true.
\end{thm}

\begin{problem}\label{problem:set-partitions-1}
For a $W$-type partition $(X,Y)$ (resp. $(V,E)$) introduced above, \textbf{find} conditions for a graph $G$ admitting a $W$-type labeling (resp. $W$-type coloring) $f$ holding $f(V(G))=X\cup Y$ (resp. $f(V(G))=X$ and $f(E(G))=E$) true.
\end{problem}

\begin{rem}\label{rem:set-partitions}
Partition $[0,n]$ into $(X,Y)$ as follows:
\begin{asparaenum}[\textrm{Exa}-1. ]
\item $X=[0,k]$, $Y=[k+1,n]$, so $[1,n]\subseteq \{|a-b|:~a\in X,b\in Y\}$ for $0\leq k\leq \frac{n}{2}$ and $\max X<\min Y$, then $(X,Y)$ is a set-ordered graceful partition.
\item $X=\{0,a_1,a_2,\dots ,a_j\}$ with $0<a_1<a_2<\dots <a_j$, $Y=[0,n]\setminus X$, so $[1,n]\subseteq \{|a-b|:~a\in X,b\in Y\}$ for $0\leq a_j\leq \frac{n}{2}$, then $(X,Y)$ is a graceful partition.
\item As $n$ is odd, $X=[0,k]$, $Y=[k+1,n]$, so $[1,n]^o\subseteq \{|a-b|:~a\in X,b\in Y\}$ for $0\leq k\leq \frac{n}{2}$, and $\max X<\min Y$, then $(X,Y)$ is a set-ordered odd-graceful partition.
\item As $n$ is odd, $X=\{0\}\cup [1,n-2]^o$, $Y=\{n\}\cup [2,n-1]^e$, so $[1,n]^o\subseteq \{|a-b|:~a\in X,b\in Y\}$, then $(X,Y)$ is an odd-graceful partition.\paralled
\end{asparaenum}
\end{rem}

\begin{problem}\label{qeu:xxxxxx}
(\cite{Yao-Chen-Yao-Cheng2013JCMCC}, Definition \ref{defn:k-lambda-magically-total-labeling}) \textbf{The $(k,\lambda )$-magically Total Partition Problem.} Let $U=[1,p(p+1)/2]$ for some integer $p\geq 2$. Let $Q$ be a $p$-set of $U$, we wish to find a $q$-set $S\subset U\setminus Q$ with respect to $p-1\leq q$ and two integers $k$ and $\lambda ~ (\neq 0)$ such that $Q\cup S=[1,p+q]$ and each $w\in S$ satisfies the following equation
\begin{equation}\label{eqa:c4-preface-problem}
x+y=k+\lambda w \text{ for some distinct }x,~ y\in Q.
\end{equation}
We call $(Q, S)$ a $(k,\lambda )$-\emph{magically total partition}. \textbf{Find} all $(k,\lambda )$-magically total partitions $(Q, S)$ for $U=[1,p(p+1)/2]$ with integer $p\geq 2$.
\end{problem}

\begin{problem}\label{qeu:xxxxxx}
($^*$ Definition \ref{defn:k-sedge-difference-magically-total-labeling}) \textbf{The $(k,\lambda )$-edge-difference Magically Total Partition Problem.} Let $U=[1,p(p+1)/2]$ for some integer $p\geq 2$. Let $X$ be a $p$-set of $U$, we wish to find a $q$-set $Y\subset U\setminus X$ with respect to $p-1\leq q$ and two integers $k$ and $\lambda~(\neq 0)$ such that $X\cup Y=[1,p+q]$ and each $w\in Y$ satisfies the following equation
\begin{equation}\label{eqa:c4-preface-problem}
|u-v|=k+\lambda w \text{ for some distinct }u, ~ v\in X.
\end{equation}
We call $(X, Y)$ a $(k,\lambda )$-\emph{edge-difference magically total partition}. \textbf{Find} all $(k,\lambda )$-edge-difference magically total partitions $(X, Y)$ for $U=[1,p(p+1)/2]$ with integer $p\geq 2$.
\end{problem}

\subsection{Problems without colorings}

\begin{problem}\label{problem:xxxxxx}
\textbf{Number of Euler graphs.} For integer $q\geq 3$, determine the number of connected Euler graphs of $q$ edges, since such connected Euler graphs can be obtained by doing a series of vertex-coinciding operations to a Hamilton cycle of $q$ edges or to a group of vertex-disjoint cycles having $q$ edges in total.
\end{problem}

\begin{problem}\label{problem:xxxxxx}
\textbf{New connectivities of graphs.} \cite{Wang-Zhang-Mei-Yao2018-Split} For a simple and connected graph $H$, $\gamma_{vs}(H)$ (resp. $\gamma_{es}(H)$) is the smallest number of $k$ for which we vertex-split $k$ vertices (resp. edge-split and leaf-split $k$ edges) of $H$ such that the resultant graph is disconnected (see Section 1). We call $\gamma_{vs}(H)$ and $\gamma_{es}(H)$ \emph{vertex-split connectivity} and \emph{edge-split connectivity} of $H$, respectively. It is not hard to show $\gamma_{vs}(H)=\kappa (H)$ and $\gamma_{es}(H)=\kappa \,'(H)$, where $\kappa (H)$ and $\kappa \,'(H)$ are the vertex-connectivity, edge-connectivity, respectively, \cite{Bondy-2008}. \textbf{Characterize} new properties of connected graphs under the vertex-split connectivity and the edge-split connectivity of graphs.
\end{problem}

\begin{problem}\label{problem:xxxxxx}
\textbf{Find particular spanning trees.} Since a Topsnut-gpw $G$ is a network, \textbf{find}: (i) all possible non-isomorphic caterpillars of $G$; (ii) all possible non-isomorphic lobsters of $G$; (iii) all possible non-isomorphic spanning trees of $G$ with the maximum number of leaves; (iv) all possible non-isomorphic unicyclic graphs of $G$.
\end{problem}

\begin{problem}\label{problem:xxxxxx}
\textbf{Lobster-pure on graphs}. If each spanning tree of a connected graph $G$ is a lobster, we call $G$ to be \emph{lobster-pure}, a \emph{lobster-pure graph}. Since lobsters admit many meaningful colorings and labelings, \textbf{find} the necessary and sufficient conditions for a graph $G$ to be lobster-pure.
\end{problem}

\begin{problem}\label{problem:xxxxxx}
\textbf{Build up algorithms for enumerating trees obtained by adding randomly leaves.} Adding randomly $m$ leaves to a given tree produces new trees, and \textbf{enumerate} these new trees.
\end{problem}

\begin{problem}\label{problem:xxxxxx}
\textbf{Meta-regular graphs for odd-integer decomposition.} A meta-regular graph $G$ with vertex set $V(G)=\{x_1,x_2,\dots ,x_{p-1},w\}$ such that degrees $\textrm{deg}_G(x_i)=k$ for $i\in [1,p-1]$ and $\textrm{deg}_G(w)=1$. \textbf{Characterize} meta-regular graphs, since such graphs are related with Integer Decomposition Problem.
\end{problem}

\begin{figure}[h]
\centering
\includegraphics[width=14cm]{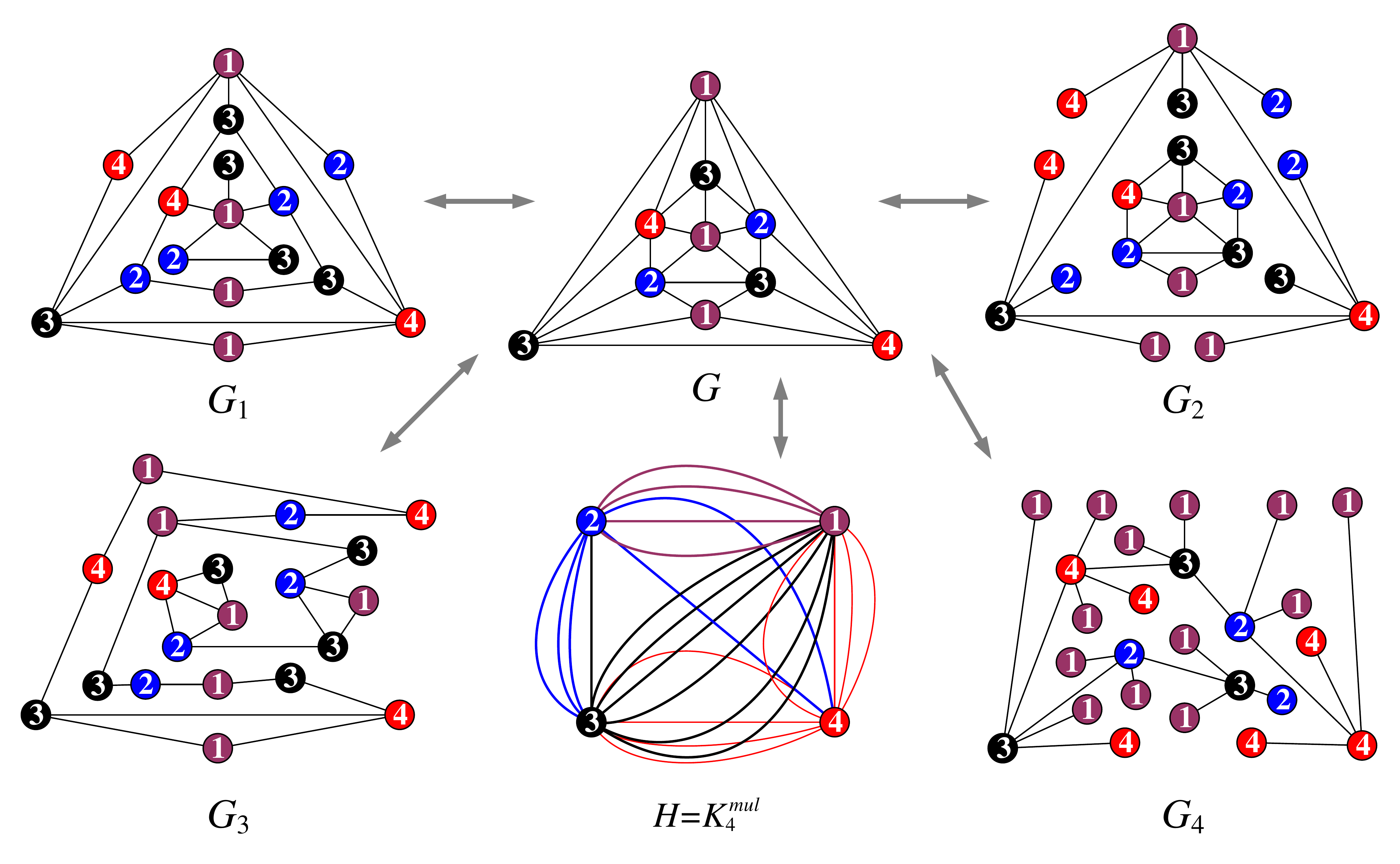}\\
\caption{\label{fig:vertex-split-problem}{\small Simple graph homomorphisms: $G_k\rightarrow G$ for $k\in [1,4]$, and a multiple graph homomorphism $G\rightarrow H=K^{mul}_4$. Conversely, $H\rightarrow_{split} G$, $G\rightarrow_{split} G_k$ for $k\in [1,4]$.}}
\end{figure}

\begin{defn} \label{defn:multiple-edge-complete-graphs}
A \emph{multiple-edge complete graph} $K^{mul}_n$ of $n$ vertices has its vertex set $V(K^{mul}_n)=\{x_1,x_2,\dots ,x_n\}$, each pair of vertices $x_i$ and $x_j$ for $i\neq j$ is joined by $a_{i,j}$ edges $e_{i,j}(k)$ with $k\in [1,a_{i,j}]$ for $a_{i,j}\geq 1$, and there exists some $a_{r,s}\geq 2$ for $r\neq s$. A multiple-edge complete graph $K^{mul}_4$ is shown in Fig.\ref{fig:vertex-split-problem}.\qqed
\end{defn}

\begin{problem}\label{problem:xxxxxx}
By Definition \ref{defn:multiple-edge-complete-graphs}, for what conditions held by $a_{i,j}$ with $i,j\in [1,n]$ and $i\neq j$, can a multiple-edge complete graph $K^{mul}_n$ (as a public-key) be vertex-split into a predicated graph $H$ (as a private-key) such that $H\rightarrow K^{mul}_n$?
\end{problem}
\begin{problem}\label{problem:xxxxxx}
By Definition \ref{defn:multiple-edge-complete-graphs}, for $n=4$, what conditions do $a_{1,2},a_{1,3},a_{1,4},a_{2,3},a_{2,4},a_{3,4}$ hold such that $K^{mul}_4$ can be vertex-split into maximal planar graphs, or out-planar graphs, or half-maximal planar graphs?
\end{problem}


\section{Hanzi-graphs and Hanzi-colorings}

Is your password secure? How to design a password that others can't guess and they can remember easily? Our Hanzi-gpws can help Chinese people to design themselves passwords by speaking and writing Hanzis without learning computer technique.

\subsection{Hanzi-graphs}

\begin{figure}[h]
\centering
\includegraphics[width=13cm]{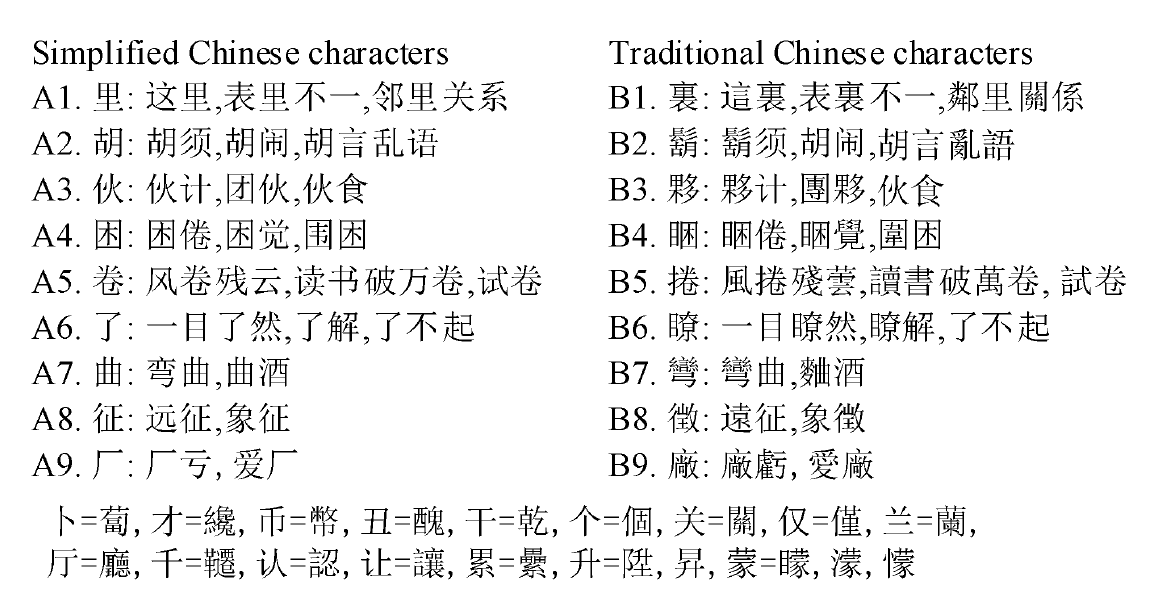}\\
\caption{\label{fig:two-types-hanzis}{\small Differences between the simplified Chinese characters (left) and the traditional Chinese characters (right).}}
\end{figure}

\begin{figure}[h]
\centering
\includegraphics[width=15.4cm]{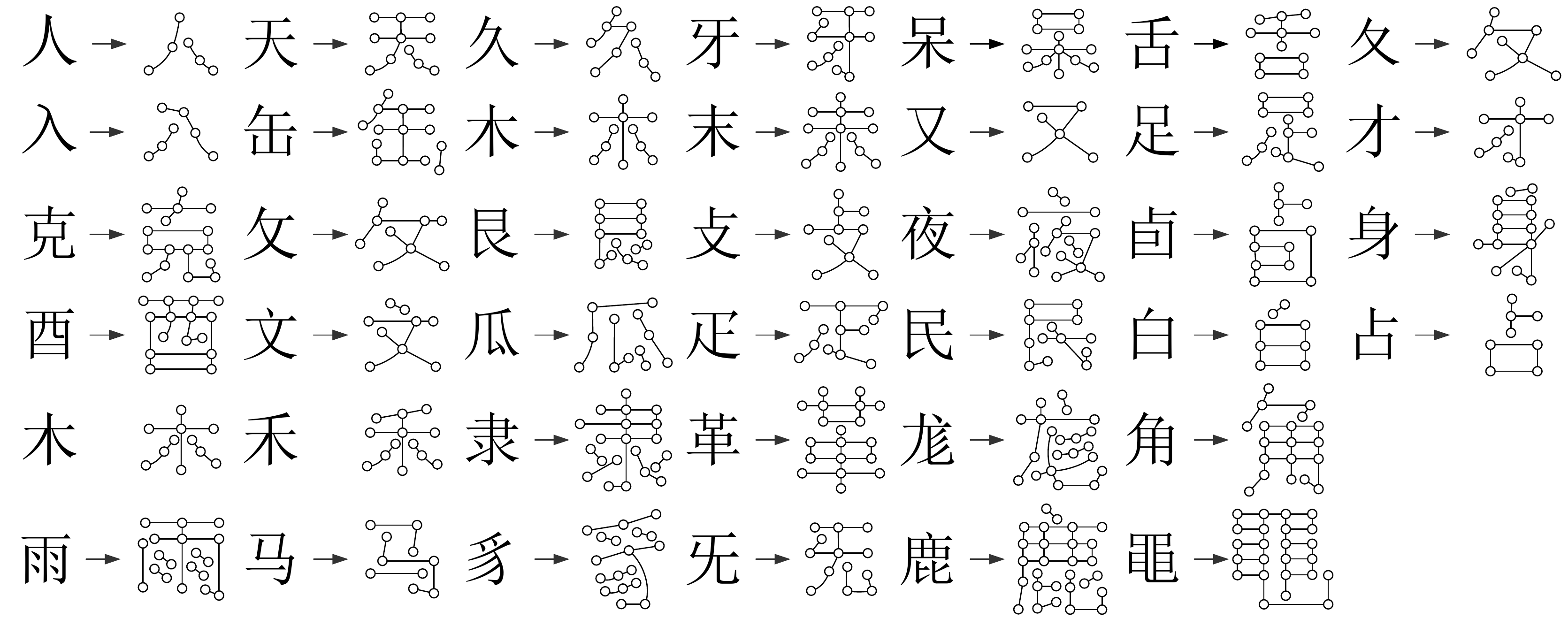}\\
\caption{\label{fig:hanzis-vs-graphs}{\small A group of mathematical models of Hanzis components and radicals.}}
\end{figure}

\begin{figure}[h]
\centering
\includegraphics[width=16.4cm]{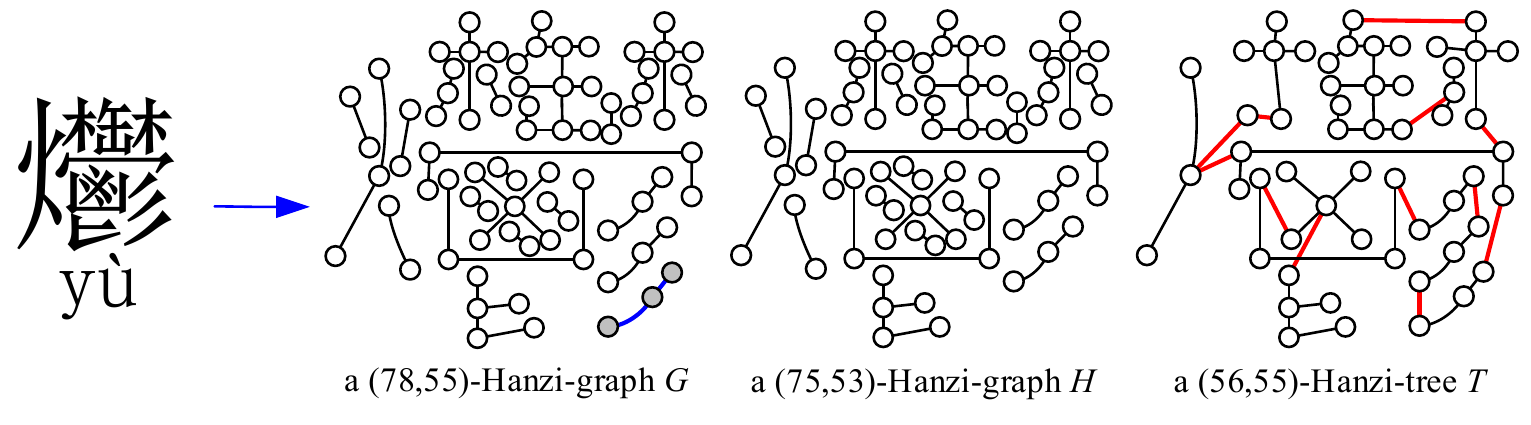}\\
\caption{\label{fig:yu-topological-structrure}{\small An example for illustrating the complex of Hanzi-graphs.}}
\end{figure}

In \cite{Yao-Mu-Sun-Sun-Zhang-Wang-Su-Zhang-Yang-Zhao-Wang-Ma-Yao-Yang-Xie2019}, the authors introduce the construction of Hanzi-graphs based on simplified Chinese characters, here, we omit it. See Fig.\ref{fig:two-types-hanzis}, Fig.\ref{fig:hanzis-vs-graphs} and Fig.\ref{fig:yu-topological-structrure} for knowing some Hanzi-graphs.

\subsection{Hanzi-idiom colorings}

\begin{defn}\label{defn:Hanzi-idiom-labelings}
\cite{Yao-Mu-Sun-Sun-Zhang-Wang-Su-Zhang-Yang-Zhao-Wang-Ma-Yao-Yang-Xie2019} Let $H_{idiom}$ be the set of Hanzi-idioms and phrases. A $(p,q)$-graph $G$ admits a \emph{Hanzi-idiom labeling} $\phi:V(G)\rightarrow H_{idiom}$, such that each edge $uv\in E(G)$ is colored with $\phi(uv)=\phi(u)(\bullet) \phi(v)$, where $(\bullet)$ is an operation on two Hanzi-idioms and phrases.\qqed
\end{defn}

\begin{figure}[h]
\centering
\includegraphics[width=16.4cm]{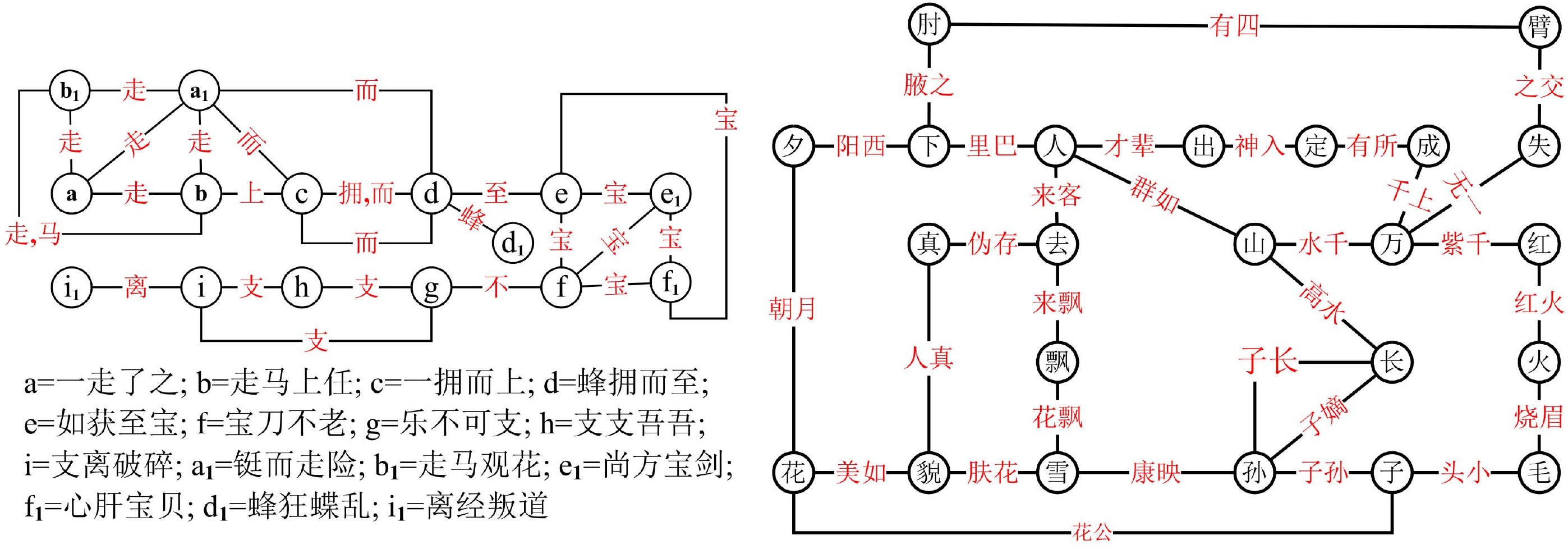}\\
\caption{\label{fig:idiom-labeling}{\small Two graphs admitting two \emph{Hanzi-idiom labelings}, cited from \cite{Yao-Mu-Sun-Sun-Zhang-Wang-Su-Zhang-Yang-Zhao-Wang-Ma-Yao-Yang-Xie2019}.}}
\end{figure}

Fig.\ref{fig:idiom-labeling} shows two graphs admitting two Hanzi-idiom labelings.

\begin{defn}\label{defn:Hanzi-idiom-phrases-labelings}
\cite{Yao-Mu-Sun-Sun-Zhang-Wang-Su-Zhang-Yang-Zhao-Wang-Ma-Yao-Yang-Xie2019} For a given graph $G$, we color its vertices or edges by Chinese phrases and idioms, so we called this labeling as a \emph{Chinese phrase/idiom labeling}, and get a new graphic password, called a \emph{Hanzi-password made by topological structure plus Chinese phrases/idioms}. \qqed
\end{defn}

The right graph $G$ in Fig.\ref{fig:idiom-labeling} admits an \emph{idiom labeling} $\phi:V(G)\rightarrow CI$, where $CI$ is the set of all Chinese idioms, such that $\phi(uv)=\phi(u)\cap \phi(v)$ for each edge $uv\in E(G)$.

\subsection{Colorings on Hanzi-graphs}

We use $H_{abcd}$ to denote a Hanzi with number-based string $abcd$ defined in \cite{GB2312-80}, and $T_{abcd}$ to be the structure (also \emph{Hanzi-graph}) of the Hanzi $H_{abcd}$, and $H^{gpw}_{abcd}\textrm{(b)}$ to be the Hanzi-gpw of the Hanzi $H_{abcd}$, namely, $H^{gpw}_{abcd}\textrm{(b)}$ is the result of coloring Hanzi-graph $T_{abcd}$ by one of the existing $W$-type colorings and the existing $W$-type labelings.

\subsubsection{Hanzi labelings and colorings}

\begin{figure}[h]
\centering
\includegraphics[width=16.4cm]{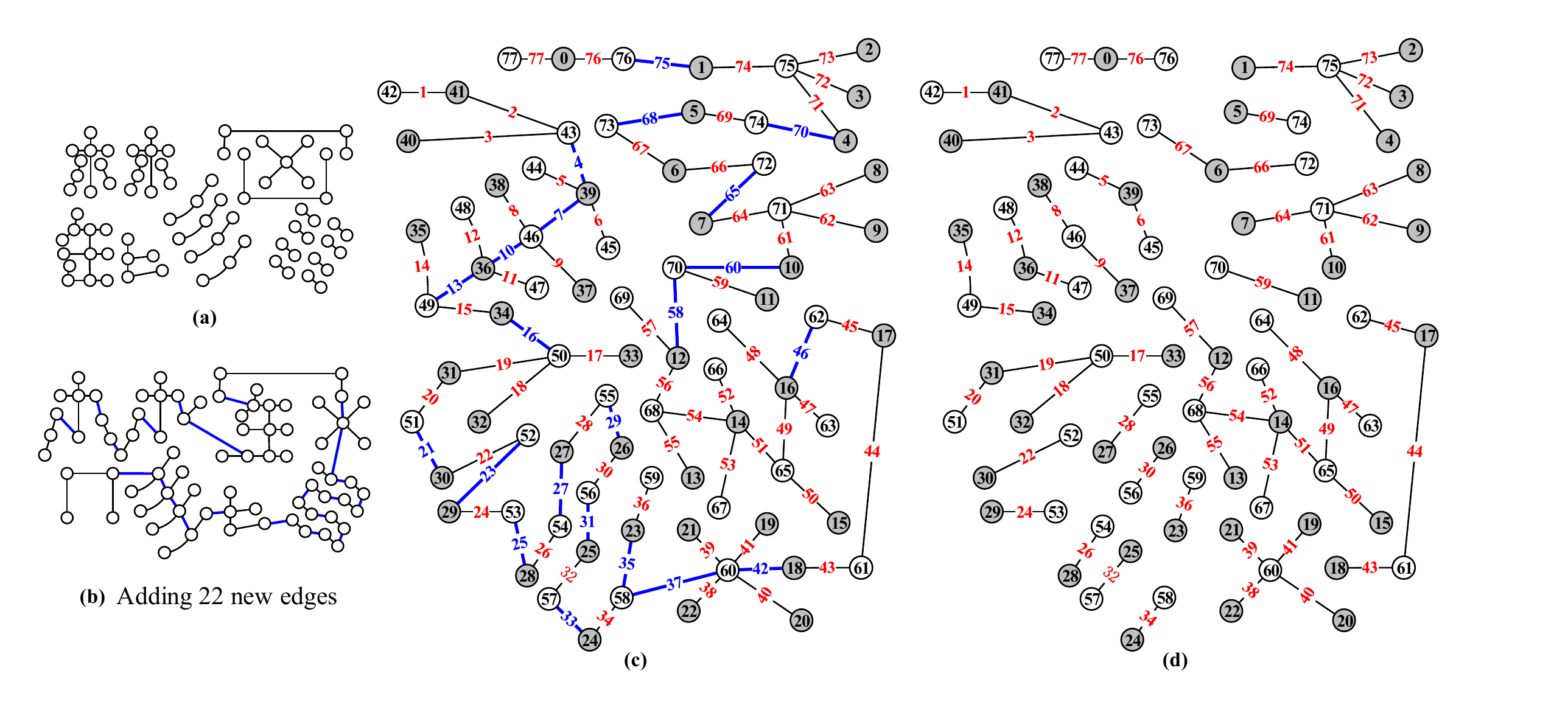}\\
\caption{\label{fig:yu-flawed-graceful}{\small A scheme for illustrating the procedure of finding a flawed set-ordered graceful labeling of the $(78,55)$-Hanzi-graph $G$ shown in Fig.\ref{fig:yu-topological-structrure}.}}
\end{figure}

\begin{figure}[h]
\centering
\includegraphics[width=16.4cm]{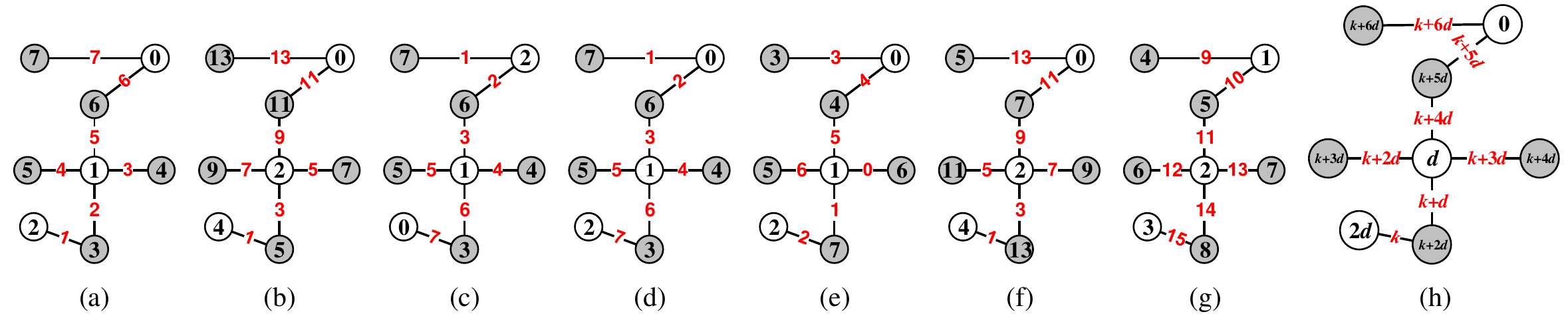}\\
\caption{\label{fig:H-5551-various-colorings}{\small Some labelings of a Hanzi-graph $T_{5551}$, cited from \cite{Zhang-Yang-Mu-Zhao-Sun-Yao-2019}.}}
\end{figure}

Fig.\ref{fig:yu-flawed-graceful} shows a $(78,55)$-Hanzi-graph $G$ admitting a \emph{flawed set-ordered graceful labeling}. In Fig.\ref{fig:H-5551-various-colorings}, a Hanzi-graph $T_{5551}$ made by a Hanzi $H_{5551}$ admits the following labelings:

(a) The Hanzi-gpw $H^{gpw}_{5551}\textrm{(a)}$ admits a \emph{graceful labeling} $\beta_1$ \cite{Gallian2020};

(b) the Hanzi-gpw $H^{gpw}_{5551}\textrm{(b)}$ admits an \emph{odd-graceful labeling} $\beta_2$ \cite{Gallian2020};

(c) the Hanzi-gpw $H^{gpw}_{5551}\textrm{(c)}$ admits an \emph{edge-magic total labeling} $\beta_3$ \cite{Gallian2020};

(d) the Hanzi-gpw $H^{gpw}_{5551}\textrm{(d)}$ admits a \emph{6C-labeling} $\beta_4$ defined in \cite{Yao-Sun-Zhang-Mu-Sun-Wang-Su-Zhang-Yang-Yang-2018arXiv};

(e) the Hanzi-gpw $H^{gpw}_{5551}\textrm{(e)}$ admits a \emph{felicitous labeling} $\beta_5$ ($\bmod~7$) \cite{Gallian2020};

(f) the Hanzi-gpw $H^{gpw}_{5551}\textrm{(f)}$ admits an \emph{odd-edge-magic matching labeling} $(\beta_6,\alpha)$ defined in \cite{YAO-SUN-WANG-SU-XU2018arXiv};

(g) the Hanzi-gpw $H^{gpw}_{5551}\textrm{(g)}$ admits an \emph{edge-magic total graceful labeling} $\beta_7$ \cite{Gallian2020}; and

(h) the Hanzi-gpw $H^{gpw}_{5551}\textrm{(h)}$ admits a \emph{$(k,d)$-graceful labeling} $\beta_8$ \cite{Gallian2020}.

\subsubsection{Hanzi group-labelings}

\begin{figure}[h]
\centering
\includegraphics[width=16.4cm]{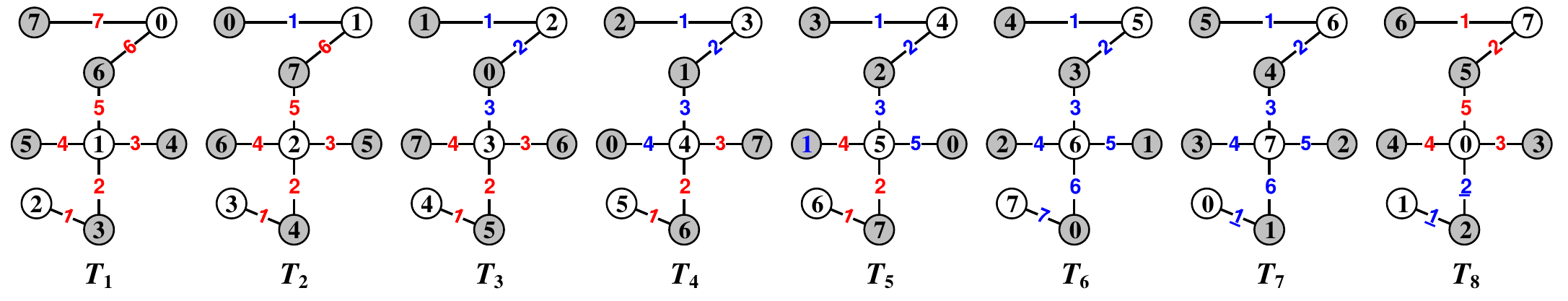}\\
\caption{\label{fig:H-5551-graceful-group}{\small An every-zero Hanzi-group made by a Hanzi-gpw $H^{gpw}_{5551}$ under modular $8$, cited from \cite{Zhang-Yang-Mu-Zhao-Sun-Yao-2019}.}}
\end{figure}

\begin{figure}[h]
\centering
\includegraphics[width=14cm]{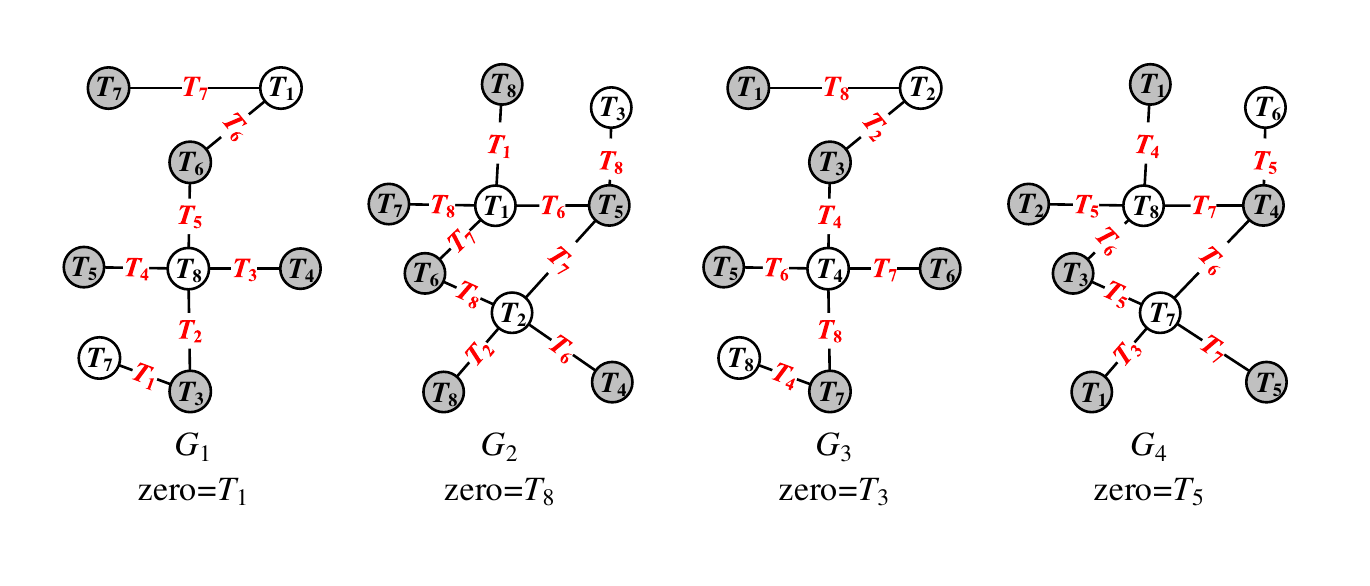}\\
\caption{\label{fig:H-5551-group-coloring}{\small Each Hanzi-graph $G_i$ admits a graphic group-labeling $\theta_i$ for $i\in [1,4]$ based on the Hanzi-group shown in Fig.\ref{fig:H-5551-graceful-group}, cited from \cite{Zhang-Yang-Mu-Zhao-Sun-Yao-2019}.}}
\end{figure}

A \emph{every-zero Hanzi-group} is shown in Fig.\ref{fig:H-5551-graceful-group} and a \emph{graphic group-labeling} is shown in Fig.\ref{fig:H-5551-group-coloring}. Observe Fig.\ref{fig:H-5551-group-coloring} carefully, we can see the following facts:
\begin{asparaenum}[\textbf{\textrm{GL}}-1 ]
\item $G_1$ admits a \emph{graceful graphic group-labeling} $\theta_1$, since $\theta_1(E(G_1))=\{T_1,T_2,\dots ,T_7\}$.
\item $G_2$ is a bipartite graph with vertex bipartition $(X,Y)$, and $\max\{i: T_i\in \theta_2(X)\}<\min\{j: T_j\in \theta_2(Y)\}$, so we say $\theta_2$ to be \emph{set-ordered}.
\item $G_3$ admits a \emph{graphic group-labeling} $\theta_3$ holding $\theta_3(x)\neq \theta_3(y)$ for any pair of vertices $x,y\in V(G_3)$.
\item The \emph{graphic group-labeling} $\theta_4$ of $G_4$ is the \emph{dual labeling} of $\theta_2$, since the indexes of $\theta_2(u)=T_i$ and $\theta_4(u)=T_{9-i}$ for $u\in V(G_2)=V(G_4)$ hold $i+(9-i)=9$.
\item Because $\theta_k(u_i)\neq \theta_k(u_j)$ for $u_i,u_j\in N(u)$ with $k\in [1,4]$, so we call each $\theta_k$ to be \emph{edge-proper graphic group-labeling}.
\item Because $\{\theta_k(u_i):~u_i\in N(u)\}\neq \{\theta_k(v_i):~v_i\in N(v)\}$ for any edge $uv\in E(G_i)$ with $k\in [1,4]$, so we call each $\theta_k$ to be \emph{adjacent edge distinguishing}.
\end{asparaenum}

We write a number-based string from $G_1$ shown in Fig.\ref{fig:H-5551-group-coloring} as follows
\begin{equation}\label{eqa:string-group-Hanzi-gpw}
{
\begin{split}
T_b(G_1)=&T_b(T_7)T_b(T_7)T_b(T_1)T_b(T_6)T_b(T_6)T_b(T_6)T_b(T_5)T_b(T_8)T_b(T_2)T_b(T_3)\\
&T_b(T_1)T_b(T_7)T_b(T_5)T_b(T_4)T_b(T_8)T_b(T_3)T_b(T_4)
\end{split}}
\end{equation}
where each number-based string $T_b(T_i)$ for $i\in [1,8]$ is:

$T_b(T_1)=$77066651231254134, $T_b(T_2)=$01167752241364235,

$T_b(T_3)=$11220033251474336, $T_b(T_4)=$21321134261504437,

$T_b(T_5)=$31422235271614550, $T_b(T_6)=$41523336607724651,

$T_b(T_7)=$51624437611034752, $T_b(T_8)=$61725550221244033.\\
Notice that $T_b(G_1)$ consists of $17\times 17=289$ numbers in $[0,9]$.

\subsubsection{Hanzi randomly labelings}

\begin{figure}[h]
\centering
\includegraphics[width=16.4cm]{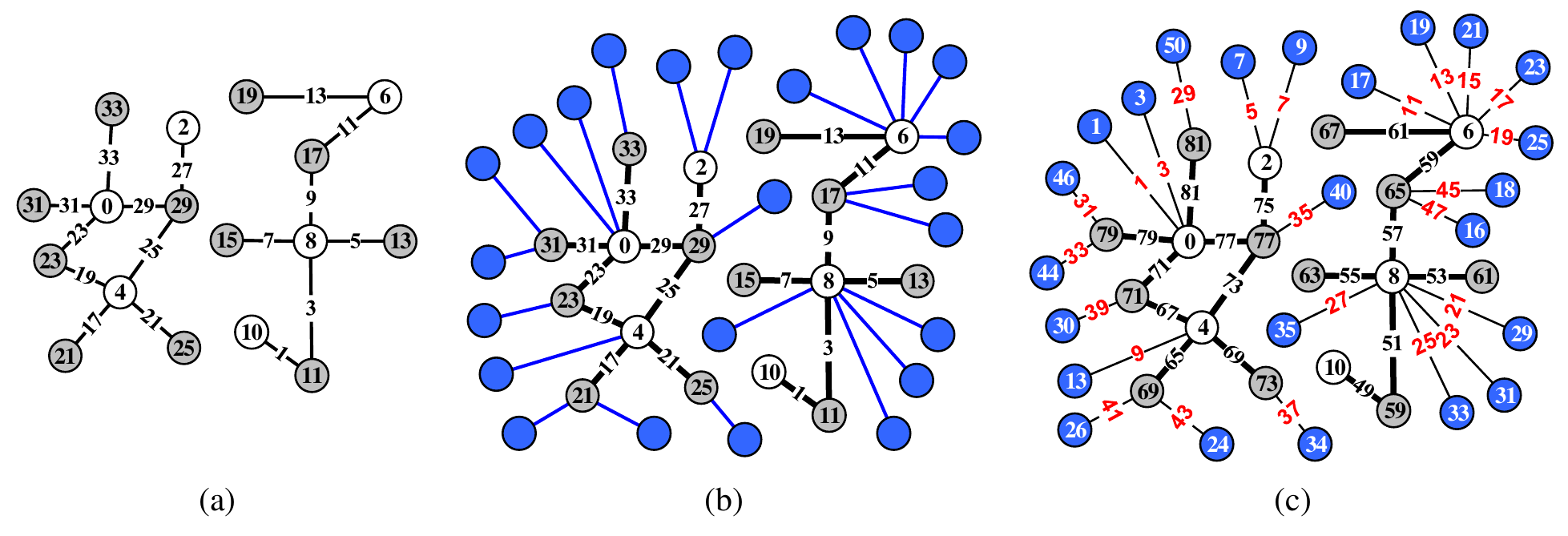}\\
\caption{\label{fig:H-2635-add-leaves}{\small (a) A Hanzi-gpw $H^{gpw}_{2635}$ admitting a flawed set-ordered odd-graceful labeling; (b) adding randomly leaves to $H^{gpw}_{2635}$ for forming a new graph $L$; (c) a flawed odd-graceful labeling of the graph $L$, cited from \cite{Zhang-Yang-Mu-Zhao-Sun-Yao-2019}.}}
\end{figure}

By \emph{adding randomly leaves} to Hanzi-gpws, we can make \emph{probabilistic number-based strings}. In Fig.\ref{fig:H-2635-add-leaves}, a Hanzi-gpw $H^{gpw}_{2635}$ admitting a flawed set-ordered odd-graceful labeling (refer to Definition \ref{defn:flawed-labeling} and Definition \ref{defn:flawed-odd-graceful-labeling}) is added randomly leaves, the resultant graph $L$ admits a flawed odd-graceful labeling, which produces a Topcode-matrix for generating probabilistic number-based strings as desired.

\begin{thm}\label{thm:xxxxxx}
\cite{Zhang-Yang-Mu-Zhao-Sun-Yao-2019} Suppose that a graphic group $G_{roup}=\{F_f(G),\oplus\}$ is equivalent to another graphic group $L_{roup}=\{F_g(L),\oplus\}$. Then a graph $R$ admits a \emph{graphic group-labeling} $\psi$ on $G_{roup}$, which is equivalent to another graphic group-labeling $\omega$ of the graph $R$ based on $L_{roup}$, that is, $\psi \sim \omega$.
\end{thm}

\subsubsection{Derivative hanzi-systems for generating Hanzi-gpws}

\begin{figure}[h]
\centering
\includegraphics[width=15.4cm]{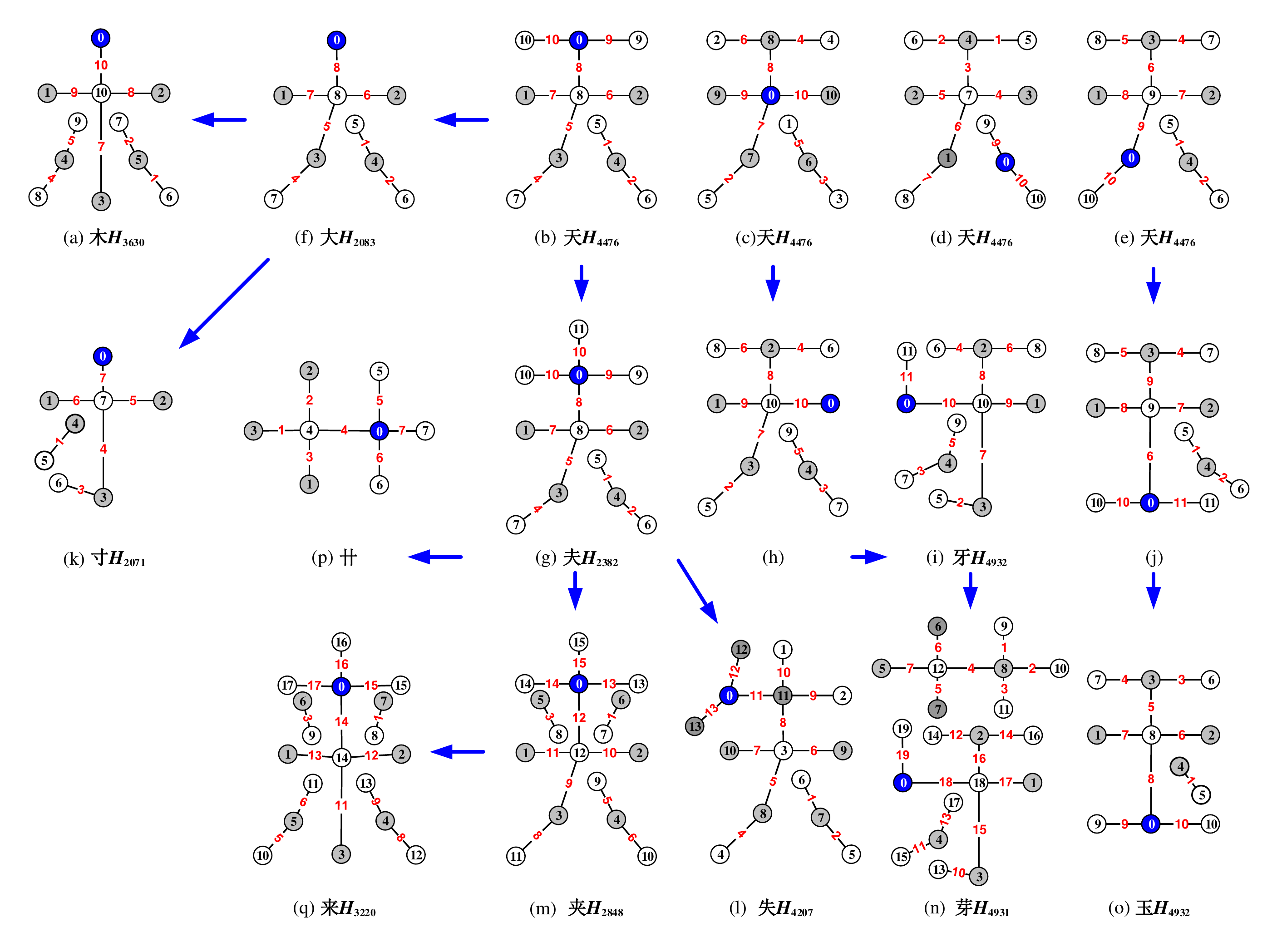}\\
\caption{\label{fig:H-4476-derivative-system}{\small A derivative hanzi-system built on a Hanzi $H_{4476}$, cited from \cite{Yao-Mu-Sun-Sun-Zhang-Wang-Su-Zhang-Yang-Zhao-Wang-Ma-Yao-Yang-Xie2019}.}}
\end{figure}

\begin{figure}[h]
\centering
\includegraphics[width=12cm]{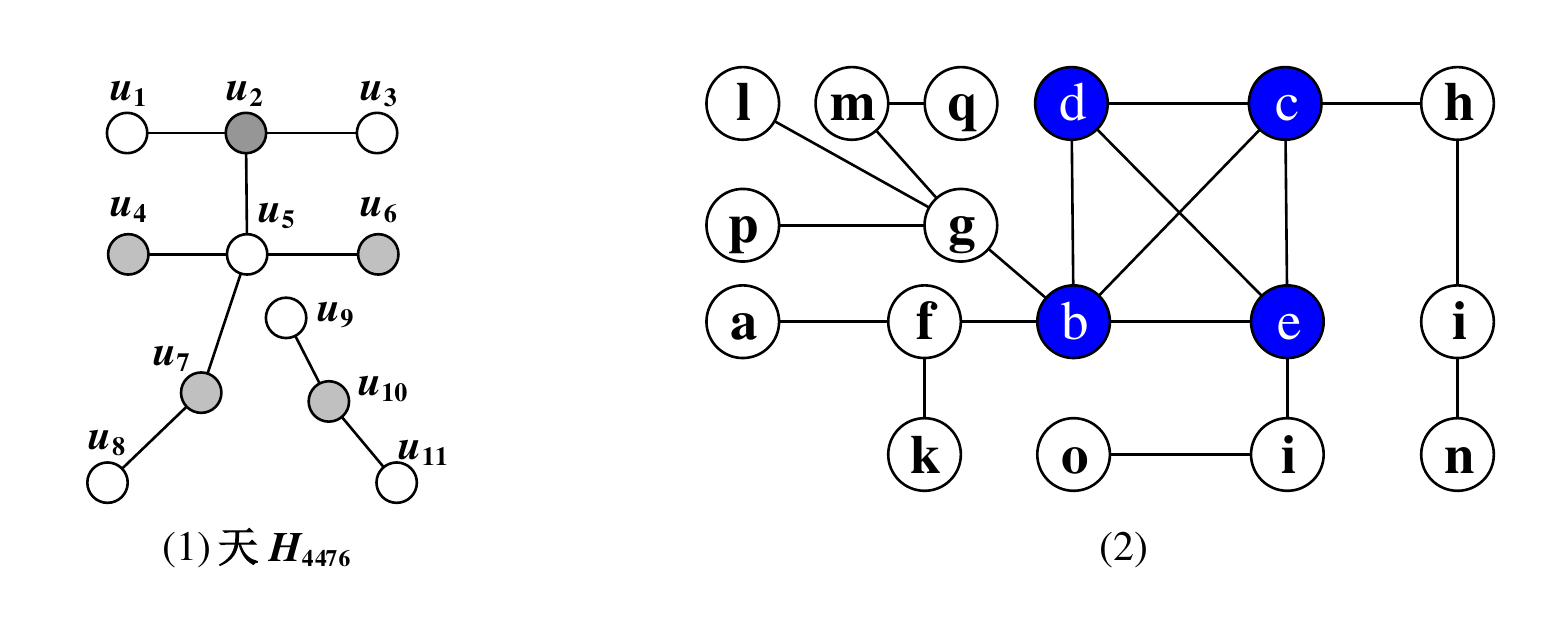}\\
\caption{\label{fig:H-4476-system-graph}{\small (1) A Hanzi-graph $T_{4476}$; (2) a hz-$\varepsilon$-graph made by the derivative Hanzi-system shown in Fig.\ref{fig:H-4476-derivative-system}, where $\varepsilon$ is the graceful labeling, cited from \cite{Yao-Mu-Sun-Sun-Zhang-Wang-Su-Zhang-Yang-Zhao-Wang-Ma-Yao-Yang-Xie2019}.}}
\end{figure}

In fact, determining the $0$-rotatable gracefulness of trees is not slight, even for caterpillars (Ref. \cite{Zhou-Yao-Cheng-2011}). A hz-$\varepsilon$-graph $G$ (see Fig.\ref{fig:H-4476-derivative-system} and Fig.\ref{fig:H-4476-system-graph}) has its vertices corresponding Hanzi-gpws, and two vertices of $G$ are adjacent to each other if the corresponding Hanzi-gpws $H^{gpw}_{u_i}$ having its $\varepsilon$-labeling $f_i$ and topological structure $T_{u_i}$ with $i=1,2$ can be transformed to each other by the following rules, where both $f_1$ and $f_2$ are the same $W$-type of labelings, for example, both $f_1$ and $f_2$ are flawed graceful labelings, and so on.

\begin{asparaenum}[\textbf{Rule}-1. ]
\item \label{item:dual-labeling} The labeling of $H^{gpw}_{u_i}$ is the \emph{dual labeling} of the labeling of $H^{gpw}_{u_{3-i}}$ with $i=1,2$, so $T_{u_1}\cong T_{u_2}$, here $T_{u_i}$ is the topological structure of the Hanzi-gpw $H^{gpw}_{u_i}$ with $i=1,2$.
\item \label{item:adding-reducing-v-e} $H^{gpw}_{u_i}$ is obtained by adding (reducing) vertices and edges to $H^{gpw}_{u_{3-i}}$, such that $f_i$ can be deduced by $f_{3-i}$ with $i=1,2$.
\item \label{item:complete-same} $H^{gpw}_{u_1}=H^{gpw}_{u_2}$, namely, $f_1=f_2$ and $T_{u_1}\cong T_{u_2}$.
\item \label{item:image} $H^{gpw}_{u_i}$ is an \emph{image} of Hanzi-gpw $H^{gpw}_{u_{3-i}}$ with $i=1,2$.
\item \label{item:inverse} $H^{gpw}_{u_i}$ is the \emph{inverse} of Hanzi-gpw $H^{gpw}_{u_{3-i}}$ with $i=1,2$.
\item \label{item:same-structure-different-lab} $T_{u_1}\cong T_{u_2}$, but $f_1\neq f_2$.
\end{asparaenum}

A \emph{generalized Hanzi-gpw system} shown in Fig.\ref{fig:H-4476-derivative-system} is based on a Hanzi-graph $T_{4476}$ shown in Fig.\ref{fig:H-4476-system-graph} (1). Based on the Hanzi-graph $T_{4476}$, four Hanzi-gpws $H^{gpw_j}_{4476}$ with $j\in[1,4]$ are (b)$=H^{gpw_1}_{4476}$, (c)$=H^{gpw_2}_{4476}$, (d)$=H^{gpw_3}_{4476}$ and (e)$=H^{gpw_4}_{4476}$ respectively, which tell us that the Hanzi-graph $T_{4476}$ admits \emph{flawed graceful $0$-rotatable labelings}, since each vertex of the Hanzi-graph $T_{4476}$ can be labelled with $0$ by some graceful labeling of $T_{4476}$ under \textbf{Rule}-\ref{item:dual-labeling}. Because each Hanzi-gpws $H^{gpw_j}_{4476}$ admit a \emph{flawed set-ordered graceful labeling} $f_j$ with $j\in[1,4]$, so $f_j(V(H^{gpw_j}_{4476}))=[0,10]$ for $j\in[1,4]$.

Clearly, our hz-$\varepsilon$-graphs are Hanzi-gpws too. In Fig.\ref{fig:H-4476-derivative-system}, we can see the following facts:

\begin{asparaenum}[(1) ]
\item Two flawed set-ordered graceful labelings in Fig.\ref{fig:H-4476-derivative-system}(c) and (h) are \emph{dual} from each other. Again, adding a new vertex and a new edge makes (i)$=H^{gpw}_{4932}$; then (n)$=H^{gpw}_{4931}$ is obtained by adding 8 new vertices and 7 new edges to (i).

\item By \textbf{Rule}-\ref{item:adding-reducing-v-e}, $H^{gpw}_{4476} \rightarrow H^{gpw}_{2328}$ from (b) to (g) after adding a vertex and an edge to $H^{gpw}_{4476}$.

\item Do \textbf{Rule}-\ref{item:dual-labeling} to (g)$=H^{gpw}_{2328}$, and adding two vertices and two edges produces (l)$=H^{gpw}_{4207}$ by \textbf{Rule}-\ref{item:adding-reducing-v-e}.

\item Deleting two vertices and two edges from (b)$=H^{gpw}_{4476}$, under \textbf{Rule}-\ref{item:adding-reducing-v-e}, so obtaining (f)$=H^{gpw}_{2083}$; and deleting a vertex and an edge of (f)$=H^{gpw}_{2083}$ produces (k)$=H^{gpw}_{2071}$; and do \textbf{Rule}-\ref{item:adding-reducing-v-e} two times for getting (a)$=H^{gpw}_{3630}$.

\item The Hanzi-gpw (q)$=H^{gpw}_{3220}$ is the result of adding four new vertices and two new edges, and deleting an old vertex and an old edge to (m)$=H^{gpw}_{2848}$.
\end{asparaenum}

\subsection{Hanzi-matrices and their operations}

A \emph{pan-matrix} has its own elements from a set $W$, for instance, $W$ is a letter set, or a Hanzi (Chinese characters) set,
or a graph set, or a poem set, or a music set, even a picture set, \emph{etc.}

\begin{defn}\label{defn:Hanzi-GB2312-80-matrix}
\cite{Yao-Mu-Sun-Sun-Zhang-Wang-Su-Zhang-Yang-Zhao-Wang-Ma-Yao-Yang-Xie2019} A \emph{GB-Hanzi-matrix} (\cite{GB2312-80}) $A_{han}(H)$ of a Hanzi-sentence $H=\{ H_{a_ib_ic_id_i}\}^m_{i=1}$ made by $m$ Hanzis $H_{a_1b_1c_1d_1}$, $H_{a_2b_2c_2d_2}$, $\dots$, $H_{a_mb_mc_md_m}$ is defined as
\begin{equation}\label{eqa:a-formula}
\centering
{
\begin{split}
A_{han}(H)= \left(
\begin{array}{ccccc}
a_{1} & a_{2} & \cdots & a_{m}\\
b_{1} & b_{2} & \cdots & b_{m}\\
c_{1} & c_{2} & \cdots & c_{m}\\
d_{1} & d_{2} & \cdots & d_{m}
\end{array}
\right)_{4\times m}=\left(\begin{array}{c}
A\\
B\\
C\\
D
\end{array} \right)=(A,B,C,D)^{T}
\end{split}}
\end{equation}\\
with $A=(a_1, a_2, \dots ,a_m)$, $B=(b_1, b_2 , \dots ,b_m)$, $C=(c_1, c_2, \dots ,c_m)$, and $D=(d_1, d_2, \dots ,d_m)$, where each Chinese code $a_ib_ic_id_i$ is defined in \cite{GB2312-80}.\qqed
\end{defn}

We have a GB-Hanzi-matrix

\begin{equation}\label{eqa:Hanzi-sentence-GB2312-80}
\centering
{
\begin{split}
A_{han}(G^*)&= \left(
\begin{array}{ccccccccc}
4 & 4 & 2 & 2 & 5 & 4 & 4 & 4 & 3\\
0 & 0 & 6 & 5 & 2 & 4 & 7 & 4 & 8\\
4 & 4 & 3 & 1 & 8 & 7 & 3 & 1 & 2\\
3 & 3 & 5 & 1 & 2 & 6 & 4 & 1 & 9
\end{array}
\right)_{4\times 9}
\end{split}}
\end{equation}
according to a Hanzi-sentence $G^*=H_{4043}$ $H_{4043}$ $H_{2635}$ $H_{2511}$ $H_{5282}$ $H_{4476}$ $H_{4734}$ $H_{4411}$ $H_{3829}$. This Hanzi-sentence $G^*$ induces a CCD-Hanzi-matrix as follows:
\begin{equation}\label{eqa:Hanzi-sentence-code}
\centering
{
\begin{split}
A_{CCD}(G^*)&= \left(
\begin{array}{ccccccccc}
4 & 4 & 5 & 5 & 5 & 5 & 4 & 5 & 5\\
E & E & 9 & 1 & 2 & 9 & E & 9 & E\\
B & B & 7 & 6 & 1 & 2 & 0 & 1 & 7\\
A & A & D & C & 9 & 9 & B & A & 3
\end{array}
\right)_{4\times 9}
\end{split}}
\end{equation} by the Chinese code of Chinese dictionary. Furthermore, in GBK Coding \cite{GBK-Coding-2010}, this Hanzi-sentence $G^*$ produces a GBK-Hanzi-matrix
\begin{equation}\label{eqa:Hanzi-sentence-GBK-Coding}
\centering
{
\begin{split}
A_{GBK}(G^*)&= \left(
\begin{array}{ccccccccc}
C & C & B & B & D & C & C & C & C\\
8 & 8 & A & 9 & 4 & C & F & C & 6\\
C & C & C & A & F & E & C & A & B\\
B & B & 4 & B & 2 & C & 2 & B & D
\end{array}
\right)_{4\times 9}.
\end{split}}
\end{equation}

Each of three Hanzi-matrices $A_{han}(G^*)$, $A_{CCD}(G^*)$ and $A_{GBK}(G^*)$ introduced above can produces a group of $36!$ number-based/text-based strings, so a number-based string $s_{han}(k)$ generated from the Hanzi-matrix $A_{han}(G^*)$ corresponds a text-based string $s_{CCD}(k)$ generated from the Hanzi-matrix $A_{CCD}(G^*)$, or a text-based string $s_{GBK}(k)$ generated from the Hanzi-matrix $A_{GBK}(G^*)$. However, it is not easy to find out the Hanzi-sentence $G^*$ from these number-based/text-based strings $s_{han}(i)$, $s_{CCD}(j)$ and $s_{GBK}(k)$ if a Hanzi-sentence made by tens of thousands of Chinese characters.

For two GB-Hanzi-matrices $A^r_{han}$ and $A^c_{han}$ defined as follows:
\begin{equation}\label{eqa:han-matrix-1}
\centering
{
\begin{split}
A^r_{han}&= \left(
\begin{array}{ccccc}
a_{1} & b_{1} & c_{1} & d_{1}\\
a_{2} & b_{2} & c_{2} & d_{2}\\
\cdots & \cdots& \cdots &\cdots\\
a_{m} & b_{m} & c_{m} & d_{m}
\end{array}
\right)_{m\times 4}=\left(\begin{array}{c}
X_1\\
X_2\\
\cdots\\
X_m
\end{array} \right)=(X_1,~X_2,~\cdots ,~X_m)^{T}
\end{split}}
\end{equation}
where $X_k=(a_{k}~ b_{k}~ c_{m}~ d_{k})$ with $k\in [1,m]$ corresponds a Hanzi $H_{a_{k}b_{k}c_{m}d_{k}}$ in \cite{GB2312-80}, and
\begin{equation}\label{eqa:han-matrix-11}
\centering
{
\begin{split}
A^c_{han}&= \left(
\begin{array}{ccccc}
x_{1} & x_{2} & \cdots & x_{n}\\
y_{1} & y_{2} & \cdots & y_{n}\\
z_{1} & z_{2} & \cdots & z_{n}\\
w_{1} & w_{2} & \cdots & w_{n}
\end{array}
\right)_{4\times n}=(Y_1,~Y_2,~\cdots ,~Y_n)
\end{split}}
\end{equation}
where $Y_j=(x_{j}~ y_{j}~ z_{j}~ w_{j})^{T}$ with $j\in [1,n]$ corresponds a Hanzi $H_{x_{j}y_{j}z_{j}w_{j}}$ in \cite{GB2312-80}, the authors in \cite{Yao-Zhao-Mu-Sun-Zhang-Zhang-Yang-IAEAC-2019} define an operation on GB-Hanzi-matrices as follows:
\begin{equation}\label{eqa:han-matrix-12}
\centering
{
\begin{split}
&\quad A^{r(\bullet) c}_{han}=A^r_{han}(\bullet) A^c_{han}=(X_1~X_2~\cdots ~X_m)^{T}_{m\times 4}(\bullet) (Y_1~Y_2~\cdots ~Y_n)_{4\times n}=(x_{i,j})_{m\times n}
\end{split}}
\end{equation}
where $A^{r\times c}_{han}=(\alpha_{i,j})_{m\times n}$ and ``$(\bullet)$'' is an \emph{abstract operation}. Two operations are defined as follows:

\vskip 0.2cm

(i) \emph{\textbf{Multiplication ``$\bullet $'' on components of two vectors}.} We define
\begin{equation}\label{eqa:han-matrix-multiplication}
{
\begin{split}
X_k\bullet Y_j&=(a_{k}~ b_{k}~ c_{k}~ d_{k})\bullet (x_{j}~ y_{j}~ z_{j}~ w_{j})^{T}=a_{k,j}b_{k,j}c_{k,j}d_{k,j}
\end{split}}
\end{equation}
where $a_{k,j}=a_{k}\cdot x_{j}~(\bmod~10)$, $b_{k,j}=b_{k}\cdot y_{j}~(\bmod~10)$, $c_{k,j}=c_{k}\cdot z_{j}~(\bmod~10)$ and $d_{k,j}=d_{k}\cdot w_{j}~(\bmod~10)$. So
\begin{equation}\label{eqa:555555}
A^{r\bullet c}_{han}=A^r_{han}\bullet A^c_{han}=(\alpha_{i,j})_{m\times n}.
\end{equation}

\begin{figure}[h]
\centering
\includegraphics[width=9cm]{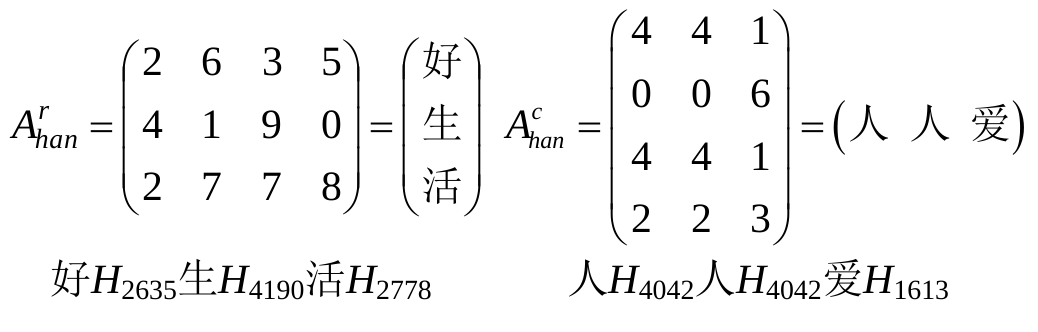}\\
\includegraphics[width=10cm]{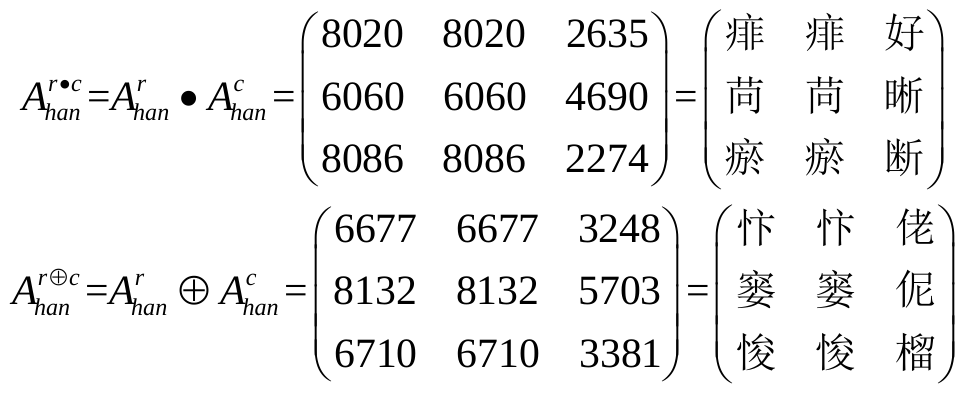}\\
\caption{\label{fig:hanzi-matrices-2}{\small Two GB-Hanzi-matrices $A^{r\bullet c}_{han}$ and $A^{r\oplus c}_{han}$ obtained by the multiplication ``$\bullet $'' and the addition ``$\oplus$'' operations to two GB-Hanzi-matrices $A^{r}_{han}$ and $A^{c}_{han}$.}}
\end{figure}

(ii) \emph{\textbf{Addition ``$\oplus$'' on components of two vectors}.} We define
\begin{equation}\label{eqa:han-matrix-addition}
{
\begin{split}
X_k\oplus Y_j&=(a_{k}~ b_{k}~ c_{m}~ d_{k})\oplus (x_{j}~ y_{j}~ z_{j}~ w_{j})^{T}=\alpha_{k,j}\beta_{k,j}\gamma_{k,j}\delta_{k,j}
\end{split}}
\end{equation}
where $\alpha_{k,j}=a_{k}+x_{j}~(\bmod~10)$, $\beta_{k,j}=b_{k}+y_{j}~(\bmod~10)$, $\gamma_{k,j}=c_{m}+z_{j}~(\bmod~10)$ and $\delta_{k,j}=d_{k}+w_{j}~(\bmod~10)$, that is
\begin{equation}\label{eqa:555555}
A^{r\oplus c}_{han}=A^r_{han}\oplus A^c_{han}=(\beta_{i,j})_{m\times n}.
\end{equation}
See examples shown in Fig.\ref{fig:hanzi-matrices-2}.

Thereby, each element of the matrix $A^{r\bullet c}_{han}=(\alpha_{i,j})_{m\times n}$, or $A^{r\oplus c}_{han}=(\beta_{i,j})_{m\times n}$ is a number of four numbers which corresponds a Hanzi in \cite{GB2312-80}, so we call both $A^{r\bullet c}_{han}$ and $A^{r\oplus c}_{han}$ two \emph{GB-Hanzi-matrices}. On the other hands, we can regard $A^{r\bullet c}_{han}$, or $A^{r\oplus c}_{han}$ as a Hanzi-matrix \emph{authentication} from the \emph{public key} $A^r_{han}$ and the \emph{private key} $A^c_{han}$, see example shown in Fig.\ref{fig:hanzi-matrices-2}.

\begin{rem}\label{rem:ABC-conjecture}
Various Hanzi-matrices differing from Topcode-matrices are new and introduce Chinese characters into mathematical regions, and will be liked by Chinese people, since Chinese characters are easily put into computer and equipments by speaking and hand-writings.\paralled
\end{rem}

\subsection{Hanzi-Lattices}

\begin{figure}[h]
\centering
\includegraphics[width=14cm]{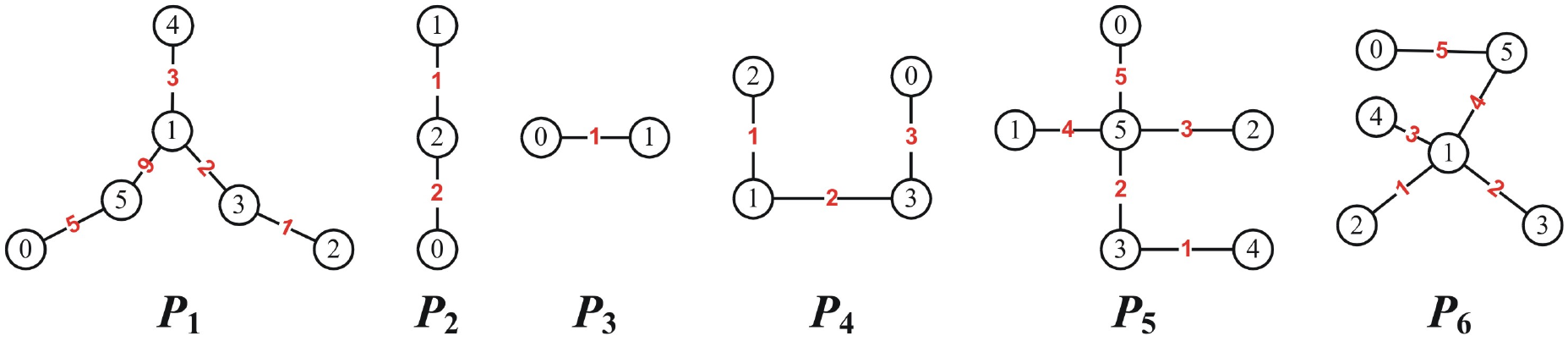}\\
\caption{\label{fig:pianpang-11}{\small Six Pianpangs in Chinese characters.}}
\end{figure}

In \cite{Wang-Yao-Su-Wanjia-Zhang-2021-IMCEC}, the authors point out that lattice not only has its important applications in information security and mathematics, but also its potential in other fields. As known, there are 250 ``Pianpang'' (see Fig.\ref{fig:pianpang-11}) and ``Bushou'' in Chinese characters (Hanzis) according to ``Chinese Character GB18030-2000'' containing 27484 Chinese characters, where ``Pianpang'' is the right and left part of a Chinese character, and ``Bushou'' can arrange Chinese characters; we have 299 ``Dutizi''; and there are 16 punctuation being the marks to clarify meaning by indicating separation of words into sentences and clauses and phrases. Let $H_{an}(1),H_{an}(2),\dots ,H_{an}(565)$ indicate all of ``Pianpang'', ``Bushou'', ``Dutizi'' and punctuation in Chinese characters, so we call $\textbf{\textrm{H}}_{an}=(H_{an}(1),H_{an}(2),\dots ,H_{an}(565))$ a \emph{Hanzi-Pianpang base}. Thereby, we can build up a \emph{Hanzi-literary lattice} $\textbf{\textrm{L}}(Z^0\textbf{\textrm{H}}_{an})$ of various Chinese writings by
\begin{equation}\label{eqa:11-Hanzi-lattice}
\textbf{\textrm{L}}(Z^0\textbf{\textrm{H}}_{an})=\left \{\bigcup ^{565}_{k=1}x_kH_{an}(k):x_k\in Z^0,H_{an}(k)\in \textbf{\textrm{H}}_{an}\right \}
\end{equation}
with $\sum ^{565}_{k=1}x_k\geq 1$. Let $C_{para}=\sum ^{565}_{k=1}x_kH_{an}(k)$ in $\textbf{\textrm{L}}(Z^0\textbf{\textrm{H}}_{an})$. We can confirm the following facts:
\begin{thm}\label{thm:Hanzi-literary-lattice}
\cite{Wang-Yao-Su-Wanjia-Zhang-2021-IMCEC} (1) Chinese paragraphs $C_{para}$ can be used in text-based passwords to make more complicated \emph{public-keys} and \emph{private-keys} for increasing the cost of decryption and attackers.

(2) Each Chinese writing limited in Chinese Character GB18030-2000 has been contained in the Hanzi-literary lattice $\textbf{\textrm{L}}(Z^0\textbf{\textrm{H}}_{an})$, such as poems, novels, essay, proses, reports \emph{et al.}
\end{thm}

\begin{figure}[h]
\centering
\includegraphics[width=16.4cm]{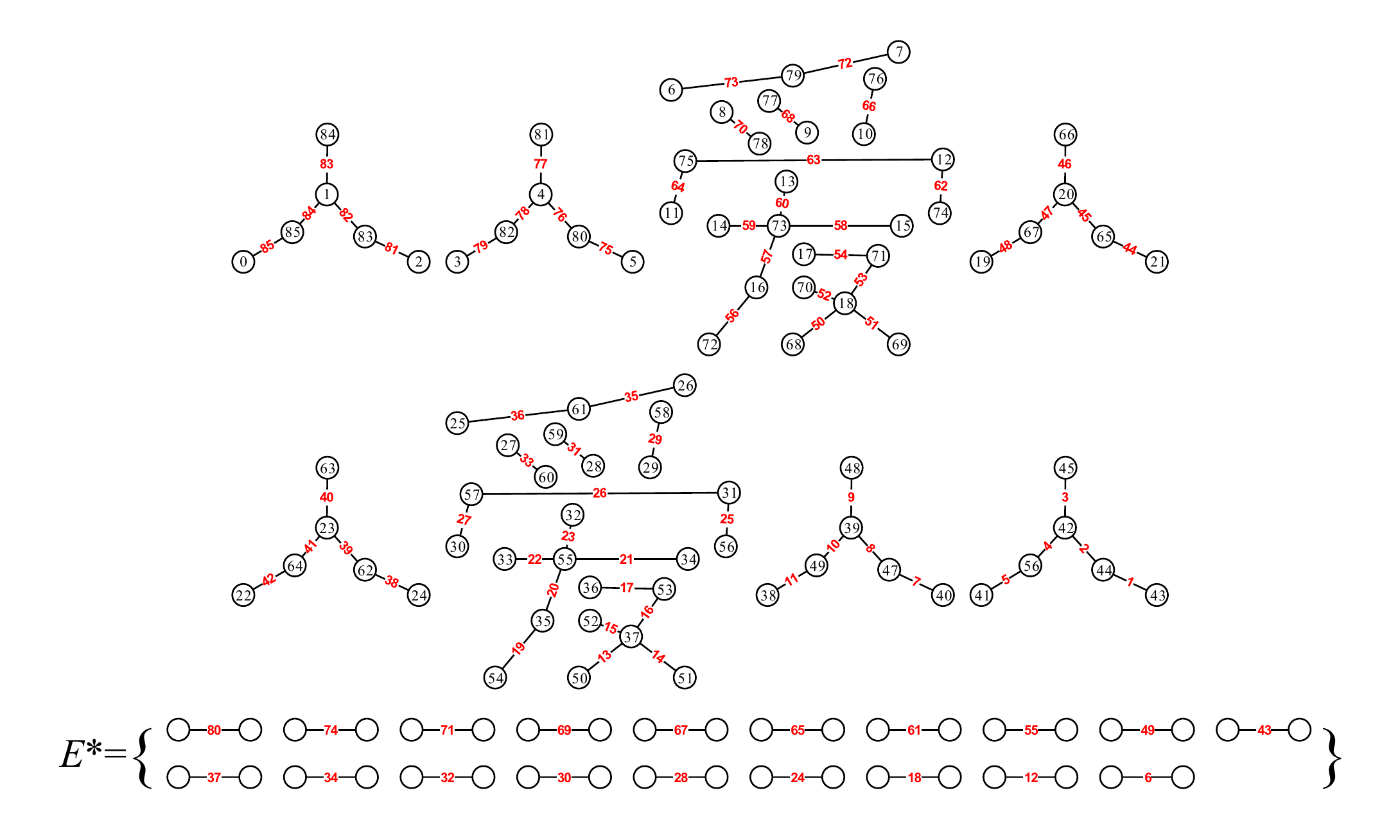}\\
\caption{\label{fig:pianpang-33}{\small A Hanzi-sentence $H_{an}(G)$ made by six Pianpangs shown in Fig.\ref{fig:pianpang-11} admits a flawed graceful labeling, cited from \cite{Wang-Yao-Su-Wanjia-Zhang-2021-IMCEC}.}}
\end{figure}

For example, Fig.\ref{fig:pianpang-33} shows us a Hanzi-sentence $H_{an}(G)=6P_1\cup 2P_2\cup 6P_3\cup 2P_4\cup 2P_5\cup 2P_6$ made by six Pianpangs shown in Fig.\ref{fig:pianpang-11}, which is a disconnected graph, and this Hanzi-sentence $H_{an}(G)$ admits a flawed graceful labeling $f$ shown in Fig.\ref{fig:pianpang-33}. And the connected graph $H_{an}(G)+E^*$ admits a graceful labeling induced by the flawed graceful labeling $f$, where $E^*$ is a set of 19 disjoint colored edges. It is not hard to see that there are many connected graph $H_{an}(G)+E^*$ admitting graceful labelings. See examples shown in Fig.\ref{fig:Hanzi-lattice-22} and Fig.\ref{fig:Hanzi-lattice-33}.

\begin{thm}\label{thm:colosed-flawed-graceful-Pianpang}
\cite{Wang-Yao-Su-Wanjia-Zhang-2021-IMCEC} If each Pianpang $H\,'_{an}(k)$ of a Hanzi-Pianpang base $\textbf{\textrm{H}}\,'_{an}=(H\,'_{an}(1),H\,'_{an}(2),\dots $, $H\,'_{an}(m))$ admits a set-ordered graceful labeling for $k\in [1,m]$, then each disconnected graph of the lattice $\textbf{\textrm{L}}(Z^0\textbf{\textrm{H}}\,'_{an})=\left \{\bigcup ^{m}_{k=1}x_kH_{an}(k):x_k\in Z^0,H\,'_{an}(k)\in \textbf{\textrm{H}}\,'_{an}\right \}$ admits a \emph{flawed graceful labeling}.
\end{thm}

\begin{rem}\label{rem:333333}
We need to consider the following problems:
\begin{asparaenum}[\textrm{Prob}-1. ]
\item \textbf{List} all Chinese paragraphs in $\textbf{\textrm{L}}(Z^0\textbf{\textrm{H}}_{an})$ with $\sum ^{565}_{k=1}x_k\leq M$ for a fixed integer $M\in Z^0\setminus \{0\}$.
\item \textbf{Judge} whether each $C_{para}$ defined in (\ref{eqa:11-Hanzi-lattice}) is meaningful or meaningless in Chinese.\paralled
\end{asparaenum}
\end{rem}

Considering the above \textrm{Prob}-1 and \textrm{Prob}-2, we can build another \emph{Hanzi-literary lattice}
\begin{equation}\label{eqa:Hanzi-Pianpang-123}
\textbf{\textrm{L}}(Z^0\textbf{\textrm{C}}_{cha})=\left \{\bigcup ^{27500}_{k=1}y_kC_{cha}(k):y_k\in Z^0,C_{cha}(k)\in \textbf{\textrm{C}}_{cha}\right \}
\end{equation}
based on a \emph{Hanzi-base} $\textbf{\textrm{C}}_{cha}=(C_{cha}(1),C_{cha}(2),\dots ,C_{cha}(27500))$ and $\sum ^{27500}_{k=17}y_k\geq 1$, where each $C_{cha}(i)$ with $i\in [1,16]$ is a punctuation, and each $C_{cha}(j)$ for $j\in [17,27500]$ is a Chinese character in Chinese Character GB18030-2000.

\begin{figure}[h]
\centering
\includegraphics[width=16.4cm]{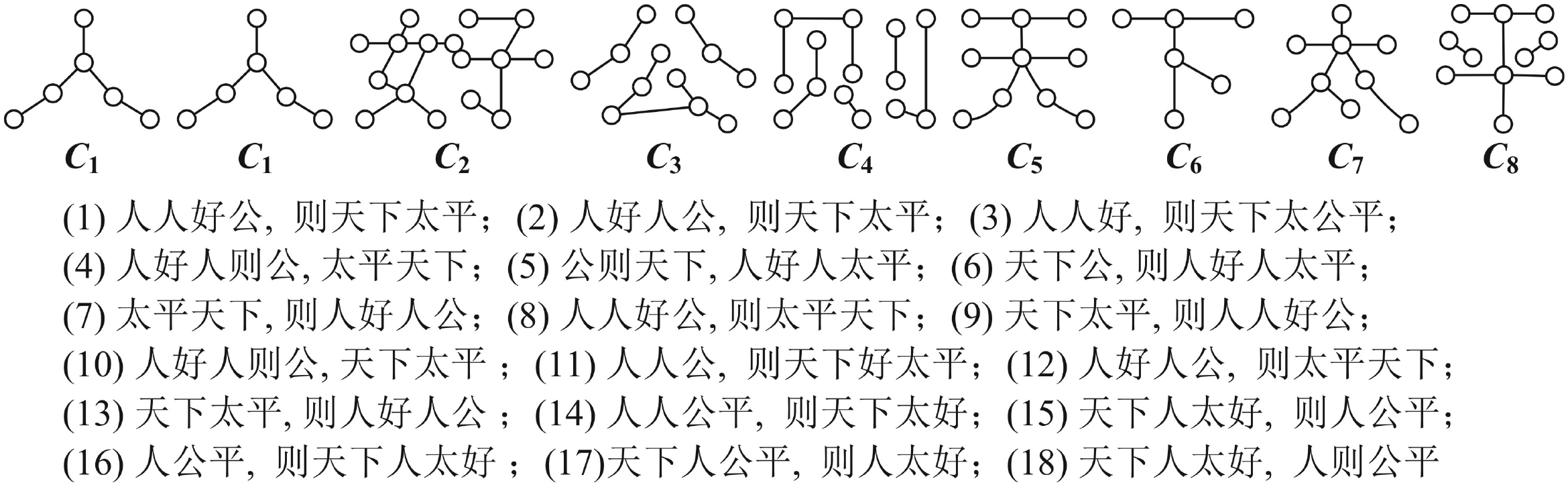}\\
\caption{\label{fig:Hanzi-lattice-11}{\small Nine Hanzis and their combinations, cited from \cite{Wang-Yao-Su-Wanjia-Zhang-2021-IMCEC}.}}
\end{figure}

\begin{figure}[h]
\centering
\includegraphics[width=16cm]{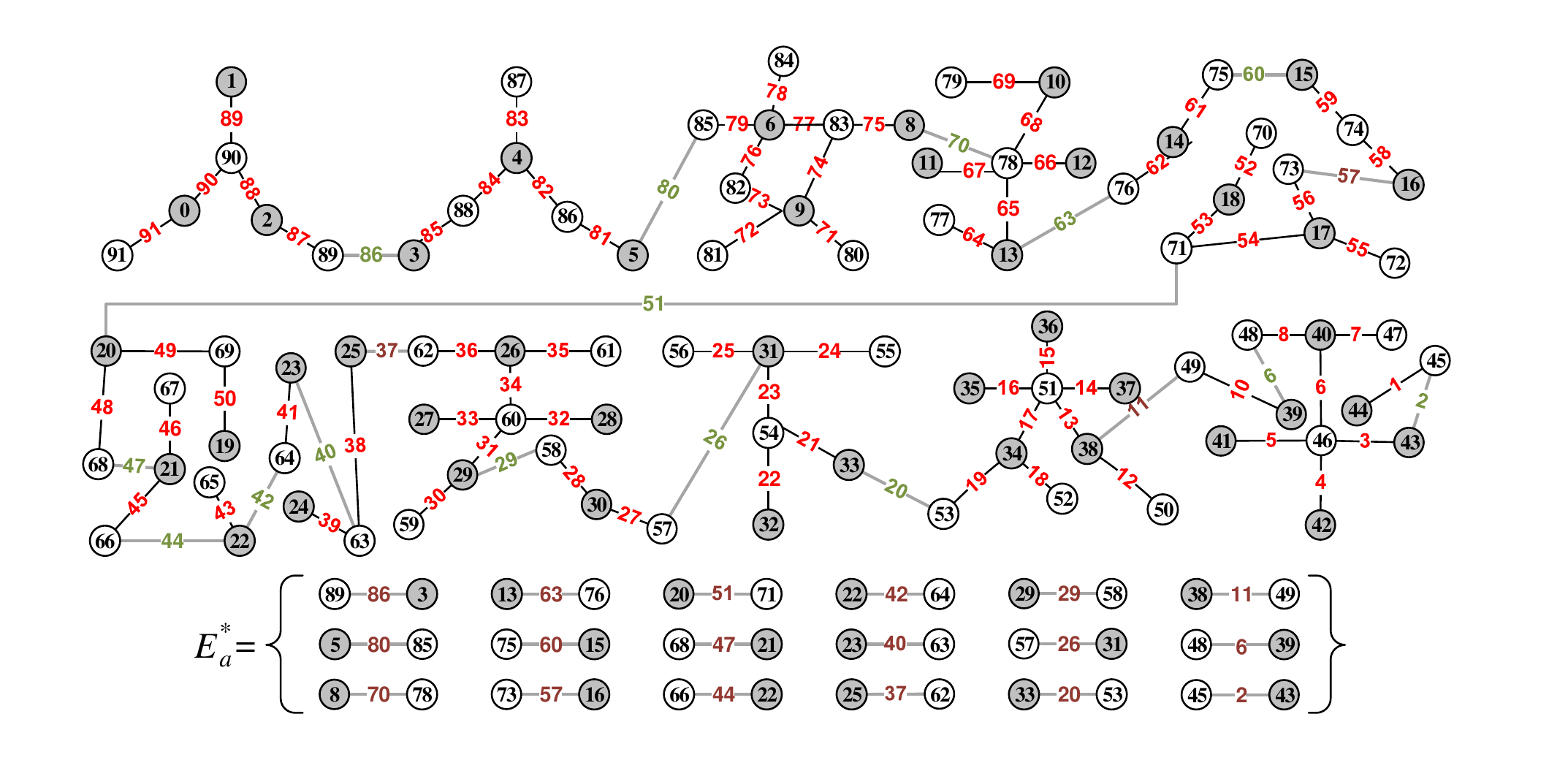}\\
\caption{\label{fig:Hanzi-lattice-22}{\small A Hanzi-sentence made by Eight Hanzis shown in Fig.\ref{fig:Hanzi-lattice-11} admits a flawed graceful labeling, cited from \cite{Yao-Mu-Sun-Sun-Zhang-Wang-Su-Zhang-Yang-Zhao-Wang-Ma-Yao-Yang-Xie2019}.}}
\end{figure}

\begin{figure}[h]
\centering
\includegraphics[width=16cm]{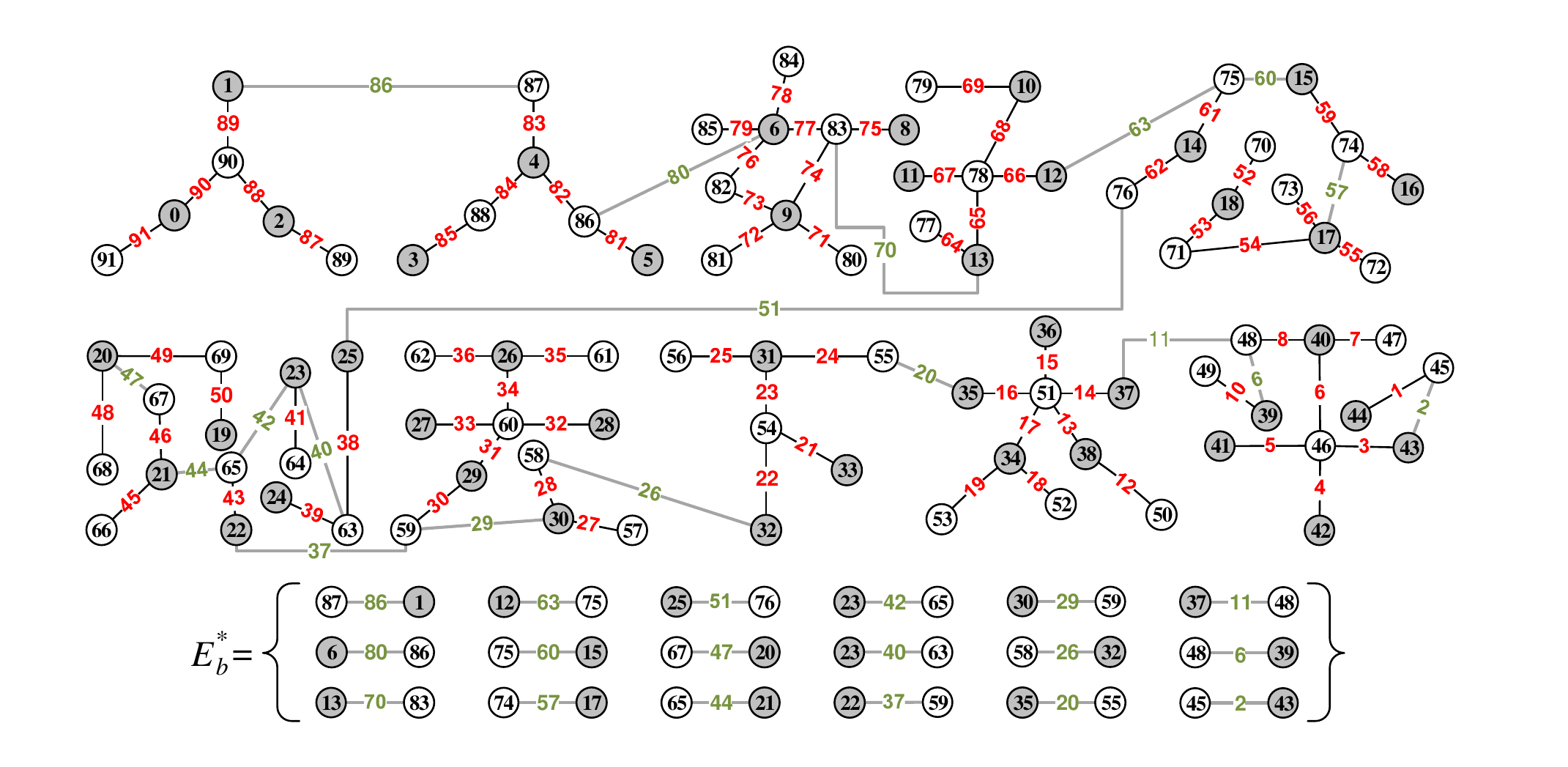}\\
\caption{\label{fig:Hanzi-lattice-33}{\small Another Hanzi-sentence made by Eight Hanzis shown in Fig.\ref{fig:Hanzi-lattice-11} admits a flawed graceful labeling, cited from \cite{Yao-Mu-Sun-Sun-Zhang-Wang-Su-Zhang-Yang-Zhao-Wang-Ma-Yao-Yang-Xie2019}.}}
\end{figure}

According to Fig.\ref{fig:Hanzi-lattice-11}, we have Hanzis $C_k$ for $k\in [1,8]$, and get the following Hanzi-sentences:

(1)$=C_1\cup C_1\cup C_2\cup C_3\cup C_4\cup C_5\cup C_6\cup C_7\cup C_8$;

(2)$=C_1\cup C_2\cup C_1\cup C_3\cup C_4\cup C_5\cup C_6\cup C_7\cup C_8$;

(3)$=C_1\cup C_2\cup C_1\cup C_3\cup C_4\cup C_5\cup C_6\cup C_7\cup C_8$;

(4)$=C_1\cup C_2\cup C_1\cup C_4\cup C_3\cup C_7\cup C_8\cup C_5\cup C_6$.


\begin{thm}\label{thm:colosed-flawed-graceful-Chinese-Character}
\cite{Wang-Yao-Su-Wanjia-Zhang-2021-IMCEC} If each Hanzi-graph $C\,'_{cha}(k)$ of a Hanzi-graph base $\textbf{\textrm{C}}\,'_{cha}=(C\,'_{cha}(1),C\,'_{cha}(2),\dots $, $C\,'_{cha}(n))$ admits a set-ordered graceful labeling (resp. set-ordered odd-graceful labeling) for $k\in [1,n]$, then each disconnected graph of the lattice
$$\textbf{\textrm{L}}(Z^0\cup \textbf{\textrm{C}}\,'_{cha})=\left \{\bigcup ^{n}_{k=1}y_kC\,'_{cha}(k):y_k\in Z^0,C\,'_{cha}(k)\in \textbf{\textrm{C}}'_{cha}\right \}
$$ admits a \emph{flawed graceful labeling} (resp. \emph{flawed odd-graceful labeling}).
\end{thm}

\begin{figure}[h]
\centering
\includegraphics[width=16cm]{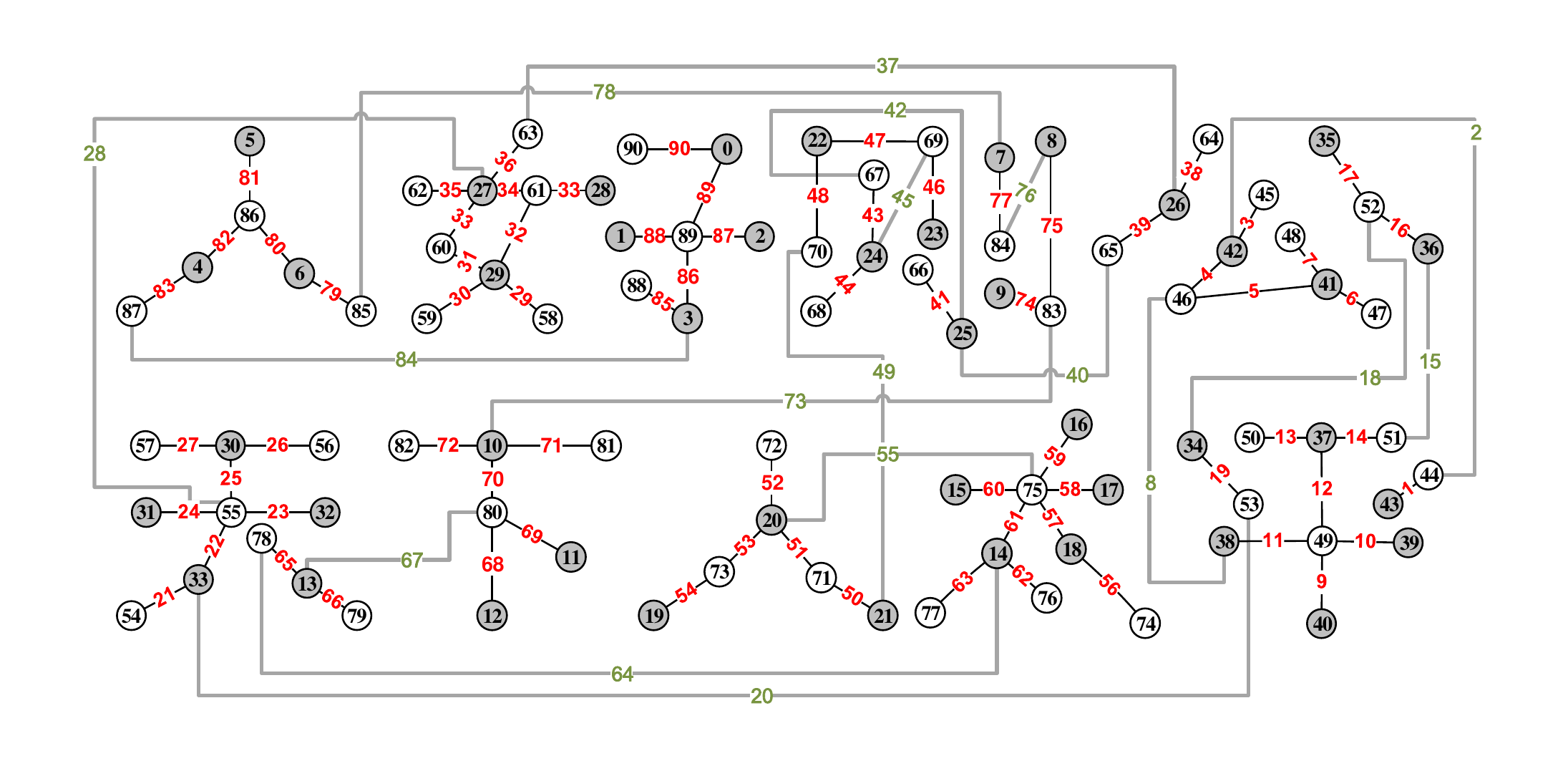}\\
\caption{\label{fig:Hanzi-lattice-44}{\small A Hanzi-sentence made by Hanzi-graphs shown in Fig.\ref{fig:Hanzi-lattice-11} admits a flawed graceful labeling, cited from \cite{Yao-Mu-Sun-Sun-Zhang-Wang-Su-Zhang-Yang-Zhao-Wang-Ma-Yao-Yang-Xie2019}.}}
\end{figure}

\subsection{Transformation between six every-zero graphic groups}

\begin{figure}[h]
\centering
\includegraphics[width=12.4cm]{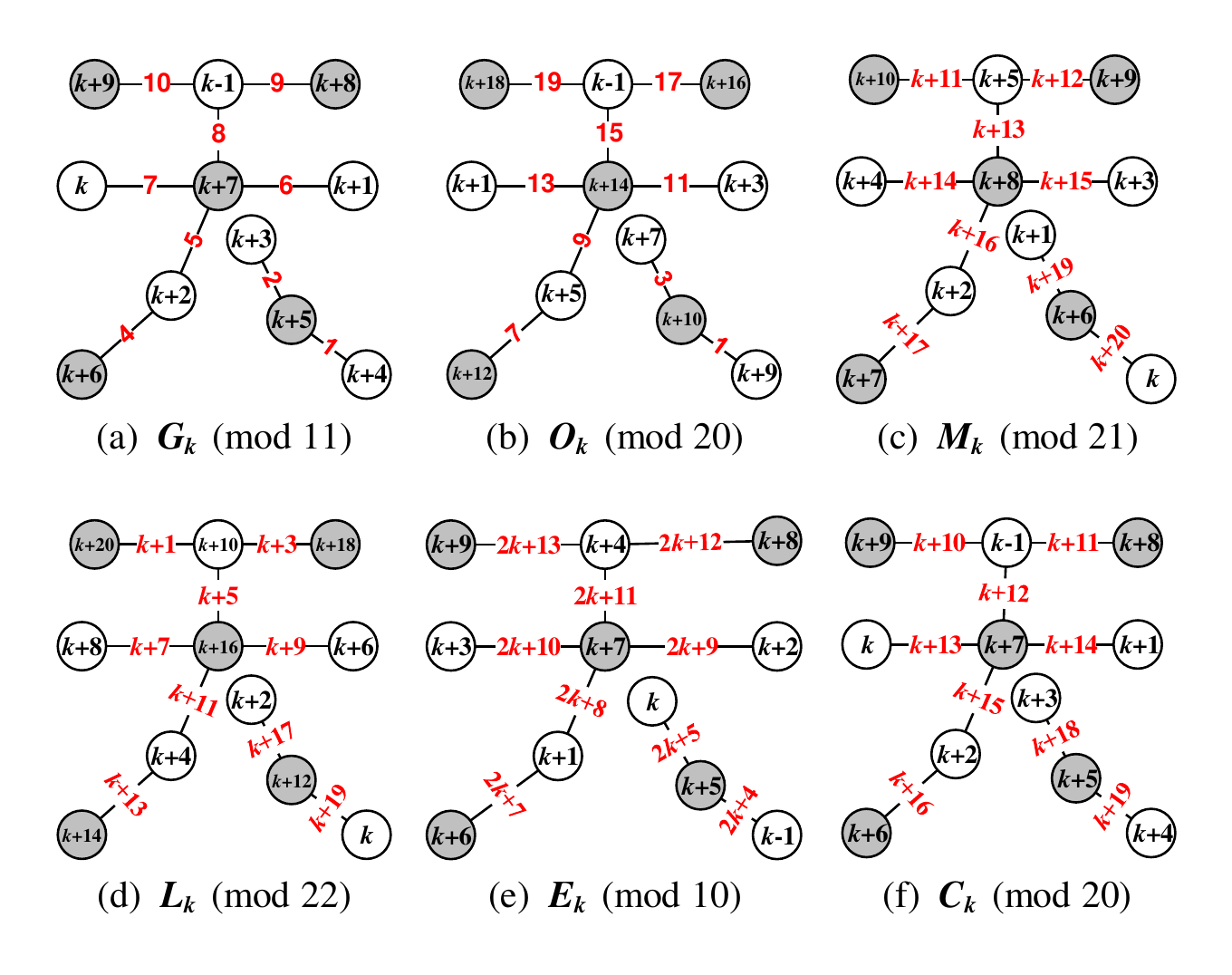}\\
\caption{\label{fig:tian-group-formulae}{\small Six every-zero graphic groups, cited from \cite{Yao-Mu-Sun-Sun-Zhang-Wang-Su-Zhang-Yang-Zhao-Wang-Ma-Yao-Yang-Xie2019}.}}
\end{figure}

Let $T_{4476}=G,O,M,L,E,C$ in the following discussion \cite{Yao-Mu-Sun-Sun-Zhang-Wang-Su-Zhang-Yang-Zhao-Wang-Ma-Yao-Yang-Xie2019}. Based on Hanzi-graph $T_{4476}$ and some graph labelings, we have the following six every-zero graphic groups $\{F_{f_X}(X),\oplus\}$ with $X=G,O,M,L,E,C$:

\begin{asparaenum}[Group-1. ]
\item $\{F_{f_G}(G),\oplus\}$ with \emph{flawed set-ordered graceful labelings} $f^{(i)}_G$ with $i\in [1,11]$, where $G=G_1$ shown in Fig.\ref{fig:tian-group-formulae} (a).
\item $\{F_{f_O}(O),\oplus\}$ with\emph{ flawed set-ordered odd-graceful labelings} $f^{(j)}_O$ with $j\in [1,21]$, where $O=O_1$ shown in Fig.\ref{fig:tian-group-formulae} (b).
\item $\{F_{f_M}(M),\oplus\}$ with \emph{flawed set-ordered edge-magic total labelings} $f^{(k)}_M$ with $k\in [1,21]$, where $M=M_1$ shown in Fig.\ref{fig:tian-group-formulae} (c).
\item $\{F_{f_L}(L),\oplus\}$ with \emph{flawed set-ordered odd-even separable edge-magic total labelings} $f^{(i)}_L$ with $i\in [1,21]$, where $L=L_1$ shown in Fig.\ref{fig:tian-group-formulae} (d).
\item $\{F_{f_E}(E),\oplus\}$ with \emph{flawed set-ordered odd-elegant labelings} $f^{(j)}_E$ with $j\in [1,11]$, where $E=E_1$ shown in Fig.\ref{fig:tian-group-formulae} (e).
\item $\{F_{f_C}(C),\oplus\}$ with \emph{flawed set-ordered odd-elegant labelings} $f^{(k)}_C$ with $k\in [1,20]$, where $C=C_1$ shown in Fig.\ref{fig:tian-group-formulae} (f).
\end{asparaenum}

\begin{figure}[h]
\centering
\includegraphics[width=16.4cm]{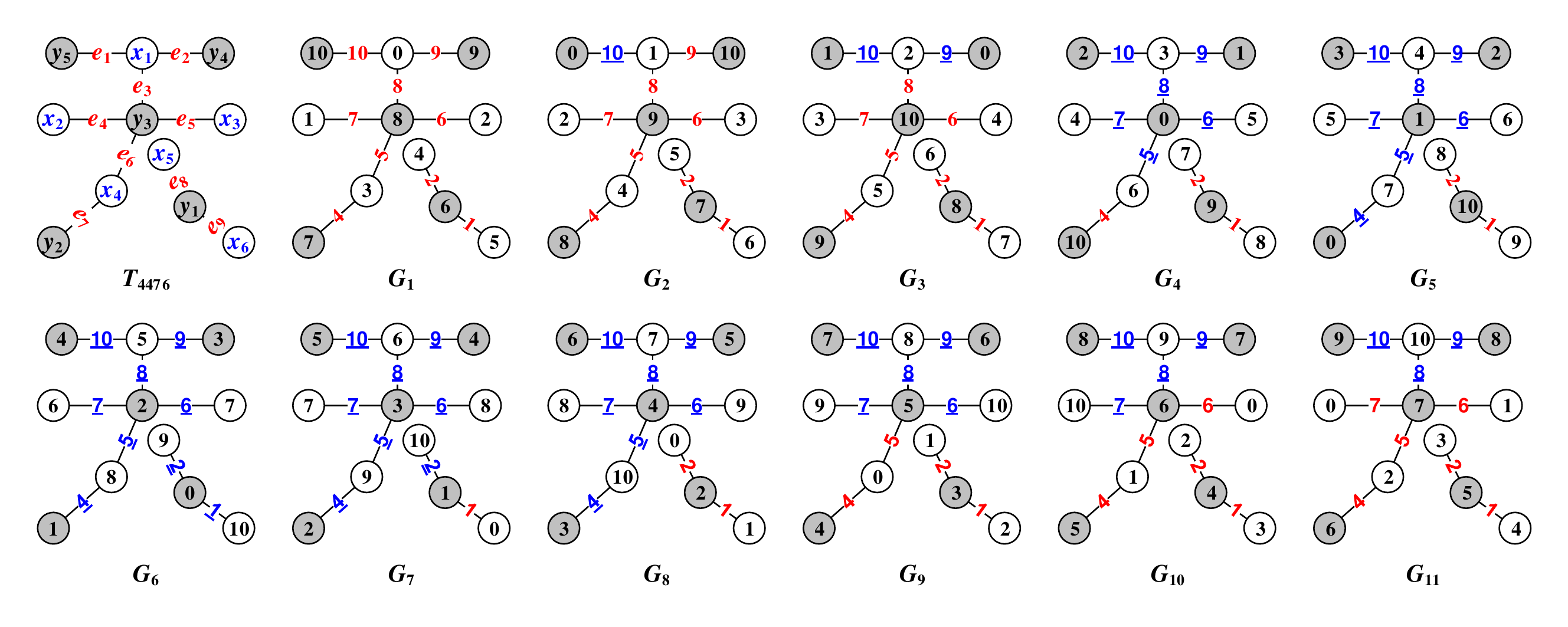}\\
\caption{\label{fig:tian-group}{\small An every-zero graphic group $\{F_{f_G}(G),\oplus\}$ made by Hanzi-graph $T_{4476}$ admitting a set-ordered graceful labeling, cited from \cite{Yao-Mu-Sun-Sun-Zhang-Wang-Su-Zhang-Yang-Zhao-Wang-Ma-Yao-Yang-Xie2019}.}}
\end{figure}

Let $V(T_{4476})=X\cup Y$ be the set of vertices of a Hanzi-graph $T_{4476}$ shown in Fig.\ref{fig:tian-group}, where
\begin{equation}\label{eqa:bipartite-sets}
{
\begin{split}
&X=\{x_1,x_2,x_3,x_4,x_5,x_6\},~Y=\{y_1,y_2,y_3,y_4,y_5\}; ~E(T_{4476})=\{e_i:~i\in [1,9]\}.
\end{split}}
\end{equation} where $e_1=x_1y_5$, $e_2=x_1y_4$, $e_3=x_1y_3$, $e_4=y_3x_2$, $e_5=y_3x_3$, $e_6=y_3x_4$, $e_7=y_2x_4$, $e_8=y_1x_5$, $e_9=y_1x_6$.

The property of ``set-ordered'' tells us:
$$f^{(j)}_G(x_i)<f^{(j)}_G(x_{i+1}),\quad f^{(j)}_G(x_6)<f^{(j)}_G(y_1),\quad f^{(j)}_G(y_j)<f^{(j)}_G(y_{j+1})
$$ for $T_{4476}=G=G_1$ shown in Fig.\ref{fig:tian-group}, and $j\in [1,11]$. Also, we have
$$
f^{(k)}_X(x_i)<f^{(k)}_X(x_{i+1}),\quad f^{(k)}_X(x_6)<f^{(k)}_X(y_1),\quad f^{(k)}_X(y_j)<f^{(k)}_X(y_{j+1})
$$
for $X=O_1,M_1,L_1,E_1,C_1$ shown in Fig.\ref{fig:tian-group-formulae}. Thereby, $G=G_1$ admits its own flawed set-ordered graceful labeling as: $f^{(1)}_G(x_i)=i-1$ for $i\in [1,6]$, and $f^{(1)}_G(y_j)=6+j-1$ for $i\in [1,5]$, and the induced edge labels $f^{(1)}_O(e_{s})=11-s$ for $s\in [1,7]$, as well as $f^{(1)}_O(e_{s})=10-s$ for $s=8,9$. We write $f^{(k)}_G(x_i)=k+i-2$ with $i\in [1,6]$ and $k\in [1,11]$, $f^{(k)}_G(y_j)=k+i-2$ with $j\in [1,5]$ and $k\in [1,11]$. Hence, $f^{(k)}_G(w)=f^{(1)}_G(w)+k-1$ for $w\in V(G)$ and $k\in [1,11]$.

\begin{figure}[h]
\centering
\includegraphics[width=16.4cm]{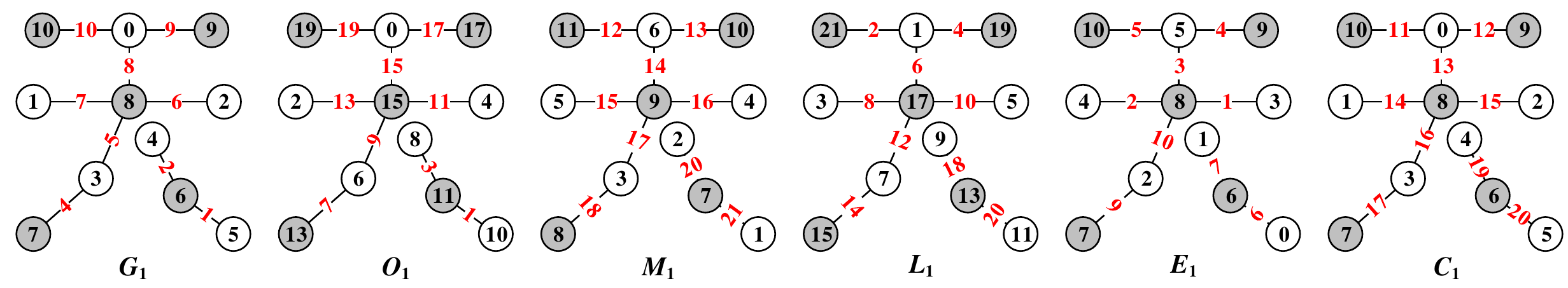}\\
\caption{\label{fig:tian-group-head}{\small The first Topsnut-gpw of six every-zero graphic groups shown in Fig.\ref{fig:tian-group-formulae}, cited from \cite{Yao-Mu-Sun-Sun-Zhang-Wang-Su-Zhang-Yang-Zhao-Wang-Ma-Yao-Yang-Xie2019}.}}
\end{figure}

We have the following equivalent relationships:
\begin{asparaenum}[(Rel-1) ]
\item $\{F_{f_G}(G),\oplus\}$ and $\{F_{f_O}(O),\oplus\}$ under $(\bmod~21)$: $f^{(1)}_O(x_i)=2f^{(1)}_G(x_i)$ for $x_i\in X$ and $f^{(1)}_O(y_i)=2f^{(1)}_G(y_i)-1$ for $y_i\in Y$, and the induced edge labels $f^{(1)}_O(e_{s})=2f^{(1)}_G(e_{s})-1$ for $s\in [1,9]$ (see Fig.\ref{fig:tian-group-head}). Thereby, we get $f^{(k)}_O(z)=f^{(1)}_O(z)+k-1$ for $z\in V(G)$, $f^{(k)}_O(e_j)=|f^{(k)}_O(y_l)-f^{(k)}_O(x_i)|$ for $e_j=x_iy_l\in E(G)$ and $k\in [1,21]$, and $G_j$ corresponds with $O_{2j-1}$ for $i\in [1,11]$.

\item $\{F_{f_G}(G),\oplus\}$ and $\{F_{f_M}(M),\oplus\}$ under $(\bmod~21)$: $f^{(1)}_M(x_i)=f^{(1)}_G(x_{6-i+1})+1$ for $i\in [1,6]$ and $f^{(1)}_M(y_i)=f^{(1)}_G(y_i)+1$ for $y_i\in Y$, and $f^{(1)}_M(e_j)=f^{(1)}_G(e_{6-j+1})+11$ for $i\in [1,9]$ (see Fig.\ref{fig:tian-group-head}). Hence, $f^{(k)}_M(z)=f^{(1)}_M(z)+k-1$ for $z\in V(G)$, $f^{(k)}_M(e_j)=f^{(1)}_M(e_j)+k-1$ for $e_j\in E(G)$ and $k\in [1,21]$.

\item $\{F_{f_G}(G),\oplus\}$ and $\{F_{f_L}(L),\oplus\}$ under $(\bmod~21)$: $f^{(1)}_L(x_i)=2f^{(1)}_G(x_{6-i+1})+1$ for $i\in [1,6]$ and $f^{(1)}_L(y_i)=2f^{(1)}_G(y_i)+1$ for $y_i\in Y$, and $f^{(1)}_L(e_j)=2f^{(1)}_G(e_{6-j+1})$ for $i\in [1,9]$ (see Fig.\ref{fig:tian-group-head}). In general, $f^{(k)}_L(z)=f^{(1)}_L(z)+k-1$ for $z\in V(G)$, $f^{(k)}_L(e_i)=f^{(1)}_L(e_i)+k-1$ for $e_i\in E(G)$ and $k\in [1,21]$.

\item $\{F_{f_G}(G),\oplus\}$ and $\{F_{f_E}(E),\oplus\}$ under $(\bmod~10)$: $f^{(1)}_E(x_i)=f^{(1)}_G(x_{6-i+1})$ for $i\in [1,6]$ and $f^{(1)}_E(y_i)=f^{(1)}_G(y_i)$ for $y_i\in Y$, and $f^{(1)}_E(e_j)=f^{(1)}_G(x_{6-i+1})+f^{(1)}_G(y_i)~(\bmod~10)$ for $i\in [1,9]$ (see Fig.\ref{fig:tian-group-head}). We obtain $f^{(k)}_E(z)=f^{(1)}_E(z)+k-1$ for $z\in V(G)$, $f^{(k)}_E(e_j)=f^{(k)}_E(x_i)+f^{(k)}_E(y_l)~(\bmod~10)$ for $e_j=x_iy_l\in E(G)$ and $k\in [1,11]$.

\item $\{F_{f_G}(G),\oplus\}$ and $\{F_{f_C}(C),\oplus\}$ under $(\bmod~20)$: $f^{(1)}_C(w)=f^{(1)}_G(w)+1$ for $w\in V(G)$, and $f^{(1)}_C(e_i)=f^{(1)}_G(e_{9-i+1})+9$ for $i\in [1,9]$ (see Fig.\ref{fig:tian-group-head}). The above transformation enables us to claim $f^{(k)}_E(z)=f^{(1)}_E(z)+k-1$ for $z\in V(G)$, $f^{(k)}_C(e_i)=f^{(1)}_C(e_i)+k$ for $e_i\in E(G)$ and $k\in [1,20]$.
\end{asparaenum}

\begin{figure}[h]
\centering
\includegraphics[width=13.4cm]{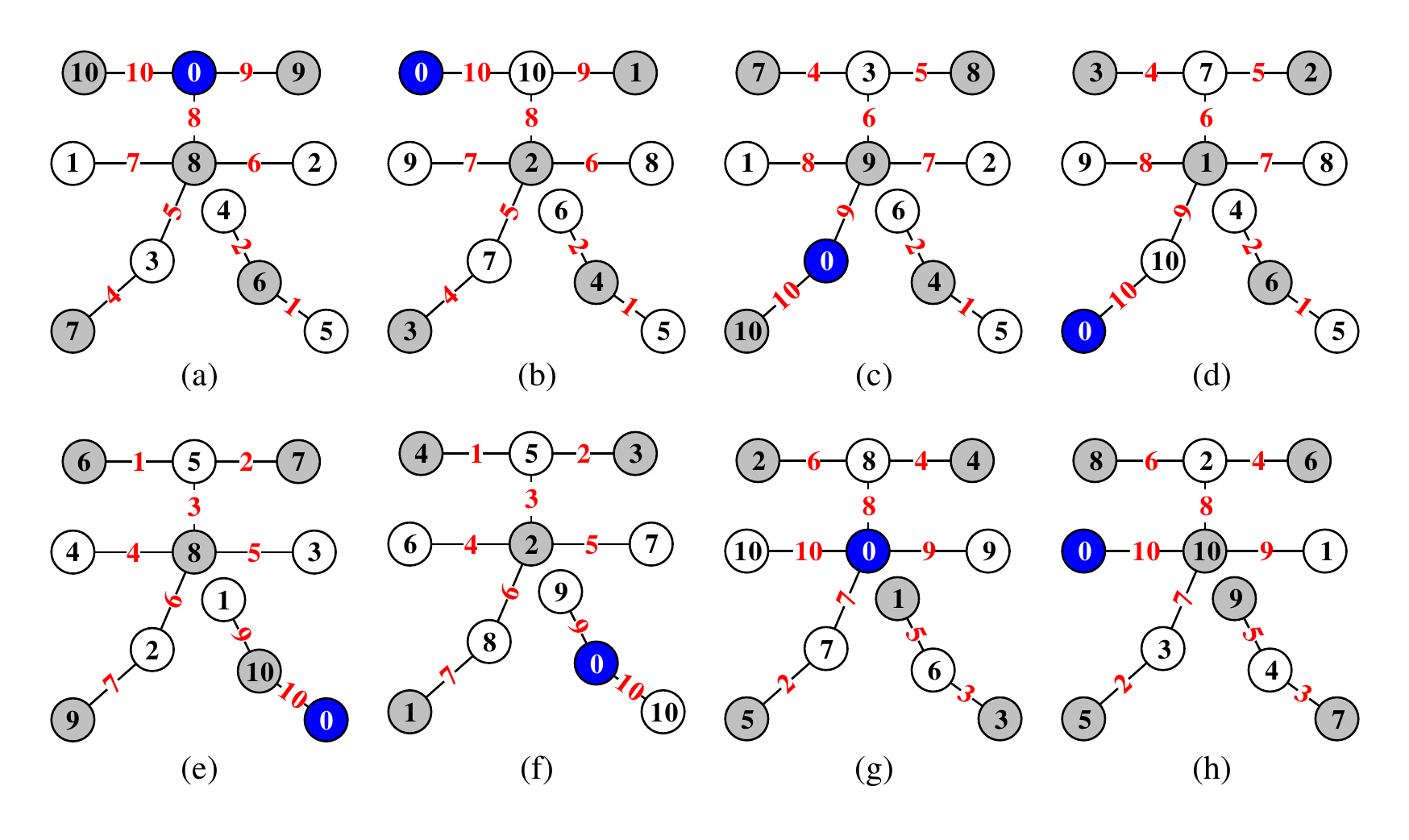}\\
\caption{\label{fig:tian-0-rotatable}{\small Hanzi-graph $T_{4476}$ admits a set-ordered $0$-rotatable system of (odd-)graceful labelings, cited from \cite{Yao-Mu-Sun-Sun-Zhang-Wang-Su-Zhang-Yang-Zhao-Wang-Ma-Yao-Yang-Xie2019}.}}
\end{figure}

Hanzi-graph $T_{4476}$, in fact, admits a \emph{flawed set-ordered $0$-rotatable system of (odd-)graceful labelings}, see Fig.\ref{fig:tian-0-rotatable}, namely, each vertex of Hanzi-graph $T_{4476}$ grows new vertices and edges. Thereby, we have at least eight every-zero graphic groups $\{F_{f_G}(G),\oplus\}$ based on $T_{4476}$.

\begin{figure}[h]
\centering
\includegraphics[width=16.4cm]{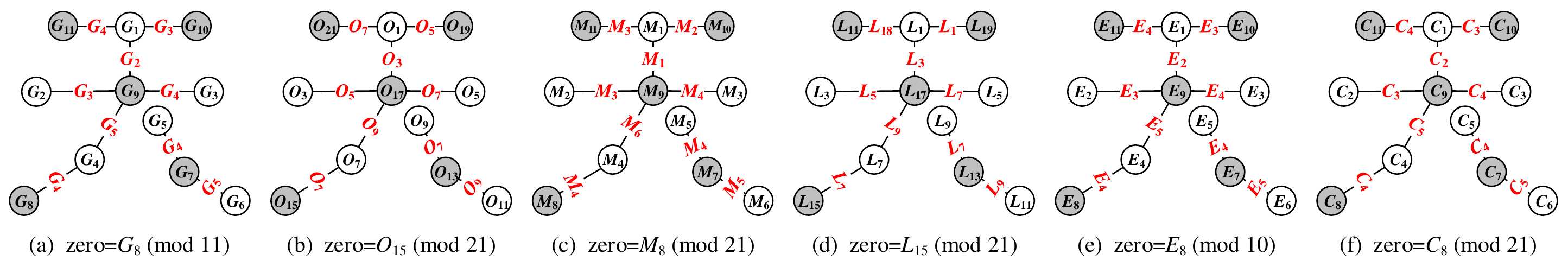}\\
\caption{\label{fig:tian-encrypted-6-groups}{\small A Hanzi-network $T_{4476}$ is encrypted by six every-zero graphic groups shown in Fig.\ref{fig:tian-group-formulae}, respectively, cited from \cite{Yao-Mu-Sun-Sun-Zhang-Wang-Su-Zhang-Yang-Zhao-Wang-Ma-Yao-Yang-Xie2019}.}}
\end{figure}

By the writing stroke order of Hanzis, a Topsnut-gpw shown in Fig.\ref{fig:tian-0-rotatable} (a) enables us to obtain a number-based string
\begin{equation}\label{eqa:t-4476-tb-paw}
{
\begin{split}
D(G_1)=101009917862088534742615
\end{split}}
\end{equation}
with $24$ numbers. Furthermore, a Hanzi-network $T_{4476}$ shown in Fig.\ref{fig:tian-encrypted-6-groups} (a), which was encrypted by an every-zero graphic group $\{F_{f_G}(G),\oplus\}$, and induces a number-based string
\begin{equation}\label{eqa:t-4476-tb-paw}
{
\begin{split}
D(T_{4476})=&D(G_{11})D(G_4)D(G_1)D(G_3)D(G_{10})D(G_{2})D(G_3)D(G_9)D(G_4)D(G_{3})D(G_{1})\\
&D(G_2)D(G_9)D(G_5)D(G_{4})D(G_4)D(G_{8})D(G_{5})D(G_4)D(G_7)D(G_5)D(G_{6})
\end{split}}
\end{equation} with at least $500$ numbers ($500 \times 8=4000$ bits). Moreover, this number-based string $D(T_{4476})$ has its own \emph{strong-rank} $H(D(T_{4476}))$ computed by
\begin{equation}\label{eqa:strong-rank}
{
\begin{split}
H(D(T_{4476}))=L(D(T_{4476}))\cdot \log_2 |X|=500\cdot \log_2~10\approx 1661 ~(\textrm{bits}),
\end{split}}
\end{equation} here $X=[0,9]$ in (\ref{eqa:strong-rank}). If $X=[0,9]\cup \{a,b,\dots ,z\}\cup \{A,B,\dots ,Z\}$, then $\log_2~|X|=\log_2 62\approx 5.9542$ bits. A string $x$ has its own strong-rank $H(x)$ computed by the strong-rank formula $H(x)=L(x)\cdot \log_2 |X|$, where $L(x)$ is the number of bytes of the string $x$ written by the letters of $X$.

We have known: Each permutation $G_{a_1}G_{a_2}\dots G_{a_{11}}$ of $G_1G_2 \dots G_{11}$ in the every-zero graphic group $\{F_{f_G}(G),\oplus\}$ can be used to label the vertices of the Hanzi-network $T_{4476}$, and each permutation $G_{a_1}G_{a_2}\dots G_{a_{11}}$ has $11$ zeros to encrypt the Hanzi-network $T_{4476}$, so we have at least $11\cdot (11)!~(=439,084,800> 2^{28})$ different approaches to encrypt the Hanzi-network $T_{4476}$. Moreover, this Hanzi-network $T_{4476}$ induces a matrix $A_{(a)}$ as follows:

\begin{equation}\label{eqa:tian-network-matrix}
\centering
A_{(a)}= \left(
\begin{array}{l}
G_1 ~ ~G_1 ~~ G_1 ~ G_3~ G_9~ G_9~ G_4~ G_5~ G_7\\
G_4 ~ ~G_3 ~ ~G_2 ~ G_2~ G_4~ G_5~ G_4~ G_4~ G_5\\
G_{11} ~ G_{10} ~ G_9 ~ G_9~ G_3~ G_4~ G_8~ G_7~ G_6
\end{array}
\right)
\end{equation}
\\So, we have $27!~(>2^{93})$ permutations obtained from the matrix $A_{(a)}$ shown in (\ref{eqa:tian-network-matrix}), and each permutation distributes us a number-based string with at least 648 numbers.

\begin{thm}\label{thm:equivalent-graphic-groups}
If a $(p,q)$-graph $G$ admits two mutually equivalent labelings $f_a$ and $f_b$, then two every-zero graphic groups $\{F_{f_a}(G),\oplus\}$ and $\{F_{f_b}(G),\oplus\}$ induced by these two labelings of $G$ are equivalent to each other.
\end{thm}

\subsection{Problems for Hanzi-graphs and Hanzi-colorings}

\begin{problem}\label{problem:xxxxxx}
\cite{Zhang-Yang-Mu-Zhao-Sun-Yao-2019} Color a graph (digraph) $G$ with Hanzis, such that there is a decomposition $G_1,G_2,\dots ,G_m$ of $G$ by the vertex-splitting operation, where each Hanzi-colored subgraph $G_i$ induces a meaningful paragraph in Chinese.
\end{problem}

\begin{problem}\label{problem:Hanzi-writing-lattice-11}
\cite{Zhang-Yang-Mu-Zhao-Sun-Yao-2019} List all Chinese paragraphs in a Hanzi-writing lattice $\textbf{\textrm{L}}(Z^0\textbf{\textrm{H}}_{an})$ defined in (\ref{eqa:Hanzi-Pianpang-123}) with $\sum ^{565}_{k=1}x_k\leq M$ for a fixed integer $M\in Z^0\setminus \{0\}$.
\end{problem}

\begin{problem}\label{problem:Hanzi-writing-lattice-22}
\cite{Zhang-Yang-Mu-Zhao-Sun-Yao-2019} Let $C_{para}=\sum ^{565}_{k=1}x_kH_{an}(k)$ in $\textbf{\textrm{L}}(Z^0\textbf{\textrm{H}}_{an})$, a Hanzi-writing lattice defined in (\ref{eqa:11-Hanzi-lattice}). Judge whether each $C_{para}$ is meaningful or meaningless in Chinese.
\end{problem}

\begin{problem}\label{problem:extrems-in-lattices-22}
\cite{Zhang-Yang-Mu-Zhao-Sun-Yao-2019} \textbf{Does} there exist a Hanzi-graphic lattice containing any Chinese essay with $m$ Chinese letters?
\end{problem}

\begin{figure}[h]
\centering
\includegraphics[width=16cm]{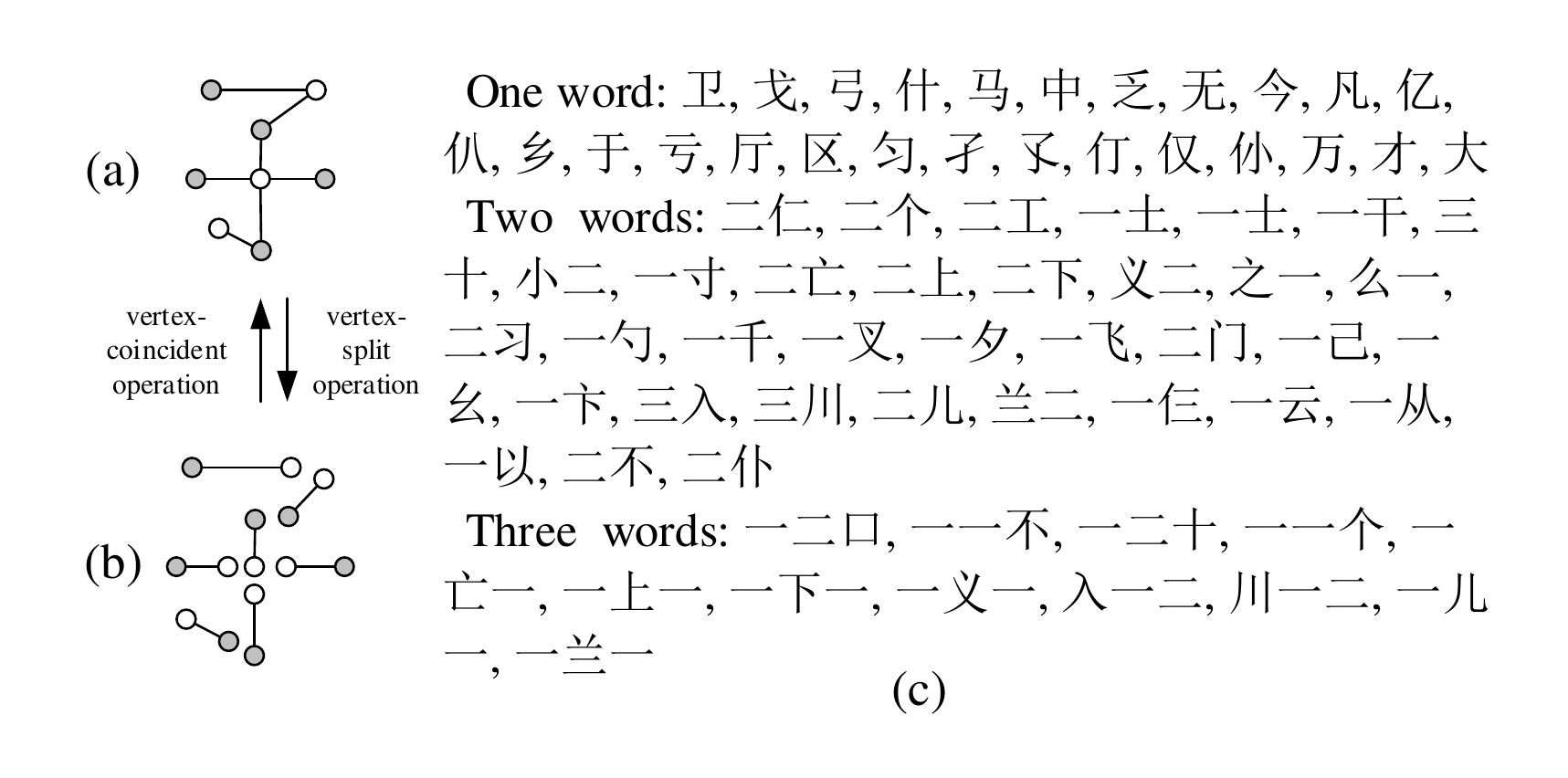}\\
\caption{\label{fig:7-edges-make-words}{\small Several Hanzi groups in (c) can be made as Hanzi-graphs of seven edges shown in (b).}}
\end{figure}

\begin{problem}\label{qeu:xxxxxx}
\cite{Zhang-Yang-Mu-Zhao-Sun-Yao-2019} \textbf{Constructing Chinese phrases.} A vertex-coinciding operation is a process of coinciding two vertices $x,u$ into one, where $x$ is the end of an edge $xy$, $u$ is the end of another edge $uv$, and $N(x)\cap N(u)=\emptyset$. Given $n$ edges $e_1,e_2,\dots ,e_n$, we, by means of the vertex-splitting operation, can use them to construct Hanzi groups $C_k=\{H_{k,1}, H_{k,2},\dots ,H_{k,d_k}\}$, find $a,b$ forming a range $a\leq d_k\leq b$. For example, we use seven edges shown in Fig.\ref{fig:7-edges-make-words}(a) and (b) to make some Chinese words and phrases shown in Fig.\ref{fig:7-edges-make-words} (c). It is not hard to see: There are horrible possibilities for this problem based on the combination of view.
\end{problem}

\begin{problem}\label{qeu:xxxxxx}
\cite{Yao-Mu-Sun-Sun-Zhang-Wang-Su-Zhang-Yang-Zhao-Wang-Ma-Yao-Yang-Xie2019}\textbf{Characterizing connected graphs admitting idiom labelings or phrase labelings.} As an example, we show a pan-Topsnut-gpw by a graph admitting a an idiom labeling shown in Fig.\ref{fig:idiom-labeling}. This problem may become a subbranch of graph labeling, since there is no result on this problems in our memory. However, this problem is restricted in Hanzis, on the other hands, this problem can be generalized to $\varepsilon$-labelings defined on essays, proses, poems, novels, reports and news \emph{etc}.
\end{problem}

\begin{problem}\label{qeu:xxxxxx}
\cite{Yao-Mu-Sun-Sun-Zhang-Wang-Su-Zhang-Yang-Zhao-Wang-Ma-Yao-Yang-Xie2019}\textbf{Estimating spaces of various Hanzi-gpws.} Clearly, the space of Hanzi-gpws is related with \emph{provable security} and \emph{computational security} in cryptography. It is not hard to see the Constructing Chinese phrases' problem induces terrible computational complex as the \emph{size} is enough large. No more research on this problem because of Hanzi-gpw does not last long.
\end{problem}

\begin{problem}\label{qeu:xxxxxx}
\cite{Yao-Mu-Sun-Sun-Zhang-Wang-Su-Zhang-Yang-Zhao-Wang-Ma-Yao-Yang-Xie2019}\textbf{Characterizing text-based string groups.} Given a group $S_{tring}$ of strings $s_1,s_2,\dots ,s_n$ with the numbers of the same length, find an operation $\theta$ between them such that $S_{tring}$ becomes a group under the operation $\theta$. An example of \emph{text-based string groups} is given by a matrix $A(\{T_b(T_i)\}^8_1)$ shown in (\ref{eqa:string-group-Hanzi-gpw}). Eight strings $T_b(T_1), T_b(T_2),\dots ,T_b(T_8)$ form a text-based string group under the the additive operation ``$\oplus$''.
\end{problem}

\begin{problem}\label{problem:Decomposing-graph-hanzi-graphs}
\cite{yarong-mu-bing-yao-IMCEC-2018} \textbf{Decomposing graphs into Hanzi-graphs (DGIHG).} Let $S_{plitHan}(G)$ be a set of all edge-disjoint Hanzi-graphs obtained by doing vertex-split operation to a given $(p,q)$-graph $G$ (as a \emph{public-key}).

(i) \textbf{How many} different groups of Hanzi-graphs $G_{i,1},G_{i,2}, \cdots,G_{i,k_n}$ (as \emph{private-keys}) having Chinese meanings are there in $S_{plitHan}(G)$?

(ii) \textbf{Compute} $\min\{k_n\}$ over all groups of Hanzi-graphs $G_{i,1},G_{i,2}, \cdots,G_{i,k_n}$ having Chinese meanings. Consider particular graphs, for example, $G$ is one of complete graphs $K_n$, Euler graphs, bipartite graphs, maximal planar graphs, and trees.

(iii) \textbf{Determine} $S_{plitHan}(G)$ and \textbf{compute} $\min\{k_n\}$.
\end{problem}

\begin{figure}[h]
\centering
\includegraphics[width=15.4cm]{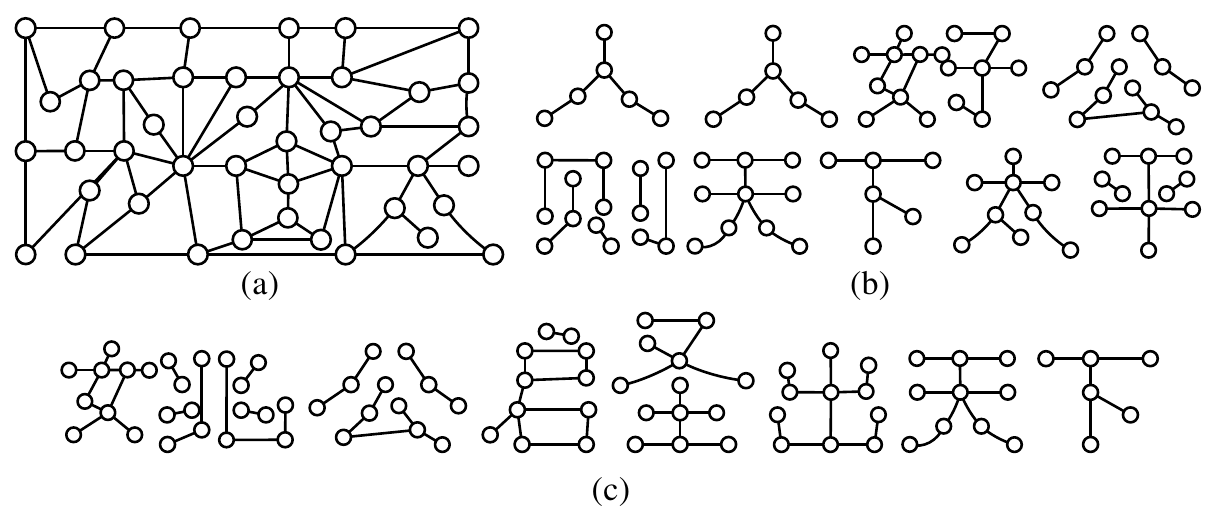}\\
\caption{\label{fig:decompose-graphs-into-Hanzos}{\small A scheme for illustrating Problem \ref{problem:Decomposing-graph-hanzi-graphs}.}}
\end{figure}

In Fig.\ref{fig:decompose-graphs-into-Hanzos}, a graph (a) is made by a group of Hanzi-graphs (b) (resp. a group of Hanzi-graphs (c)) through a series of vertex-coinciding operations. Conversely, the graph (a) can be vertex-split into the group of Hanzi-graphs (b) (resp. another group of Hanzi-graphs (c)). Thereby, we have (b)$\rightarrow$(a) and (c)$\rightarrow$(a), as well as (a)$\rightarrow_{split}$(b) and (a)$\rightarrow_{split}$(c).

\begin{rem}\label{rem:ABC-conjecture}
In \cite{Zhang-Yang-Mu-Zhao-Sun-Yao-2019}, the authors pointed that Hanzis have the solid advantages as the following:

1. \emph{Reading and thinking speeds.} Chinese people uses 1200 sounds for reading Hanzis, and each Hanzi is a voice. The more kinds of voices there are, the faster the thinking speed is. English has 20 vowels and 20 consonants, so the types of English voices will not exceed $20\times 20 = 400$.

2. \emph{Language stability and remembering efficiency.} Chinese does not make new words for new things. The emergence of new things is always accompanied by English new words. For example, ``rocket'' is not ``fire-driven-arrow'' in English, there is no connection between arrow and fire in English. However, we have ``rocket~$=H_{2780}H_{2893}$'' in Chinese, since ``arrow~$=H_{2893}$'' driven by ``fire~$=H_{2780}$'' (Ref. \cite{GB2312-80}). So, Hanzis keep the inheritance of language and civilization, only with such stable characters we can understand our ancestors' civilization and make continuous succession and progress.

3. \emph{Topological figures.} Hanzis are naturally topological structures, that is, ``two-dimensional'' on paper and ``image'' language making Chinese people's thinking curve, thereby, the biggest advantage of Hanzis is that the information density is very high.

4. \emph{Vast space.} China Online Dictionary includes Xinhua Dictionary, Modern Chinese Dictionary, Modern Idiom Dictionary, Ancient Chinese Dictionary in 12 dictionaries total about 20950 Hanzis; 360,000 words (28,770 commonly used words); 520,000 word groups; 31920 idioms; 4320 synonyms; 7690 antonyms; 14000 allegorical sayings; 28070 riddles; 19420 aphorisms; and countless essays, poems, novels, poems, reports, news, and so on.

5. \emph{Convenience for Chinese people.} Hanzis are easily impute into computer, mobile equipments with touch screen by speaking, writing and key-input.\paralled
\end{rem}


\section*{Acknowledgment}
The author, \emph{Bing Yao}, was supported by the National Natural Science Foundation of China under grant No. 61163054, No. 61363060 and No. 61662066. The author, \emph{Hongyu Wang}, thanks gratefully the National Natural Science Foundation of China under grants No. 61902005, and China Postdoctoral Science Foundation Grants No. 2019T120020 and No. 2018M641087.

My students have done a lot of works on graph labelings and graph colorings, the main members are of Dr. Xiangqian Zhou (School of Mathematics and Statistics, Huanghuai University, Zhu Ma Dian, Henan); Dr. Hongyu Wang, Dr. Xiaomin Wang, Dr. Fei Ma, Dr. Jing Su, Dr. Hui Sun (School of Electronics Engineering and Computer Science, Peking University, Beijing); Dr. Xia Liu (School of Mathematics and Statistics, Beijing Institute of Technology, Beijing); Dr. Chao Yang (School of Mathematics, Physics and Statistics, Shanghai University of Engineering Science, Shanghai); Inst. Meimei Zhao (College of Science, Gansu Agricultural University, Lanzhou); Inst. Sihua Yang (School of Information Engineering, Lanzhou University of Finance and Economics, Lanzhou); Inst. Jingxia Guo (Lanzhou University of technology, Lanzhou); Inst. Wanjia Zhang (College of Mathematics and Statistics, Hotan Teachers College, Hetian); Dr. Xiaohui Zhang (College of Mathematics and Statistics, Jishou University, Jishou, Hunan); Dr. Lina Ba (School of Mathematics and Statistics, Lanzhou University, Lanzhou); Lingfang Jiang, Haixia Tao, Jiajuan Zhang, Xiyang Zhao, Tianjiao Dai, Yaru Wang, Yichun Li, Yarong Mu.

Thanks for the teachers of Graph Cloring/labeling Symposium: Prof. Mingjun Zhang (School of Information Engineering, Lanzhou University of Finance and Economics, Lanzhou); Prof. Ming Yao (College of Information Process and Control Engineering, Lanzhou Petrochemical Polytechnic, Lanzhou); Inst. Yirong Sun (College of Mathematics and Statistics, Northwest Normal University, Lanzhou); Prof. Lijuan Qi (Department of basic courses, Lanzhou Institute of Technology, Lanzhou); Prof. Jianmin Xie (College of Mathematics, Lanzhou City University, Lanzhou).

\end{CJK}

\end{document}